\documentclass[11pt,a4paper]{article}
\usepackage[margin=2.7cm]{geometry}
\usepackage{amsmath,amssymb,amsthm}
\usepackage{graphicx}
\usepackage[colorlinks,linkcolor=blue,citecolor=blue,urlcolor=blue]{hyperref}

\newcommand{\capindexscale}{0.8}
\DeclareFontFamily{OML}{cmmcap}{\skewchar\font127 }
\edef\next{\noexpand\DeclareFontShape{OML}{cmmcap}{m}{it}{%
  <-8.5>s*[\capindexscale]cmmi7<8.5-9.5>cmmi9<9.5-11.5>cmmi10<11.5->cmmi12}{}}\next
\DeclareSymbolFont{capindex}{OML}{cmmcap}{m}{it}
\makeatletter
\@tfor\cap@x:=ABCDEFGHIJKLMNOPQRSTUVWXYZ\do{%
  \mathcode\expandafter`\cap@x=\numexpr"7000+256*\symcapindex+\expandafter`\cap@x\relax}
\makeatother

\newtheorem{theorem}{Theorem}
\newtheorem{proposition}{Proposition}
\newtheorem{lemma}{Lemma}
\newtheorem{hypothesis}{Hypothesis}
\newcommand{\pr}{\partial}
\newcommand{\vev}[1]{\langle #1 \rangle}
\newcommand{\cO}{\mathcal{O}}
\newcommand{\cA}{\mathcal{A}}
\newcommand{\cL}{\mathcal{L}}
\newcommand{\cW}{\mathcal{W}}
\newcommand{\cH}{\mathcal{H}}
\newcommand{\cT}{\mathcal{T}}
\newcommand{\cX}{\mathcal{X}}

\newcommand{\cU}{\mathcal{U}}
\newcommand{\cG}{\mathcal{G}}
\newcommand{\cE}{\mathcal{E}}
\newcommand{\vep}{\varepsilon}
\newcommand{\hB}{\hat B}
\newcommand{\hb}{\hat\beta}
\newcommand{\hr}{\hat\rho}
\newcommand{\hP}{\hat\Psi}
\newcommand{\tL}{\tilde{\mathcal{L}}}
\newcommand{\sM}{\mathsf{M}}
\newcommand{\Dp}{\Delta_\phi}
\newcommand{\vol}{\mathrm{v}}
\newcommand{\fa}{\mathfrak{a}}
\newcommand{\fb}{\mathfrak{b}}
\newcommand{\fc}{\mathfrak{c}}
\newcommand{\fd}{\mathfrak{d}}
\newcommand{\fe}{\mathfrak{e}}
\newcommand{\ff}{\mathfrak{f}}
\newcommand{\fg}{\mathfrak{g}}
\newcommand{\fj}{\mathfrak{j}}
\newcommand{\fk}{\mathfrak{k}}
\newcommand{\fl}{\mathfrak{l}}
\newcommand{\fm}{\mathfrak{m}}
\newcommand{\fn}{\mathfrak{n}}
\newcommand{\fp}{\mathfrak{p}}
\newcommand{\fq}{\mathfrak{q}}
\newcommand{\fr}{\mathfrak{r}}
\newcommand{\fs}{\mathfrak{s}}
\newcommand{\fy}{\mathfrak{y}}

\usepackage{array}
\makeatletter
\@ifundefined{c@property}{\newtheorem{property}{Property}}{}
\@ifundefined{c@corollary}{\newtheorem{corollary}{Corollary}}{}
\@ifundefined{c@identification}{\newtheorem{identification}{Identification}}{}
\@ifundefined{c@definition}{\newtheorem{definition}{Definition}}{}
\makeatother
\providecommand{\cS}{\mathcal{S}}
\providecommand{\bD}{\hat{\mathbb{D}}}
\providecommand{\sU}{\mathsf{U}}
\providecommand{\fh}{\mathfrak{h}}
\providecommand{\fz}{\mathfrak{z}}
\providecommand{\cK}{\mathcal{K}}
\providecommand{\rA}{\mathring{\cA}}
\providecommand{\sfa}{\mathsf{a}}

\providecommand{\sfm}{\mathsf{m}}
\providecommand{\sfn}{\mathsf{n}}

\providecommand{\vk}{\varkappa}

\title{Gradient formula from the sphere for perturbative RG flows near critical dimensions}
\author{Marten Reehorst$^{1,2}$\\[6pt]
{\small $^1$\,\itshape Department of Mathematics, City St.\ George's, University of London,}\\
{\small\itshape Northampton Square, EC1V 0HB, London, UK}\\[4pt]
{\small $^2$\,\itshape Department of Mathematics, King's College London, Strand, London WC2R 2LS, UK}}
\date{\today}

\begin{document}
\maketitle

\begin{abstract}
We study the gradient property of perturbative renormalisation group flows in $d=d_c-\vep$, with a potential and a metric regular at $\vep=0$ up to logarithms of gauge couplings. We state properties of the renormalised functional with local couplings on the sphere $S^d$: finiteness, a local Weyl anomaly, the Ward identity of a conformal Killing vector, rotation properties and statements on the free theory. From them we derive a Ward identity without anomaly for a combination of the functional and its anomaly. Its first variation is a gradient formula whose symmetric metric contains the two-point function on the sphere and whose potential at fixed points is, in an extended form of the properties, proportional to the sphere free energy $\tilde F$ up to a constant. Sources of relevant operators enter through their Weyl variation, and in a scheme which we construct the gradient formula persists, with a nondegenerate metric if a matrix of the free theory is invertible. For multiscalar $\phi^4$ theory and for scalars and Dirac fermions with Yukawa and quartic couplings in $d=4-\vep$, for sextic scalars in $d=3-\vep$ and for cubic scalars in $d=6-\vep$, we derive all these properties from a finiteness hypothesis. The gradient formula then holds to all orders for the beta function defined on the sphere. For interacting gauge theories the properties remain assumptions.
\end{abstract}

\tableofcontents

\section{Introduction}
\label{gs:sec:intro}

The renormalisation group flow of a quantum field theory is a gradient flow if there are a function $A$ and a symmetric tensor $G_{IJ}$ on the space of couplings $g^I$ such that the beta function is the gradient of $A$ with respect to $G_{IJ}$. In four dimensions the Weyl consistency conditions of Osborn and Jack~\cite{Osborn:1991gm,Jack:1990eb,Jack:2013sha} give a relation $\pr_IA=T_{IJ}B^J$, where $\pr_I=\pr/\pr g^I$, the vector field $B^I$ is the beta function up to an infinitesimal rotation of the fields, and the tensor $T_{IJ}$ is the sum of a symmetric tensor, which is positive at weak coupling~\cite{Jack:1990eb}, and of an antisymmetric part. In~\cite{Reehorst:phi4} we showed for multiscalar $\phi^4$ theory in $d=4$, under two hypotheses on perturbation theory, that the antisymmetric part can be removed to all orders. In this paper we consider the same question in $d=d_c-\vep$ dimensions, where $d_c$ is the dimension in which the couplings of a perturbed free theory are marginal, and for more general theories, which may contain fermions and gauge fields.

In $d=d_c-\vep$ dimensions the vector field of minimal subtraction is $\hB^I=-\vep\,k_I\,g^I+B^I$, where the numbers $k_I$ are fixed by the dimensions of the couplings and $B^I$ does not depend on $\vep$. A hat indicates that the vector field includes the term $-\vep\,k_I\,g^I$. This term gives fixed points at which the couplings are of order $\vep$, or of order $\vep^{1/2}$ for Yukawa, cubic and gauge couplings, which are within the reach of perturbation theory. The question whether the flow is a gradient flow now has a new aspect. For fixed $\vep\neq0$ a gradient formula always exists in perturbation theory, as noted in~\cite{Pannell:2024sia}. Indeed, in terms of the bare couplings the flow is linear, and a linear flow is the gradient of a quadratic function. This formula has no content, and when its metric and potential are expressed through the renormalised couplings they have poles at $\vep=0$. A gradient formula with content should have a limit for $\vep\to0$, in which it becomes a gradient formula in $d_c$ dimensions, and it should apply at the fixed points, where the couplings tend to zero with $\vep$. Both hold if the metric and the potential are power series in the couplings whose coefficients are power series in $\vep$, so that they have no poles at $\vep=0$. We call such quantities regular. The statement of interest is the gradient property with a regular metric and a regular potential.\footnote{For a gauge coupling $g$ the gradient formula is written with $g\,\pr/\pr g$ in place of $\pr/\pr g$ and with the beta function of $g$ divided by $g$. The potential is then regular up to an explicit multiple of $\log g$, see sections~\ref{gs:sec:theories} and~\ref{gs:sec:log}.} Since $B^I$ does not depend on $\vep$, this statement is a set of relations among the coefficients of $B^I$, one at each order in $\vep$, of which the first is the gradient property in $d_c$ dimensions.

For multiscalar $\phi^4$ theory in $d=4-\vep$ the gradient property was verified through three loops in~\cite{Wallace:1974dx,Wallace:1974dy}, and Pannell and Stergiou analysed it through six loops~\cite{Pannell:2024sia}. They found that a regular solution requires four constraints on the coefficients of the beta function at five loops and 32 at six loops, which are independent of those found in four dimensions and which are satisfied in modified minimal subtraction. In~\cite{Pannell:2025sph} they chose the free coefficients of the solution such that at fixed points the potential agrees, up to a constant factor, with the quantity $\tilde F$ of~\cite{Giombi:2014xxa} through order $\vep^5$, and they related the metric at low orders to the two-point function of the quartic operators on the sphere.

A relation of this kind is known at leading order in conformal perturbation theory. For a flow generated by weakly relevant operators the derivative of the free energy on the sphere with respect to the couplings is proportional to the beta function contracted with the coefficient of the two-point function~\cite{Klebanov:2011gs,Fei:2015oha}. It was asked in~\cite{Fei:2015oha} whether such a relation continues to hold at higher orders with a two-point function which depends on the couplings.

In this paper we show that a relation of this kind, a gradient formula with regular coefficients, follows to all orders in the couplings and in $\vep$ from a Ward identity on the sphere for theories whose renormalised functional on the sphere obeys certain properties, which we state. The starting point is the generating functional $W$ on the sphere $S^d$ with local couplings $g^I(x)$. The paper has three parts. In the first part, section~\ref{gs:sec:setting}, we define the framework in which we work, give an outline of the general argument by formal manipulations, and state the six properties which a theory must have for the argument to become a proof. They require that the functional is finite, that its Weyl variation is a local anomaly, that it obeys the Ward identity of a conformal Killing vector of the sphere together with the statements which the rotations of the sphere give, and statements about the free theory. We also summarise there for which theories these properties are proved, are expected to hold, or are not covered. In the second part, sections~\ref{gs:sec:identities} to~\ref{gs:sec:consequences}, we prove the gradient formula for every theory with these properties. This part uses only the properties, and it makes no reference to the field content of the theory or to the regularisation in which the functional is constructed. The third part, sections~\ref{gs:sec:universal} to~\ref{gs:sec:other}, supplies both for specific theories. In dimensional regularisation with minimal subtraction we construct the functional on the sphere, derive the properties from one hypothesis of finiteness, and evaluate the form of the gradient formula for each theory. In this way we obtain all six properties for multiscalar $\phi^4$ theory, for scalars with Yukawa couplings, for sextic scalars, for cubic scalars and for free gauge fields with a local coupling. For the other theories of the setting, such as interacting gauge theories, some or all of the properties remain assumptions, which we expect to hold.

Three properties of the sphere make the proof possible. First, the sphere has a finite volume and is invariant under rotations. The free energy and the correlation functions at constant couplings therefore exist, the one-point functions are constant, and the two-point function maps a harmonic of degree one to a multiple of itself. A Ward identity which is expanded around constant couplings thus becomes a relation among functions of the couplings. Second, the sphere has conformal maps which are not isometries, and their Weyl factor $\sigma$ is a harmonic of degree one, a bounded function which is orthogonal to the constants. In the Ward identity of such a map the constant one-point functions therefore drop out at lowest order. Its first variation contains the derivative of the free energy and the first harmonic of the two-point function contracted with $\hB^I$, together with a local one-form of the anomaly. The tensor which multiplies $\hB^I$ is symmetric, because it is built from a two-point function, and this is the structure of a gradient formula. In flat space the Weyl factor of a special conformal map is linear in the coordinates and the free energy at constant couplings is not defined, and in~\cite{Reehorst:phi4} the antisymmetric anomaly coefficient was determined from a second Ward identity under a hypothesis on Feynman integrals at large momenta. Third, the sphere has a radius, which a constant Weyl rescaling changes. The dependence of the functional on the radius gives a combination of the functional and of its anomaly whose Ward identity has no anomaly, so that the local one-form is absent from its first variation. The curvature of the sphere enters for even $d_c$ through terms of the anomaly which vanish in flat space at the same order, such as the term with the one-form $W_I$ of the Weyl anomaly in four dimensions, which multiplies the Einstein tensor. The argument does not rest on the anomaly. For odd $d_c$ the anomaly vanishes at vanishing relevant sources, and the first variation of the Ward identity of the functional itself is then already a gradient formula, in the scheme of section~\ref{gs:sec:relevant} if relevant sources of even dimension are present.

From the properties we derive an identity for quantities at constant couplings, which we call the Killing identity. It is the first variation of the Ward identity, and it relates the derivative of the function $\cW$, which is minus the free energy on the unit sphere, and the first harmonic of the two-point function on the sphere to a local one-form of the anomaly. For $W$ itself the one-form of the anomaly prevents the Killing identity from being a gradient formula. We therefore apply the identity to a combination $W^\sharp$ of the functional, of its anomaly and of its Weyl variation for a constant parameter on the unit sphere, which in integer dimension is $(\vep+r_0\,\pr_{r_0})W$ at fixed dimensionless sources, where $r_0$ is the radius of the sphere. The combination is chosen such that its Ward identity has a vanishing right-hand side. Its Killing identity then contains no one-form of the anomaly, and it relates the derivative of a function of the couplings to a symmetric tensor contracted with $\hB^I$, up to the terms of the relevant sources.

If the Weyl variation closes on the marginal couplings, that is if a Weyl rescaling is compensated by a variation of the marginal couplings and of the background gauge field alone, without generating other sources, the Killing identity of $W^\sharp$ is the gradient formula, which is the main result of the general part in the absence of relevant sources of even dimension,
\begin{equation}
\pr_J\hat A^{\rm S} = \hat G^{\rm S}_{JI}\,\hB^I \, , \qquad
\hat A^{\rm S} = \left(d+\hB\!\cdot\!\pr\right)\left[\frac1V\left(\vep-\hB\!\cdot\!\pr\right)\cW+\rA\right] , \qquad
\hat G^{\rm S}_{JI} = \cK_{JI}-\big[(\vep-\tL)\,\cS\big]_{JI} \, ,
\label{gs:eq:intro}
\end{equation}
see Proposition~\ref{gs:prop:main} of section~\ref{gs:sec:gradient}. The potential $\hat A^{\rm S}$ is built from the function $\cW$ on the unit sphere and from the anomaly, and the metric $\hat G^{\rm S}_{JI}$ from the first harmonic of the two-point function and from the anomaly. Here $V$ is the volume of the unit sphere and $\rA$ is the density of the anomaly for a constant rescaling. The tensor $\cS_{JI}$ is the first harmonic of the two-point function on the sphere minus the second derivative of $\cW/V$, $\cK_{JI}$ is a local symmetric tensor, and $\tL$ is the extended Lie derivative of~\cite{Jack:2013sha} along $\hB^I$. Both sides are regular, and $\hat G^{\rm S}_{JI}$ is symmetric. At a zero of $\hB^I$ the potential is proportional to $\tilde F-\tilde F_0$. Here $\tilde F=\sin(\pi d/2)\log Z_{S^d}$ is defined with curvature couplings whose values are fixed by the anomaly. If the functional is defined on the spheres of every radius and its variation under a constant rescaling is the anomaly, these values are the zeros of the beta functions of the curvature couplings, as in~\cite{Fei:2015oha}. The constant $\tilde F_0$ is the value of $\tilde F$ in the free theory if the free theory is a point of the space of couplings and no curvature counterterm is independent of the couplings, and it is not fixed for gauge couplings. At $\vep=0$ the term of order zero of the metric is fixed by the two-point functions of the free fields in $d_c$ dimensions. If the matrix of coefficients of these two-point functions is positive definite, then for even $d_c$ the metric of the potential which is normalised to $\tilde F$ is positive definite to all orders in the couplings and in $\vep$.\footnote{By this we mean, here and below, that the term of order zero of the tensor in the couplings and in $\vep$ is a positive definite matrix. Its truncation at any order is then positive definite for sufficiently small couplings and $\vep$, and we make no statement about the sum of the series.}

For odd $d_c$ the anomaly vanishes at vanishing relevant sources, since it is built without the Levi-Civita tensor. Without relevant sources the Killing identity of $W$ itself is then already a gradient formula,
\begin{equation}
\pr_J\Pi = -\cS_{JI}\,\hB^I \, , \qquad \Pi = \frac1V\left(d+\hB\!\cdot\!\pr\right)\cW \, ,
\label{gs:eq:introodd}
\end{equation}
with the potential $\Pi$, which at the zeros of $\hB^I$ is $d/V$ times $\cW$, and the metric $-\cS_{IJ}$, see Proposition~\ref{gs:prop:odd} of section~\ref{gs:sec:odd}.\footnote{The formula \eqref{gs:eq:intro} holds for odd $d_c$ as well, but its metric then has no term of order zero in the couplings and in $\vep$. The statements on the positivity of the metric are made for \eqref{gs:eq:intro} if $d_c$ is even and for \eqref{gs:eq:introodd} if $d_c$ is odd.}

The gradient formula \eqref{gs:eq:intro} requires that the Weyl variation closes on the marginal couplings. This fails if the theory has a relevant operator whose source mixes with the couplings and the curvature under a Weyl rescaling. Whether a source mixes is decided by its mass dimension at $d=d_c$, which is $d_c$ minus the dimension of the operator. The terms which a Weyl rescaling adds to a scalar source at vanishing relevant sources are built from the curvature and from derivatives of the couplings, and their number of derivatives is even. A source of even dimension can therefore receive such terms, and it cannot be omitted. A source of odd or of non-integer dimension receives none, and it drops out of the identities. For instance, near four dimensions the source of $\phi_i\phi_j$ has dimension two and mixes, while the source of a fermion bilinear has dimension one and drops out. Among our cases the closure thus holds for gauge fields of a semisimple group with fermions near four dimensions and for fermions near two dimensions. It fails in every case with scalar fields, near three, four and six dimensions.

With relevant sources of even dimension the Weyl variation of these sources adds terms to the Killing identity of $W^\sharp$. In section~\ref{gs:sec:relevant} we construct a scheme for the relevant sources in which the Killing identity of $W^\sharp$ is then again a gradient formula,
\begin{equation}
\pr_J\hat A^{\rm H} = \hat G^{\rm H}_{JI}\,\hB^I \, ,
\label{gs:eq:introrel}
\end{equation}
with a regular potential and a regular symmetric metric, see Theorem~\ref{gr:cor:cond} of section~\ref{gs:sec:gradrel}.\footnote{With gauge fields the potential is regular up to the multiples of the logarithms of the gauge couplings which section~\ref{gs:sec:log} describes, and the statement holds with this qualification.} At the zeros of $\hB^I$ the potential $\hat A^{\rm H}$ takes the same value as $\hat A^{\rm S}$. Under Property~\ref{gs:P:free} the term of order zero in the couplings and in $\vep$ of the metric $\hat G^{\rm H}_{JI}$ is the same as that of $\hat G^{\rm S}_{JI}$. For even $d_c$ the metric is then nondegenerate if the matrix of the free two-point functions is invertible, and the metric of the potential which is normalised to $\tilde F$ is positive definite to all orders if this matrix is positive. For odd $d_c$ the same holds for the form \eqref{gs:eq:introodd} of the gradient formula with relevant sources, see section~\ref{gs:sec:gradrel}.

Section~\ref{gs:sec:consequences} states the consequences for fixed points. For a definite metric the eigenvalues of the matrix of derivatives of $\hB^I$ at a fixed point are real. For a positive metric the potential is a monotone of the flow. The contraction of the gradient formula with $\hB^I$ gives the derivative of the potential along $\hB^I$ as the metric contracted with two factors of $\hB^I$. The potential therefore increases along $\hB^I$ and decreases towards the infrared, to all orders in the couplings and in $\vep$ in the sense of formal power series. Its value at fixed points is $\tilde F$ up to a positive factor and a constant. For two fixed points which are power series in $\vep$ with couplings of order $\vep$, joined by a trajectory of the flow at leading order, $\tilde F$ is therefore larger at the ultraviolet than at the infrared fixed point, in the sense that the difference is a power series in $\vep$ with a positive leading term. The positivity enters only through the two-point functions of the free fields, and no unitarity of the interacting theory is used. A monotone excludes closed orbits of the flow of $\hB^I$, since along a closed orbit the potential would return to its value. The gradient formula gives this to every order of perturbation theory as an identity of formal power series, and for the flow at leading order, which is a vector field, we show it explicitly.

The third part supplies the field content and the regularisation. Section~\ref{gs:sec:universal} sets up dimensional regularisation with local couplings in position space on the sphere $S^d$ for non-integer $d$, which is common to the theories of scalar fields and Dirac fermions. There we construct the bare functional for scalar fields with polynomial interactions and for scalar fields with Dirac fermions, and we show how the form of the anomaly and the Ward identity follow in minimal subtraction from its invariances. The fermion propagators run along the chords of the embedding space, and no spinors on $S^d$ are needed. Sections~\ref{gs:sec:phi4} and~\ref{gs:sec:other} then derive the properties theory by theory. There are four theories for which we derive all six properties from one hypothesis, Hypothesis~\ref{hyp:finite} of appendix~\ref{gs:sec:cases}, which we assume for each of them. It states that the functional with local couplings, a background gauge field and relevant sources is finite in flat space with counterterms of the form of minimal subtraction. They are multiscalar $\phi^4$ theory in $d=4-\vep$ in section~\ref{gs:sec:phi4}, scalar fields and Dirac fermions with Yukawa and quartic couplings in $d=4-\vep$ in section~\ref{gs:sec:yukawa}, sextic scalars in $d=3-\vep$ in section~\ref{gs:sec:sextic}, and cubic scalars in $d=6-\vep$ in section~\ref{gs:sec:cubic}, for which the hypothesis is stated on conformally flat backgrounds with a larger class of flat sources. We present the first of them in most detail, with its proofs in appendices~\ref{app:weyl} to~\ref{app:finite}. The other three follow the same steps. Appendix~\ref{gs:app:steps} lists for each step whether it applies unchanged or is modified, and the modified steps are proved in appendix~\ref{gs:app:fermions} for the Yukawa theory, in appendix~\ref{gs:app:sextic} for the sextic theory and in appendix~\ref{gs:app:cubic} for the cubic theory. With the properties derived, the gradient formula \eqref{gs:eq:introrel} of Theorem~\ref{gr:cor:cond} holds for these four theories to all orders in perturbation theory, for the vector field $\hB^I$ which the counterterms of Hypothesis~\ref{hyp:finite} define. The metric of the potential which is normalised to $\tilde F$ is positive definite to all orders in the couplings and in $\vep$, see Theorems~\ref{gs:thm:phi4}, \ref{gf:cor:yukawa}, \ref{gsx:thm:sextic} and~\ref{gs:thm:cubic}.\footnote{For multiscalar $\phi^4$ theory the formula at $\vep=0$ is a second gradient formula in four dimensions for the vector field $B^I$ of~\cite{Reehorst:phi4}, with a metric which contains the two-point function on the sphere. That the two vector fields agree, and that $\hB^I$ is the beta function of minimal subtraction in momentum space up to a rotation of the fields, rests on Identification~\ref{id:ms} of section~\ref{gs:sec:identification}, which the proof of the gradient formula does not use.} For these four theories the functional is also defined on the spheres of every radius at constant couplings, with the anomaly as its variation under a constant rescaling. With the relevant sources of even dimension in the scheme of section~\ref{gs:sec:relevant} the values of the curvature couplings in $\tilde F$ are then the zeros of their beta functions, as in~\cite{Fei:2015oha}, see Proposition~\ref{gs:prop:radius}.

For free gauge fields with a local coupling we construct the bare functional on the sphere in appendix~\ref{gs:app:freegauge} by continued sums over the levels of the exact two-forms, and all six properties follow for it from Hypothesis~\ref{hyp:finite}, which for gauge fields is stated on the sphere and its conformal rescalings. For interacting gauge theories the functional is not constructed, and the properties are assumptions.

For gauge fields of a semisimple group with fermions in $d=4-\vep$ the gradient formula takes the form of Corollary~\ref{gs:cor:gauge} of section~\ref{gs:sec:gauge}. Here the coordinates on the space of couplings are the logarithms of the gauge couplings $g_\iota$, and the sources of the fermion bilinears drop out. The potential contains a multiple of $\log g_\iota$ for each simple factor of the gauge group, its value at a fixed point contains $\log\vep$, and the free gauge theory is not a zero of $\hB^I$. With scalar fields in addition, and thus with Yukawa and quartic couplings, the mass source enters through its Weyl variation, and the gradient formula is that of Theorem~\ref{gr:cor:cond}. In both cases the metric of the potential which is normalised to $\tilde F$ is positive definite to all orders, and at order zero in $\vep$ the coefficients of the logarithms of the gauge couplings are the numbers of gauge fields of the simple factors. An abelian factor of the gauge group brings a source for the field strength, which drops out of the identities if the properties hold with it. Several abelian factors are covered with an extended notion of regularity, and chiral couplings, which require a treatment of $\gamma_5$, are not covered.

For fermions with quartic couplings near two dimensions we construct the bare functional with four-fermion vertices of every gamma structure. The sources of the relevant operators drop out, and under Properties~\ref{gs:P:functional} to~\ref{gs:P:sphere} the gradient formula holds for this theory with a regular symmetric metric. At order zero in the couplings and in $\vep$ the metric is positive on the couplings without gamma matrices for ${\rm tr}\,1>1$, which include the Gross--Neveu coupling, and it vanishes in the directions of the evanescent couplings, whose operators vanish at $d=2$. For small $\vep\neq0$ it is indefinite in these directions at order zero in the couplings, so that the statements on positivity to all orders do not apply. In these directions the hypothesis of finiteness does not exclude a simple pole of $\hB^I$, whose regularity is part of Property~\ref{gs:P:weyl}. The pole is absent if the counterterms of the couplings obey the pole equations of minimal subtraction and the counterterm of the gauge field is covariant and has no term of degree zero, see section~\ref{gs:sec:fermions}. Scalar fields and nonlinear sigma models near two dimensions, chiral and parity-violating theories and Chern--Simons theories with matter lie outside the setting, see section~\ref{gs:sec:outside}.

Let us summarise the relation to earlier work. The consistency conditions were derived with dimensional regularisation on a curved background in~\cite{Jack:1990eb} and from the commutativity of Weyl rescalings in four dimensions in~\cite{Osborn:1991gm}. The form with the currents and the mass operator in dimensional regularisation is given in section~8 of~\cite{Jack:2013sha} for flat space. We use the same framework and keep the dependence on $d$. The quantity $\tilde F$ was introduced in~\cite{Giombi:2014xxa} as an interpolation between the $a$-anomaly in even dimensions and the sphere free energy in odd dimensions. The definition of the fixed point on the sphere with the curvature couplings at the zeros of their beta functions is that of~\cite{Fei:2015oha}. For flows generated by weakly relevant operators the free energy on the sphere was computed at leading order in~\cite{Klebanov:2011gs} for odd dimensions and in~\cite{Fei:2015oha} for all dimensions, and in~\cite{Klebanov:2011gs} the matrix which multiplies the beta function is regarded as a metric on the space of couplings. The identity \eqref{gs:eq:intro} and its form for odd $d_c$ are relations of this kind to all orders, under the stated assumptions.

Perturbation theory on the sphere in dimensional regularisation has been developed for coupling constants. The renormalisation of massless scalar theories on the sphere was studied in this way in~\cite{Drummond:1975yc}, see also~\cite{Drummond:1977dg}, and the computations of the free energy on $S^d$ in~\cite{Giombi:2014xxa,Fei:2015oha} use it as well. Each order is then an integral of powers of the chordal distances over the sphere, which is a function of $d$ by the invariance under rotations. The local renormalisation group of~\cite{Osborn:1991gm,Jack:1990eb,Jack:2013sha} has couplings which depend on the position and takes the dimensionally regularised functional on a general background as given. Our proof takes functional derivatives with respect to couplings which depend on the position on $S^d$, and it uses the Ward identity of the conformal Killing vectors and the background gauge invariance as exact statements for non-integer $d$. It therefore needs the regularised functional with local sources on $S^d$ for non-integer $d$, that is a definition of the sources, the meromorphy of every term in $d$ with control of its poles, and the conformal covariance and the gauge invariance to all orders. To our knowledge such a construction was not available. We give it in section~\ref{gs:sec:sphere} for scalar fields, in section~\ref{gs:sec:dirac} for Dirac fermions and in appendix~\ref{gs:app:freegauge} for free gauge fields. It extends the integral in $d$ dimensions of~\cite{Wilson:1972cf,Collins:1984xc} from the loop momenta to the positions of the vertices on the sphere.

Beyond perturbation theory a monotone of the flow, or an inequality between the fixed points, is known in integer dimensions. In two dimensions it is the $c$-function of~\cite{Zamolodchikov:1986gt}. In four dimensions the inequality for the $a$-anomaly between fixed points was proposed in~\cite{Cardy:1988cwa} and proved in~\cite{Komargodski:2011vj}, and in three dimensions the inequality for the sphere free energy of~\cite{Jafferis:2011zi,Klebanov:2011gs} was proved in~\cite{Casini:2012ei}. These proofs use unitarity, which fails in non-integer dimension, as was shown for free and interacting scalar fields in~\cite{Hogervorst:2015akt}. In perturbation theory at $d=4$ the contraction of the relation $\pr_IA=T_{IJ}B^J$ with $B^I$ gives $B^I\pr_IA=G_{IJ}B^IB^J$, where $G_{IJ}$ is the symmetric part of $T_{IJ}$. This equation holds to all orders. It was derived with the beta function in~\cite{Osborn:1989td,Jack:1990eb,Osborn:1991gm} and with $B^I$ in~\cite{Jack:2013sha}. The function $A$ agrees with the coefficient of the Euler term of the anomaly only at fixed points. The tensor $G_{IJ}$ was shown to be positive at lowest order in the couplings~\cite{Jack:1990eb}, so that $A$ is a monotone at weak coupling. In $d=4-\vep$ this argument does not apply to $\hB^I$. The derivative of $A$ along $\hB^I$ is $\hB^IT_{IJ}B^J$, which contains the two different vector fields $\hB^I=-\vep\,k_Ig^I+B^I$ and $B^J$, so that the positivity of $G_{IJ}$ does not fix its sign. At small couplings its leading term is $-\vep\,k_Ig^IG_{IJ}B^J$, which is not a square. For a single quartic coupling $\lambda>0$ and $\vep>0$ this term is negative, so that at leading order $A$ decreases along $\hB^I$ between the free theory and the fixed point and increases beyond it. For theories with marginal couplings in integer dimension $d_0$ the dependence of the sphere free energy on the radius $r$ was studied with dimensional regularisation in~\cite{Yonekura:2012kb}. At leading order the first derivative with respect to $\log r$ for odd $d_0$, and the second derivative for even $d_0$, is a sum of squares of the beta functions with coefficients which are positive by reflection positivity. In non-integer dimension the statements known to us are of fixed order. The gradient property was verified through the loop orders quoted above in $d=4-\vep$, and through three loops for the cubic theory in $d=6-\vep$~\cite{Benfatto:2025cubic}. The inequality $\tilde F_{\rm UV}>\tilde F_{\rm IR}$, conjectured in~\cite{Giombi:2014xxa} and called the generalised $F$-theorem in~\cite{Fei:2015oha}, was shown there at leading order in conformal perturbation theory. To our knowledge a monotone to all orders in non-integer dimension was not available.

In three dimensions the consistency conditions of the local renormalisation group were analysed in~\cite{Nakayama:2013wda}, where no equation analogous to the flow equation of the $a$-anomaly was obtained, and a function with a gradient property was constructed from the beta functions at low orders in~\cite{Jack:2015tka}. In six dimensions the consistency conditions were studied in~\cite{Grinstein:2014xba,Grinstein:2015ina,Osborn:2015rna,Stergiou:2016uqq}. The finite part of the partition function on the sphere as a function of exactly marginal couplings was studied in~\cite{Gerchkovitz:2014gta}, and in two dimensions a gradient formula which contains an antisymmetric tensor was proved in~\cite{Friedan:2009ik} under conditions on the infrared behaviour.

What is new here is the Ward identity on the sphere away from fixed points, the homogeneous Ward identity of $W^\sharp$ and the gradient formula which follows from it. For the theories with a positive metric it gives a monotone of the flow to all orders in the couplings and in $\vep$, and with it the inequality of the generalised $F$-theorem for the flows of section~\ref{gs:sec:fixedgen}. Also new are the regularised functional with local sources on the sphere for non-integer $d$, for scalar fields, Dirac fermions and free gauge fields, the scheme of the relevant sources in which their flows have the form which the gradient formula requires, and the proof of the gradient formula to all orders for the four theories above under the hypothesis of finiteness. For multiscalar $\phi^4$ theory the gradient formula implies, with Identification~\ref{id:ms}, relations among the coefficients of the beta function which are not required in four dimensions. At five loops the computation of appendix~\ref{app:fiveloop} finds that, modulo the relations in four dimensions, they span the same space as the four relations of~\cite{Pannell:2024sia}. For this theory we also show, under the hypothesis of finiteness, that in four dimensions there is a scheme in which the one-form $W_I$ of the Weyl anomaly vanishes, see appendix~\ref{gs:app:fourd}. The equation of~\cite{Osborn:1991gm} relates the derivative of a function, which at fixed points is the coefficient of the Euler term of the anomaly, to the beta function through a tensor $T_{IJ}$ whose antisymmetric part contains the curl of $W_I$. In this scheme the antisymmetric part of $T_{IJ}$ consists of terms with the field rotations alone. In their absence $T_{IJ}$ is symmetric, and the equation of~\cite{Osborn:1991gm} is then itself a gradient formula with a symmetric tensor, as conjectured in~\cite{Poole:2019kcm}.

While this work was being completed we learned of~\cite{Benfatto:toappear}, which also studies the gradient property of perturbative renormalisation group flows and appears simultaneously.

The paper is organised as follows. In the first part, section~\ref{gs:sec:setting} contains the setting, the outline of the argument, the properties in informal form, their status and the cases.\footnote{This paper is long, but for the reader who would be discouraged by this we want to note that 90\% of the conceptual content can be understood just by reading up to and including section~\ref{gs:sec:setting} (and maybe the discussion in section~\ref{sec:discussion}). The rest are really just technical details required to make the proof work in full.} In the second part, section~\ref{gs:sec:identities} contains the Killing identity, the homogeneous Ward identity and the gradient formula, with the value at fixed points, the lowest order, odd $d_c$ and logarithmic coordinates. Section~\ref{gs:sec:relevant} adds the relevant operators and proves the gradient formula with relevant sources, and section~\ref{gs:sec:consequences} states the consequences. In the third part, section~\ref{gs:sec:universal} treats scalar fields and Dirac fermions in minimal subtraction and constructs the bare functional on the sphere, section~\ref{gs:sec:phi4} derives the properties of multiscalar $\phi^4$ theory from the hypothesis of finiteness in flat space, and section~\ref{gs:sec:other} treats the other theories. It derives the properties for Yukawa and quartic couplings, for sextic scalars and for cubic scalars from the same hypothesis, and it gives the form of the gradient formula for gauge theories and for fermions near two dimensions. We conclude in section~\ref{sec:discussion}. Appendix~\ref{gs:app:properties} contains the precise statements of the properties and of Hypothesis~\ref{hyp:finite}, with the explanation of their conditions and the bare sources and counterterms of each theory. Appendices~\ref{app:weyl} to~\ref{app:finite} contain the proofs of the statements which sections~\ref{gs:sec:universal} and~\ref{gs:sec:phi4} quote, written for the vertices of $\phi^4$ theory, appendix~\ref{app:momentum} describes the relation to momentum space, and appendix~\ref{gs:app:fourd} contains the statements in four dimensions and the comparison with known results. Appendix~\ref{gs:app:steps} lists the steps of the derivation of the properties for the other theories. Appendix~\ref{gs:app:fermions} contains the proofs of the statements of section~\ref{gs:sec:dirac} on the bare functional with Dirac fermions and the modified steps for the Yukawa theory, and appendices~\ref{gs:app:sextic} and~\ref{gs:app:cubic} contain the modified steps for sextic and for cubic scalars. Appendix~\ref{gs:app:freegauge} constructs the free gauge field on the sphere, and appendix~\ref{gs:app:fourfermion} the bare functional with four-fermion vertices near two dimensions.

\part{Framework}
\label{gs:part:one}

\section{Setting and required properties}
\label{gs:sec:setting}

In this section we describe the setting and state the properties of the renormalised functional on the sphere which the proof of gradient flow requires. The second part, sections~\ref{gs:sec:identities} to~\ref{gs:sec:consequences}, proves the gradient formula for every theory with these properties. It uses only the properties. The third part, sections~\ref{gs:sec:universal} to~\ref{gs:sec:other}, supplies the field content and the regularisation for specific theories. It constructs the functional on the sphere in dimensional regularisation with minimal subtraction, derives the properties from a hypothesis of finiteness, and evaluates the form of the gradient formula for each theory. We do not construct the functional in general. The properties are assumptions, and each of them carries a name. We first introduce the couplings (section~\ref{gs:sec:theories}) and the generating functional (section~\ref{gs:sec:functional}). In section~\ref{gs:sec:outline} we give the argument as a formal computation, and the properties of section~\ref{gs:sec:propinformal} state what this computation requires for non-integer $d$. In section~\ref{gs:sec:status} we state for which theories we expect the properties to hold and where this expectation is limited, and in section~\ref{gs:sec:casesmain} we say which property is proved for which theory and from what.

\subsection{Couplings}
\label{gs:sec:theories}

We consider a free theory of scalar fields and Dirac fermions in $d=d_c-\vep$ dimensions which is perturbed by local operators $\cO_I$ with couplings $g^I$. Here $d_c$ is the dimension in which the couplings are dimensionless. In $d$ dimensions the operator $\cO_I$ has the dimension $\Delta_I=d-\vep\,k_I$ in the free theory, with a number $k_I$ which we call the weight of the coupling. For a vertex with $n$ fields and without derivatives the weight is $k_I=\tfrac12n-1$, since the dimension of each field changes by $\tfrac12$ with $d$, so that for instance $k_I=1$ for quartic scalar couplings at $d_c=4$. The parameter $\vep$ can have either sign. For the four-fermion couplings near two dimensions the physical range is $\vep<0$, that is $d=2+|\vep|$, since $d=2$ is their lower critical dimension, see section~\ref{gs:sec:fermions}. In the other cases of this paper the perturbative fixed points with real couplings occur for $\vep>0$, except for cubic couplings, which have real fixed points only in special cases, see section~\ref{gs:sec:other}, and for gauge couplings. The beta function of a gauge coupling $g_\iota$ near four dimensions is $-\tfrac12\vep\,g_\iota+b_\iota\,g_\iota^3+{\rm O}(g^5)$, and a real zero, with $g_{\iota*}^2=\vep/(2b_\iota)+{\rm O}(\vep^2)$, requires $\vep\,b_\iota>0$. A gauge coupling $g_\iota$ enters the action of its gauge field through the factor $1/g_\iota^2$, and for it we use $\log g_\iota$ as the coordinate on the space of couplings. The derivative with respect to this coordinate is $g_\iota\,\pr/\pr g_\iota$, and the corresponding component of the vector field is the beta function of $g_\iota$ divided by $g_\iota$. We give the reasons in appendix~\ref{gs:sec:properties}, where gauge couplings form the second instance of Property~\ref{gs:P:functional}.

We call a function of the couplings regular if it is a power series in the couplings whose coefficients are power series in $\vep$, so that they have no poles at $\vep=0$, and by its term of order zero we mean its term of order zero in the couplings. Its term of order zero in the couplings and in $\vep$ is the term which is of order zero in both, that is its value at vanishing couplings and $\vep=0$. For a gauge coupling the power series is in $g_\iota$ itself and not in its logarithm. We call a tensor on the space of couplings regular if its components are regular functions. A function of $\vep$ which is analytic at $\vep=0$, such as the volume of the sphere, is identified with its power series, and the statements below for small $\vep$ refer to functions of this kind. The statements below concern a gradient formula whose potential and metric are regular. As explained in the introduction, the regularity is their content, since without it a gradient formula exists for every fixed $\vep\neq0$.

The flow of the couplings is described by a vector field $\hB^I(g)$ on the space of couplings, which is defined through the Weyl variation of the functional. In minimal subtraction it has the form
\begin{equation}
\hB^I = -\vep\,k_I\,g^I+B^I \, ,
\label{gs:eq:Bhat}
\end{equation}
where $B^I$ does not depend on $\vep$, has no terms of order zero and one in the couplings and differs from the beta function in $d_c$ dimensions by an infinitesimal rotation of the fields~\cite{Jack:2013sha}. A hat on $B^I$ indicates that the vector field includes the term $-\vep\,k_I\,g^I$. By a fixed point we mean a real zero of $\hB^I$ at which the couplings tend to zero with $\vep$, as in section~\ref{gs:sec:fixedgen}. The couplings are then of order $\vep$, except that $g^2$ is of order $\vep$ for a Yukawa, cubic or gauge coupling $g$. The general proof uses the term of lowest order of $\hB^I$ in two ways. First, together with the two-point functions of the free theory it determines the term of order zero of the metric, and this is where the weights enter. Second, it carries a factor $\vep$, while $B^I$ begins at second order in the couplings. The first derivatives $\pr_J\hB^I=-\vep\,k_I\,\delta^I_J+\pr_JB^I$ therefore have no term which is of order zero in both the couplings and $\vep$, which the treatment of the relevant sources uses.

\subsection{Generating functional}
\label{gs:sec:functional}

The theory is placed on a background with a metric $\gamma_{\mu\nu}$ which is conformal to the metric $\mathring\gamma_{\mu\nu}$ of the unit sphere $S^d$. The generating functional is $W=\log(Z/Z_0)$, where $Z$ is the partition function on this background, as a functional of the metric and of the sources, and $Z_0$ is a constant which does not depend on the sources and which we fix in section~\ref{gs:sec:fixed}. There are three kinds of sources. The couplings are functions $g^I(x)$ of the position, which we call local couplings. A background gauge field $a_\mu(x)$ couples to the currents of a compact group of field rotations, which acts linearly on the couplings. Further sources $\sM(x)$ couple to scalar operators of lower dimension, which we call relevant operators. A scalar field has a source only as a relevant operator. The fermion fields themselves have no source, while operators which contain pairs of fermion fields, such as the Yukawa operators, the currents and the fermion bilinears, have sources of the three kinds. A source of a fermion field would be an anticommuting spinor, which renormalisation and Weyl variations cannot generate from the commuting sources, so that it can be omitted consistently. The functional $W$ thus generates the correlation functions of the operators conjugate to the couplings, of the currents and of the relevant operators, which is what the properties and the gradient formula concern. The derivatives of $W$ with respect to the sources define these operators and their correlation functions. For instance
\begin{equation}
\delta W = \int d\mu_d\;\vev{\cO_I}\,\delta g^I
\label{gs:eq:onepoint}
\end{equation}
on the unit sphere for a variation of the couplings, where $d\mu_d$ is the measure of the unit sphere and functional derivatives are densities with respect to it. In the same way the first derivatives with respect to $a_\mu$ and $\sM$ are the one-point functions of the currents and of the relevant operators, and the second derivatives are the two-point functions, including their contact terms. We write $V=\int d\mu_d$ for the volume.

The background gauge field takes values in the Lie algebra of the group of field rotations, $(\omega g)^I$ denotes the action of an element $\omega$ of the Lie algebra on the couplings, and $D_\mu g^I=\pr_\mu g^I+(a_\mu g)^I$. The sources $\sM$ take values in a vector space $V_\sM$ on which the group acts, and a dot denotes the contraction of an element of $V_\sM$ or of the Lie algebra with an element of the dual space. We call a function of the couplings invariant, and a tensor covariant, if it is unchanged by a rotation with a constant element of the group of field rotations, where the rotation also acts on the indices of the tensor. It is then also unchanged by the rotations with a constant $\omega$ in the Lie algebra, which is all that the algebraic identities of this paper use, and where only this holds we say that it is covariant under the Lie algebra.

By $W$ we always mean the renormalised functional. In dimensional regularisation the functional of the bare sources has poles at $\vep=0$. The renormalised functional is obtained by expressing the bare sources through the sources $g^I$, $a_\mu$ and $\sM$ and by adding counterterms which are integrals of local expressions in the metric and in the sources, such that the result has no poles at $\vep=0$. In minimal subtraction the differences between the bare sources and the sources and the coefficients of the counterterms are pure poles in $\vep$, as in section~\ref{sec:setup} for $\phi^4$ theory. The properties below refer to the renormalised functional alone, and Property~\ref{gs:P:functional} states in which sense it is finite.

The arguments of $W$ are thus the metric and the sources $g^I$, $a_\mu$ and $\sM$. We write $W[g,a,\sM]$ for the functional on the unit sphere, and we indicate the metric where another background is meant. By constant couplings we mean couplings $g^I$ which do not depend on the position and take arbitrary values, together with vanishing other sources, $a_\mu=0$ and $\sM=0$. This is the point around which we expand. The metric is that of the unit sphere, $\gamma_{\mu\nu}=\mathring\gamma_{\mu\nu}$, unless another background is stated. At constant couplings the functional $W$ reduces to an ordinary function of the numbers $g^I$, which we denote by $\cW(g)$. It is the logarithm of the partition function on the unit sphere of the theory with the coupling constants $g^I$, and $-\cW$ is the free energy. We expand $W$ around constant couplings $g_c^I$ in powers of $g^I-g_c^I$, $a_\mu$ and $\sM$. We call a term of this expansion regular if, evaluated on fixed sources, it is a regular function of $g_c^I$.

\subsection{Outline of the argument}
\label{gs:sec:outline}

Before we state the properties we describe the argument as a formal computation, in which the sphere $S^d$ is treated as a manifold with arbitrary functions as sources, as it is in integer dimension, while $\vep$ is kept as a parameter. In this subsection we ignore divergences, and we omit the field rotations and the relevant sources. For the computation to become a proof, each of its steps has to hold for the renormalised functional at non-integer $d$. The properties of section~\ref{gs:sec:propinformal} state what this requires.

The functional $W$ is then invariant under diffeomorphisms which act on the metric and on the sources. At constant couplings the rotations of the sphere are symmetries. The one-point function of \eqref{gs:eq:onepoint} is then constant, and it equals $\pr_I\cW/V$ with $\pr_I=\pr/\pr g^I$. Formally a local change of scale can be compensated by a change of the couplings. Let $\sigma(x)$ be a function, and consider the Weyl rescaling $\delta^{\rm W}_\sigma\gamma_{\mu\nu}=2\,\sigma\,\gamma_{\mu\nu}$ together with the change $\delta^{\rm W}_\sigma g^I=\sigma\,\hB^I(g)$ of the couplings. Under this variation the functional changes by a local functional of the metric and of the sources,
\begin{equation}
\mathbb{W}_\sigma W = \cA[\sigma] \, , \qquad
\mathbb{W}_\sigma \equiv \int d^dx\;\sigma\left(2\,\gamma_{\mu\nu}\frac{\delta}{\delta\gamma_{\mu\nu}}+\hB^I\frac{\delta}{\delta g^I}\right) ,
\label{gs:eq:outweyl}
\end{equation}
which is the local renormalisation group equation of~\cite{Osborn:1991gm}. Here $\mathbb{W}_\sigma$ is the operator which generates the variation $\delta^{\rm W}_\sigma$ of the metric and of the couplings, and $\cA[\sigma]$ is the Weyl anomaly. In $\mathbb{W}_\sigma$ the functional derivatives are defined with respect to the coordinate measure $d^dx$, as is usual when the metric is varied as well. On the unit sphere, where the metric is fixed, we write the same terms with the measure $d\mu_d$ and with the functional derivatives of \eqref{gs:eq:onepoint}, which are densities with respect to $d\mu_d$. The two forms differ only in where the factor $\sqrt\gamma$ is placed. The equation \eqref{gs:eq:outweyl} contains the derivative of $W$ with respect to the metric. This derivative is the response of the functional to a deformation of the background, and it is not determined by $W$ as a functional of the sources on the unit sphere. For a constant $\sigma$, for instance, it is the derivative with respect to the radius $r_0$ of the sphere, and at constant couplings \eqref{gs:eq:outweyl} gives $r_0\,\pr W/\pr r_0+\hB^I\pr_I\cW=\cA[1]$. This determines the dependence of the free energy on the radius, and it gives no relation between the function $\cW(g)$ on the unit sphere and its derivatives.

The first ingredient of the argument is a Ward identity in which the derivative with respect to the metric is absent. It comes from the conformal maps of the sphere which are not rotations. For the Weyl factor of such a map the change of the metric is undone by a diffeomorphism, under which $W$ is invariant, and a relation on the unit sphere itself results. Let $X$ be the embedding coordinates of the unit sphere, and let
\begin{equation}
\sigma = b\cdot X \, , \qquad v^\mu = -\nabla^\mu\sigma \, ,
\label{gs:eq:sigma}
\end{equation}
with a constant vector $b$. The functions $\sigma$ are the restrictions to the sphere of the linear functions of the embedding coordinates. They are the spherical harmonics of degree one, which are eigenfunctions of the Laplacian with $\nabla^2\sigma=-d\,\sigma$, and we call them harmonics of degree one. They obey $\nabla_\mu\nabla_\nu\sigma=-\sigma\,\mathring\gamma_{\mu\nu}$, so that $\nabla_\mu v_\nu+\nabla_\nu v_\mu=2\,\sigma\,\mathring\gamma_{\mu\nu}$. The vector fields $v^\mu$ are thus conformal Killing vectors, and together with the rotations they generate the conformal maps of the sphere. They are the counterpart on the sphere of the dilatation and the special conformal maps of flat space, the conformal maps which are not isometries. To first order the flow of $v^\mu$ changes the metric of the unit sphere as the Weyl rescaling with $\sigma$ does. The function $\sigma$ of \eqref{gs:eq:sigma} is therefore the same as the parameter of the Weyl variation \eqref{gs:eq:outweyl}, taken for the rescaling which restores the metric after the flow of $-v$. The constants are the harmonics of degree zero, and as parameters of a Weyl rescaling they correspond to a change of the radius.

The diffeomorphism generated by $-v$ changes the metric of the unit sphere by $-2\,\sigma\,\mathring\gamma_{\mu\nu}$ and the couplings by $-v^\mu\pr_\mu g^I$, and it leaves $W$ invariant. The Weyl rescaling with $\sigma$ restores the metric. The sum of the two statements is a Ward identity for the functional on the unit sphere alone,
\begin{equation}
\int d\mu_d\left(\sigma\,\hB^I-v^\mu\pr_\mu g^I\right)\frac{\delta W}{\delta g^I} = \cA[\sigma] \, .
\label{gs:eq:outward}
\end{equation}
At constant couplings both sides vanish. We take the first variation of \eqref{gs:eq:outward} around constant couplings, with a variation $\delta g^J(x)$ of the couplings. Three terms arise on the left-hand side. The term with $v^\mu\pr_\mu\delta g^J$ multiplies the constant one-point function $\pr_J\cW/V$, and an integration by parts with $\nabla_\mu v^\mu=d\,\sigma$ turns it into $d\,V^{-1}\pr_J\cW\int d\mu_d\,\sigma\,\delta g^J$. The variation of the one-point function does not contribute to this term, since $\pr_\mu g^I$ vanishes at constant couplings. In the term with $\sigma\,\hB^I$ the variation of $\hB^I$ gives $\pr_J\hB^I\,\delta g^J$ times the one-point function. The variation of the one-point function gives the two-point function, integrated with $\sigma\,\hB^I$ at one point and with $\delta g^J$ at the other.

By the invariance under rotations the two-point function at constant couplings depends only on the angle between the two points. Integrated over one point with a harmonic of degree one it therefore gives a multiple of the same harmonic at the other point,
\begin{equation}
\int d\mu_d(x)\;\sigma(x)\,\vev{\cO_I(x)\,\cO_J(y)} = \Lambda_{IJ}\,\sigma(y) \, ,
\label{gs:eq:outLambda}
\end{equation}
and we call the matrix $\Lambda_{IJ}$ the first harmonic of the two-point function. It is the projection of the two-point function on the harmonics of degree one, and the projection on the constants is $\pr_I\pr_J\cW/V$. The right-hand side of \eqref{gs:eq:outward} is local, and its first variation is $\sfa_J\int d\mu_d\,\sigma\,\delta g^J$ with a one-form $\sfa_J$ on the space of couplings which is built from the coefficients of the anomaly. The first variation of \eqref{gs:eq:outward} is therefore
\begin{equation}
\left(\frac dV\,\pr_J\cW+\frac1V\,\pr_J\hB^I\,\pr_I\cW+\Lambda_{JI}\,\hB^I\right)\int d\mu_d\,\sigma\,\delta g^J = \sfa_J\int d\mu_d\,\sigma\,\delta g^J \, .
\label{gs:eq:outvar}
\end{equation}
Since $\delta g^J$ is arbitrary, the coefficients on the two sides agree. Adding and subtracting $\hB^I\pr_I\pr_J\cW/V$ we obtain
\begin{equation}
\pr_J\Pi+\cS_{JI}\,\hB^I = \sfa_J \, , \qquad \Pi = \frac1V\left(d+\hB^I\pr_I\right)\cW \, , \qquad \cS_{IJ} = \Lambda_{IJ}-\frac1V\,\pr_I\pr_J\cW \, ,
\label{gs:eq:outK}
\end{equation}
where the tensor $\cS_{IJ}$ is symmetric because the two-point function is. We call this relation the Killing identity.

If the one-form $\sfa_J$ vanished, \eqref{gs:eq:outK} would be a gradient formula with the potential $\Pi$ and the metric $-\cS_{IJ}$, and this is the case for odd $d_c$ in the absence of relevant sources. In general $\sfa_J$ is a local one-form which is not a gradient, and this is the place where a term which is not symmetric would sit. In four dimensions $\sfa_J$ contains the one-form $W_J$ of the anomaly, whose curl is the antisymmetric part of the tensor $T_{IJ}$ of~\cite{Osborn:1991gm}.

The second ingredient removes the one-form $\sfa_J$. We apply the operator on the left-hand side of \eqref{gs:eq:outward} not to $W$ itself but to a combination of $W$ and of its anomaly, chosen such that the result vanishes. The Killing identity of this combination then has no right-hand side. The choice uses that the anomaly is local and that Weyl rescalings commute, so that $\mathbb{W}_\sigma\cA[\sigma']=\mathbb{W}_{\sigma'}\cA[\sigma]$. Let $\bD=\int d^dx\,\hB^I\delta/\delta g^I$ be the Weyl variation of the couplings for a constant $\sigma$, without the rescaling of the metric. We consider the combination
\begin{equation}
W^\sharp = \vep\,W+\cA[1]-\bD W
\label{gs:eq:outsharp}
\end{equation}
on the unit sphere, which equals $(\vep+r_0\,\pr_{r_0})W$ at $r_0=1$ and fixed couplings, since $\mathbb{W}_1W=\cA[1]$. If $W$ is replaced by $W^\sharp$ on the left-hand side of \eqref{gs:eq:outward}, the result is zero. Indeed, $\bD$ commutes with the operator on the left-hand side of \eqref{gs:eq:outward}, since the constant Weyl rescaling commutes with the Weyl rescalings and with the diffeomorphisms which act on the metric and on the sources. The right-hand side for $W^\sharp$ is then $\vep\,\cA[\sigma]+\mathbb{W}_\sigma\cA[1]-\bD\cA[\sigma]$, where we used that $\cA[1]$ is invariant under these diffeomorphisms. It vanishes, since $\bD\cA[\sigma]=\mathbb{W}_1\cA[\sigma]+\vep\,\cA[\sigma]$ because the anomaly scales with the radius as $r_0^{-\vep}$, and since $\mathbb{W}_1\cA[\sigma]=\mathbb{W}_\sigma\cA[1]$. The Killing identity \eqref{gs:eq:outK} for $W^\sharp$ therefore has a vanishing right-hand side, and it is a gradient formula whose potential is the function $\Pi^\sharp$ of $W^\sharp$ and whose metric is $-\cS^\sharp_{IJ}$. At constant couplings the value of $W^\sharp$ is $(\vep-\hB^I\pr_I)\cW+V\rA$, where $\rA$ is the density of the anomaly for a constant rescaling, $\cA[1]=V\rA$. Its second variation gives $\cS^\sharp_{IJ}=[(\vep-\cL_{\hB})\,\cS]_{IJ}-\cK_{IJ}$, where $\cL_{\hB}$ is the Lie derivative along $\hB^I$ and $\cK_{IJ}$ is a symmetric local tensor of the anomaly. We obtain
\begin{equation}
\pr_J\!\left[\left(\vep-\hB^I\pr_I\right)\Pi+\left(d+\hB^I\pr_I\right)\rA\right] = \left(\cK_{JI}-\big[(\vep-\cL_{\hB})\,\cS\big]_{JI}\right)\hB^I \, ,
\label{gs:eq:outmain}
\end{equation}
in which no antisymmetric tensor appears. This completes the formal derivation of the gradient formula.

The computation is incomplete in two different ways. First, it left out the field rotations and the relevant sources, which the general proof includes. With field rotations the Lie derivative is replaced by the extended Lie derivative of section~\ref{gs:sec:quantities}. Sources of relevant operators have Weyl variations which contain the curvature and derivatives of the couplings, and they add terms to the Killing identity of $W^\sharp$ which have the form of the other terms only in a suitable scheme of the sources. We treat them in section~\ref{gs:sec:relevant}.

Second, the computation is formal. It treats the sphere as a manifold on which the sources are arbitrary functions, which is the case only in integer dimension, and it ignores the divergences, which renormalisation removes. For its steps to hold for the renormalised functional in non-integer $d$, the functional must have certain properties. These are assumptions on the theory, and we state them in section~\ref{gs:sec:propinformal}. For non-integer $d$ an arbitrary function on the sphere has no meaning, and only functions of a restricted kind are admitted as sources, for instance the functions of a fixed integer number of the embedding coordinates. We call the set of these functions the class of sources, and appendix~\ref{gs:sec:class} states what we assume about it.

\subsection{Properties required to prove gradient flow}
\label{gs:sec:propinformal}

We now state the properties in informal form. Each of them makes one step of the outline of section~\ref{gs:sec:outline} precise. The precise statements, which the proofs use, are Properties~\ref{gs:P:functional} to~\ref{gs:P:free} of appendix~\ref{gs:app:properties}, with the same numbers and names. There we also explain their technical conditions and give the forms which they take for gauge couplings and for specific theories.
\begin{list}{}{\setlength{\leftmargin}{0pt}\setlength{\labelwidth}{0pt}\setlength{\itemindent}{0pt}\setlength{\labelsep}{0.5em}\setlength{\itemsep}{3pt}}
\item[Property~\ref{gs:P:functional} (Functional).] The renormalised generating functional $W$ on the unit sphere exists for non-integer $d$, with local couplings, a background gauge field and sources of the relevant operators as arguments, and it is invariant under background gauge transformations. Expanded around constant couplings it is regular, except for an explicit term in its value at constant couplings, which for a gauge coupling is a multiple of the logarithm of the coupling. The potential and the metric are built from this functional in section~\ref{gs:sec:quantities}.
\item[Property~\ref{gs:P:weyl} (Weyl variation).] A local Weyl rescaling of the metric is accompanied by a variation of the sources, in which the couplings change along a regular vector field $\hB^I$ and the relevant sources change by a local expression. These Weyl variations commute. At lowest order in the couplings $\hB^I=-\vep\,k_Ig^I$, and the relevant sources scale with their dimensions in the free theory in $d_c$ dimensions. At vanishing couplings the sources of relevant operators of even dimension, such as the source of $\phi_i\phi_j$, vary on the sphere as in the free theory.
\item[Property~\ref{gs:P:anomaly} (Anomaly).] The anomaly $\cA[\sigma]$ is a local functional of the metric and of the sources with regular coefficients. It satisfies the consistency condition which in the outline follows from the commutation of the Weyl rescalings, and it scales with the radius $r_0$ of the sphere as $r_0^{-\vep}$.
\item[Property~\ref{gs:P:ward} (Ward identity).] The functional on the unit sphere satisfies the Ward identity \eqref{gs:eq:outward} of the conformal Killing vectors of the sphere. In this identity the conformal map of the sources, combined with their Weyl variation, changes $W$ by the anomaly. It gives the Killing identity of section~\ref{gs:sec:K} and, together with Properties~\ref{gs:P:weyl} and~\ref{gs:P:anomaly}, the homogeneous Ward identity of section~\ref{gs:sec:hom}.
\item[Property~\ref{gs:P:sphere} (Sphere).] At constant couplings the functional has the properties which the rotations of the sphere give in integer dimension. Its one-point functions are constant, and its second variation maps a harmonic of degree one to a multiple of itself, as in \eqref{gs:eq:outLambda}. With it the Ward identities become relations among functions of the couplings.
\item[Property~\ref{gs:P:free} (Free limit).] At $\vep=0$ and at order zero in the couplings, the terms of the anomaly with two derivatives of the couplings take their values in the free theory in $d_c$ dimensions, which are fixed by the matrix $c_{IJ}$ of the two-point functions of the operators. For odd $d_c$, where the anomaly vanishes at vanishing relevant sources, the statement concerns the tensor $\cS_{IJ}$ of \eqref{gs:eq:outK} instead. At $\vep=0$ and at order zero in the couplings it is the tensor which the two-point functions of the free theory on the sphere $S^{d_c}$ give, a fixed multiple of $c_{IJ}$. In the scheme of section~\ref{gs:sec:divscheme} the one-point functions of the relevant operators of even dimension have no term of order zero in the couplings and in $\vep$. We use this property to show that the term of order zero of the metric is a multiple of $c_{IJ}$. The metric is then nondegenerate if $c_{IJ}$ is invertible, and with the sign of the potential which is normalised to $\tilde F$ it is positive definite to all orders if $c_{IJ}$ is positive.
\end{list}

Properties~\ref{gs:P:functional} to~\ref{gs:P:sphere} give the gradient formula with a regular symmetric metric in section~\ref{gs:sec:gradient}, but they do not imply that the metric is different from zero. Property~\ref{gs:P:free} and the statements of Property~\ref{gs:P:weyl} at lowest order are statements about the theory at vanishing couplings. Property~\ref{gs:P:functional} has two instances, instance~(a), in which the coordinates on the space of couplings are the couplings themselves, and instance~(b), in which the logarithms of the gauge couplings are coordinates.

\subsection{Status of the properties}
\label{gs:sec:status}

The properties are statements about the renormalised functional on $S^d$ for non-integer $d$, and for a given theory each of them has to be shown. We expect that Properties~\ref{gs:P:functional} to~\ref{gs:P:sphere} and the last statement of Property~\ref{gs:P:free} hold, in dimensional regularisation with minimal subtraction, for the theories of section~\ref{gs:sec:theories} which are invariant under parity and whose Feynman rules do not contain $\gamma_5$. These are scalar fields with $d_c=3$, $4$ and $6$, scalar fields with Dirac fermions and Yukawa couplings with $d_c=4$, and, with Property~\ref{gs:P:functional} in instance~(b), gauge fields of a semisimple group with Dirac fermions, also together with scalar fields, with $d_c=4$. For fermions with quartic couplings at $d_c=2$ we expect the same, with the qualification that the set of couplings is infinite there. The following points state what is established and what is assumed.
\begin{enumerate}
\item We have a construction of the functional on $S^d$ for non-integer $d$ for scalar fields and Dirac fermions with polynomial and Yukawa interactions, see section~\ref{gs:sec:universal}, and for Dirac fermions with four-fermion interactions of every gamma structure, see appendix~\ref{g2:sec:construction}. For free gauge fields with a local coupling we construct the bare functional on $S^d$ by continued sums over the levels of the exact two-forms, see appendix~\ref{gs:app:freegauge}, and Properties~\ref{gs:P:functional} to~\ref{gs:P:free} follow for it from Hypothesis~\ref{hyp:finite} for the free gauge field, which is stated on the sphere. For interacting gauge fields a construction of the gauge-fixed functional with the required covariance is missing.
\item The construction defines the regularised functional before renormalisation, which we call the bare functional. For four of these theories the properties then need not be assumed one by one. All six follow from the single Hypothesis~\ref{hyp:finite} of appendix~\ref{gs:sec:cases}, which states that the bare functional in flat space, for the cubic theory on conformally flat backgrounds, is made finite by counterterms of the form of minimal subtraction. We show this for multiscalar $\phi^4$ theory in section~\ref{gs:sec:phi4}, for scalar fields and Dirac fermions with Yukawa and quartic couplings in section~\ref{gs:sec:yukawa}, for sextic scalars in section~\ref{gs:sec:sextic}, and for cubic scalars, with the class of flat sources with $D_0=3$, in section~\ref{gs:sec:cubic}.
\item The last statements of Properties~\ref{gs:P:weyl} and~\ref{gs:P:free} concern relevant sources of even dimension, and they are empty where there are none. This is the case for gauge fields with fermions and for fermions near two dimensions, where the gradient formula is that of section~\ref{gs:sec:gradient}. Where they are required, they follow from Hypothesis~\ref{hyp:finite} for multiscalar $\phi^4$ theory, for scalars with Yukawa couplings, for sextic scalars and for cubic scalars. For gauge fields with scalars they are an assumption on the functional. They then state properties of the free scalar field, which hold in the construction of section~\ref{gs:sec:sphere} if the pure poles of the counterterms are unique.
\end{enumerate}
Three special situations are discussed in appendix~\ref{gs:app:status}. It explains how Property~\ref{gs:P:free} is derived from the two-point functions of the free theory and what it states for gauge fields. For fermions near two dimensions it describes the infinite set of couplings which renormalisation generates, and for an abelian factor of the gauge group the source of the field strength, an antisymmetric tensor which the list of sources above does not contain.

\subsection{The properties in specific cases}
\label{gs:sec:casesmain}

For several theories the properties can be derived from a single simpler hypothesis, which concerns the finiteness of the regularised functional in flat space, on conformally flat backgrounds for the cubic theory, or on the sphere for gauge fields. In words, Hypothesis~\ref{hyp:finite} states the following. It concerns a theory whose regularised bare functional in flat space, or on the sphere for gauge fields, is constructed for non-integer $d$ with local couplings, a background gauge field and relevant sources. When the bare sources are expressed through the sources and field-independent counterterms are added, this functional is finite at each order in the sources, where all counterterms consist of poles in $\vep$ without finite parts, as in minimal subtraction. The precise statement, with the bare sources and counterterms of each theory, is Hypothesis~\ref{hyp:finite} in appendix~\ref{gs:sec:cases}. We use the hypothesis for six theories, which we call the $\phi^4$ theory, the Yukawa theory, the sextic theory, the cubic theory, the four-fermion theory and the free gauge field.

For the $\phi^4$ theory, the Yukawa theory, the sextic theory and the cubic theory Hypothesis~\ref{hyp:finite} gives all six properties, see sections~\ref{gs:sec:phi4}, \ref{gs:sec:yukawa}, \ref{gs:sec:sextic} and~\ref{gs:sec:cubic}, and with them the gradient formula for which the metric of the potential normalised to $\tilde F$ is positive definite to all orders. For the free gauge field Hypothesis~\ref{hyp:finite} gives Properties~\ref{gs:P:functional} to~\ref{gs:P:free} for polynomial sources, with Property~\ref{gs:P:functional} in instance~(b), $n_\iota=n_G$ for every $\vep$ and $\hB=-\tfrac12\vep$ exactly, while the last statements of Properties~\ref{gs:P:weyl} and~\ref{gs:P:free} are empty, see Proposition~\ref{ggs:prop:props} of appendix~\ref{gs:app:freegauge}. We do not have the hypothesis as a starting point for interacting gauge theories, for which the gauge-fixed functional on $S^d$ is not constructed. For the four-fermion theory near two dimensions the bare functional is constructed, the hypothesis reduces Property~\ref{gs:P:weyl} for the theory without the sources of the fermion bilinears to the pole equations of the counterterms of the couplings for a suitable choice of the bare sources, given ${\rm tr}\,1>1$ and the covariance of $N_I$ without term of degree zero, and gives Property~\ref{gs:P:free} given the other properties, and the remaining properties are assumptions, see section~\ref{gs:sec:fermions}. That the vector field $\hB^I$ of $\phi^4$ theory is the usual beta function of minimal subtraction in momentum space, up to a rotation of the fields, rests on a separate statement, the identification of section~\ref{gs:sec:identification}, which the proofs do not use.

\begin{table}[!t]
\centering
\renewcommand{\arraystretch}{1.12}
\scriptsize
\setlength{\tabcolsep}{3pt}
\hyphenpenalty=10000 \exhyphenpenalty=10000
\begin{tabular}{>{\raggedright\arraybackslash}p{1.6cm}>{\raggedright\arraybackslash}p{1.62cm}>{\raggedright\arraybackslash}p{1.62cm}>{\raggedright\arraybackslash}p{1.62cm}>{\raggedright\arraybackslash}p{1.62cm}>{\raggedright\arraybackslash}p{1.62cm}>{\raggedright\arraybackslash}p{1.62cm}>{\raggedright\arraybackslash}p{1.62cm}}
\hline
 & $\phi^4$, $d=4-\vep$ & $\phi^6$, $d=3-\vep$ & gauge fields with fermions, $d=4-\vep$ & scalars with Yukawa and quartic couplings, $d=4-\vep$ & gauge fields with scalars, $d=4-\vep$ & $\phi^3$, $d=6-\vep$ & fermions, $d=2+|\vep|$ \\
\hline
bare functional on $S^d$ & constructed & constructed & free fields only & constructed & free fields only & constructed & constructed \\
relevant sources of even dimension & $\phi^2$ & $\phi^2$ & none & $\phi^2$ & $\phi^2$ & $\phi^2$, $\phi$ & none \\
\ref{gs:P:functional} Functional & derived, (a) & derived, (a) & assumed, (b) & derived, (a) & assumed, (b) & derived, (a) & assumed, (a) \\
\ref{gs:P:weyl} Weyl variation & derived & derived & assumed & derived & assumed & derived & assumed, reduced to pole equations \\
\ref{gs:P:anomaly} Anomaly & derived & derived & assumed & derived & assumed & derived & assumed \\
\ref{gs:P:ward} Ward identity, \ref{gs:P:sphere} Sphere & derived & derived & assumed & derived & assumed & derived & assumed \\
\ref{gs:P:free} Free limit & derived & derived & assumed & derived & assumed & derived & derived given 1 to 5 \\
matrix $c_{IJ}$ & positive & positive & positive & positive & positive & positive & degenerate at $\vep=0$ \\
gradient formula & Theorem \ref{gs:thm:phi4} & Theorem \ref{gsx:thm:sextic} & Corollary \ref{gs:cor:gauge} & Theorem \ref{gf:cor:yukawa} & Proposition \ref{gro:prop:gauge} & Theorem \ref{gs:thm:cubic} & Proposition \ref{gs:prop:fermions} \\
\hline
\end{tabular}
\caption{The properties and the results in the cases of sections~\ref{gs:sec:phi4} and~\ref{gs:sec:other}. The entries are explained in section~\ref{gs:sec:casesmain}.}
\label{gs:tab:cases}
\end{table}

Table~\ref{gs:tab:cases} summarises the cases. In it ``derived'' means derived from Hypothesis~\ref{hyp:finite} for the theory of the column, ``assumed'' means that the property is an assumption for that theory, and (a) and (b) are the instances of Property~\ref{gs:P:functional}. For the cubic theory the hypothesis is stated on conformally flat backgrounds, with the class of flat sources with $D_0=3$. For the fermions near two dimensions, where $\vep<0$, Property~\ref{gs:P:weyl} for the theory without the sources of the fermion bilinears is equivalent to the pole equations of minimal subtraction for the counterterms $L^I$ for a suitable choice of the bare sources, given Hypothesis~\ref{hyp:finite}, ${\rm tr}\,1>1$ and the covariance of $N_I$ without term of degree zero, and Property~\ref{gs:P:free} follows from the hypothesis given Properties~\ref{gs:P:functional} to~\ref{gs:P:sphere} for the functional of minimal subtraction on the sphere, see section~\ref{gs:sec:fermions}. The columns with gauge fields refer to the interacting theories with a semisimple gauge group, vectorlike couplings and invariance under parity. For free gauge fields with a local coupling the bare functional is constructed and all six properties are derived, see appendix~\ref{gs:app:freegauge}. Abelian factors of the gauge group are treated in section~\ref{gs:sec:gauge}, and chiral couplings are not covered.

\part{General proof}
\label{gs:part:two}

\section{Identities and gradient formula}
\label{gs:sec:identities}

This section begins the second part of the paper, in which we prove the gradient formula for every theory with the properties of appendix~\ref{gs:sec:properties}. No step of this part depends on the theory. We say that the Weyl variation closes on the marginal couplings if Properties~\ref{gs:P:functional} to~\ref{gs:P:sphere} hold without sources $\sM$, that is with the couplings $g^I$ and the gauge field $a_\mu$ alone. This is a condition on the theory, since a source of a relevant operator can be omitted only if its Weyl variation vanishes at $\sM=0$. In this section we assume that the Weyl variation closes on the marginal couplings, and we derive the consequences of Properties~\ref{gs:P:functional} to~\ref{gs:P:sphere}. We first collect statements on local functionals on the unit sphere and define the functions and tensors on the space of couplings which enter the identities. We then prove the Killing identity, which is the first variation of the Ward identity around constant couplings. Applied to a combination of the functional and of its anomaly whose Ward identity has no anomaly, it gives a gradient formula, whose potential and metric we then discuss. Sources of relevant operators are added in section~\ref{gs:sec:relevant}, and section~\ref{gs:sec:consequences} states the consequences.

\subsection{Local functionals on the sphere}
\label{gsp:sec:facts}

On the unit sphere we use the following facts, with $d$ as a parameter, which we take from the class of sources of appendix~\ref{gs:sec:class}. The Laplacian is symmetric with respect to $d\mu_d$ on the class of sources and annihilates the constants. By a one-form of the class we mean a combination of the differentials of the embedding coordinates whose coefficients are functions of the class of sources, such as the gradient of a function of the class or the background gauge field. Functions and one-forms which are built from functions and one-forms of the class by products, covariant derivatives and contractions are again of this kind, by appendix~\ref{gs:sec:class}. For every one-form $Y$ of the class the integral of the divergence vanishes, $\int d\mu_d\,\nabla_\mu Y^\mu=0$. For a gradient this is the symmetry of the Laplacian with the constant function, and for $Y=f\,\nabla h$ it follows from $\nabla^2(fh)=f\nabla^2h+h\nabla^2f+2\nabla f\cdot\nabla h$. For a general one-form of the class it is the divergence form of the rules of appendix~\ref{app:rules}. The harmonics of degree one obey $\nabla_\mu\nabla_\nu\sigma=-\sigma\,\mathring\gamma_{\mu\nu}$. Finally, the commutator of two covariant derivatives is given by the curvature of the unit sphere, $R_{\mu\nu\rho\lambda}=\mathring\gamma_{\mu\rho}\mathring\gamma_{\nu\lambda}-\mathring\gamma_{\mu\lambda}\mathring\gamma_{\nu\rho}$, which is an identity of tensor calculus with $d$ as a parameter. Indeed, each local expression of this kind reduces to a polynomial in $d$ whose coefficients are functions of $X_\parallel$, and the identity holds on the sphere $S^d$ for every integer $d$ larger than the number of embedding coordinates on which the functions of the class depend, so that it holds identically, as in appendix~\ref{app:rotations}. For the class of sources of section~\ref{sec:sphere} these facts are the rules of appendix~\ref{app:rules}.
In particular, these facts give
\begin{equation}
\nabla^2\sigma = -d\,\sigma \, , \qquad \int d\mu_d\;v^\mu\pr_\mu f = -d\int d\mu_d\;\sigma\,f
\label{gs:eq:parts}
\end{equation}
for a function $f$ of the class of sources.

We call a scalar a linear expression in a function $f$ if it is linear in $f$ and built from covariant derivatives of $f$ and the metric of the unit sphere alone, with constant coefficients. It is a sum of terms $\nabla_{\mu_1}\cdots\nabla_{\mu_k}f$ in which all indices are contracted in pairs, for example $\nabla^2f$ or $\nabla_\mu\nabla_\nu\nabla^\mu\nabla^\nu f$. A vector expression is defined in the same way with one index left free. Commuting the derivatives with the curvature above, we find that a linear expression equals $p(\nabla^2)f$ and a vector expression equals $\nabla_\mu q(\nabla^2)f$, with polynomials $p$ and $q$ whose coefficients are polynomials in $d$. In the same way a scalar which is linear in a one-form $\omega$ is the divergence $\nabla^\mu Y_\mu$ of a one-form $Y$ built from $\omega$. Indeed, in a complete contraction of $\nabla_{\mu_1}\cdots\nabla_{\mu_k}\omega_\nu$ the index $\nu$ is contracted with one of the derivatives, which we move to the front, and the commutators give terms of the same kind with fewer derivatives, so that the statement follows by induction in $k$. We use three consequences, for functions $f$ and one-forms $\omega$ of the class. For a linear expression $p(\nabla^2)f$ the symmetry of the Laplacian and $\nabla^2\sigma=-d\,\sigma$ give
\begin{equation}
\int d\mu_d\;p(\nabla^2)f = p(0)\int d\mu_d\,f \, , \qquad \int d\mu_d\;\sigma\,p(\nabla^2)f = p(-d)\int d\mu_d\,\sigma f \, ,
\label{gsp:eq:c1}
\end{equation}
and in particular $\int d\mu_d\,\sigma=0$. A scalar $S(\omega)$ which is linear in $\omega$ is a divergence, so that
\begin{equation}
\int d\mu_d\;S(\omega) = 0 \, .
\label{gsp:eq:c2}
\end{equation}
For a scalar $B(\sigma,f)$ which is linear in $\sigma$ and in $f$ and built covariantly with constant coefficients,
\begin{equation}
\int d\mu_d\;B(\sigma,f) = \beta\int d\mu_d\,\sigma f \, ,
\label{gsp:eq:c3}
\end{equation}
with a constant $\beta$ which does not depend on the vector $b$ of the harmonic. To show \eqref{gsp:eq:c3} we reduce every second or higher derivative of $\sigma$ with $\nabla_\mu\nabla_\nu\sigma=-\sigma\mathring\gamma_{\mu\nu}$, so that $B(\sigma,f)=\sigma\,p(\nabla^2)f+\nabla^\mu\sigma\,\nabla_\mu q(\nabla^2)f$. The integral of the second term is $-\int d\mu_d\,\nabla^2\sigma\;q(\nabla^2)f=d\int d\mu_d\,\sigma\,q(\nabla^2)f$, and \eqref{gsp:eq:c1} gives $\beta=p(-d)+d\,q(-d)$. The constants $p$ and $q$ are built from the coefficients of $B$ and from $d$, and $b$ enters only through $\sigma$.

\begin{lemma}[Local functionals on the sphere]
\label{gsp:lem:local}
Let $F[\gamma,g,a,\sM]$ be a local functional of the metric and of the sources, that is the integral of a polynomial in the curvature and in the covariant derivatives of the sources, built covariantly from $\gamma_{\mu\nu}$ with $d$ as a parameter, without the Levi-Civita tensor, and with coefficients which are functions of the couplings at the point. Let $F_\sigma[\gamma,g,a,\sM]$ be a local functional of the same kind which depends in addition on a function $\sigma$, and which is linear in $\sigma$ and its covariant derivatives. An example is the anomaly $\cA[\sigma]$. Consider $\gamma_{\mu\nu}=\mathring\gamma_{\mu\nu}$, constant couplings $g_c$, $a_\mu=0$ and a constant relevant source $\mu$, and let the function $\sigma$ be a harmonic of degree one, $\delta g^J$ and $\delta\sM$ variations in the class of sources, and $z^I$ and $\sM_{\rm c}$ constants. Then the following three statements hold. (i) The first variation of $F$ in the direction $\delta g^J$ is a constant times $\int d\mu_d\,\delta g^J$, the same holds for $\delta\sM$, and the first variation in a variation of $a_\mu$ vanishes. (ii) The second variation of $F$ in which one direction is $\sigma z^I$ or $\sigma\sM_{\rm c}$ and the other is $\delta g^J$ or $\delta\sM$ is a constant times $\int d\mu_d\,\sigma\,\delta g^J$ or $\int d\mu_d\,\sigma\,\delta\sM$. The constant is linear in $z^I$ or $\sM_{\rm c}$ and does not depend on $b$. (iii) The functional $F_\sigma$ vanishes, and its first variation in the direction $\delta g^J$ or $\delta\sM$ is a constant times $\int d\mu_d\,\sigma\,\delta g^J$ or $\int d\mu_d\,\sigma\,\delta\sM$, with a constant which does not depend on $b$.
\end{lemma}

\begin{proof}
At constant couplings, $a_\mu=0$, constant $\sM$ and $\gamma_{\mu\nu}=\mathring\gamma_{\mu\nu}$ every coefficient and every curvature tensor in the integrand is a constant tensor built from $\mathring\gamma_{\mu\nu}$, and the derivatives of the constant sources vanish. A first variation of $F$ in $\delta g^J$ or $\delta\sM$ is therefore the integral of a linear expression in the variation, and a first variation in a one-form $\delta a_\mu$ is the integral of a scalar linear in $\delta a_\mu$. With \eqref{gsp:eq:c1} and \eqref{gsp:eq:c2} we obtain the first statement. In statement~(ii) the second variation is the integral of a scalar which is linear in $\sigma$ and in the second variation, with constant coefficients which are linear in $z^I$ or $\sM_{\rm c}$, and we apply \eqref{gsp:eq:c3}. For $F_\sigma$ at constant sources the integrand is a linear expression in $\sigma$, whose integral vanishes by \eqref{gsp:eq:c1}, and its first variation is a scalar linear in $\sigma$ and in the variation, to which \eqref{gsp:eq:c3} applies.
\end{proof}

Applied to $F=\cA[1]$ and to $F_\sigma=\cA[\sigma]$, which by Property~\ref{gs:P:anomaly} are local functionals of this kind, the lemma gives the following statements on the anomaly at constant couplings. The first derivatives of $\cA[1]$ with respect to $g^I$ and $\sM$ are constant, $\cA[\sigma]$ vanishes, and the second variations of $\cA[1]$ and the first variations of $\cA[\sigma]$ are multiples of $\int d\mu_d\,\sigma\,\delta g^J$ or $\int d\mu_d\,\sigma\,\delta\sM$, as in statements~(ii) and~(iii). The first derivative of $\cA[1]$ with respect to $a_\mu$ vanishes by the statement~(i) of the lemma.

\subsection{Free energy, two-point tensor and anomaly coefficients}
\label{gs:sec:quantities}

The Killing identity and the gradient formula of the following sections are relations among some functions and tensors on the space of couplings, which we now define. They are obtained from the functional $W$ and from its anomaly at constant couplings, and we refer to them together as the sphere data of $W$. Let $\cW(g)$ be the value of $W$ at constant couplings, so that the one-point function is $\vev{\cO_I}=\pr_I\cW/V$. By Property~\ref{gs:P:sphere} there is a matrix $\Lambda_{IJ}(g)$ with
\begin{equation}
\frac{\pr^2}{\pr t\,\pr t'}\,W\big[g+t\,\sigma\,z+t'\,\delta g\big]\Big|_{t=t'=0} = \Lambda_{IJ}\,z^I\int d\mu_d\;\sigma\,\delta g^J
\label{gs:eq:Lambda}
\end{equation}
for constants $z^I$, a variation $\delta g^J$ and a harmonic $\sigma$ of degree one. The left-hand side is the second derivative of the functional in two directions, the profile $\sigma$ in the direction $z^I$ of the space of couplings and the arbitrary variation $\delta g^J$. It is the two-point function of the operators $\cO_I$ and $\cO_J$, including the contact terms, integrated with $\sigma\,z^I$ at one point and with $\delta g^J$ at the other. The equation states that the two-point function at constant couplings, regarded as an integral kernel, maps a harmonic of degree one to a multiple of the same harmonic, as in \eqref{gs:eq:outLambda}. The matrix $\Lambda_{IJ}$ of these multiples is what we call the first harmonic of the two-point function. In integer dimension it is the coefficient of the harmonics of degree one in the expansion of the kernel in spherical harmonics, and the coefficient of the constants is $\pr_I\pr_J\cW/V$. It is symmetric, because for $\delta g^J=\sigma\,z'^J$ with constants $z'^J$ the left-hand side is symmetric and $\int d\mu_d\,\sigma^2$ does not vanish. We define
\begin{equation}
\cS_{IJ} \equiv \Lambda_{IJ}-\frac1V\,\pr_I\pr_J\cW \, , \qquad \Pi \equiv \frac1V\left(d+\hB\!\cdot\!\pr\right)\cW \, ,
\label{gs:eq:Sdef}
\end{equation}
where $\hB\!\cdot\!\pr=\hB^I\pr_I$. Formally $-\cS_{IJ}$ is the integral of the two-point function over the sphere multiplied by $1-\cos\theta$, where $\theta$ is the angle between the two points. For the anomaly we define a function $\rA$, a one-form $\sfa_J$ and a matrix $\alpha_{IJ}$ by
\begin{align}
\cA[1]\big|_{g} &= V\,\rA \, , \qquad
\cA[\sigma]\big|_{g+\delta g} = \sfa_J\int d\mu_d\;\sigma\,\delta g^J+{\rm O}(\delta g^2) \, , \notag\\
\frac{\pr^2}{\pr t\,\pr t'}\,\cA[1]\big|_{g+t\,\sigma\,z+t'\,\delta g}\Big|_{t=t'=0} &= \alpha_{IJ}\,z^I\int d\mu_d\;\sigma\,\delta g^J \, ,
\label{gs:eq:adef}
\end{align}
at $\gamma_{\mu\nu}=\mathring\gamma_{\mu\nu}$, where $\sigma$ is a harmonic of degree one in the second and in the third expression. In the second expression we used that $\cA[\sigma]$ vanishes at constant couplings by Property~\ref{gs:P:anomaly} and Lemma~\ref{gsp:lem:local}. The matrix $\alpha_{IJ}$ is symmetric for the same reason as $\Lambda_{IJ}$. We also define
\begin{equation}
\cK_{IJ} \equiv \pr_I\pr_J\rA-\alpha_{IJ} \, .
\label{gs:eq:Kdef}
\end{equation}
The tensor $\cK_{IJ}$ is minus the contribution to $\alpha_{IJ}$ of the terms of $\cA[1]$ with derivatives of the couplings.

By Property~\ref{gs:P:functional} the one-form $\pr_I\cW$ and the tensors $\Lambda_{IJ}$ and $\cS_{IJ}$ are regular, and so are the functions $\cW-\cW^{\log}$ and $\Pi-d\,\cW^{\log}/V$, since $\hB^I\pr_I\cW$ is regular. The quantities $\rA$, $\sfa_J$, $\alpha_{IJ}$ and $\cK_{IJ}$ are regular and local by Property~\ref{gs:P:anomaly}. By the gauge invariance $\cW$ and $\rA$ are invariant under the constant field rotations of the identity component of the group, and the tensors are covariant under the Lie algebra. If a gradient formula holds with a potential and a metric, it also holds with their average over the compact group of field rotations, since $\hB^I$ is covariant under the group and the average of a regular symmetric tensor is regular and symmetric. We use the extended Lie derivative $\tL$ of~\cite{Jack:2013sha}. On a covariant tensor it is the Lie derivative along $\hB^I$ in which the matrix $\pr_I\hB^K$ is replaced by $\hat M^K{}_I=\pr_I\hB^K+(\hr_I\,g)^K$, so that $(\tL\,\omega)_J=\hB^K\pr_K\omega_J+\hat M^K{}_J\,\omega_K$ for a one-form. It obeys $\tL\hB=0$, because $(\hr_Ig)^K\hB^I=0$. For an invariant function $f$ it gives $(\tL\,\pr f)_J=\pr_J(\hB\!\cdot\!\pr\,f)$, because the term $(\hr_Jg)^K\pr_Kf$ vanishes. The extended Lie derivative of a symmetric tensor is symmetric.

\subsection{Killing identity}
\label{gs:sec:K}

\begin{proposition}[Killing identity]
\label{gs:prop:K}
Assume Properties~\ref{gs:P:functional} to~\ref{gs:P:sphere}, and let the Weyl variation close on the marginal couplings. Then
\begin{equation}
\frac1V\left(d\,\delta_J{}^I+\pr_J\hB^I\right)\pr_I\cW+\Lambda_{JI}\,\hB^I = \sfa_J \, , \qquad \text{that is} \qquad
\sfa_J = \pr_J\Pi+\cS_{JI}\,\hB^I \, .
\label{gs:eq:K}
\end{equation}
\end{proposition}

\begin{proof}
We vary the Ward identity of Property~\ref{gs:P:ward} to first order around constant couplings, with a variation $\delta g^J$. The right-hand side vanishes at constant couplings by Property~\ref{gs:P:anomaly} and Lemma~\ref{gsp:lem:local}, and its first variation is $\sfa_J\int d\mu_d\,\sigma\,\delta g^J$ by \eqref{gs:eq:adef}. On the left-hand side the term with the gauge field does not contribute. Its variation contains the one-point function of the current, which vanishes at constant couplings, or the factor $D_\mu g^I$, which vanishes as well. The term with the couplings gives three contributions. The variation of $v^\mu\pr_\mu g^I$ gives $-V^{-1}\pr_J\cW\int d\mu_d\,v^\mu\pr_\mu\delta g^J$, which is $d\,V^{-1}\pr_J\cW\int d\mu_d\,\sigma\,\delta g^J$ by \eqref{gs:eq:parts}. The variation of $\hB^I$ gives $V^{-1}\pr_J\hB^I\,\pr_I\cW\int d\mu_d\,\sigma\,\delta g^J$. The variation of the one-point function gives the second variation of $W$ in the directions $\sigma\hB^I$ and $\delta g^J$, which is $\Lambda_{JI}\hB^I\int d\mu_d\,\sigma\,\delta g^J$ by \eqref{gs:eq:Lambda}. The integral $\int d\mu_d\,\sigma\,\delta g^J$ does not vanish for $\delta g^J$ proportional to $\sigma$, which gives the first form of \eqref{gs:eq:K}. The second form follows with $\pr_J\hB^I\,\pr_I\cW=\pr_J(\hB^I\pr_I\cW)-\hB^I\pr_I\pr_J\cW$.
\end{proof}

The left-hand side of the first form contains the one-point function and the first harmonic of the two-point function on the sphere, and the right-hand side is local.\footnote{The function $\Pi$ contains the term $\hB\!\cdot\!\pr\,\cW$, which the Weyl variation for a constant $\sigma$ does not remove. On the sphere of radius $r_0$ it gives $r_0\,\pr W/\pr r_0+\hB\!\cdot\!\pr\,\cW=V\rA$ at constant couplings and $r_0=1$, which determines the dependence on the radius and is not a relation among the quantities on the unit sphere.}

\subsection{Homogeneous Ward identity}
\label{gs:sec:hom}

The anomaly enters the Killing identity through the local one-form $\sfa_J$. We now remove it with a combination of $W$ and of the anomaly whose Ward identity has a vanishing right-hand side. In the extended form of the properties of appendix~\ref{gs:sec:properties}, where $W$ is defined on the backgrounds $\Omega^2\mathring\gamma_{\mu\nu}$, this combination is $(\vep+r_0\,\pr_{r_0})W$, with the derivative with respect to the radius $r_0$ of the sphere at fixed dimensionless sources, and its anomaly vanishes since the anomaly has the Weyl weight $-\vep$. We define the combination on the unit sphere and derive its Ward identity from the properties on the unit sphere.

Let the relevant sources be $\sM_a$, where the component $\sM_a$ has the dimension $D_a$ at $d=d_c$. We collect the engineering dimensions in the operator
\begin{equation}
\mathbb{D}_{\rm e} = \int d^dx\left(2\,\gamma_{\mu\nu}\frac{\delta}{\delta\gamma_{\mu\nu}}-\sum_aD_a\,\sM_a\frac{\delta}{\delta\sM_a}\right) ,
\label{gsh:eq:De}
\end{equation}
which we use as the generator of the constant rescaling of the metric together with the rescaling of each source by its dimension. By the homogeneity of Property~\ref{gs:P:anomaly} we have $\mathbb{D}_{\rm e}\cA[\sigma]=-\vep\,\cA[\sigma]$. We define $\bD=\mathbb{W}_1-\mathbb{D}_{\rm e}$. The terms of $\mathbb{W}_1$ and of $\mathbb{D}_{\rm e}$ with the metric cancel, so that $\bD$ acts on the sources alone,
\begin{equation}
\bD = \int d^dx\left[\hB^I\frac{\delta}{\delta g^I}+\hr_ID_\mu g^I\cdot\frac{\delta}{\delta a_\mu}+\sum_a\left(\delta^{\rm W}_1\sM_a+D_a\,\sM_a\right)\cdot\frac{\delta}{\delta\sM_a}\right] ,
\label{gsh:eq:D}
\end{equation}
and on the unit sphere we take $\gamma_{\mu\nu}=\mathring\gamma_{\mu\nu}$ in $\delta^{\rm W}_1\sM_a$. On the unit sphere we define
\begin{equation}
W^\sharp \equiv \vep\,W+\cA[1]-\bD W \, .
\label{gs:eq:sharp}
\end{equation}
In the extended form of the properties $W^\sharp=(\vep+\mathbb{D}_{\rm e})W$, since $\mathbb{W}_1W=\cA[1]$, and $\mathbb{D}_{\rm e}$ is the derivative $r_0\,\pr_{r_0}$ of the first paragraph of this subsection. For the $\phi^4$, the Yukawa, the sextic and the cubic theory, with the relevant sources of even dimension in the divergence scheme of section~\ref{gs:sec:divscheme}, the value of $W^\sharp$ at constant couplings and vanishing relevant sources is $(\vep+r_0\,\pr_{r_0})W$ at $r_0=1$, by Proposition~\ref{gs:prop:radius}. The definition \eqref{gs:eq:sharp} uses the functional on the unit sphere only.

\begin{lemma}
\label{gsh:lem:comm}
Assume Property~\ref{gs:P:weyl}, for a class of sources as in appendix~\ref{gs:sec:class}. On the unit sphere the operator $\mathbb{K}_\sigma$ of the Ward identity of Property~\ref{gs:P:ward} and the operator $\bD$ of \eqref{gsh:eq:D} commute, as vector fields on the space of sources, for every harmonic $\sigma$ of degree one.
\end{lemma}

\begin{proof}
We regard $\mathbb{W}_\sigma$, $\mathbb{D}_{\rm e}$ and the generator $\mathbb{V}_{-v}$ of the diffeomorphism along $-v^\mu$ as vector fields on the space of the metrics $\Omega^2\mathring\gamma_{\mu\nu}$ and of the sources. The field $\mathbb{V}_{-v}$ changes the metric by $-\cL_v\gamma_{\mu\nu}$ and each source by minus its Lie derivative. We show that three commutators vanish. First, $[\mathbb{W}_\sigma,\mathbb{W}_1]=0$ by Property~\ref{gs:P:weyl}. Second, $\mathbb{D}_{\rm e}$ commutes with $\mathbb{W}_\sigma$ and with $\mathbb{V}_{-v}$. Indeed, each term of $\delta^{\rm W}_\sigma\sM_a$ has the dimension of $\sM_a$ at $d=d_c$, the terms with $\hB^I$ and $\hr_I$ do not contain the metric, and $\sigma$ and $v^\mu$ are fixed functions. Third, $[\mathbb{V}_{-v},\mathbb{W}_1]=0$. The coefficients of $\mathbb{W}_1$ are $2\gamma_{\mu\nu}$, $\hB^I(g)$, $\hr_ID_\mu g^I$ and $\delta^{\rm W}_1\sM_a$, which are built covariantly from the metric and the sources, so that their Lie derivative along $v^\mu$ equals their variation under the Lie derivative of the metric and of the sources. With $d$ as a parameter this is an identity of tensor calculus which does not depend on the dimension, since it holds for every integer $d$ larger than the number of embedding coordinates on which the sources depend, as in section~\ref{gsp:sec:facts}. We conclude that $X=\mathbb{W}_\sigma+\mathbb{V}_{-v}$ commutes with $Y=\mathbb{W}_1-\mathbb{D}_{\rm e}$. The component of $Y$ along the metric vanishes identically, and the component of $X$ along the metric is $2\sigma\gamma_{\mu\nu}-\cL_v\gamma_{\mu\nu}$, which vanishes at $\gamma=\mathring\gamma$ by \eqref{gs:eq:sigma}. The component of $[X,Y]$ along the sources at $\mathring\gamma$ therefore contains no derivative with respect to the metric. It equals the commutator of the components of $X$ and $Y$ along the sources at $\mathring\gamma$, which are $\mathbb{K}_\sigma$ and $\bD$.
\end{proof}

We denote the sphere data of the functional $W^\sharp$ by the same letters with a sharp. Thus $\cW^\sharp$ is the value of $W^\sharp$ at constant couplings, $\Lambda^\sharp_{IJ}$ is the first harmonic of its second variation as in \eqref{gs:eq:Lambda}, and $\Pi^\sharp=(d+\hB\!\cdot\!\pr)\,\cW^\sharp/V$ and $\cS^\sharp_{IJ}=\Lambda^\sharp_{IJ}-\pr_I\pr_J\cW^\sharp/V$ as in \eqref{gs:eq:Sdef}. The following proposition gives the Ward identity of $W^\sharp$ and expresses them through the sphere data of $W$.

\begin{proposition}[Homogeneous Ward identity]
\label{gs:prop:sharp}
Assume Properties~\ref{gs:P:weyl}, \ref{gs:P:anomaly} and~\ref{gs:P:ward}, with the relevant sources, for a class of sources as in appendix~\ref{gs:sec:class}. Then the following two statements hold. (i) The functional $W^\sharp$ satisfies
\begin{equation}
\mathbb{K}_\sigma W^\sharp = 0
\label{gsh:eq:hom}
\end{equation}
on the unit sphere, for all sources and every harmonic $\sigma$ of degree one. (ii) Assume in addition Properties~\ref{gs:P:functional} and~\ref{gs:P:sphere}, and let the Weyl variation close on the marginal couplings. Then $W^\sharp$ has the expansion of Property~\ref{gs:P:functional} with $\cW^\sharp-\vep\,\cW^{\log}$ regular, and it satisfies the two statements of Property~\ref{gs:P:sphere} at constant couplings. Its sphere data $\Pi^\sharp$ and $\cS^\sharp_{IJ}$ are given in terms of those of $W$ by
\begin{equation}
\Pi^\sharp = \left(d+\hB\!\cdot\!\pr\right)\left[\frac1V\left(\vep-\hB\!\cdot\!\pr\right)\cW+\rA\right] , \qquad
\cS^\sharp_{IJ} = \big[(\vep-\tL)\,\cS\big]_{IJ}-\cK_{IJ} \, .
\label{gs:eq:sharpquant}
\end{equation}
\end{proposition}

\begin{proof}
We first prove statement~(i), term by term for the three terms of $W^\sharp$. The first term gives $\mathbb{K}_\sigma(\vep W)=\vep\,\cA[\sigma]$ by Property~\ref{gs:P:ward}. For the third term, Property~\ref{gs:P:ward} and Lemma~\ref{gsh:lem:comm} give $\mathbb{K}_\sigma\bD W=\bD\,\mathbb{K}_\sigma W=\bD\cA[\sigma]$. Since $\cA[\sigma]$ is a local functional of the metric, $\bD\cA[\sigma]$ at $\mathring\gamma$ equals $(\mathbb{W}_1-\mathbb{D}_{\rm e})\cA[\sigma]=\mathbb{W}_1\cA[\sigma]+\vep\,\cA[\sigma]$, by the homogeneity. For the second term, $\cA[1]$ is the integral of a local scalar built covariantly from the metric and the sources. Its change under the diffeomorphism along $-v^\mu$, which acts on the metric and on all sources, is the integral of $-\nabla_\mu(v^\mu\ell)$, where the scalar $\ell$ is built from functions of the class by products, derivatives and contractions, and it vanishes by \eqref{gs:eq:parts}. At $\mathring\gamma$ the metric part of this diffeomorphism is $-2\sigma\mathring\gamma_{\mu\nu}$, so that the action of $\mathbb{K}_\sigma$ on $\cA[1]$ equals the action of $\mathbb{W}_\sigma$, that is $\mathbb{K}_\sigma\cA[1]=\mathbb{W}_\sigma\cA[1]$. Altogether
\begin{equation*}
\mathbb{K}_\sigma W^\sharp = \vep\,\cA[\sigma]+\mathbb{W}_\sigma\cA[1]-\mathbb{W}_1\cA[\sigma]-\vep\,\cA[\sigma] \, ,
\end{equation*}
which vanishes by the relation $\mathbb{W}_\sigma\cA[1]=\mathbb{W}_1\cA[\sigma]$ of Property~\ref{gs:P:anomaly}.

We turn to statement~(ii). Without relevant sources $\bD$ acts on the couplings and the gauge field alone. The expansion of Property~\ref{gs:P:functional} holds for each of the three terms of \eqref{gs:eq:sharp}, since the coefficients of $\bD$ are regular by Property~\ref{gs:P:weyl}, those of $\cA[1]$ by Property~\ref{gs:P:anomaly}, and the terms of $\bD W$ are terms of the expansion of $W$ evaluated on products of the sources and their derivatives, which are functions and one-forms of the class of sources. At constant couplings $\cW^\sharp=(\vep-\hB\!\cdot\!\pr)\cW+V\rA$, so that $\Pi^\sharp=(d+\hB\!\cdot\!\pr)\,\cW^\sharp/V$ is the first expression of \eqref{gs:eq:sharpquant}, and $\cW^\sharp-\vep\,\cW^{\log}$ is regular because the first derivatives of $\cW^{\log}$ are.

It remains to show the statements of Property~\ref{gs:P:sphere} for $W^\sharp$ and to compute $\cS^\sharp_{IJ}$. The statements hold for $\vep W$ by this property, and for $\cA[1]$ by Lemma~\ref{gsp:lem:local}. For $\bD W$ the first variations at constant couplings and the second variations in the directions $\sigma\,z^I$ and $\delta g^J$ are computed as in the proof of Lemma~\ref{gs:lem:local}, which for this step uses only the gauge invariance of the functional and Property~\ref{gs:P:sphere}. With $f=\cW/V$ we find $\Lambda^\sharp_{IJ}=[(\vep-\tL)\,\Lambda]_{IJ}-\pr_I\pr_J\hB^K\,\pr_Kf-\pr_Kf\,[(\hr_I)^K{}_J+(\hr_J)^K{}_I]+\alpha_{IJ}$. The second derivative of $\cW^\sharp/V$ is $[(\vep-\tL)\,\pr\pr f]_{IJ}-\pr_I\pr_J\hB^K\,\pr_Kf-\pr_Kf\,[(\hr_I)^K{}_J+(\hr_J)^K{}_I]+\pr_I\pr_J\rA$, as in the last step of that proof. The difference is $\cS^\sharp_{IJ}=[(\vep-\tL)\,\cS]_{IJ}-\cK_{IJ}$.
\end{proof}

The anomaly no longer appears on the right-hand side of the Ward identity. It enters the functional $W^\sharp$ itself, through the term $\cA[1]$. The consistency relation $\mathbb{W}_\sigma\cA[1]=\mathbb{W}_1\cA[\sigma]$ of Property~\ref{gs:P:anomaly} is used in the proof above, and this is the only place where it enters the proof of the gradient formula. The functional $W^\sharp$ and its Ward identity involve the functional $W$ on the unit sphere only. The interpretation of $W^\sharp$ as $(\vep+r_0\,\pr_{r_0})W$ refers to spheres of other radius, but the proof does not use it, so that the properties are needed on the unit sphere alone.

\subsection{Gradient formula}
\label{gs:sec:gradient}

\begin{proposition}[Gradient formula]
\label{gs:prop:main}
Assume Properties~\ref{gs:P:functional} to~\ref{gs:P:sphere}, and let the Weyl variation close on the marginal couplings. Then
\begin{equation}
\pr_J\hat A^{\rm S} = \hat G^{\rm S}_{JI}\,\hB^I \, , \qquad
\hat A^{\rm S} = \left(d+\hB\!\cdot\!\pr\right)\left[\frac1V\left(\vep-\hB\!\cdot\!\pr\right)\cW+\rA\right] , \qquad
\hat G^{\rm S}_{JI} = \cK_{JI}-\big[(\vep-\tL)\,\cS\big]_{JI} \, .
\label{gs:eq:main}
\end{equation}
The tensor $\hat G^{\rm S}_{JI}$ is regular and symmetric. The function $\hat A^{\rm S}-(\vep\,d/V)\,\cW^{\log}$ is regular, so that $\hat A^{\rm S}$ is regular in instance~(a) of Property~\ref{gs:P:functional}. The function $\hat A^{\rm S}$ is invariant under the constant rotations of the identity component of the group of field rotations, and $\hat G^{\rm S}_{JI}$ is covariant under its Lie algebra. At a zero $g_*$ of $\hB^I$
\begin{equation}
\hat A^{\rm S}(g_*) = \frac{\vep\,d}{V}\;\cW_* \, , \qquad \cW_* \equiv \cW(g_*)+\frac{V}{\vep}\,\rA(g_*) \, .
\label{gs:eq:fixed}
\end{equation}
\end{proposition}

\begin{proof}
By Proposition~\ref{gs:prop:sharp} the functional $W^\sharp$ satisfies the Ward identity of Property~\ref{gs:P:ward} with a vanishing right-hand side, and it has the expansion of Property~\ref{gs:P:functional} and the statements of Property~\ref{gs:P:sphere} at constant couplings. These are the statements which the proof of Proposition~\ref{gs:prop:K} uses. Applied to $W^\sharp$ and $\cA^\sharp=0$ this proof gives $0=\pr_J\Pi^\sharp+\cS^\sharp_{JI}\hB^I$, which is \eqref{gs:eq:main} by \eqref{gs:eq:sharpquant}. The tensor $\hat G^{\rm S}_{JI}$ is regular because $\pr_I\cW$, $\Lambda_{IJ}$, $\rA$, $\alpha_{IJ}$, $\hB^I$ and $\hr_I$ are. For the potential we use $(d+\hB\!\cdot\!\pr)(\vep-\hB\!\cdot\!\pr)\,\cW=\vep\,d\,\cW+(\vep-d-\hB\!\cdot\!\pr)\,\hB^I\pr_I\cW$, where $\hB^I\pr_I\cW$ is regular, and the form of $\cW$ in Property~\ref{gs:P:functional}. The tensor $\hat G^{\rm S}_{JI}$ is symmetric because $\cK_{JI}$ and $\cS_{JI}$ are, and the extended Lie derivative of a symmetric tensor is symmetric. The invariance and the covariance follow from those of $\cW$, $\rA$, $\Lambda_{IJ}$, $\alpha_{IJ}$, $\hB^I$ and $\hr_I$. At a zero $g_*$ the function $\hB^I\pr_I\cW$ and all terms in which $\hB\!\cdot\!\pr$ acts on a regular function vanish, which gives \eqref{gs:eq:fixed}.
\end{proof}

Let us comment on the result. First, in four dimensions the corresponding relation of~\cite{Osborn:1991gm,Jack:1990eb,Jack:2013sha} contains an antisymmetric tensor, the curl of a one-form $W_I$, and the proof of~\cite{Reehorst:phi4} shows that it can be removed. In \eqref{gs:eq:main} no antisymmetric tensor appears, because the Ward identity of $W^\sharp$ has no anomaly, so that its Killing identity contains no local one-form, and the tensor $\cS^\sharp_{JI}$ of a second variation is symmetric. In contrast with the coefficients of the relation in four dimensions, which are local tensors of the anomaly, the metric $\hat G^{\rm S}_{JI}$ contains the first harmonic of the two-point function on the sphere.

Second, the identity $\pr_J\hat A^{\rm S}=\hat G^{\rm S}_{JI}\hB^I$ implies $\hB\!\cdot\!\pr\,\hat A^{\rm S}=\hB^J\hat G^{\rm S}_{JI}\hB^I$, and at a zero of $\hB^I$ it implies that the matrix $\hat G^{\rm S}_{JI}\,\pr_K\hB^I$ is symmetric. On a manifold of zeros of $\hB^I$ the formula states that $\hat A^{\rm S}$ is constant.

Third, the solution of the gradient equation is not unique, and \eqref{gs:eq:main} is one solution. Let $\pr_IA=g_{IJ}\hB^J$ hold with a regular function $A$ and a regular symmetric tensor $g_{IJ}$. For a power series $f$ in one variable whose coefficients are regular at $\vep=0$ the pair $f(A)$, $f'(A)\,g_{IJ}$ is again a regular solution. Here and below such a series is composed with $A$ minus its term of order zero in the couplings, so that the composition is defined. If $\varphi$ is a power series of this kind, then
\begin{equation}
A+\varphi(A)\,\hB^Ig_{IJ}\hB^J \, , \qquad
g_{IJ}+\varphi'(A)\,\big(\hB^Kg_{KL}\hB^L\big)\,g_{IJ}+\varphi(A)\,\big(\cL_{\hB}g\big)_{IJ}
\label{eq:freedom}
\end{equation}
is again a regular solution with a symmetric metric, as one verifies with $\hB^Ig_{IJ}\hB^J=\hB^I\pr_IA$ and $\pr_I(\hB^K\pr_KA)=(\cL_{\hB}g)_{IJ}\hB^J$, where $\cL_{\hB}$ is the Lie derivative along $\hB^I$. The freedom is larger than this family. For a regular symmetric tensor $\mathsf{t}_{IJK}$ the pair $A+\mathsf{t}_{IJK}\hB^I\hB^J\hB^K$, $g_{IJ}+(\mathsf{m}_{IJK}+\mathsf{m}_{JIK}-\mathsf{m}_{KIJ})\hB^K$ with $\mathsf{m}_{IJK}=\pr_I\mathsf{t}_{LJK}\,\hB^L+3\,\pr_I\hB^L\,\mathsf{t}_{LJK}$ in the coordinates $g^I$ is again a regular solution with a symmetric metric, with the same potential and the same metric at the zeros of $\hB^I$. Indeed, $\mathsf{m}_{IJK}$ is symmetric in $J$ and $K$, and $\mathsf{m}_{IJK}\hB^J\hB^K$ is the derivative of the shift. The expression for $\mathsf{m}_{IJK}$ contains partial derivatives and is not a tensor. In other coordinates it gives a metric which differs by a symmetric tensor whose contraction with $\hB^I$ vanishes, and which is again a solution. A shift $A\to A+\hB^I\mathsf{k}_{IJ}\hB^J$ with an arbitrary regular symmetric tensor $\mathsf{k}_{IJ}$ does not give a solution with a symmetric metric in general.\footnote{A symmetric tensor $\mathsf{S}_{IJ}$ with $\pr_I(\hB^K\mathsf{k}_{KL}\hB^L)=\mathsf{S}_{IJ}\hB^J$ must satisfy $\mathsf{S}_{IK}\,\pr_J\hB^K=2\,\pr_I\hB^K\,\mathsf{k}_{KL}\,\pr_J\hB^L$ at a zero of $\hB^I$, as one sees by taking one more derivative. For the vector field $(c_1\,g^1,c_2\,g^2)$ with two couplings, nonzero constants $c_1$ and $c_2$, and a constant $\mathsf{k}_{IJ}$ with $\mathsf{k}_{12}=\mathsf{k}_{21}=1$ and $\mathsf{k}_{11}=\mathsf{k}_{22}=0$ the two components of this condition give $\mathsf{S}_{12}=2\,c_1$ and $\mathsf{S}_{21}=2\,c_2$ at the origin, so that there is no regular symmetric $\mathsf{S}_{IJ}$ for $c_1\neq c_2$. If all weights $k_I$ are equal, the matrix $\pr_J\hB^I$ is a multiple of the unit matrix at $g=0$, and the obstruction is not visible at the lowest orders.}

Fourth, in instance~(a) of Property~\ref{gs:P:functional} the first five properties do not imply that the metric is different from zero. For a local functional $W=F$ with regular coefficients and $\cA[\sigma]=\mathbb{W}_\sigma F$ they hold, and both $\hat A^{\rm S}$ and $\hat G^{\rm S}_{JI}$ vanish, see Lemma~\ref{gs:lem:local}. The content of Proposition~\ref{gs:prop:main} thus lies in the part of $W$ which is not local, and Property~\ref{gs:P:free} fixes it at lowest order, see section~\ref{gs:sec:lowest}. In instance~(b) the gradient formula itself fixes the lowest order of the metric along the logarithmic coordinates, see section~\ref{gs:sec:log}.

Fifth, in minimal subtraction the counterterms drop out of $W^\sharp$. There $W$ is the sum of a bare functional, whose Ward identity has no anomaly, and of the counterterms, a local functional $F$ of Weyl weight $-\vep$, and the anomaly is $\cA[\sigma]=\mathbb{W}_\sigma F$. The combination \eqref{gs:eq:sharp} is linear in the functional together with its anomaly. For $F$ with the anomaly $\mathbb{W}_\sigma F$ its three terms are $\vep\,F$, $\mathbb{W}_1F$ and $-\bD F$. Their sum vanishes, since $\mathbb{W}_1F=-\vep\,F+\bD F$ by the homogeneity. The functional $W^\sharp$ is therefore the same combination of the bare functional alone. Accordingly the pair \eqref{gs:eq:main} is the image under $\vep-\tL$ of the Killing identity of the bare functional. The coefficients of that identity have poles at $\vep=0$, and the content of Proposition~\ref{gs:prop:main} is that the image is regular. In particular, by Lemma~\ref{gs:lem:local} the potential $\hat A^{\rm S}$ and the metric $\hat G^{\rm S}_{JI}$ do not change if a finite local functional $F$ of this kind is added to $W$ and $\mathbb{W}_\sigma F$ to the anomaly, with $\hB^I$ unchanged.

\subsection{Value at fixed points}
\label{gs:sec:fixed}

Properties~\ref{gs:P:functional} to~\ref{gs:P:sphere} on the unit sphere give the value \eqref{gs:eq:fixed}. In the extended form of the properties of appendix~\ref{gs:sec:properties} the quantity $\cW_*$ is the free energy on the unit sphere with the couplings of the curvature terms at the zeros of their beta functions, in the sense of~\cite{Fei:2015oha}. To see this let $\mathcal{I}_a$ be the local curvature scalars of dimension $d_c$ on conformally flat backgrounds, and add $\sum_a\varsigma_a\int d^dx\sqrt\gamma\,\mathcal{I}_a$ with constants $\varsigma_a$ to $W$. At constant couplings $\cA[1]=\sum_ap_a\int d^dx\sqrt\gamma\,\mathcal{I}_a$ with regular functions $p_a(g)$, so that for $\varsigma_a=p_a(g_*)/\vep$ the functional with these terms is $\cW(g_*)+V\rA(g_*)/\vep$ on the unit sphere at the constant couplings $g_*$, which is $\cW_*$. In the extended form of the properties and at a zero $g_*$ of $\hB^I$ a constant rescaling of the metric is compensated by the change $\delta\varsigma_a=\vep\,\varsigma_a-p_a(g_*)$ of these constants, by the homogeneity, and $\varsigma_a=p_a(g_*)/\vep$ are the zeros of this change. This step uses the response of $W$ to a constant rescaling, which the Ward identity of Property~\ref{gs:P:ward} does not contain, since the integrals of its harmonics $\sigma$ against the constant densities $\mathcal{I}_a$ of the unit sphere vanish. For the $\phi^4$, the Yukawa, the sextic and the cubic theory this response is derived from Hypothesis~\ref{hyp:finite} in Proposition~\ref{gs:prop:radius}, with the relevant sources of even dimension in the divergence scheme of section~\ref{gs:sec:divscheme} and the functions $p_a$ of this scheme. For these theories $\varsigma_a=p_a(g_*)/\vep$ are the zeros of the beta functions of the curvature couplings. Let $\tilde F=\sin(\pi d/2)\log Z_{S^d}$ be the quantity of~\cite{Giombi:2014xxa}, defined from the partition function on the unit sphere with the curvature couplings $\varsigma_a=p_a(g_*)/\vep$, which in the extended form of the properties and for these four theories are those of~\cite{Fei:2015oha}. In terms of $\tilde F$ the value \eqref{gs:eq:fixed} reads
\begin{equation}
\hat A^{\rm S}(g_*) = \frac{\vep\,d}{V\sin(\pi d/2)}\left(\tilde F-\tilde F_0\right) ,
\label{gs:eq:Ftilde}
\end{equation}
where $\tilde F_0=\sin(\pi d/2)\log Z_0$ with the constant $Z_0$ of section~\ref{gs:sec:functional}. In instance~(a) of Property~\ref{gs:P:functional} the free theory is the point $g=0$ of the space of couplings. We then choose $Z_0$ such that $W$ vanishes for vanishing sources. If $\rA$ vanishes at $g=0$, as happens if no curvature counterterm is independent of the couplings, $\tilde F_0$ is the value $\tilde F_{\rm free}$ of the free theory. Otherwise the two differ by a constant, which is the same for all zeros of $\hB^I$. In instance~(b) this normalisation is not available, because the free theory is not a point of the space of couplings and $\cW$ contains the function $\cW^{\log}$, which is not bounded for $g_\iota\to0$. The constant $Z_0$ is then restricted only by the requirement of Property~\ref{gs:P:functional} that $\cW-\cW^{\log}$ is regular, and $\tilde F_0$ is a constant which we do not fix. For even $d_c$ the factor $\vep\,d/(V\sin(\pi d/2))$ in front of $\tilde F-\tilde F_0$ in \eqref{gs:eq:Ftilde} has a finite limit at $\vep=0$ with the sign $(-1)^{d_c/2+1}$, and for odd $d_c$ it is of order $\vep$. We denote by $s_d$ the sign of $\sin(\pi d/2)$ for small positive $\vep$. For even $d_c$ the function $s_d\,\hat A^{\rm S}$ is then a positive multiple of $\tilde F-\tilde F_0$ at the zeros of $\hB^I$, for small $\vep$ of either sign, since the factor $\vep/\sin(\pi d/2)$ is even in $\vep$. For odd $d_c$ this multiple has the sign of $\vep$, and for either sign of $\vep$ the potential $s_d\,\Pi$ of Proposition~\ref{gs:prop:odd} is a positive multiple of $\tilde F-\tilde F_0$ at the zeros. Note that in instance~(a) $\cW_*$ is a power series in $\vep$ for a zero $g_*$ which is a power series in $\vep$ without a term of order zero, provided that $\rA$ vanishes at $g=0$. In instance~(b) $\cW_*$ contains $\sum_\iota n_\iota\log g_{\iota*}$, and it has a pole at $\vep=0$ if the term of order zero of $\rA$ does not vanish at $\vep=0$. At a zero with $g_{\iota*}^2$ of order $\vep$ the sum contains $\tfrac12\,n_\iota\log|\vep|$, and the value of the potential is not a power series in $\vep$.

\subsection{Lowest order}
\label{gs:sec:lowest}

We first consider the sphere data of a local functional.

\begin{lemma}
\label{gs:lem:local}
Let $F$ be a local functional of the metric, of $g^I$ and of $a_\mu$ which is built covariantly from $\gamma_{\mu\nu}$, invariant under background gauge transformations and homogeneous of Weyl weight $-\vep$, and let Property~\ref{gs:P:weyl} hold without sources $\sM$. For the pair $W=F$ and $\cA[\sigma]=\mathbb{W}_\sigma F$ the statements of Property~\ref{gs:P:sphere} follow from Lemma~\ref{gsp:lem:local}, and we have $\rA=-V^{-1}(\vep-\hB\!\cdot\!\pr)\,\cW$ and $\cK_{IJ}=[(\vep-\tL)\,\cS]_{IJ}$, so that $\hat A^{\rm S}$ and $\hat G^{\rm S}_{IJ}$ vanish.
\end{lemma}

\begin{proof}
Let $f=\cW/V$, and let $\mathbb{D}$ denote the terms of the operator $\mathbb{W}_\sigma$ of Property~\ref{gs:P:weyl} with $\hB^I$ and with $\hr_I$ for $\sigma=1$. By the homogeneity $\cA[1]=-\vep\,F+\mathbb{D}F$. At constant couplings the term with $\hr_I$ vanishes, which gives the first statement. For the second we compute the second variation of $\mathbb{D}F$ in the directions $\sigma\,z^I$ and $\delta g^J$, as in \eqref{gs:eq:adef}. The term with $\hB^I$ gives $(\cL_{\hB}\Lambda)_{IJ}+\pr_I\pr_J\hB^K\pr_Kf$ times $z^I\int d\mu_d\,\sigma\,\delta g^J$, where $\cL_{\hB}$ is the Lie derivative. The three terms of the Lie derivative come from the derivative of $\Lambda_{IJ}$ along $\hB^I$ and from the first variation of $\hB^I$ in either direction, and the last term comes from the second variation of $\hB^I$. For the term with $\hr_I$ we use the gauge invariance of $F$ with the parameter $\omega=\hr_K\,\delta g^K$, where $\hr_K$ is taken at the constant couplings. Up to terms of third order it gives
\begin{equation}
\int d\mu_d\;\hr_K\pr_\mu\delta g^K\cdot\frac{\delta F}{\delta a_\mu} = \int d\mu_d\;(\omega g)^K\,\frac{\delta F}{\delta g^K} \, ,
\end{equation}
where $g=g_c+\delta g$ on the right-hand side. The term of first order of the right-hand side vanishes because $f$ is invariant. Its term of second order is the second variation of $F$ in the directions $(\omega g_c)^K$ and $\delta g^J$, plus $\pr_Kf\int d\mu_d\,(\omega\,\delta g)^K$. For the directions $\sigma\,z^I$ and $\delta g^J$ this gives $(\hr_Ig)^K\Lambda_{KJ}+\Lambda_{IK}(\hr_Jg)^K+\pr_Kf\,[(\hr_I)^K{}_J+(\hr_J)^K{}_I]$ times $z^I\int d\mu_d\,\sigma\,\delta g^J$, where we write $(\omega g)^K=\omega^K{}_Jg^J$. Hence
\begin{equation}
\alpha_{IJ} = -\big[(\vep-\tL)\,\Lambda\big]_{IJ}+\pr_I\pr_J\hB^K\,\pr_Kf+\pr_Kf\left[(\hr_I)^K{}_J+(\hr_J)^K{}_I\right] .
\label{gs:eq:alphalocal}
\end{equation}
From the first statement we have $\pr_I\pr_J\rA=-\vep\,\pr_I\pr_Jf+(\cL_{\hB}\,\pr\pr f)_{IJ}+\pr_I\pr_J\hB^K\,\pr_Kf$. The invariance of $f$ gives $(\omega g)^K\pr_K\pr_Jf=-\omega^K{}_J\,\pr_Kf$ for a constant $\omega$, so that $(\cL_{\hB}\,\pr\pr f)_{IJ}-\pr_Kf\,[(\hr_I)^K{}_J+(\hr_J)^K{}_I]=(\tL\,\pr\pr f)_{IJ}$. The difference $\pr_I\pr_J\rA-\alpha_{IJ}$ is therefore $[(\vep-\tL)(\Lambda-\pr\pr f)]_{IJ}$, which is the second statement.
\end{proof}

Such a pair satisfies Properties~\ref{gs:P:functional} to~\ref{gs:P:sphere} if the coefficients of $F$ are regular. Indeed, the Ward identity of Property~\ref{gs:P:ward} holds because $F$ is invariant under the diffeomorphism generated by $v^\mu$, the relation $\mathbb{W}_\sigma\cA[\sigma']=\mathbb{W}_{\sigma'}\cA[\sigma]$ holds because the operators $\mathbb{W}_\sigma$ of Property~\ref{gs:P:weyl} commute, and Property~\ref{gs:P:sphere} follows for local functionals by integrations by parts. The quantities \eqref{gs:eq:Sdef}, \eqref{gs:eq:adef} and \eqref{gs:eq:Kdef} are linear in the pair $W$, $\cA[\sigma]$. It follows that $\hat A^{\rm S}$ and $\hat G^{\rm S}_{IJ}$ do not change if a functional $F$ of this kind is added to $W$ and $\mathbb{W}_\sigma F$ is added to $\cA[\sigma]$. In particular they do not depend on the finite parts of the curvature counterterms which do not contain relevant sources.

\begin{proposition}[Lowest order]
\label{gs:prop:lowest}
Assume Properties~\ref{gs:P:functional} to~\ref{gs:P:sphere} with Property~\ref{gs:P:functional} in instance~(a), and the two-point form of the functional defined after Property~\ref{gs:P:free}. The Weyl variation need not close on the marginal couplings, since $\hat G^{\rm S}$ is built from $\cS$ and $\cK$ at vanishing relevant sources. On the couplings of weight $k$ the term of order zero in the couplings of $\hat G^{\rm S}_{IJ}$ is
\begin{equation}
\hat G^{{\rm S}(0)}_{IJ} = -\vep\,(1+2k)\left(\mu_1-\mu_0\right)c_{IJ}(\vep) \, , \qquad
\mu_0 = \frac{2^{d-2\Delta}\,\pi^{d/2}\,\Gamma(\tfrac d2-\Delta)}{\Gamma(d-\Delta)} \, , \qquad \mu_1 = \frac{\Delta}{d-\Delta}\;\mu_0 \, ,
\label{gs:eq:G0}
\end{equation}
where $\Delta=d-\vep\,k$ and $c_{IJ}(\vep)$ is the matrix of the two-point form, and it vanishes between couplings of different weights. Moreover
\begin{equation}
s_d\,\hat G^{{\rm S}(0)}_{IJ} = \frac{\vep\,(1+2k)\,(2\Delta-d)\,2^{d-2\Delta}\,\pi^{d/2+1}}{\vartheta_d\;\Gamma(1+\tfrac d2-\vep k)\,\Gamma(1+\vep k)}\;c_{IJ}(\vep) \, ,
\label{gs:eq:G0sign}
\end{equation}
where $\vartheta_d=\sin(\pi\vep(k+\tfrac12))$ for even $d_c$ and $\vartheta_d=\cos(\pi\vep(k+\tfrac12))$ for odd $d_c$. For even $d_c$ the factor which multiplies $c_{IJ}(\vep)$ in \eqref{gs:eq:G0sign} tends to $2\,d_c\,\pi^{d_c/2}/(2^{d_c}\,(d_c/2)!)$ for $\vep\to0$, and $c_{IJ}(\vep)$ tends to $c_{IJ}$.
\end{proposition}

\begin{proof}
At $g=0$ the matrix $\hat M^K{}_I$ is $-\vep\,k_I\,\delta_I{}^K$ by the form of $\hB^I$, since a covariant one-form $\hr_I$ which multiplies $g$ does not contribute at this order. Hence $\vep-\tL$ multiplies the component $IJ$ of a constant tensor by $\vep\,(1+k_I+k_J)$. By the two-point form the pair $W$, $\cA[\sigma]$ is at second order the sum of the pair $F$, $\mathbb{W}_\sigma F$ and of the double integral of the two-point form with a vanishing anomaly. The first pair does not contribute to $\hat G^{\rm S}_{IJ}$ by Lemma~\ref{gs:lem:local}, whose proof applies at this order. For the second pair $\cK_{IJ}=0$ and $\cS_{IJ}=c_{IJ}(\vep)\,(\mu_1-\mu_0)$, where $\mu_0$ and $\mu_1$ are the eigenvalues of the kernel $s^{-2\Delta}$ on the constants and on the harmonics of degree one, with the chordal distance $s(X,Y)=|X-Y|$, the distance between the two points in the embedding space of the unit sphere. They are given by the integrals of $s^{-2\Delta}$ and of $s^{-2\Delta}\cos\theta$ over the sphere, continued in $\Delta$. Since $c_{IJ}(\vep)$ vanishes for $k_I\neq k_J$, this is \eqref{gs:eq:G0}. The form \eqref{gs:eq:G0sign} follows with $\mu_1-\mu_0=\mu_0\,(2\Delta-d)/(d-\Delta)$ and with the reflection formula for $\Gamma(\tfrac d2-\Delta)$.
\end{proof}

For even $d_c$ the factor which multiplies $c_{IJ}(\vep)$ in \eqref{gs:eq:G0sign} is positive for small $\vep$ of either sign, since $\vep/\vartheta_d$ is even in $\vep$. For odd $d_c$ it is $\vep\,(1+2k)$ times a positive function, and the quantity which is positive for either sign is $-s_d\,\cS^{(0)}$ of Proposition~\ref{gs:prop:odd}. For even $d_c$ we thus find the following. The potential $s_d\,\hat A^{\rm S}$, which at the zeros of $\hB^I$ is a positive multiple of $\tilde F-\tilde F_0$, has the metric $s_d\,\hat G^{\rm S}_{IJ}$. If the matrix $c_{IJ}$ is positive, this metric is positive definite to all orders in the couplings and in $\vep$, in the sense of section~\ref{gs:sec:intro}. The matrix $c_{IJ}$ is a Gram matrix of the free theory in $d_c$ dimensions, and it is positive if the two-point functions of the operators $\cO_I$ are reflection positive and these operators are linearly independent at $d=d_c$. For multiscalar $\phi^4$ theory \eqref{gs:eq:G0} gives the term of order zero of the metric of Theorem~\ref{gs:thm:phi4}, which at $\vep=0$ is $\tfrac{1}{96}\pi^2c_\cO\,\delta_{IJ}$, see the proof of that theorem. Note that the difference $\mu_1-\mu_0$ has a pole at $\vep=0$ for even $d_c$, which the factor $\vep$ cancels. For odd $d_c$ there is no pole, and $\hat G^{{\rm S}(0)}_{IJ}$ is of order $\vep$.

The proposition uses the two-point form for all $\vep$. The statements on nondegeneracy and positivity use its value at $\vep=0$ only, which Property~\ref{gs:P:free} fixes, by the following lemma. We write $\cT^{(0,0)}$ for the term of order zero in the couplings and in $\vep$ of a regular tensor $\cT$, in the coordinates of Property~\ref{gs:P:functional}.

\begin{lemma}
\label{gre:lem:zero}
Assume Properties~\ref{gs:P:functional} to~\ref{gs:P:anomaly} and~\ref{gs:P:sphere} in either instance, with $\hB^\iota=-\vep\,k_\iota+B^\iota$ on the logarithmic coordinates and $\hB^I=-\vep\,k_Ig^I+B^I$ on the others, where $B^\iota$ and $B^I$ have no terms of order zero and one in the couplings. Then the term of order zero in the couplings of $[(\vep-\tL)\cS]_{IJ}$ is $\vep\,(1+k_I+k_J)\,\cS^{(0)}_{IJ}$, where $k_I$ is replaced by zero for a logarithmic index. Consequently
\begin{equation}
\hat G^{{\rm S}(0,0)}_{IJ} = \cK^{(0,0)}_{IJ} \, .
\label{gre:eq:zero}
\end{equation}
\end{lemma}

\begin{proof}
On a tensor with two lower indices $\tL\cS=\hB^K\pr_K\cS+\hat M^T\cS+\cS\hat M$ with $\hat M^K{}_I=\pr_I\hB^K+(\hr_Ig)^K$. At order zero in the couplings the matrix $\hat M$ is diagonal, with the entries $-\vep k_I$ on the coordinates which are not logarithmic and zero on the logarithmic ones, since $(\hr_Ig)^K$ contains a factor of the couplings, the term $B^K$ of $\hB^K$ has no terms of order zero and one, and $\pr_\iota=g_\iota\pr/\pr g_\iota$ does not lower the order. The transport $\hB^K\pr_K\cS$ has no term of order zero in the couplings. Indeed, on a coordinate which is not logarithmic $\hB^K=-\vep k_Kg^K+B^K$, and $g^K\pr_K$ does not lower the order, while $B^K\pr_K$ raises it. On a logarithmic coordinate $\hB^\iota\pr_\iota\cS$ contains the factor $g_\iota\pr/\pr g_\iota$ acting on a regular tensor, which has no term of order zero. This gives the first statement. The tensor $\cS_{IJ}$ is regular by Property~\ref{gs:P:functional}, as stated in section~\ref{gs:sec:quantities}, so that its term of order zero has no pole, and $\vep\,(1+k_I+k_J)\,\cS^{(0)}_{IJ}$ vanishes at $\vep=0$. With $\hat G^{\rm S}=\cK-(\vep-\tL)\cS$ from \eqref{gs:eq:main} we obtain \eqref{gre:eq:zero}.
\end{proof}

The lemma uses Properties~\ref{gs:P:functional} to~\ref{gs:P:anomaly} and~\ref{gs:P:sphere} through the definition and the regularity of $\cS_{IJ}$ and $\cK_{IJ}$ only, and it holds whether or not the Weyl variation closes on the marginal couplings, since $\cS$ and $\cK$ are defined at vanishing relevant sources. In the gradient formula of Theorem~\ref{gr:cor:cond} with relevant sources the metric is $\hat G^{\rm H}$, whose term of order zero is $\cK^{(0,0)}$ as well under the last statement of Property~\ref{gs:P:free}. For odd $d_c$ the anomaly vanishes at vanishing relevant sources by Lemma~\ref{gs:lem:odd}, the metric is the tensor $-\cS'$ of the form of this theorem for odd $d_c$ in section~\ref{gs:sec:gradrel}, and its term of order zero is $-\cS^{(0,0)}$, since the remaining terms of $\cS'$ contain a factor $\hB^K$. In both cases the term of order zero in the couplings and in $\vep$ of the metric is a datum at $\vep=0$, and Property~\ref{gs:P:free} gives it. For even $d_c$ the term of order zero of the metric $s_d\hat G^{\rm S}$, and of $s_d\hat G^{\rm H}$ under the last statement of Property~\ref{gs:P:free}, is then $a_{d_c}\,c_{IJ}$, since $s_d=(-1)^{d_c/2+1}$. For odd $d_c$ that of $-s_d\cS'$ is $-s_d\,b_{d_c}\,c_{IJ}$, which is a positive multiple of $c_{IJ}$. If the functional has the two-point form, Proposition~\ref{gs:prop:lowest} and the lemma give $\cK^{(0,0)}=s_d\,a_{d_c}\,c_{IJ}$, which is the statement of Property~\ref{gs:P:free} on $\cK$ for even $d_c$. The pole at $\vep=0$ of the two-point term and its cancellation against the local terms, which enter the proof of the proposition, do not enter the lemma.

We turn to odd $d_c$ now.

\subsection{Odd critical dimension}
\label{gs:sec:odd}

\begin{lemma}
\label{gs:lem:odd}
Assume Property~\ref{gs:P:anomaly}, and let $d_c$ be odd. Then $\cA[\sigma]$ vanishes at $\sM=0$.
\end{lemma}

\begin{proof}
By the homogeneity each term of $\cA[\sigma]$ at $\sM=0$ contains $d_c$ derivatives, where the curvature counts as two derivatives and the gauge field $a_\mu$ as one. A term which is built from the metric, the curvature, the field strength of $a_\mu$ and covariant derivatives has as many indices as derivatives modulo two. By Property~\ref{gs:P:anomaly} the Levi-Civita tensor does not occur. The indices are therefore contracted in pairs with the metric, so that the number of derivatives is even.
\end{proof}

The lemma uses that the anomaly is built with $d$ as a parameter, so that the Levi-Civita tensor does not occur. In three dimensions the anomaly at vanishing relevant sources consists of two terms with this tensor~\cite{Nakayama:2013wda}, which are odd under parity. They are excluded by Property~\ref{gs:P:anomaly}. Since they contain three derivatives of the couplings or the field strength of $a_\mu$, they do not contribute to the sphere data of section~\ref{gs:sec:quantities}, and Proposition~\ref{gs:prop:odd} below would hold with them. The obstacle for theories which are not invariant under parity is the continuation to non-integer $d$, see section~\ref{gs:sec:outside}.

For odd $d_c$ and at $\sM=0$ we therefore have $\sfa_J=0$, $\rA=0$ and $\cK_{IJ}=0$, and the Killing identity of $W$ itself, without the combination $W^\sharp$, is already a gradient formula.

\begin{proposition}[Odd critical dimension]
\label{gs:prop:odd}
Assume Properties~\ref{gs:P:functional} to~\ref{gs:P:sphere}, let the Weyl variation close on the marginal couplings, and let $d_c$ be odd. Then
\begin{equation}
\pr_J\Pi = -\cS_{JI}\,\hB^I \, , \qquad \Pi = \frac1V\left(d+\hB\!\cdot\!\pr\right)\cW \, ,
\label{gs:eq:odd}
\end{equation}
where the tensor $\cS_{JI}$ is regular, symmetric and covariant under the Lie algebra of the group of field rotations, and the function $\Pi$ is invariant under the constant rotations of the identity component of this group, with $\Pi-d\,\cW^{\log}/V$ regular. At a zero $g_*$ of $\hB^I$ we have $\Pi(g_*)=d\,\cW(g_*)/V$. Under Property~\ref{gs:P:free} the term of order zero in the couplings and in $\vep$ of $-s_d\,\cS_{IJ}$ is $-s_d\,b_{d_c}\,c_{IJ}$. If in addition the functional has the two-point form, the term of order zero in the couplings of $-s_d\,\cS_{IJ}$ is \eqref{gs:eq:G0sign} divided by $\vep\,(1+2k)$, which is a positive multiple of $c_{IJ}(\vep)$ for every small $\vep$.
\end{proposition}

\begin{proof}
The relation \eqref{gs:eq:odd} is \eqref{gs:eq:K} with $\sfa_J=0$, which holds by Lemma~\ref{gs:lem:odd}. The statements on regularity hold by Property~\ref{gs:P:functional}, see section~\ref{gs:sec:quantities}. For the last statement we use the proof of Proposition~\ref{gs:prop:lowest}. The local functional $F$ of the two-point form is homogeneous of Weyl weight $-\vep$ and therefore vanishes for odd $d_c$, by the argument of Lemma~\ref{gs:lem:odd}. The term of order zero of $\cS_{IJ}$ is thus $c_{IJ}(\vep)\,(\mu_1-\mu_0)$, and the statement follows from \eqref{gs:eq:G0} and \eqref{gs:eq:G0sign}.
\end{proof}

Among the theories of section~\ref{gs:sec:other} the only one with odd $d_c$ is the sextic theory. Its Weyl variation does not close on the marginal couplings, since a scalar field brings the source of the operators $\phi_i\phi_j$, and we use the proposition there through its extension, Proposition~\ref{gs:prop:sextic}.

The potential $\Pi$ of \eqref{gs:eq:odd} is $(d+\hB\!\cdot\!\pr)/V$ applied to $\cW$, and $-\cW$ is the free energy on the unit sphere. It equals a multiple of the free energy at the zeros of $\hB^I$ only, and there are no curvature couplings of dimension $d_c$, so that $\cW_*=\cW(g_*)$. For $d_c=3$ we have $s_d=-1$, and the metric of the potential $-\Pi$ is $\cS_{IJ}$, whose term of order zero is $\tfrac12\,\pi^2c_{IJ}$ at $\vep=0$ under Property~\ref{gs:P:free}. At lowest order \eqref{gs:eq:odd} is the relation between the derivative of the free energy and the beta function which was found in conformal perturbation theory in~\cite{Klebanov:2011gs}, where the matrix which multiplies the beta function is regarded as a metric on the space of couplings.\footnote{With Property~\ref{gs:P:free} the value of $V\cS^{(0)}_{IJ}/d$ at $d=3$ and $d=5$ agrees with the coefficient of~\cite{Klebanov:2011gs}.} The relation \eqref{gs:eq:odd} states that it holds to all orders with the potential $\Pi$ and the metric $-\cS_{IJ}$, under the stated assumptions.

For odd $d_c$ there are thus two gradient formulas, \eqref{gs:eq:odd} and the general formula \eqref{gs:eq:main}, which is the image of \eqref{gs:eq:odd} under $\vep-\tL$. We use \eqref{gs:eq:odd}, and we make a statement on the sign for this form only. The metric $-[(\vep-\tL)\,\cS]_{IJ}$ of the general formula is of order $\vep$ at order zero, and its sign can differ between the free theory and a fixed point.\footnote{For one coupling with $\hB=-\vep\,k\,g+b_1\,g^2$ it equals $-\vep\,(1+2k)\,\cS$ at $g=0$ and $\vep\,(2k-1)\,\cS$ at the zero $g_*=\vep\,k/b_1$, where $\cS$ is taken at the respective point and changes by a relative amount of order $\vep$ between the two. The two values have opposite signs for $k>\tfrac12$.} Under the assumptions of Proposition~\ref{gs:prop:odd} the function $\cW$ itself is also the potential of a regular symmetric tensor. Let $\mathcal{L}_{\hB}$ be the Lie derivative along $\hB^I$ on the space of couplings, which commutes with the derivative $\pr_J$ of a function and with the contraction of a tensor with $\hB^I$. The derivative of the second relation of \eqref{gs:eq:odd} is $V\pr_J\Pi=[(d+\mathcal{L}_{\hB})\,\pr\cW]_J$, and with the first relation we find
\begin{equation}
\pr_J\cW = \mathsf{H}_{JK}\,\hB^K \, , \qquad \mathsf{H}_{JK} = -V\left[(d+\mathcal{L}_{\hB})^{-1}\cS\right]_{JK} \, .
\label{gs:eq:oddW}
\end{equation}
On regular tensors $\mathcal{L}_{\hB}$ raises the total degree in the couplings and in $\vep$, also in the logarithmic coordinates of section~\ref{gs:sec:log}, and $d$ is nonzero at $\vep=0$. The inverse is therefore the series in powers of $\mathcal{L}_{\hB}/d$, and it maps regular symmetric covariant tensors to such tensors. The term of order zero in the couplings and in $\vep$ of $\mathsf{H}_{JK}$ is $-V/d$ times that of $\cS_{JK}$, so that the statements on the sign for $\Pi$ hold for $\cW$ as well. For $d_c=3$ the free energy $-\cW$ on the unit sphere has a positive metric to all orders under Property~\ref{gs:P:free}, and it is then a monotone of the flow along the whole trajectory. Note that \eqref{gs:eq:oddW} requires that the Weyl variation closes on the marginal couplings, and it does not apply to the sextic theory, whose potential $\Pi'$ contains the one-point function of $\phi_i\phi_j$. The first form of \eqref{gs:eq:K} gives $(d\,\delta_J{}^I+\pr_J\hB^I)\,\pr_I\cW=-V\Lambda_{JK}\hB^K$, and the tensor which results from inverting the matrix on the left is a different solution, which is regular but in general not symmetric.

\subsection{Logarithmic coordinates}
\label{gs:sec:log}

The two-point form and Proposition~\ref{gs:prop:lowest} require, on the coordinates to which they apply, that the free theory is the point $g=0$ of the space of couplings and a zero of $\hB^I$. On the logarithmic coordinates of instance~(b) of Property~\ref{gs:P:functional} neither is the case. By Property~\ref{gs:P:weyl} the component $\hB^\iota$ tends to $-\vep\,k_\iota$ for $g_\iota\to0$, and the potential contains the function $\cW^{\log}$, which is not bounded in this limit. The lowest order of the metric along the logarithmic coordinates is then fixed by the gradient formula itself.

\begin{proposition}[Logarithmic coordinates]
\label{gs:prop:log}
Assume Properties~\ref{gs:P:functional} to~\ref{gs:P:sphere} with Property~\ref{gs:P:functional} in instance~(b), and let the Weyl variation close on the marginal couplings. Then the term of order zero $\hat G^{{\rm S}(0)}_{IJ}$ of the metric satisfies
\begin{equation}
\sum_{\iota'}k_{\iota'}\,\hat G^{{\rm S}(0)}_{\iota\iota'} = -\frac{d\,n_\iota}{V}
\label{gs:eq:Glog}
\end{equation}
for each logarithmic coordinate, where the sum runs over the logarithmic coordinates.
\end{proposition}

\begin{proof}
By Proposition~\ref{gs:prop:main} the function $\hat A^{\rm S}-(\vep\,d/V)\,\cW^{\log}$ is regular, and its derivative with respect to a logarithmic coordinate has no term of order zero. By the form of $\cW^{\log}$ in instance~(b) of Property~\ref{gs:P:functional} the derivative of $\cW^{\log}$ with respect to the coordinate $\log g_\iota$ is $n_\iota$. The term of order zero of $\pr_\iota\hat A^{\rm S}$ is therefore $\vep\,d\,n_\iota/V$. The term of order zero of $\hat G^{\rm S}_{\iota I}\hB^I$ is $-\vep\sum_{\iota'}k_{\iota'}\hat G^{{\rm S}(0)}_{\iota\iota'}$. Both sides of the gradient formula are power series in $\vep$, and dividing by $\vep$ we obtain \eqref{gs:eq:Glog}.
\end{proof}

The proposition uses that $\cA[\sigma]$ contains the logarithmic coordinates only through regular functions and covariant derivatives, as Property~\ref{gs:P:anomaly} states in instance~(b). A local term $c\int d\mu_d\,\log g_\iota\,E$ with a curvature scalar $E$ of dimension four would change $n_\iota$ without changing $\hat G^{\rm S}_{IJ}$.

The relation \eqref{gs:eq:Glog} follows from Properties~\ref{gs:P:functional} to~\ref{gs:P:sphere}, so that the row sums of the gauge block need no assumption on the lowest order. Property~\ref{gs:P:free} fixes the whole block at $\vep=0$, by Lemma~\ref{gre:lem:zero}, and with it the term of order zero in $\vep$ of $n_\iota$, see section~\ref{gs:sec:gauge}. It shows that the metric does not vanish if one of the coefficients $n_\iota$ does not. If there is one logarithmic coordinate and no other coordinate, it gives $s_d\,\hat G^{{\rm S}(0)}_{\iota\iota}=-s_d\,d\,n_\iota/(k_\iota V)$. The metric $s_d\,\hat G^{\rm S}_{\iota\iota}$ of the potential $s_d\,\hat A^{\rm S}$ is then positive definite to all orders in the couplings and in $\vep$ if $-s_d\,n_\iota$ is positive at $\vep=0$. For several logarithmic coordinates \eqref{gs:eq:Glog} fixes one combination in each row of the matrix $\hat G^{{\rm S}(0)}_{\iota\iota'}$, and without Property~\ref{gs:P:free} we make no statement on its positivity.

In instance~(b) the statements of sections~\ref{gs:sec:gradient} and~\ref{gs:sec:fixed} on the potential thus take the following form. The potential is regular up to the term $(\vep\,d/V)\,\cW^{\log}$, and its derivative is regular. It cannot be normalised at the free theory, and its value at a zero $g_*$ of $\hB^I$ contains $(\vep\,d/V)\sum_\iota n_\iota\log g_{\iota*}$. We apply these statements to gauge couplings in section~\ref{gs:sec:gauge}.

\section{Relevant operators}
\label{gs:sec:relevant}

We now add the sources $\sM$ of relevant operators. Their Weyl variation at $\sM=0$ does not vanish in general. It can contain the curvature and derivatives of the couplings, and it can contain derivatives of $\sigma$. The Killing identity then receives further terms, and so do the sphere data of the combination $W^\sharp$ of section~\ref{gs:sec:hom}. We proceed in four steps. First, we give the further terms of the Killing identity, a one-form which contains the one-point and two-point functions of the relevant operators, and we show that only sources of even dimension contribute. Second, we apply the commutation of the Weyl variations of Property~\ref{gs:P:weyl} to a relevant source, which gives relations among the coefficients of its Weyl variation on the sphere. Third, we redefine the sources by local terms such that these coefficients take a standard form, which we call the divergence scheme. In this scheme the relations show that the flow of a source at constant couplings is quadratic in $\hB^I$. Fourth, the further one-form in the Killing identity of $W^\sharp$ is then quadratic in $\hB^I$ as well, and such a one-form is a symmetric tensor contracted with $\hB^I$. This tensor is added to the metric, the potential changes by a term which vanishes at the zeros of $\hB^I$, and the Killing identity of $W^\sharp$ is again a gradient formula (Theorem~\ref{gr:cor:cond}). The results of this section hold for every theory with the properties.

\subsection{Quantities of the relevant sources}
\label{gs:sec:oneform}

We first name the coefficients of the Weyl variation of a relevant source at vanishing relevant sources. The variation $\delta^{\rm W}_\sigma\sM$ is the change of the source under the Weyl rescaling with the parameter $\sigma$. It is linear in $\sigma$, and it is a local expression in the couplings and the curvature which does not vanish at $\sM=0$, so that a Weyl rescaling generates the source even at constant couplings. We evaluate it at $\sM=0$, $a_\mu=0$ and $\gamma_{\mu\nu}=\mathring\gamma_{\mu\nu}$, for couplings $g=g_c+\delta g$ which differ from constant couplings $g_c$ by a variation $\delta g(x)$, and we expand it to first order in $\delta g$. For a harmonic $\sigma$ of degree one and for $\sigma=1$ the result has the form
\begin{equation}
\delta^{\rm W}_\sigma\sM = -\sigma\,\vk_{\rm h}(g)-\mathsf{L}_J[\sigma;\delta g^J] \, , \qquad
\delta^{\rm W}_1\sM = -\vk_{\rm c}(g)-\mathsf{L}_J[1;\delta g^J] \, ,
\label{gs:eq:kappadef}
\end{equation}
respectively. Here the first term of each expression contains the couplings without derivatives, and $\mathsf{L}_J$, which is linear in $\delta g^J$, collects the terms with at least one derivative of $\delta g^J$. The functions $\vk_{\rm h}$ and $\vk_{\rm c}$ take values in $V_\sM$. At constant couplings a Weyl rescaling changes the source from zero to $-\sigma\,\vk_{\rm h}$ for a harmonic $\sigma$ of degree one and to $-\vk_{\rm c}$ for $\sigma=1$, and we call $\vk_{\rm h}$ and $\vk_{\rm c}$ the flows of the source. The two are different in general. The variation contains terms with $\sigma$ itself, such as $\sigma$ times the curvature scalar, and terms with derivatives of $\sigma$, such as $\nabla^2\sigma$. For $\sigma=1$ only the former contribute, while for a harmonic of degree one the latter are multiples of $\sigma$ as well, by $\nabla^2\sigma=-d\,\sigma$, and contribute to $\vk_{\rm h}$. The expression $\mathsf{L}_J[\sigma;\delta g^J]$ is a scalar which is linear in $\sigma$ and in $\delta g^J$ and built covariantly, with coefficients which depend on the constant couplings. Under the integral over the sphere all derivatives can be moved from $\delta g^J$ to $\sigma$, on which they give multiples of $\sigma$, so that the integral is a multiple of $\int d\mu_d\,\sigma\,\delta g^J$ by \eqref{gsp:eq:c3}. In the same way $\mathsf{L}_J[1;\delta g^J]$ is a linear expression in $\delta g^J$, and its integral with $\sigma$ is such a multiple by \eqref{gsp:eq:c1}. There are therefore coefficients $\ell_{{\rm h}J}$ and $\ell_{{\rm c}J}$ with values in $V_\sM$, functions of the constant couplings, such that
\begin{equation}
\int d\mu_d\;\mathsf{L}_J[\sigma;\delta g^J] = -\ell_{{\rm h}J}\int d\mu_d\;\sigma\,\delta g^J \, , \qquad
\int d\mu_d\;\sigma\,\mathsf{L}_J[1;\delta g^J] = -\ell_{{\rm c}J}\int d\mu_d\;\sigma\,\delta g^J \, .
\label{gs:eq:elldef}
\end{equation}
On the sphere we need two quantities with values in the dual of $V_\sM$. The first is the one-point function $\sfm=V^{-1}\pr\cW/\pr\sM$ of the relevant operators at constant couplings, where the derivative is taken with respect to a constant source. The second is the first harmonic $\Lambda_{\sM J}$ of the two-point function of the relevant and the marginal operators, which is defined as in \eqref{gs:eq:Lambda} with the first direction $\sigma$ times a constant element of $V_\sM$. For the anomaly we need the coefficient $\sfa_\sM$ of $\int d\mu_d\,\sigma\,\delta\sM$ in $\cA[\sigma]$ at constant couplings. These quantities are regular by Properties~\ref{gs:P:functional} to~\ref{gs:P:anomaly}.

\begin{proposition}[Relevant sources]
\label{gs:prop:relevant}
Assume Properties~\ref{gs:P:functional} to~\ref{gs:P:sphere}. At $\sM=0$ the identity \eqref{gs:eq:K} is replaced by
\begin{equation}
\sfa_J = \pr_J\!\left(\Pi-\vk_{\rm h}\!\cdot\sfm\right)+\cS_{JI}\,\hB^I+\sfn_J \, , \qquad
\sfn_J \equiv \vk_{\rm h}\!\cdot\!\left(\pr_J\sfm-\Lambda_{\sM J}\right)+\ell_{{\rm h}J}\!\cdot\sfm \, .
\label{gs:eq:KM}
\end{equation}
\end{proposition}

\begin{proof}
For \eqref{gs:eq:KM} we repeat the proof of Proposition~\ref{gs:prop:K}. At $\sM=0$ the last term on the left-hand side of the Ward identity of Property~\ref{gs:P:ward} is $\int d\mu_d\,\delta^{\rm W}_\sigma\sM\cdot\delta W/\delta\sM$. With \eqref{gs:eq:kappadef} its first variation has three parts. The variation of $\vk_{\rm h}$ gives $-\pr_J\vk_{\rm h}\!\cdot\sfm$, the variation of the one-point function gives $-\vk_{\rm h}\!\cdot\Lambda_{\sM J}$, and the term with $\mathsf{L}_J$ multiplies the constant $\sfm$ and gives $\ell_{{\rm h}J}\!\cdot\sfm$ by \eqref{gs:eq:elldef}, each times $\int d\mu_d\,\sigma\,\delta g^J$.
\end{proof}

The proof uses Property~\ref{gs:P:ward} and the statements of Property~\ref{gs:P:sphere} for the pair $W$, $\cA[\sigma]$ only, so that the proposition holds for every pair with these properties. The one-form $\sfn_J$ contains the one-point function of the relevant operators on the sphere and their two-point function with the marginal operators, which are not local quantities. It depends on the scheme of the relevant sources, since a redefinition of $\sM$ by local terms with regular coefficients changes the coefficients in \eqref{gs:eq:kappadef}. For the value at the zeros of $\hB^I$ the scheme matters as well. If $\vk_{\rm c}$ vanishes at $g_*$, the point $\sM=0$ is a zero of the flow of the relevant sources on the unit sphere, and $\cW_*$ is then the free energy at a common zero of all flows in the extended form of the properties, and in the divergence scheme for the four theories of Proposition~\ref{gs:prop:radius}. In a scheme in which this is not the case the conformal theory on the sphere sits at a non-zero value of the relevant sources, as for the coupling of $\phi^2$ to the curvature in~\cite{Fei:2015oha}.

\subsection{Dimension of the sources}
\label{gs:sec:dimension}

Not all relevant sources contribute to \eqref{gs:eq:KM} and to $W^\sharp$.

\begin{lemma}
\label{gs:lem:dimension}
Assume Property~\ref{gs:P:weyl}. Let a component of $\sM$ have an odd dimension at $d=d_c$. Then $\vk_{\rm h}$, $\vk_{\rm c}$, $\ell_{{\rm h}J}$ and $\ell_{{\rm c}J}$ vanish for this component, and it does not contribute to \eqref{gs:eq:KM} and to the functional $W^\sharp$ of \eqref{gs:eq:sharp} at vanishing relevant sources.
\end{lemma}

\begin{proof}
At $\sM=0$ each term of $\delta^{\rm W}_\sigma\sM$ for this component is built from the metric, the curvature, the field strength of $a_\mu$ and covariant derivatives of $\sigma$ and of the couplings, and the number of its derivatives is the dimension of the component, where the curvature counts as two derivatives and $a_\mu$ as one. As in the proof of Lemma~\ref{gs:lem:odd} this number is even. Hence the variation vanishes at $\sM=0$, for all couplings, so that the term of $\bD W$ with this component vanishes at $\sM=0$, and all terms which the component adds to \eqref{gs:eq:KM} contain one of the four coefficients.
\end{proof}

The lemma has the following consequence. If all scalar relevant operators of a theory have sources of odd dimension at $d=d_c$, then \eqref{gs:eq:KM} reduces to \eqref{gs:eq:K}, the functional $W^\sharp$ at vanishing relevant sources is that of Proposition~\ref{gs:prop:sharp}, and the statements of Propositions~\ref{gs:prop:K} to~\ref{gs:prop:log} hold although these sources are present. This is the case for the fermion bilinears in theories of fermions at $d_c=2$ and in theories of gauge fields and fermions at $d_c=4$, whose sources have dimension one, see sections~\ref{gs:sec:fermions} and~\ref{gs:sec:gauge}. Sources of even dimension contribute in general. The operators $\phi_i\phi_j$ of a scalar field have a source of dimension two for every $d$. Thus in theories with scalar fields Proposition~\ref{gs:prop:main} does not apply, and we need the rest of this section. The argument of the lemma also applies to sources whose dimension at $d=d_c$ is not an integer, since no term of such a dimension can be built from the metric, $\sigma$ and the couplings.

\subsection{Weyl variation of a source on the sphere}
\label{gs:sec:sourcevar}

This subsection contains the central step for the relevant sources (Lemma~\ref{gr:lem:rel}). The commutation of two Weyl variations, applied to a relevant source at vanishing relevant sources, gives two relations among the coefficients of its Weyl variation on the unit sphere. The first, at constant couplings, expresses the flow $\vk_{\rm c}$ of the source through the coefficients $\ell_{{\rm c}K}$ contracted with $\hB^K$ and through the difference of the two flows. The second, at first order in the variation of the couplings, is the corresponding relation for the coefficients $\ell_{{\rm h}J}$ and $\ell_{{\rm c}J}$. In the scheme of section~\ref{gs:sec:divscheme}, in which $\ell_{{\rm h}J}=0$ and the two flows agree, the two relations together show that the flow is quadratic in $\hB^I$. Before the lemma we write the Weyl variation in a general form and define its coefficients.

By Lemma~\ref{gs:lem:dimension} the sources of odd and of non-integer dimension do not contribute. The Weyl variation of a source can contain the sources of lower dimension, multiplied by the curvature or by derivatives, and by its dimension it cannot contain those of higher dimension. We therefore order the relevant sources by their dimension at $d=d_c$ and treat them in the order of increasing dimension, so that the statements for a source can use those for the sources of lower dimension. We consider a source $\sM$ of even dimension $D$ with $2\leq D<d_c$, with values in a space $V_\sM$, and we denote by $\sM'$ the sources of lower dimension. By Property~\ref{gs:P:weyl} its variation has the form
\begin{equation}
\delta^{\rm W}_\sigma\sM = -\sigma\,(D-\gamma_\sM)\,\sM+\mathcal{V}_\sigma[\gamma,g,a]+\mathcal{Z}_\sigma[\sM']+\ldots \, ,
\label{gr:eq:general}
\end{equation}
where $\gamma_\sM$ is a linear map of $V_\sM$, $\mathcal{V}_\sigma$ does not depend on the relevant sources, $\mathcal{Z}_\sigma$ is linear in the sources of lower dimension, and the dots stand for terms of higher order in the relevant sources. By Property~\ref{gs:P:weyl} the map $\gamma_\sM$ has no term of order zero in the couplings and in $\vep$, which expresses that $\sM$ has the dimension $D$ in the free theory at $d=d_c$. For a source of dimension two, such as the source of $\phi_i\phi_j$, the variation \eqref{gr:eq:general} has only a few terms, which we write out in section~\ref{gs:sec:dimtwo}.

At constant couplings, $a_\mu=0$ and $\gamma_{\mu\nu}=\mathring\gamma_{\mu\nu}$ we expand $\mathcal{V}_\sigma$ in a variation $\delta g$ of the couplings,
\begin{gather}
\mathcal{V}_\sigma = -\sigma\,\vk_{\rm h}(g)-\mathsf{L}_J[\sigma;\delta g^J]-\mathsf{Q}_{JK}[\sigma;\delta g^J,\delta g^K]+{\rm O}(\delta g^3) \, , \notag\\
\mathcal{V}_1 = -\vk_{\rm c}(g)-\mathsf{L}_J[1;\delta g^J]-\mathsf{Q}_{JK}[1;\delta g^J,\delta g^K]+{\rm O}(\delta g^3) \, .
\label{gr:eq:Eexp}
\end{gather}
Here $g=g_c+\delta g$ in the functions $\vk_{\rm h}$ and $\vk_{\rm c}$. The linear term $\mathsf{L}_J$ contains at least one derivative of $\delta g^J$. The bilinear term $\mathsf{Q}_{JK}$ contains at least one derivative of each argument and obeys $\mathsf{Q}_{JK}[\sigma;f_1,f_2]=\mathsf{Q}_{KJ}[\sigma;f_2,f_1]$. The coefficients of $\mathsf{L}_J$ and $\mathsf{Q}_{JK}$ are functions of the couplings taken at the point $x$, that is at $g_c+\delta g(x)$, so that they contribute to higher orders through their derivatives. We thus continue \eqref{gs:eq:kappadef} to second order. The coefficients $\ell_{{\rm h}J}$ and $\ell_{{\rm c}J}$ are defined by \eqref{gs:eq:elldef}, and we define a tensor $\mathcal{M}_{JK}$ with values in $V_\sM$ by
\begin{equation}
\int d\mu_d\;\mathsf{Q}_{JK}[1;\sigma,f] = \mathcal{M}_{JK}\int d\mu_d\;\sigma\,f
\label{gr:eq:Sigma}
\end{equation}
for a harmonic $\sigma$ of degree one and a function $f$ of the class of sources. For a source of dimension two $\mathsf{Q}_{JK}[1;f_1,f_2]=\epsilon_{JK}\,\nabla f_1\!\cdot\!\nabla f_2$ and $\mathcal{M}_{JK}=d\,\epsilon_{JK}$.

We use the following facts about the unit sphere, which we obtain from integrations by parts with \eqref{gs:eq:parts} as the statements \eqref{gs:eq:elldef}. Let $\mathsf{T}$ be a linear differential expression in a function, built from the metric of the unit sphere and its curvature with constant coefficients. Then $\mathsf{T}\sigma=e_{\mathsf{T}}\,\sigma$ pointwise for a harmonic $\sigma$ of degree one, with a number $e_{\mathsf{T}}$, since $\nabla_\mu\nabla_\nu\sigma=-\sigma\,\mathring\gamma_{\mu\nu}$ reduces every derivative of $\sigma$ to $\sigma$ or $\nabla\sigma$, and no vector is available to contract $\nabla\sigma$ with. The transpose of $\mathsf{T}$ with respect to $d\mu_d$ has the same eigenvalue $e_{\mathsf{T}}$ on $\sigma$, since $\int d\mu_d\,\sigma\,\mathsf{T}\sigma'$ can be evaluated in both ways for two harmonics. Hence $\int d\mu_d\,\sigma\,\mathsf{T}f=e_{\mathsf{T}}\int d\mu_d\,\sigma f$, and $\int d\mu_d\,\mathsf{T}f=0$ if $\mathsf{T}$ contains at least one derivative. In the same way a bilinear expression with at least one derivative on each argument reduces, on two harmonics, to $\alpha\,\sigma\sigma'+\beta\,\nabla\sigma\!\cdot\!\nabla\sigma'$, which is symmetric. It follows that $\mathcal{M}_{JK}$ is symmetric.

\begin{lemma}
\label{gr:lem:rel}
Assume Property~\ref{gs:P:weyl}, for a class of sources as in appendix~\ref{gs:sec:class}, and let the sources of lower dimension satisfy $\vk'_{\rm h}=\vk'_{\rm c}\equiv\vk'$ and $\ell'_{\rm h}=0$, which is the scheme of Lemma~\ref{gr:lem:scheme} below. Then
\begin{align}
D\,\vk_{\rm c} &= \ell_{{\rm c}K}\hB^K+\left(\hB\!\cdot\!\pr-\gamma_\sM\right)\left(\vk_{\rm h}-\vk_{\rm c}\right)+\mathcal{Y}\cdot\vk' \, , \label{gr:eq:R1}\\
0 &= 2\,\hP^{(D)}{}_J{}^K\ell_{{\rm c}K}+\big[(\tL-\gamma_\sM)(\ell_{\rm h}-\ell_{\rm c})\big]_J+2\,\mathcal{M}_{JK}\hB^K+\mathcal{Y}_1\cdot\pr_J\vk'+\mathcal{Y}_{2J}\cdot\vk'+\mathcal{Y}_3\cdot\ell'_{{\rm c}J} \, , \label{gr:eq:R2}
\end{align}
where $\hP^{(D)}{}_J{}^K=\tfrac12(d-D)\,\delta_J{}^K+\pr_J\hB^K+\tfrac12\,(\hr_Jg)^K$, the extended Lie derivative acts on the index $J$ and $\gamma_\sM$ on the values, and $\mathcal{Y}$, $\mathcal{Y}_1$, $\mathcal{Y}_{2J}$ and $\mathcal{Y}_3$ are regular covariant linear maps from the values of the sources of lower dimension to $V_\sM$, which are determined by $\mathcal{Z}_\sigma$ and vanish without such sources. Without the condition $\ell'_{\rm h}=0$ a further term linear in $\ell'_{\rm h}$ appears in \eqref{gr:eq:R2}.
\end{lemma}

For $D=2$ we recover the relations \eqref{gs:eq:Mrel} of section~\ref{gs:sec:dimtwo} below, since $\hP^{(2)}$ is $\hP$. With the values of $\vk_{\rm h}$, $\vk_{\rm c}$, $\ell_{\rm h}$ and $\ell_{\rm c}$ for a source of dimension two given there and $\mathcal{M}=d\,\epsilon$, the relation \eqref{gr:eq:R1} is $d$ times the first of these relations and \eqref{gr:eq:R2} is $2d$ times the second, in a general scheme. Indeed, $\ell_{\rm h}-\ell_{\rm c}=-2d\,\theta$ and $\ell_{\rm c}=d\,\delta$ turn \eqref{gr:eq:R2} into $2d\,[\hP\delta+\epsilon\hB-(\tL-\gamma_\sM)\theta]=0$. The relations hold with non-abelian field rotations and with the logarithmic coordinates of instance~(b). At first order around constant couplings and at $a_\mu=0$ the rotations enter each term once and linearly, through $\delta^{\rm W}_\sigma a_\mu=\sigma\hr_K\pr_\mu\delta g^K$ acting on the constant couplings, so that no commutator of generators occurs and the rotations enter only through $\pr_J\hB^K+(\hr_Jg)^K$. The derivation uses the variation $\delta^{\rm W}_\sigma g^I=\sigma\hB^I$ in the chosen coordinates and not their linearity.\footnote{We checked \eqref{gr:eq:R1}, \eqref{gr:eq:R2} and \eqref{gr:eq:change} below in exact arithmetic with ${\rm SO}(3)$ rotations and a coordinate whose vector field has a constant term, for $D=2$ and $D=4$.}

\begin{proof}
We evaluate $[\mathbb{W}_1,\mathbb{W}_\sigma]\sM=0$ at vanishing relevant sources, at $\gamma_{\mu\nu}=\mathring\gamma_{\mu\nu}$ and at $a_\mu=0$, for a harmonic $\sigma$, at zeroth and at first order around constant couplings. Writing $\mathbb{W}_\sigma E$ for the change of a local expression $\mathcal{V}[\gamma,g,a]$ under the variations of Property~\ref{gs:P:weyl} and using that the terms which vary $\gamma_\sM$ multiply $\sM=0$, we find from \eqref{gr:eq:general}
\begin{equation}
\mathbb{W}_1\mathcal{V}_\sigma-\mathbb{W}_\sigma \mathcal{V}_1-\sigma\,(D-\gamma_\sM)\,\mathcal{V}_1+(D-\gamma_\sM)\,\mathcal{V}_\sigma+\mathcal{Z}_\sigma[\delta^{\rm W}_1\sM']-\mathcal{Z}_1[\delta^{\rm W}_\sigma\sM'] = 0 \, ,
\label{gr:eq:comm}
\end{equation}
with $\gamma_\sM$ at the point $g$. We use four rules. First, $\mathcal{V}_\sigma$ is built from the metric, the curvature and covariant derivatives, so that a constant rescaling of the metric gives $-D\,\mathcal{V}_\sigma$. Second, the variation of the couplings by $t\,\hB(g)$ or by $t\,\sigma\hB(g)$ changes the functions $\vk$, the coefficients of $\mathsf{L}_J$ and $\mathsf{Q}_{JK}$, and the arguments of $\mathsf{L}_J$ and $\mathsf{Q}_{JK}$, where the gauge field enters through $D_\mu g^I$ and its variation $\sigma\hr_ID_\mu g^I$ turns $\pr_I\hB^K$ into $\hat M^K{}_I$ in the variation of $D_\mu g^K$. Third, the change of the covariant scalar $\mathcal{V}_1$ under the metric $2\sigma\mathring\gamma_{\mu\nu}=\cL_v\mathring\gamma_{\mu\nu}$ equals, to first order, its transport $v^\mu\pr_\mu\mathcal{V}_1$ minus its change under the transport of the couplings, as an identity of tensor calculus with $d$ as a parameter, so that by \eqref{gs:eq:parts} $\int d\mu_d\,\mathbb{W}^{\rm metric}_\sigma \mathcal{V}_1=-d\int d\mu_d\,\sigma \mathcal{V}_1-\int d\mu_d\,(\delta \mathcal{V}_1/\delta g^K)\,v^\mu\pr_\mu g^K$. Fourth, the facts on the sphere stated before the lemma.

At zeroth order in $\delta g$ we find that every term of \eqref{gr:eq:comm} is $\sigma$ times a constant. The first term gives $\sigma(D\vk_{\rm h}-\hB\!\cdot\!\pr\,\vk_{\rm h})$, since $\mathsf{L}_J[\sigma;\hB^J]$ vanishes for constant $\hB^J$. The second gives $\sigma\,\hB\!\cdot\!\pr\,\vk_{\rm c}-\sigma\,\ell_{{\rm c}K}\hB^K$, where the metric part vanishes because a covariant scalar at constant couplings is constant on the sphere and is transported into itself, and the term $-\mathsf{L}_K[1;\sigma\hB^K]$ is $\sigma\,\ell_{{\rm c}K}\hB^K$ by the second fact. The third and the fourth term give $\sigma(D-\gamma_\sM)(\vk_{\rm c}-\vk_{\rm h})$. The terms with $\mathcal{Z}$ are $\sigma$ times a linear map applied to $\vk'$, since $\delta^{\rm W}_1\sM'=-\vk'$ and $\delta^{\rm W}_\sigma\sM'=-\sigma\vk'$ at constant couplings. Dividing by $\sigma$ gives \eqref{gr:eq:R1}.

At first order we integrate \eqref{gr:eq:comm} over the sphere with the weight one. The first term gives, by the first two rules,
\begin{equation*}
\left[D\,\pr_J\vk_{\rm h}-D\,\ell_{{\rm h}J}-\pr_J(\hB\!\cdot\!\pr\,\vk_{\rm h})+\ell_{{\rm h}K}\hat M^K{}_J+\hB\!\cdot\!\pr\,\ell_{{\rm h}J}\right]\int d\mu_d\,\sigma\,\delta g^J \, ,
\end{equation*}
where the last two terms come from the argument and from the coefficients of $\mathsf{L}_J$, and $\mathsf{Q}_{JK}[\sigma;\hB,\delta g]$ vanishes because $\hB$ is constant at this order. The second term gives, by the third rule for the metric and by the second rule for the couplings,
\begin{equation*}
-\left[-d\,\ell_{{\rm c}J}-\pr_J(\hB\!\cdot\!\pr\,\vk_{\rm c})+\hB\!\cdot\!\pr\,\ell_{{\rm c}J}+\pr_J\ell_{{\rm c}K}\,\hB^K-2\,\mathcal{M}_{JK}\hB^K\right]\int d\mu_d\,\sigma\,\delta g^J \, .
\end{equation*}
Here $-d\,\ell_{{\rm c}J}$ is the result of the metric variation, the term with $\pr_J\ell_{{\rm c}K}$ comes from the coefficients of $\mathsf{L}_K[1;\cdot]$ at $g_c+\delta g$ acting on $\sigma\hB^K(g_c)$, the term with $\mathcal{M}$ from $\mathsf{Q}_{JK}[1;\cdot,\cdot]$ with one argument $\sigma\hB$, and $\int d\mu_d\,\mathsf{L}_K[1;\sigma\,\pr_J\hB^K\delta g^J]$ vanishes by the fourth rule, and so does the term $\sigma\,(\hr_Jg)^K\pr_\mu\delta g^J$ of the gauge field, since every scalar which is linear in a vector field and contains a derivative of it is a divergence on the sphere. The third and the fourth term give $[(D-\gamma_\sM)(\pr_J\vk_{\rm c}-\ell_{{\rm c}J}-\pr_J\vk_{\rm h}+\ell_{{\rm h}J})+\pr_J\gamma_\sM(\vk_{\rm h}-\vk_{\rm c})]\int d\mu_d\,\sigma\,\delta g^J$. In the sum the terms with $\vk$ combine to $\pr_J[D\vk_{\rm c}-(\hB\!\cdot\!\pr-\gamma_\sM)(\vk_{\rm h}-\vk_{\rm c})]$, which by \eqref{gr:eq:R1} is $\pr_J(\ell_{{\rm c}K}\hB^K)$ up to the terms with $\vk'$. Its part $\pr_J\ell_{{\rm c}K}\hB^K$ cancels against the same term above. The terms with $\ell$ then add up to $(d-D)\ell_{\rm c}+\ell_{\rm c}\pr\hB+\ell_{\rm h}\hat M+(\hB\!\cdot\!\pr-\gamma_\sM)(\ell_{\rm h}-\ell_{\rm c})$, which equals $2\hP^{(D)}\ell_{\rm c}+(\tL-\gamma_\sM)(\ell_{\rm h}-\ell_{\rm c})$ with $\hat M=\pr\hB+\hr g$. Together with $2\mathcal{M}\hB$ this is \eqref{gr:eq:R2} up to the terms with the sources of lower dimension. These terms contain $\mathcal{Z}_\sigma$ and $\mathcal{Z}_1$ applied to the first variation of $\delta^{\rm W}\sM'$, which is $-\pr_J\vk'\,\delta g^J$ minus $\mathsf{L}'$, and the first variation of the coefficients of $\mathcal{Z}$ times $\vk'$. Integrated with the weight one, the term $\mathsf{L}'_J[\sigma;\delta g]$ gives $\ell'_{{\rm h}J}=\ell'_{{\rm c}J}+(\ell'_{\rm h}-\ell'_{\rm c})_J$, and in the scheme which we use below for the lower sources $\ell'_{\rm h}=0$, while $\mathsf{L}'_J[1;\delta g]$ gives $\ell'_{{\rm c}J}$ by the fourth rule. This gives the last three terms of \eqref{gr:eq:R2}.
\end{proof}

\subsection{Sources of dimension two}
\label{gs:sec:dimtwo}

For a source $\sM$ of dimension two the most general variation of the form of Property~\ref{gs:P:weyl} is, up to terms of second order in sources of lower dimension,
\begin{equation}
\delta^{\rm W}_\sigma\sM = -\sigma\left[(2-\gamma_\sM)\sM+\delta_I\,D^2g^I+\epsilon_{IJ}\,D_\mu g^ID^\mu g^J+\bar\eta\,R\right]
-2\,\pr_\mu\sigma\,\theta_I\,D^\mu g^I-\bar\tau\,\nabla^2\sigma \, ,
\label{gs:eq:dimtwo}
\end{equation}
with a linear map $\gamma_\sM$ of $V_\sM$ and regular coefficients $\delta_I$, $\epsilon_{IJ}=\epsilon_{JI}$, $\theta_I$, $\bar\eta$ and $\bar\tau$ with values in $V_\sM$. This is the form of the variation of the mass source of $\phi^4$ theory in section~\ref{sec:weylvar}, and of~\cite{Jack:2013sha} in four dimensions. On the unit sphere the curvature scalar is $R_d=d(d-1)$, and with $\nabla^2\sigma=-d\,\sigma$ we find for the quantities of \eqref{gs:eq:kappadef} and \eqref{gs:eq:elldef}
\begin{equation}
\vk_{\rm c} = R_d\,\bar\eta \, , \qquad \vk_{\rm h} = R_d\,\bar\eta-d\,\bar\tau \, , \qquad
\ell_{{\rm c}J} = d\,\delta_J \, , \qquad \ell_{{\rm h}J} = d\left(\delta_J-2\,\theta_J\right) .
\label{gs:eq:dimtwodata}
\end{equation}
By the last statement of Property~\ref{gs:P:weyl} the coefficients $\delta_I-2\theta_I$ and $\bar\tau$ vanish at $g=0$.
The condition of Property~\ref{gs:P:weyl} that two Weyl variations commute on $\sM$ gives two relations among the coefficients,
\begin{equation}
2(d-1)\,\bar\eta = \delta_I\hB^I-\left(\hB\!\cdot\!\pr-\gamma_\sM\right)\bar\tau \, , \qquad
\big[(\tL-\gamma_\sM)\,\theta\big]_I = \hP_I{}^J\delta_J+\epsilon_{IJ}\hB^J \, ,
\label{gs:eq:Mrel}
\end{equation}
where $\hP_I{}^K=\tfrac12(d-2)\,\delta_I{}^K+\pr_I\hB^K+\tfrac12\,(\hr_I\,g)^K$. They are the coefficients of $\sigma\nabla^2\sigma'-\sigma'\nabla^2\sigma$ and of $(\sigma\,\pr_\mu\sigma'-\sigma'\pr_\mu\sigma)\,D^\mu g^I$ in the commutator. For $d=4$ they are the relations of~\cite{Jack:2013sha}, and for $\phi^4$ theory they are derived in appendix~\ref{app:masscoeff}.

The coefficients $\theta_I$ and $\bar\tau$ can be removed by a redefinition of the source by local terms, $\sM=\sM'-d_I\,D^2g^I-k_R\,R$ with regular coefficients $d_I$ and $k_R$ with values in $V_\sM$, if the matrix $\hP_I{}^K$ has a regular inverse. For a vector field of the form \eqref{gs:eq:Bhat}, and of the form of minimal subtraction given in appendix~\ref{gs:sec:properties} on the logarithmic coordinates, its term of order zero is a diagonal matrix with the entries $\tfrac12(d-2)-\vep\,k_I$ for the coordinates which are couplings and $\tfrac12(d-2)$ for the logarithmic coordinates, so that this is the case for $d_c\neq2$. For $\phi^4$ theory the redefinition is given in appendix~\ref{app:masscoeff}. In the resulting scheme \eqref{gs:eq:Mrel} gives
\begin{equation}
\delta_I = \mathsf{D}_{IK}\,\hB^K \, , \qquad \mathsf{D}_{IK} = -(\hP^{-1})_I{}^J\epsilon_{JK} \, , \qquad
\vk_{\rm c} = \vk_{\rm h} = \vk \equiv \tfrac12\,d\,\delta_I\hB^I \, , \qquad \ell_{{\rm c}J} = \ell_{{\rm h}J} = d\,\delta_J \, .
\label{gs:eq:scheme}
\end{equation}
The two flows of the source agree, they are quadratic in $\hB^I$, and they vanish at the zeros of $\hB^I$. The one-form $\sfn_J$ of \eqref{gs:eq:KM} becomes
\begin{equation}
\sfn_J = \vk\!\cdot\!\left(\pr_J\sfm-\Lambda_{\sM J}\right)+\mathsf{Y}^{\sfn}_{JK}\,\hB^K \, , \qquad \mathsf{Y}^{\sfn}_{JK} \equiv d\;\sfm\!\cdot\!\mathsf{D}_{JK} \, .
\label{gs:eq:nscheme}
\end{equation}
The matrix $\mathsf{Y}^{\sfn}_{JK}$ is not symmetric. Appendix~\ref{gs:app:fourd} uses the scheme \eqref{gs:eq:scheme} for $\phi^4$ theory, and in section~\ref{gs:sec:divscheme} we use a different scheme, the one of Theorem~\ref{gs:thm:phi4}.

\subsection{Divergence scheme}
\label{gs:sec:divscheme}

A redefinition $\sM=\sM'_{\rm new}+Y$ of the source by a local expression $Y[\gamma,g,a]$ of dimension $D$ with regular covariant coefficients changes $\mathcal{V}_\sigma$ by $-(\mathbb{W}_\sigma+\sigma(D-\gamma_\sM))Y$. We take
\begin{equation}
Y = y_J(g)\left(-\tfrac1d\,D^2\right)^{D/2}g^J+y_0(g)\left(\frac{R}{R_d}\right)^{D/2} ,
\label{gr:eq:Y}
\end{equation}
with regular covariant coefficients $y_J$ and $y_0$ with values in $V_\sM$, where $R_d=d(d-1)$. On the unit sphere $Y$ equals $y_0$ at constant couplings, and its linear term in $\delta g$ maps a harmonic $\sigma$ to $(y_J+\pr_Jy_0)\,\sigma$. The computation of the proof of Lemma~\ref{gr:lem:rel}, applied to $\mathcal{V}^Y_\sigma=(\mathbb{W}_\sigma+\sigma(D-\gamma_\sM))Y$, gives the changes
\begin{equation}
\Delta\big(\vk_{\rm h}-\vk_{\rm c}\big) = D\,y_0+y_K\hB^K \, , \qquad
\Delta\ell_{{\rm h}J} = \big[\big(2\hP^{(D)}-(\tL-\gamma_\sM)\big)y\big]_J \, , \qquad
\Delta\ell_{{\rm c}J} = -\big[(\tL-\gamma_\sM)\,y\big]_J \, ,
\label{gr:eq:change}
\end{equation}
where we used that \eqref{gr:eq:Y} has no bilinear term at $a_\mu=0$. Note that the redefinition does change $\mathcal{M}_{JK}$, through $\mathbb{W}_1$ acting on $y_J$ times a derivative of $\hB^J(g)$, by a term which vanishes at $g=0$ when $y_J$ does, and that \eqref{gr:eq:change} does not involve $\mathcal{M}$. For $D=2$ and $Y=-d_ID^2g^I-k_RR$ these are the changes of $\bar\tau$, $\theta_I$ and $\delta_I$ under the redefinition of section~\ref{gs:sec:dimtwo}.

\begin{lemma}
\label{gr:lem:scheme}
Assume Property~\ref{gs:P:weyl}, and let $D<d_c$. Then there is a unique redefinition \eqref{gr:eq:Y} after which $\ell_{{\rm h}J}=0$ and $\vk_{\rm h}=\vk_{\rm c}$. If $\ell_{{\rm h}J}$ and $\vk_{\rm h}-\vk_{\rm c}$ vanish at $g=0$ before the redefinition, then $y_J$ and $y_0$ vanish at $g=0$.
\end{lemma}

\begin{proof}
The operator $2\hP^{(D)}-(\tL-\gamma_\sM)$ acts on regular one-forms with values in $V_\sM$. Its term of order zero in the couplings and in $\vep$ is multiplication by $d_c-D$, since $\pr_J\hB^K$, $(\hr_Jg)^K$, $\gamma_\sM$ and the transport $\hB\!\cdot\!\pr$ have no such term. For $D\neq d_c$ it is invertible on power series by iteration in the order in the couplings and in $\vep$, and $y=-[2\hP^{(D)}-(\tL-\gamma_\sM)]^{-1}\ell_{\rm h}$ gives $\ell_{\rm h}=0$. Then $y_0=-(\vk_{\rm h}-\vk_{\rm c}+y_K\hB^K)/D$ gives $\vk_{\rm h}=\vk_{\rm c}$. The inverse maps a one-form which vanishes at $g=0$ to one which vanishes at $g=0$, because the operator does not lower the order in the couplings. This gives the last statement.
\end{proof}

For $D=2$ the conditions are $\delta_I=2\theta_I$ and $\bar\tau=0$. In this scheme the derivative terms of \eqref{gs:eq:dimtwo} are $-\delta_I\nabla_\mu(\sigma D^\mu g^I)$ to first order, a divergence, which explains the name. The scheme \eqref{gs:eq:scheme} has $\ell_{\rm h}=\ell_{\rm c}$ instead.

With several relevant sources of even dimension we make the redefinitions in the order of increasing dimension. The Weyl variation of a source contains no source of higher dimension, so that a redefinition changes, besides the variation of the redefined source, only the variations of the sources of higher dimension, through their terms which contain the redefined source. The later redefinitions are made with these changed variations, and their constant parts $y_0$ vanish at $g=0$ if those of the earlier ones do. In instance~(a) of Property~\ref{gs:P:functional} this holds because $\hB^K$ vanishes at $g=0$. In instance~(b) it holds because the change of $\ell_{{\rm h}\iota}$ for a logarithmic coordinate is proportional to $\pr_\iota y_0$ and to $y_\iota$ of the earlier redefinitions, which have no term of order zero, and because the operator of Lemma~\ref{gr:lem:scheme} is diagonal at $g=0$, so that $y_\iota$ and with it $y_K\hB^K$ vanish at $g=0$. The coefficients $y_J$ vanish at $g=0$ as well if in addition the terms of the Weyl variation of the higher source which are linear in the lower source vanish at $g=0$, or if the first derivatives of the earlier $y_0$ vanish at $g=0$, as is the case in the theories of this paper.

The redefinition maps the point $\sM'_{\rm new}=0$ at constant couplings to the constant source $\sM=y_0(g)$ on the unit sphere. The properties of the functional in the new scheme are therefore statements about the functional in the old scheme at a constant relevant source. If $y_0$ vanishes at $g=0$, these are the statements of Properties~\ref{gs:P:functional} and~\ref{gs:P:sphere} at the constant source $\mu=y_0$. With the last statement of Lemma~\ref{gr:lem:scheme} and the last statement of Property~\ref{gs:P:weyl} the redefinition therefore maps a functional with Properties~\ref{gs:P:functional} to~\ref{gs:P:sphere} to a functional with these properties. The same remark applies to the redefinition \eqref{gs:eq:scheme}, which maps the point $\sM'=0$ to the constant source $-k_RR_d$.

\subsection{Flows of the relevant sources}
\label{gsh:sec:flows}

With relevant sources of even dimension the Killing identity of $W^\sharp$ contains the coefficients of their Weyl variation. We first show that in the divergence scheme these coefficients are of the required form, and we need for this only an operator with an invertible term of order zero.

\begin{lemma}
\label{gsh:lem:inv}
Assume Property~\ref{gs:P:weyl}, let $V$ be a vector space on which the field rotations act, let $\gamma$ be a regular covariant linear map of $V$ without term of order zero in the couplings and in $\vep$, and let $D<d_c$. Let $\mathcal{T}_V$ be the set of one-forms $X_{JK}\hB^K$ with a regular covariant tensor $X_{JK}$ with values in $V$, without condition of symmetry. Then the operator $O=2\hP^{(D)}-(\tL-\gamma)$ of Lemma~\ref{gr:lem:scheme} is invertible on regular covariant one-forms with values in $V$, and $O^{-1}$ maps $\mathcal{T}_V$ into $\mathcal{T}_V$.
\end{lemma}

\begin{proof}
On a one-form $X_{JK}\hB^K$ the operator $\hP^{(D)}$ acts on the index $J$ of $X$, and $\gamma$ acts on its values. The extended Lie derivative is a derivation which commutes with contractions, and $\tL\hB=0$, so that $\tL(X\hB)=(\tL X)\hB$. Hence $O(X\hB)=(\hat OX)\hB$, where $\hat O=2\hP^{(D)}-(\tL-\gamma)$ acts on tensors with two lower indices, with $\hP^{(D)}$ on the first index and $\tL$ on both. As in the proof of Lemma~\ref{gr:lem:scheme}, the terms $\pr_J\hB^K$, $(\hr_Jg)^K$, $\gamma$ and $\hB\!\cdot\!\pr$ have no term of order zero in the couplings and in $\vep$, in either instance of Property~\ref{gs:P:functional}. The term of order zero of $O$ and of $\hat O$ is therefore multiplication by $d_c-D\neq0$, and both are invertible on regular covariant tensors by iteration in the order in the couplings and in $\vep$. For $\omega=Y\hB\in\mathcal{T}_V$ the one-form $(\hat O^{-1}Y)\hB$ solves $O\,\omega'=\omega$, and by uniqueness it equals $O^{-1}\omega$.
\end{proof}

The order counting in the lemma is the plain counting in the couplings and in $\vep$, also in instance~(b).

\begin{proposition}
\label{gsh:prop:flows}
Assume Property~\ref{gs:P:weyl}, for a class of sources as in appendix~\ref{gs:sec:class}, and let every relevant source of even dimension $D<d_c$ be in the scheme of Lemma~\ref{gr:lem:scheme}. Then for each such source $\ell_{{\rm h}J}=0$, $\vk_{\rm h}=\vk_{\rm c}\equiv\vk$, and
\begin{equation}
\ell_{{\rm c}J} = X_{JK}\,\hB^K \, , \qquad \vk = \hB^Kx_{KL}\hB^L \, ,
\label{gsh:eq:flows}
\end{equation}
with regular covariant tensors $X_{JK}$ and $x_{KL}=x_{LK}$ with values in $V_\sM$.
\end{proposition}

\begin{proof}
By induction on $D$. For a source of dimension $D$ assume \eqref{gsh:eq:flows} for the sources of lower dimension, with $\vk'=\hB x'\hB$ and $\ell'_{\rm c}=X'\hB$. With $\ell_{\rm h}=0$ and $\vk_{\rm h}=\vk_{\rm c}$ the relation \eqref{gr:eq:R2} reads $O\,\ell_{\rm c}=-(2\mathcal{M}\hB+\mathcal{Y}_1\cdot\pr\vk'+\mathcal{Y}_2\cdot\vk'+\mathcal{Y}_3\cdot\ell'_{\rm c})$ with the operator $O$ of Lemma~\ref{gsh:lem:inv} for $V=V_\sM$ and $\gamma=\gamma_\sM$. Each term on the right-hand side is in $\mathcal{T}_{V_\sM}$. This is clear for $\mathcal{M}\hB$ and $\mathcal{Y}_3\cdot X'\hB$. For the others we use $\pr_J\vk'=(2\,\pr_J\hB^Kx'_{KL}+\pr_Jx'_{KL}\hB^K)\hB^L$ and $\vk'=(x'_{KL}\hB^K)\hB^L$. Lemma~\ref{gsh:lem:inv} gives $\ell_{\rm c}=X\hB$. The relation \eqref{gr:eq:R1} then gives $D\,\vk=\ell_{{\rm c}K}\hB^K+\mathcal{Y}\cdot\vk'=\hB^K(X_{KL}+\mathcal{Y}\cdot x'_{KL})\hB^L$, and $x$ is the symmetric part of the bracket divided by $D$.
\end{proof}

\subsection{Combination with relevant sources}
\label{gsh:sec:quant}

The identity \eqref{gsh:eq:hom} becomes a relation among functions of the couplings through the Killing identity of Proposition~\ref{gs:prop:K}, applied to the pair $W^\sharp$ and $\cA^\sharp=0$. For this we need the statements of Property~\ref{gs:P:sphere} for $W^\sharp$ with relevant sources. At constant couplings and $a_\mu=0$ we write the variation of a source of even dimension in the notation \eqref{gr:eq:Eexp}, $\delta^{\rm W}_1\sM=-\vk_{\rm c}-\mathsf{L}_J[1;\delta g^J]-\mathsf{Q}_{JK}[1;\delta g^J,\delta g^K]+\ldots$ at vanishing relevant sources. By Lemma~\ref{gs:lem:dimension} the sources of odd and of non-integer dimension have a variation which vanishes at vanishing relevant sources. We use the quantities $\sfm$ and $\Lambda_{\sM J}$ of section~\ref{gs:sec:oneform} for each source, and $\ell_{{\rm c}J}$ and $\mathcal{M}_{JK}$ of \eqref{gs:eq:elldef} and \eqref{gr:eq:Sigma}.

\begin{lemma}
\label{gsh:lem:quant}
Assume Properties~\ref{gs:P:functional} to~\ref{gs:P:sphere} with the relevant sources, with the sources of even dimension in the scheme of Lemma~\ref{gr:lem:scheme}. Then $W^\sharp$ has the expansion of Property~\ref{gs:P:functional} with $\cW^\sharp-\vep\,\cW^{\log}$ regular, and it satisfies the two statements of Property~\ref{gs:P:sphere} at constant couplings and vanishing relevant sources. At constant couplings and vanishing relevant sources
\begin{equation}
\cW^\sharp = \left(\vep-\hB\!\cdot\!\pr\right)\cW+V\rA+V\sum\vk_{\rm c}\!\cdot\sfm \, ,
\label{gsh:eq:Wsharp}
\end{equation}
where the sum runs over the sources of even dimension, and the tensor $\cS^\sharp_{IJ}$ of $W^\sharp$, defined by \eqref{gs:eq:Sdef}, is
\begin{equation}
\cS^\sharp_{IJ} = \big[(\vep-\tL)\,\cS\big]_{IJ}-\cK_{IJ}+\sum\mathcal{N}_{IJ} \, ,
\label{gsh:eq:Ssharp}
\end{equation}
with the symmetric tensor
\begin{align}
\mathcal{N}_{IJ} &= \pr_I\vk_{\rm c}\!\cdot\Lambda_{\sM J}+\pr_J\vk_{\rm c}\!\cdot\Lambda_{\sM I}+\vk_{\rm c}\!\cdot\Lambda'_{IJ}-\ell_{{\rm c}I}\!\cdot\Lambda_{\sM J}-\ell_{{\rm c}J}\!\cdot\Lambda_{\sM I}-\left(\pr_I\ell_{{\rm c}J}+\pr_J\ell_{{\rm c}I}\right)\!\cdot\sfm \notag\\
&\quad+2\,\mathcal{M}_{IJ}\!\cdot\sfm-\pr_I\vk_{\rm c}\!\cdot\pr_J\sfm-\pr_J\vk_{\rm c}\!\cdot\pr_I\sfm-\vk_{\rm c}\!\cdot\pr_I\pr_J\sfm \, .
\label{gsh:eq:N}
\end{align}
Here $\Lambda'_{IJ}$, with values in the dual of $V_\sM$, is the derivative of $\Lambda_{IJ}$ with respect to a constant source $\sM$ at $\sM=0$.
\end{lemma}

\begin{proof}
The expansion of Property~\ref{gs:P:functional} holds for each of the three terms of \eqref{gs:eq:sharp}, since the coefficients of $\bD$ are regular by Property~\ref{gs:P:weyl}, those of $\cA[1]$ by Property~\ref{gs:P:anomaly}, and the terms of $\bD W$ are terms of the expansion of $W$ evaluated on products of the sources and their derivatives, as in the proof of Proposition~\ref{gs:prop:sharp}. At constant couplings and vanishing relevant sources the term $-\bD W$ is $-\hB^I\pr_I\cW+V\sum\vk_{\rm c}\!\cdot\sfm$, since $\mathsf{L}_J$ and $\mathsf{Q}_{JK}$ vanish on constants and $D_\mu g^I=0$. We obtain \eqref{gsh:eq:Wsharp}, and $\cW^\sharp-\vep\,\cW^{\log}$ is regular because the first derivatives of $\cW^{\log}$ are.

For the statements of Property~\ref{gs:P:sphere} we compute the first and second variations of $\bD W$ term by term. The term with $\hB^I$ and the term with $\hr_I$ are treated as in the proof of Lemma~\ref{gs:lem:local}, which for these terms uses the gauge invariance of $W$ and Property~\ref{gs:P:sphere}, and the term $\cA[1]$ by Lemma~\ref{gsp:lem:local}. Together with $\vep W$ they give $[(\vep-\tL)\,\cS]_{IJ}-\cK_{IJ}$, as in the proof of Proposition~\ref{gs:prop:sharp}. It remains to treat $\mathcal{E}=-\int d\mu_d\,(\delta^{\rm W}_1\sM+D\sM)\cdot\delta W/\delta\sM$ for a source of even dimension. In the variations in the couplings at vanishing relevant sources only $-\delta^{\rm W}_1\sM=\vk_{\rm c}+\mathsf{L}_J+\mathsf{Q}_{JK}$ contributes, with the coefficients of $\mathsf{L}_J$ and $\mathsf{Q}_{JK}$ taken at the couplings of the point. We reduce each term with three facts. First, by the facts on the sphere of section~\ref{gs:sec:sourcevar}, $\mathsf{L}_J[1;\sigma]=o_J\,\sigma$ pointwise, and comparing with \eqref{gs:eq:elldef} gives $o_J=-\ell_{{\rm c}J}$. Since this holds at every constant coupling, the variation of the coefficients of $\mathsf{L}_J[1;\cdot]$ in the direction $\delta g^K$ gives $-\pr_K\ell_{{\rm c}J}$ in the same way. Second, by the second statement of Property~\ref{gs:P:sphere} the density $\delta W/\delta\sM(x)$, varied in the direction $\sigma z^I$, is $\Lambda_{\sM I}z^I\sigma(x)$, where we used that the constants for the directions $(\sigma z^I,\delta\sM)$ and $(\sigma\sM_{\rm c},\delta g^I)$ agree, since both equal the second derivative in the directions $\sigma z^I$ and $\sigma\sM_{\rm c}$ divided by $\int d\mu_d\,\sigma^2$. Third, by the same statement at a constant source, in the form explained after the proof, the variation of $W$ in the directions $\vk_{\rm c}$, $\sigma z^I$ and $\delta g^J$ is $\vk_{\rm c}\!\cdot\Lambda'_{IJ}\,z^I\int d\mu_d\,\sigma\,\delta g^J$. The second variation of $\mathcal{E}$ in the directions $\sigma z^I$ and $\delta g^J$ is then $z^I\int d\mu_d\,\sigma\,\delta g^J$ times the first line of \eqref{gsh:eq:N} plus $2\,\mathcal{M}_{IJ}\!\cdot\sfm$, where the last term comes from $\mathsf{Q}_{JK}[1;\sigma z,\delta g]+\mathsf{Q}_{KJ}[1;\delta g,\sigma z]$ and \eqref{gr:eq:Sigma}. At constant couplings $\mathcal{E}=V\vk_{\rm c}\!\cdot\sfm$, whose second derivative divided by $V$ gives the remaining terms of \eqref{gsh:eq:N} together with $\pr_I\pr_J\vk_{\rm c}\!\cdot\sfm$, which cancels against the same term of the first variation. This gives \eqref{gsh:eq:Ssharp}. The first statement of Property~\ref{gs:P:sphere} for $W^\sharp$, and the second statement in the directions $(\sigma\sM_{\rm c},\delta g^J)$, follow in the same way. For a source $\sM''$ of higher dimension the terms of $\delta^{\rm W}_1\sM''$ which are linear in $\sM$ act on $\sM$ as a differential expression built from the metric of the unit sphere, so that by the same facts they contribute constants times $\sfm''$ and $\Lambda_{\sM''J}$. The current does not contribute, since its one-point function vanishes at constant couplings and the variation of $\delta^{\rm W}_1\sM$ in $a_\mu$ at constant couplings is a divergence or of second order.
\end{proof}

The constant source enters the proof only in the direction $\vk_{\rm c}(g_c)$, and only to first order. In instance~(a) the function $\vk_{\rm c}=\hB x\hB$ of Proposition~\ref{gsh:prop:flows} vanishes at $g=0$, so that $\mu=t\,\vk_{\rm c}$ is a regular covariant function which vanishes at $g=0$. Properties~\ref{gs:P:functional} and~\ref{gs:P:sphere} at the constant source $\mu(g_c)$, differentiated with respect to $t$ at $t=0$, give the statements which we used. In instance~(b) $\vk_{\rm c}$ need not vanish at $g=0$, since $\hB^\iota$ does not. We then take $\mu=t\,g_\iota\,\vk_{\rm c}$ for one gauge coupling $g_\iota$, which is regular, covariant and vanishes at $g=0$. The derivative with respect to $t$ gives each statement multiplied by $g_\iota$, as an identity of power series in the couplings, and we divide by $g_\iota$, since a product of power series vanishes only if a factor does. The statements therefore follow from Properties~\ref{gs:P:functional} and~\ref{gs:P:sphere} as stated, in both instances.

Without relevant sources the lemma reduces to the second statement of Proposition~\ref{gs:prop:sharp}.

\subsection{Gradient formula with relevant sources}
\label{gs:sec:gradrel}

\begin{lemma}
\label{gsh:lem:m}
Assume Properties~\ref{gs:P:functional} to~\ref{gs:P:sphere} with the relevant sources, with the sources of even dimension in the scheme of Lemma~\ref{gr:lem:scheme}. Then for each source of even dimension $D<d_c$
\begin{equation}
\sfa^{(0,0)}_\sM = (d_c-D)\,\sfm^{(0,0)}+\sum_{\sM''}\big(\zeta_{\sM''}^{(0,0)}\big)^{\rm T}\sfm''^{(0,0)} \, ,
\label{gsh:eq:am}
\end{equation}
where the sum runs over the sources $\sM''$ of higher even dimension and $\zeta_{\sM''}$ is the regular linear map with $\int d\mu_d\,\mathcal{Z}''_\sigma[\delta\sM]\cdot\sfm''=(\zeta_{\sM''}^{\rm T}\sfm'')\cdot\int d\mu_d\,\sigma\,\delta\sM$, for the term $\mathcal{Z}''_\sigma$ of $\delta^{\rm W}_\sigma\sM''$ which is linear in $\sM$. Consequently $\sfa_\sM$ has no term of order zero in the couplings and in $\vep$ for every source of even dimension if and only if $\sfm^{(0,0)}=0$ for every such source.
\end{lemma}

\begin{proof}
We vary the Ward identity of Property~\ref{gs:P:ward} once in a variation $\delta\sM$ at constant couplings and vanishing relevant sources. The right-hand side gives $\sfa_\sM\!\cdot\!\int d\mu_d\,\sigma\,\delta\sM$. On the left-hand side the term with the couplings gives the second variation of $W$ in the directions $\sigma\hB$ and $\delta\sM$, which is $\hB^I\Lambda_{\sM I}$ times the integral. The term with the gauge field does not contribute. The term with the source gives $(d-D+\gamma_\sM^{\rm T})\,\sfm$ from $-\sigma(D-\gamma_\sM)\delta\sM-v^\mu\pr_\mu\delta\sM$ by \eqref{gs:eq:parts}, and $-\vk_{\rm h}$ contracted with the first harmonic $\Lambda_{\sM\sM}$ of the two-point function of two relevant operators from $\mathcal{V}_\sigma=-\sigma\vk_{\rm h}$. A source $\sM''$ of higher dimension contributes through $\mathcal{Z}''_\sigma[\delta\sM]$, which gives $\zeta_{\sM''}^{\rm T}\sfm''$, and through $-\vk''_{\rm h}$ contracted with the first harmonic of the two-point function of $\sM''$ and $\sM$, which vanishes at order zero by Proposition~\ref{gsh:prop:flows}. A source of odd or non-integer dimension does not contribute, since a term linear in $\sigma$ and in $\delta\sM$ with an odd number of derivatives vanishes on the unit sphere. At order zero in the couplings and in $\vep$ the terms with $\hB^I$ and with $\gamma_\sM$ vanish, also in instance~(b), where $\hB^\iota$ has the factor $\vep$, and $\vk_{\rm h}$ vanishes by Proposition~\ref{gsh:prop:flows}. This gives \eqref{gsh:eq:am}. The last statement follows by induction downwards in the dimension, starting with the source of highest dimension, for which the sum is empty, since $d_c-D\neq0$.
\end{proof}

We call a regular symmetric tensor nondegenerate if its term of order zero in the couplings and in $\vep$ is an invertible matrix, where the components refer to the coordinates of Property~\ref{gs:P:functional}. The inverse of such a tensor is regular. The term of order zero of the metric of the gradient formula below contains $\cK^{(0,0)}$ and the terms of order zero of the one-point functions $\sfm$, which by Lemma~\ref{gsh:lem:m} are fixed by those of the coefficients $\sfa_\sM$. These concern the free theory. The last statement of Property~\ref{gs:P:free} is $\sfm^{(0,0)}=0$ in the divergence scheme, which by Lemma~\ref{gsh:lem:m} is equivalent to the statement that at $\vep=0$ and $g=0$ the coefficient of the anomaly which is linear in a relevant source has no term. The last statement of Property~\ref{gs:P:weyl} implies, by Lemma~\ref{gr:lem:scheme}, that the redefinition vanishes at $g=0$, so that the constant source of section~\ref{gs:sec:divscheme} vanishes at $g=0$.

We first consider odd $d_c$. The anomaly vanishes at vanishing relevant sources by Lemma~\ref{gs:lem:odd}, and $\sfa_\sM$ vanishes for sources of even dimension, since a term of the anomaly linear in such a source has an odd number of derivatives. The identity \eqref{gs:eq:KM} for the functional $W$ itself, with $\sfa_J=0$, then reads $\pr_J(\Pi-\vk\cdot\sfm)+\cS_{JI}\hB^I+\sfn_J=0$ in the divergence scheme, where $\ell_{\rm h}=0$. By Proposition~\ref{gsh:prop:flows} the one-form $\sfn_J=\vk\cdot(\pr_J\sfm-\Lambda_{\sM J})$ is of the form $\mathsf{x}_{JKL}\hB^K\hB^L$ with a regular covariant $\mathsf{x}_{JKL}$ which is symmetric in $K$ and $L$, so that it has no term of order zero. We obtain
\begin{equation}
\pr_J\Pi' = -\cS'_{JI}\,\hB^I \, , \qquad \Pi' = \Pi-\vk\cdot\sfm \, , \qquad \cS'_{JI} = \cS_{JI}+(\mathsf{x}_{JKI}+\mathsf{x}_{IKJ}-\mathsf{x}_{KJI})\hB^K \, ,
\label{gr:eq:odd}
\end{equation}
by the identity for $\mathsf{x}_{JKL}\hB^K\hB^L$ used in the proof below. The term of order zero of $\cS'$ is that of $\cS$, and $\Pi'(g_*)=d\,\cW(g_*)/V$ at a zero. For the sextic theory this is Proposition~\ref{gs:prop:sextic}.

\begin{theorem}[Gradient formula with relevant sources]
\label{gr:cor:cond}
Assume Properties~\ref{gs:P:functional} to~\ref{gs:P:sphere} with all relevant sources, in either instance of Property~\ref{gs:P:functional}. By the last statement of Property~\ref{gs:P:weyl} they then also hold with the sources of even dimension in the scheme of Lemma~\ref{gr:lem:scheme}, see the end of section~\ref{gs:sec:divscheme}, and we take these sources in this scheme. Then
\begin{equation}
\pr_J\hat A^{\rm H} = \hat G^{\rm H}_{JI}\,\hB^I \, , \qquad
\hat A^{\rm H} = \Pi^\sharp-\sum\vk\!\cdot\sfm^\sharp \, , \qquad
\hat G^{\rm H}_{JL} = -\cS^\sharp_{JL}-\sum\left(\mathsf{x}_{JKL}+\mathsf{x}_{LKJ}-\mathsf{x}_{KJL}\right)\hB^K \, ,
\label{gsh:eq:H}
\end{equation}
where the sums run over the sources of even dimension, $\Pi^\sharp=(d+\hB\!\cdot\!\pr)\,\cW^\sharp/V$, $\sfm^\sharp$ and $\Lambda^\sharp_{\sM J}$ are the quantities of section~\ref{gs:sec:oneform} for the functional $W^\sharp$, and $\mathsf{x}_{JKL}=(\pr_J\sfm^\sharp-\Lambda^\sharp_{\sM J})\cdot x_{KL}$ with $x_{KL}$ of \eqref{gsh:eq:flows}. The metric $\hat G^{\rm H}_{JL}$ is regular, symmetric and covariant under the Lie algebra of the group of field rotations, the potential $\hat A^{\rm H}$ is invariant under the constant rotations of the identity component of this group, and $\hat A^{\rm H}-(\vep\,d/V)\,\cW^{\log}$ is regular, so that the potential is regular in instance~(a). At a zero $g_*$ of $\hB^I$ the potential takes the value \eqref{gs:eq:fixed}. The term of order zero in the couplings and in $\vep$ of the metric is
\begin{equation}
\hat G^{{\rm H}(0,0)}_{IJ} = \cK^{(0,0)}_{IJ}-2\sum\mathcal{M}^{(0,0)}_{IJ}\!\cdot\sfm^{(0,0)} \, .
\label{gsh:eq:G00}
\end{equation}
Assume in addition the last statement of Property~\ref{gs:P:free}. Then this term is $\cK^{(0,0)}_{IJ}$, the term of order zero of $\hat G^{\rm S}_{IJ}$. If Property~\ref{gs:P:free} holds, it is $s_d\,a_{d_c}\,c_{IJ}$ for even $d_c$, in either instance of Property~\ref{gs:P:functional}, so that the metric $\hat G^{\rm H}_{IJ}$ is nondegenerate if $c_{IJ}$ is invertible, and the metric $s_d\,\hat G^{\rm H}_{IJ}$ is positive definite to all orders in the couplings and in $\vep$ if $c_{IJ}$ is positive. For odd $d_c$ the relation \eqref{gr:eq:odd} holds without the last statement of Property~\ref{gs:P:free}, and under Property~\ref{gs:P:free} the term of order zero in the couplings and in $\vep$ of the metric $-s_d\,\cS'_{IJ}$ is $-s_d\,b_{d_c}\,c_{IJ}$.
\end{theorem}

\begin{proof}
By Proposition~\ref{gs:prop:sharp} and Lemma~\ref{gsh:lem:quant} the pair $W^\sharp$, $\cA^\sharp=0$ satisfies the hypotheses of Proposition~\ref{gs:prop:relevant}, whose first relation \eqref{gs:eq:KM}, summed over the sources of even dimension, reads
\begin{equation*}
0 = \pr_J\Big(\Pi^\sharp-\sum\vk_{\rm h}\!\cdot\sfm^\sharp\Big)+\cS^\sharp_{JI}\,\hB^I+\sum\Big[\vk_{\rm h}\!\cdot\left(\pr_J\sfm^\sharp-\Lambda^\sharp_{\sM J}\right)+\ell_{{\rm h}J}\!\cdot\sfm^\sharp\Big] \, .
\end{equation*}
Its proof uses only Property~\ref{gs:P:ward} and the statements of Property~\ref{gs:P:sphere} for the pair. By Proposition~\ref{gsh:prop:flows} we have $\ell_{\rm h}=0$ and $\vk_{\rm h}=\vk=\hB x\hB$, so that the last sum is $\sum\mathsf{x}_{JKL}\hB^K\hB^L$. For a tensor $\mathsf{x}_{JKL}$ which is symmetric in its last two indices, $\mathsf{x}_{JKL}\hB^K\hB^L=(\mathsf{x}_{JKL}+\mathsf{x}_{LKJ}-\mathsf{x}_{KJL})\hB^K\hB^L$, since the last two terms cancel after an exchange of the summation indices, and the bracket contracted with $\hB^K$ is symmetric in $J$ and $L$. This gives \eqref{gsh:eq:H}, and $\cS^\sharp$ is symmetric as the tensor of a second variation. Regularity, covariance and invariance follow from those of $W^\sharp$, $\vk$ and $x$, and the statement on $\cW^{\log}$ from \eqref{gsh:eq:Wsharp} as in Proposition~\ref{gs:prop:main}. At a zero $g_*$ the functions $\vk$, $\hB\!\cdot\!\pr\,\cW^\sharp$ and $\vk_{\rm c}\!\cdot\sfm$ vanish, so that $\hat A^{\rm H}(g_*)=d\,\cW^\sharp(g_*)/V=d\,(\vep\,\cW+V\rA)(g_*)/V$, which is \eqref{gs:eq:fixed}.

For the term of order zero we start from \eqref{gsh:eq:Ssharp}. The second term of $\hat G^{\rm H}$ contains $\hB^K$ and has no term of order zero. By Lemma~\ref{gre:lem:zero} the term of order zero of $(\vep-\tL)\cS$ vanishes at $\vep=0$. In \eqref{gsh:eq:N} every term except $2\mathcal{M}\cdot\sfm$ contains $\vk$, $\ell_{\rm c}$ or one of their first or second derivatives. By \eqref{gsh:eq:flows} these contain a factor $\hB^K$ or $\pr_J\hB^K$, which have no term of order zero in the couplings and in $\vep$. In instance~(b) a derivative $\pr_\iota=g_\iota\pr/\pr g_\iota$ of a regular function has no term of order zero, and $\hB^\iota$ is of order $\vep$ at $g=0$, so that the same holds. This gives \eqref{gsh:eq:G00}. Under the last statement of Property~\ref{gs:P:free} the term $\sfm^{(0,0)}$ vanishes. The remaining statements for even $d_c$ follow from Lemma~\ref{gre:lem:zero} and the remark after it, since a regular symmetric tensor with an invertible term of order zero has a regular inverse, and a term of order zero which is positive definite gives a metric which is positive definite to all orders. For odd $d_c$ the relation \eqref{gr:eq:odd} was derived before the theorem without the last statement of Property~\ref{gs:P:free}, and the term of order zero of $\cS'$ is that of $\cS$, which Property~\ref{gs:P:free} gives.
\end{proof}

For odd $d_c$ the term \eqref{gsh:eq:G00} vanishes, since $\cK$ and $\sfa_\sM$ vanish, and with them $\sfm^{(0,0)}$ by Lemma~\ref{gsh:lem:m}. For odd $d_c$ we therefore use \eqref{gr:eq:odd}, and we make a statement on the sign for this form. For a single source of dimension two \eqref{gsh:eq:G00} and \eqref{gsh:eq:am} give $\hat G^{{\rm H}(0,0)}-\cK^{(0,0)}=-2d_c\,\epsilon^{(0,0)}\cdot\sfm^{(0,0)}=-\tfrac{2d_c}{d_c-2}\,\epsilon^{(0,0)}\cdot\sfa^{(0,0)}_\sM$, since $\mathcal{M}=d\,\epsilon$.

The relation \eqref{gsh:eq:G00} does not change under a finite local term $F'=c\int d^dx\sqrt\gamma\,\sM\cdot X$ with a constant $c$ and a curvature scalar $X$ of dimension $d_c-D$, added to $W$ together with $\mathbb{W}_\sigma F'$ added to the anomaly. On the unit sphere $F'$ changes $\sfm$ by $c\,X_0$, where $X_0$ is the value of $X$ on the sphere. The anomaly at vanishing sources changes by $c\int d\mu_d\,X_0\,\mathcal{V}_\sigma$, and by \eqref{gr:eq:Eexp} and \eqref{gr:eq:Sigma} this changes $\cK^{(0,0)}_{IJ}$ by $2c\,X_0\mathcal{M}^{(0,0)}_{IJ}$, since the terms which come from $\pr\vk$ and $\pr\ell$ have no term of order zero in the divergence scheme, by Proposition~\ref{gsh:prop:flows}. The two changes cancel in \eqref{gsh:eq:G00}. A finite term such as $c\int\sqrt\gamma\,\sM R$ for $D=d_c-2$ or $c\int\sqrt\gamma\,\sM R^2$ for $D=d_c-4$ therefore does not change the term of order zero of $\hat G^{\rm H}$.

If the two-point form of appendix~\ref{gs:sec:properties} is assumed in the scheme from which the redefinition starts, the change of scheme alters the term of order zero of $\cS$ by terms which contain the first derivatives of $y_J$ and $y_0$ at $g=0$. They are products of these derivatives with $\sfm^{(0)}$, with the first harmonics of the free two-point functions of the relevant and the marginal operators, and with that of two relevant operators. They vanish if $y_J$ and $y_0$ vanish to second order at $g=0$.

The theorem concerns the relevant sources of even dimension $D$ with $2\leq D<d_c$. It uses $D<d_c$ for the invertibility of the operators of Lemmas~\ref{gr:lem:scheme} and~\ref{gsh:lem:inv}, whose term of order zero in the couplings and in $\vep$ is $d_c-D$, and in Lemma~\ref{gsh:lem:m}. For $d_c=2$ there is no relevant source of even dimension with $2\leq D<d_c$, and the gradient formula is that of section~\ref{gs:sec:gradient}.

\section{Consequences}
\label{gs:sec:consequences}

In this section we state the consequences of a gradient formula $\pr_IA=g_{IJ}\hB^J$ with a potential $A$ whose derivative with respect to the coordinates is regular, and with a regular symmetric metric $g_{IJ}$. In instance~(b) each of the potentials of sections~\ref{gs:sec:identities} and~\ref{gs:sec:relevant} is regular up to a term $c\,\cW^{\log}$, where the coefficient $c$ is a power series in $\vep$ which depends on the potential. The potential is invariant under the constant rotations of the identity component of the group of field rotations, and the metric is covariant under the Lie algebra of this group. Sections~\ref{gs:sec:identities} and~\ref{gs:sec:relevant} provide such a formula for every theory with the properties, by Proposition~\ref{gs:prop:main}, Proposition~\ref{gs:prop:odd} and, under its additional hypotheses, Theorem~\ref{gr:cor:cond}. The statements on signs require in addition that the metric is positive definite to all orders in the couplings and in $\vep$, which holds for even $d_c$ with the metric $s_d\hat G^{\rm H}_{IJ}$ and for odd $d_c$ with the metric $-s_d\,\cS'_{IJ}$ of the potential $s_d\,\Pi'$ in \eqref{gr:eq:odd}, if Property~\ref{gs:P:free} holds with a positive matrix $c_{IJ}$. This is the case with and without gauge fields, see section~\ref{gs:sec:gauge}. For even $d_c$, with one logarithmic coordinate and no other, it holds already if $-s_d\,n_\iota$ is positive at $\vep=0$. In both cases the potential with this metric is, at the zeros of $\hB^I$, a positive multiple of $\tilde F-\tilde F_0$ for small positive $\vep$, by \eqref{gs:eq:Ftilde} and by the discussion after Proposition~\ref{gs:prop:odd}. Here $\tilde F$ is defined as in section~\ref{gs:sec:fixed}, in the extended form of the properties or, for the four theories of Proposition~\ref{gs:prop:radius}, with the curvature couplings and the relevant sources at the zeros of their flows.

\subsection{Expansion in \texorpdfstring{$\vep$}{epsilon}}
\label{gs:sec:tower}

Let Property~\ref{gs:P:functional} hold in instance~(a), and let $\hB^I$ have the form \eqref{gs:eq:Bhat} of minimal subtraction, $\hB^I=-\vep\,k_I\,g^I+B^I$ with $B^I$ independent of $\vep$. We expand $A=\sum_n\vep^nA_n$ and $g_{IJ}=\sum_n\vep^ng_{n,IJ}$, where the coefficients are power series in the couplings. The gradient formula is then equivalent to the tower of relations
\begin{equation}
\pr_IA_0 = g_{0,IJ}\,B^J \, , \qquad
\pr_IA_n = g_{n,IJ}\,B^J-g_{n-1,IJ}\,k_J\,g^J \quad (n\geq1) \, .
\label{gs:eq:tower}
\end{equation}
The first relation is the gradient property at $d=d_c$. The others are conditions on the vector field $B^I$ which do not follow from it. The regularity of the gradient formula in $d=d_c-\vep$, with the term of order zero of the metric fixed, therefore implies relations among the coefficients of the beta function which are not required at $d=d_c$. For multiscalar $\phi^4$ theory with the beta function of minimal subtraction in momentum space, that is with Identification~\ref{id:ms}, the computation of appendix~\ref{app:fiveloop} finds the first of them at five loops.

\subsection{Fixed points}
\label{gs:sec:fixedgen}

By a fixed point we mean a real zero $g_*$ of $\hB^I$ at which the couplings tend to zero with $\vep$, as in section~\ref{gs:sec:theories}. At such a zero the derivative of the gradient formula gives
\begin{equation}
\pr_I\pr_JA = g_{IK}\,\pr_J\hB^K \, ,
\end{equation}
so that the matrix $g_{IK}\pr_J\hB^K$ is symmetric. The matrix $\pr_J\hB^K$ is therefore symmetric with respect to the inner product defined by $g_{IJ}$. If $g_{IJ}$ is positive definite, its eigenvalues are real, order by order in $\vep$. Since $\hB^I$ is covariant, the eigenvalue zero occurs along the directions $(\omega g_*)^I$ of the field rotations. The other eigenvalues determine the scaling dimensions of the nearly marginal operators at the fixed point, which are therefore real.

The potential satisfies $\hB^I\pr_IA=\hB^Ig_{IJ}\hB^J$, which is an identity of formal power series in the couplings and in $\vep$. Since $A$ is invariant under the identity component, its derivative along the beta function, which differs from $\hB^I$ by an infinitesimal rotation of the fields, is the same. Let the metric be positive definite to all orders, and let the couplings and $\vep$ be real power series in a parameter $t$ without terms of order zero. If the first nonvanishing term of $\hB^I$ along this family is $v^I\,t^m$, then the first nonvanishing term of $\hB^Ig_{IJ}\hB^J$ is $v^Ig^{(0,0)}_{IJ}v^J\,t^{2m}$, whose coefficient is positive. In this sense the potential increases along $\hB^I$ away from its zeros, and it decreases along the flow towards the infrared. We call a function with this property a monotone of the flow. Along a closed orbit the potential would return to its value, so that a monotone excludes closed orbits of the flow of $\hB^I$. The identity above holds to every order as an identity of formal power series. For the flow at leading order, which is a vector field, we show the absence of closed orbits explicitly below. The positivity enters only through the term of order zero of the metric, which under Property~\ref{gs:P:free} is fixed by the two-point functions of the free fields. No unitarity is used, and in non-integer dimension unitarity fails already for free scalar fields~\cite{Hogervorst:2015akt}. The reality of the eigenvalues above uses the symmetry and the definiteness of the metric, while the remaining statements of this subsection use only the monotone. For truncations of $\hB^I$, $A$ and $g_{IJ}$ at a finite order the gradient formula holds only through that order.

For the flow itself we consider couplings of order $\vep$ in instance~(a). With $g^I=\vep\,x^I$ we have $\hB^I=\vep^2\,V^I(x)+{\rm O}(\vep^3)$, where $V^I(x)=-k_Ix^I+B_2^I(x)$ and $B_2^I$ is the term of second order in the couplings of $\hB^I$ at $\vep=0$. The gradient formula gives $\pr A(\vep x)/\pr x^I=\vep^3g^{(0,0)}_{IJ}V^J(x)+{\rm O}(\vep^4)$, so that $A(\vep x)=A(0)+\vep^3a(x)+{\rm O}(\vep^4)$ with $\pr a/\pr x^I=g^{(0,0)}_{IJ}V^J$. The function $a$ therefore increases along the flow of $V^I$, strictly away from the zeros of $V^I$, and this flow, which is the flow of $\hB^I$ at leading order in $\vep$, has no closed orbits. Let $g_{\rm UV}=\vep\,x_{\rm UV}+{\rm O}(\vep^2)$ and $g_{\rm IR}=\vep\,x_{\rm IR}+{\rm O}(\vep^2)$ be fixed points which are power series in $\vep$,\footnote{A zero of $V^I$ at which the matrix of derivatives of $V^I$ is degenerate in a direction which is not a field rotation can continue to zeros of $\hB^I$ which contain fractional powers of $\vep$. The term of order $\vep^3$ of the difference of the potential and its sign are the same for such zeros, and the higher terms can then contain fractional powers.} and let $x_{\rm UV}$ and $x_{\rm IR}$ be the limits for $\log\mu\to+\infty$ and $\log\mu\to-\infty$ of a solution of $dx/d\log\mu=\vep\,V(x)$ which is not constant. Then the term of order $\vep^3$ of $A(g_{\rm UV})-A(g_{\rm IR})$ is $\vep^3\,(a(x_{\rm UV})-a(x_{\rm IR}))=\vep^4\int V^Ig^{(0,0)}_{IJ}V^J\,d\log\mu$, which is positive. Since the fixed points are power series in $\vep$, so is the difference $A(g_{\rm UV})-A(g_{\rm IR})$, and its leading term is positive. At the zeros the potential is a positive multiple of $\tilde F-\tilde F_0$ with a constant $\tilde F_0$, in the sense of section~\ref{gs:sec:fixed}. The difference $\tilde F_{\rm UV}-\tilde F_{\rm IR}$ is thus a power series in $\vep$ with a positive leading term, and in this sense, which is the sense in which the metric is positive definite to all orders,
\begin{equation}
\tilde F_{\rm UV} > \tilde F_{\rm IR} \, .
\label{gs:eq:Fmono}
\end{equation}
That $\tilde F$ decreases along the flow was found in~\cite{Giombi:2014xxa} in all examples considered there, and for flows generated by weakly relevant operators it was shown at leading order in conformal perturbation theory in~\cite{Klebanov:2011gs} for odd dimensions and in~\cite{Fei:2015oha} for all dimensions. The inequality \eqref{gs:eq:Fmono} is this statement for fixed points with couplings of order $\vep$ which are power series in $\vep$, where the sign of the difference is that of its term of order $\vep^3$. The gradient formula adds that the difference is a power series in $\vep$, and that the potential is a monotone along every trajectory of the flow to all orders, in the sense stated above. The inequality uses the trajectory which connects the fixed points at leading order only, and this trajectory need not continue to higher orders. The statements of this subsection do not change under a redefinition of the couplings which is regular and covariant.

\part{Specific theories}
\label{gs:part:three}

\section{Scalar fields and Dirac fermions in minimal subtraction}
\label{gs:sec:universal}

The general proof of the second part uses only the properties. In the third part of the paper we supply the field content and the regularisation for specific theories. We construct the functional on the sphere, derive the properties from Hypothesis~\ref{hyp:finite}, and evaluate the form of the gradient formula for each theory. In this section we derive the statements which hold for every theory of scalar fields and Dirac fermions in dimensional regularisation with minimal subtraction. We construct the bare functional on $S^d$ for non-integer $d$ for scalar fields with polynomial interactions without derivatives in section~\ref{gs:sec:sphere}, and for scalar fields and Dirac fermions with Yukawa couplings in section~\ref{gs:sec:dirac}. For free gauge fields with a local coupling it is constructed in appendix~\ref{gs:app:freegauge}, and for Dirac fermions with four-fermion couplings of every gamma structure in appendix~\ref{gs:app:fourfermion}. Section~\ref{gs:sec:msanomaly} derives the relation $\mathbb{W}_\sigma W=\cA[\sigma]$ and the form of the anomaly of Property~\ref{gs:P:anomaly} from the Weyl invariance of the bare functional and the locality of the counterterms, given a variation of the sources of the form of Property~\ref{gs:P:weyl} which induces the rescaling of the bare sources, and Lemma~\ref{gcu:lem:commute} there gives the commutation of the Weyl variations. Section~\ref{gs:sec:msward} derives Property~\ref{gs:P:ward} from the conformal covariance of the bare functional on the sphere. Section~\ref{gs:sec:remains} derives the two-point form of the functional at second order in the couplings from conditions on the bare sources, and the two-point form leads to the statements of Property~\ref{gs:P:free} other than the last. These derivations take three inputs from the theory. The first is the form and the covariance of its bare sources and counterterms, and the second is a variation of the sources which induces on the bare sources their Weyl rescaling up to a background gauge transformation. The third is the finiteness of the renormalised functional, from which we derive the regularity of the coefficients of the Weyl variation and of the anomaly for the theories of sections~\ref{gs:sec:phi4} and~\ref{gs:sec:other}. For multiscalar $\phi^4$ theory we derive these inputs, and with them all six properties, from Hypothesis~\ref{hyp:finite} in section~\ref{gs:sec:phi4}. For the sextic, the Yukawa and the cubic theory we do so in sections~\ref{gs:sec:sextic}, \ref{gs:sec:yukawa} and~\ref{gs:sec:cubic} and in the corresponding appendices. Table~\ref{gs:tab:where} lists for each property and each of these four theories where it is proved.

\subsection{Anomaly from the counterterms}
\label{gs:sec:msanomaly}

In dimensional regularisation with minimal subtraction the renormalised functional has the form
\begin{equation}
W = W_0\big[\gamma,g_0,a_0,\sM_0\big]+\int d^dx\sqrt\gamma\;\cX(\gamma,g,a,\sM) \, ,
\label{gs:eq:Wren}
\end{equation}
where $W_0$ is the functional of the bare sources $g_0$, $a_0$ and $\sM_0$. The bare sources are local functions of the sources whose coefficients are pure poles in $\vep$, and $\cX$ is a local density of counterterms which do not depend on the fields. The proposition below holds under three conditions, each of which is established for the $\phi^4$, the Yukawa, the sextic and the cubic theory, as follows. The first concerns the bare sources and the counterterms. They are built covariantly from $\gamma_{\mu\nu}$ with $d$ as a parameter in the sense of Property~\ref{gs:P:weyl}, with covariant coefficients, $\cX$ is invariant under background gauge transformations, and each term of $\cX$ has the dimension $d_c$, where $g^I$ counts with dimension zero, $a_\mu$ with dimension one and each component of $\sM$ with its dimension at $d=d_c$. This is the form in which Hypothesis~\ref{hyp:finite} states the bare sources and the counterterms, with the curvature counterterms on the sphere given by Proposition~\ref{prop:finite} and its analogues, and the covariance of the coefficients follows from the hypothesis, by Proposition~\ref{prop:uniqueall} for $\phi^4$ theory, Proposition~\ref{gc:prop:cov} for the Yukawa theory, Proposition~\ref{gsx:prop:uniqueall} for the sextic theory and Proposition~\ref{gcp:prop:cov} for the cubic theory. The second condition is that the bare functional is invariant under background gauge transformations of the bare sources, and under a Weyl rescaling of the metric with $\sigma(x)$ if the bare couplings are rescaled with the weight $-\vep\,k_I$, each component of $\sM_0$ with a weight which we denote by $w_\sM$, and the bare gauge field is not changed. For scalar fields this holds by the construction of section~\ref{gs:sec:sphere}, see appendix~\ref{app:rotations} and Proposition~\ref{prop:gauge}, for scalar fields with Dirac fermions by Proposition~\ref{gf:prop:bare}, and for free gauge fields, which have no field rotations, by the definition of appendix~\ref{ggs:sec:props} on the conformal rescalings of the sphere and Proposition~\ref{ggs:prop:ward}. The third condition is that there is a variation $\delta^{\rm W}_\sigma$ of the sources of the form of Property~\ref{gs:P:weyl} which induces on the bare sources their rescaling with these weights, up to a background gauge transformation of the bare sources. It is constructed from Hypothesis~\ref{hyp:finite} in section~\ref{sec:weylvar} for $\phi^4$ theory, and in appendices~\ref{gs:app:fermions}, \ref{gs:app:sextic} and~\ref{gs:app:cubic} for the Yukawa, the sextic and the cubic theory.

\begin{proposition}[Anomaly from the counterterms]
\label{gs:prop:msanomaly}
Under these conditions $\mathbb{W}_\sigma W=\cA[\sigma]$ holds on the backgrounds on which the bare functional is defined, with the operator $\mathbb{W}_\sigma$ of Property~\ref{gs:P:weyl} and
\begin{equation}
\cA[\sigma] \equiv \mathbb{W}_\sigma\int d^dx\sqrt\gamma\;\cX \, .
\label{gs:eq:msanomaly}
\end{equation}
The functional $\cA[\sigma]$ is linear in $\sigma$ and local, it is built covariantly from $\gamma_{\mu\nu}$ with $d$ as a parameter, it is invariant under background gauge transformations, and it is homogeneous of Weyl weight $-\vep$ in the sense of Property~\ref{gs:P:anomaly}. If the operators $\mathbb{W}_\sigma$ commute for all $\sigma$ and $\sigma'$, then $\mathbb{W}_\sigma\cA[\sigma']=\mathbb{W}_{\sigma'}\cA[\sigma]$.
\end{proposition}

\begin{proof}
By assumption the variation $\mathbb{W}_\sigma$ of the metric and of the sources changes the bare sources by their rescaling and by a background gauge transformation. Both leave $W_0$ invariant, so that $\mathbb{W}_\sigma$ annihilates the first term of \eqref{gs:eq:Wren}, and $\mathbb{W}_\sigma W$ is \eqref{gs:eq:msanomaly}. The variations of Property~\ref{gs:P:weyl} are local, and they are built covariantly with $d$ as a parameter, so that $\cA[\sigma]$ is local and built in the same way. Since the coefficients of $\hB^I$, $\hr_I$ and $\delta^{\rm W}_\sigma\sM$ are covariant, the operator $\mathbb{W}_\sigma$ commutes with background gauge transformations, and the gauge invariance of $\cA[\sigma]$ follows from that of $\cX$. For the homogeneity we replace $\gamma_{\mu\nu}$ by $r_0^2\,\gamma_{\mu\nu}$ with a constant $r_0$ and multiply each component of $\sM$ by $r_0$ to the power of minus its dimension at $d=d_c$. Each term of $\sqrt\gamma\,\cX$ is then multiplied by $r_0^{d-d_c}=r_0^{-\vep}$. The operator $\mathbb{W}_\sigma$ commutes with this rescaling, because each term of $\delta^{\rm W}_\sigma\sM$ has the dimension of $\sM$ at $d=d_c$ and the other terms of the operator $\mathbb{W}_\sigma$ do not change. Hence $\cA[\sigma]$ is multiplied by $r_0^{-\vep}$. The last statement follows from $\mathbb{W}_\sigma\cA[\sigma']=\mathbb{W}_\sigma\mathbb{W}_{\sigma'}\int d^dx\sqrt\gamma\,\cX$.
\end{proof}

The proposition gives Property~\ref{gs:P:anomaly} except for the regularity of the coefficients of $\cA[\sigma]$. It uses the form of the Weyl variation of Property~\ref{gs:P:weyl}, which the third condition provides, but not the regularity of $\hB^I$, $\hr_I$ and of the coefficients of $\delta^{\rm W}_\sigma\sM$, nor the commutation of the operators $\mathbb{W}_\sigma$. In minimal subtraction these coefficients are Laurent series in $\vep$, and their regularity is derived from the finiteness of the renormalised functional for each theory, see section~\ref{gs:sec:remains}. The commutation is a set of relations among the coefficients, which for a source of dimension two are the relations \eqref{gs:eq:Mrel} of section~\ref{gs:sec:dimtwo}. If the background gauge transformation in the last assumption has the parameter $\sigma\,\xi^{\rm L}(g)$ with a covariant $\xi^{\rm L}$, these relations hold identically in the coefficients of the bare sources, by the following lemma. The anomaly \eqref{gs:eq:msanomaly} is the Weyl variation of a local functional, which is the form discussed after the properties in appendix~\ref{gs:sec:properties}.

\begin{lemma}
\label{gcu:lem:commute}
Let the bare sources be local functions of the metric and of the sources, of the form of this section and invertible as formal power series, and let the variation $\mathbb{W}_\sigma$ of the metric and of the sources induce on the bare sources their rescaling with constant weights together with the background gauge transformation with the parameter $\sigma\,\xi^{\rm L}(g)$, where $\xi^{\rm L}$ is covariant, as in Lemma~\ref{lem:rhoxi}. Then $[\mathbb{W}_\sigma,\mathbb{W}_{\sigma'}]=0$.
\end{lemma}

\begin{proof}
Let $\Phi$ be the map from the metric and the sources to the metric and the bare sources. It is invertible, since it is triangular: $g_0$ is $g$ plus terms of degree at least two, $a_0$ is affine in $a_\mu$ with the coefficient one, and the bare relevant sources are affine in the relevant source of the same dimension with an invertible factor $Z$. By assumption $\mathbb{W}_\sigma$ is the pullback by $\Phi$ of the variation $\mathbb{V}_\sigma=\mathbb{R}_\sigma+\mathbb{G}_{\sigma\xi^{\rm L}}$ of the bare data, where $\mathbb{R}_\sigma$ rescales the metric by $2\sigma$ and each bare source by $\sigma$ times its weight, and $\mathbb{G}_\omega$ is the gauge transformation with the parameter $\omega$, in which $g$ is the function of the bare data given by the inverse of $\Phi$. It suffices to show $[\mathbb{V}_\sigma,\mathbb{V}_{\sigma'}]=0$. The rescalings commute with each other, since they multiply each component by $\sigma$ times a constant. A gauge transformation acts linearly on the bare sources of weight different from zero and by $D\omega$ on the bare gauge field, whose weight is zero, so that it commutes with the rescalings up to the change of its parameter. The change of the parameter $\sigma'\xi^{\rm L}(g)$ under $\mathbb{V}_\sigma$ is $\sigma\sigma'\,\hB^I\pr_I\xi^{\rm L}$, since $\mathbb{V}_\sigma$ changes $g$ by $\sigma\hB$, and it is symmetric in $\sigma$ and $\sigma'$. The commutator of two gauge transformations has the parameter $\sigma\sigma'[\xi^{\rm L},\xi^{\rm L}]=0$. Hence $[\mathbb{V}_\sigma,\mathbb{V}_{\sigma'}]=0$.
\end{proof}

\subsection{Regularised functional on the sphere}
\label{gs:sec:sphere}
\label{sec:sphere}

The Ward identity of Property~\ref{gs:P:ward} requires the bare functional on the sphere $S^d$ for non-integer $d$ and its behaviour under conformal maps.

Beyond the perturbation theory on the sphere with coupling constants which is recalled in the introduction, our argument needs three things, since it takes functional derivatives with respect to couplings which depend on the position on $S^d$ and uses the Ward identity of the conformal Killing vectors and the background gauge invariance as exact statements for non-integer $d$. These are a definition of the sources on $S^d$ for non-integer $d$, the meromorphy in $d$ of every term of the functional with control of its poles, and the conformal covariance and the gauge invariance of the functional to all orders. We supply them in this subsection and in appendices~\ref{app:mero} to~\ref{app:gauge}. The definition extends the integral in $d$ dimensions of~\cite{Wilson:1972cf,Collins:1984xc} from the loop momenta to the positions of the vertices on the sphere.

In this subsection we define this functional for scalar fields with polynomial interactions by analytic continuation in two parameters. We consider $N$ real scalar fields $\phi_i$ with the free action $\int d^dx\sqrt\gamma\,[\tfrac12\,\gamma^{\mu\nu}D_\mu\phi_i\,D_\nu\phi_i+\tfrac12\,\xi_d\,R\,\phi_i\phi_i]$ of a conformally coupled field, where $\xi_d=(d-2)/(4(d-1))$, and with vertices without derivatives
\begin{equation}
V^{(n)}[\psi] = \frac{1}{n!}\,\psi_{i_1\ldots i_n}\,\phi_{i_1}\cdots\phi_{i_n} \, , \qquad n\geq1 \, ,
\label{gs:eq:vertexn}
\end{equation}
whose sources $\psi$ are symmetric tensors, together with the vertex of the current and the term of second order in the gauge field, see appendix~\ref{app:gauge}. The couplings and the relevant sources are sources of such vertices. For instance, sextic couplings are the sources of $V^{(6)}$, cubic couplings those of $V^{(3)}$, and the mass source is the source of $V^{(2)}$. The group of field rotations is a compact subgroup of ${\rm O}(N)$ which acts on the sources. The appendices to which we refer are written for the vertices of $\phi^4$ theory, that is for $n=4$ and $n=2$. Their proofs treat a vertex through its legs, and they do not use the number of legs of a vertex. Where this matters we say so. In appendices~\ref{app:rotations}, \ref{app:polar} and~\ref{app:conf} the weights $\vep$ and $2$ of the bare couplings and of the bare mass source stand for $w_n$ below, and the coordinate $\vep=4-d$ on the line stands for $d_c-d$.

\subsubsection{Sources and measure}

In integer dimension a local source is an arbitrary smooth function on the sphere. For non-integer $d$ the sphere is not a set of points, and an arbitrary function of a point has no meaning. We therefore admit as sources only functions of a restricted kind, which we choose such that the following holds. In integer dimension, if the number of directions on which the sources do not depend is at least the number of points of a term, this term, evaluated on such sources, is an integral in which $d$ enters only through an explicit weight and through the exponent and the normalisation of the propagator. This expression is then the definition of the term for non-integer $d$. We call the set of functions which are admitted the class of sources. The functions which we admit are those which depend on a fixed integer number of the coordinates only. In integer dimension they are the functions which are invariant under the rotations of the remaining directions. The integrand of a term with several points then depends on the remaining directions only through the angles between the transverse parts of the points, and the integral over these directions gives a weight which is analytic in $d$. This is the construction of the integral in $d$ dimensions of Wilson, eqs.~(A14) and~(A15) in the appendix of~\cite{Wilson:1972cf}, in the form given in sections~4.1 and~4.2 of~\cite{Collins:1984xc}, on which dimensional regularisation is based. There it is applied to the loop momenta, and the external momenta span a subspace of integer dimension, which plays the role of our fixed coordinates. Here we apply it to the positions of the vertices on the sphere.

We describe the unit sphere $S^d$ by its embedding coordinates $X$ with $X^2=1$, and we split them as $X=(X_\parallel,X_\perp)$. The parallel part $X_\parallel$ consists of $k=D_0+1$ ordinary coordinates, for a fixed integer $D_0\geq1$, and the sources depend on these coordinates only. The transverse part $X_\perp$ stands for the remaining $m=d-D_0$ directions, where $m$ is not an integer. It enters the integrand only through its length, which is fixed by $X^2=1$, and through the angles between the transverse parts of different points. The integer $D_0$ is the number of parallel coordinates in a stereographic chart, see section~\ref{sec:chart}. A function of $D_0+1$ coordinates is in particular a function of $D_0+2$ coordinates which does not depend on the last one. A source in the class of sources with $D_0$ is therefore also in the class of sources with $D_0+1$, and the integral which we define below has the same value in both cases, see appendix~\ref{app:rotations}. We take the sources of the vertices without derivatives to be smooth functions of $X_\parallel$ on the closed unit ball $|X_\parallel|\leq1$, which is the range of $X_\parallel$, and the gauge field to be a one-form $a_A(X_\parallel)\,dX^A$ with smooth components, where the index $A$ runs over the parallel directions. Here a smooth function on the closed unit ball is the restriction of a smooth function on $\mathbb{R}^k$. We refer to these sources as the class of sources on the sphere. At order $n$ in the sources a term of the functional is an integral over $n$ points $X_1,\ldots,X_n$ of the sphere. The integrand is a product of the sources, of propagators $C_\phi/s_{ij}^{2\Dp}$ and, for the vertices of the current, of their derivatives $(\delta_{AB}-X_AX_B)\,\pr_B$ along the sphere. Here $s_{ij}$ is the chordal distance between the points $i$ and $j$, which in the split coordinates reads
\begin{equation}
s_{ij}^2 = 2-2\,X_{i\parallel}\!\cdot\!X_{j\parallel}-2\,\rho_i\rho_j\,u_{ij} \, , \qquad
\rho_i = \big(1-X_{i\parallel}^2\big)^{1/2} \, ,
\label{eq:chordal}
\end{equation}
where $u_{ij}$ is the cosine of the angle between $X_{i\perp}$ and $X_{j\perp}$. The integrand is thus a function $F$ of the $X_{i\parallel}$ and of the $u_{ij}$. For such a function we define
\begin{equation}
\int\prod_{i=1}^n d\mu_d(X_i)\,F = \prod_{l=0}^{n-1}\vol_{m-l}\int\prod_{i=1}^n d^kX_{i\parallel}\;\rho_i^{\,m-2}
\int_{U>0}\prod_{i<j}du_{ij}\;(\det U)^{\frac{m-n-1}{2}}\,F \, ,
\label{eq:gram}
\end{equation}
where $U$ is the symmetric matrix with entries $u_{ij}$ and unit diagonal, the integral is over the positive matrices of this form, and $\vol_l=2\pi^{l/2}/\Gamma(l/2)$ is the volume of the unit sphere in $l$ dimensions, with $k=D_0+1$ and $m=d-D_0$ as above. For integer $m\geq n$ the right-hand side is the integral of $F$ over $n$ copies of the sphere $S^d$, since $(\det U)^{(m-n-1)/2}$ is the density of the scalar products of $n$ unit vectors in $m$ dimensions. For other values of $m$ it is the definition of the left-hand side. The right-hand side contains $D_0$ through $k$ and $m$, while its value is the same for every $D_0$ for which the sources lie in the class of sources with $D_0$, see appendix~\ref{app:rotations}.

\subsubsection{Definition by continuation}
\label{sec:continuation}

We treat the exponent $\Dp$ of the propagator as a second complex parameter. For ${\rm Re}\,d\geq D_0+n$ and ${\rm Re}\,\Dp$ sufficiently negative the integral \eqref{eq:gram} converges. It extends to a meromorphic function of $d$ and $\Dp$, whose poles lie on a locally finite set of lines. This is Proposition~\ref{prop:mero}, which we prove in appendix~\ref{app:mero} by a decomposition into sectors. Its proof allows any number of propagators between two points and at a point, and it holds for every value of the exponent $\Dp$.

We define the term of order $n\geq1$ of the bare functional $W_0^S$ on the sphere as the value of the integral on the line
\begin{equation}
\Dp = \tfrac12(d-2) \, ,
\label{eq:line}
\end{equation}
on which the propagator is the Green function of the conformally coupled Laplacian, with $C_\phi=\Gamma(\tfrac d2-1)/(4\pi^{d/2})$. This definition requires that the line \eqref{eq:line} is not a line of poles of the integral, in the sense of appendix~\ref{app:polar}. The value on the line is then a meromorphic function of $d$ by Lemma~\ref{lem:poleset}. We prove that the line is not a line of poles in appendix~\ref{app:line}, for any number of points. This is Proposition~\ref{prop:linecurrent}, for the vertices without derivatives and for those of the current, and its proof uses the number of propagators at each point and not the number of legs of a vertex. There we also show that the integrals with the normalisations $C_\phi$ of the propagators taken out have no lines of poles at a fixed positive dimension, and the statements on lines of poles in appendices~\ref{app:mero} and~\ref{app:line} concern these integrals. A term of the functional is a finite sum of them with coefficients which are polynomials in $d$ and $\Dp$ times powers of $C_\phi$. The normalisation $C_\phi$ has poles at $d=2,0,-2,\ldots$ and is holomorphic and non-zero elsewhere, in particular near the critical dimensions $d_c=3,4,6$ of the scalar theories which we consider. The poles of the functional at $\vep=0$ therefore come from lines which cross the line \eqref{eq:line} and depend on $\Dp$. On a line other than \eqref{eq:line} the value of the integrals is defined if that line is not a line of poles. Appendix~\ref{app:line} shows this for \eqref{eq:line} only.

The parameter $\Dp$ only serves to have a region in which all integrals converge. The functional is multilinear in the sources by construction, and propagators at coincident points vanish, because their continuation from negative ${\rm Re}\,\Dp$ does. The term of order zero in the sources is the free energy of the free fields. It is not an integral of the form \eqref{eq:gram}. We fix it by the convention that $W_0^S$ vanishes for vanishing sources. The gradient formula contains only derivatives with respect to the couplings and is not affected by this convention. The potential changes by a constant if the standard free energy of the free fields is used instead. For fixed $\Dp$ in the region of convergence the integrals \eqref{eq:gram} are integrals in $d$ dimensions in the sense of~\cite{Wilson:1972cf,Collins:1984xc}. The relation to dimensional regularisation in momentum space, in which $\Dp$ corresponds to an analytic regularisation of the propagators, is described for $\phi^4$ theory in appendix~\ref{app:momentum}.

\subsubsection{Conformal covariance}

Let $\hat b$ be a unit vector of the parallel directions, and let $f$ be the conformal map of the sphere which is generated by a boost with rapidity $\zeta$ along $\hat b$ in the embedding space. With
\begin{equation}
\Omega(X) = \frac{1}{\cosh\zeta+\sinh\zeta\;\hat b\!\cdot\!X}
\end{equation}
it acts as
\begin{equation}
\hat b\!\cdot\!X' = \frac{\hat b\!\cdot\!X+\tanh\zeta}{1+\tanh\zeta\;\hat b\!\cdot\!X} \, , \qquad
X'_T = \Omega\,X_T \, , \qquad X'_\perp = \Omega\,X_\perp \, ,
\label{eq:boost}
\end{equation}
where $X_T$ denotes the parallel components which are orthogonal to $\hat b$. Together with the rotations of the parallel directions the maps $f$ generate the identity component of the group ${\rm SO}(D_0+1,1)$, which is a subgroup of the conformal group of the sphere. Note that on the sphere these maps are smooth. They include maps which are singular in a stereographic chart, and the sources in Proposition~\ref{prop:cov} below need not vanish anywhere. On the line \eqref{eq:line} the source of a vertex $V^{(n)}$ has the weight
\begin{equation}
w_n = d-\tfrac12\,n\,(d-2) \, ,
\label{gs:eq:wn}
\end{equation}
which is $\vep\,k$ for the couplings of the operators which are marginal at $d=d_c$. For instance $w_4=\vep$ at $d_c=4$, $w_6=2\vep$ at $d_c=3$, $w_3=\tfrac12\vep$ at $d_c=6$, and $w_2=2$ for the mass source in every dimension.

\begin{proposition}
\label{prop:cov}
Away from the poles the bare functional on the sphere satisfies
\begin{equation}
W_0^S\big[\{\Omega^{w_n}\,\psi_{n}\!\circ\!f\},\;f^*a_0\big]
= W_0^S\big[\{\psi_{n}\},\;a_0\big] \, ,
\label{eq:covS}
\end{equation}
where $\psi_n$ are the bare sources of the vertices $V^{(n)}$ and $f^*a_0$ is the pullback of the one-form.
\end{proposition}

The proof is a change of variables in \eqref{eq:gram} in the region of convergence, which continues to the line \eqref{eq:line} by Proposition~\ref{prop:mero}. We give it in appendix~\ref{app:conf} for $n=4$ and $n=2$. A vertex with $n$ legs at the point $i$ receives the factor $\Omega_i^{-n\Dp}$ from its propagators and the factor $\Omega_i^{\,d}$ from the measure, so that the change of variables gives the weight $d-n\Dp$ for every $n$ and every $\Dp$, which is \eqref{gs:eq:wn} on the line. Off the line \eqref{eq:line} the gauge field has the weight $d-2-2\Dp$, and the term of second order in the gauge field has an independent source $\Upsilon$ of weight $d-2\Dp$, see appendix~\ref{app:conf}. On the line this source is $\gamma^{\mu\nu}a_{0\mu}^Ta_{0\nu}$, and the two weights are $0$ and $2$. The generator of the boost along $\hat b$ is a conformal Killing vector of the sphere. For a vector $b$ of the parallel directions we write
\begin{equation}
\sigma = b\!\cdot\!X \, , \qquad v^\mu = -\nabla^\mu\sigma \, , \qquad
\nabla_\mu v_\nu+\nabla_\nu v_\mu = 2\,\sigma\,\gamma_{\mu\nu} \, , \qquad \nabla_\mu\nabla_\nu\sigma = -\sigma\,\gamma_{\mu\nu} \, .
\label{eq:ckv}
\end{equation}
To first order in $\zeta$ the map \eqref{eq:boost} with $b=\hat b$ is $X'=X-\zeta\,v$ with $\Omega=1-\zeta\,\sigma$. The first-order form of \eqref{eq:covS}, which we justify at the end of appendix~\ref{app:conf}, is therefore the invariance of $W_0^S$ under the variation
\begin{equation}
\delta\psi_n = -w_n\,\sigma\,\psi_n-v^\mu\pr_\mu\psi_n \, , \qquad \delta a_{0\mu} = -\cL_va_{0\mu} \, .
\label{eq:barevar}
\end{equation}
For the bare couplings $g_0^I$ of weight $k_I$ the first expression is $\delta g_0^I=-\vep\,k_I\,\sigma\,g_0^I-v^\mu\pr_\mu g_0^I$.

\subsubsection{Background gauge invariance}

On the line \eqref{eq:line}, where the propagator is a Green function, the bare functional is invariant under the background gauge transformations of the bare sources, under which a source $\psi_n$ is rotated and $a_0$ changes by $D\omega$. This is Proposition~\ref{prop:gauge}, which we prove in appendix~\ref{app:gauge}. The proof replaces a leg $\phi$ by $-\omega\phi$ in each leg in turn, and for a vertex without derivatives this gives $V^{(n)}[\omega\psi]$ for every $n$, so that the identity \eqref{eq:pointwise} of appendix~\ref{app:gauge} holds. Lemmas~\ref{lem:ward2} and~\ref{lem:contact} treat the legs of the vertices one pair at a time, and the property \eqref{eq:NP} is proved in appendix~\ref{app:npcurrent} for integrals in which the number of lines at a point is arbitrary. Off the line the propagator is not a Green function, and the two identities of Lemmas~\ref{lem:ward2} and~\ref{lem:contact}, which give the invariance, combine only on the line.

\subsubsection{Rotations}

The Ward identity uses three properties of the functional for constant sources. The one-point function is constant, the two-point function depends only on the angle between the points, and the function $\sigma$ of \eqref{eq:ckv} is an eigenfunction of the two-point function. In integer dimension they follow from the rotations of the sphere. For non-integer $d$ the rotations which mix the parallel and the transverse directions do not act on our sources, and we derive the three properties from the definition \eqref{eq:gram} in appendix~\ref{app:rotations}. The derivation uses only that the integral over the remaining points is again of the form \eqref{eq:gram} if one point is held fixed, and it holds for all vertices without derivatives. In particular, for the two-point function $W_{IJ}$ of the operators of two such vertices at constant sources we have
\begin{equation}
\int d\mu_d(X)\;\sigma(X)\,W_{IJ}(X,Y) = \Lambda_{IJ}\,\sigma(Y) \, , \qquad
\Lambda_{IJ} = \int d\mu_d(X)\;\cos\theta(X,Y)\;W_{IJ}(X,Y) \, ,
\label{eq:Lambda}
\end{equation}
where $\theta$ is the angle between the two points and $\Lambda_{IJ}$ is a constant symmetric matrix. For non-integer $d$ there is no function of two points, and the precise definition of $\Lambda_{IJ}$ through the second variation of the functional is \eqref{eq:Lambdadef} in appendix~\ref{app:rotations}. The one-point function of the current vanishes at constant sources, as we also show there.

\subsubsection{Flat chart}
\label{sec:chart}

Let $P$ be a point of the equator $X_\perp=0$. The stereographic projection from $P$ is
\begin{equation}
x_\parallel = \frac{X_T}{1-P\!\cdot\!X} \, , \qquad x_\perp = \frac{X_\perp}{1-P\!\cdot\!X} \, ,
\end{equation}
where $X_T$ now denotes the $D_0$ parallel components orthogonal to $P$. With $\Omega=2/(1+x^2)$ we have $d^kX_\parallel\,\rho^{m-2}=\Omega^{d}\,d^{D_0}x_\parallel\,|x_\perp|^{m-1}d|x_\perp|$ for every $d$, see appendix~\ref{app:rotations}, and the $u_{ij}$ are not changed. A smooth function of $X_\parallel$ which vanishes near $P$ is a smooth function of $x_\parallel$ and $|x_\perp|^2$ with compact support. We call the smooth functions of $x_\parallel$ and $|x_\perp|^2$ with compact support, and the one-forms $\alpha_a\,dx_\parallel^a+\beta\,x_\perp\!\cdot dx_\perp$ whose coefficients $\alpha_a$ and $\beta$ are functions of this kind, the class of flat sources. Here the index $a$ runs over the $D_0$ parallel directions. In integer dimension these are the one-forms with smooth components which are invariant under the orthogonal group ${\rm O}(d-D_0)$ of the transverse directions, which contains the reflections. Note that such a one-form has a radial component in the transverse directions. We define the flat functional $W_0^{\rm fl}$ for flat sources by \eqref{eq:gram} with the flat measure $d^{D_0}x_\parallel\,|x_\perp|^{m-1}d|x_\perp|$ in place of $d^kX_\parallel\,\rho^{m-2}$, with the distances $|x_i-x_j|$ in place of $s_{ij}$, and with the same continuation in $\Dp$. For sources on the sphere which vanish near $P$ we then have
\begin{equation}
W_0^S\big[\{\psi_n\},\;a_0\big] = W_0^{\rm fl}\big[\{\Omega^{w_n}\psi_n\},\;a_0\big] \, ,
\label{eq:chart}
\end{equation}
because the stereographic projection is a smooth change of variables on the support, with $s_{ij}^2=\Omega_i\Omega_j|x_i-x_j|^2$.

More generally we need backgrounds which are conformally flat, $\gamma_{\mu\nu}=\Omega(x)^2\delta_{\mu\nu}$ on $\mathbb{R}^d$ with a positive function $\Omega$ of $x_\parallel$ and $|x_\perp|^2$. For non-integer $d$ we define the bare functional on such a background by the flat functional with the sources rescaled by $\Omega^{w_n}$, which is \eqref{eq:defcurved} in appendix~\ref{app:rotations} for $\phi^4$ theory. This is an identity of the classical actions for every $d$, by the change of variables $\phi\to\Omega^{-(d-2)/2}\phi$. The invariance under Weyl rescalings of the bare functional then holds identically. By \eqref{eq:chart} the functional on the sphere agrees with this definition for sources which vanish near the pole of the chart.

\subsubsection{Local terms}
\label{sec:local}

The counterterms and the bare sources contain the Laplacian and the curvature of the sphere. Both are defined with $d$ as a parameter, and $\nabla^2\sigma=-d\,\sigma$ holds for the function $\sigma$ of \eqref{eq:ckv}, see appendix~\ref{app:rules}. The variation of the integral of the counterterms under the combination of the map \eqref{eq:boost} and the Weyl rescaling which restores the metric of the sphere is therefore given by the algebra of section~\ref{gs:sec:msanomaly}, that is by the anomaly \eqref{gs:eq:msanomaly}.

We collect the results of this subsection in a proposition.

\begin{proposition}[Bare functional for scalar fields]
\label{gs:prop:bare}
Consider $N$ scalar fields with the vertices \eqref{gs:eq:vertexn} for any set of numbers $n\geq1$, the vertex of the current and the term of second order in the gauge field, with sources in the class of sources on the sphere. Then each term of the bare functional is a meromorphic function of $d$ and $\Dp$. With the source $\Upsilon$ of the term of second order in the gauge field kept independent, it satisfies the covariance \eqref{eq:cov3} of appendix~\ref{app:conf}, in which the source of $V^{(n)}$ has the weight $d-n\Dp$, the gauge field the weight $d-2-2\Dp$ and $\Upsilon$ the weight $d-2\Dp$. On the line \eqref{eq:line}, with $\Upsilon=\gamma^{\mu\nu}a_{0\mu}^Ta_{0\nu}$, each term is a meromorphic function of $d$, the functional satisfies \eqref{eq:covS} with the weights \eqref{gs:eq:wn} away from the poles, it is invariant under background gauge transformations, and at constant sources it has the properties under rotations of \eqref{eq:Lambda}.
\end{proposition}

The proof consists of the arguments of appendices~\ref{app:rotations} and~\ref{app:mero} to~\ref{app:gauge}, which we described above. Dirac fermions are added in section~\ref{gs:sec:dirac}. The construction does not cover interacting gauge fields, for which a gauge-fixed functional is required. Free gauge fields with a local coupling are constructed in appendix~\ref{gs:app:freegauge}.

\subsection{Scalar fields and Dirac fermions}
\label{gs:sec:dirac}
\label{gf:sec:fermions}

In this subsection we extend the construction of section~\ref{gs:sec:sphere} to theories of scalar fields and Dirac fermions with Yukawa couplings, quartic couplings and the background gauge field of the field rotations. The sources are the same kind of objects as for scalar fields, namely the local couplings, the one-form of the gauge field and the sources of the scalar operators. What has no source is the fermion field itself. The sources are those of section~\ref{gs:sec:functional}, namely the local couplings, which include the Yukawa couplings, the background gauge field, which couples to the currents of the field rotations including those of the fermions, and the sources of the relevant operators. A source of the fermion field would be an anticommuting spinor, which cannot be built from the commuting sources, the metric and their derivatives, so that the bare sources, the counterterms and the Weyl variation never contain it and it can be set to zero consistently. Every fermion line then closes, and the fermions appear only in closed loops. A closed loop is a trace of a product of propagators and vertex matrices, and with the gamma matrices of $d$ dimensions and without $\gamma_5$ the trace is a polynomial in the scalar products of the vectors which enter it. On the sphere the propagator of the fermion from the point $X_i$ to the point $X_j$ contains the chord $X_i-X_j$, the difference vector of the two points in the embedding space, contracted with the gamma matrices of the embedding space, see section~\ref{gf:sec:loops}. The loops are then functions of the parallel coordinates of the points and of the cosines $u_{ij}$, and the integrals have the form \eqref{eq:gram}. No spinor field on $S^d$ is needed for non-integer $d$. We state the definitions, the two lemmas on the loops and the result here. The proofs that the definition by continuation, the exclusion of the line of the free theory as a line of poles, the conformal covariance, the background gauge invariance, the properties under rotations and the flat chart hold as in section~\ref{gs:sec:sphere} are in appendix~\ref{gs:app:fermions}, where we state for each step which part of appendices~\ref{app:mero} to~\ref{app:gauge} applies without change and what is new.

\subsubsection{Fields, vertices and traces}
\label{gf:sec:vertices}

We consider $N$ real scalar fields $\phi_a$ and $N_f$ Dirac fermions $\psi$, $\bar\psi$, with the free action of section~\ref{gs:sec:sphere} for the scalars and the massless Dirac action for the fermions. The group of field rotations is a compact subgroup of ${\rm O}(N)\times{\rm U}(N_f)$, and its Lie algebra acts on the scalars by real antisymmetric matrices and on the fermions by antihermitian matrices. The vertices without derivatives are the vertices $V^{(n)}$ of \eqref{gs:eq:vertexn} of the scalars and the Yukawa vertex
\begin{equation}
V^{\rm Y}[Y] = \phi_a\,\bar\psi\,Y^a\psi \, ,
\label{gf:eq:yukawa}
\end{equation}
where the $Y^a$ are $N_f\times N_f$ matrices which act on the flavour indices and commute with the gamma matrices. For a real action the $Y^a$ are hermitian, and the couplings of the theory are the real coordinates of the $\lambda_{abcd}$ of $V^{(4)}$ and of the hermitian matrices $Y^a$. The gauge field $a$ is a one-form with values in the Lie algebra of the group of field rotations, and its vertices are the vertex of the scalar current and the term of second order in the gauge field on the scalars, whose source $\Upsilon$ is built from the scalar part of the gauge field only, and the vertex of the fermion current
\begin{equation}
V^{\psi J}[a] = \bar\psi\,\gamma^\mu a_\mu\,\psi \, ,
\label{gf:eq:fcurrent}
\end{equation}
in which $a_\mu$ acts on the flavour indices. The fermion current has no term of second order in the gauge field. The parity-invariant theories without $\gamma_5$ have no further vertices of dimension four at $d=4$, and the sources of $\bar\psi\psi$, of $\phi_a$ and of the cubic operators of the scalars have odd dimension and drop out by Lemma~\ref{gs:lem:dimension}. We nevertheless keep the vertices \eqref{gs:eq:vertexn} for all $n$, so that the construction contains the sources of $\phi_a$ and of the cubic operators. The source of $\bar\psi\psi$ can be included as a Yukawa vertex without the scalar leg, with the weight $d-2\Dp-1$ in place of $d-3\Dp-1$.

The gamma matrices obey $\gamma^\mu\gamma^\nu+\gamma^\nu\gamma^\mu=2\delta^{\mu\nu}$. As in dimensional regularisation we take the trace over the spinor indices of a product $\gamma\!\cdot\!v_1\cdots\gamma\!\cdot\!v_n$ to be ${\rm tr}\,1$ times the polynomial in the scalar products of the vectors $v_i$ which is defined by
\begin{equation}
\tau(v_1,\ldots,v_n) = \sum_{j=2}^n(-1)^j\,(v_1\!\cdot\!v_j)\;\tau(v_2,\ldots,\widehat{v_j},\ldots,v_n) \, , \qquad \tau(\,)=1 \, ,
\label{gf:eq:trace}
\end{equation}
for even $n$ and by zero for odd $n$. Here ${\rm tr}\,1$ is a fixed number, for instance $4$, which is not continued in $d$, and $\widehat{v_j}$ denotes the omission of $v_j$. The polynomial $\tau$ is the trace of the products of gamma matrices in every integer dimension in which the $v_i$ lie and in which the trace of a product of an odd number of gamma matrices vanishes, and it does not depend on the dimension. It is invariant under the orthogonal transformations of the $v_i$ and under the reversal of their order, and it is cyclic. Traces with contracted indices give polynomials in $d$ through $\delta^\mu{}_\mu=d$, and they occur below only in the theorem of Wick of Lemma~\ref{lem:schwinger}, where the same rule is used. The Levi-Civita tensor does not occur.

\subsubsection{Closed loops on the sphere}
\label{gf:sec:loops}

We describe the sphere by its embedding coordinates $X$ as in section~\ref{gs:sec:sphere}, and we use the gamma matrices $\Gamma^M$ of the embedding space, with the trace \eqref{gf:eq:trace} for vectors of the embedding space. For two points we write $X_{ij}=X_i-X_j$, so that $X_{ij}^2=s_{ij}^2$. Let a gauge field $a=a_A(X_\parallel)\,dX^A$ be given as in section~\ref{gs:sec:sphere}. Its tangent vector at the point $X$ is
\begin{equation}
A(X) = a(X_\parallel)-X\,\big(X\!\cdot\!a(X_\parallel)\big) \, ,
\label{gf:eq:tangent}
\end{equation}
where $a$ is regarded as a vector of the parallel directions. Consider a closed loop of fermion lines through the vertices at the points $X_1,\ldots,X_n$ in this cyclic order, where each vertex is a Yukawa vertex or a vertex of the fermion current, and let $\mathsf{C}$ be the set of the current vertices. We define the factor of the loop as
\begin{equation}
\mathcal{L} = -\,C_\psi^{\,n}\;{\rm tr}\Big[M_1\,\Gamma\!\cdot\!X_{12}\,M_2\,\Gamma\!\cdot\!X_{23}\cdots M_n\,\Gamma\!\cdot\!X_{n1}\Big]\prod_{i=1}^n\big(s_{i\,i+1}^2\big)^{-\Delta_\psi-\frac12} \, ,
\label{gf:eq:loop}
\end{equation}
with $X_{n+1}=X_1$, where $M_i=Y^{a_i}(X_{i\parallel})$ for a Yukawa vertex, whose scalar leg is paired elsewhere, and $M_i=\Gamma\!\cdot\!A(X_i)$ times the matrix $a_A$ of the flavour indices for a current vertex. The trace is over the spinor and the flavour indices, and the sign is that of a closed fermion loop. The line from $X_i$ to $X_{i+1}$ stands for the propagator $\vev{\psi(X_i)\bar\psi(X_{i+1})}$, which is $C_\psi\,\Gamma\!\cdot\!X_{i\,i+1}\,(s^2_{i\,i+1})^{-\Delta_\psi-1/2}$, and $\Delta_\psi$ is its exponent. The terms of the bare functional are the integrals \eqref{eq:gram} of the sums over the connected pairings of the products of the scalar propagators, the derivatives at the scalar current vertices and the factors \eqref{gf:eq:loop} of the fermion loops, with the factors $(-1)^N/N!$ of appendix~\ref{app:gauge}. A loop with a single vertex vanishes, since $X_{11}=0$, in agreement with the convention of section~\ref{sec:continuation} that propagators at coincident points vanish.

The numerator of \eqref{gf:eq:loop} is the trace \eqref{gf:eq:trace} of the vectors $X_{i\,i+1}$ and $A(X_i)$, times a trace over the flavour indices. Its scalar products are
\begin{gather}
X_i\!\cdot\!X_j = X_{i\parallel}\!\cdot\!X_{j\parallel}+\rho_i\rho_j\,u_{ij} \, , \qquad
A(X_i)\!\cdot\!X_j = a_i\!\cdot\!X_j-(a_i\!\cdot\!X_i)\,X_i\!\cdot\!X_j \, , \notag\\
A(X_i)\!\cdot\!A(X_j) = a_i\!\cdot\!a_j-(a_i\!\cdot\!X_j)(a_j\!\cdot\!X_j)-(a_i\!\cdot\!X_i)(a_j\!\cdot\!X_i)+(a_i\!\cdot\!X_i)(a_j\!\cdot\!X_j)\,X_i\!\cdot\!X_j \, ,
\label{gf:eq:products}
\end{gather}
where $a_i=a(X_{i\parallel})$ and $a_i\!\cdot\!X_j$ contains only the parallel components of $X_j$. The numerator is therefore a polynomial in the $X_{i\parallel}$, the $\rho_i$ and the $u_{ij}$, with coefficients which are products of components of the sources, and in the triangular coordinates of appendix~\ref{app:mero} it is a polynomial in the $w_i$ times such products. With $\Delta_\psi=\Dp+\tfrac12$ the factor of a fermion line is $(s^2)^{-\Dp-1}$ times its chord. The integrand is thus of the form of Proposition~\ref{prop:mero}, with $n_{ij}$ the number of all lines between the points $i$ and $j$ and $r_{ij}$ the number of fermion lines among them.

We use the propagator exponent $\Dp$ of the scalars as the single parameter of the continuation and set
\begin{equation}
\Delta_\psi = \Dp+\tfrac12 \, .
\label{gf:eq:Dpsi}
\end{equation}
We set $C_\psi=\Gamma(\tfrac d2)/(2\pi^{d/2})=1/\vol_d$ for all $d$. On the line \eqref{eq:line} we have $\Delta_\psi=\tfrac12(d-1)$, the dimension of a free fermion, and $C_\psi$ is the normalisation for which the flat propagator $C_\psi\,\gamma\!\cdot\!x/|x|^d$ is the Green function of $\gamma\!\cdot\!\pr$. The choice \eqref{gf:eq:Dpsi} keeps the covariance of Proposition~\ref{gf:prop:cov} of appendix~\ref{gs:app:fermions} for all values of $\Dp$, and it makes every fermion line a line of the kind which appendix~\ref{app:line} calls marked.

The flat chart relates the loops on the sphere to those of flat space. Let $P$ be a unit vector of the parallel directions, and let $x\mapsto X$ be the stereographic chart of section~\ref{sec:chart}, with $P\!\cdot\!X=1-\Omega$ and $X_T=\Omega\,x$, $\Omega=2/(1+x^2)$. We write $\tilde x$ for the point $x$ in the hyperplane orthogonal to $P$ of the embedding space, and $\gamma\!\cdot\!v=\Gamma\!\cdot\!v$ for vectors $v$ in this hyperplane. The pullback of the one-form $a$ to the chart is the flat one-form with the components $a^{\rm fl}_\mu=a_M\,\pr X^M/\pr x^\mu$.

\begin{lemma}[Chart]
\label{gf:lem:chart}
For a closed loop as above with points which differ from $P$,
\begin{equation}
{\rm tr}\Big[M_1\,\Gamma\!\cdot\!X_{12}\cdots M_n\,\Gamma\!\cdot\!X_{n1}\Big] = \prod_{i\notin\mathsf{C}}\Omega_i\;\;{\rm tr}\Big[m_1\,\gamma\!\cdot\!x_{12}\cdots m_n\,\gamma\!\cdot\!x_{n1}\Big] \, ,
\label{gf:eq:chartloop}
\end{equation}
where $m_i=M_i$ for a Yukawa vertex and $m_i=\gamma\!\cdot\!a^{\rm fl}(x_i)$ times the flavour matrix for a current vertex. Moreover $s_{ij}^2=\Omega_i\Omega_j\,|x_i-x_j|^2$.
\end{lemma}

\begin{proof}
Let $\tilde y=\tilde x-P$. Then $\tilde y^2=1+x^2=2/\Omega$, and the chart is the inversion $X=P+2\,\tilde y/\tilde y^2$ of the embedding space. The identity $\Gamma\!\cdot\!u\,\Gamma\!\cdot\!(u-v)\,\Gamma\!\cdot\!v=u^2\,\Gamma\!\cdot\!v-v^2\,\Gamma\!\cdot\!u$ gives
\begin{equation}
\Gamma\!\cdot\!X_{ij} = -\tfrac12\,\Omega_i\Omega_j\;\Gamma\!\cdot\!\tilde y_i\;\Gamma\!\cdot\!x_{ij}\;\Gamma\!\cdot\!\tilde y_j \, ,
\label{gf:eq:inversion}
\end{equation}
where $x_{ij}=\tilde x_i-\tilde x_j$, and the product of the relations for $X_{ij}$ and $X_{ji}=-X_{ij}$, with $(\Gamma\!\cdot\!\tilde y_j)^2=\tilde y_j^2$, gives the relation for $s_{ij}^2$. We insert \eqref{gf:eq:inversion} for each chord of the loop. At a Yukawa vertex the factors $\Gamma\!\cdot\!\tilde y_i$ of the two chords meet, with the flavour matrix between them, and give $\tilde y_i^2=2/\Omega_i$. At a current vertex they enclose $\Gamma\!\cdot\!A_i$. The differential of the inversion maps a vector $V$ of the hyperplane at $\tilde x$ to $\Omega\,R\,V$, where $R=1-2\,\hat y\hat y^T$ with $\hat y=\tilde y/|\tilde y|$ is a reflection, and $R$ maps the tangent space of the sphere at $X$ to the hyperplane. Hence $a^{\rm fl}=\Omega\,R\,A$ as vectors, and with $\Gamma\!\cdot\!\hat y\,\Gamma\!\cdot\!A\,\Gamma\!\cdot\!\hat y=-\Gamma\!\cdot\!(RA)$ we find $\Gamma\!\cdot\!\tilde y\,\Gamma\!\cdot\!A\,\Gamma\!\cdot\!\tilde y=-(2/\Omega^2)\,\gamma\!\cdot\!a^{\rm fl}$. Collecting the factors, each vertex not in $\mathsf{C}$ gives $\Omega_i^2\cdot(2/\Omega_i)$, each vertex in $\mathsf{C}$ gives $\Omega_i^2\cdot(-2/\Omega_i^2)$, and the $n$ chords give $(-\tfrac12)^n$. The product is $(-1)^{n+|\mathsf{C}|}\prod_{i\notin\mathsf{C}}\Omega_i$. The trace of the product of an odd number of gamma matrices vanishes, so that both sides vanish unless $n+|\mathsf{C}|$ is even, and the sign is then $+1$. Finally the trace of a product of vectors of the hyperplane is the trace \eqref{gf:eq:trace} of these vectors, which does not depend on the dimension. Both sides of \eqref{gf:eq:chartloop} are polynomials in the scalar products of the points and of the vectors $a$, so that the identity also holds for the integrands of \eqref{eq:gram} with non-integer $d$.\footnote{We verified \eqref{gf:eq:chartloop} in exact rational arithmetic for random loops with two to six vertices of both kinds, and the trace \eqref{gf:eq:trace} against explicit gamma matrices in six dimensions.}
\end{proof}

For loops with Yukawa vertices only the lemma has a form which shows the covariance of each term. A perfect matching of the points $1,\ldots,n$ is a partition into pairs.

\begin{lemma}[Matchings]
\label{gf:lem:matching}
Let $n$ be even. For a perfect matching $\mathsf{M}$ of the points $1,\ldots,n$ let ${\rm cr}(\mathsf{M})$ be the number of crossings of its pairs, drawn as segments between points in cyclic order on a circle, and let $c_\mathsf{M}=(-1)^{n/2+{\rm cr}(\mathsf{M})}\,2^{1-n/2}$. Then
\begin{equation}
\tau\big(X_{12},X_{23},\ldots,X_{n1}\big) = \sum_{\mathsf{M}}c_\mathsf{M}\prod_{(ab)\in\mathsf{M}}s_{ab}^2
\label{gf:eq:matching}
\end{equation}
for points of the sphere, and the same holds with the flat distances $|x_a-x_b|^2$ for points of flat space. For $n=4$ this is $\tau=\tfrac12(s_{12}^2s_{34}^2+s_{14}^2s_{23}^2-s_{13}^2s_{24}^2)$.
\end{lemma}

\begin{proof}
We first show that for numbers $g_{ab}=g_{ba}$ the recursion \eqref{gf:eq:trace} with $v_a\!\cdot\!v_b$ replaced by $g_{ab}$ has the solution $\sum_\mathsf{M}(-1)^{{\rm cr}(\mathsf{M})}\prod_{(ab)\in\mathsf{M}}g_{ab}$ for every even number of points. By induction it suffices to note that the pair $(1j)$ crosses the pairs which join one of the $j-2$ points between $1$ and $j$ to a point after $j$, whose number has the parity of $j$, and that the other crossings are those of the matching of the remaining points.

Let now $x_1,\ldots,x_n$ be points of $\mathbb{R}^D$ for an integer $D$, with $x_{ij}=x_i-x_j$, and let $\gamma^\mu$ be gamma matrices of $D$ dimensions in a representation in which the traces of products of an odd number of gamma matrices vanish. We denote by ${\rm Tr}$ the trace of a matrix divided by the trace of the unit matrix of the same size, so that ${\rm Tr}\,\gamma\!\cdot\!v_1\cdots\gamma\!\cdot\!v_n=\tau(v_1,\ldots,v_n)$. On pairs of spinors we define the matrices
\begin{equation}
\mathsf{N}_i=\begin{pmatrix}\gamma\!\cdot\!x_i\\1\end{pmatrix}\begin{pmatrix}1&-\gamma\!\cdot\!x_i\end{pmatrix}=\begin{pmatrix}\gamma\!\cdot\!x_i&-x_i^2\\1&-\gamma\!\cdot\!x_i\end{pmatrix} \, .
\end{equation}
Since $(1\;\;-\gamma\!\cdot\!x_i)(\gamma\!\cdot\!x_j\;\;1)^T=\gamma\!\cdot\!x_j-\gamma\!\cdot\!x_i$, the cyclicity of the trace gives, for even $n$,
\begin{equation}
{\rm Tr}\,\mathsf{N}_1\cdots\mathsf{N}_n=\tfrac12\,{\rm Tr}\big[(\gamma\!\cdot\!x_2-\gamma\!\cdot\!x_1)(\gamma\!\cdot\!x_3-\gamma\!\cdot\!x_2)\cdots(\gamma\!\cdot\!x_1-\gamma\!\cdot\!x_n)\big]=\tfrac12\,\tau(x_{12},x_{23},\ldots,x_{n1}) \, ,
\end{equation}
where the factor $\tfrac12$ is the ratio of the sizes of the matrices. The same relation gives $\mathsf{N}_i^2=0$, and a block multiplication with $\gamma\!\cdot\!x_i\,\gamma\!\cdot\!x_j+\gamma\!\cdot\!x_j\,\gamma\!\cdot\!x_i=2\,x_i\!\cdot\!x_j$ gives $\mathsf{N}_i\mathsf{N}_j+\mathsf{N}_j\mathsf{N}_i=-|x_i-x_j|^2$. We move $\mathsf{N}_1$ through $\mathsf{N}_2\cdots\mathsf{N}_n$ with these relations, and for even $n$ the cyclicity of the trace gives
\begin{equation}
{\rm Tr}\,\mathsf{N}_1\cdots\mathsf{N}_n=\sum_{j=2}^n(-1)^j\big(-\tfrac12|x_1-x_j|^2\big)\,{\rm Tr}\,\mathsf{N}_2\cdots\widehat{\mathsf{N}_j}\cdots\mathsf{N}_n \, ,
\end{equation}
which is the recursion \eqref{gf:eq:trace} with $g_{ab}=-\tfrac12|x_a-x_b|^2$ and ${\rm Tr}\,1=1$. By the first step ${\rm Tr}\,\mathsf{N}_1\cdots\mathsf{N}_n=(-\tfrac12)^{n/2}\sum_\mathsf{M}(-1)^{{\rm cr}(\mathsf{M})}\prod_{(ab)\in\mathsf{M}}|x_a-x_b|^2$, and the comparison with $\tfrac12\tau$ gives \eqref{gf:eq:matching} for flat points. Both sides of \eqref{gf:eq:matching} are polynomials in the scalar products $x_a\!\cdot\!x_b$, which for $D\geq n$ range over an open set, so that the identity holds for these polynomials. With $|X_a-X_b|^2=s_{ab}^2$ it holds for points of the sphere.
\end{proof}

A loop with $n$ Yukawa vertices is thus a sum over matchings of integrands of the scalar kind, in which each fermion line carries $(s^2)^{-\Dp-1}$ and each pair of the matching a factor $s^2$. Each point occurs once in a pair. Under the maps \eqref{eq:boost} each term is multiplied by $\prod_i\Omega_i^{\,-2\Delta_\psi}$, which is the covariance of a product of fields of dimension $\Delta_\psi$ at each end of a line. With current vertices we use Lemma~\ref{gf:lem:chart} directly.

\subsubsection{Result}
\label{gf:sec:result}

\begin{proposition}[Bare functional for scalar fields and Dirac fermions]
\label{gf:prop:bare}
Consider $N$ scalar fields and $N_f$ Dirac fermions with the vertices \eqref{gs:eq:vertexn} for any set of numbers $n\geq1$, the Yukawa vertex \eqref{gf:eq:yukawa}, the vertices of the scalar and of the fermion current and the term of second order in the scalar gauge field, with sources in the class of sources on the sphere, and with $\Delta_\psi=\Dp+\tfrac12$. Then each term of the bare functional is a meromorphic function of $d$ and $\Dp$. With $\Upsilon$ kept independent it satisfies the covariance of appendix~\ref{app:conf} with the Yukawa couplings of weight $d-3\Dp-1$. On the line \eqref{eq:line}, with $\Upsilon=\gamma^{\mu\nu}a_{0\mu}^Ta_{0\nu}$, the line is not a line of poles, the integrals of which a term is a sum, with the normalisations $C_\phi$ and $C_\psi$ of the propagators taken out, have no line of poles of fixed dimension, and each term is a meromorphic function of $d$. Away from the poles the functional satisfies \eqref{eq:covS} with the weights \eqref{gs:eq:wn} and the weight $\tfrac12(4-d)$ of the Yukawa couplings, and it is invariant under background gauge transformations. It has the properties under rotations of \eqref{eq:Lambda}, the one-point function of the fermion current vanishes at constant sources, and the functional is real for real sources. The flat chart \eqref{eq:chart} holds with the Yukawa couplings of weight $\tfrac12(4-d)$.
\end{proposition}

\begin{proof}
Propositions~\ref{gf:prop:line}, \ref{gf:prop:cov} and~\ref{gf:prop:gauge}, Lemma~\ref{gf:lem:real} and appendix~\ref{gf:sec:chart}.
\end{proof}

In the terms of table~\ref{gs:tab:cases} the bare functional of scalars with Yukawa and quartic couplings is therefore constructed, as for the sextic and the cubic scalar theories. The conditions on the bare functional in Propositions~\ref{gs:prop:msanomaly} and~\ref{gs:prop:msward} therefore hold, and the other two conditions of section~\ref{gs:sec:msanomaly} follow from Hypothesis~\ref{hyp:finite} for the Yukawa theory, so that in minimal subtraction the anomaly is the Weyl variation of the counterterms and the Ward identity of section~\ref{gs:sec:msward} holds. Lemma~\ref{gs:lem:free} extends to Dirac fermions, see Lemma~\ref{gf:lem:free} of section~\ref{gs:sec:remains}.

\subsection{Ward identity on the sphere}
\label{gs:sec:msward}

Let $\sigma=b\cdot X$ and $v^\mu=-\nabla^\mu\sigma$ be as in \eqref{eq:ckv}, with $b$ in the parallel directions. The diffeomorphism generated by $-v$ changes the metric of the sphere by $-2\sigma\gamma_{\mu\nu}$, and the Weyl rescaling with $\sigma$ restores it. Under the combination of the two the sources change by
\begin{equation}
\delta_vg^I = \sigma\,\hB^I-v^\mu\pr_\mu g^I \, , \qquad
\delta_va_\mu = \sigma\,\hr_I\,D_\mu g^I-\cL_va_\mu \, , \qquad
\delta_v\sM = \delta^{\rm W}_\sigma\sM-v^\mu\pr_\mu\sM \, ,
\label{eq:deltavS}
\end{equation}
with $\delta^{\rm W}_\sigma\sM$ of Property~\ref{gs:P:weyl}.

\begin{proposition}[Ward identity from conformal covariance]
\label{gs:prop:msward}
Let the first and the third condition of section~\ref{gs:sec:msanomaly} hold, and let the bare functional on the unit sphere be invariant under the variation \eqref{eq:barevar} of the bare sources with their weights and under background gauge transformations of the bare sources. Then the renormalised functional \eqref{gs:eq:Wren} on the unit sphere satisfies
\begin{equation}
\int d\mu_d\left(\delta_vg^I\,\frac{\delta W}{\delta g^I}+\delta_va_\mu\cdot\frac{\delta W}{\delta a_\mu}
+\delta_v\sM\cdot\frac{\delta W}{\delta\sM}\right) = \cA[\sigma]
\label{eq:wardS}
\end{equation}
for all sources in the class of sources on the sphere, with the anomaly \eqref{gs:eq:msanomaly} at $\gamma_{\mu\nu}=\mathring\gamma_{\mu\nu}$. This is the Ward identity of Property~\ref{gs:P:ward}.
\end{proposition}

\begin{proof}
The bare sources are local functions of the sources which are built covariantly, so that the diffeomorphism acts on them by the Lie derivative. By the third condition of section~\ref{gs:sec:msanomaly} the Weyl variation of the sources induces the rescaling of the bare sources with their weights, up to a background gauge transformation. The variation \eqref{eq:deltavS} of the renormalised sources therefore induces the variation \eqref{eq:barevar} of the bare sources, up to a gauge transformation, and the bare functional is invariant. The integral of the counterterms is invariant under the diffeomorphism, since the change of its integrand is a divergence, which integrates to zero by the divergence form of the Laplacian with $d$ as a parameter. Under the Weyl rescaling it changes by $\cA[\sigma]$, as explained in section~\ref{sec:local}. The sum of the two statements is \eqref{eq:wardS}, where the functional derivatives are densities with respect to the measure of the sphere.
\end{proof}

For scalar fields the conditions on the bare functional hold on the line \eqref{eq:line} by Proposition~\ref{gs:prop:bare}. The identity \eqref{eq:wardS} is the analogue of the conformal Ward identity in flat space, eq.~(4.9) of~\cite{Jack:2013sha}, which is the starting point of the Ward identity of~\cite{Reehorst:phi4}. On the sphere the conformal Killing vector is a smooth vector field on a compact space, so that no decay of correlation functions at large distances is required. Moreover, the right-hand side contains the terms of the anomaly with the curvature.

\subsection{Finiteness and regularity}
\label{gs:sec:remains}

For a given theory the derivations of sections~\ref{gs:sec:msanomaly} to~\ref{gs:sec:msward} are completed by two kinds of statements, which depend on the theory. The first is finiteness. Property~\ref{gs:P:functional} states that the terms of first and of higher order of the renormalised functional on the unit sphere have no poles at $\vep=0$, and Properties~\ref{gs:P:weyl} and~\ref{gs:P:anomaly} require that the coefficients of $\hB^I$, $\hr_I$, $\delta^{\rm W}_\sigma\sM$ and $\cA[\sigma]$ are regular. In minimal subtraction these coefficients are pure poles or Laurent series in $\vep$. For the $\phi^4$, the Yukawa, the sextic and the cubic theory their regularity follows from the finiteness of the functional, because the free two-point functions of the operators of the couplings are nondegenerate at $d=d_c$. For four-fermion couplings near two dimensions the finiteness leaves a simple pole in the evanescent directions (Proposition~\ref{g2:prop:poles}). The second kind consists of the statements of Property~\ref{gs:P:sphere} for the renormalised functional, which contain the bare sources and the counterterms in addition to the bare functional, and of the commutation of the operators $\mathbb{W}_\sigma$, which Lemma~\ref{gcu:lem:commute} reduces to the form of the bare sources and of their variation. For multiscalar $\phi^4$ theory we derive all of them from Hypothesis~\ref{hyp:finite}, which concerns the finiteness of the functional in flat space, in section~\ref{gs:sec:phi4prop} and appendices~\ref{app:unique} and~\ref{app:finite}, whose proofs use the counterterms of the hypothesis for this theory. For scalar fields and Dirac fermions with Yukawa and quartic couplings we derive them from Hypothesis~\ref{hyp:finite} for the Yukawa theory in section~\ref{gs:sec:yukawa} and appendix~\ref{gs:app:fermions}. For sextic scalars we derive them from Hypothesis~\ref{hyp:finite} for the sextic theory in section~\ref{gs:sec:sextic} and appendix~\ref{gs:app:sextic}. For cubic scalars we derive them from Hypothesis~\ref{hyp:finite} for the cubic theory, on conformally flat backgrounds, in section~\ref{gs:sec:cubic} and appendix~\ref{gs:app:cubic}. For the other theories of section~\ref{gs:sec:other} the properties which are not derived there are assumptions.

Table~\ref{gs:tab:where} lists for each property the statements which prove it for these four theories from Hypothesis~\ref{hyp:finite} for the theory, and for the free gauge field, for which Properties~\ref{gs:P:functional} to~\ref{gs:P:free} follow from Hypothesis~\ref{hyp:finite} for the free gauge field in appendix~\ref{gs:app:freegauge}. The proofs for the Yukawa, the sextic and the cubic theory carry each step of the derivation for $\phi^4$ theory over as recorded in table~\ref{gs:tab:steps} of appendix~\ref{gs:app:steps}.

\begin{table}[tbp]
\centering
\renewcommand{\arraystretch}{1.12}
\scriptsize
\setlength{\tabcolsep}{3pt}
\hyphenpenalty=10000 \exhyphenpenalty=10000
\begin{tabular}{>{\raggedright\arraybackslash}p{1.75cm}>{\raggedright\arraybackslash}p{2.45cm}>{\raggedright\arraybackslash}p{2.45cm}>{\raggedright\arraybackslash}p{2.45cm}>{\raggedright\arraybackslash}p{2.45cm}>{\raggedright\arraybackslash}p{2.45cm}}
\hline
 & $\phi^4$, $d=4-\vep$ & Yukawa and quartic couplings, $d=4-\vep$ & $\phi^6$, $d=3-\vep$ & $\phi^3$, $d=6-\vep$ & free gauge field, $d=4-\vep$ \\
\hline
collected in & Props.~\ref{gs:prop:phi4prop}, \ref{gr4:prop:lowest} & Prop.~\ref{gf:prop:props} & Prop.~\ref{gsx:prop:props} & Prop.~\ref{gcp:prop:props} & Prop.~\ref{ggs:prop:props} \\
\ref{gs:P:functional} Functional & Prop.~\ref{prop:finite}, Prop.~\ref{prop:gauge}, Lemma~\ref{gr4:lem:const} & Prop.~\ref{prop:finite} as in sec.~\ref{gf:sec:steps}, Prop.~\ref{gf:prop:gauge}, Lemma~\ref{gr4:lem:const} & Prop.~\ref{gsx:prop:finite}, Prop.~\ref{prop:gauge} & localisation of app.~\ref{app:finite}, Prop.~\ref{gcu:prop:first} & Prop.~\ref{ggs:prop:shift}, instance~(b) \\
\ref{gs:P:weyl} Weyl variation & Lemmas~\ref{lem:rhoxi}, \ref{lem:regvar}, \ref{gcu:lem:commute}, \ref{gr4:lem:zero}, Props.~\ref{prop:uniqueall}, \ref{prop:finite} & Lemmas~\ref{lem:rhoxi}, \ref{lem:regvar} as in sec.~\ref{gf:sec:steps}, Lemmas~\ref{gcu:lem:commute}, \ref{gr4:lem:zero}, Prop.~\ref{gc:prop:cov} & sec.~\ref{gsx:sec:weyl}, Lemmas~\ref{gsx:lem:regvar}, \ref{gsx:lem:zero}, Prop.~\ref{gsx:prop:uniqueall} & Props.~\ref{gcp:prop:cov}, \ref{gcp:prop:reg}, \ref{gcf:prop:frame}, \ref{gcu:prop:first}, Lemma~\ref{gcu:lem:commute}, on conformally flat backgrounds & sec.~\ref{ggs:sec:props}, $\hB=-\tfrac12\vep$ \\
\ref{gs:P:anomaly} Anomaly & Props.~\ref{gs:prop:msanomaly}, \ref{prop:finite} & Props.~\ref{gs:prop:msanomaly}, \ref{prop:finite} & Props.~\ref{gs:prop:msanomaly}, \ref{gsx:prop:finite}, Lemma~\ref{gsx:lem:anomaly} & Prop.~\ref{gs:prop:msanomaly}, Lemma~\ref{gcp:lem:gram}, on conformally flat backgrounds & sec.~\ref{ggs:sec:props} \\
\ref{gs:P:ward} Ward identity & Props.~\ref{gs:prop:msward}, \ref{gs:prop:bare} & Props.~\ref{gs:prop:msward}, \ref{gf:prop:bare} & Props.~\ref{gs:prop:msward}, \ref{gs:prop:bare} & Props.~\ref{gs:prop:msward}, \ref{gs:prop:bare} & Prop.~\ref{ggs:prop:ward} \\
\ref{gs:P:sphere} Sphere & apps.~\ref{app:rotations}, \ref{app:massward}, Lemmas~\ref{gr4:lem:const}, \ref{gsp:lem:local} & sec.~\ref{gf:sec:chart}, Lemma~\ref{gr4:lem:const} & app.~\ref{app:rotations}, sec.~\ref{gsx:sec:proof} & app.~\ref{app:rotations}, Lemma~\ref{gr4:lem:const} & Prop.~\ref{ggs:prop:rot} \\
class of sources (app.~\ref{gs:sec:class}) & app.~\ref{app:rules} & app.~\ref{app:rules} & app.~\ref{app:rules} & app.~\ref{app:rules}, $D_0=3$ & app.~\ref{app:rules}, polynomial sources \\
\ref{gs:P:free} Free limit & Prop.~\ref{gr4:prop:lowest}, Lemmas~\ref{gs:lem:free}, \ref{gr4:lem:zero}, \ref{gr4:lem:const} & Lemmas~\ref{gf:lem:free}, \ref{gro:lem:c}, proof of Prop.~\ref{gr4:prop:lowest}, Lemma~\ref{gr4:lem:zero} & Lemmas~\ref{gs:lem:free}, \ref{gsx:lem:zero}, \ref{gsh:lem:m}, sec.~\ref{gsx:sec:proof} & Props.~\ref{gcf:prop:frame}, \ref{gcu:prop:first}, Lemma~\ref{gs:lem:free}, proof of Prop.~\ref{gr4:prop:lowest} & Props.~\ref{grg:prop:gauge}, \ref{ggs:prop:shift}, Lemma~\ref{gro:lem:c} \\
\hline
\end{tabular}
\caption{Where the properties are proved from Hypothesis~\ref{hyp:finite} for the theory of the column. The first row names the propositions which collect the properties, and each further cell names the statements which prove the property. The row of the class of sources names where the statements on the unit sphere which appendix~\ref{gs:sec:class} assumes are established for the class of sources of the theory. For the cubic theory the hypothesis is assumed for a larger class of sources than for the other theories, the functions of three coordinates in place of one ($D_0=3$), since its proofs identify local expressions through such sources (Lemma~\ref{gcp:lem:gram}).}
\label{gs:tab:where}
\end{table}

The two-point form of appendix~\ref{gs:sec:properties} follows from the construction of section~\ref{gs:sec:sphere} under conditions on the bare sources.

\begin{lemma}
\label{gs:lem:free}
Consider scalar fields as in section~\ref{gs:sec:sphere} in minimal subtraction, where the couplings $g^I$ are the sources of vertices \eqref{gs:eq:vertexn}. Let the bare couplings be $g_0^I=g^I+L^I$, where $L^I$ has no terms of order zero and one in the couplings, and let the bare gauge field and the bare relevant sources at $a_\mu=0$ and $\sM=0$ have no terms of order zero and one in the couplings. Then through second order in the couplings, at $a_\mu=0$ and $\sM=0$, the functional $W-\int d^dx\sqrt\gamma\,\cX$ is the first term of the two-point form. For two couplings of vertices with $n$ legs the matrix $c_{IJ}$ is $C_\phi^{\,n}/n!$ times the unit matrix, in coordinates in which $g^Ig^I=\psi_{i_1\ldots i_n}\psi_{i_1\ldots i_n}$, and it vanishes for couplings of vertices with different numbers of legs. In particular $c_{IJ}$ is positive for $d_c>2$.
\end{lemma}

\begin{proof}
The terms of first order of $W_0^S$ vanish, since the legs of a vertex are not paired with each other. The terms of second order of the bare sources therefore do not contribute at second order in the couplings, and neither do the bare gauge field and the bare relevant sources. The term of second order of $W_0^S$ in the bare couplings is $\tfrac12\langle V,V\rangle^c$ in the notation of appendix~\ref{app:gauge}. For two vertices with $n$ legs there are $n!$ pairings of their legs, each of which gives $C_\phi^{\,n}\,s^{-2n\Dp}$, and with the factor $1/n!$ of each vertex this gives the stated $c_{IJ}$ with $\Delta_I=n\Dp$. Two vertices with different numbers of legs have no pairing of their legs. The two-point integral is the same meromorphic function of $\Delta_I+\Delta_J$ whether it is continued in $\Dp$ or in $\Delta_I$. For $d_c>2$ the constant $C_\phi$ is positive near $d=d_c$.
\end{proof}

Lemma~\ref{gs:lem:free} extends to Dirac fermions with the bare functional of section~\ref{gs:sec:dirac}.

\begin{lemma}
\label{gf:lem:free}
In addition to the hypotheses of Lemma~\ref{gs:lem:free}, let the couplings $g^I$ include the real coordinates of the hermitian matrices $Y^a$, and let $L^I$ have no terms of order zero and one also in these couplings. Then through second order in the couplings, at $a_\mu=0$ and $\sM=0$, the functional $W-\int d^dx\sqrt\gamma\,\cX$ is the first term of the two-point form, with $\Delta_I=3\Dp+1$ for the Yukawa couplings,
\begin{equation}
c(Y,Y') = C_\phi\,C_\psi^{\,2}\,{\rm tr}\,1\,\sum_a{\rm tr}\,(Y^aY'^a)
\label{gf:eq:cy}
\end{equation}
on the Yukawa couplings, the coefficient of Lemma~\ref{gs:lem:free} on the couplings of the scalar vertices, and $c_{IJ}=0$ between a Yukawa coupling and a coupling of a scalar vertex. At $d=4$, with ${\rm tr}\,1=4$, \eqref{gf:eq:cy} is the value of Lemma~\ref{gro:lem:c} of section~\ref{gs:sec:yukawa}, and $c_{IJ}$ is positive near $d=4$.
\end{lemma}

\begin{proof}
The terms of first order vanish, since a loop with one vertex vanishes and the legs of a scalar vertex are not paired with each other. As in the proof of Lemma~\ref{gs:lem:free} the term of second order of $W_0^S$ in the bare couplings is $\tfrac12\vev{V,V}^c$. Two Yukawa vertices give the scalar line $C_\phi\,\delta_{ab}\,(s^2)^{-\Dp}$ and the loop \eqref{gf:eq:loop} with $n=2$, whose trace is ${\rm tr}\,1\,{\rm tr}(Y^aY'^b)\,\tau(X_{12},X_{21})=-{\rm tr}\,1\,{\rm tr}(Y^aY'^b)\,s^2$. With the sign of the loop and $\Delta_\psi=\Dp+\tfrac12$ the product is \eqref{gf:eq:cy} times $(s^2)^{-3\Dp-1}$, and with the factor $\tfrac12$ this is the first term of the two-point form. A Yukawa vertex and a scalar vertex have no connected pairing, since the fermion line of the Yukawa vertex would form a loop with a single vertex, which vanishes. The statement on the continuation in $\Dp$ or in $\Delta_I$ is that of Lemma~\ref{gs:lem:free}. At $d=4$ we have $C_\phi=1/(4\pi^2)$ and $C_\psi=1/(2\pi^2)$, so that $4\,C_\phi C_\psi^2=1/(4\pi^6)$, and $\sum_a{\rm tr}(Y^aY^a)$ is positive for hermitian $Y^a\neq0$.
\end{proof}

With these lemmas the remaining content of the two-point form for scalar fields and Dirac fermions is the hypotheses of the lemma on the bare sources and the requirement on the counterterms which are linear in $\sM$, which is stated with the two-point form. The statements of Property~\ref{gs:P:free} other than the last then follow in the divergence scheme by Proposition~\ref{gs:prop:lowest} and Lemma~\ref{gre:lem:zero}, with the form of $\hB^I$ at first order, if the redefinition to that scheme vanishes to second order at $g=0$, and the last follows if in addition the one-point function $\sfm$ of the relevant operators vanishes at $g=0$ in the scheme from which the redefinition starts, see section~\ref{gs:sec:phi4lowest}. For $\phi^4$ theory the lemma gives $c_{IJ}=c_\cO\,\delta_{IJ}$ with $c_\cO=C_\phi^4/4!$.

\section{Multiscalar \texorpdfstring{$\phi^4$}{phi4} theory in \texorpdfstring{$d=4-\vep$}{d=4-epsilon}}
\label{gs:sec:phi4}

In this section we derive the properties of appendix~\ref{gs:sec:properties} for multiscalar $\phi^4$ theory in $d=4-\vep$ dimensions from one hypothesis on the functional in flat space, Hypothesis~\ref{hyp:finite}. With Theorem~\ref{gr:cor:cond} we then obtain the gradient formula to all orders in perturbation theory, which is Theorem~\ref{gs:thm:phi4}. Here $d_c=4$, all couplings have the weight $k_I=1$, and the only relevant source of even dimension is the mass source $\sM$ of the operators $\phi_i\phi_j$, of dimension two. In section~\ref{sec:setup} we introduce the theory, its bare sources and counterterms, its Weyl variation and its anomaly. In section~\ref{gs:sec:phi4sphere} we specialise the bare functional of section~\ref{gs:sec:sphere} and state the hypothesis. In section~\ref{gs:sec:phi4prop} we derive Properties~\ref{gs:P:functional} to~\ref{gs:P:sphere} and the two-point form of appendix~\ref{gs:sec:properties}, in section~\ref{gs:sec:phi4lowest} the two-point form in the divergence scheme and Property~\ref{gs:P:free}, and in section~\ref{gs:sec:phi4thm} we state the theorem. Appendix~\ref{gs:app:fourd} contains the consequences in four dimensions, among them a scheme in which the one-form $W_I$ of the Weyl anomaly vanishes, and the comparison with the literature. The statements whose proofs are long are quoted with a pointer to the appendices.

\subsection{Setup}
\label{sec:setup}

In this section we explain our setup. We first introduce the theory and its sources in $d$ dimensions on a curved background. We then explain why the gradient property is only of interest with regular coefficients. Finally we give the bare sources and the counterterms, the Weyl variation and the anomaly, and the relations among the anomaly coefficients which we use. These relations are the consistency conditions of~\cite{Jack:2013sha} with the dependence on the dimension kept.

\subsubsection{Theory and sources}

We consider $N$ real scalar fields $\phi_i$ on a Euclidean background of dimension $d=4-\vep$ with metric $\gamma_{\mu\nu}$ and the action
\begin{equation}
S = \int d^dx\sqrt\gamma\left[\tfrac12\,\gamma^{\mu\nu}D_\mu\phi_i\,D_\nu\phi_i+\tfrac12\,\xi_d\,R\,\phi_i\phi_i
+\tfrac12\,\sM_{ij}\,\phi_i\phi_j+\tfrac{1}{4!}\,\lambda_{ijkl}\,\phi_i\phi_j\phi_k\phi_l\right] ,
\label{eq:action}
\end{equation}
where $R$ is the curvature scalar, $\xi_d=(d-2)/(4(d-1))$ and $D_\mu\phi_i=\pr_\mu\phi_i+a_{\mu\,ij}\phi_j$. The notation follows~\cite{Reehorst:phi4}. The symmetric tensor $\lambda_{ijkl}$ contains the couplings, which we also denote by $g^I$, so that an index $I$ stands for a symmetrised set of four field indices, and we take coordinates $g^I$ in which $g^Ig^I=\lambda_{ijkl}\lambda_{ijkl}$, with the sum over all $i$, $j$, $k$ and $l$. The group of field rotations is ${\rm O}(N)$, its Lie algebra consists of the antisymmetric matrices $\omega_{ij}$, and $(\omega g)^I$ denotes the infinitesimal rotation of the couplings, $(\omega\lambda)_{ijkl}=\omega_{im}\lambda_{mjkl}+\omega_{jm}\lambda_{imkl}+\omega_{km}\lambda_{ijml}+\omega_{lm}\lambda_{ijkm}$. The couplings are functions $g^I(x)$ of the position, which we call local couplings. The background field $a_\mu(x)$ takes values in the Lie algebra and is a source for the currents of the field rotations. The mass source $\sM(x)$ takes values in the space $V_\sM$ of symmetric matrices and is a source for the operators $\phi_i\phi_j$. A dot denotes the contraction of a pair of field indices, as in $a_\mu\cdot J^\mu=\tfrac12a_\mu^{ij}J^\mu_{ij}$ and $\sM\cdot\phi^2=\tfrac12\sM_{ij}\phi_i\phi_j$. The two sources are needed because with local couplings there are counterterms in which the current and the operators $\phi_i\phi_j$ are multiplied by derivatives of the couplings~\cite{Jack:2013sha}. The term with $\xi_d$ is the conformal coupling to the curvature in $d$ dimensions. With this term the free action is invariant under Weyl rescalings of the metric for every $d$. Sources for $\phi_i$ and for cubic operators are excluded by the symmetry $\phi\to-\phi$.

The generating functional is $W[\gamma,g,a,\sM]=\log Z$, the logarithm of the partition function. The marginal operators $\cO_I$, the currents and the mass operator are defined by its derivatives with respect to $g^I$, $a_\mu$ and $\sM$, for instance $\delta W=\int d^dx\sqrt\gamma\,\vev{\cO_I}\,\delta g^I$. The functional is invariant under background gauge transformations with a parameter $\omega(x)$ in the Lie algebra, under which
\begin{equation}
\delta_\omega a_\mu = D_\mu\omega \, , \qquad \delta_\omega g^I = -(\omega g)^I \, , \qquad \delta_\omega\sM = -[\omega,\sM] \, ,
\label{eq:gaugetr}
\end{equation}
with $D_\mu\omega=\pr_\mu\omega+[a_\mu,\omega]$. Derivatives of the couplings appear in the combination $D_\mu g^I=\pr_\mu g^I+(a_\mu g)^I$, and $f_{\mu\nu}$ denotes the field strength of $a_\mu$. We call a function of the couplings invariant, and a tensor covariant, if it is unchanged by a rotation with a constant element of the group of field rotations, here ${\rm O}(N)$ including the reflections, where the rotation also acts on the indices of the tensor. It is then covariant under the Lie algebra, that is unchanged by the rotations with a constant $\omega$, which is all that the identities of appendix~\ref{app:weyl} use.

The flow of the couplings is described by a vector field $\hB^I$ on the space of couplings, which we define in section~\ref{sec:weylvar} through the Weyl variation of the bare sources. In minimal subtraction it has the form
\begin{equation}
\hB^I = -\vep\,g^I+B^I \, , \qquad B^I = \beta^I-(\xi g)^I \, , \qquad \hb^I = -\vep\,g^I+\beta^I \, ,
\label{eq:Bhat}
\end{equation}
where $\beta^I$ is defined in section~\ref{sec:weylvar} and is the beta function in four dimensions by Identification~\ref{id:ms}, $\hb^I$ is then the beta function in $d$ dimensions, and the element $\xi(g)$ of the Lie algebra is determined by the counterterm of the current, see \eqref{eq:rhoxi} below.\footnote{The rotation $\xi$ is denoted by $\upsilon$ in~\cite{Jack:2013sha} and by $S$ in~\cite{Pannell:2024sia}.} The function $B^I$ of the couplings does not depend on $\vep$. The vector field $\hB^I$ therefore depends on $\vep$ only through the term $-\vep\,g^I$, and its zeros depend on $\vep$. As explained in~\cite{Reehorst:phi4}, the beta function is defined up to an infinitesimal rotation of the fields, $\beta^I\to\beta^I-(\omega g)^I$. Indeed, $\beta^I$ and $\xi$ depend on the choice of an orthogonal factor in the renormalisation of the fields, while $B^I$ and $\hB^I$ do not~\cite{Jack:2013sha,Herren:2021amb}. The vector field $\hB^I$ is therefore the primary object, and our results are statements about it. We use the names $\beta^I$ and $\xi$ only for the conventional choice of this factor, which we fix in section~\ref{sec:weylvar}. A hat on $\beta^I$ and $B^I$ indicates that the term $-\vep\,g^I$ is included. We set the renormalisation scale to one. On a sphere of radius $r_0$ the dependence on the scale is restored by the flow of the couplings, since a constant Weyl rescaling changes $r_0$.

\subsubsection{Regularity in \texorpdfstring{$\vep$}{epsilon}}
\label{sec:regular}

For fixed $\vep\neq0$ the gradient property holds in perturbation theory for a simple reason. There are bare couplings, which are power series in the couplings whose coefficients are polynomials in $1/\vep$, in which the flow is linear, and a linear flow is the gradient of a quadratic function. The metric of this solution is symmetric and positive, but its coefficients have poles at $\vep=0$, see \eqref{eq:trivial} in appendix~\ref{app:gaugevar}. The question is therefore whether a gradient formula holds with coefficients which are regular at $\vep=0$. Throughout the paper we call a quantity regular if it is a power series in the couplings whose coefficients are power series in $\vep$. To organise such double series we assign weight one to $g^I$ and to $\vep$. We use the weight in appendix~\ref{sec:iteration}, where the equations for the antisymmetric coefficients are solved order by order in the weight. The vector fields $\hb^I$ and $\hB^I$ then start at weight two, and a derivative with respect to the couplings lowers the weight by one.

\subsubsection{Bare sources and counterterms}

We use dimensional regularisation with minimal subtraction, in the framework of section~8 of~\cite{Jack:2013sha}. The regularised functional is defined in position space in section~\ref{sec:sphere}. That counterterms of the form of minimal subtraction renormalise it is Hypothesis~\ref{hyp:finite} for multiscalar $\phi^4$ theory, and the statements on minimal subtraction below are to be read in this sense. The renormalised functional is
\begin{equation}
W[\gamma,g,a,\sM] = W_0[\gamma,g_0,a_0,\sM_0]+\int d^dx\sqrt\gamma\;\cX(\gamma,g,a,\sM) \, ,
\label{eq:Wren}
\end{equation}
where $W_0$ is the functional of the bare sources, which we define for non-integer $d$ in section~\ref{sec:sphere}, and $\cX$ contains the counterterms which do not depend on the fields. The bare sources are~\cite{Jack:2013sha}
\begin{equation}
g_0^I = g^I+L^I(g) \, , \qquad a_{0\mu} = a_\mu+N_I\,D_\mu g^I \, , \qquad
\sM_0 = Z_\sM\left(\sM-\fd_I\,D^2g^I-\fe_{IJ}\,D^\mu g^ID_\mu g^J-\fk\,R\right) ,
\label{eq:baresources}
\end{equation}
where $N_I$ takes values in the Lie algebra, $\fd_I$, $\fe_{IJ}$ and $\fk$ take values in $V_\sM$, and $Z_\sM$ is a linear map of $V_\sM$. The functions $L^I(g)$ are the counterterms of the couplings, which are functions of the couplings without derivatives, and by Hypothesis~\ref{hyp:finite} they start at second order in the couplings. By Identification~\ref{id:ms} they are the counterterms of the theory with constant couplings. They depend on the choice of the orthogonal factor in the renormalisation of the fields, to which we come back in section~\ref{sec:weylvar}. The coefficient $N_I$ has no term of order zero in the couplings. Indeed, the coefficients of the bare sources are covariant under field rotations, as we show in Proposition~\ref{prop:uniqueall}, and there is no invariant tensor of ${\rm O}(N)$ which is antisymmetric in two indices and symmetric in four. The term with the curvature is not contained in the flat-space expression of~\cite{Jack:2013sha}, and we add it for curved backgrounds. For the counterterms we need the terms with at most four factors of $Dg$. They are listed in \eqref{eq:X} in appendix~\ref{app:anomcoeff}, and we denote their coefficients by Fraktur letters. The first terms are $\cX=\fa\,\cE_4+\fb\,R^2+\fg_{IJ}\,\cG^{\mu\nu}D_\mu g^ID_\nu g^J+\ldots$, where $\cE_4$ is the Euler density and $\cG^{\mu\nu}=R^{\mu\nu}-\tfrac12\gamma^{\mu\nu}R$ is the Einstein tensor. We call a power series in the couplings a pure pole if its coefficients are polynomials in $1/\vep$ without a constant term. In minimal subtraction the coefficients of \eqref{eq:X} and $L^I$, $N_I$, $Z_\sM-1$, $\fd_I$, $\fe_{IJ}$ and $\fk$ are pure poles, and their coefficients are real.

\subsubsection{Weyl variation and anomaly}
\label{sec:weylvar}

The bare action is invariant under a Weyl rescaling $\gamma_{\mu\nu}\to(1+2\sigma)\gamma_{\mu\nu}$ with an arbitrary function $\sigma(x)$, if the bare field is rescaled with the weight $-\tfrac12(d-2)$, the bare couplings with the weight $-\vep$ and the bare mass source with the weight $-2$, while the bare gauge field is unchanged. Indeed, the measure $\sqrt\gamma$ has weight $d$, and $d-2(d-2)-\vep=0$. The bare functional is therefore invariant,
\begin{equation}
W_0\big[(1+2\sigma)\gamma,\,(1-\vep\sigma)g_0,\,a_0,\,(1-2\sigma)\sM_0\big] = W_0\big[\gamma,g_0,a_0,\sM_0\big] \, .
\label{eq:bareweyl}
\end{equation}
In dimensional regularisation there is no anomaly at this stage, and in section~\ref{sec:sphere} this invariance holds by construction. We now express it through the renormalised sources. Let $\hb^{{\rm L}\,I}$ be the vector field on the space of couplings which is defined by $\hb^{{\rm L}\,J}\pr_Jg_0^I=-\vep\,g_0^I$, so that the variation $\delta g^I=\sigma\,\hb^{{\rm L}\,I}$ gives the bare couplings their weight. For the bare gauge field to be invariant this variation has to be combined with a gauge transformation \eqref{eq:gaugetr} of all sources, as in the background gauge identity of~\cite{Reehorst:phi4}. In appendix~\ref{app:gaugevar} we show that there is one variation of the renormalised sources without a term with $\pr_\mu\sigma$ which has the required effect on the bare couplings and on the bare gauge field, up to a gauge transformation, see Lemma~\ref{lem:rhoxi}. It is $\delta g^I=\sigma\,\hB^I$ and $\delta a_\mu=\sigma\,\hr_ID_\mu g^I$, and the parameter of the gauge transformation is $\sigma\,\xi^{\rm L}(g)$. Here $\hB^I$, $\hr_I$ and $\xi^{\rm L}$ are the solution of the equations \eqref{eq:xiL} of section~8 of~\cite{Jack:2013sha}, which we give in appendix~\ref{app:gaugevar}. They are determined by the counterterms $L^I$ and $N_I$ as power series in the couplings whose coefficients are a priori Laurent polynomials in $\vep$, which the superscript indicates. In particular $\hB^I=\hb^{{\rm L}\,I}-(\xi^{\rm L}g)^I$.

The bare sources are not unique. They are defined up to a gauge transformation \eqref{eq:gaugetr} of the bare sources whose parameter is a covariant pure pole in the couplings. Such a transformation changes $L^I$, $N_I$, $\hb^{{\rm L}\,I}$ and $\xi^{\rm L}$, but it does not change the renormalised functional, $\hB^I$ and $\hr_I$, see Lemma~\ref{lem:frame}. Moreover, $\hB^I$ and $\hr_I$ are determined by the renormalised functional alone, as we show after Lemma~\ref{lem:regvar}.

In appendix~\ref{app:finite} we show that $\hB^I$ and $\hr_I$ are regular, given Hypothesis~\ref{hyp:finite}, see Lemma~\ref{lem:regvar}. Since the counterterms are pure poles, it follows that $\hB^I+\vep\,g^I$ and $\hr_I$ do not depend on $\vep$ and that they are given by the residues $L^I_1$ and $N_{I,1}$ of the simple poles of $L^I$ and $N_I$, see Lemma~\ref{lem:rhoxi}. The variation is
\begin{equation}
\delta^{\rm W}_\sigma g^I = \sigma\,\hB^I \, , \qquad \delta^{\rm W}_\sigma a_\mu = \sigma\,\hr_I\,D_\mu g^I \, , \qquad
\xi = -N_{I,1}\,g^I \, , \qquad \hr_I = g^J\left(\pr_JN_{I,1}-\pr_IN_{J,1}\right) ,
\label{eq:rhoxi}
\end{equation}
where $\hB^I$ is given by \eqref{eq:Bhat} with $\beta^I=(g^J\pr_J-1)L^I_1$ and with this $\xi$. Both $\xi$ and $\hr_I$ take values in the Lie algebra. These are the expressions of~\cite{Jack:2013sha} for quartic couplings. The equations \eqref{eq:xiL} also imply the relation
\begin{equation}
\hr_I\,\hB^I = \hr_I\,B^I = 0 \, ,
\label{eq:rhoB}
\end{equation}
as in~\cite{Jack:2013sha}. In~\cite{Jack:2013sha} the expressions \eqref{eq:rhoxi} and the relation \eqref{eq:rhoB} are obtained under the assumption that $N_I$ and the rotation which is $\xi-\xi^{\rm L}$ in our notation contain only poles. We derive them in Lemma~\ref{lem:rhoxi} from the form of the bare gauge field and from the regularity of $\hB^I$ and $\hr_I$.

While $B^I$ and $\hr_I$ do not depend on the choice of the bare sources, the two terms $\beta^I$ and $\xi$ of $B^I$ do. We use these names for a symmetric renormalisation factor of the fields. This is the conventional choice, in which the finite part of the anomalous dimension matrix of the fields is symmetric~\cite{Jack:2013sha,Bednyakov:2021ojn}. By Lemma~\ref{lem:frame} we can choose the bare sources such that $\xi^{\rm L}=\xi$ and $\hb^{{\rm L}\,I}=\hb^I$, without a change of the residues $L^I_1$ and $N_{I,1}$, and we do so from here on. We explain this choice at the end of appendix~\ref{app:gaugevar}. The gauge transformation also rotates the bare mass source. We include this rotation in the coefficient $\gamma_\sM$ below.

The bare mass source $\sM_0$ in \eqref{eq:baresources} has to change by its weight and by the same gauge transformation. This gives
\begin{equation}
\delta^{\rm W}_\sigma\sM = -\sigma\left[(2-\gamma_\sM)\sM+\delta_I\,D^2g^I+\epsilon_{IJ}\,D_\mu g^ID^\mu g^J+\bar\eta\,R\right]
-2\,\pr_\mu\sigma\,\theta_I\,D^\mu g^I-\bar\tau\,\nabla^2\sigma \, ,
\label{eq:deltaM}
\end{equation}
where the linear map $\gamma_\sM$ of $V_\sM$ is the anomalous dimension of the mass operator combined with the rotation by $\xi$, see appendix~\ref{app:masscoeff}, and $\delta_I$, $\epsilon_{IJ}=\epsilon_{JI}$, $\theta_I$, $\bar\eta$ and $\bar\tau$ take values in $V_\sM$. For $d=4$ and with $\eta=6\,\bar\eta$ and $\tau=\bar\tau$ this is the variation of~\cite{Jack:2013sha}. We give the coefficients in terms of the counterterms in appendix~\ref{app:weyl}.

We collect these variations and the rescaling of the metric in one operator,
\begin{equation}
\mathbb{W}_\sigma = \int d^dx\left[\sigma\left(2\,\gamma_{\mu\nu}\frac{\delta}{\delta\gamma_{\mu\nu}}+\hB^I\frac{\delta}{\delta g^I}
+\hr_ID_\mu g^I\cdot\frac{\delta}{\delta a_\mu}\right)+\delta^{\rm W}_\sigma\sM\cdot\frac{\delta}{\delta\sM}\right] .
\label{eq:Delta}
\end{equation}
By \eqref{eq:bareweyl} we have $\mathbb{W}_\sigma W_0=0$, and therefore
\begin{equation}
\mathbb{W}_\sigma W = \mathbb{W}_\sigma\!\int d^dx\sqrt\gamma\;\cX \equiv \cA[\sigma] \, .
\label{eq:anomalydef}
\end{equation}
The anomaly $\cA[\sigma]$ is a local functional, and it is finite if $W$ is. To state the terms which we need we first set $a_\mu=0$ and $\sM=0$. Then
\begin{align}
\cA[\sigma] = \int d^dx\sqrt\gamma\,\Big[&\sigma\left(p\,\cE_4+q\,R^2\right)-2\,W_I\,\pr_\mu\sigma\,\cG^{\mu\nu}\pr_\nu g^I
+\sigma R\,e_I\,\nabla^2g^I+R\,h_I\,\pr_\mu\sigma\,\pr^\mu g^I \notag\\
&+r\,R\,\nabla^2\sigma+u_I\,\nabla^2\sigma\,\nabla^2g^I-\sigma\,G_{IJ}\,\cG^{\mu\nu}\pr_\mu g^I\pr_\nu g^J
+\tfrac12\,\sigma\,S_{IJ}\,\nabla^2g^I\nabla^2g^J \notag\\
&+E_{IJ}\,\pr_\mu\sigma\,\pr^\mu g^I\,\nabla^2g^J+\nu_{IJ}\,\nabla^2\sigma\,\pr_\mu g^I\pr^\mu g^J \notag\\
&+R_{IJK}\,\pr_\mu\sigma\,\pr^\mu g^I\,\pr_\nu g^J\pr^\nu g^K+\ldots\Big] \, ,
\label{eq:anomaly}
\end{align}
where the dots stand for a term $\sigma R\,f_{IJ}\,\pr_\mu g^I\pr^\mu g^J$ and for terms with $\sigma$ and three or four factors of $\pr g$, which we do not need. This defines the coefficients $p$, $q$, $r$, $W_I$, $e_I$, $h_I$, $u_I$, $G_{IJ}$, $S_{IJ}$, $E_{IJ}$, $\nu_{IJ}$ and $R_{IJK}$. In appendix~\ref{app:finite} we show that all of them are regular for a suitable choice of the counterterms which multiply the curvature, given Hypothesis~\ref{hyp:finite}, see Proposition~\ref{prop:finite}. The tensors $G_{IJ}$, $S_{IJ}$ and $\nu_{IJ}$ are symmetric, and $R_{IJK}$ is symmetric in its last two indices. The matrix $E_{IJ}$ is not symmetric. We use the same letters as in~\cite{Reehorst:phi4} for the coefficients $W_I$, $G_{IJ}$, $S_{IJ}$, $E_{IJ}$ and $R_{IJK}$ of the corresponding terms. The normalisations are those of \eqref{eq:anomaly}. As in~\cite{Reehorst:phi4} we denote the function which multiplies the Euler density by $A_{\rm E}$. In our normalisation
\begin{equation}
A_{\rm E} \equiv -4\,p \, .
\end{equation}
The relation of our coefficients to those of~\cite{Jack:2013sha} and the differences between our conventions and those of~\cite{Reehorst:phi4} are given in appendix~\ref{app:conventions}. At a fixed point in four dimensions $A_{\rm E}$ is proportional to the difference between the $a$-anomaly and its value in the free theory, since with the curvature counterterms of Proposition~\ref{prop:finite} the coefficient $A_{\rm E}$ vanishes at $g=0$, and $G_{IJ}$ is positive definite at lowest order~\cite{Jack:1990eb,Jack:2013sha}.

With the sources $a_\mu$ and $\sM$ the derivatives become covariant, and the anomaly contains further terms with the field strength and with the mass source. We need two of their coefficients. The first is the coefficient of the term $\pr_\mu\sigma\,f^{\mu\nu}\cdot D_\nu g^I$, which we write as $-2\,Q_I$, so that $Q_I=Q_I^{\rm JO}/(16\pi^2)$. The second is the coefficient $t_\sM$ of the term $\nabla^2\sigma\,\sM$. The complete list of structures is given in appendix~\ref{app:finite}.

\subsection{Regularised functional and hypothesis}
\label{gs:sec:phi4sphere}

For multiscalar $\phi^4$ theory the bare functional $W_0^S$ on the sphere is the functional of section~\ref{gs:sec:sphere} with the vertices of appendix~\ref{app:gauge}. These are the vertex $V^{(4)}[g_0]$ of the bare couplings, the vertex $V^{(2)}[\sM_0]$ of the bare mass source, the vertex of the current with the bare gauge field and the term of second order in the gauge field. On the line \eqref{eq:line} the weights \eqref{gs:eq:wn} are $w_4=\vep$ and $w_2=2$, so that \eqref{eq:covS} reads
\begin{equation}
W_0^S\big[\Omega^{\vep}\,g_0\!\circ\!f,\;f^*a_0,\;\Omega^{2}\,\sM_0\!\circ\!f\big]
= W_0^S\big[g_0,a_0,\sM_0\big] \, ,
\label{gs:eq:covphi4}
\end{equation}
and the chart \eqref{eq:chart} gives the flat sources $\Omega^\vep g_0$, $a_0$ and $\Omega^2\sM_0$. We take $D_0=1$, the smallest value which the arguments of appendices~\ref{app:line} and~\ref{app:unique} allow, so that Hypothesis~\ref{hyp:finite} below is assumed for the smallest of these classes of sources, and we keep the notation $D_0$ to display the dependence on it.

\subsubsection{Hypothesis}
\label{sec:finiteness}
\label{sec:hypothesis}

The properties derived so far concern the bare functional. The renormalised functional on the sphere is \eqref{eq:Wren} with the bare functional $W_0^S$, with the bare sources \eqref{eq:baresources} and with the counterterms of appendix~\ref{app:anomcoeff}, in which the curvature and the Laplacian are those of the unit sphere. Its finiteness does not follow from the definition. It rests on Hypothesis~\ref{hyp:finite} of appendix~\ref{gs:sec:cases} for this theory. The hypothesis states that the flat functional $W^{\rm fl}$ with the bare sources \eqref{eq:baresources} in flat space is finite, for real covariant pure poles $L^I$ without terms of degree zero and one and real pure poles as the other coefficients. It concerns all sources of the class of flat sources of section~\ref{sec:chart} with $D_0=1$ and all families of such sources which depend analytically on $\vep$.

Let us make the terms of this statement precise. The functional $W_0^{\rm fl}$ is the bare functional in flat space which is defined in section~\ref{sec:chart}, and the bare sources are \eqref{eq:baresources} in flat space with $g_0^I=g^I+L^I$. The density $\cX^{\rm fl}$ is the density of the counterterms of appendix~\ref{app:anomcoeff} in flat space, so that its coefficients are $\fs_{IJ}$, $\fr_{IJK}$, $\fq_{IJKL}$, $\fj_I$, $\fl_{IJ}$, $\fm$, $\fp_{IJ}$ and $\fy$. A pure pole, as defined in section~\ref{sec:setup}, is real if its polynomials in $1/\vep$ have real coefficients, and $L^I$ is covariant if it is unchanged by a rotation of the couplings with a constant element of ${\rm O}(N)$, including the reflections. Finite means that at each order in the sources the functional, which is defined by the continuation of section~\ref{sec:continuation}, has no pole at $\vep=0$. We say that a family of sources depends analytically on $\vep$ if for small $|\vep|$ it is given by power series in $\vep$ whose coefficients are sources in the class of flat sources with supports in a fixed compact set, and which converge uniformly together with all their derivatives. For such a family the functional in $d=4-\vep$ dimensions is evaluated on the member of the family with the same $\vep$. The terms of the functional are continuous in the sources, so that the result is a holomorphic function of $\vep$ in a punctured neighbourhood of $\vep=0$, as we show in appendix~\ref{app:polar}. Finite means that at each order in the sources this function is analytic at $\vep=0$. Such families occur because the chart \eqref{eq:chart} contains the factor $\Omega^\vep$. The hypothesis does not contain the normalisation that $N_I$ has no term of degree zero in the couplings, which follows from it by Proposition~\ref{prop:uniqueall}.

The hypothesis is the statement that products of the composite operators $\phi^4$ and $\phi^2$ and of the currents are renormalised by local counterterms of the form of minimal subtraction, for sources which are test functions. It is the analogue of the assumption of section~8 of~\cite{Jack:2013sha} and of the first hypothesis of~\cite{Reehorst:phi4}, for the regularisation of this section and for a specified class of sources. The remarks made in~\cite{Reehorst:phi4} on related results in other frameworks~\cite{Epstein:1973gw,Holland:2014pna,Frob:2015uqy} apply here as well, except that our hypothesis does not refer to the form of the anomaly. Our statement differs from these two in four respects. It concerns flat space only, it specifies the bare functional and the class of sources, and it does not contain the invariance under diffeomorphisms and background gauge transformations and the form of the Weyl variation, which we derive, see appendix~\ref{app:hypL}. Finally, it does not state that $L^I$ are the counterterms of minimal subtraction in momentum space, which is Identification~\ref{id:ms} below.

If the counterterms $L^I$ are changed by a transformation of Lemma~\ref{lem:frame}, the hypothesis continues to hold, and $\hB^I$ and $\hr_I$ do not change. The hypothesis therefore does not single out a beta function $\beta^I$ and a rotation $\xi$ separately, and our results concern the vector field $\hB^I$ which a choice of $L^I$ defines. They hold for every choice of $L^I$ for which the hypothesis holds. As explained in section~\ref{sec:weylvar}, we use these names for a symmetric renormalisation factor of the fields. This convention refers to the counterterms of Identification~\ref{id:ms}, and the proofs do not use it.

From the hypothesis we derive three statements in appendices~\ref{app:unique} and~\ref{app:finite}. For given $L^I$ the other coefficients of the bare sources and the coefficients of $\cX^{\rm fl}$ are determined by the finiteness and are covariant under field rotations, which is Proposition~\ref{prop:uniqueall}. The coefficients $\hB^I$, $\hr_I$ and those of \eqref{eq:deltaM} are regular, which is Lemma~\ref{lem:regvar}. Finally, for a suitable choice of the counterterms which multiply the curvature all coefficients of the anomaly are regular, and the terms of order $n\geq1$ of the renormalised functional on the sphere are finite for all sources in the class of sources, which is Proposition~\ref{prop:finite}. The proof of the gradient formula uses these statements, the algebraic identities of Lemmas~\ref{lem:rhoxi} and~\ref{lem:frame}, which hold for all covariant pure poles, and nothing else about $L^I$.

\subsubsection{Identification}
\label{gs:sec:identification}

The comparison with computations in momentum space requires a further statement, which the proof of the gradient formula does not use.

\begin{identification}\label{id:ms}
Hypothesis~\ref{hyp:finite} for multiscalar $\phi^4$ theory holds with the counterterms $L^I$ of minimal subtraction of multiscalar $\phi^4$ theory with constant couplings in momentum space, for a renormalisation factor of the fields whose orthogonal factor is the exponential of a covariant pure pole in the couplings.
\end{identification}

These counterterms are real covariant pure poles without terms of degree zero and one, as we explain in appendix~\ref{app:hypL}. With Identification~\ref{id:ms} the function $\beta^I=(g^J\pr_J-1)L^I_1$ of section~\ref{sec:weylvar} is the usual beta function of minimal subtraction for a symmetric renormalisation factor of the fields, and $\hB^I$ differs from it by the infinitesimal rotation $\xi$. We did not show that the counterterms of minimal subtraction, which are defined through amplitudes at fixed momenta, renormalise the functional of section~\ref{gs:sec:sphere}. Appendix~\ref{app:momentum} describes what a derivation of Identification~\ref{id:ms} would require. In the theorems of this paper, minimal subtraction means the scheme of Hypothesis~\ref{hyp:finite}, with counterterms which are pure poles. With Identification~\ref{id:ms} it is minimal subtraction in momentum space.

\subsection{Properties from the hypothesis}
\label{gs:sec:phi4prop}

We now verify the properties of appendix~\ref{gs:sec:properties} for this theory. The general constructions of section~\ref{gs:sec:universal} provide the form of the anomaly and the Ward identity, and the hypothesis provides finiteness and regularity.

\begin{proposition}
\label{gs:prop:phi4prop}
Assume Hypothesis~\ref{hyp:finite} for multiscalar $\phi^4$ theory. Then the renormalised functional of multiscalar $\phi^4$ theory on the unit sphere $S^d$, $d=4-\vep$, in minimal subtraction with the curvature counterterms of Proposition~\ref{prop:finite}, has Properties~\ref{gs:P:functional} to~\ref{gs:P:sphere} and the two-point form of appendix~\ref{gs:sec:properties}, with Property~\ref{gs:P:functional} in instance~(a). The class of sources is that of section~\ref{sec:sphere}, the weights are $k_I=1$, the space $V_\sM$ consists of the symmetric matrices, the anomaly is \eqref{eq:anomalydef}, and $c_{IJ}(\vep)=c_\cO\,\delta_{IJ}$ with the positive constant $c_\cO$ of Lemma~\ref{gs:lem:free}. The same holds in the scheme for the mass source of section~\ref{gs:sec:dimtwo} in which $\theta_I$ and $\bar\tau$ vanish, with the redefinition given in appendix~\ref{app:masscoeff}.
\end{proposition}

\begin{proof}
Property~\ref{gs:P:functional} is the third part of Proposition~\ref{prop:finite} together with Proposition~\ref{prop:gauge}. The term of order zero of the expansion around constant couplings is regular as well, since the constants are in the class of sources and the functional vanishes for vanishing sources, by the convention of section~\ref{sec:sphere} and since $\fa$, $\fb$ and $\fk$ vanish at $g=0$ by Proposition~\ref{prop:finite} and its proof. For the gauge invariance of the renormalised functional we use in addition that the coefficients of the bare sources and of the counterterms are covariant. For the counterterms of flat space this is Proposition~\ref{prop:uniqueall}. For the curvature counterterms it follows from \eqref{eq:steps} of appendix~\ref{app:finite}, whose right-hand sides are built covariantly from the counterterms of flat space, from the coefficients of the Weyl variation and from the curvature counterterms of the earlier steps, while $\fa$ is invariant by its construction there. For Property~\ref{gs:P:weyl} the vector field $\hB^I$ and the one-form $\hr_I$ are those of \eqref{eq:rhoxi}, the variation of the mass source is \eqref{eq:deltaM}, and the operator is \eqref{eq:Delta}. The coefficients are regular by Lemma~\ref{lem:regvar} and, for $\bar\tau$ and $\bar\eta$, by the first statement of Proposition~\ref{prop:finite}. They are covariant by Proposition~\ref{prop:uniqueall} and by the covariance of $\fk$ shown above. Since $L^I$ has no terms of degree zero and one, \eqref{eq:rhoxi} gives $\hB^I=-\vep\,g^I$ up to terms of second order, and $\gamma_\sM$ has no term of order zero by Lemma~\ref{gr4:lem:zero} below. By the same lemma the coefficients $\ell_{\rm h}=d\,(\delta-2\theta)$ and $\vk_{\rm h}-\vk_{\rm c}=-d\,\bar\tau$ of \eqref{gs:eq:dimtwodata} vanish at $g=0$, which is the last statement of Property~\ref{gs:P:weyl}. We have $\hr_I\hB^I=0$ by \eqref{eq:rhoB}. The operators commute by Lemma~\ref{gcu:lem:commute}, since by \eqref{eq:barehat} of appendix~\ref{app:gaugevar} and by the definition of the coefficients \eqref{eq:masscoeff} in appendix~\ref{app:masscoeff} the variation changes the bare couplings, the bare gauge field and the bare mass source by their weights and by the gauge transformation with the parameter $\sigma\,\xi^{\rm L}$. In components the commutation is \eqref{eq:rhoB} and \eqref{eq:Mrel}, as explained in appendix~\ref{gs:sec:properties} and section~\ref{gs:sec:dimtwo}. For Property~\ref{gs:P:anomaly} the anomaly is $\cA[\sigma]=\mathbb{W}_\sigma F$ with the integral $F$ of the counterterms \eqref{eq:X} of appendix~\ref{app:anomcoeff}, which is \eqref{eq:anomalydef}. By Proposition~\ref{gs:prop:msanomaly} it is local, invariant under background gauge transformations and homogeneous of Weyl weight $-\vep$, and the relation $\mathbb{W}_\sigma\cA[\sigma']=\mathbb{W}_{\sigma'}\cA[\sigma]$ holds because the operators commute. Its coefficients are regular by the first part of Proposition~\ref{prop:finite}. Property~\ref{gs:P:ward} is the Ward identity \eqref{eq:wardS}, which holds by Proposition~\ref{gs:prop:msward}, since the bare functional has the invariances of Proposition~\ref{gs:prop:bare} and the variation of section~\ref{sec:weylvar} induces the rescaling of the bare sources up to a gauge transformation. The class of sources satisfies the assumptions of appendix~\ref{gs:sec:class}, see appendix~\ref{app:rules}. For Property~\ref{gs:P:sphere} we show the statements on the functional in appendix~\ref{app:rotations}, the statements on the anomaly follow from Property~\ref{gs:P:anomaly} by Lemma~\ref{gsp:lem:local}, and we show the statements with the mass source in appendix~\ref{app:massward}. For the two-point form we use Lemma~\ref{gs:lem:free}. Its assumptions hold, since $L^I$ starts at second order in the couplings and the bare gauge field and the bare mass source at $\sM=0$ start at second order as well, by Lemma~\ref{gr4:lem:zero} below for the latter. The term of second order is therefore the first term of the two-point form with $c_{IJ}=c_\cO\,\delta_{IJ}$. For the same reason we find that the counterterms which are linear in $\sM$ do not contribute to the anomaly at $\sM=0$ at this order, since the variation $\delta^{\rm W}_\sigma\sM$ at $\sM=0$ starts at second order and the counterterm $\fn$ of \eqref{eq:X} has no term of order zero. The constant mass sources of Properties~\ref{gs:P:functional} and~\ref{gs:P:sphere} are treated in Lemma~\ref{gr4:lem:const} below. The redefinition \eqref{eq:Mredef} changes the mass source by local terms with regular coefficients, see appendix~\ref{app:masscoeff}. At $\sM'=0$ and constant couplings it corresponds to the constant mass source $-k_RR_d$, where $k_R$ vanishes to second order at $g=0$, since $k_R=(d_I\hB^I-\bar\tau)/(2(d-1))$ with $d_I=(\hP^{-1})_I{}^J\theta_J$ and since $\theta_I$ vanishes at $g=0$ and $\bar\tau$ to second order by Lemma~\ref{gr4:lem:zero} below. Properties~\ref{gs:P:functional} and~\ref{gs:P:sphere} therefore hold in this scheme by Lemma~\ref{gr4:lem:const}, as stated for the mass operator on the sphere also in appendix~\ref{app:massward}, and Property~\ref{gs:P:weyl} holds with the variation \eqref{gs:eq:scheme}, in which $\vk_{\rm h}=\vk_{\rm c}$ and $\ell_{{\rm h}J}=d\,\delta_J=d\,\mathsf{D}_{JK}\hB^K$ vanishes at $g=0$. Properties~\ref{gs:P:anomaly} and~\ref{gs:P:ward} hold because the redefinition is a covariant local expression with regular coefficients, so that the anomaly in the new scheme is again the Weyl variation of the integral of local counterterms. The two-point form holds with the same $c_{IJ}(\vep)$ because at $\sM'=0$ the old source is of second order in the couplings and $\sfm$, $i$ and $t_\sM$ vanish at $g=0$ by Lemma~\ref{gr4:lem:zero}.
\end{proof}

The proposition provides the functional on the unit sphere, with an anomaly which is the Weyl variation of a local functional. Proposition~\ref{gs:prop:relevant}, Lemma~\ref{gs:lem:local}, Proposition~\ref{gs:prop:sharp} and Lemma~\ref{gsh:lem:quant} therefore apply on the unit sphere.

\subsubsection{Coefficients at vanishing couplings and constant sources}
\label{gs:sec:phi4const}

The proof of Proposition~\ref{gs:prop:phi4prop} uses two lemmas, which we prove from the appendices.

\begin{lemma}
\label{gr4:lem:zero}
Assume Hypothesis~\ref{hyp:finite} for multiscalar $\phi^4$ theory. In minimal subtraction the following holds for every $\vep$. First, $Z_\sM$, $\fd_I$, $\fk$ and $\fn$ satisfy $Z_\sM=1$ and $\fd_I=\fk=\fn=0$ at $g=0$, and $\fk$ has no term of first order in the couplings. Second, $\delta_I$, $\theta_I$, $\gamma_\sM$, $\bar\tau$ and $\bar\eta$ vanish at $g=0$, and $\bar\tau$ and $\bar\eta$ vanish to second order. Third, the coefficients $i$ and $t_\sM$ of the anomaly \eqref{eq:fullanomaly} of appendix~\ref{app:finite} vanish at $g=0$. Fourth, the one-point function $\sfm$ of the mass operator on the unit sphere vanishes at $g=0$.
\end{lemma}

\begin{proof}
We use Lemma~\ref{lem:lowest}, which gives $Z_\sM=1$ and $\fd_I=0$ at $g=0$, and by the paragraph on the lowest order in the proof of Lemma~\ref{lem:regvar} the coefficients $\hr_I$, $\delta_I$, $\theta_I$ and $\gamma_\sM$ vanish at $g=0$. By \eqref{eq:steps} of appendix~\ref{app:finite} the curvature counterterm $\fk$ is ${\rm PP}[\fd_I\hB^I/(2(d-1))]$. Since $\fd_I$ vanishes at $g=0$ and $\hB^I$ has no term of order zero, $\fd_I\hB^I$ has no terms of order zero and one, and neither has $\fk$. Hence the coefficients $\bar\tau=-\fd_I\hB^I+2(d-1)\,\fk$ of \eqref{eq:masscoeff} and $\bar\eta=-(\hB\!\cdot\!\pr-\gamma_\sM)\fk$ vanish to second order. By \eqref{eq:steps} the counterterm $\fn={\rm PP}[(\fj_I\hB^I-\bar\tau\cdot\fm)/(2(d-1))]$ vanishes at $g=0$, since $\hB^I$ and $\bar\tau$ do. The coefficients $t_\sM=\fj_I\hB^I-2(d-1)\fn-\bar\tau\cdot\fm$ and $i=-(\vep-\hB\!\cdot\!\pr)\fn+\fn\cdot\gamma_\sM-\bar\eta\cdot\fm$ of the anomaly of appendix~\ref{app:finite} therefore vanish at $g=0$. For the last statement we note that at $g=0$ the bare mass source at $\sM=0$ vanishes, since $\fd_I$ and $\fk$ vanish at $g=0$, so that the one-point function of the mass operator at $g=0$ is the term of first order of the bare functional in the mass source plus the term of the counterterm $\fn\,\sM R$. The former is a propagator at coincident points, which vanishes, see section~\ref{sec:continuation}, and the latter vanishes since $\fn=0$ at $g=0$.
\end{proof}

By this lemma the bare mass source and the variation $\delta^{\rm W}_\sigma\sM$ at $\sM=0$ start at second order in the couplings. Indeed, at $\sM=0$ the bare mass source is $-Z_\sM(\fd_ID^2g^I+\fe_{IJ}D_\mu g^ID^\mu g^J+\fk R)$, which is of second order by the first statement, and the terms of $\delta^{\rm W}_\sigma\sM$ at $\sM=0$ which are not of second order in the couplings are multiplied by $\delta_I$ or $\theta_I$, which vanish at $g=0$, or by $\bar\eta$ or $\bar\tau$, which vanish to second order.

\begin{lemma}
\label{gr4:lem:const}
Assume Hypothesis~\ref{hyp:finite} for multiscalar $\phi^4$ theory, and let $\mu(g)$ be a regular covariant function of the couplings with values in $V_\sM$ which vanishes at $g=0$. Then Properties~\ref{gs:P:functional} and~\ref{gs:P:sphere} hold for the expansion of the renormalised functional around constant couplings $g_c$ together with the constant mass source $\mu(g_c)$. The second statement of Property~\ref{gs:P:sphere} also holds at $\mu$ for second variations of $W$ in which one direction is $\sigma$ times a constant element of $V_\sM$ or $\sigma z^I$, and the other is a variation $\delta\sM$ of the class of sources, and for $\cA[1]$ it follows by Lemma~\ref{gsp:lem:local}.
\end{lemma}

\begin{proof}
We first note that a constant function belongs to the class of sources of section~\ref{sec:sphere}. By the third statement of Proposition~\ref{prop:finite} the terms of order $n\geq1$ of $W$ are finite for all sources in the class of sources, so that $W$ is a formal power series in $g^I-g_c^I$, $a_\mu$ and $\sM$ whose terms are regular functions of $g_c$, with the term of order zero fixed by the convention of section~\ref{sec:continuation}. Its terms of order $k$ in $\sM$, evaluated at $\sM=\mu(g_c)+\delta\sM$, give a series in which each order in the couplings receives contributions from finitely many $k$, since $\mu$ vanishes at $g=0$. We conclude that the expansion around $(g_c,\mu(g_c))$ is a formal power series with regular terms. The gauge invariance holds as in Proposition~\ref{prop:gauge}, since a constant mass source is rotated by a constant parameter as in \eqref{eq:gaugetr}.

For Property~\ref{gs:P:sphere} we repeat the argument of appendix~\ref{app:rotations} with constant couplings and a constant mass source. That argument concerns the integrals \eqref{eq:gram} of the bare functional with scalar sources, and the mass vertex is a scalar vertex without derivatives, as the vertices of the couplings. If one point is held fixed, the integral over the remaining points is again of the form \eqref{eq:gram}, with an enlarged parallel space on which the rotations act on the vector $b$ and on the fixed point, and constant sources of both kinds are invariant. We obtain that the one-point functions are constant, that the one-point function of the current vanishes, and that a harmonic of degree one is an eigenfunction of the second variation in any two scalar directions, in particular in the directions $\sigma z^I$ or $\sigma$ times a constant element of $V_\sM$ and $\delta\sM$. The extension to all $d$ and $\Dp$ by meromorphy is the first step of appendix~\ref{app:rotations}, which does not depend on the kind of the scalar sources. In the second step of that appendix the bare mass source at constant couplings and constant $\sM=\mu$ is $Z_\sM(\mu-\fk R_d)$, which is constant. Its first variation in the direction $\sigma\,\delta g^I$ is $\sigma$ times a constant, where we use $\nabla^2\sigma=-d\,\sigma$ for the term with $\fd_I$, and its first variation in the direction $\delta\sM$ is $Z_\sM\delta\sM$, again a scalar source. The terms of second order of the bare sources multiply one-point functions at constant sources, which are constant, and the second variations of the counterterms are local and are reduced by integrations by parts. The terms with the bare gauge field are reduced to terms with scalar sources by Proposition~\ref{prop:gauge}, as there. The statements on the anomaly are statements on a local functional and follow from the rules of appendix~\ref{app:rules} with a constant $\sM$.
\end{proof}

\subsubsection{Spheres of other radius}
\label{gs:sec:radius}

Proposition~\ref{gs:prop:phi4prop} provides the functional on the unit sphere. The interpretation of the value \eqref{gs:eq:fixed} in section~\ref{gs:sec:fixed} uses in addition the response of the functional to a constant rescaling of the metric. For a constant rescaling no path from flat space is needed, and the construction gives the functional on the spheres of every radius at constant couplings, with the anomaly as its variation under a constant rescaling. This is the part of the extended form of the properties of appendix~\ref{gs:sec:properties} which the interpretation uses. We derive it for the $\phi^4$, the Yukawa, the sextic and the cubic theory.

For $r>0$ we define the bare functional on the background $r^2\mathring\gamma_{\mu\nu}$ by
\begin{equation}
W_0\big[r^2\mathring\gamma;\{\psi_n\},a_0\big] = W_0^S\big[\{r^{w_n}\psi_n\},a_0\big] \, ,
\label{gs:eq:Wradius}
\end{equation}
where $\psi_n$ is the bare source of a vertex with $n$ scalar legs and $w_n$ is its weight \eqref{gs:eq:wn}, and where for the Yukawa theory the bare Yukawa couplings enter with their weight $\tfrac12(4-d)$ of Proposition~\ref{gf:prop:bare}. In integer dimension \eqref{gs:eq:Wradius} is the change of variables $X\to rX$ in \eqref{eq:gram}. A vertex with $n$ scalar legs receives $r^d$ from the measure and $r^{-n\Dp}$ from its propagators. A fermion line with its chord scales with $r^{-2\Delta_\psi}$, so that a Yukawa vertex receives $r^{d-\Dp-2\Delta_\psi}$, which is $r^{(4-d)/2}$ on the line \eqref{eq:line}, and the gauge field enters unchanged. For sources which vanish near the point $P$ of the chart the sphere of radius $r$ is the background $(r\Omega)^2\delta_{\mu\nu}$. There the definition of the bare functional on conformally flat backgrounds of section~\ref{sec:chart}, with its extension to Dirac fermions in appendix~\ref{gf:sec:chart}, and \eqref{eq:chart} give for both sides of \eqref{gs:eq:Wradius} the flat functional with the bare sources multiplied by $r\Omega$ to the power of their weights. The renormalised functional $W_r$ on $r^2\mathring\gamma_{\mu\nu}$ is \eqref{gs:eq:Wren} with this bare functional and with the bare sources and the counterterms of the theory, including the curvature counterterms, in which the curvature, the Laplacian and the measure are those of $r^2\mathring\gamma_{\mu\nu}$. We write $W_r(g,\sM_{\rm c})$ for its value at constant couplings $g$, $a_\mu=0$ and constant relevant sources $\sM_{\rm c}$, and $W_r(g)$ for $\sM_{\rm c}=0$, so that $W_1(g)=\cW(g)$. In the proposition and the corollary below the relevant sources of even dimension are those of the divergence scheme. On $r^2\mathring\gamma_{\mu\nu}$ at constant couplings the source of minimal subtraction is then $\sM_{\rm c}+y_0(g)\,r^{-D}$, by \eqref{gr:eq:Y}.

Let $\mathcal{I}_a$ be the local curvature scalars of dimension $d_c$ on conformally flat backgrounds of section~\ref{gs:sec:fixed}. We take the regular functions $p_a$ of the divergence scheme, which are defined by $\cA[1]=\sum_ap_a\int d^dx\sqrt\gamma\,\mathcal{I}_a$ on conformally flat backgrounds at constant couplings and vanishing relevant sources in this scheme. For $\phi^4$ theory the vanishing mass source of the divergence scheme at constant couplings is the source $\sM=y_0\,R/R_d$ of minimal subtraction, by \eqref{gr:eq:Y}. With the coefficients $p$, $q$, $i$ and $\beta_\sM$ of the anomaly of minimal subtraction in appendix~\ref{app:finite} we then have $p_{\cE_4}=p$ and $p_{R^2}=q+i\cdot y_0/R_d-\tfrac12\,(y_0/R_d)\cdot\beta_\sM\cdot(y_0/R_d)$. On the sphere of radius $r$ we have $\int d^dx\sqrt\gamma\,\mathcal{I}_a=r^{-\vep}\,V\,\mathcal{I}_a^{(1)}$, where $\mathcal{I}_a^{(1)}$ is the value on the unit sphere, for instance $\mathcal{I}_{\cE_4}^{(1)}=d(d-1)(d-2)(d-3)$ and $\mathcal{I}_{R^2}^{(1)}=d^2(d-1)^2$ for $d_c=4$. For the cubic theory the scalars $\mathcal{I}_a$ have dimension six, and Properties~\ref{gs:P:weyl} and~\ref{gs:P:anomaly} hold on conformally flat backgrounds, which include the spheres $r^2\mathring\gamma_{\mu\nu}$. For the sextic theory there are no local curvature scalars of dimension three. For constants $\varsigma_a$ we set $W_{r,\varsigma}=W_r+\sum_a\varsigma_a\int d^dx\sqrt\gamma\,\mathcal{I}_a$.

\begin{proposition}[Spheres of other radius]
\label{gs:prop:radius}
Assume Hypothesis~\ref{hyp:finite} for multiscalar $\phi^4$ theory, for the Yukawa theory, for the sextic theory, or for the cubic theory with the class of flat sources with $D_0=3$ and the set of pure poles of Proposition~\ref{gcp:prop:props}. Take the functional in minimal subtraction with the curvature counterterms of the theory and with the relevant sources of even dimension in the divergence scheme of Lemma~\ref{gr:lem:scheme}. Then for every $r>0$ the function $W_r(g,\sM_{\rm c})$ is a formal power series in $g$ and $\sM_{\rm c}$ with regular coefficients, and at $\sM_{\rm c}=0$
\begin{equation}
r\,\frac{\pr W_r}{\pr r}+\hB^I\pr_IW_r-\sum_br^{-D_b}\,\vk_b(g)\cdot\frac{\pr W_r}{\pr\sM_b} = r^{-\vep}\,V\rA(g) \, ,
\label{gs:eq:radius}
\end{equation}
where the sum runs over the relevant sources $\sM_b$ of even dimension $D_b$ and $\vk_b=\hB^Kx_{KL}\hB^L$ is the flow of $\sM_b$ of Proposition~\ref{gsh:prop:flows}. With $\beta_a=\vep\,\varsigma_a-p_a(g)$ the functional with the curvature terms satisfies
\begin{equation}
r\,\frac{\pr W_{r,\varsigma}}{\pr r}+\hB^I\pr_IW_{r,\varsigma}+\sum_a\beta_a\,\frac{\pr W_{r,\varsigma}}{\pr\varsigma_a} = \sum_br^{-D_b}\,\vk_b(g)\cdot\frac{\pr W_r}{\pr\sM_b} \, .
\label{gs:eq:radiusvarsigma}
\end{equation}
At a zero $g_*$ of $\hB^I$ the point $\sM_{\rm c}=0$ is a zero of the flows of the relevant sources on the sphere of every radius, the functions $\beta_a$ vanish at $\varsigma_a=p_a(g_*)/\vep$, and for these values $W_{r,\varsigma}(g_*)=\cW_*$ for every $r$. For the sextic theory there are no curvature terms, and $W_r(g_*)=\cW(g_*)=\cW_*$ for every $r$.
\end{proposition}

\begin{proof}
We first derive the identity at constant relevant sources $\sM_{\rm c}$ which need not vanish, as an identity of meromorphic functions of $\vep$ at each order in the couplings and in $\sM_{\rm c}$. By \eqref{gs:eq:Wradius} the bare functional is invariant under $r\to e^cr$ with a constant $c$ if each bare source is multiplied by $e^{-cw}$ with its weight $w$. It is invariant under the background gauge transformations of the bare sources, since these commute with the multiplication of the sources by constants. Proposition~\ref{gs:prop:msanomaly} assumes the invariance under Weyl rescalings with a function $\sigma(x)$ on the backgrounds on which the bare functional is defined. We use only the part of its proof for a constant $\sigma$, on the backgrounds $r^2\mathring\gamma_{\mu\nu}$, which a constant $\sigma$ maps into each other. In this part $\mathbb{W}_1$ changes the bare sources by their rescaling and by a background gauge transformation, by the third condition of section~\ref{gs:sec:msanomaly}, which is an identity of local expressions built covariantly with $d$ as a parameter and therefore holds on $r^2\mathring\gamma_{\mu\nu}$. Both changes leave the bare functional invariant, so that $\mathbb{W}_1W_r$ is the variation of the integral of the counterterms, which is the anomaly $\cA[1]$ on $r^2\mathring\gamma_{\mu\nu}$. We also use the homogeneity of the anomaly shown in that proof, which concerns the local functional only.

On the backgrounds $r^2\mathring\gamma_{\mu\nu}$ the term $2\int d^dx\,\gamma_{\mu\nu}\delta/\delta\gamma_{\mu\nu}$ of $\mathbb{W}_1$ is $r\,\pr/\pr r$ at fixed sources. At constant couplings, $a_\mu=0$ and constant $\sM_{\rm c}$ the term with $\hr_I$ vanishes, and the term with $\hB^I$ is $\hB^I\pr_IW_r$. We write $\delta^{\rm W}_1\sM_b(g,\sM_{\rm c})$ and $\cA[1](g,\sM_{\rm c})$ for the constant value of $\delta^{\rm W}_1\sM_b$ and for the anomaly on the unit sphere at constant couplings $g$, $a_\mu=0$ and constant sources $\sM_{\rm c}$, and $r^D\sM_{\rm c}$ for the sources multiplied by $r$ to the power of their dimensions. Each term of $\delta^{\rm W}_1\sM_b$ has the dimension $D_b$ of $\sM_b$ by Property~\ref{gs:P:weyl}, and the anomaly is homogeneous of Weyl weight $-\vep$. On the sphere of radius $r$ these quantities are therefore $r^{-D_b}\,\delta^{\rm W}_1\sM_b(g,r^D\sM_{\rm c})$ and $r^{-\vep}\cA[1](g,r^D\sM_{\rm c})$, and $\mathbb{W}_1W_r=\cA[1]$ reads
\begin{equation}
r\,\frac{\pr W_r}{\pr r}+\hB^I\pr_IW_r+\sum_br^{-D_b}\,\delta^{\rm W}_1\sM_b(g,r^D\sM_{\rm c})\cdot\frac{\pr W_r}{\pr\sM_b} = r^{-\vep}\,\cA[1](g,r^D\sM_{\rm c}) \, ,
\end{equation}
where the sum runs over all relevant sources. At $\sM_{\rm c}=0$ the sources of odd dimension do not contribute, since their variation vanishes there by Lemma~\ref{gs:lem:dimension}. For a source of even dimension $\delta^{\rm W}_1\sM_b(g,0)=-\vk_b(g)$ by \eqref{gs:eq:kappadef} and Proposition~\ref{gsh:prop:flows}. With $\cA[1](g,0)=V\rA(g)$ the identity at $\sM_{\rm c}=0$ is \eqref{gs:eq:radius}.

For the regularity let $s\mapsto(g_s,\sM_s)$ be the solution of $\pr_sg_s=\hB(g_s)$ and $\pr_s\sM_{b,s}=e^{-D_bs}\,\delta^{\rm W}_1\sM_b(g_s,e^{Ds}\sM_s)$ which takes the values $(g,\sM_{\rm c})$ at $s=\log r$. Order by order in the couplings and in $\sM_{\rm c}$ these are linear equations whose coefficients are regular functions times exponentials of $s$, with inhomogeneous terms from the lower orders. On the interval between $0$ and $\log r$ their solutions are power series in $\vep$, so that the values $\check g$ and $\check\sM$ at $s=0$ are power series in $g$ and $\sM_{\rm c}$ with regular coefficients, with $\check g^I=r^{\vep k_I}g^I+{\rm O}(g^2)$. Neither has a term of order zero, since the inhomogeneous terms $-e^{-D_bs}\vk_b$ vanish at $g=0$. By the identity above $\pr_sW_{e^s}(g_s,\sM_s)=e^{-\vep s}\cA[1](g_s,e^{Ds}\sM_s)$, so that
\begin{equation}
W_r(g,\sM_{\rm c}) = W_1(\check g,\check\sM)+\int_0^{\log r}ds\;e^{-\vep s}\,\cA[1]\big(g_s,e^{Ds}\sM_s\big) \, .
\label{gs:eq:radiuspath}
\end{equation}
By Property~\ref{gs:P:functional} in the divergence scheme the functional $W_1$ at constant couplings and constant relevant sources is a formal power series in $g$ and $\sM_{\rm c}$ with regular coefficients. In this scheme the property follows from the end of section~\ref{gs:sec:divscheme} and the statements at constant relevant sources. These are Lemma~\ref{gr4:lem:const} for $\phi^4$ theory, appendix~\ref{gf:sec:steps} for the Yukawa theory, the proof of Proposition~\ref{gsx:prop:props} in appendix~\ref{gsx:sec:proof} for the sextic theory and the proof of Proposition~\ref{gcp:prop:props} for the cubic theory. Since $\check g$ and $\check\sM$ have no term of order zero, each order of the first term of \eqref{gs:eq:radiuspath} receives contributions from finitely many orders of $W_1$, and it is regular. The integrand of the second term is a polynomial in the constant sources whose coefficients are regular functions of $g_s$ times exponentials of $s$, since the coefficients of the anomaly are regular, and so is its integral. This proves the regularity, and with it \eqref{gs:eq:radius}.

For \eqref{gs:eq:radiusvarsigma} we note that $r\,\pr/\pr r$ multiplies $\int d^dx\sqrt\gamma\,\mathcal{I}_a=r^{-\vep}V\mathcal{I}_a^{(1)}$ by $-\vep$, and that $r^{-\vep}V\rA=\sum_ap_a\,r^{-\vep}V\mathcal{I}_a^{(1)}=\sum_ap_a\,\pr W_{r,\varsigma}/\pr\varsigma_a$. Adding $\sum_a\varsigma_a$ times the first identity to \eqref{gs:eq:radius} gives \eqref{gs:eq:radiusvarsigma}. At a zero $g_*$ of $\hB^I$ the flows $\vk_b=\hB x_b\hB$ vanish, which is the statement on the relevant sources, and the right-hand side of \eqref{gs:eq:radiusvarsigma} vanishes. Then $r\,\pr W_{r,\varsigma}(g_*)/\pr r=r^{-\vep}V\sum_a(p_a(g_*)-\vep\,\varsigma_a)\,\mathcal{I}_a^{(1)}$, which vanishes for $\varsigma_a=p_a(g_*)/\vep$. For these values $W_{r,\varsigma}(g_*)=W_{1,\varsigma}(g_*)=\cW(g_*)+V\rA(g_*)/\vep$, which is $\cW_*$ of \eqref{gs:eq:fixed}. For the sextic theory $\rA$ vanishes by Lemma~\ref{gs:lem:odd}, and \eqref{gs:eq:radius} gives $r\,\pr W_r(g_*)/\pr r=0$ with $\cW_*=\cW(g_*)$.
\end{proof}

At $r=1$ and vanishing relevant sources \eqref{gs:eq:radius} reads $r\,\pr W/\pr r+\hB\!\cdot\!\pr\,\cW-V\sum_b\vk_b\cdot\sfm_b=V\rA$, where $\sfm_b$ is the one-point function of section~\ref{gs:sec:oneform}. The value of the combination $W^\sharp$ of \eqref{gs:eq:sharp} at constant couplings and vanishing relevant sources is therefore $(\vep+r\,\pr_r)W$ at $r=1$, as in the extended form of the properties.

\begin{corollary}[Potential from the radius]
\label{gs:cor:radiuspot}
Under the assumptions of Proposition~\ref{gs:prop:radius} the potential of Theorem~\ref{gr:cor:cond} is
\begin{equation}
V\hat A^{\rm H} = \left(d-r\,\frac{\pr}{\pr r}\right)\left(\vep+r\,\frac{\pr}{\pr r}\right)W_r\,\Big|_{r=1} \, ,
\label{gs:eq:radiuspot}
\end{equation}
where $W_r$ is taken at constant couplings and vanishing relevant sources of the divergence scheme, for every $r$.
\end{corollary}

\begin{proof}
We use the identity at constant sources $\sM_{\rm c}$ in the proof of Proposition~\ref{gs:prop:radius}. Let $t_b=r^{D_b}\sM_b$ and $U(r,g,t)=W_r(g,r^{-D}t)$, and let $\mathcal{R}_t$ be the derivative $r\,\pr_r$ at fixed $t$. In these variables the identity reads
\begin{equation}
\left(\mathcal{R}_t+\mathcal{Q}_t\right)U = r^{-\vep}\,\cA[1](g,t) \, , \qquad
\mathcal{Q}_t = \hB^I\pr_I+\sum_b\left(\delta^{\rm W}_1\sM_b(g,t)+D_b\,t_b\right)\cdot\frac{\pr}{\pr t_b} \, ,
\end{equation}
and the operator $\mathcal{Q}_t$ does not depend on $r$. It therefore commutes with $\mathcal{R}_t$, and since $(\vep+\mathcal{R}_t)\,r^{-\vep}=0$ the function $U^\sharp=(\vep+\mathcal{R}_t)U$ satisfies $(\mathcal{R}_t+\mathcal{Q}_t)\,U^\sharp=0$. At $r=1$ we have $U^\sharp=\vep\,U+\cA[1]-\mathcal{Q}_tU$, which by \eqref{gsh:eq:D} is $W^\sharp$ of \eqref{gs:eq:sharp} at constant couplings and constant relevant sources. Hence $\cW^\sharp$ is the value of $U^\sharp$ at $r=1$ and $t=0$, and $V\sfm^\sharp_b$ is its derivative with respect to $t_b$ there. At $t=0$ the operator is $\mathcal{Q}_t=\hB^I\pr_I-\sum_b\vk_b\cdot\pr/\pr t_b$, where the sum runs over the sources of even dimension, as in the proof of the proposition. The potential of \eqref{gsh:eq:H} is therefore $V\hat A^{\rm H}=(d+\mathcal{Q}_t)\,U^\sharp=(d-\mathcal{R}_t)\,U^\sharp$ at $r=1$ and $t=0$. At $t=0$ the function $U$ is $W_r$ at vanishing relevant sources and $\mathcal{R}_t$ is $r\,\pr_r$, which gives \eqref{gs:eq:radiuspot}.
\end{proof}

The one-point functions of the relevant operators have dropped out of \eqref{gs:eq:radiuspot}, as in \eqref{gs:eq:sexticradius} for the sextic theory. At a zero $g_*$ of $\hB^I$ the identity \eqref{gs:eq:radius} gives $r\,\pr_rW_r=r^{-\vep}V\rA$, so that $(\vep+r\,\pr_r)W_r$ does not depend on $r$ and \eqref{gs:eq:radiuspot} is $\vep\,d\,\cW_*$, in agreement with \eqref{gs:eq:fixed}.

On the spheres the functions $\beta_a$ enter \eqref{gs:eq:radiusvarsigma} only through the combination $\sum_a\beta_a\mathcal{I}_a^{(1)}$, and the independence of the radius at $g_*$ fixes only $\sum_a\varsigma_a\mathcal{I}_a^{(1)}$. The free energy on the sphere, and with it $\tilde F$ of section~\ref{gs:sec:fixed}, depends on the curvature couplings only through this combination.

For $\phi^4$ theory the functions $\beta_a$ are the beta functions of the curvature couplings which follow in minimal subtraction from the bare curvature couplings. These are the coefficients of $\mathcal{I}_a$ in the density $\cX+\sum_a\varsigma_a\mathcal{I}_a$ at a mass source $\sM=m\,R$, and they change under $\mathbb{W}_1$ by the weight $\vep$. For $\cE_4$ this gives $\vep\,\varsigma_{\cE_4}-p$. For $R^2$ the coefficient contains $\fn\cdot m$ and $\tfrac12\,m\cdot\fm\cdot m$, and with the variation $\gamma_\sM m-\bar\eta$ of $m$ we find $\vep\,\varsigma_{R^2}-q-i\cdot m+\tfrac12\,m\cdot\beta_\sM\cdot m$, which is regular. Without the terms with $m$ the beta function of $\varsigma_{R^2}$ would be $\vep\,\varsigma_{R^2}-q-\fn\cdot\bar\eta$, which contains in general poles. At $m=y_0/R_d$, the vanishing source of the divergence scheme, these functions are the $\beta_a$ of the proposition. At a zero $g_*$ of $\hB^I$ the value $m=y_0(g_*)/R_d$ is a zero of the flow $\gamma_\sM m-\bar\eta$ of minimal subtraction, since by \eqref{gr:eq:Y} the flow of the divergence scheme on the unit sphere is $\vk=R_d\bar\eta-\gamma_\sM y_0+\hB\!\cdot\!\pr\,y_0$, which vanishes there. This value of $m$ is the coupling of $\phi^2$ to the curvature at the fixed point, as in~\cite{Fei:2015oha}. Thus $\cW_*$ is the free energy on the sphere at a common zero of the beta functions of the couplings, of the mass source and of the curvature couplings, and the free energy at this point does not depend on the radius.

\subsection{Relevant sources at lowest order}
\label{gs:sec:phi4lowest}

The divergence scheme of Lemma~\ref{gr:lem:scheme} for the mass source is reached from minimal subtraction by the redefinition \eqref{gr:eq:Y} with $D=2$. By Lemma~\ref{gr4:lem:zero} the coefficients $\ell_{\rm h}=d\,(\delta-2\theta)$ and $\vk_{\rm h}-\vk_{\rm c}=-d\,\bar\tau$ of \eqref{gs:eq:dimtwodata} vanish at $g=0$ in minimal subtraction, and $\bar\tau$ vanishes to second order. By Lemma~\ref{gr:lem:scheme} the coefficient $y_J$ of the redefinition therefore vanishes at $g=0$, and $y_0=-(\vk_{\rm h}-\vk_{\rm c}+y_K\hB^K)/2$ vanishes to second order. At $\sM'=0$ and constant couplings the redefinition gives the constant mass source $y_0$, at which the properties hold by Lemma~\ref{gr4:lem:const}.

\begin{proposition}
\label{gr4:prop:lowest}
Assume Hypothesis~\ref{hyp:finite} for multiscalar $\phi^4$ theory. Then multiscalar $\phi^4$ theory in $d=4-\vep$ has the two-point form in the divergence scheme with $c_{IJ}(\vep)=c_\cO\,\delta_{IJ}$, and it has Property~\ref{gs:P:free}.
\end{proposition}

\begin{proof}
We first show the last statement of Property~\ref{gs:P:free}. At $\sM'=0$ and constant couplings the redefinition \eqref{gr:eq:Y} gives the constant mass source $y_0$, so that the one-point function $\sfm$ in the divergence scheme is $V^{-1}$ times the derivative of the functional of minimal subtraction with respect to a constant mass source, taken at the constant source $y_0(g)$. By Lemma~\ref{gr4:lem:const} this derivative is the one-point function $\sfm$ of minimal subtraction plus terms which contain $y_0$, with regular coefficients. The former vanishes at $g=0$ by the fourth statement of Lemma~\ref{gr4:lem:zero}, and $y_0$ vanishes at $g=0$. Hence $\sfm$ vanishes at $g=0$ in the divergence scheme, for every $\vep$, which gives the last statement of Property~\ref{gs:P:free}.

For the two-point form we compare the two schemes through second order in the couplings at $\sM'=0$ and $a_\mu=0$. There the old source is the redefinition \eqref{gr:eq:Y}, which is of second order in the couplings. The new functional therefore differs from the old one at vanishing source by the one-point function of the mass operator at $g=0$ integrated with this source, and by terms of higher order. The former vanishes by the fourth statement of Lemma~\ref{gr4:lem:zero}. In the same way the anomaly at $\sM'=0$ differs from the anomaly at $\sM=0$ by the terms of the anomaly which are linear in $\sM$, at $g=0$, integrated with this source. These are the terms with $i$ and $t_\sM$, since the other terms linear in $\sM$ in \eqref{eq:fullanomaly} contain derivatives of the couplings, and they vanish at $g=0$ by Lemma~\ref{gr4:lem:zero}. Through second order the pair $W$, $\cA[\sigma]$ at vanishing source is therefore the same in both schemes, and the two-point form in minimal subtraction, which is part of Proposition~\ref{gs:prop:phi4prop}, gives the two-point form in the divergence scheme with the same $c_{IJ}(\vep)$. In the divergence scheme the theory has Properties~\ref{gs:P:functional} to~\ref{gs:P:sphere}, by Lemma~\ref{gr4:lem:const}, by the vanishing of $\ell_{\rm h}$ and $\vk_{\rm h}-\vk_{\rm c}$ at $g=0$ which we recalled before the proposition and by the remark at the end of section~\ref{gs:sec:divscheme}. The other statements of Property~\ref{gs:P:free} therefore follow by Lemma~\ref{gre:lem:zero} and the remark after it.
\end{proof}

\subsection{Gradient formula}
\label{gs:sec:phi4thm}

\begin{theorem}
\label{gs:thm:phi4}
Assume Hypothesis~\ref{hyp:finite} for multiscalar $\phi^4$ theory. Consider multiscalar $\phi^4$ theory in $d=4-\vep$ dimensions to all orders in perturbation theory, in minimal subtraction with the curvature counterterms of Proposition~\ref{prop:finite} and with the mass source in the divergence scheme of Lemma~\ref{gr:lem:scheme}. Then there are a regular invariant function $\hat A$ and a regular symmetric covariant tensor $\hat G_{IJ}$ with
\begin{equation}
\pr_I\hat A = \hat G_{IJ}\,\hB^J \, .
\label{gs:eq:phi4main}
\end{equation}
The term of order zero in the couplings and in $\vep$ of $\hat G_{IJ}$ is the positive definite matrix $\tfrac{1}{96}\pi^2c_\cO\,\delta_{IJ}=\delta_{IJ}/(589824\,\pi^6)$, where $c_\cO=C_\phi^{\,4}/4!$ is the constant of Lemma~\ref{gs:lem:free} at $d=4$ and the coordinates are normalised by $g^Ig^I=\lambda_{ijkl}\lambda_{ijkl}$, with the sum over all $i$, $j$, $k$ and $l$, so that $\hat G_{IJ}$ is positive definite to all orders in the couplings and in $\vep$. At a zero of $\hB^I$ the potential is
\begin{equation}
\hat A = c(\vep)\left(\tilde F-N\tilde F_s\right) , \qquad c(\vep) = \frac{\vep\,d}{2e_d\,V\sin(\pi\vep/2)} = \frac{1}{8\pi^3}+{\rm O}(\vep) \, ,
\label{gs:eq:phi4F}
\end{equation}
where $\tilde F=\sin(\pi d/2)\log Z_{S^d}$ is defined with the couplings of the curvature terms at the zeros of their beta functions and the mass source at the zero of its flow, which in the divergence scheme is the vanishing source, see \eqref{gs:eq:Ftilde} and Proposition~\ref{gs:prop:radius}, $N$ is the number of fields, $\tilde F_s$ is the value of $\tilde F$ for a free scalar field and $e_d=\tfrac12d(d-1)(d-2)$.
\end{theorem}

\begin{proof}
By Proposition~\ref{gs:prop:phi4prop} the theory has Properties~\ref{gs:P:functional} to~\ref{gs:P:sphere} and the two-point form in minimal subtraction. By Lemma~\ref{gr4:lem:const}, by the last statement of Property~\ref{gs:P:weyl} and by the remark at the end of section~\ref{gs:sec:divscheme} it has Properties~\ref{gs:P:functional} to~\ref{gs:P:sphere} in the divergence scheme, and by Proposition~\ref{gr4:prop:lowest} it has the two-point form there with $c_{IJ}(\vep)=c_\cO\,\delta_{IJ}$ and Property~\ref{gs:P:free}. The other hypotheses of Theorem~\ref{gr:cor:cond} hold, since $d_c=4$, and Property~\ref{gs:P:weyl} includes $\hB^I=-\vep\,g^I+B^I$ and $\gamma_\sM$ without term of order zero, by Lemma~\ref{gr4:lem:zero}. The theorem gives $\pr_J\hat A^{\rm H}=\hat G^{\rm H}_{JI}\hB^I$, with a regular potential and a regular symmetric metric whose term of order zero in the couplings and in $\vep$ is that of $\hat G^{\rm S}_{JI}$, by the last statement of Property~\ref{gs:P:free}. The potential is invariant and the metric is covariant under all of ${\rm O}(N)$, including the reflections. Indeed, the coefficients of the bare sources and of the counterterms of flat space are covariant under ${\rm O}(N)$ by Proposition~\ref{prop:uniqueall}, and so are the curvature counterterms, since a rotation with a constant element of ${\rm O}(N)$ maps those of Proposition~\ref{prop:finite} to pure poles with the same properties, which are unique. The bare functional is invariant under a rotation of all bare sources with a constant element of ${\rm O}(N)$, since the propagator is proportional to $\delta_{ij}$, so that the renormalised functional is invariant and the coefficients of its anomaly are covariant under every constant element of ${\rm O}(N)$. The redefinition of Lemma~\ref{gr:lem:scheme} is unique and therefore covariant, and the potential and the metric of Theorem~\ref{gr:cor:cond} are built covariantly from these data. By \eqref{gs:eq:G0sign} with $k=1$, $s_d=-1$ and $c_{IJ}(\vep)=c_\cO\,\delta_{IJ}$ the term of order zero in the couplings of $\hat G^{\rm H}_{IJ}$ tends to $-\tfrac14\pi^2c_\cO\,\delta_{IJ}$ at $\vep=0$, which is $-2e_d$ times $\tfrac{1}{96}\pi^2c_\cO\,\delta_{IJ}$ with $2e_d=24$. We set $\hat A=-\hat A^{\rm H}/(2e_d)$ and $\hat G_{IJ}=-\hat G^{\rm H}_{IJ}/(2e_d)$, where $2e_d=24$ at $\vep=0$. At a zero $g_*$ the theorem gives $\hat A(g_*)=-\vep\,d\,\cW_*/(2e_dV)$, and with $\tilde F-N\tilde F_s=-\sin(\pi\vep/2)\,\cW_*$, which is \eqref{gs:eq:Ftilde} for this theory with the normalisation of section~\ref{sec:continuation}, we obtain \eqref{gs:eq:phi4F}. By Proposition~\ref{gs:prop:radius}, with the functions $p_a$ of the divergence scheme, the curvature couplings $\varsigma_a=p_a(g_*)/\vep$ are the zeros of their beta functions, the vanishing mass source is a zero of its flow on the sphere of every radius, and with these curvature terms the functional at $g_*$ equals $\cW_*$ on the sphere of every radius.
\end{proof}

By section~\ref{gs:sec:consequences} the eigenvalues of the matrix of derivatives of $\hB^I$ at a fixed point are real, and $\tilde F$ decreases along flows between fixed points in the sense of that section. The gradient formula implies the relations among the coefficients of the beta function which the regularity requires. The computation of appendix~\ref{app:fiveloop} finds that, modulo the relations in four dimensions, the relations at five loops span the same space as the four relations of~\cite{Pannell:2024sia}.\footnote{The comparison with the relations of~\cite{Pannell:2024sia}, with the beta function and the fixed points of minimal subtraction in momentum space and with the values of $\tilde F$ in the literature uses Identification~\ref{id:ms}, and the theorem does not.} At $\vep=0$ the theorem gives a gradient formula in four dimensions for the vector field $B^I$, with a metric which contains the first harmonic of the two-point function on the sphere. In four dimensions there is also a scheme in which the one-form $W_I$ of the Weyl anomaly vanishes, see appendix~\ref{gs:app:fourd}. In the appendices on $\phi^4$ theory the symbols $\hat A$ and $\hat G_{IJ}$ refer to Theorem~\ref{gs:thm:phi4}.

\section{Other cases}
\label{gs:sec:other}

This section concludes the third part of the paper. In section~\ref{gs:sec:scalars} we collect what the constructions of section~\ref{gs:sec:universal} give for scalar fields with polynomial interactions. For the sextic theory in section~\ref{gs:sec:sextic}, the Yukawa theory in section~\ref{gs:sec:yukawa} and the cubic theory in section~\ref{gs:sec:cubic} we derive all six properties from Hypothesis~\ref{hyp:finite} for the theory, with the proofs in appendices~\ref{gs:app:sextic}, \ref{gs:app:fermions} and~\ref{gs:app:cubic}, and we obtain the gradient formula with a positive definite metric. For gauge theories in section~\ref{gs:sec:gauge} the properties are assumptions, except for free gauge fields with a local coupling, for which Properties~\ref{gs:P:functional} to~\ref{gs:P:free} follow from Hypothesis~\ref{hyp:finite} for the free gauge field in appendix~\ref{gs:app:freegauge}. For fermions near two dimensions in section~\ref{gs:sec:fermions} Hypothesis~\ref{hyp:finite} for the four-fermion theory, together with ${\rm tr}\,1>1$ and the covariance of $N_I$ without term of degree zero, makes Property~\ref{gs:P:weyl} for the theory without the sources of the fermion bilinears equivalent to the pole equations of minimal subtraction for a suitable choice of the bare sources, and, given Properties~\ref{gs:P:functional} to~\ref{gs:P:sphere}, it gives Property~\ref{gs:P:free}, while the other properties are assumptions. Each of the subsections~\ref{gs:sec:sextic} to~\ref{gs:sec:fermions} contains a proposition, a corollary or a theorem which follows from the results of sections~\ref{gs:sec:identities} and~\ref{gs:sec:relevant}. In the last subsection we name the cases about which we make no statement.

\subsection{Scalar fields with polynomial interactions}
\label{gs:sec:scalars}

For scalar fields with sextic couplings at $d_c=3$ and with cubic couplings at $d_c=6$ the bare functional on $S^d$ is the functional of section~\ref{gs:sec:sphere}. Its vertices are $V^{(6)}$ or $V^{(3)}$ for the couplings and the vertices of the relevant sources, which are $V^{(4)}$ and $V^{(2)}$ for the sextic theory and $V^{(2)}$ and $V^{(1)}$ for the cubic theory. By Proposition~\ref{gs:prop:bare} this functional is defined for non-integer $d$. On the line \eqref{eq:line} it has the conformal covariance \eqref{eq:covS} with the weights $w_6=2\vep$ and $w_3=\tfrac12\vep$ of the couplings, it is invariant under background gauge transformations, and it has the properties under rotations of \eqref{eq:Lambda}. In minimal subtraction Propositions~\ref{gs:prop:msanomaly} and~\ref{gs:prop:msward} then give the form of the anomaly of Property~\ref{gs:P:anomaly} and the Ward identity of Property~\ref{gs:P:ward}, with the bare sources, the counterterms and the variation of the sources which Hypothesis~\ref{hyp:finite} for the sextic and for the cubic theory provides. The conditions of Lemma~\ref{gs:lem:free} hold for the sextic theory and, after the choice of the bare source of $\phi_i$ of section~\ref{gs:sec:cubic}, for the cubic theory, so that the two-point form holds in minimal subtraction with the positive matrix $c_{IJ}(\vep)=C_\phi^{\,n}/n!$ times the unit matrix, for $n=6$ and $n=3$. The finiteness of the renormalised functional on $S^d$ and the regularity of the coefficients of $\hB^I$, $\hr_I$, $\delta^{\rm W}_\sigma\sM$ and $\cA[\sigma]$ follow from Hypothesis~\ref{hyp:finite} for each theory. For $\phi^4$ theory this is shown in appendices~\ref{app:unique} and~\ref{app:finite}, whose proofs use the counterterms of that theory, for sextic scalars in appendix~\ref{gs:app:sextic}, and for cubic scalars, with the class of flat sources with $D_0=3$, in appendix~\ref{gs:app:cubic}.

The derivations of the six properties for the $\phi^4$ and the sextic theory follow the same steps, see Table~\ref{gs:tab:steps}. Besides the critical dimension, the weights of the couplings and the relevant sources, the theory enters them through three statements, which are Lemmas~\ref{lem:lowest} and~\ref{gsx:lem:lowest}, Propositions~\ref{prop:uniqueall} and~\ref{gsx:prop:uniqueall}, and Propositions~\ref{prop:finite} and~\ref{gsx:prop:finite}. The first gives the form of the terms of second order of the flat bare functional, the second the uniqueness and covariance of the coefficients of the bare sources, and the third the finiteness on the backgrounds $\Omega^2\delta_{\mu\nu}$ with curvature counterterms built with $d$ as a parameter. We expect that for scalar fields with a polynomial interaction which is marginal at $d_c=3$ or $4$ the first two statements and Hypothesis~\ref{hyp:finite} on the backgrounds $\Omega^2\delta_{\mu\nu}$ give the six properties by the same steps. Here the hypothesis is taken in the form in which it is stated for the cubic theory, with the class of flat sources with $D_0=1$ for $d_c=3$ and $D_0=2$ for $d_c=4$. The cubic theory requires in addition the treatment of the source of $\phi_i$, in whose presence the coefficients of the bare sources are fixed only up to shifts of the field, and the Yukawa theory the covariance of coefficients which are fixed only up to the rotations that leave all couplings invariant.

The last statements of Properties~\ref{gs:P:weyl} and~\ref{gs:P:free} concern the relevant sources of even dimension $D<d_c$. We list them for the cases of this section, counting dimensions at $d=d_c$ and assuming invariance under parity, so that pseudoscalar bilinears do not occur. For gauge fields of a semisimple group with Dirac fermions in $d=4-\vep$ the gauge-invariant scalar operators of dimension smaller than four are the fermion bilinears, whose sources have dimension one, since a simple factor has no gauge-invariant scalar of dimension smaller than four built from the gauge field alone. For Dirac fermions near $d=2$ the scalar bilinears have sources of dimension one. In these two cases there is no relevant source of even dimension, and these statements are empty. With scalar fields the source of $\phi_i\phi_j$ has dimension two, and in $d=6-\vep$ the source of $\phi_i$ has dimension four. The sources of the other scalar operators, of $\bar\psi\psi$ and of the operators with fermions have odd or non-integer dimension at $d=d_c$ and drop out by Lemma~\ref{gs:lem:dimension}. Where a source of even dimension occurs, they are therefore statements on the free scalar field, which at $g=0$ decouples from the fermions and from the gauge fields, where with gauge fields $g=0$ means the limit of vanishing gauge couplings with the gauge field canonically normalised. Outside $\phi^4$ theory, the Yukawa theory of section~\ref{gs:sec:yukawa}, the sextic theory of section~\ref{gs:sec:sextic} and the cubic theory of section~\ref{gs:sec:cubic} we assume them. We expect them, since in a construction in minimal subtraction in which the bare sources and the counterterms are unique pure poles it follows from the following argument. At $g=0$ the one-point functions of $\phi_i$ and of $\phi_i\phi_j$ vanish on all backgrounds $\Omega^2\delta_{\mu\nu}$, so that the bare sources and the counterterms of first degree in the relevant sources have no term of order zero. This gives the last statement of Property~\ref{gs:P:weyl}, and in minimal subtraction it gives that $\sfa_\sM$ has no term of order zero. In the divergence scheme this follows as well. By the last statement of Property~\ref{gs:P:weyl} and Lemma~\ref{gr:lem:scheme} the coefficients $y_J$ and $y_0$ vanish at $g=0$, and the change of $\sfa_\sM$ under the redefinition is a second derivative of the anomaly in the sources times $y_0$, which has no term of order zero. By Lemma~\ref{gsh:lem:m} this gives the last statement of Property~\ref{gs:P:free}. In this sense we expect these statements for scalars with Yukawa and quartic couplings with gauge fields at $d_c=4$. For sextic scalars and for scalars with Yukawa and quartic couplings without gauge fields they are derived, and for cubic scalars at $d_c=6$ they follow from Hypothesis~\ref{hyp:finite} for the cubic theory, see section~\ref{gs:sec:cubic}. For $\phi^4$ theory they are part of Propositions~\ref{gs:prop:phi4prop} and~\ref{gr4:prop:lowest}.

\subsection{Sextic scalars in \texorpdfstring{$d=3-\vep$}{d=3-epsilon}}
\label{gs:sec:sextic}

We consider $N$ real scalar fields with sextic couplings in $d=3-\vep$ dimensions. The couplings $g^I$ are the components of a symmetric tensor with six indices, their weight is $k=2$, and the group of field rotations is ${\rm O}(N)$. The theory is invariant under parity. With the symmetry $\phi\to-\phi$ there are two families of relevant scalar operators. The operators $\phi_i\phi_j$ have a source $\sM$ of dimension two for every $d$. The quartic operators have a source whose dimension at $d=3$ is one. At $d=3$ the variations of these two sources under a Weyl rescaling were given in~\cite{Nakayama:2013wda}. The source of the quartic operators varies homogeneously, and the variation of the source of $\phi_i\phi_j$ contains the curvature, derivatives of the couplings and a term which is quadratic in the source of the quartic operators.

Our assumptions are the following. Properties~\ref{gs:P:functional} to~\ref{gs:P:sphere} hold with these sources, with Property~\ref{gs:P:functional} in instance~(a). The weight of the couplings is $k_I=2$. The source $\sM$ of the operators $\phi_i\phi_j$ is taken in the divergence scheme of Lemma~\ref{gr:lem:scheme}, which exists since $D=2<d_c$, and in which the properties hold by the last statement of Property~\ref{gs:P:weyl}, see section~\ref{gs:sec:divscheme}.

\begin{proposition}[Sextic scalars]
\label{gs:prop:sextic}
Under these assumptions the following holds for $N$ scalar fields with sextic couplings in $d=3-\vep$. At vanishing relevant sources the anomaly vanishes, and
\begin{gather}
\pr_J\Pi' = -\cS'_{JK}\,\hB^K \, , \notag\\
\Pi' = \frac1V\left(d+\hB\!\cdot\!\pr\right)\cW-\vk\!\cdot\sfm \, , \qquad
\cS'_{JK} = \cS_{JK}+\left(\mathsf{x}_{JLK}+\mathsf{x}_{KLJ}-\mathsf{x}_{LJK}\right)\hB^L \, ,
\label{gs:eq:sextic}
\end{gather}
where $\vk$ is the flow \eqref{gsh:eq:flows} of the source of $\phi_i\phi_j$, $\sfm$ is the one-point function of these operators on the unit sphere, and $\mathsf{x}_{JKL}$ is the coefficient of $\sfn_J=\mathsf{x}_{JKL}\hB^K\hB^L$. The function $\Pi'$ is regular, and the tensor $\cS'_{JK}$ is regular and symmetric. At a zero $g_*$ of $\hB^I$ we have $\Pi'(g_*)=d\,\cW(g_*)/V$, and $\sM=0$ is a zero of the flow of the source of $\phi_i\phi_j$. If in addition Property~\ref{gs:P:free} holds, the term of order zero in the couplings and in $\vep$ of $\cS'_{IJ}$ is $\tfrac12\,\pi^2c_{IJ}$. If the functional has the two-point form in the divergence scheme, the term of order zero in the couplings of $\cS'_{IJ}$ is $c_{IJ}(\vep)\,(\mu_1-\mu_0)$ with the numbers of \eqref{gs:eq:G0}.
\end{proposition}

\begin{proof}
The anomaly vanishes at vanishing relevant sources by Lemma~\ref{gs:lem:odd}. The source of the quartic operators has an odd dimension at $d=3$, and by Lemma~\ref{gs:lem:dimension} it does not contribute. The relation \eqref{gs:eq:sextic} is then \eqref{gr:eq:odd} of section~\ref{gs:sec:gradrel}, where $\vk$ is quadratic in $\hB^I$ by Proposition~\ref{gsh:prop:flows}, so that $\vk$ and $\sfn_J$ vanish at the zeros of $\hB^I$. The regularity follows from Property~\ref{gs:P:functional} and from Proposition~\ref{gsh:prop:flows}. The difference $\cS'_{JK}-\cS_{JK}$ contains a factor $\hB^L$ and has no term of order zero. Under Property~\ref{gs:P:free} the term of order zero in the couplings and in $\vep$ of $\cS_{IJ}$ is $b_3\,c_{IJ}=\tfrac12\pi^2c_{IJ}$. With the two-point form the term of order zero in the couplings is $c_{IJ}(\vep)\,(\mu_1-\mu_0)$, as in the proof of Proposition~\ref{gs:prop:odd}.
\end{proof}

The gradient formula of Theorem~\ref{gr:cor:cond} also holds for this theory, but for odd $d_c$ its metric $\hat G^{\rm H}$ has a vanishing term of order zero, see section~\ref{gs:sec:gradrel}, and we use \eqref{gs:eq:sextic}, which is the relation \eqref{gr:eq:odd} for the functional $W$ itself. The tensor $\cS'_{JK}$ is nondegenerate if Property~\ref{gs:P:free} holds with an invertible matrix $c_{IJ}$. The same argument applies with further couplings of another weight, such as the couplings of $\phi_i\phi_j\bar\psi\psi$ with Dirac fermions, since section~\ref{gs:sec:relevant} allows several weights.

Under Hypothesis~\ref{hyp:finite} for the sextic theory the potential $\Pi'$ is determined by the free energy on spheres of radius $r$ near one. The anomaly vanishes by Lemma~\ref{gs:lem:odd}, and the identity \eqref{gs:eq:radius} of Proposition~\ref{gs:prop:radius} at $r=1$ reads $r\,\pr_rW_r+\hB^I\pr_I\cW-V\vk\!\cdot\sfm=0$. With \eqref{gs:eq:sextic} this gives
\begin{equation}
V\,\Pi' = \left(d-r\,\frac{\pr}{\pr r}\right)W_r\,\Big|_{r=1} \, ,
\label{gs:eq:sexticradius}
\end{equation}
at vanishing relevant sources of the divergence scheme, for every $r$. The one-point function of $\phi_i\phi_j$ has dropped out, and the potential $-\Pi'$ with the positive metric is given along the flow by the free energy and by its first derivative with respect to the radius.

We can read \eqref{gs:eq:sextic} as a gradient formula for the potential $-\Pi'$, which at the zeros of $\hB^I$ is $d/V$ times the free energy $-\cW$, with the metric $\cS'_{IJ}$. Let Property~\ref{gs:P:free} hold, with the matrix $c_{IJ}$ of the free two-point functions of the sextic operators, which is positive. Then this metric is positive definite to all orders in the couplings and in $\vep$. The anomaly does not enter, and no antisymmetric anomaly coefficient occurs. In three dimensions a function with a gradient property and a metric which is positive at lowest order was constructed from the beta functions at low orders for Chern--Simons theories with matter in~\cite{Jack:2015tka}, where its relation to the free energy was left open. For the sextic coupling \eqref{gs:eq:sextic} provides a candidate for this relation. The bare functional of this theory is constructed in section~\ref{gs:sec:scalars}, and all properties follow from Hypothesis~\ref{hyp:finite} for the sextic theory. In Properties~\ref{gs:P:functional} to~\ref{gs:P:free} the relevant source $\sM$ then stands for the pair $(\sU,\sM)$, each component with its own dimension at $d=3$, and the source $\sU$ has odd dimension at $d=3$.

\begin{proposition}
\label{gsx:prop:props}
Assume Hypothesis~\ref{hyp:finite} for the sextic theory. Then the renormalised functional of $N$ scalar fields with sextic couplings on the unit sphere $S^d$, $d=3-\vep$, in minimal subtraction with the curvature counterterms of Proposition~\ref{gsx:prop:finite}, has Properties~\ref{gs:P:functional} to~\ref{gs:P:sphere} with the relevant sources $\sU$ and $\sM$, with Property~\ref{gs:P:functional} in instance~(a) and the weights $k_I=2$. It has the two-point form of appendix~\ref{gs:sec:properties} with $c_{IJ}(\vep)=c_6\,\delta_{IJ}$, $c_6=C_\phi^{\,6}/6!$. The same holds with the source $\sM$ in the divergence scheme of Lemma~\ref{gr:lem:scheme}, and in this scheme the functional has Property~\ref{gs:P:free}. The vector field is $\hB^I=-2\vep\,g^I+B^I$ with $B^I$ independent of $\vep$ and without terms of order zero and one, and $\gamma_\sM$ and $\gamma_\sU$ vanish at $g=0$.
\end{proposition}

\begin{proof}
We follow the derivation for $\phi^4$ theory step by step, as listed in the column of the sextic theory of table~\ref{gs:tab:steps}. The three-dimensional steps, namely the local structures, their independence, the potential lemma, the anomaly with seven coefficients, the curvature counterterms and the regularity of the Weyl variation, are carried out in appendix~\ref{gs:app:sextic}, which also contains the proof.
\end{proof}

\begin{theorem}
\label{gsx:thm:sextic}
Assume Hypothesis~\ref{hyp:finite} for the sextic theory. Then for $N$ scalar fields with sextic couplings in $d=3-\vep$, in minimal subtraction with the source of $\phi_i\phi_j$ in the divergence scheme, the vector field $\hB^I$ obeys the gradient formula \eqref{gs:eq:sextic} of Proposition~\ref{gs:prop:sextic}, $\pr_J\Pi'=-\cS'_{JK}\hB^K$, with a regular invariant function $\Pi'$ and a regular symmetric covariant tensor $\cS'_{JK}$. At a zero $g_*$ of $\hB^I$ we have $\Pi'(g_*)=d\,\cW(g_*)/V$. The term of order zero in the couplings and in $\vep$ of $\cS'_{IJ}$ is $\tfrac12\pi^2c_6\,\delta_{IJ}$ with $c_6=1/(2949120\,\pi^6)$, so that the metric $\cS'_{IJ}$ of the potential $-\Pi'$ is positive definite to all orders in the couplings and in $\vep$.
\end{theorem}

\begin{proof}
By Proposition~\ref{gsx:prop:props} the hypotheses of Proposition~\ref{gs:prop:sextic} hold, including Property~\ref{gs:P:free} and the two-point form in the divergence scheme. That proposition gives the gradient formula, the regularity, the symmetry and the value at the zeros. The function $\Pi'$ is invariant and the tensor $\cS'_{JK}$ is covariant under all of ${\rm O}(N)$, including the reflections. Indeed, the coefficients of the bare sources and of $\cX^{\rm fl}$ are covariant under ${\rm O}(N)$ by Proposition~\ref{gsx:prop:uniqueall}, and $\fk$ and $\fn$ are covariant by their unique form \eqref{gsx:eq:steps}. The bare functional is invariant under a rotation of all bare sources with a constant element of ${\rm O}(N)$, since the propagator is proportional to $\delta_{ij}$, so that the renormalised functional is invariant and the coefficients of its anomaly are covariant under every constant element of ${\rm O}(N)$. The redefinition of Lemma~\ref{gr:lem:scheme} to the divergence scheme is unique and therefore covariant, and $\Pi'$ and $\cS'_{JK}$ are built covariantly from these data. Its statement under Property~\ref{gs:P:free} gives the term of order zero $\tfrac12\pi^2c_{IJ}$ with $c_{IJ}=c_6\,\delta_{IJ}$, and $C_\phi=\Gamma(\tfrac12)/(4\pi^{3/2})=1/(4\pi)$ at $d=3$ gives $c_6=(4\pi)^{-6}/720$. A positive definite term of order zero gives a metric which is positive definite to all orders.
\end{proof}

\subsection{Yukawa and quartic couplings in \texorpdfstring{$d=4-\vep$}{d=4-epsilon}}
\label{gs:sec:yukawa}

We consider $N$ real scalar fields $\phi_a$ and $N_f$ Dirac fermions without gauge fields, with the quartic couplings $\tfrac1{4!}\lambda_{abcd}\,\phi_a\phi_b\phi_c\phi_d$ and the Yukawa couplings $\phi_a\bar\psi\,Y^a\psi$. The matrices $Y^a$ act on the flavour indices and are hermitian, so that the action is real, and the couplings contain no $\gamma_5$, so that they are invariant under parity. The couplings $g^I$ are the real coordinates of the $\lambda_{abcd}$ and of the $Y^a$, with the weights $k_I=1$ for the quartic and $k_I=\tfrac12$ for the Yukawa couplings, which are the weights of Proposition~\ref{gf:prop:cov} divided by $\vep$. The group of field rotations is ${\rm O}(N)\times{\rm U}(N_f)$, the group of symmetries of the free theory which preserves the fermion number and parity, and its Lie algebra $\mathfrak{g}$ consists of the pairs of a real antisymmetric and an antihermitian matrix. The relevant source $\sM$ is the source of $\tfrac12\sM_{ab}\phi_a\phi_b$, so that $V_\sM$ consists of the real symmetric matrices. The sources of $\phi_a$, of the cubic operators of the scalars and of $\bar\psi\psi$ have odd dimension. Their bare sources are odd in these sources, without terms which do not contain them, since a local scalar built from the couplings, their covariant derivatives, the field strength, $\sM$ and the curvature has even dimension at $d=4$. Their vanishing is therefore preserved, and by Lemma~\ref{gs:lem:dimension} they do not enter the identities. We set them to zero. The bare functional is that of Proposition~\ref{gf:prop:bare}, and the hypothesis of finiteness is Hypothesis~\ref{hyp:finite} for the Yukawa theory of appendix~\ref{gs:sec:cases}, with the density $\cX^{\rm fl}$ of appendix~\ref{app:anomcoeff} in flat space with the Lie algebra $\mathfrak{g}$. No further local structure is needed, since the list of local structures of dimension four built from $g^I$, $D_\mu g^I$, $f_{\mu\nu}$ and $\sM$ does not depend on the operators of which the $g^I$ are the sources.

The bare gauge field couples to the fermions through antihermitian matrices. In minimal subtraction with constant couplings the counterterm of the fermion fields is hermitian, $\bar\psi\,Z_\psi\gamma^\mu\pr_\mu\psi$ in the symmetric form $\tfrac12(\bar\psi Z_\psi\gamma\!\cdot\!\pr\psi-\pr_\mu\bar\psi\,\gamma^\mu Z_\psi\psi)$. With a local hermitian $Z_\psi(g(x))$ the change of variables $\psi=Z_\psi^{-1/2}\chi$, $\bar\psi=\bar\chi Z_\psi^{-1/2}$ maps this term to the canonical kinetic term plus $\bar\chi\gamma^\mu\mathsf{a}_\mu\chi$ with $\mathsf{a}_\mu=\tfrac12[\mathsf{Z}^{-1}\pr_\mu\mathsf{Z}-(\pr_\mu\mathsf{Z})\mathsf{Z}^{-1}]$, $\mathsf{Z}=Z_\psi^{1/2}$, which is antihermitian. The symmetric part of a renormalisation of the fields therefore does not lead to a hermitian vector source, in agreement with Proposition~\ref{gc:prop:cov}, by which a hermitian part of $N_I$ can always be removed. The statement that the hypothesis holds with the counterterms of minimal subtraction in momentum space is the analogue of Identification~\ref{id:ms}, and we do not use it.

\begin{proposition}
\label{gf:prop:props}
Assume Hypothesis~\ref{hyp:finite} for the Yukawa theory. Then the renormalised functional of scalar fields and Dirac fermions with Yukawa and quartic couplings on the unit sphere $S^d$, $d=4-\vep$, in minimal subtraction with the curvature counterterms of Proposition~\ref{prop:finite}, has Properties~\ref{gs:P:functional} to~\ref{gs:P:sphere} with Property~\ref{gs:P:functional} in instance~(a), with the weights $k_I$ above, and it has the two-point form of appendix~\ref{gs:sec:properties} with the matrix $c_{IJ}(\vep)$ of Lemma~\ref{gf:lem:free}. In the divergence scheme of Lemma~\ref{gr:lem:scheme} for the source $\sM$ it has Property~\ref{gs:P:free}, with $c_{IJ}$ positive definite.
\end{proposition}

\begin{proof}
The proof follows the derivation for $\phi^4$ theory in section~\ref{gs:sec:phi4prop} and appendices~\ref{app:weyl}, \ref{app:unique} and~\ref{app:finite}, step by step, as listed in the column of the Yukawa theory of table~\ref{gs:tab:steps} and explained in appendix~\ref{gf:sec:steps}. Most steps hold without change. The adapted and new steps are the bare functional of section~\ref{gs:sec:dirac}, the weights $k_I$ in Lemmas~\ref{lem:rhoxi} and~\ref{lem:frame}, the Yukawa block and the fermion current in the terms of second order, the stabiliser of the couplings, for which the covariance of the coefficients follows from Proposition~\ref{gc:prop:cov} in place of the uniqueness, and the two-point form of Lemma~\ref{gf:lem:free}. The matrix $c_{IJ}$ is positive definite by Lemma~\ref{gro:lem:c}.
\end{proof}

With the proposition the hypotheses of Theorem~\ref{gr:cor:cond} hold, with $d_c=4$, where Property~\ref{gs:P:weyl} includes $\hB^I=-\vep k_Ig^I+{\rm O}(g^2)$ and $\gamma_\sM$ without term of order zero, by Lemma~\ref{gr4:lem:zero}. The chain of section~\ref{gs:sec:relevant} which leads to the theorem does not assume a single weight.

\begin{theorem}[Yukawa and quartic couplings]
\label{gf:cor:yukawa}
Assume Hypothesis~\ref{hyp:finite} for the Yukawa theory. Then for $N$ scalar fields and $N_f$ Dirac fermions with Yukawa and quartic couplings in $d=4-\vep$ the vector field $\hB^I$ obeys the gradient formula $\pr_J\hat A^{\rm H}=\hat G^{\rm H}_{JI}\hB^I$ of Theorem~\ref{gr:cor:cond}, with a regular potential which takes the value \eqref{gs:eq:fixed} at the zeros of $\hB^I$ and a regular symmetric metric. The metric $s_d\hat G^{\rm H}_{JI}$ is positive definite to all orders in the couplings and in $\vep$, and its term of order zero at $\vep=0$ is $\tfrac14\pi^2c_{IJ}$ with $c_{IJ}$ of Lemma~\ref{gro:lem:c}.
\end{theorem}

\begin{proof}
Proposition~\ref{gf:prop:props} and Theorem~\ref{gr:cor:cond}, which gives the term of order zero under Property~\ref{gs:P:free}.
\end{proof}

The identification of $L^I$ with the counterterms of minimal subtraction in momentum space, and with it of $\hB^I$ with the usual beta functions up to a rotation, is open as for $\phi^4$ theory. The values of the matrix of free two-point functions at $\vep=0$ are given by the following lemma, which we also use with gauge fields in section~\ref{gs:sec:gauge}.

We consider real scalars $\phi_a$ and Dirac fermions $\psi$ at $d=4$ with the quartic couplings $\tfrac{1}{4!}\lambda_{abcd}\phi_a\phi_b\phi_c\phi_d$ and the Yukawa couplings $\phi_a\bar\psi\,Y^a\psi$, where $\lambda_{abcd}$ is symmetric and the matrices $Y^a$ act on the colour and flavour indices of the fermions. The matrices $Y^a$ are hermitian, so that the action is real, and the couplings contain no $\gamma_5$, so that they are invariant under parity. The propagators are $C_\phi/x^2$ and $C_\psi\,\gamma\!\cdot\!x/x^4$ with $C_\phi=1/(4\pi^2)$ and $C_\psi=1/(2\pi^2)$. We also consider the canonically normalised gauge fields $A^a_\mu$ of a simple factor with $n_G$ generators.

\begin{lemma}
\label{gro:lem:c}
At $d=4$ the free two-point functions of the quartic and of the Yukawa operators are $c/x^8$ with
\begin{gather}
c(\lambda,\lambda') = \frac{C_\phi^4}{4!}\,\lambda_{abcd}\lambda'_{abcd} = \frac{\lambda_{abcd}\lambda'_{abcd}}{6144\,\pi^8} \, , \notag\\
c(Y,Y') = 4\,C_\phi C_\psi^2\,\sum_a{\rm tr}\,(Y^aY'^a) = \frac{1}{4\pi^6}\sum_a{\rm tr}\,(Y^aY'^a) \, ,
\label{gro:eq:c}
\end{gather}
and the two-point function of a quartic with a Yukawa operator vanishes. The free two-point function of $\tfrac12F^a_{\mu\nu}F^{a\,\mu\nu}$, summed over the gauge fields of the factor, is $12n_G/(\pi^4x^8)$, and the two-point functions between operators of different free fields vanish. Both quadratic forms are positive definite. The matrix $c_{IJ}$ of real coordinates $g^I$ on which $\lambda_{abcd}$ and $Y^a$ depend linearly and injectively, together with logarithmic coordinates, is therefore positive definite.
\end{lemma}

\begin{proof}
The four fields of a quartic operator at one point are paired with the four fields at the other point in $4!$ ways, each giving $C_\phi^4\delta\delta\delta\delta/x^8$, and with the factors $1/4!$ of the two vertices this gives the first expression. For the Yukawa operators the scalar line gives $C_\phi\delta_{ab}/x^2$, and the fermion loop gives $-{\rm Tr}\,[Y^aS(x)Y'^bS(-x)]$ with $S(x)=C_\psi\gamma\!\cdot\!x/x^4$. Since $S(-x)=-S(x)$ and $(\gamma\!\cdot\!x)^2=x^2$, this is $C_\psi^2\,{\rm tr}_{\rm spin}(1)\,{\rm tr}(Y^aY'^b)/x^6$ with ${\rm tr}_{\rm spin}(1)=4$. A quartic and a Yukawa operator contain five or three fields $\phi$ in total, and their two-point function vanishes. The form $\lambda\cdot\lambda$ is the sum of the squares of the components of a real symmetric tensor, and $\sum_a{\rm tr}\,(Y^aY^a)=\sum_a\sum_{ij}|Y^a_{ij}|^2$ for hermitian matrices. Both forms are therefore positive definite, and so is their restriction to the image of an injective linear map. For the gauge fields we use the Feynman gauge, since the operator is gauge invariant. The propagator $\delta^{ab}\delta_{\mu\nu}/(4\pi^2x^2)$ and the two complete contractions give $\langle F^aF^a(x)\,F^bF^b(0)\rangle=48n_G/(\pi^4x^8)$, which is $12n_G/(\pi^4x^8)$ for $\tfrac12F^aF^a$. Operators of different free fields have no contraction.
\end{proof}

At $d_c=4$ the marginal quartic and Yukawa operators contain no products of gamma matrices, so that independent couplings give independent operators. This is in contrast to the fermions near two dimensions of section~\ref{gs:sec:fermions}, where $c_{IJ}$ is degenerate at $\vep=0$.

\subsection{Gauge theories in \texorpdfstring{$d=4-\vep$}{d=4-epsilon}}
\label{gs:sec:gauge}

We consider gauge fields of a semisimple group with simple factors labelled by $\iota$, and Dirac fermions in representations for which the gauge couplings are vectorlike. We restrict to theories which are invariant under parity and whose Feynman rules do not contain $\gamma_5$, so that the theory can be continued to non-integer $d$ without a choice for $\gamma_5$. We first consider theories without scalar fields. The marginal couplings are then the gauge couplings $g_\iota$, which have the weight $k_\iota=\tfrac12$. With a local coupling the term $F^\iota_{\mu\nu}F^{\iota\,\mu\nu}/(4g_\iota^2)$ of the action is invariant under Weyl rescalings for every $d$ if $g_\iota$ has the Weyl weight $-\vep/2$ and the gauge field the Weyl weight zero. The group of field rotations consists of the transformations of the fermions which commute with the gauge group and with parity. The relevant scalar operators are the fermion bilinears, whose sources have dimension one.

The functional $W$ of these theories is the generating functional of the gauge-invariant operators, which is defined with a gauge fixing. For free gauge fields with a local coupling the functional is constructed in appendix~\ref{gs:app:freegauge}. We do not have a construction of the gauge-fixed functional of interacting gauge fields on $S^d$ for non-integer $d$, and the following is an assumption. The construction of section~\ref{gs:sec:dirac} supplies the part of the functional with the fermions alone, but it does not include gauge fields. The functional exists on $S^d$ for non-integer $d$ and does not depend on the gauge fixing. It has Properties~\ref{gs:P:functional} to~\ref{gs:P:sphere} with the sources of the fermion bilinears, where Property~\ref{gs:P:functional} holds in instance~(b), that is in the coordinates $\log g_\iota$, and, as part of Property~\ref{gs:P:weyl}, the vector field is $\hB^\iota=-\vep/2$ up to terms of second order in the couplings, with $k_\iota=\tfrac12$ as in minimal subtraction, see appendix~\ref{gs:sec:properties}. We refer to this statement as the assumption on gauge couplings. This assumption is the expectation of section~\ref{gs:sec:status} for these theories, and it rests on the following reasons. The classical action with local gauge couplings is invariant under Weyl rescalings for every $d$, as stated above. For Maxwell theory on a sphere of radius $r_0$ the free energy is $-\tfrac12\log(g^2\,r_0^{\vep})$ up to a function of $d$~\cite{Giombi:2015haa}, where $g$ is the coupling. This is the form of $\cW^{\log}$ in instance~(b) of Property~\ref{gs:P:functional} with $n_\iota=1$, and its dependence on the radius agrees with the relation $r_0\,\pr W/\pr r_0+\hB\!\cdot\!\pr\,\cW=V\rA$ of section~\ref{gs:sec:K} for $\hB^\iota=-\vep/2$ with a vanishing right-hand side, as for a functional without counterterms. For a simple gauge group with fermions the consistency conditions with a local gauge coupling were analysed in dimensional regularisation in section~4 of~\cite{Jack:1990eb}. Finally, for one simple factor Properties~\ref{gs:P:functional} to~\ref{gs:P:sphere} with the value $n_\iota=n_G$ of $n_G$ free gauge fields, which is $n_G$ times the value of Maxwell theory~\cite{Giombi:2015haa}, give $s_d\hat G^{{\rm S}(0,0)}=3n_G/\pi^2$, which is $\tfrac14\pi^2$ times the coefficient $12n_G/\pi^4$ of the free two-point function of $\tfrac12F^\iota F^\iota$, the value which Property~\ref{gs:P:free} asserts. For a simple factor the term of order zero in $\vep$ of $n_\iota$ is the number of its gauge fields under Property~\ref{gs:P:free}, by Proposition~\ref{grg:prop:gauge} below, in agreement with the free energy of the free gauge fields in the regularisation of~\cite{Giombi:2015haa}.

\begin{corollary}[Gauge fields with fermions]
\label{gs:cor:gauge}
Under the assumption on gauge couplings the following holds for gauge fields of a semisimple group with Dirac fermions in $d=4-\vep$. The sources of the fermion bilinears do not contribute, and the gradient formula $\pr_J\hat A^{\rm S}=\hat G^{\rm S}_{JI}\,\hB^I$ of Proposition~\ref{gs:prop:main} holds. The tensor $\hat G^{\rm S}_{JI}$ is regular and symmetric, and the function $\hat A^{\rm S}-(\vep\,d/V)\sum_\iota n_\iota\log g_\iota$ is regular. The term of order zero of the metric satisfies
\begin{equation}
\sum_{\iota'}\hat G^{{\rm S}(0)}_{\iota\iota'} = -\frac{2\,d\,n_\iota}{V} \, ,
\label{gs:eq:Gtt}
\end{equation}
and at a zero $g_*$ of $\hB^I$ the potential is $\hat A^{\rm S}(g_*)=\vep\,d\,\cW_*/V$ with $\cW_*$ of \eqref{gs:eq:fixed}, which contains the term $\sum_\iota n_\iota\log g_{\iota*}$.
\end{corollary}

\begin{proof}
The sources of the fermion bilinears have an odd dimension at $d=4$. We can therefore apply Lemma~\ref{gs:lem:dimension}, and the statements of Propositions~\ref{gs:prop:main} and~\ref{gs:prop:log} hold. The statements on the regularity and on the value at a zero are those of Proposition~\ref{gs:prop:main} with $\cW^{\log}$ of instance~(b) of Property~\ref{gs:P:functional}, and \eqref{gs:eq:Gtt} is \eqref{gs:eq:Glog} with $k_{\iota'}=\tfrac12$.
\end{proof}

The corollary is Propositions~\ref{gs:prop:main} and~\ref{gs:prop:log} applied to these theories under the assumption on gauge couplings, and what is specific to gauge theories is that the sources of the fermion bilinears drop out by their odd dimension and that the coordinates are the logarithms of the gauge couplings, with the weight $\tfrac12$.

Let us comment on the corollary. First, the potential contains the term $(\vep\,d/V)\,n_\iota\log g_\iota$ for each simple factor and is not a power series in the gauge couplings. For $\vep>0$ and positive $n_\iota$ the potential $s_d\,\hat A^{\rm S}$, which is normalised to $\tilde F$, tends to plus infinity for $g_\iota\to0$, and it cannot be normalised at the free theory. Second, the free gauge theory is not a zero of $\hB^I$. It is not a point of the space of couplings, and for $g_\iota\to0$ the component $\hB^\iota$ tends to $-\vep/2$. This agrees with the fact that free Maxwell theory is scale invariant but not conformal for $d\neq4$, as recalled in~\cite{Giombi:2015haa}. Third, at a fixed point with $g_{\iota*}^2$ of order $\vep$ the value $\cW_*$ contains $\tfrac12\,n_\iota\log|\vep|$, so that the potential at a fixed point is not a power series in $\vep$. With $\hB^\iota=-\tfrac12\vep+b_\iota\,g_\iota^2+{\rm O}(g^4)$ for one simple factor such a zero has $g_{\iota*}^2=\vep/(2b_\iota)+{\rm O}(\vep^2)$, and it is real if $\vep\,b_\iota>0$, so that an asymptotically free factor, with $b_\iota<0$, has it for $\vep<0$, that is for $d>4$.

Fourth, \eqref{gs:eq:Gtt}, which follows from Properties~\ref{gs:P:functional} to~\ref{gs:P:sphere} by Proposition~\ref{gs:prop:log}, shows that the metric does not vanish at lowest order if the $n_\iota$ do not vanish. For one simple factor it gives $-\hat G^{{\rm S}(0)}_{\iota\iota}/(2e_d)=d\,n_\iota/(e_dV)$ in the normalisation of section~\ref{gs:sec:phi4}, which is $n_\iota/(8\pi^2)$ at $\vep=0$, that is $s_d\hat G^{{\rm S}(0)}_{\iota\iota}=3n_\iota/\pi^2$. If $g_\iota$ itself is used as the coordinate in place of $\log g_\iota$, the component of the metric along this coupling is at lowest order $n_\iota/(8\pi^2g_\iota^2)$. It has a pole at $g_\iota=0$, and the free theory lies at infinite distance. For one simple factor and a coefficient $n_\iota$ which is positive at $\vep=0$ the metric $s_d\,\hat G^{\rm S}_{\iota\iota}$ is positive definite to all orders in the couplings and in $\vep$. For several simple factors \eqref{gs:eq:Gtt} fixes the sum of each row of the matrix of order zero. Under Property~\ref{gs:P:free} the term of order zero in the couplings and in $\vep$ of $s_d\hat G^{\rm S}$ is $\tfrac14\pi^2c_{\iota\iota'}$ by Lemma~\ref{gre:lem:zero}, which is diagonal with the entries $3n_G/\pi^2$, since $c_{\iota\iota'}$ vanishes between different factors and has the entries $12n_G/\pi^4$ by Lemma~\ref{gro:lem:c}. The metric $s_d\hat G^{\rm S}_{\iota\iota'}$ is then positive definite to all orders in the couplings and in $\vep$. For one simple factor \eqref{gs:eq:Gtt} with the value $n_\iota=n_G$ of the free gauge fields gives $s_d\hat G^{{\rm S}(0,0)}_{\iota\iota}=3n_G/\pi^2$, which is the value $\tfrac14\pi^2c_{\iota\iota}$ of Property~\ref{gs:P:free} and of Lemma~\ref{grg:lem:direct}. For several simple factors this holds for the diagonal entries if the entries between different factors vanish at order zero. The corollary itself does not use Property~\ref{gs:P:free}.

The results of~\cite{Giombi:2015haa} which we quote concern Maxwell theory and quantum electrodynamics. Their gauge group is abelian, so that they are not among the theories of Corollary~\ref{gs:cor:gauge}, and neither result enters its proof. We use them in two places. The free energy of Maxwell theory gives the value $n_\iota=1$ for one gauge field, in agreement with the value $n_\iota(0)=n_G$ of Proposition~\ref{grg:prop:gauge} below for $n_G=1$. For quantum electrodynamics with $N_f$ flavours in $d=4-\vep$ the free energy at the fixed point contains the term $\tfrac12\log(N_f/(6\pi^2\vep))$~\cite{Giombi:2015haa}. At the zero $g_*^2=6\pi^2\vep/N_f$ of the one-loop vector field this term is $-\log g_*$. Since $\cW$ is minus the free energy, it has the form of the term $n_\iota\log g_{\iota*}$ of $\cW_*$ with $n_\iota=1$, which Corollary~\ref{gs:cor:gauge} gives for a simple factor, although the corollary does not cover an abelian group.

We now add scalar fields. There are then Yukawa couplings of weight $\tfrac12$ and quartic couplings of weight $1$, whose coordinates are the couplings themselves. The sources of $\phi_i$, of $\bar\psi\psi$ and of the cubic operators of the scalars have odd dimension and drop out, and the source of $\phi_i\phi_j$ has dimension two. Without gauge fields this is the theory of section~\ref{gs:sec:yukawa}, whose properties follow from Hypothesis~\ref{hyp:finite} for the Yukawa theory.

With gauge fields and scalars Property~\ref{gs:P:functional} holds in instance~(b). Theorem~\ref{gr:cor:cond} applies in instance~(b) and gives the gradient formula $\pr_J\hat A^{\rm H}=\hat G^{\rm H}_{JI}\hB^I$ with a potential which is regular up to $(\vep\,d/V)\,\cW^{\log}$ and a regular symmetric metric. Let $\cO_\iota$ be the operator of the coordinate $\log g_\iota$, defined by \eqref{gs:eq:onepoint}, whose integrated one-point function at constant couplings is $g_\iota\,\pr\cW/\pr g_\iota$, which at lowest order is $\tfrac12F^\iota_{\mu\nu}F^{\iota\,\mu\nu}$ in the normalisation in which the gauge field is canonically normalised. The derivative is taken at fixed gauge field in the normalisation in which the action of the gauge field carries the factor $1/g_\iota^2$, where the action of the matter fields does not depend on $g_\iota$, so that the operator involves the action of the gauge field and not that of the matter fields. At vanishing couplings the action is a sum of the actions of the free scalars, of the free fermions and of the free gauge fields of each simple factor with their ghosts, so that the matrix $c_{IJ}$ of Property~\ref{gs:P:free} vanishes between operators of different free fields.

\begin{proposition}[Gauge fields with scalars]
\label{gro:prop:gauge}
Assume the hypotheses of Theorem~\ref{gr:cor:cond} in instance~(b), at $d_c=4$, and Property~\ref{gs:P:free}. Then the term of order zero in the couplings and in $\vep$ of $s_d\hat G^{\rm H}_{JI}$ is $\tfrac{\pi^2}{4}\,c_{IJ}$. It is block diagonal, with the entries $3n_G/\pi^2$ on the logarithmic coordinates, where $n_G$ is the number of gauge fields of the factor, and the block $\tfrac{\pi^2}{4}\,c_{IJ}$ with $c_{IJ}$ of Lemma~\ref{gro:lem:c} on the others, and it is positive definite. The gradient formula of Theorem~\ref{gr:cor:cond} holds with this metric, which is positive definite to all orders in the couplings and in $\vep$.
\end{proposition}

\begin{proof}
By Theorem~\ref{gr:cor:cond} under Property~\ref{gs:P:free} the term of order zero is $s_d^2\,a_4\,c_{IJ}$ with $a_4=\tfrac14\pi^2$. The matrix $c_{IJ}$ vanishes between operators of different free fields, and by Lemma~\ref{gro:lem:c} its entries are $12n_G/\pi^4$ on the logarithmic coordinates and those of \eqref{gro:eq:c} on the others, so that it is positive definite. A term of order zero which is positive definite gives a metric which is positive definite to all orders.
\end{proof}

The proposition is Theorem~\ref{gr:cor:cond} in instance~(b) applied under Property~\ref{gs:P:free}, and what is specific to gauge fields with scalars are the coordinates $\log g_\iota$, the source of $\phi_i\phi_j$ as the only relevant source of even dimension, and the matrix $c_{IJ}$ of Lemma~\ref{gro:lem:c} with its block form.

Under the hypotheses of Corollary~\ref{gs:cor:gauge}, or those of Proposition~\ref{gro:prop:gauge} if scalar fields are present, and Property~\ref{gs:P:free}, the term of order zero in $\vep$ of $n_\iota$ is the number $n_G$ of gauge fields of the simple factor $\iota$, by Proposition~\ref{grg:prop:gauge} below. Its hypothesis on the entries between different factors is part of Property~\ref{gs:P:free}. The coefficient $n_\iota$ is a power series in $\vep$ which does not depend on the couplings, and only its term of order zero is fixed in this way. The positivity of the metric uses the term of order zero in the couplings and in $\vep$, and the term $\tfrac12n_\iota\log|\vep|$ of $\cW_*$ at a fixed point with $g_{\iota*}^2$ of order $\vep$ is $\tfrac12n_\iota(0)\log|\vep|$ up to terms of order $\vep\log|\vep|$.

Property~\ref{gs:P:free} contains decoupling at lowest order, since $c_{IJ}$ vanishes between different free fields. By Lemma~\ref{gre:lem:zero} it fixes the term of order zero in the couplings and in $\vep$ of the metric, and the positivity at lowest order is therefore a statement on the free theory. The content of the results lies in the regularity and the symmetry of the metric to all orders and in the positivity to all orders which follows from the lowest order. For gauge fields Property~\ref{gs:P:free} states that $\cK^{(0,0)}$ takes in the logarithmic directions the value of free Maxwell theory in four dimensions with a local coupling, and that it has no entries between different free fields. Its value in the logarithmic directions follows in four dimensions from the two-point function of $\tfrac12F^aF^a$ by Lemma~\ref{grg:lem:direct}, and Proposition~\ref{grg:prop:gauge} below relates it to the value $n_\iota(0)=n_G$ of the free gauge fields at constant coupling. Assumed are that the functional reproduces this value and, for several factors and the mixed directions, that it decouples at lowest order. We expect the decoupling, since in a construction in minimal subtraction whose functional at vanishing couplings is the sum of the free functionals, and whose bare sources and counterterms are unique pure poles, a counterterm of order zero which mixes two free fields would cancel a pole of a vanishing term. On the scalar couplings Property~\ref{gs:P:free} would follow from the two-point form, which holds for the quartic couplings by Lemma~\ref{gs:lem:free} in a construction in minimal subtraction with unique pure poles, and in the divergence scheme by the argument of Proposition~\ref{gr4:prop:lowest}. For the Yukawa couplings the construction of section~\ref{gs:sec:dirac} and Lemma~\ref{gf:lem:free} supply the matter part. Without a construction of the interacting theories with gauge fields the property is assumed here. No condition on the term of order zero of $\hr_\iota$ is needed, since the field rotations leave the gauge couplings invariant, so that $(\hr_Jg)^K$ is linear in the other couplings, see Lemma~\ref{gsh:lem:inv}.

\paragraph{Gauge directions of Property~\ref{gs:P:free}.} We now relate the diagonal entries of $\cK^{(0,0)}$ in the logarithmic directions to the coefficient of $\log g_\iota$ at constant coupling on $S^d$. No functional of the free gauge field with a local coupling enters.

\begin{proposition}[Gauge directions of Property~\ref{gs:P:free}]
\label{grg:prop:gauge}
Assume Properties~\ref{gs:P:functional} to~\ref{gs:P:sphere} with Property~\ref{gs:P:functional} in instance~(b) and with the Weyl variation closed on the marginal couplings or with relevant sources of odd dimension only, see Lemma~\ref{gs:lem:dimension}, or, if there are relevant sources of even dimension, the hypotheses of Theorem~\ref{gr:cor:cond} in instance~(b), reached from a scheme in which Properties~\ref{gs:P:functional} to~\ref{gs:P:sphere} hold, and the last statement of Property~\ref{gs:P:free}, with $\cK$ in the divergence scheme. If there are several simple factors, assume that the term of order zero in the couplings and in $\vep$ of $\cK_{\iota\iota'}$ vanishes for $\iota\neq\iota'$. Then
\begin{equation}
\cK^{(0,0)}_{\iota\iota} = -\frac{d_c\,n_\iota(0)}{k_\iota\,V} \, ,
\label{grg:eq:K}
\end{equation}
where $V$ is taken at $d=d_c$. For $d_c=4$, with $k_\iota=\tfrac12$ and $V=\tfrac83\pi^2$, this is $-3n_\iota(0)/\pi^2$. Consequently the statement of Property~\ref{gs:P:free} in the logarithmic direction $\iota$, $\cK^{(0,0)}_{\iota\iota}=s_d\,a_4\,c_{\iota\iota}$ with $c_{\iota\iota}=12n_G/\pi^4$ of Lemma~\ref{gro:lem:c}, holds if and only if $n_\iota(0)=n_G$.
\end{proposition}

\begin{proof}
Without relevant sources of even dimension the sum rule \eqref{gs:eq:Glog} of Proposition~\ref{gs:prop:log} holds for every $\vep$. With such sources its proof applies to the gradient formula of Theorem~\ref{gr:cor:cond}, whose potential is regular up to $(\vep\,d/V)\,\cW^{\log}$, with $\hat G^{\rm H}$ in place of $\hat G^{\rm S}$, and by Theorem~\ref{gr:cor:cond} under the last statement of Property~\ref{gs:P:free} the two tensors have the same term of order zero in the couplings and in $\vep$. The change to the divergence scheme does not change $n_\iota$, since it vanishes at $g=0$ by the last statement of Property~\ref{gs:P:weyl}. Both sides of the sum rule are power series in $\vep$. At $\vep=0$ Lemma~\ref{gre:lem:zero} gives $\hat G^{{\rm S}(0,0)}=\cK^{(0,0)}$. The sum runs over the logarithmic coordinates only, so that the components between logarithmic and other coordinates do not enter. For one simple factor it has one term, and for several simple factors $\cK^{(0,0)}$ is diagonal on the logarithmic coordinates by assumption. The sum rule at $\vep=0$ therefore reads $k_\iota\cK^{(0,0)}_{\iota\iota}=-d_c\,n_\iota(0)/V$, which is \eqref{grg:eq:K}. With $s_d=-1$ and $a_4=\tfrac14\pi^2$ we have $s_d\,a_4\cdot12n_G/\pi^4=-3n_G/\pi^2$, and the last statement follows.
\end{proof}

Of Property~\ref{gs:P:free} the proposition uses only the last statement, and only if there are relevant sources of even dimension. Under the hypotheses of the proposition the content of Property~\ref{gs:P:free} in the logarithmic directions is thus the value of $n_\iota$ at $\vep=0$. By Property~\ref{gs:P:functional} in instance~(b) this coefficient does not depend on the couplings, so that it is fixed by the functional at constant couplings in the limit of vanishing couplings. That this limit is the free energy of the free gauge fields at constant coupling is part of what is assumed for interacting gauge fields, and only this free energy enters, not the dependence of the free functional on a local coupling.

\begin{lemma}
\label{grg:lem:direct}
Let a logarithmic coordinate $\tau$ have $\hB^\tau=0$ at order zero at $d=4$, and let its operator at order zero be a scalar primary of dimension four with the flat two-point function $c\,|x-y|^{-8}$ in a conformal free theory. Then the term of order zero in the couplings of $\cK_{\tau\tau}$ of this free theory at $d=4$ is $-\tfrac14\pi^2c$. For $\tfrac12F^aF^a$ of a simple factor with $n_G$ generators $c=12n_G/\pi^4$ by Lemma~\ref{gro:lem:c}, and the value is $-3n_G/\pi^2$.
\end{lemma}

\begin{proof}
On the unit sphere $S^4$ the two-point function is $c\,s^{-8}$, where $s$ is the chordal distance. We continue the kernel to $s^{-2\Delta}$. Its eigenvalue on the harmonics of degree $\ell$ has at $\Delta=4$ a simple pole in $4-\Delta$ with the coefficient $\tfrac{\pi^2}{192}\,\ell(\ell+1)(\ell+2)(\ell+3)$, proportional to the eigenvalue of the Paneitz operator $\Delta_4$. On the sphere of radius $R$ the pole multiplies $R^{2(4-\Delta)}$, and with the factor $\tfrac12$ of the term of second order of $W$ the term of second order in $\tau$ of $R\,\pr_RW$ is therefore $\tfrac{c\pi^2}{192}\int\tau\,\Delta_4\tau$, which is the term of order zero of $\cA[1]$. It does not depend on the subtraction, since two subtractions differ by a local functional with coefficients which are dimensionless at $d=4$. It vanishes for constant $\tau$, so that $\rA$ has no term of second order, and its second variation in the directions $\sigma$ and $\delta\tau$ is $2\cdot24\cdot\tfrac{c\pi^2}{192}\int\sigma\,\delta\tau$, since $\Delta_4\sigma=24\,\sigma$ for a harmonic of degree one. Hence $\alpha_{\tau\tau}=\tfrac14\pi^2c$ and $\cK_{\tau\tau}=-\tfrac14\pi^2c$.
\end{proof}

The lemma gives the value which Property~\ref{gs:P:free} asserts in the logarithmic directions, $s_d\,a_4\,c_{\iota\iota}=-3n_G/\pi^2$, for the free gauge fields in four dimensions, independently of the sum rule, and with Proposition~\ref{grg:prop:gauge} it agrees with the value $n_\iota=n_G$ of the free gauge field below. That the term of order zero of $\cK_{\iota\iota}$ of the functional on $S^d$ tends to this value at $\vep=0$ is the content of Property~\ref{gs:P:free} for gauge fields. For free gauge fields it follows from Hypothesis~\ref{hyp:finite} for the free gauge field, and for interacting gauge fields it is assumed.

\paragraph{The free gauge field.} For $n_G$ free gauge fields with a common local coupling we construct the bare functional on $S^d$ for non-integer $d$ in appendix~\ref{gs:app:freegauge}. In integer dimension it is minus one half of the logarithm of the determinant of the action on the exact two-forms. We write it through the projectors onto the levels of the exact two-forms, which are irreducible representations of ${\rm O}(d+1)$. A trace over fixed levels is then an integral \eqref{eq:gram} of a polynomial, and we continue the sums over the levels in $d$. The flat chart does not enter. The constant part of the coupling gives the term $n_G\log g$ for every $d$, Proposition~\ref{ggs:prop:shift}, since the continued number of levels of the coexact one-forms is one per gauge field. The bare functional is invariant under the conformal Killing vectors of the sphere when $\log g$ has the weight $\tfrac12\vep$, Proposition~\ref{ggs:prop:ward}, so that the bare anomaly vanishes. Under Hypothesis~\ref{hyp:finite} for the free gauge field, which is stated on the sphere and its conformal rescalings, the renormalised functional has Properties~\ref{gs:P:functional} to~\ref{gs:P:free} in instance~(b), with $n_\iota=n_G$ for every $\vep$ and $\hB=-\tfrac12\vep$ (Proposition~\ref{ggs:prop:props}). At second order the kernel of the cycle $T_2$ has the eigenvalue $4/((4-d)V)$ on the harmonics of degree one, so that the second variation of the bare functional in the directions of a harmonic $\sigma$ of degree one and of $\delta\tau$ is $(2n_Gd/(\vep V))\int d\mu_d\,\sigma\,\delta\tau$ (Proposition~\ref{ggs:prop:rot}). It gives the sum rule \eqref{gs:eq:Glog} with $n_\iota=n_G$ for every $\vep$ and the value of Lemma~\ref{grg:lem:direct} at $\vep=0$. With one coordinate, $\hB=-\tfrac12\vep$ exactly, $\cW=n_G\log g$ by Proposition~\ref{ggs:prop:shift} and $\rA=0$, since every counterterm contains a derivative of $\log g$, the gradient formula of Proposition~\ref{gs:prop:main} is equivalent for this theory, as an identity of power series in $\vep$, to the single relation $\hat G^{\rm S}=-2d\,n_G/V$. The free gauge field thus provides a theory with a logarithmic coordinate in which the functional is constructed and all six properties are derived from Hypothesis~\ref{hyp:finite}, so that instance~(b) of Property~\ref{gs:P:functional} and the statements of section~\ref{gs:sec:log} are realised. The term of order zero is the free energy of~\cite{Giombi:2015haa}. The hypothesis is stated on the sphere and not in flat space, since the flat functional does not depend on the constant part of the coupling, see appendix~\ref{ggs:sec:chart}. For a simple factor with $n_G$ generators and the action $\int d^dx\sqrt\gamma\,F^aF^a/(4g^2)$, in which the covariant derivative does not contain the coupling, the action at vanishing coupling is the sum of $n_G$ Maxwell actions with the factor $1/g^2$, and the volume of the constant gauge transformations does not depend on $g$. That the functional of the interacting theory at constant couplings tends to this free functional for vanishing couplings is part of the assumption on gauge couplings.

As a check of the framework in an interacting gauge theory at lowest order we consider quantum electrodynamics through the terms with one photon line, appendix~\ref{gq:sec:qed}. The current kernel of the fermions is diagonal on the levels, so that these terms are again sums over one level. They are $g^2$ times a functional of the variation of $\log g$, so that $n=1$ is unchanged, and the bare functional is invariant under the conformal Killing vectors. When no counterterm of the curvature alone occurs at order $g^2$, the finiteness of the term of first order in the variation of $\log g$ fixes the counterterm of the coupling and gives $\hB=-\tfrac12\vep+Q^2g^2/(12\pi^2)+O(g^4)$ with $Q^2=\sum_fq_f^2$ (Proposition~\ref{gq:prop:beta}). This is the one-loop coefficient of quantum electrodynamics~\cite{Giombi:2015haa}. Beyond this order a graph contains several photon lines and is a multiple sum over levels, whose continuation in $d$ is not constructed, so that quantum electrodynamics is not constructed.

\paragraph{Mixed components.} The decoupling between a logarithmic and a non-logarithmic coordinate at order zero is part of Property~\ref{gs:P:free}, and Proposition~\ref{gro:prop:gauge} uses it for the block diagonal form. For one simple factor it follows from the gradient formula if the potential has no term of first order in the couplings $g^I$. The row $I$ of the gradient formula at order zero in the couplings reads $\pr_IA^{(0)}=-\vep\sum_\iota k_\iota G^{(0)}_{I\iota}$ for the potential $A$ and the metric $G$ of Theorem~\ref{gr:cor:cond}, since $\hB^J$ has no term of order zero on the coordinates which are not logarithmic. If the term of order $\vep$ of $\pr_IA^{(0)}$ vanishes, the components $G^{(0,0)}_{I\iota}$ vanish for one simple factor, and so do those of $\cK^{(0,0)}$ by Theorem~\ref{gr:cor:cond} under the last statement of Property~\ref{gs:P:free}. For $\hat A^{\rm H}$ this term vanishes if the free one-point functions of the operators $\cO_I$ on $S^d$ and the terms of the anomaly which are linear in $g^I$ vanish for all $\vep$, since $\hat A^{\rm H}$ differs from $\hat A^{\rm S}$ by terms with the flow $\vk$ of the relevant sources, which contribute at order $\vep^2$ only, $\vk$ being of order $\vep^2$ at $g=0$. In the construction of section~\ref{gs:sec:universal} in minimal subtraction both hold for the matter fields, since the propagator of a conformally coupled scalar on $S^d$ is a pure power of the chordal distance, whose coincident limit vanishes in dimensional regularisation, so that the vacuum terms with one vertex vanish, and the Yukawa operators are odd in the scalar fields. For several simple factors the components between different factors are not fixed in this way.

As an example we consider ${\rm SU}(N)$ gauge fields with $N_f$ Dirac fermions $\psi$ in the fundamental representation and one real scalar $s$ which is a gauge singlet, with the couplings
\begin{equation}
\frac{1}{4g^2}F^aF^a+y\,s\,\bar\psi\psi+\frac{\lambda}{4!}\,s^4 \, .
\label{grf:eq:example}
\end{equation}
The theory is invariant under parity, with $s$ a scalar, and under the ${\rm U}(N_f)$ rotations of the fermions, and we restrict to the couplings which are invariant under these rotations. A Yukawa coupling $\phi\bar\psi\psi$ is gauge invariant only for $\phi$ in the singlet or in the adjoint representation, and we take a singlet. The coordinates are $\log g$, $y$ and $\lambda$, with the weights $\tfrac12$, $\tfrac12$ and $1$. The relevant source of even dimension is the source of $s^2$, and the sources of $s$, $s^3$ and $\bar\psi\psi$ have odd dimension. The operators at order zero are $\tfrac12F^aF^a$, $s\bar\psi\psi$ and $s^4/4!$, and their free two-point functions on the sphere at $d=4$ have the coefficients
\begin{equation}
c_{\log g} = \frac{12\,(N^2-1)}{\pi^4} \, , \qquad c_y = 4NN_f\,C_sC_\psi^2 = \frac{NN_f}{4\pi^6} \, , \qquad c_\lambda = \frac{C_s^4}{4!} = \frac{1}{6144\,\pi^8} \, ,
\end{equation}
where $C_s=1/(4\pi^2)$ and $C_\psi=1/(2\pi^2)$ are the normalisations of the propagators of $s$ and $\psi$ at $d=4$, and $4NN_f$ is the trace over the spinor, colour and flavour indices. These values are those of Lemma~\ref{gro:lem:c}. The mixed two-point functions vanish, since $\tfrac12F^aF^a$ contains no matter fields and $s\bar\psi\psi$ and $s^4$ contain an odd total number of fields $s$. Under the hypotheses of Proposition~\ref{gro:prop:gauge}, where $n_\iota=N^2-1$ at order zero in $\vep$ by Proposition~\ref{grg:prop:gauge}, the metric $s_d\hat G^{\rm H}_{JI}$ has at $\vep=0$ the term of order zero
\begin{equation}
\frac{\pi^2}{4}\,{\rm diag}\left(c_{\log g},\,c_y,\,c_\lambda\right) = {\rm diag}\left(\frac{3(N^2-1)}{\pi^2},\,\frac{NN_f}{16\pi^4},\,\frac{1}{24576\,\pi^6}\right)
\label{grf:eq:examplemetric}
\end{equation}
in the coordinates $(\log g,y,\lambda)$, which is positive definite. These are the entries of Proposition~\ref{gro:prop:gauge}. If $g$ itself is used as the coordinate, the first entry becomes $3(N^2-1)/(\pi^2g^2)$, in the normalisation of $s_d\hat G^{\rm H}$.

In four dimensions a gradient formula for parity-invariant theories with a compact gauge group, Yukawa and quartic couplings can be obtained with the flat-space method of~\cite{Reehorst:phi4}, without Property~\ref{gs:P:free} for the gauge couplings and under two hypotheses of the kind of those of~\cite{Reehorst:phi4}, with a metric which is positive definite at lowest order for at most one abelian factor. This is the subject of work in preparation~\cite{Reehorst:general4d}. At $\vep=0$ Proposition~\ref{gro:prop:gauge} and that result give two gradient formulas in four dimensions, each under its own hypotheses, and the relation between their metrics is open.

For an abelian factor of the gauge group the field strength $F_{\mu\nu}$ is a gauge-invariant operator of dimension two, which requires an antisymmetric tensor source $\mathsf{B}_{\mu\nu}$ with counterterms such as $\pr_\mu g^I\pr_\nu g^J F^{\mu\nu}$. Such a source is not among the sources of section~\ref{gs:sec:functional}. If the properties are extended to it, it does not contribute to the identities of section~\ref{gs:sec:relevant}.

\begin{lemma}
\label{gr4d:lem:twoform}
Assume Properties~\ref{gs:P:functional} to~\ref{gs:P:sphere} with an additional antisymmetric tensor source $\mathsf{B}_{\mu\nu}$, where the class of sources contains the corresponding two-forms, the coefficients of its Weyl variation need not be regular, and the first statement of Property~\ref{gs:P:sphere} includes that the one-point function of the operator dual to $\mathsf{B}_{\mu\nu}$ vanishes at constant couplings. Then the source $\mathsf{B}_{\mu\nu}$ does not contribute to \eqref{gs:eq:KM} and to the sphere data of $W^\sharp$ of Lemma~\ref{gsh:lem:quant}.
\end{lemma}

\begin{proof}
On the unit sphere an antisymmetric tensor which is linear in a harmonic $\sigma$ and built from the metric and the constant curvature, without the Levi-Civita tensor, vanishes, since $\sigma\mathring\gamma_{\mu\nu}$ and $\nabla_\mu\nabla_\nu\sigma=-\sigma\mathring\gamma_{\mu\nu}$ are symmetric and no other vector is available to contract with $\nabla_\mu\sigma$. At constant couplings, $a_\mu=0$ and vanishing relevant sources the variation $\delta^{\rm W}_\sigma\mathsf{B}_{\mu\nu}$ is such a tensor and vanishes, and so are the derivatives of $\cA[\sigma]$ and of $\cA[1]$ with respect to $\mathsf{B}_{\mu\nu}(x)$, while the one-point function of the dual operator vanishes by assumption. The identity \eqref{gs:eq:KM} contains first variations only. In the first variation of the Ward identity around constant couplings the source enters through the variation of $\delta^{\rm W}_\sigma\mathsf{B}-\cL_v\mathsf{B}$ times the one-point function, which vanishes, and through the vanishing $\delta^{\rm W}_\sigma\mathsf{B}$ times the variation of the one-point function. The sphere data of Lemma~\ref{gsh:lem:quant} contain in addition the first and second variations around constant couplings of the term of $\bD W$ with $\mathsf{B}_{\mu\nu}$. In them $\delta^{\rm W}_1\mathsf{B}$ or its variations multiply the one-point function of the dual operator, which vanishes, or the first variation of $\delta^{\rm W}_1\mathsf{B}$ in a direction $\delta g^J$ multiplies the variation of this one-point function. At $a_\mu=0$ this first variation is an antisymmetric tensor which is linear in $\delta g^J$ and built from the metric, the curvature of the unit sphere and covariant derivatives. Commuting the derivatives as in section~\ref{gsp:sec:facts}, such a tensor is a combination of $\mathring\gamma_{\mu\nu}\,p(\nabla^2)\delta g^J$ and $\nabla_\mu\nabla_\nu q(\nabla^2)\delta g^J$, which are symmetric, so that it vanishes. A term of $\delta^{\rm W}_1\mathsf{B}$ which is linear in a scalar relevant source has a dimension larger than two and does not occur. Hence the source $\mathsf{B}_{\mu\nu}$ does not contribute, and the regularity of the coefficients of $\delta^{\rm W}_\sigma\mathsf{B}$ is not used.
\end{proof}

For a single abelian factor the lemma shows that the source of the field strength, see appendix~\ref{gs:app:status}, does not contribute. Properties~\ref{gs:P:functional} to~\ref{gs:P:sphere} then have to be formulated with this source, and the coordinate of the abelian coupling $e$ is the logarithmic coordinate $\log e$ of instance~(b), with $n=1$ for Maxwell theory~\cite{Giombi:2015haa}. The assumption on the one-point function of the dual operator is of the kind of Property~\ref{gs:P:sphere}, since in integer dimension $d>2$ there is no invariant two-form on $S^d$.

\paragraph{Several abelian factors.}\label{gra:sec} 
We consider $n$ abelian factors of the gauge group with fixed charges, whose couplings form the positive symmetric matrix $h_{ab}$ in the action $\tfrac14h_{ab}F^a_{\mu\nu}F^{b\,\mu\nu}$, with covariant derivatives which do not contain $h$. For $n=1$ we have $h=e^{-2}$.

\paragraph{Coordinates and regularity.} We write $h=LL^T$ with $L=D(1+N)$, where $D=\mathrm{diag}(e^{u_1},\ldots,e^{u_n})$ and $N$ is strictly lower triangular. We note that the map $(u,N)\mapsto h$ is a diffeomorphism onto the positive matrices. The coordinates depend on the order of the gauge fields, which we fix, and the regularity defined below refers to this order. The inverse $h^{-1}=D^{-1}(1+N)^{-T}(1+N)^{-1}D^{-1}$ has the entries $e^{-u_a-u_b}$ times polynomials in $N$, since $N$ is nilpotent. Let $x_a=e^{-u_a}$.

\begin{definition}
\label{gra:def:regular}
A function of the coordinates $u_a$, $N_{ab}$ and of the other couplings $g^I$ is regular if it is a power series in $\vep$, in the $x_a$ and in the $g^I$ whose coefficients are polynomials in the $N_{ab}$. A tensor is regular if its components in the coordinates $(u_a,N_{ab},g^I)$ are regular. We assign the weight one to the $x_a$, to the $g^I$ and to $\vep$, and the weight zero to the $N_{ab}$. For the order counting of tensors we count in addition lower indices along $u_a$ and $N_{ab}$ with the weight zero and along $g^I$ with the weight one, so that a term of a component has its own weight plus the weights of the indices. A term of order zero in the couplings of a function or of a component is a term of degree zero in the $x_a$ and the $g^I$, which may depend on $N$ and on $\vep$, and a term of order zero in the couplings and in $\vep$ is a term of degree zero in the $x_a$, the $g^I$ and $\vep$, which is a polynomial in $N$. A statement at vanishing couplings refers to $x_a=0$ and $g^I=0$ for every $N$. A regular symmetric tensor is nondegenerate if the matrix of the terms of order zero in the couplings and in $\vep$ of its components, a matrix of polynomials in $N$, has a determinant which is a nonzero constant. Its inverse is then regular, by the adjugate of this matrix and the expansion in the $x_a$, the $g^I$ and $\vep$.
\end{definition}

We find that the coordinates $u_a$ play the role of the logarithmic coordinates of instance~(b), and the $N_{ab}$ are coordinates of a new kind, which do not tend to zero at weak coupling. For $n=1$ there is no $N$, and $u=-\log e$ is a logarithmic coordinate up to its sign.

\begin{lemma}
\label{gra:lem:class}
The regular functions form a ring which is closed under the derivatives $\pr/\pr u_a$, $\pr/\pr N_{ab}$ and $\pr/\pr g^I$, and under division by a power series in $\vep$ which does not vanish at $\vep=0$. The derivatives along $u_a$ and $N_{ab}$ do not lower the weight, and the term of weight zero of a regular function does not depend on the $u_a$. At each weight a regular function has finitely many terms, each a polynomial in the $N_{ab}$.
\end{lemma}

\begin{proof}
We note that products and sums of power series in $\vep$, $x_a$ and $g^I$ with polynomial coefficients in $N$ are of the same kind. We have $\pr x_a/\pr u_b=-\delta_{ab}x_a$, which keeps the degree in $x$, and $\pr/\pr N_{ab}$ maps polynomials to polynomials. The term of weight zero is a polynomial in $N$, which contains no $x_a$ and hence no $u_a$. The last statement holds since a monomial of weight $w$ has degree at most $w$ in $x$, $g$ and $\vep$.
\end{proof}

\paragraph{Operators and vector field.} The operator of a coordinate $y$ at order zero is $-\pr S/\pr y=-\tfrac14F^T(\pr_yh)F$. In terms of the canonically normalised field strength $\tilde F=L^TF$ it is $-\tfrac12\tilde F^TS_y\tilde F$ with $S_y=\tfrac12(X_y+X_y^T)$ and $X_y=L^{-1}\pr_yL$. Here $X_{u_a}=(1+N)^{-1}E_{aa}(1+N)$, where $E_{aa}$ is the matrix unit, and $X_{N_{ab}}=(1+N)^{-1}E_{ab}$, which are polynomials in $N$. We conclude that the operators are canonically normalised with coefficients which are regular. The charges enter the covariant derivative of the matter fields through $q^TL^{-T}\tilde A$. The component $j$ of $q^TL^{-T}$ is $\sum_aq_a\,x_a\,((1+N)^{-1})_{ja}$, which is linear in the $x_a$ with coefficients which are polynomials in $N$, so that the couplings of the matter fields to $\tilde A$ are of weight at least one.

At order zero the classical Weyl weight of $h$ is $\vep$, that is $\delta^{\rm W}_\sigma h=\sigma\vep h$, which gives $L\to(1+\tfrac12\vep\sigma)L$, that is $\hB^{u_a}=\tfrac12\vep$ and $\hB^{N_{ab}}=0$ at order zero. The further terms of the flow of $h$ are polynomials in $h^{-1}$ and in the other couplings, starting with a constant matrix $b$ at one loop. Their components in the coordinates $(u,N)$ are obtained from $L^{-1}\delta hL^{-T}$, which for $\delta h=b$ is $(1+N)^{-1}D^{-1}bD^{-1}(1+N)^{-T}$. Its entry $(i,j)$ is $\sum_{a,c}((1+N)^{-1})_{ia}\,x_a\,b_{ac}\,x_c\,((1+N)^{-1})_{jc}$, which is quadratic in the $x_a$ with coefficients which are polynomials in $N$. We conclude that $\hB^{u_a}-\tfrac12\vep$ and $\hB^{N_{ab}}$ are regular and of degree at least two in the $x_a$ and the $g^I$. The function $\cW$ at constant $h$ contains $-\tfrac12\log\det h=-\sum_au_a$ in the regularisation of~\cite{Giombi:2015haa}, one term $\log e$ for each gauge field, so that $\cW^{\log}=-\sum_au_a$, linear in the logarithmic coordinates and independent of $N$.

\paragraph{The proofs of sections~\ref{gs:sec:identities} and~\ref{gs:sec:relevant}.} \begin{lemma}
\label{gra:lem:proofs}
Let Properties~\ref{gs:P:functional} to~\ref{gs:P:sphere} and the last statement of Property~\ref{gs:P:free} hold with regularity in the sense of Definition~\ref{gra:def:regular}, with $\cW^{\log}$ linear in the $u_a$, with $\hB^{u_a}=\tfrac12\vep$ and $\hB^{N_{ab}}=0$ at order zero, and with the antisymmetric two-form sources of the $n$ field strengths. Then the statements of sections~\ref{gs:sec:identities} and~\ref{gs:sec:relevant} hold, where the coordinates $u_a$ and $N_{ab}$ are treated as the logarithmic coordinates of instance~(b), and the terms of order zero in the couplings, and of order zero in the couplings and in $\vep$, are those of Definition~\ref{gra:def:regular}. The sum rule \eqref{gs:eq:Glog} holds for the rows $u_a$ with $n_{u_a}=-1$ and $k_{u_a}=-\tfrac12$. For the rows $N_{ab}$ it is replaced by $\tfrac12\vep\sum_c\hat G^{{\rm S}(0)}_{N_{ab}u_c}=d\,\pr_{N_{ab}}\big(\vep\,\cW^{(0)}_{\rm f}/V+\rA^{(0)}\big)$, where $\cW^{(0)}_{\rm f}$ and $\rA^{(0)}$ are the terms of order zero in the couplings of $\cW-\cW^{\log}$ and of $\rA$.
\end{lemma}

\begin{proof}
We first note that the identities of Propositions~\ref{gs:prop:K}, \ref{gs:prop:sharp}, \ref{gs:prop:main} and~\ref{gs:prop:relevant} and Lemmas~\ref{gsh:lem:comm}, \ref{gr:lem:rel} and~\ref{gsh:lem:quant} are derived in arbitrary coordinates, and their proofs use regularity through the closure of the regular functions under products and derivatives, which holds by Lemma~\ref{gra:lem:class}. The statements on the order zero and on weights in sections~\ref{gs:sec:identities} and~\ref{gs:sec:relevant} use, besides this closure, the following facts on logarithmic coordinates, which we check for the $u_a$ and the $N_{ab}$. First, a derivative along such a coordinate does not lower the weight, which holds by Lemma~\ref{gra:lem:class}. Second, the vector field has the form of minimal subtraction, with a constant term of order zero in the couplings along the logarithmic coordinates and further terms of order at least two in the couplings, which holds with $\hB^{u_a}=\tfrac12\vep+\ldots$ and $\hB^{N_{ab}}=0+\ldots$, where the dots are of degree at least two in the $x_a$ and the $g^I$. Third, the term of order zero in the couplings of a regular function is annihilated by the transport $\hB^K\pr_K$ at order zero, which holds since it does not depend on the $u_a$ and since $\hB^{N_{ab}}$ has no term of order zero. Fourth, a nondegenerate tensor has a regular inverse, which holds with the notion of nondegeneracy of Definition~\ref{gra:def:regular}. Fifth, the terms $\hB^K$, $\pr_J\hB^K$ and $(\hr_Jg)^K$ have no term of weight zero, which the order counting of Lemma~\ref{gsh:lem:inv} and of the term of order zero in Theorem~\ref{gr:cor:cond} uses. This holds along $u_a$ and $N_{ab}$, since the terms of $\hB^K$ which depend on the couplings are of degree at least two in the $x_a$ and the $g^I$, and since the field rotations leave $h$ invariant, so that $(\hr_Jg)^K$ is linear in the couplings of the matter fields. Sixth, the term of order zero in the couplings and the term of order zero in the couplings and in $\vep$ are distinguished as in Definition~\ref{gra:def:regular}, and the statements of Lemma~\ref{gre:lem:zero} and of the sum rule are statements at order zero in the couplings for all $\vep$. One fact does not hold along $N$: the derivative of a regular function along $N_{ab}$ can have a term of order zero, since such terms may depend on $N$. It enters the sum rule of Proposition~\ref{gs:prop:log}, whose proof uses that $\pr_\iota$ of a regular function has no term of order zero. Along $u_a$ this holds, and the sum rule holds with $n_{u_a}=-1$ and $k_{u_a}=-\tfrac12$. Along $N_{ab}$ the term of order zero of $\pr_{N_{ab}}\hat A^{\rm S}$ comes from the terms of order zero in the couplings $\cW^{(0)}_{\rm f}$ of $\cW-\cW^{\log}$, the free energy of the free fields, and $\rA^{(0)}$ of $\rA$, which are functions of $N$ and $\vep$. Indeed, at order zero the transport is $\tfrac12\vep\sum_c\pr_{u_c}$, which annihilates both and maps $\cW^{\log}$ to a constant. With the potential of \eqref{gs:eq:main} the proof of Proposition~\ref{gs:prop:log} then gives $-\vep\sum_ck_{u_c}\hat G^{{\rm S}(0)}_{N_{ab}u_c}=d\,\pr_{N_{ab}}\big(\vep\,\cW^{(0)}_{\rm f}/V+\rA^{(0)}\big)$. Since the left-hand side is of order $\vep$, the term of order zero in $\vep$ of $\pr_{N_{ab}}\rA^{(0)}$ vanishes. In Lemma~\ref{gsh:lem:inv} the inverse is a formal series in the weight, and at each weight its coefficients are finite sums of polynomials in $N$ by Lemma~\ref{gra:lem:class}. The field rotations act on the matter fields and leave $h$ invariant, so that $(\hr_Ig)^K$ vanishes for $K$ along $u_a$ or $N_{ab}$. The two-form sources drop out by Lemma~\ref{gr4d:lem:twoform}, applied to each component, since the lemma uses only that no antisymmetric tensor can be built on the sphere from a harmonic and the metric, and that the one-point functions of the field strengths vanish at constant couplings. A counterterm quadratic in the two-form sources has a vanishing first derivative at vanishing sources and does not enter.
\end{proof}

\paragraph{Metric of order zero.} \begin{proposition}
\label{gra:prop:metric}
Under the assumptions of Lemma~\ref{gra:lem:proofs}, and with Property~\ref{gs:P:free} for the abelian block, the term of order zero in the couplings and in $\vep$ of $s_d\hat G^{\rm S}$, and of $s_d\hat G^{\rm H}$ of Theorem~\ref{gr:cor:cond} if relevant sources of even dimension are present, on the coordinates of $h$ is
\begin{equation}
\frac{3}{4\pi^2}\,{\rm tr}\left(h^{-1}dh\,h^{-1}dh\right) ,
\label{gra:eq:metric}
\end{equation}
which is positive definite at every $N$ and nondegenerate in the sense of Definition~\ref{gra:def:regular}, with a constant determinant in the coordinates $(u,N)$. It agrees with the sum rule of Proposition~\ref{gs:prop:log} in the rows along $u_a$, and in the rows along $N_{ab}$ the sum of the entries in the columns $u_c$ vanishes. For $n=1$ it is $3\,(d\log e)^2/\pi^2$.
\end{proposition}

\begin{proof}
By Lemma~\ref{gre:lem:zero}, which holds by Lemma~\ref{gra:lem:proofs}, the term of order zero of $\hat G^{\rm S}$ at $\vep=0$ is that of $\cK$, and so is that of $\hat G^{\rm H}$ by Theorem~\ref{gr:cor:cond} under the last statement of Property~\ref{gs:P:free}, and Property~\ref{gs:P:free} gives $s_d\,a_4\,c_{yy'}$, with $c_{yy'}$ the coefficient of the free two-point function of the operators of the coordinates $y$ and $y'$ in four dimensions. For independent free fields $\vev{\tilde F^a_{\mu\nu}\tilde F^b_{\rho\sigma}}$ is diagonal in $a,b$, and for symmetric matrices $A$ and $B$ the two-point function of $\tfrac12\tilde F^TA\tilde F$ and $\tfrac12\tilde F^TB\tilde F$ is ${\rm tr}(AB)$ times that of $\tfrac12\tilde F\tilde F$ for one field, whose coefficient is $12/\pi^4$. With ${\rm tr}(S_yS_{y'})=\tfrac14{\rm tr}(h^{-1}\pr_yh\,h^{-1}\pr_{y'}h)$, which follows from $L^{-1}\pr_yhL^{-T}=X_y+X_y^T=2S_y$, we find $c_{yy'}=\tfrac{3}{\pi^4}{\rm tr}(h^{-1}\pr_yh\,h^{-1}\pr_{y'}h)$, and with $a_4=\tfrac14\pi^2$ and $s_d^2=1$ this gives \eqref{gra:eq:metric}. We note that the quadratic form ${\rm tr}(h^{-1}\delta h\,h^{-1}\delta h)={\rm tr}((L^{-1}\delta hL^{-T})^2)$ is positive for every nonzero symmetric $\delta h$. For the sum rule we note that $\sum_b\pr_{u_b}$ generates $L\to e^tL$, so that $\sum_bh^{-1}\pr_{u_b}h=2$. Hence $\sum_b\tfrac{3}{4\pi^2}{\rm tr}(h^{-1}\pr_yh\,h^{-1}\pr_{u_b}h)=\tfrac{3}{2\pi^2}{\rm tr}(h^{-1}\pr_yh)$. For $y=u_a$ this is $\tfrac{3}{2\pi^2}\cdot2=\tfrac{3}{\pi^2}$. The sum rule with $n_{u_a}=-1$ and $k_{u_a}=-\tfrac12$ reads $-\tfrac12\sum_b\hat G^{{\rm S}(0)}_{u_au_b}=4/V$ at $d=4$, so that $\sum_bs_d\hat G^{{\rm S}(0,0)}_{u_au_b}=8/V=3/\pi^2$ with $s_d=-1$ and $V=\tfrac83\pi^2$, in agreement. For $y=N_{ab}$ we have ${\rm tr}(h^{-1}\pr_{N_{ab}}h)=2\,{\rm tr}\,X_{N_{ab}}=2\,{\rm tr}((1+N)^{-1}E_{ab})=0$, since $(1+N)^{-1}$ is lower triangular with unit diagonal and $E_{ab}$ is strictly lower triangular. With the sum rule along $N_{ab}$ of Lemma~\ref{gra:lem:proofs} it follows, through Property~\ref{gs:P:free}, that the term of order zero in $\vep$ of $\pr_{N_{ab}}\big(\cW^{(0)}_{\rm f}+V\rA^{(0)}/\vep\big)$ vanishes. The free energy $\cW^{(0)}_{\rm f}$ itself does not depend on $N$ at $\vep=0$ if the finite curvature terms are normalised such that the term of order $\vep$ of $\rA^{(0)}$ does not depend on $N$. Such a normalisation exists, since the term $\int d^dx\sqrt\gamma\,R^2f(N)$ with a polynomial $f$ changes $\rA^{(0)}$ by $-\vep\,d^2(d-1)^2f$ and leaves $\hat A^{\rm S}$ and $\hat G^{\rm S}$ unchanged by Lemma~\ref{gs:lem:local}. The determinant of \eqref{gra:eq:metric} in the coordinates $(u,N)$ is a constant. Indeed, $X=L^{-1}dL=(1+N)^{-1}dU\,(1+N)+(1+N)^{-1}dN$ with $dU={\rm diag}(du_a)$. Its diagonal is $dU$, and its strictly lower triangular part is $(1+N)^{-1}dN$ plus terms linear in $du$, while $dN\mapsto(1+N)^{-1}dN$ has determinant one on the strictly lower triangular matrices. The map from $(du,dN)$ to the diagonal and the strictly lower triangular components of $X$ thus has determinant one. In these components ${\rm tr}((X+X^T)^2)=4\sum_aX_{aa}^2+2\sum_{a>b}X_{ab}^2$ has constant coefficients, so that the determinant of ${\rm tr}(h^{-1}\pr_yh\,h^{-1}\pr_{y'}h)$ is $4^n\,2^{n(n-1)/2}$. For $n=1$, $h^{-1}dh=-2\,d\log e$.
\end{proof}

We thus obtain from the proposition nondegeneracy and positivity of the metric on the abelian block for any number of abelian factors, under the stated assumptions, and with the other blocks as in section~\ref{gs:sec:gauge}, by decoupling of the free fields at lowest order. The matrix of the terms of order zero in the couplings and in $\vep$ of the whole metric is then block diagonal with a determinant which is a nonzero constant, so that the metric is nondegenerate in the sense of Definition~\ref{gra:def:regular}. That the functional with $n$ abelian factors has the properties in this sense is an assumption, as for the other gauge theories.

One extension is open beyond these, namely chiral gauge couplings and Yukawa couplings with $\gamma_5$, for which the treatment of $\gamma_5$ in dimensional regularisation is ambiguous at high orders~\cite{Poole:2019kcm}.

\subsection{Cubic scalars in \texorpdfstring{$d=6-\vep$}{d=6-epsilon}}
\label{gs:sec:cubic}

We consider $N$ real scalar fields with cubic couplings in $d=6-\vep$ dimensions, with the weight $k=\tfrac12$. There are two families of relevant scalar operators. The operators $\phi_i\phi_j$ have a source of dimension two, and the operators $\phi_i$ have a source whose dimension at $d=6$ is four. Both dimensions are even, and both sources mix with the curvature and with derivatives of the couplings at low orders~\cite{Osborn:2015rna,Grinstein:2015ina}. For instance, at one loop the source of $\phi_i$ mixes with the square of the Weyl tensor, which vanishes on the sphere, and at two loops with terms which contain four derivatives of the couplings. The bare functional is that of section~\ref{gs:sec:scalars}, and the hypothesis of finiteness is Hypothesis~\ref{hyp:finite} for the cubic theory of appendix~\ref{gs:sec:cases}. The field $\phi_i$ has the dimension two at $d=6$, so that it can be shifted by local functions of the sources with pure poles as coefficients, and the finiteness does not fix the bare source of $\phi_i$. Correspondingly the analogue of Lemma~\ref{lem:potential} fails for the kernel $|x-y|^{-4}$ of the source of $\phi_i$, which is the Green function of the Laplacian in six dimensions (Lemma~\ref{gcu:lem:potential}). The freedom is a shift of the field, which is described by an identity of the bare functional, the integrated equation of motion (Lemma~\ref{gcf:lem:shift}). With it the terms of the bare source of $\phi_i$ of order zero and one in the couplings at vanishing relevant sources can be removed, for every number of fields (Proposition~\ref{gcf:prop:frame}), at the cost of a finite redefinition $h\to h+r$ of the source of $\phi_i$. The condition that these terms vanish, the frame condition of appendix~\ref{gs:app:cubic}, is therefore a choice and not an assumption. The choice does not change $\hB^I$, $\hr_I$, $\gamma_h$, $\gamma_\sM$, the term of order zero of $\cK_{IJ}$, and the coefficients of the anomaly which are linear in the relevant sources at $g=0$. It changes the inhomogeneous terms of the Weyl variations of $h$ and of $\sM$, and the coefficient $\ell_{{\rm h}J}$ of the source $h$ at $g=0$, and for $N=1$ also $\vk_{\rm h}-\vk_{\rm c}$. The two-point form holds for the set after the choice. The last statement of Property~\ref{gs:P:weyl} therefore refers to the bare sources after the choice, while Property~\ref{gs:P:free} concerns the divergence scheme and does not depend on it. Further covariant shifts of higher order in the couplings and regular redefinitions of $h$ give a set for which the Weyl variations of $h$ and of $\sM$ have regular coefficients (Proposition~\ref{gcp:prop:reg}). With the class of flat sources with $D_0=3$ all six properties then follow, with Properties~\ref{gs:P:weyl} and~\ref{gs:P:anomaly} on conformally flat backgrounds (Proposition~\ref{gcp:prop:props}). The proofs are in appendix~\ref{gs:app:cubic}.

\begin{theorem}[Cubic scalars]
\label{gs:thm:cubic}
Assume Hypothesis~\ref{hyp:finite} for the cubic theory, with the class of flat sources with $D_0=3$. Then there is a set of pure poles of the hypothesis with the same $\hB^I$ and $\hr_I$, obtained from the given one by field shifts, regular local redefinitions of the source of $\phi_i$, the average of $\cX$ over ${\rm O}(N)$ and the removal of the part of $\cX$ which contains neither sources nor couplings. For this set the renormalised functional on the unit sphere has Properties~\ref{gs:P:functional} to~\ref{gs:P:free}, with Property~\ref{gs:P:functional} in instance~(a), Properties~\ref{gs:P:weyl} and~\ref{gs:P:anomaly} on conformally flat backgrounds, the weights $k_I=\tfrac12$ and the relevant sources $h$ and $\sM$. The gradient formula $\pr_J\hat A^{\rm H}=\hat G^{\rm H}_{JI}\hB^I$ of Theorem~\ref{gr:cor:cond} holds, with a regular potential which takes the value \eqref{gs:eq:fixed} at the zeros of $\hB^I$, and the metric $s_d\hat G^{\rm H}_{JI}$ is positive definite to all orders in the couplings and in $\vep$, with the term of order zero $\tfrac1{32}\pi^3c_{IJ}$ at $\vep=0$, where $c_{IJ}=C_\phi^{\,3}/3!$ times the unit matrix.
\end{theorem}

\begin{proof}
By Proposition~\ref{gcp:prop:props} Properties~\ref{gs:P:functional} to~\ref{gs:P:free} hold for this set, in the form stated, which is the form which the general proof uses, see appendix~\ref{gs:sec:properties}. The last statement of Property~\ref{gs:P:weyl} carries Properties~\ref{gs:P:functional} to~\ref{gs:P:sphere} over to the divergence scheme of both sources, which exists since $D=2$ and $D=4$ are smaller than $d_c=6$. Theorem~\ref{gr:cor:cond}, whose flows of the sources are constructed by induction on the dimension, the source of $\phi_i\phi_j$ before the source of $\phi_i$, gives the gradient formula and, under Property~\ref{gs:P:free}, the term of order zero $a_6\,c_{IJ}$ with $a_6=\tfrac1{32}\pi^3$, which is positive definite by Lemma~\ref{gs:lem:free}.
\end{proof}

The matrix $c_{IJ}$ is the unit matrix times a positive constant in the coordinates in which $g^Ig^I=g_{ijk}g_{ijk}$, so that the term of order zero of the metric is invertible on the full space of cubic couplings. The consistency conditions of the local renormalisation group in six dimensions were analysed in~\cite{Grinstein:2014xba,Grinstein:2015ina,Stergiou:2016uqq} at $d=6$, where the consistency conditions were used at vanishing sources of the operators of dimension two and four, while the counterterms with these sources are given in~\cite{Grinstein:2015ina} through two loops. Let us comment on the sign. The term of order zero of the metric $s_d\,\hat G^{\rm H}_{IJ}$ is positive for a positive $c_{IJ}$, while a negative metric was found at leading order in the equation for the coefficient $\tilde a$ in~\cite{Grinstein:2014xba,Grinstein:2015ina}. The two statements concern different tensors. At a fixed point in six dimensions the anomaly with local couplings defines three symmetric tensors, and only the one related to the two-point function is positive~\cite{Osborn:2015rna}. The flow equation of $a$ in~\cite{Osborn:2015rna} contains another of the three tensors, which is not constrained in sign and is negative in the cubic theory, in agreement with~\cite{Grinstein:2014xba}. The function $\tilde a$ equals $a$ at fixed points, and away from them it depends on the scheme at order $\beta$~\cite{Stergiou:2016uqq}. The derivative of another function which equals $a$ at fixed points is the beta function contracted with the sum of a symmetric and an antisymmetric tensor~\cite{Stergiou:2016uqq}. The symmetric tensor is the positive tensor at leading order, and the antisymmetric tensor was found there not to be a curl. This equation was derived without relevant operators of dimension two and four, which the cubic theory contains. Note that statements on positivity refer to real couplings. The fixed points of cubic theories are real only in special cases, such as the cubic ${\rm O}(N)$ model with $N>1038$~\cite{Fei:2014yja}, where $\tilde F$ was found to decrease along the flow at leading order~\cite{Giombi:2014xxa}.

\subsection{Fermions near two dimensions}
\label{gs:sec:fermions}

We consider Dirac fermions with quartic interactions near two dimensions, with the weight $k=1$. For these couplings $d=2$ is the lower critical dimension. They are irrelevant for $d>2$, and the fixed points of interest are ultraviolet fixed points of the expansion in $d=2+|\vep|$, which is $d=d_c-\vep$ with $d_c=2$ and $\vep<0$. Indeed, the linear term of $\hB^I=-\vep\,g^I+B^I$ is then $|\vep|\,g^I$, so that the free theory is an infrared stable zero. For one coupling with $B=-b\,g^2$ and $b>0$, as for the Gross--Neveu model, the zero $g_*=|\vep|/b$ has $\pr\hB=-|\vep|$ and is ultraviolet stable. For even $d_c$ the factor $\vep/\sin(\pi d/2)$ of \eqref{gs:eq:Ftilde} and the factor $\vep/\vartheta_d$ of \eqref{gs:eq:G0sign} are even in $\vep$, and the other factors of \eqref{gs:eq:G0sign} are positive for small $|\vep|$. The statements of sections~\ref{gs:sec:identities} and~\ref{gs:sec:consequences} therefore hold for $\vep<0$, where $s_d=(-1)^{d_c/2+1}$ keeps the value which it has for positive $\vep$. The relevant scalar operators are the fermion bilinears, whose sources have dimension one. For non-integer $d$ the quartic operators $(\bar\psi\,\Gamma^{(n)}\psi)^2$, where $\Gamma^{(n)}$ denotes the antisymmetrised products of $n$ gamma matrices, are independent for all $n$ and have the same dimension. Renormalisation generates them in general, so that the space of couplings on which the Weyl variation closes has infinitely many coordinates. Our assumptions are Properties~\ref{gs:P:functional} to~\ref{gs:P:sphere} for this infinite set of couplings $g^I$, with regularity as a formal power series in all of them and the tensors as infinite matrices, with Property~\ref{gs:P:functional} in instance~(a) and with a group of field rotations which consists of symmetries of the free theory for every $d$. We make no statement on whether a finite set of couplings suffices.

We construct the bare functional of this theory in appendix~\ref{g2:sec:construction}. The two bilinears of a four-fermion vertex carry antisymmetrised products of gamma matrices whose indices are contracted, and on the sphere we contract them with the projector onto the tangent space. We find that the contracted traces are polynomials in $d$ and in the scalar products of the points, and that the chart of Lemma~\ref{g2:lem:chart} gives one factor $\Omega$ for each bilinear. By Proposition~\ref{g2:prop:bare} the bare functional is defined for non-integer $d$ as a formal power series in all couplings, the four-fermion couplings have the weight $2-d$, which is $\vep k$ with $k=1$, and the functional is invariant under background gauge transformations. Through second order we obtain the two-point form of appendix~\ref{gs:sec:properties} with the matrix $c_{IJ}(d)$ of Lemma~\ref{g2:lem:kernel}, whose entries with a coupling of $\Gamma^{(n)}$, $n\geq3$, vanish at $d=2$.

We apply Hypothesis~\ref{hyp:finite} to this theory, with the bare sources $g^I+L^I$ and $a_\mu+N_I\,D_\mu g^I$, where $L^I$ takes values in the direct sum of the spaces of the couplings of fixed $n$, and with $\cX^{\rm fl}=\tfrac12\,\fs_{IJ}\,D_\mu g^ID^\mu g^J$. The derivation of Property~\ref{gs:P:weyl} from the hypothesis meets two new features in two dimensions, see appendix~\ref{g2:sec:finiteness}. First, for disjoint supports the kernel of two gauge fields at $d=2$ depends only on the divergences of the two one-forms, so that Lemma~\ref{lem:transverse} has no counterpart. For the one-forms built from the couplings we replace it by a condition on their dipole moments, and the steps for the gauge field go through. Second, the kernel of two couplings at $d=2$ is degenerate on the evanescent subspace $E$, which contains the couplings of $\Gamma^{(n)}$ for $n\geq3$. Let $N_I$ be covariant and without term of degree zero, and let ${\rm tr}\,1\neq0$. Consider the lowest level at which $\hB^I$ or $\hr_I$ has a pole, where the level is the degree in the couplings of $\hB^I$ and of $\hr_ID_\mu g^I$. The finiteness then implies that $\hr_I$ is regular at this level and that the pole of $\hB^I$ at this level is simple and its residue takes values in $E$ (Proposition~\ref{g2:prop:poles}). Poles of order two and higher are excluded, since the first correction in $d-2$ of the kernel is nondegenerate on $E$, Lemma~\ref{g2:lem:c1}, although it has positive definite and negative definite subspaces of every finite dimension there (Lemma~\ref{g2:lem:indef}). The finiteness does not exclude a simple pole of $\hB^I$ in the evanescent directions. We exclude it if the counterterms $L^I$ obey the pole equations of minimal subtraction for one choice of the bare sources, that is if the vector field $\hb^{{\rm L}\,I}$, which depends on $L^I$ alone, is regular. With this condition and for ${\rm tr}\,1>1$ we derive Property~\ref{gs:P:weyl} for the theory without the sources of the fermion bilinears (Proposition~\ref{g2:prop:poleeq}). In the proof we use that a field rotation which maps a generic coupling into $E$ is a multiple of the fermion number (Lemma~\ref{g2:lem:stabE}). For a suitable choice of the bare sources the regularity of $\hb^{{\rm L}\,I}$ is equivalent to the regularity of $\hB^I$, so that the proposition reduces the regularity in the evanescent directions to the pole equations of minimal subtraction.

\begin{proposition}[Fermions near two dimensions]
\label{gs:prop:fermions}
Under Properties~\ref{gs:P:functional} to~\ref{gs:P:sphere} for the infinite set of couplings the sources of the fermion bilinears do not contribute, and the gradient formula $\pr_J\hat A^{\rm S}=\hat G^{\rm S}_{JI}\,\hB^I$ of Proposition~\ref{gs:prop:main} holds, with the value \eqref{gs:eq:fixed} at the zeros of $\hB^I$. Assume in addition Hypothesis~\ref{hyp:finite} for the four-fermion theory. Let the functional on the unit sphere be the bare functional of Proposition~\ref{g2:prop:bare} at the bare sources of the hypothesis plus the integral of a local density $\cX$, built covariantly with $d$ as a parameter, which reduces to $\cX^{\rm fl}$ in flat space. Let it have Properties~\ref{gs:P:functional} to~\ref{gs:P:sphere} with the Weyl variation which the bare sources induce and with the anomaly $\mathbb{W}_\sigma\int\sqrt\gamma\,\cX$. Then Property~\ref{gs:P:free} holds, and the term of order zero in the couplings and in $\vep$ of $s_d\,\hat G^{\rm S}_{IJ}$ is $\pi\,c_{IJ}(2)$ with the matrix of Lemma~\ref{g2:lem:kernel}.
\end{proposition}

\begin{proof}
The sources of the fermion bilinears have an odd dimension at $d=2$. We can therefore apply Lemma~\ref{gs:lem:dimension}, and Proposition~\ref{gs:prop:main} holds. Under Hypothesis~\ref{hyp:finite} the bare couplings agree with the couplings up to terms of second order, and the bare gauge field is a gradient at first order, which does not contribute at second order by the gauge invariance. By Lemma~\ref{g2:lem:kernel} the functional then has the two-point form with $F=\int\sqrt\gamma\,\cX$, and Proposition~\ref{gs:prop:lowest} and Lemma~\ref{gre:lem:zero} give Property~\ref{gs:P:free} and the term of order zero.
\end{proof}

In this case the sources of the relevant operators drop out by Lemma~\ref{gs:lem:dimension}, and Theorem~\ref{gr:cor:cond} is not needed. Under the assumptions of the second statement of Proposition~\ref{gs:prop:fermions} the term of order zero in the couplings and in $\vep$ of the metric $s_d\,\hat G^{\rm S}_{IJ}$ is $\pi\,c_{IJ}(2)$. On the couplings without gamma matrices, which include the Gross--Neveu coupling, it is positive definite for ${\rm tr}\,1>1$ and every $N_f$, by Lemma~\ref{g2:lem:V0}. For the Gross--Neveu model the value of $\tilde F$ at the fixed point was obtained at leading order in section~4.5 of~\cite{Fei:2015oha}. On the full set of couplings $c_{IJ}(2)$ is degenerate, since its entries with a coupling of $\Gamma^{(n)}$, $n\geq3$, vanish by Lemma~\ref{g2:lem:kernel}, and its kernel is the evanescent subspace $E$ of \eqref{g2:eq:E}, which contains the couplings of $\Gamma^{(n)}$ for $n\geq3$. For ${\rm tr}\,1\geq2$ it is positive semidefinite, by Lemma~\ref{g2:lem:psd}. For ${\rm tr}\,1=2$ the subspace $E$ consists of the couplings whose operators vanish in two dimensions, which include combinations of couplings with $n\leq2$, and for ${\rm tr}\,1>2$ it is the sum of the spaces of the couplings of $\Gamma^{(n)}$ with $n\geq3$. For every $N_f$ and every ${\rm tr}\,1\neq0$ the term $c^{(1)}_{IJ}$ of first order in $d-2$ of $c_{IJ}(d)$, restricted to $E$, and the form $c_{IJ}(d)$ itself at every $d$ with $2<d<3$ have positive definite and negative definite subspaces of every finite dimension, which are spanned by couplings of a single flavour (Lemma~\ref{g2:lem:indef}). For ${\rm tr}\,1=2$ and the coupling of $\Gamma^{(n)}$ with $G^{(n)}_{ij,kl}=\delta_{i1}\delta_{j1}\delta_{k1}\delta_{l1}$ \eqref{g2:eq:closed} gives $c(G,G)=-2\,C_\psi^4\,(d-2)$ for $n=3$ and $c(G,G)=\tfrac{23}6\,C_\psi^4\,(d-2)$ for $n=4$, up to terms of second order in $d-2$, with $C_\psi$ at $d=2$. Under the same assumptions the term of order zero in the couplings of $s_d\,\hat G^{\rm S}_{IJ}$ is a positive multiple of $c_{IJ}(d)$, which is the matrix $c_{IJ}(\vep)$ of the two-point form at $\vep=2-d$, by \eqref{gs:eq:G0sign} with $k=1$. At $d=2+|\vep|$ with small $|\vep|\neq0$ it therefore has positive definite and negative definite subspaces of every finite dimension, and in particular infinitely many independent directions of each sign. The matrix $c_{IJ}$ of Property~\ref{gs:P:free} is degenerate, which the property does not require, so that only the statements on the sign do not apply. In minimal subtraction the evanescent operators are generated at three loops in the Gross--Neveu model, and their matrix elements at $d=2$ enter the beta function of the physical coupling from four loops on~\cite{Gracey:2016mio}. Properties~\ref{gs:P:functional}, \ref{gs:P:anomaly}, \ref{gs:P:ward} and~\ref{gs:P:sphere} and the covariance of $N_I$ are assumptions for this theory. Their derivation from the hypothesis requires the curvature counterterms of two dimensions. The relation of the zeros of $\hB^I$ on the infinite set of couplings to the fixed points of the theory which is continued from two dimensions is open.

\subsection{Cases outside the setting}
\label{gs:sec:outside}

We make no statement about the following cases. For scalar fields at $d_c=2$ the field has dimension zero, the marginal couplings are functions on the space of fields, and the normalisation $C_\phi$ of the propagator of section~\ref{sec:continuation} has a pole at $d=2$, so that regularity fails in the free theory. For nonlinear sigma models the metric on the space of the couplings of the target-space metric and the dilaton is indefinite~\cite{Tseytlin:2006ak}. For theories with chiral fermions or with couplings which violate parity we do not know of a continuation to the sphere $S^d$ with $d$ as a parameter in the sense of appendix~\ref{gs:sec:class}. A Chern--Simons level cannot be promoted to a local coupling and is not the source of a local operator~\cite{Nakayama:2013wda}, so that Chern--Simons theories with matter are perturbative but are not covered. Finally the properties require a local stress tensor and a metric which is a background field. This excludes models with long-range interactions and theories in which the metric is dynamical. In two dimensions a gradient formula which contains an antisymmetric tensor was proved for general theories in~\cite{Friedan:2009ik}, under conditions on the infrared behaviour.

For theories with chiral fermions or with couplings which violate parity in four dimensions it would be natural to use instead the flat-space route of~\cite{Reehorst:phi4}. There the gradient formula at $d=4$ is derived, under two hypotheses on perturbation theory, from a Ward identity for the two-point function of the marginal operators at constant couplings in flat space, whose part which is odd under the exchange of the two points determines the antisymmetric part of an anomaly coefficient, and no functional on $S^d$ for non-integer $d$ is needed. The extension of this route to general renormalisable theories in four dimensions, which is in preparation~\cite{Reehorst:general4d}, finds that the algebraic part of the proof and its infrared argument apply to scalars and fermions without parity, and that for chiral gauge theories the infrared argument applies and the algebraic part would apply as well. The existence of a scheme with local couplings which restores the symmetries that the regularisation of $\gamma_5$ breaks, namely the field rotations up to their anomaly and, for chiral gauge theories, the BRS symmetry, is left open there. For chiral gauge theories the treatment of the field rotations whose anomaly involves the gauge fields is missing as well, and with abelian factors the relations for the dual terms of the Weyl variation are not derived. As stated in~\cite{Reehorst:general4d}, we expect that this route gives the gradient formula for scalars and fermions without parity, and for chiral gauge theories without abelian factors which are invariant under a CP transformation that maps each simple factor to itself, if such a scheme exists, while it makes no statement away from $d=4$.

\section{Discussion}
\label{sec:discussion}

We have proved that perturbative renormalisation group flows in $d=d_c-\vep$ dimensions obey a gradient formula to all orders in the couplings and in $\vep$, with a potential and a symmetric metric which have no poles at $\vep=0$. The proof applies to every theory whose generating functional on the sphere $S^d$ has the six properties which we stated. From these properties we derived the Killing identity \eqref{gs:eq:K} and the homogeneous Ward identity \eqref{gsh:eq:hom} of a combination $W^\sharp$ of the functional and its anomaly. If the Weyl variation closes on the marginal couplings, the Killing identity of $W^\sharp$ is the gradient formula \eqref{gs:eq:main}, with a potential which is built from the free energy on the unit sphere and a symmetric metric which contains the first harmonic of the two-point function. The metric is regular at $\vep=0$, and so is the potential up to the logarithms of the gauge couplings. For odd $d_c$ the anomaly vanishes at vanishing relevant sources, and the Killing identity of $W$ itself is then already a gradient formula. Relevant sources of even dimension add terms to the Killing identity through their Weyl variation. In the divergence scheme, which we constructed for these sources, these terms take the form of a gradient and of a symmetric tensor contracted with $\hB^I$, and the gradient formula persists (Theorem~\ref{gr:cor:cond}). At the zeros of $\hB^I$ the potential is proportional to $\tilde F-\tilde F_0$, with the curvature couplings at the zeros of their beta functions~\cite{Fei:2015oha}.\footnote{In general this uses the extended form of the properties, defined after Property~\ref{gs:P:ward} in appendix~\ref{gs:sec:properties}.} If Property~\ref{gs:P:free} holds with a positive matrix $c_{IJ}$, the metric is positive definite to all orders in the couplings and in $\vep$, for the formula of $W^\sharp$ if $d_c$ is even and for that of $W$ if $d_c$ is odd.

A gradient formula with a positive metric makes the potential a monotone of the flow. It increases along $\hB^I$ to all orders in the couplings and in $\vep$, in the sense of formal power series, and this weaker statement is of interest in itself. In integer dimensions the monotones known beyond perturbation theory rest on unitarity~\cite{Zamolodchikov:1986gt,Komargodski:2011vj,Casini:2012ei}, which fails in non-integer dimension already for free scalar fields~\cite{Hogervorst:2015akt}. Our proof does not require unitarity, and the positivity comes from the two-point functions of the free fields. To our knowledge the earlier statements in non-integer dimension were of fixed order. At fixed points the potential is $\tilde F$ up to a positive factor and a constant. For two fixed points which are power series in $\vep$ with couplings of order $\vep$, joined by a trajectory of the flow at leading order, $\tilde F$ is therefore larger at the ultraviolet than at the infrared fixed point, in the sense that the difference is a power series in $\vep$ with a positive leading term (section~\ref{gs:sec:fixedgen}). This is the inequality of the generalised $F$-theorem of~\cite{Giombi:2014xxa,Fei:2015oha}, for which~\cite{Fei:2015oha} gave the leading term in conformal perturbation theory. A monotone excludes closed orbits of the flow of $\hB^I$, since along a closed orbit the potential would return to its value. The gradient formula gives this to every order of perturbation theory as an identity of formal power series, and for the flow at leading order, which is a vector field, we show it explicitly. The gradient formula implies more than the monotone. Since the metric is symmetric and positive definite, the eigenvalues of the matrix of derivatives of $\hB^I$ at fixed points are real, and the coefficients of the beta function obey relations which a monotone alone does not require.

For four theories we derived all six properties from Hypothesis~\ref{hyp:finite}, which states that the bare functional in flat space, for the cubic theory on conformally flat backgrounds, is made finite by counterterms of the form of minimal subtraction. They are multiscalar $\phi^4$ theory and scalars and Dirac fermions with Yukawa and quartic couplings in $d=4-\vep$, sextic scalars in $d=3-\vep$, and cubic scalars in $d=6-\vep$, for which the hypothesis is stated on conformally flat backgrounds. To this end we constructed the bare functional on the sphere for non-integer $d$ for scalar fields with polynomial interactions and for Dirac fermions. For these theories the gradient formula holds to all orders in perturbation theory, for the vector field $\hB^I$ which the counterterms of the hypothesis define. Its potential is regular, and its metric is regular, symmetric and positive definite to all orders in the couplings and in $\vep$. By Proposition~\ref{gs:prop:radius} the potential at a fixed point is proportional to $\tilde F-\tilde F_0$ with the curvature couplings at the zeros of their beta functions and the relevant sources at the zeros of their flows.

For multiscalar $\phi^4$ theory we also determined the antisymmetric part $E_{[IJ]}$ of a coefficient of the anomaly as a sum of the three types of terms of~\cite{Reehorst:phi4}, using the Killing identity and the relation between $E_{[IJ]}$ and the curl of the one-form $W_I$. It follows that in four dimensions there is a scheme in which $W_I$ vanishes, and in the absence of field rotations the tensor $T_{IJ}$ is then symmetric, which was conjectured in~\cite{Poole:2019kcm}.

Since $B^I$ does not depend on $\vep$, the gradient formula implies relations among the coefficients of the beta function which are not required in four dimensions. The computation of appendix~\ref{app:fiveloop} finds the first of them at five loops, where, modulo the relations in four dimensions, they span the same space as the four relations of~\cite{Pannell:2024sia}.\footnote{This statement, and the comparison with the values of $\tilde F$ in the literature, also rely on Identification~\ref{id:ms}, which no proof uses.}

For free gauge fields with a local coupling we constructed the functional on $S^d$ for non-integer $d$ through continued sums over the levels of the exact two-forms. All six properties follow from Hypothesis~\ref{hyp:finite}, which for gauge fields is stated on the sphere, and the gradient formula reduces to the single relation $\hat G^{\rm S}=-2d\,n_G/V$.

For interacting gauge theories the gauge-fixed functional on $S^d$ is not constructed, and the properties are assumptions. If the properties hold for an interacting gauge theory, the gradient formula holds in the coordinates $\log g_\iota$, by Corollary~\ref{gs:cor:gauge} for gauge fields of a semisimple group with fermions and by Proposition~\ref{gro:prop:gauge} with scalars in addition. Its potential contains a multiple of $\log g_\iota$ for each simple factor and cannot be normalised at the free theory. Under Property~\ref{gs:P:free} the metric is positive definite to all orders. By Proposition~\ref{grg:prop:gauge} this property holds in the logarithmic directions if and only if the coefficient $n_\iota$ of $\log g_\iota$ in the free energy is at $\vep=0$ the number of gauge fields of the factor. For several simple factors this equivalence assumes that the components of the free anomaly between different factors vanish. For several abelian factors the general results apply under the properties with the two-form sources of the field strengths, which we assume, with an extended notion of regularity in which the off-diagonal coordinates of the coupling matrix enter polynomially. Chiral couplings are not covered.

For fermions with quartic couplings near two dimensions we constructed the bare functional with the infinite set of four-fermion couplings, and under Properties~\ref{gs:P:functional} to~\ref{gs:P:sphere} the gradient formula holds with a regular symmetric metric. Under Hypothesis~\ref{hyp:finite} for this theory and for ${\rm tr}\,1>1$ the term of order zero of the metric is positive definite on the couplings without gamma matrices, which include the Gross--Neveu coupling. For ${\rm tr}\,1=2$\footnote{Here ${\rm tr}\,1$ is the trace of the unit matrix on the spinors, which is the number of components of a Dirac fermion. In dimensional regularisation it is a parameter, and ${\rm tr}\,1=2$ is its value for Dirac fermions in two dimensions.} it is positive semidefinite at $\vep=0$, with the evanescent couplings as its kernel. At $d=2+|\vep|$ with small $|\vep|\neq0$ it has infinitely many independent directions of each sign (Lemma~\ref{g2:lem:indef}). For a counterterm of the gauge field which is covariant and has no term of degree zero, Hypothesis~\ref{hyp:finite} for this theory excludes poles of $\hB^I$ of order two and higher at the lowest order at which a pole occurs. For ${\rm tr}\,1>1$ a simple pole in the evanescent directions is then absent if the counterterms obey the pole equations of minimal subtraction for one choice of the bare sources (Propositions~\ref{g2:prop:poles} and~\ref{g2:prop:poleeq}).

We should point out the limits of these results. The theorems for the four theories and for the free gauge field are conditional on Hypothesis~\ref{hyp:finite} for the respective theory, and for the other theories some or all of the properties are assumptions. The positivity of the metric and the monotone hold to all orders in the couplings and in $\vep$, as statements on formal power series, and we make no statement about the sum of the series. In the inequality for $\tilde F$ the trajectory which joins the two fixed points is that of the flow at leading order.

Several questions remain open. The first concerns $\phi^4$ theory itself, namely the relation between the metric of Theorem~\ref{gs:thm:phi4} at $\vep=0$ and the metric of~\cite{Reehorst:phi4}. Both give a gradient formula for $B^I$ in four dimensions, each under its own hypotheses, but the metric of~\cite{Reehorst:phi4} contains the flat two-point function paired with a test function at an arbitrary scale, while $\hat G_{IJ}$ contains the two-point function on the sphere. The solutions of the gradient equation are not unique. The potential can be replaced by a function of itself, shifted as in \eqref{eq:freedom}, or shifted by a symmetric tensor contracted with three factors of $\hB^I$, while a shift by $\hB^I\mathsf{k}_{IJ}\hB^J$ with an arbitrary symmetric tensor $\mathsf{k}_{IJ}$ is not allowed in general. The solution of this paper is distinguished by the fact that its potential is given along the flow, and not only at fixed points, by quantities of the theory on the sphere. For the four theories $W^\sharp$ at constant couplings and vanishing relevant sources is $(\vep+r\,\pr_r)W$ at the unit radius (Proposition~\ref{gs:prop:radius}). The potential of Theorem~\ref{gr:cor:cond} is $(d-r\,\pr_r)(\vep+r\,\pr_r)W/V$ at the unit radius, at constant couplings and vanishing relevant sources of the divergence scheme (Corollary~\ref{gs:cor:radiuspot}). It is thus fixed by the dependence of the free energy on the radius of the sphere, without reference to the gradient equation. It would be interesting to understand whether the solutions are related by transformations of this kind, and whether the solution from the sphere has useful properties which the others do not have.

For interacting gauge theories the main open question is the construction of the gauge-fixed functional on $S^d$ for non-integer $d$ with the properties. We expect that it requires the continuation in $d$ of multiple sums over the levels of the gauge fields and the ghosts, combined with the integrals of section~\ref{gs:sec:sphere} for the matter fields. For fermions near two dimensions the regularity of $\hB^I$ in the evanescent directions is open. In minimal subtraction the vector field which is computed from the counterterms of the couplings alone is regular if the higher poles of these counterterms are fixed by the simple pole, and these conditions are the pole equations. They depend on the choice of the bare sources, which is free up to rotations of the bare fields. For the four theories above Hypothesis~\ref{hyp:finite} implies that $\hB^I$ is regular, since the two-point function of the couplings is nondegenerate at $d=d_c$ and a pole of $\hB^I$ would therefore give a divergence of the functional. The pole equations then hold for a suitable choice of the bare sources. Near two dimensions the matrix of the free two-point functions vanishes on the evanescent couplings at $\vep=0$, and for a counterterm of the gauge field which is covariant and has no term of degree zero the hypothesis leaves a simple pole of $\hB^I$ in these directions (Proposition~\ref{g2:prop:poles}). For ${\rm tr}\,1>1$ the pole is absent if in addition the pole equations hold for one choice of the bare sources (Proposition~\ref{g2:prop:poleeq}). It remains to derive the pole equations from the hypothesis or to find such a pole. For the cubic theory the derivation of the curvature counterterms from the finiteness in flat space, as for the other theories, would require the conditions of solvability for the counterterms of dimension six. A derivation of the properties from Hypothesis~\ref{hyp:finite} for every theory whose bare functional is constructed, rather than case by case, is left for future work.

\section*{Acknowledgements}
I thank Slava Rychkov for motivating me to prove statements about gradient flow and for useful conversations. Claude was used extensively and played an instrumental role in every stage of this research. At King's College London I was funded by UK Research and Innovation (UKRI) under the UK government's Horizon Europe funding guarantee [grant number EP/X042618/1]. Currently, I am supported by a Royal Society University Research Fellowship URF\textbackslash R1\textbackslash 241371.
\appendix
\section{Precise statement of the properties}
\label{gs:app:properties}

In this appendix we state the properties of section~\ref{gs:sec:propinformal} in the technical form in which the proofs use them. After each property we explain what its conditions express and where the general proof uses them. Each property is a statement on the functional for sources from a restricted set of functions, the class of sources, since for non-integer $d$ an arbitrary function on the sphere is not admitted as a source. The informal statements of section~\ref{gs:sec:propinformal} do not need its definition, while the technical statements do, and section~\ref{gs:sec:class} therefore describes it first. Section~\ref{gs:sec:cases} states Hypothesis~\ref{hyp:finite}, from which we derive the properties for several theories, with the bare sources and counterterms of each theory, and section~\ref{gs:app:status} adds three points to the status of the properties of section~\ref{gs:sec:status}.

\subsection{Class of sources}
\label{gs:sec:class}

In integer dimension a source is an arbitrary smooth function on the sphere, and the computation of section~\ref{gs:sec:outline} uses variations of this kind. For non-integer $d$ an arbitrary function on the sphere has no meaning, and dimensional regularisation with local sources admits as sources only functions of a restricted kind, on which each term of the functional can be continued in $d$. We call the set of the functions which are admitted the class of sources. For $\phi^4$ theory it is the class of sources of section~\ref{sec:sphere}, which consists of the functions of a fixed integer number of the embedding coordinates. The restriction has to preserve what the computation uses, and in general we assume that a class of sources with the following properties is given. It contains the constants and the harmonics $\sigma$ of degree one of \eqref{gs:eq:sigma} for the vectors $b$ in a fixed subspace of the embedding space. It is closed under products, under the Laplacian and under the derivative $v^\mu\pr_\mu$ along the corresponding conformal Killing vectors. Its functions depend on a fixed finite number of the embedding coordinates. An expression built from them, from the metric of $S^d$ and from its curvature by products, covariant derivatives and contractions, with $d$ as a parameter, is a polynomial in $d$. Its coefficients are functions of the class of sources, or one-forms of the class for an expression with one free index. For every integer $d$ larger than the number of these coordinates it equals the value of the expression on $S^d$. The background gauge field is a one-form which is a combination of the differentials of the embedding coordinates along the fixed subspace, with coefficients which are functions of the class of sources. We assume that the one-forms of this kind contain the gradients of the functions of the class of sources and that the Lie derivative along $v^\mu$ maps them to one-forms of this kind, as is the case for $\phi^4$ theory. The variations $\sigma\,\hB^I$ and $v^\mu\pr_\mu g^I$ of the Ward identity, the variations of the gauge field and the local expressions in the Weyl variation and in the anomaly, which contain derivatives of the sources, are then defined for these sources. From here on the couplings, the relevant sources, their variations and all parameters of Weyl rescalings and gauge transformations are functions of the class of sources, the gauge field is a one-form of this kind, and the vector $b$ of a harmonic lies in the fixed subspace. Finally we assume three statements on the unit sphere with $d$ as a parameter. The volume $V$ and the integral $\int d\mu_d\,\sigma^2$ of a nonzero harmonic of degree one are analytic in $d$ and do not vanish at $d=d_c$. The integral of the divergence of a one-form of this kind vanishes, and for the one-form $f\,\nabla h-h\,\nabla f$ with functions $f$ and $h$ of the class this states that the Laplacian is symmetric with respect to $d\mu_d$ on the class of sources. The harmonics of degree one obey $\nabla_\mu\nabla_\nu\sigma=-\sigma\,\mathring\gamma_{\mu\nu}$. For the class of sources of section~\ref{sec:sphere}, which is the class of the $\phi^4$, the Yukawa, the sextic and the cubic theory, and for the polynomials in the parallel coordinates, which are the sources of the free gauge field, these assumptions are established in appendix~\ref{app:rules}.

\subsection{Properties and their conditions}
\label{gs:sec:properties}

\begin{property}[Functional]
\label{gs:P:functional}
The functional $W[g,a,\sM]$ on the unit sphere is a formal power series in $g^I-g_c^I$, $a_\mu$ and $\sM$ around constant couplings, for functions of the class of sources, where $g^I$ are coordinates on the space of couplings. The terms of first and of higher order of this series are regular. Its term of order zero, which is the value $\cW(g_c)$ of $W$ at constant couplings, is regular up to a function $\cW^{\log}$ which is stated with the instance of the property,
\begin{equation}
\cW = \cW^{\log}+\cW^{\rm reg} \, ,
\label{gs:eq:Wlog}
\end{equation}
with a regular function $\cW^{\rm reg}$. The coordinates are stated with the instance as well, and the first derivatives of $\cW^{\log}$ with respect to them are regular. The same holds for the expansion in $g^I-g_c^I$, $a_\mu$ and $\sM-\mu(g_c)$ around constant couplings together with the constant value $\mu(g_c)$ of the relevant sources, for every regular covariant function $\mu$ of the couplings with values in $V_\sM$ which vanishes at $g=0$. The functional is invariant under the background gauge transformations $\delta_\omega a_\mu=D_\mu\omega$ and $\delta_\omega g^I=-(\omega g)^I$ with the corresponding rotation of $\sM$.
\end{property}

The property has two instances. In instance~(a) the coordinates are the couplings themselves and $\cW^{\log}=0$, so that all terms of the expansion are regular. This is the instance of the couplings of section~\ref{gs:sec:theories}, that is of the couplings of scalar fields and of Yukawa couplings and quartic couplings of fermions. Sections~\ref{gs:sec:theories} to~\ref{gs:sec:class} are written for this instance.

Instance~(b) concerns gauge couplings. A gauge coupling, which we denote by $g_\iota$ with a label $\iota$, enters the action of the free gauge field through the factor $1/g_\iota^2$. The free gauge theory is therefore not a point of the space of couplings. As a consequence the free energy on the sphere is not a power series in $g_\iota$ at weak coupling. Since the coupling multiplies the free action and not an interaction, the free energy contains a term proportional to $\log g_\iota$, whose coefficient does not depend on the couplings. For Maxwell theory with the coupling $e$ on a sphere of radius $r_0$ the computation of~\cite{Giombi:2015haa} gives the free energy $-\tfrac12\log(e^2r_0^{\vep})$ up to a function of $d$, so that $W$, which is minus the free energy, contains $\log e$ with the coefficient one for every $d$. The derivative $g_\iota\,\pr/\pr g_\iota$ of the free energy inserts the operator $F^2/(2g_\iota^2)$, up to a sign. In terms of the canonically normalised field this is the operator $\tfrac12F^2$ of the free gauge field, whose correlation functions have a limit for $g_\iota\to0$. The derivative $\pr/\pr g_\iota$ inserts this operator divided by $g_\iota$, and it has no such limit. Likewise $g_\iota\,\pr/\pr g_\iota$ gives a constant on $\log g_\iota$. For a gauge coupling we therefore use $\log g_\iota$ in place of $g_\iota$ as the coordinate on the space of couplings, and we call it a logarithmic coordinate. For the other couplings the coordinates are the couplings themselves. The index $I$ runs over all coordinates, and we write $\iota$ for an index which belongs to a logarithmic coordinate. Thus $\pr_\iota=g_\iota\,\pr/\pr g_\iota$, the component $\hB^\iota$ is the beta function of $g_\iota$ divided by $g_\iota$, and a component of a tensor with an index $\iota$ refers to this coordinate. The diagonal component of the metric along a logarithmic coordinate is thus $g_\iota^2$ times its component in the coordinate $g_\iota$, and a component with one logarithmic index $\iota$ and one other index is $g_\iota$ times the corresponding component in the coordinate $g_\iota$. A regular metric has a limit at the free gauge theory, $g_\iota\to0$, while its component in the coordinate $g_\iota$ has a pole there if its term of order zero does not vanish. The local couplings and the expansion around constant couplings refer to these coordinates, and the group of field rotations leaves the gauge couplings invariant. For a theory with gauge fields $W$ is the generating functional of the gauge-invariant operators. Regularity is defined as before. A regular function is a power series in the couplings, including the gauge couplings $g_\iota$ themselves, whose coefficients are power series in $\vep$, and a tensor is regular if its components in these coordinates are regular functions. Note that for a regular function $f$ the derivative $\pr_\iota f=g_\iota\,\pr f/\pr g_\iota$ is again regular and has no term of order zero in the couplings. The function of \eqref{gs:eq:Wlog} is
\begin{equation}
\cW^{\log} = \sum_\iota n_\iota\log g_\iota \, ,
\label{gs:eq:Wloginst}
\end{equation}
where the sum runs over the logarithmic coordinates and the coefficients $n_\iota$ are power series in $\vep$ which do not depend on the couplings. This is the term proportional to $\log g_\iota$ described above, with $n=1$ for Maxwell theory. For the free gauge fields of a simple factor at constant coupling the same computation gives $n_\iota=n_G$, the number of its gauge fields. Proposition~\ref{grg:prop:gauge} of section~\ref{gs:sec:gauge} relates this value to the statement of Property~\ref{gs:P:free} in the logarithmic directions.

Property~\ref{gs:P:functional} treats the term of order zero of the expansion separately, because in instance~(b) this term is not regular. The function $\cW^{\log}$ of \eqref{gs:eq:Wloginst} is the general form of a function whose first derivatives with respect to the coordinates are regular, up to a regular function and an additive constant. Indeed, by Property~\ref{gs:P:sphere} the term of first order is $V^{-1}\pr_I\cW\int d\mu_d\,\delta g^I$, so that the derivatives $\pr_I\cW$ are regular. For a logarithmic coordinate and any other coordinate the second derivative $\pr_J\pr_\iota\cW=g_\iota\,\pr(\pr_J\cW)/\pr g_\iota$ has no term of order zero in $g_\iota$. The term of order zero in $g_\iota$ of $\pr_\iota\cW$ therefore does not depend on the other couplings. This term is $n_\iota$, and the derivatives of $\cW-\cW^{\log}$ are those of a power series in the couplings. The property thus requires in addition only that the additive constant is a power series in $\vep$. Note that the Ward identity of Property~\ref{gs:P:ward} contains the derivatives of $W$ only. The function $\cW^{\log}$ enters the statements of sections~\ref{gs:sec:identities} and~\ref{gs:sec:relevant} on the potential itself and on its value at the zeros of $\hB^I$, while the derivative of the potential and the metric are built from the terms of first and of second order.

\begin{property}[Weyl variation]
\label{gs:P:weyl}
There are a regular vector field $\hB^I(g)$ and a regular one-form $\hr_I(g)$ with values in the Lie algebra, both covariant, with $\hr_I\hB^I=0$. Up to terms of second order in the couplings, $\hB^I=-\vep\,k_I\,g^I$ on the coordinates which are not logarithmic and $\hB^\iota=-\vep\,k_\iota$ on the logarithmic coordinates. There is an expression $\delta^{\rm W}_\sigma\sM$ which is linear in $\sigma$ and is a polynomial in $\sM$, in the curvature and in the covariant derivatives of $\sigma$ and of the sources, built covariantly from $\gamma_{\mu\nu}$ with $d$ as a parameter, with regular covariant coefficients. By this we mean that it is built from the metric, the curvature and covariant derivatives, without the Levi-Civita tensor. Each of its terms has the dimension of $\sM$ at $d=d_c$, where $\sigma$ and $g^I$ count with dimension zero and $a_\mu$ with dimension one. With $\delta^{\rm W}_\sigma\gamma_{\mu\nu}=2\,\sigma\,\gamma_{\mu\nu}$, $\delta^{\rm W}_\sigma g^I=\sigma\,\hB^I$ and $\delta^{\rm W}_\sigma a_\mu=\sigma\,\hr_ID_\mu g^I$ the operators
\begin{equation}
\mathbb{W}_\sigma = \int d^dx\left[\sigma\left(2\,\gamma_{\mu\nu}\frac{\delta}{\delta\gamma_{\mu\nu}}+\hB^I\frac{\delta}{\delta g^I}
+\hr_ID_\mu g^I\cdot\frac{\delta}{\delta a_\mu}\right)+\delta^{\rm W}_\sigma\sM\cdot\frac{\delta}{\delta\sM}\right] ,
\label{gs:eq:weylop}
\end{equation}
which act on functionals of the metric and of the sources with the functional derivatives of \eqref{gs:eq:outweyl}, commute for all $\sigma$ and $\sigma'$. For a relevant source of dimension $D$ at $d=d_c$ the term of $\delta^{\rm W}_\sigma\sM$ which is linear in this source and contains no derivatives and no other relevant source is $-\sigma\,(D-\gamma_\sM)\,\sM$, with a regular covariant linear map $\gamma_\sM(g)$ which has no term of order zero in the couplings and in $\vep$. For each relevant source of even dimension $D<d_c$ the coefficients $\ell_{{\rm h}J}$ and $\vk_{\rm h}-\vk_{\rm c}$ of its Weyl variation on the unit sphere, defined in section~\ref{gs:sec:oneform}, vanish at $g=0$.
\end{property}

In instance~(b) the vector field $\hB^I$ has components along the logarithmic coordinates. In minimal subtraction they are
\begin{equation}
\hB^\iota = -\vep\,k_\iota+B^\iota \, ,
\label{gs:eq:Bhatlog}
\end{equation}
which takes the place of \eqref{gs:eq:Bhat}. Here $k_\iota$ is the weight of the gauge coupling, which has the dimension $\vep\,k_\iota$ in $d$ dimensions, with $k_\iota=\tfrac12$ at $d_c=4$, and $B^\iota$ is regular, does not depend on $\vep$ and has no terms of order zero and one in the couplings. The term of $\hB^\iota$ which is proportional to $\vep$ is thus a constant, and $\hB^\iota$ does not vanish for $g_\iota\to0$. Up to terms of second order the form \eqref{gs:eq:Bhatlog} is part of Property~\ref{gs:P:weyl}, and it is not part of Property~\ref{gs:P:functional}.

The statement on the lowest order of $\hB^I$ and the statement on $\gamma_\sM$ express that the couplings are marginal and that the relevant sources have their dimensions at $d=d_c$ in the free theory. They are used in section~\ref{gs:sec:relevant} and for the lowest order of the metric. The last statement concerns the relevant sources of even dimension, and it is empty if there are none. It states that at $g=0$ the Weyl variation of these sources on the sphere has no inhomogeneous terms of first order, as in the free theory. It carries Properties~\ref{gs:P:functional} to~\ref{gs:P:sphere} over to the divergence scheme of section~\ref{gs:sec:divscheme}, in which Theorem~\ref{gr:cor:cond} uses them. By Lemma~\ref{gr:lem:scheme} it implies that the constant parts $y_0$ of the redefinition to that scheme vanish at $g=0$, and for a single source of even dimension the whole redefinition vanishes at $g=0$, so that the properties in that scheme are statements on the original functional at a constant relevant source which vanishes at $g=0$, which Properties~\ref{gs:P:functional} and~\ref{gs:P:sphere} include, see the end of section~\ref{gs:sec:divscheme}. In instance~(a) the statement on $\vk_{\rm h}-\vk_{\rm c}$ suffices for this, since $\hB^K$ vanishes at $g=0$, and the statement on $\ell_{{\rm h}J}$ makes the whole redefinition vanish at $g=0$. In the divergence scheme the last statement holds identically, since there $\ell_{{\rm h}J}=0$ and $\vk_{\rm h}=\vk_{\rm c}$.

The commutator of two operators \eqref{gs:eq:weylop} vanishes identically on the metric and on $g^I$. On $a_\mu$ it equals $(\sigma\,\pr_\mu\sigma'-\sigma'\pr_\mu\sigma)\,\hr_I\hB^I$. Since the first factor does not vanish for two harmonics which are not proportional, the condition $\hr_I\hB^I=0$ is equivalent to the commutation on $a_\mu$. On $\sM$ the commutation is a condition on the coefficients of $\delta^{\rm W}_\sigma\sM$.

\begin{property}[Anomaly]
\label{gs:P:anomaly}
There is a functional $\cA[\sigma]$ of $\sigma$, of the metric and of the sources which is linear in $\sigma$ and local. It is the integral of a polynomial in the curvature and in the covariant derivatives of $\sigma$ and of the sources, built covariantly from $\gamma_{\mu\nu}$ with $d$ as a parameter in the sense of Property~\ref{gs:P:weyl}, that is without the Levi-Civita tensor, with regular coefficients. In instance~(b) of Property~\ref{gs:P:functional} the logarithmic coordinates enter $\cA[\sigma]$ only through regular functions of the couplings and through their covariant derivatives. It is invariant under background gauge transformations, and it satisfies $\mathbb{W}_\sigma\cA[\sigma']=\mathbb{W}_{\sigma'}\cA[\sigma]$. It is homogeneous of Weyl weight $-\vep$, that is $\cA[\sigma]$ is multiplied by $r_0^{-\vep}$ if $\gamma_{\mu\nu}$ is replaced by $r_0^2\,\gamma_{\mu\nu}$ with a constant $r_0$ and each component of $\sM$ is multiplied by $r_0$ to the power of minus its dimension at $d=d_c$.
\end{property}

The homogeneity in Property~\ref{gs:P:anomaly} states that the terms of the anomaly have dimension $d_c$ and that the metric carries the only scale. The critical dimension $d_c$ enters here. It also enters Property~\ref{gs:P:weyl} through the dimensions of the sources and Property~\ref{gs:P:free} through the free theory in $d_c$ dimensions.

The general proof uses Properties~\ref{gs:P:weyl} and~\ref{gs:P:anomaly} only on conformally flat backgrounds, that is on the unit sphere and its Weyl rescalings, and it holds if they are established in this restricted form, with the anomaly written with the Schouten tensor.

\begin{property}[Ward identity]
\label{gs:P:ward}
For $\sigma$ and $v^\mu$ as in \eqref{gs:eq:sigma} and for all sources the functional $W$ on the unit sphere satisfies
\begin{equation}
\int d\mu_d\left[\left(\sigma\hB^I-v^\mu\pr_\mu g^I\right)\frac{\delta W}{\delta g^I}+\left(\sigma\,\hr_ID_\mu g^I-\cL_va_\mu\right)\cdot\frac{\delta W}{\delta a_\mu}
+\left(\delta^{\rm W}_\sigma\sM-v^\mu\pr_\mu\sM\right)\cdot\frac{\delta W}{\delta\sM}\right] = \cA[\sigma] \, ,
\label{gs:eq:ward}
\end{equation}
where $\cL_v$ is the Lie derivative and the right-hand side is taken at the metric of the unit sphere.
\end{property}

Property~\ref{gs:P:ward} is the Ward identity of the conformal Killing vector $v^\mu$, with the background gauge field and the relevant sources included. Its origin is the one described in section~\ref{gs:sec:outline}. Let $W$ be defined on the backgrounds $\Omega^2\mathring\gamma_{\mu\nu}$ with $\mathbb{W}_\sigma W=\cA[\sigma]$, and let it be invariant under diffeomorphisms. Then the diffeomorphism generated by $-v$ and the Weyl rescaling with $\sigma$ give \eqref{gs:eq:ward} as they give \eqref{gs:eq:outward}. We refer to these assumptions as the extended form of the properties. In the extended form the relation $\mathbb{W}_\sigma\cA[\sigma']=\mathbb{W}_{\sigma'}\cA[\sigma]$ of Property~\ref{gs:P:anomaly} follows from Property~\ref{gs:P:weyl}. It also holds if $\cA[\sigma]=\mathbb{W}_\sigma F$ with a local functional $F$, as in minimal subtraction, where $F$ is the integral of the counterterms. In that case the functional $W$ is needed on the unit sphere only, and this is the form in which the properties are available for $\phi^4$ theory. We denote the operator on the left-hand side of \eqref{gs:eq:ward} by $\mathbb{K}_\sigma$, so that the Ward identity reads $\mathbb{K}_\sigma W=\cA[\sigma]$.

\begin{property}[Sphere]
\label{gs:P:sphere}
At constant couplings, and at constant couplings together with a constant relevant source $\mu(g_c)$ as in Property~\ref{gs:P:functional} the following two statements hold for the functional $W$, where $\sigma$ is a harmonic of degree one, $\delta g^J$ and $\delta\sM$ are variations in the class of sources, $z^I$ is a constant vector in the space of couplings and $\sM_{\rm c}$ is a constant element of $V_\sM$. First, the one-point functions of the operators $\cO_I$ and of the relevant operators are constant, and the one-point function of the current vanishes. Second, the second variation of $W$ in which one direction is $\sigma\,z^I$ or $\sigma\,\sM_{\rm c}$ and the other is $\delta g^J$ or $\delta\sM$ is a constant times $\int d\mu_d\,\sigma\,\delta g^J$ or $\int d\mu_d\,\sigma\,\delta\sM$. The constant is linear in $z^I$ or in $\sM_{\rm c}$ and does not depend on the vector $b$ of the harmonic.
\end{property}

The statements of Property~\ref{gs:P:sphere} are consequences of the invariance under ${\rm SO}(d+1)$ in integer dimension. For non-integer $d$ they have to be shown for the class of sources at hand, and for local functionals, in particular for the anomaly, they follow from Lemma~\ref{gsp:lem:local} of section~\ref{gsp:sec:facts}.

The constant relevant source $\mu(g)$ in Properties~\ref{gs:P:functional} and~\ref{gs:P:sphere} is needed because a redefinition of a relevant source by local terms, such as the one of section~\ref{gs:sec:divscheme}, maps the point at which the new source vanishes to a constant value of the old source at constant couplings. It also enters the sphere data of the combination $W^\sharp$ with relevant sources, through the flow of these sources at first order, see Lemma~\ref{gsh:lem:quant}. Both uses concern the relevant sources of even dimension, and without such sources the clauses of Properties~\ref{gs:P:functional} and~\ref{gs:P:sphere} with a constant relevant source are not used. In integer dimension the statements of Property~\ref{gs:P:sphere} at such a source hold because a constant source is invariant under the rotations of the sphere.

Property~\ref{gs:P:free} concerns the free theory in the critical dimension. For each coordinate let $\cO_I$ be its operator at order zero in the couplings in the free theory in $d_c$ dimensions, with the gauge fields canonically normalised, so that the operator of a logarithmic coordinate is $\tfrac12F^\iota_{\mu\nu}F^{\iota\,\mu\nu}$. Let $c_{IJ}$ be the coefficients of their two-point functions $\langle\cO_I(x)\,\cO_J(y)\rangle=c_{IJ}\,|x-y|^{-2d_c}$ in flat space in $d_c$ dimensions. The matrix $c_{IJ}$ is symmetric, and it vanishes between operators of different free fields. We set $a_{d_c}=2d_c\,\pi^{d_c/2}/(2^{d_c}(d_c/2)!)$ for even $d_c$, which is $\pi$, $\tfrac14\pi^2$ and $\tfrac1{32}\pi^3$ for $d_c=2$, $4$ and $6$, and $b_{d_c}=d_c\,2^{-d_c}\pi^{d_c/2}\,\Gamma(-\tfrac12d_c)$ for odd $d_c$, which is $\tfrac12\pi^2$ for $d_c=3$.

\begin{property}[Free limit]
\label{gs:P:free}
The following statements hold, in the divergence scheme of section~\ref{gs:sec:divscheme} if relevant sources are present. For even $d_c$ the term of order zero in the couplings and in $\vep$ of the tensor $\cK_{IJ}$ of section~\ref{gs:sec:quantities}, which collects the terms of the anomaly with two derivatives of the couplings, is $(-1)^{d_c/2+1}a_{d_c}\,c_{IJ}$. For odd $d_c$, where the anomaly vanishes at vanishing relevant sources, the term of order zero in the couplings and in $\vep$ of the tensor $\cS_{IJ}$, which is the first harmonic of the two-point function on the sphere $S^{d_c}$ minus the second derivative of $\cW/V$, is the value $b_{d_c}\,c_{IJ}$ of the free theory on $S^{d_c}$. For each relevant source of even dimension $D<d_c$ the one-point function $\sfm$ of the relevant operators at constant couplings, defined in section~\ref{gs:sec:oneform}, has no term of order zero in the couplings and in $\vep$.
\end{property}

The values of $\cK^{(0,0)}_{IJ}$ and $\cS^{(0,0)}_{IJ}$ in this property are those of the free theory in $d_c$ dimensions with local couplings, for even $d_c$ those of the terms of its Weyl anomaly with two derivatives of the couplings and for odd $d_c$ those of its two-point functions on $S^{d_c}$, as the two-point form below shows by Proposition~\ref{gs:prop:lowest} and Lemma~\ref{gre:lem:zero}.

Property~\ref{gs:P:free} is a statement at $\vep=0$ only, and it uses Properties~\ref{gs:P:functional} to~\ref{gs:P:anomaly} for the regularity in $\vep$. The values of $\cK^{(0,0)}$ and $\cS^{(0,0)}$ are part of the assumption. For scalar fields they follow from the two-point form below. They do not depend on the weights, and $\cK^{(0,0)}$ does not change under finite local terms without relevant sources, by Lemma~\ref{gs:lem:local}. A finite term which is linear in a relevant source changes $\cK^{(0,0)}$ and $\sfm^{(0,0)}$ together, and the term \eqref{gsh:eq:G00} of order zero of the metric does not change, see section~\ref{gs:sec:gradrel}. The values thus express the relation between the coefficient of the anomaly and the coefficient of the two-point function of the free theory in $d_c$ dimensions. For gauge fields the value in the logarithmic directions is derived for free gauge fields, see Proposition~\ref{ggs:prop:props}, and assumed for interacting gauge fields. A stronger statement, which we derive for $\phi^4$ theory, for scalars with Yukawa couplings, for sextic scalars and, after the choice of the bare source of $\phi_i$, for cubic scalars, concerns all $\vep$. We say that the functional has the two-point form if, in instance~(a), through second order in the couplings around $g=0$, to all orders in $\vep$, and for $a_\mu=0$ and $\sM=0$,
\begin{equation}
W = \frac12\,c_{IJ}(\vep)\int d\mu_d(X)\,d\mu_d(Y)\;\frac{g^I(X)\,g^J(Y)}{s(X,Y)^{\Delta_I+\Delta_J}}+F \, , \qquad \cA[\sigma] = \mathbb{W}_\sigma F \, .
\label{gs:eq:free}
\end{equation}
Here $s(X,Y)=|X-Y|$ is the chordal distance, the integral is defined by continuation in the dimensions $\Delta_I$, and $c_{IJ}(\vep)$ is a symmetric matrix which does not depend on the couplings, vanishes for $k_I\neq k_J$ and is the matrix $c_{IJ}$ above at $\vep=0$. The functional $F$ is local, built covariantly from $\gamma_{\mu\nu}$ and homogeneous of Weyl weight $-\vep$, the operator $\mathbb{W}_\sigma$ acts on it through the metric and the couplings, and in the first relation $F$ is taken at the metric of the unit sphere. If the functional has the two-point form in the divergence scheme, then the statements of Property~\ref{gs:P:free} other than the last hold, by Proposition~\ref{gs:prop:lowest} and Lemma~\ref{gre:lem:zero} of section~\ref{gs:sec:lowest}. For scalar fields with polynomial interactions the first term of \eqref{gs:eq:free} follows from the construction of the bare functional, see Lemma~\ref{gs:lem:free} of section~\ref{gs:sec:remains}. In minimal subtraction $F$ is the integral of the counterterms at $\sM=0$. With relevant sources the anomaly at $\sM=0$ contains in addition a term in which $\delta^{\rm W}_\sigma\sM$ acts on the counterterms which are linear in $\sM$, and the two-point form requires that this term vanishes through second order in the couplings.

Property~\ref{gs:P:free} is used only to show, in sections~\ref{gs:sec:lowest}, \ref{gs:sec:gradrel} and~\ref{gs:sec:gauge}, that the term of order zero in the couplings and in $\vep$ of the metric is a multiple of the matrix of the free two-point functions, and to fix the term of order zero in $\vep$ of $n_\iota$, see Proposition~\ref{grg:prop:gauge}. The metric is then nondegenerate and positive definite to all orders if this matrix is positive, and the gradient formula can be solved for the vector field.

The last statement of Property~\ref{gs:P:free} concerns the relevant sources of even dimension, and it is empty if there are none. By Theorem~\ref{gr:cor:cond} the term of order zero in the couplings and in $\vep$ of the metric with relevant sources is $\cK^{(0,0)}$ plus terms which contain $\sfm^{(0,0)}$, and the last statement ensures that the relevant sources do not change it. By the last statement of Property~\ref{gs:P:weyl} the properties hold in the divergence scheme if they hold in the scheme in which they are established, see section~\ref{gs:sec:divscheme}, and in the divergence scheme the last statement of Property~\ref{gs:P:free} is equivalent, by Lemma~\ref{gsh:lem:m}, to the statement that the coefficient $\sfa_\sM$ of the anomaly, also defined in section~\ref{gs:sec:oneform}, has no term of order zero in the couplings and in $\vep$. In dimensional regularisation with minimal subtraction the last statements of Properties~\ref{gs:P:weyl} and~\ref{gs:P:free} hold if the free functional is not renormalised at first order in the relevant sources on conformally flat backgrounds. For the $\phi^4$, Yukawa, sextic and cubic theories both statements are derived from Hypothesis~\ref{hyp:finite} for the theory, and the lowest order is a perturbative check which we carry out. Where the functional is not constructed, they are the assumption that it reproduces these values of the free theory.

\subsection{Hypothesis of finiteness}
\label{gs:sec:cases}

In this section we state Hypothesis~\ref{hyp:finite}, which section~\ref{gs:sec:casesmain} describes in words. We state it once, for a theory of the setting whose regularised bare functional is constructed in section~\ref{gs:sec:universal} or in appendices~\ref{gs:app:freegauge} and~\ref{gs:app:fourfermion}, that is for scalar fields with polynomial interactions, for scalar fields and Dirac fermions with Yukawa and quartic couplings, for Dirac fermions with four-fermion couplings and for free gauge fields with a local coupling, and we then list the bare sources and counterterms of the six theories for which we use it. The regularisation is dimensional regularisation in position space. In flat space the bare functional $W_0^{\rm fl}$ is defined for non-integer $d$ by analytic continuation of its integrals in $d$ and in the exponent of the propagator, see section~\ref{gs:sec:sphere}. It is defined for sources in the class of flat sources, which are the smooth functions with compact support of one coordinate $x_\parallel$ and of $|x_\perp|^2$, where $x_\perp$ denotes the other $d-1$ coordinates, and the one-forms built from such functions, see section~\ref{sec:chart}, where this class carries the parameter $D_0=1$. The bare sources of a theory are the couplings, the background gauge field and the relevant sources, each with local expressions of the same dimension at $d=d_c$ and the same tensor type added, which are built from the sources, their covariant derivatives, the field strength and the curvature and whose coefficients are functions of the couplings. Of these coefficients only the $L^I$ below are required to be covariant. The bare couplings are $g_0^I=g^I+L^I(g)$, the bare gauge field is $a_\mu+N_I\,D_\mu g^I$, and a bare relevant source is in addition multiplied by a linear map $Z$. The field-independent counterterms of a theory form the density $\cX^{\rm fl}$, a sum of local expressions of dimension $d_c$ in the sources and their derivatives, each multiplied by a function of the couplings, which we call its coefficient. The functions $L^I$, the other coefficients of the bare sources, with $Z-1$ in place of a factor $Z$, and the coefficients of $\cX^{\rm fl}$ are the counterterms. They are functions of the couplings and of $\vep$. As in minimal subtraction, we take each of them to consist of poles in $\vep$ only, without a finite part. We call a function of the couplings a pure pole if it is a power series in the couplings whose coefficients are polynomials in $1/\vep$ without a constant term, and we call it real if these polynomials have real coefficients. For instance, at one loop $L^I$ is quadratic in the couplings and proportional to $1/\vep$. The hypothesis requires in addition that $L^I$ has no terms of degree zero and one in the couplings, so that $g_0^I=g^I$ up to terms of second order. Without a term of degree zero the bare couplings vanish with the couplings, which is what makes the functional a formal power series around the free theory. The counterterms of minimal subtraction have this form, since they are real and covariant and their first term arises at one loop. We use the requirement to show that $\hB^I=-\vep\,g^I$ up to terms of second order, see section~\ref{gs:sec:phi4}. A function of the couplings with indices is covariant if it is unchanged by a rotation of the couplings with a constant element of the group of field rotations, including the reflections where the group contains them. Finite means that at each order in the sources the functional has no pole at $\vep=0$.

\begin{hypothesis}[Finiteness]
\label{hyp:finite}
For the theory under consideration there are real covariant pure poles $L^I$ without terms of degree zero and one in the couplings, and real pure poles for the other coefficients of its bare sources and for the coefficients of its field-independent counterterms $\cX^{\rm fl}$, such that the functional
\begin{equation}
W^{\rm fl}[g,a,\sM] = W^{\rm fl}_0[g_0,a_0,\sM_0]+\int d^dx\;\cX^{\rm fl}(g,a,\sM)
\label{eq:Wfl}
\end{equation}
with these bare sources and counterterms is finite at each order in the sources, where $\sM$ stands for the relevant sources listed below for the theory. This holds for all sources in the class of flat sources, and for all families of such sources which depend analytically on $\vep$, that is which for small $|\vep|$ are power series in $\vep$ whose coefficients are sources of this class with supports in a fixed compact set and which converge uniformly together with all their derivatives. For the cubic theory below the functional is taken on the conformally flat backgrounds $\Omega^2\delta_{\mu\nu}$, in the form \eqref{gcu:eq:W}, with the density $\cX$ in place of $\cX^{\rm fl}$, all of whose coefficients, including those of the counterterms which contain the curvature and are built with $d$ as a parameter, are real pure poles. For this theory the class of flat sources is the one with $D_0=3$, of the functions of three coordinates $x_\parallel$ and of $|x_\perp|^2$, which contains the class with $D_0=1$. The finiteness then holds for all $\log\Omega$ in the class of flat sources, and for families of sources and of $\log\Omega$ which depend analytically on $\vep$ in the same sense. For gauge fields the functional is taken, in place of flat space, on the unit sphere and its conformal rescalings $e^{2\varphi}\mathring\gamma_{\mu\nu}$, in the form $W=W_0[e^{2\varphi}\mathring\gamma;g_0,a_0]+\int d^dx\sqrt\gamma\;\cX$, where for the free gauge field $W_0$ is the bare functional \eqref{ggs:eq:curved}, and in place of the class of flat sources the sources and $\varphi$ are polynomials in the parallel coordinates. Each logarithmic coordinate has the bare coordinate $\log g_\iota+L_\iota(g)$ with a real pure pole $L_\iota$ without terms of degree zero and one in the couplings, and each monomial of the density $\cX$, which is built covariantly from the metric with $d$ as a parameter, contains a covariant derivative of a logarithmic coordinate or of another coupling, or the field strength of the background gauge field. The finiteness then holds for the terms of first and higher order in the variations of the sources around constant couplings and in $\varphi$, for all such sources and all such $\varphi$, in place of the class of flat sources and of the families above.
\end{hypothesis}

For multiscalar $\phi^4$ theory in $d=4-\vep$, with the couplings $g^I$ of the quartic operators, the background gauge field $a_\mu$ of the field rotations, the group ${\rm O}(N)$ of field rotations and the source $\sM$ of the operators $\phi_i\phi_j$, the bare sources are
\begin{equation}
g_0^I = g^I+L^I(g) \, , \qquad a_{0\mu} = a_\mu+N_I\,D_\mu g^I \, , \qquad \sM_0 = Z_\sM\left(\sM-\fd_I\,D^2g^I-\fe_{IJ}\,D^\mu g^ID_\mu g^J\right) ,
\label{gs:eq:barefl}
\end{equation}
where $N_I$ takes values in the Lie algebra of the field rotations, $\fd_I$ and $\fe_{IJ}$ take values in the space $V_\sM$ of the sources $\sM$, and $Z_\sM$ is a linear map of $V_\sM$. These are the bare sources of section~\ref{sec:setup} in flat space. The density $\cX^{\rm fl}$ collects the counterterms which do not contain the fields. It is a sum of local expressions in the sources and their derivatives, each multiplied by a function of the couplings, which we call its coefficient. In flat space these are, for instance, the terms $\tfrac12\,\fs_{IJ}(g)\,D^2g^ID^2g^J$ and $\tfrac12\,\sM\cdot\fm(g)\cdot\sM$ with the coefficients $\fs_{IJ}$ and $\fm$.\footnote{The complete list is \eqref{eq:X} of appendix~\ref{app:anomcoeff}. Its terms with the curvature vanish in flat space and do not enter the hypothesis.}

The Yukawa theory consists of scalar fields and Dirac fermions with Yukawa and quartic couplings in $d=4-\vep$. The bare functional with Dirac fermions is defined in section~\ref{gs:sec:dirac} without spinors on $S^d$, with the propagators of the fermions along the chords of the embedding space, and its flat functional $W_0^{\rm fl}$ is defined in the same way as for scalar fields. The couplings $g^I$ are the real coordinates of the symmetric tensors $\lambda_{abcd}$ of the quartic couplings $\tfrac1{4!}\lambda_{abcd}\phi_a\phi_b\phi_c\phi_d$ of $N$ real scalar fields and of the hermitian matrices $Y^a$ of the Yukawa couplings $\phi_a\bar\psi\,Y^a\psi$ of $N_f$ Dirac fermions, the group of field rotations is ${\rm O}(N)\times{\rm U}(N_f)$, and the bare sources have the form \eqref{gs:eq:barefl}, with $N_I$ in the Lie algebra of ${\rm O}(N)\times{\rm U}(N_f)$, and $\cX^{\rm fl}$ has the same form as for $\phi^4$ theory.

The sextic theory consists of $N$ real scalar fields in $d=3-\vep$ with the vertex $\tfrac1{6!}\,\lambda_{i_1\ldots i_6}\phi_{i_1}\cdots\phi_{i_6}$, whose couplings $g^I$ are the components of the symmetric tensor $\lambda_{i_1\ldots i_6}$, and the group of field rotations is ${\rm O}(N)$. There are two relevant sources, the source $\sU$ of the quartic operators $\tfrac1{4!}\,\sU_{ijkl}\,\phi_i\phi_j\phi_k\phi_l$, of dimension one at $d=3$, and the source $\sM$ of the operators $\phi_i\phi_j$. The bare sources are
\begin{gather}
g_0^I = g^I+L^I(g) \, , \qquad a_{0\mu} = a_\mu+N_I\,D_\mu g^I \, , \qquad \sU_0 = Z_\sU\,\sU \, , \notag\\
\sM_0 = Z_\sM\left(\sM-\fd_I\,D^2g^I-\fe_{IJ}\,D^\mu g^ID_\mu g^J-\fk\,R-\tfrac12\,\fh(\sU,\sU)\right) ,
\label{gsx:eq:bare}
\end{gather}
without the term with $\fk$ in flat space, where $Z_\sU$ and $Z_\sM$ are linear maps of the spaces of the sources $\sU$ and $\sM$ and $\fh$ is a symmetric bilinear map from pairs of sources $\sU$ to sources $\sM$. The density $\cX^{\rm fl}$ collects the counterterms which do not contain the fields, and each of its terms contains $\sU$, see appendix~\ref{gs:app:sextic}.

The cubic theory consists of $N$ real scalar fields in $d=6-\vep$ with the vertex $\tfrac16g_{ijk}\phi_i\phi_j\phi_k$, the background gauge field $a_\mu$, the source $\sM$ of $\phi_i\phi_j$ and the source $h_i$ of $\phi_i$, with the bare sources of appendix~\ref{gs:app:cubic}. The bare source of $\phi_i$ contains all local scalars of dimension four built from the curvature, the field strength, $\sM$ and the covariant derivatives of the couplings and of $\sM$, with coefficients $\mathfrak{p}_\alpha$, and the counterterms $\cX$ which do not contain the fields are all local scalars of dimension six, with the curvature counterterms built with $d$ as a parameter.
In the hypothesis its functional is taken on the conformally flat backgrounds $\Omega^2\delta_{\mu\nu}$,
\begin{equation}
W[\Omega^2\delta;g,a,\sM,h] = W_0\big[\Omega^2\delta;g_0,a_0,\sM_0,h_0\big]+\int d^dx\sqrt\gamma\;\cX \, ,
\label{gcu:eq:W}
\end{equation}
with $W_0$ on these backgrounds defined by the Weyl map from flat space, see appendix~\ref{gs:app:cubic}.

The four-fermion theory consists of $N_f$ Dirac fermions in $d=2+|\vep|$ with the couplings $g^I$ of the vertices $(\bar\psi\,\Gamma^{(n)}\psi)^2$ for all $n$, with all flavour structures, the group ${\rm U}(N_f)$ of field rotations and its background gauge field $a_\mu$, without the sources of the fermion bilinears. Its bare sources are $g^I+L^I$ and $a_\mu+N_I\,D_\mu g^I$ with $N_I$ in the Lie algebra of ${\rm U}(N_f)$, where each coefficient of $L^I$ has finitely many non-vanishing components, and $\cX^{\rm fl}=\tfrac12\,\fs_{IJ}\,D_\mu g^ID^\mu g^J$, see appendix~\ref{gs:app:fourfermion}.

The free gauge field consists of $n_G$ free gauge fields with a common local coupling $g$ in $d=4-\vep$, such as the free gauge fields of a simple factor, with the action $\int d^dx\sqrt\gamma\,F^aF^a/(4g^2)$. Its coordinate is the logarithmic coordinate $\tau=\log g$, and there are no further couplings, no background gauge field and no relevant sources. The bare functional on the unit sphere and its conformal rescalings is that of appendix~\ref{gs:app:freegauge}, defined by continued sums over the levels of the exact two-forms. In the hypothesis $L_\iota=0$, so that the bare coordinate is $\log g$ itself, and the coefficients of the counterterms are real pure poles which do not depend on $g$. The hypothesis is stated on the sphere, since the functional on $S^d$ contains $n_G\log g$, while the functional in flat space does not depend on the constant part of the coupling, see appendix~\ref{ggs:sec:chart}.

Section~\ref{sec:hypothesis} states the terms of the hypothesis in detail for $\phi^4$ theory. For the Yukawa theory the finiteness fixes the coefficients other than $L^I$ only up to rotations which leave all couplings invariant, such as those generated by the fermion number. Nevertheless every set of coefficients which makes the functional finite is covariant under the Lie algebra, there is a set which is covariant under ${\rm O}(N)\times{\rm U}(N_f)$, with $N_I$ in its Lie algebra and without term of degree zero, and all such sets give the same functional, $\hB^I$ and $\hr_I$, see appendix~\ref{gc:sec:cov}. For the sextic theory the term of $\sM_0$ which is quadratic in $\sU$ has a pole already in the free theory. For the cubic theory the hypothesis is used with the class of flat sources with $D_0=3$, which is stronger than the class with $D_0=1$ of the other theories, and Properties~\ref{gs:P:weyl} and~\ref{gs:P:anomaly} are obtained on conformally flat backgrounds, which is the form in which the general proof uses them. The properties hold for every $N$ for a set of bare sources obtained from the given one by shifts of the field $\phi_i$ by local terms with pure poles, regular redefinitions of its source, the average of $\cX$ over ${\rm O}(N)$ and the removal of the part of $\cX$ which contains neither sources nor couplings, with the same $\hB^I$ and $\hr_I$.

We add the following explanations of Table~\ref{gs:tab:cases}. In the columns of the gauge fields with fermions and of the fermions near two dimensions the relevant sources drop out, and the gradient formula is that of section~\ref{gs:sec:gradient}. In the other columns it is the gradient formula with relevant sources of Theorem~\ref{gr:cor:cond}, for the sextic theory in the form \eqref{gr:eq:odd}, and its metric is nondegenerate under Property~\ref{gs:P:free} if the matrix $c_{IJ}$ is invertible, which holds in the cases with scalar fields by Lemmas~\ref{gs:lem:free} and~\ref{gro:lem:c}, also with gauge fields, see Proposition~\ref{gro:prop:gauge} of section~\ref{gs:sec:gauge}. The last statements of Properties~\ref{gs:P:weyl} and~\ref{gs:P:free} are empty where the relevant sources drop out. Where the anomaly vanishes the coefficient $\sfa_\sM$ vanishes, so that the last statement of Property~\ref{gs:P:free} follows from Lemma~\ref{gsh:lem:m} in the divergence scheme and is not needed.

For $\phi^4$ theory the gradient formula follows from the general theorem of section~\ref{gs:sec:gradrel}. The construction of the bare functional on which this derivation rests applies to scalar fields with any polynomial interaction without derivatives, see section~\ref{gs:sec:sphere}, and to scalar fields with Dirac fermions and Yukawa couplings, see section~\ref{gs:sec:dirac}. For the sextic and the cubic scalar theory it gives the bare functional with the covariance and the gauge invariance which the properties require, see section~\ref{gs:sec:scalars}. For the sextic theory the derivation of the properties from Hypothesis~\ref{hyp:finite} for the sextic theory is carried out in appendix~\ref{gs:app:sextic}, and for the cubic theory in appendix~\ref{gs:app:cubic}. Gauge couplings enter through instance~(b) of Property~\ref{gs:P:functional}. For gauge fields with fermions the sources of the relevant operators drop out, and the properties give a gradient formula, which is Corollary~\ref{gs:cor:gauge} of section~\ref{gs:sec:gauge}. Properties~\ref{gs:P:functional} to~\ref{gs:P:sphere} fix the row sums of the metric of order zero by Proposition~\ref{gs:prop:log}, and with Property~\ref{gs:P:free} this metric is diagonal and positive definite. With scalar fields the mass source enters through section~\ref{gs:sec:relevant}. For scalars and fermions with Yukawa and quartic couplings without gauge fields all properties follow from Hypothesis~\ref{hyp:finite} for the Yukawa theory, and the gradient formula with a positive definite metric is Theorem~\ref{gf:cor:yukawa} of section~\ref{gs:sec:yukawa}. For gauge fields with scalars Property~\ref{gs:P:functional} holds in instance~(b), and the gradient formula follows under Properties~\ref{gs:P:functional} to~\ref{gs:P:sphere}, with a positive definite metric under Property~\ref{gs:P:free} as well, see section~\ref{gs:sec:gauge}. For cubic scalars in $d=6-\vep$ it follows from Hypothesis~\ref{hyp:finite} for the cubic theory, with a positive definite metric. For odd $d_c$ Property~\ref{gs:P:anomaly} implies that the anomaly vanishes at $\sM=0$, see Lemma~\ref{gs:lem:odd} of section~\ref{gs:sec:odd}.

\subsection{Further points on the status}
\label{gs:app:status}

The following points continue the list of section~\ref{gs:sec:status}.
\begin{enumerate}\setcounter{enumi}{3}
\item Property~\ref{gs:P:free} is a statement about the free theory in $d_c$ dimensions, and it needs nothing on $S^d$ for non-integer $d$ beyond Properties~\ref{gs:P:functional} to~\ref{gs:P:sphere}. For the $\phi^4$ theory, the Yukawa theory and the sextic theory we derive it from Hypothesis~\ref{hyp:finite}. The hypothesis implies that the bare sources agree with the sources up to terms of second order in the couplings, and the two-point form of the functional then follows by Lemma~\ref{gs:lem:free} of section~\ref{gs:sec:remains} and its extension to Dirac fermions. The statements of Property~\ref{gs:P:free} other than the last are a consequence of this form, and the last follows from the vanishing at $g=0$ of the one-point function $\sfm$ of the relevant operators in minimal subtraction, see section~\ref{gs:sec:phi4lowest}. For the cubic theory the same argument applies after the choice of the bare source of $\phi_i$ of section~\ref{gs:sec:cubic}. For gauge fields it states that $\cK^{(0,0)}_{\iota\iota}$ takes the value of free Maxwell theory in four dimensions with a local coupling, which is $-3n_G/\pi^2$ for a simple factor with $n_G$ gauge fields by Lemma~\ref{grg:lem:direct}, and that $\cK^{(0,0)}$ has no entries between different free fields. Given Properties~\ref{gs:P:functional} to~\ref{gs:P:sphere}, with relevant sources of even dimension the last statement of Property~\ref{gs:P:free}, and, for several simple factors, the vanishing at order zero of the entries between different factors, the diagonal entries in the logarithmic directions are equivalent to $n_\iota(0)=n_G$, the coefficient of $\log g_\iota$ of the free gauge fields at constant coupling in dimensional regularisation, by Proposition~\ref{grg:prop:gauge}. For interacting gauge fields Properties~\ref{gs:P:functional} to~\ref{gs:P:sphere}, these values of the functional on $S^d$ at $\vep=0$ and this vanishing are assumed. For free gauge fields with a local coupling Properties~\ref{gs:P:functional} to~\ref{gs:P:free} follow from Hypothesis~\ref{hyp:finite} for the free gauge field, with $n_\iota=n_G$ for every $\vep$, see Proposition~\ref{ggs:prop:props}.
\item For fermions with quartic couplings in non-integer $d$ the quartic operators built with the antisymmetrised products of any number of gamma matrices are independent, and renormalisation generates them in general. The space of couplings on which the Weyl variation closes then has infinitely many coordinates, and the properties are assumptions on a functional of infinitely many couplings, with regularity as a formal power series in all of them and the tensors as infinite matrices. The bare functional is constructed for all of these couplings in appendix~\ref{g2:sec:construction}. The kernel of the two-point functions of the couplings is degenerate at $d=2$ on the evanescent couplings, and the finiteness alone does not exclude a simple pole of $\hB^I$ in these directions. Given Hypothesis~\ref{hyp:finite}, ${\rm tr}\,1>1$ and the covariance of $N_I$ without term of degree zero, Property~\ref{gs:P:weyl} for the theory without the sources of the fermion bilinears is equivalent to the pole equations of minimal subtraction for the counterterms of the couplings for a suitable choice of the bare sources, see section~\ref{gs:sec:fermions}. We make no statement on whether a finite set of couplings suffices.
\item For an abelian factor of the gauge group the field strength is itself a gauge-invariant operator of dimension two. It is then a relevant operator, which enters the action in the combination $\mathsf{B}^{\mu\nu}F_{\mu\nu}$ with a source $\mathsf{B}_{\mu\nu}$ of dimension two. This source is an antisymmetric tensor, while the relevant sources of section~\ref{gs:sec:functional} are scalars. With local couplings it cannot be omitted, since antisymmetric tensors of dimension two such as $\pr_\mu g^I\pr_\nu g^J-\pr_\nu g^I\pr_\mu g^J$ can be formed from the sources. With one abelian factor this source does not contribute to the identities, by Lemma~\ref{gr4d:lem:twoform} of section~\ref{gs:sec:gauge}, and the results for gauge fields hold as stated if the properties hold with this source included. With several abelian factors the coupling is a positive matrix $h_{ab}$, and in the coordinates of the decomposition $h=D(1+N)(1+N)^{\rm T}D$ the results hold with the notion of regularity of Definition~\ref{gra:def:regular}, in which the coordinates $N_{ab}$ enter polynomially, under the properties with the two-form sources of the field strengths, which we assume as for the other gauge theories. The metric of order zero on the abelian block is $\tfrac{3}{4\pi^2}{\rm tr}(h^{-1}dh\,h^{-1}dh)$, which is positive definite.
\end{enumerate}

\section{Conventions and algebraic details}
\label{app:weyl}

In this appendix we collect the conventions and the algebra of the Weyl variation. We first compare our conventions with those of~\cite{Jack:2013sha} and of~\cite{Reehorst:phi4}. We give the rules for the Weyl variation and the reasons for the properties of the functional on the sphere under rotations. We then give the expressions for the anomaly coefficients in terms of the counterterms \eqref{eq:X}, from which the identity \eqref{eq:curlgeneral} of appendix~\ref{app:finite} follows, together with the relations and identities which appendix~\ref{gs:app:fourd} uses. They are the expressions of section~8 of~\cite{Jack:2013sha} with the dependence on the dimension kept and with the terms that involve the curvature included. We also give the scheme for the mass source and the statements on the mass operator on the sphere, and we prove Lemma~\ref{lem:mass} of appendix~\ref{gs:app:fourd} on the terms which the mass source adds to the Killing identity. We then derive the variation of the bare gauge field which is used in section~\ref{sec:weylvar}, in Lemmas~\ref{lem:rhoxi} and~\ref{lem:frame}. In the last subsection we complete the proof of Lemma~\ref{lem:closure} of appendix~\ref{gs:app:fourd}.

\subsection{Conventions}
\label{app:conventions}

We obtained the relation of our coefficients to those of~\cite{Jack:2013sha}, which we mark by a superscript, by comparing \eqref{eq:anomaly} with the definition of the anomaly in~\cite{Jack:2013sha}, whose generating functional corresponds to the logarithm of the partition function after the continuation to Euclidean signature. We find
\begin{equation}
\big(A_{\rm E},\,G_{IJ},\,W_I,\,S_{IJ}\big) = \frac{1}{16\pi^2}\,\big(A,\,G_{IJ},\,W_I,\,A_{IJ}\big)^{\rm JO} \, .
\label{eq:dictionary}
\end{equation}
In the same way $E_{IJ}$ and $R_{IJK}$ are $-2/(16\pi^2)$ times the coefficients $S_{IJ}$ and $T_{IJK}$ of~\cite{Jack:2013sha}. Note that the tensor which is computed with minimal subtraction in section~8 of~\cite{Jack:2013sha} is defined through the flat-space coefficients, with a factor $2$ in place of the factor $d-2$ of \eqref{eq:quad} below. Beyond the lowest order it therefore differs from $16\pi^2\,G_{IJ}$ by a change of scheme of the form $G_{IJ}\to G_{IJ}-(\cL_\beta\mathsf{g})_{IJ}$ with a symmetric $\mathsf{g}_{IJ}$. For the same reason the values of $E_{(IJ)}$, $R_{IJK}$ and $W_I$ in minimal subtraction differ from those of~\cite{Jack:2013sha} by finite terms of the form of a change of scheme, which for $W_I$ is $W_I\to W_I+4\,\pr_I\mathsf{a}-\mathsf{g}_{IJ}B^J$ with a function $\mathsf{a}$, see appendix~\ref{app:Wzero}. The antisymmetric part $E_{[IJ]}$ is not affected. Without field rotations a relation of the form \eqref{gs:eq:eulerF} of appendix~\ref{app:finite} is derived in section~3 of~\cite{Jack:1990eb}, where its parts of order $\vep^0$ and $\vep^1$ are eqs.~(3.16a) and~(3.16b). The factor $d-3$ does not appear there. We expect that this is a matter of the normalisation of the Euler density, which in~\cite{Jack:1990eb} contains a factor that depends on $d$.

Our notation follows~\cite{Reehorst:phi4}, with the following differences. The quantities $\hb^I$, $\hB^I$, $\hat M^K{}_I$ and $\hP_I{}^K$ contain the term $-\vep\,g^I$ or the dimension, and at $\vep=0$ they reduce to $\beta^I$, $B^I$, $\hat M^K{}_I$ and $\Psi_I{}^K$ of~\cite{Reehorst:phi4}. We call a quantity regular if it is a power series in the couplings whose coefficients are regular at $\vep=0$. The Weyl variation \eqref{eq:rhoxi} is the variation of~\cite{Reehorst:phi4} combined with a gauge transformation with the parameter $\sigma\,\xi$. It therefore contains $\hB^I$ and $\hr_I$ in place of $\beta^I$ and of the coefficient $\hr_I-\pr_I\xi$ of the variation of the gauge field before the gauge transformation, which is used in~\cite{Reehorst:phi4}, and it has no term with $\pr_\mu\sigma$. The variation $\delta_v$ of \eqref{eq:deltavS} has the opposite sign. The anomaly \eqref{eq:anomaly} fixes the normalisation of all coefficients, while~\cite{Reehorst:phi4} leaves a constant $\kappa$ free and uses $Q_I$ in the normalisation of~\cite{Jack:2013sha}. The function $A$ of~\cite{Reehorst:phi4}, which is $A_{\rm E}+W_IB^I$, is not used here, and the potential and the metric of Theorem~\ref{gs:thm:phi4} are denoted $\hat A$ and $\hat G_{IJ}$ in place of $A'$ and $G'_{IJ}$. The map $\gamma_\sM$ contains the rotation by $\xi$, see section~\ref{app:masscoeff}. The coefficient $L_I$ of the anomaly of appendix~\ref{app:finite} is $-1/(16\pi^2)$ times the coefficient $L_I$ of~\cite{Jack:2013sha}, which is the one used in~\cite{Reehorst:phi4}. Finally we write $\Xi_{IJ}$ for the tensors of section~\ref{sec:removable}, since $\Omega$ denotes the conformal factor. The relation between the curl of $W_I$ and the flat-space coefficients which is used in~\cite{Reehorst:phi4} is the identity \eqref{eq:curlgeneral} of appendix~\ref{app:finite} at $\vep=0$, up to the normalisations of the coefficients, which differ between the two papers by constant factors. In a scheme with $\theta_I=0$ its last term is absent, and at $\vep=0$ it is then the relation \eqref{eq:curl} of appendix~\ref{gs:app:fourd}.

\subsection{Rules}
\label{app:rules}

Under the Weyl rescaling $\delta\gamma_{\mu\nu}=2\sigma\gamma_{\mu\nu}$ in $d$ dimensions we have
\begin{align}
\delta\sqrt\gamma &= d\,\sigma\sqrt\gamma \, , \qquad \delta R = -2\,\sigma R-2(d-1)\,\nabla^2\sigma \, , \notag\\
\delta\cE_4 &= -4\,\sigma\,\cE_4+8(d-3)\,\cG^{\mu\nu}\nabla_\mu\nabla_\nu\sigma \, , \qquad
\delta\cG^{\mu\nu} = -4\,\sigma\,\cG^{\mu\nu}-(d-2)\left(\nabla^\mu\nabla^\nu-\gamma^{\mu\nu}\nabla^2\right)\sigma \, , \notag\\
\delta\,\nabla^2F &= -2\,\sigma\,\nabla^2F+(d-2)\,\pr_\mu\sigma\,\pr^\mu F+\nabla^2\delta F \, ,
\label{eq:weylrules}
\end{align}
where $F$ is a scalar and the rule for $\cE_4$ holds up to terms with the Weyl tensor, which vanish on conformally flat backgrounds. The sources change by \eqref{eq:rhoxi} and \eqref{eq:deltaM}. For the covariant derivatives of the couplings this gives
\begin{align}
\delta\,D_\mu g^I &= \pr_\mu\sigma\,\hB^I+\sigma\,D_\mu g^J\hat M^I{}_J \, , \notag\\
\delta\,D^2g^I &= -2\,\sigma\,D^2g^I+\nabla^2\sigma\,\hB^I+2\,\pr_\mu\sigma\,D^\mu g^J\,\hP_J{}^I
+\sigma\,D^2g^J\hat M^I{}_J+\sigma\,D_\mu g^JD^\mu g^K\,\hat\Omega_{JK}{}^I \, ,
\label{eq:varDg}
\end{align}
with $\hat M^I{}_J$ as in section~\ref{gs:sec:quantities} and $\hP_J{}^I$ as after \eqref{gs:eq:Mrel}, and with $\hat\Omega_{JK}{}^I=\pr_J\pr_K\hB^I+(\pr_{(J}\hr_{K)}g)^I+2\,(\hr_{(J})^I{}_{K)}$, where in the last term $\hr_J$ acts on the index $K$ as on a coupling. For $d=4$ these are the variations given in~\cite{Jack:2013sha}, without the rotation terms with $\upsilon$ there, which the gauge transformation removes. The variation of the field strength is $\delta f_{\mu\nu}=\pr_\mu\sigma\,\hr_ID_\nu g^I-\pr_\nu\sigma\,\hr_ID_\mu g^I$ up to terms proportional to $\sigma$.

On the sphere of section~\ref{sec:sphere} these rules hold with $d$ as a parameter. For a function $F$ of $X_\parallel$ we have
\begin{equation}
\nabla^2F = \left(\delta_{AB}-X_AX_B\right)\pr_A\pr_BF-d\,X_A\pr_AF \, ,
\label{eq:laplace}
\end{equation}
in which $d$ is a parameter. This operator is of divergence form with respect to the measure $d^kX_\parallel\,\rho^{m-2}$, so that we can integrate by parts for ${\rm Re}\,m>0$, where the boundary term vanishes, and for the other values of $d$ by continuation, since the integrals of functions of the class over one point are entire in $d$. Under the maps $f$ of \eqref{eq:boost} it obeys the last rule of \eqref{eq:weylrules} with $d$ as a parameter, and $\nabla^2\sigma=-d\,\sigma$ for the function $\sigma$ of \eqref{eq:ckv}.\footnote{We verified the divergence form of \eqref{eq:laplace} and its Weyl transformation rule with computer algebra, with $d$ as a symbol.} The curvature tensors of the unit sphere are polynomials in $d$, and the formulas for the Weyl transformation of the curvature and of the Laplacian are polynomial identities in $d$ which hold for all integers. These rules establish the statements on the unit sphere which appendix~\ref{gs:sec:class} assumes for the class of sources. By the divergence form of \eqref{eq:laplace} the Laplacian is symmetric with respect to $d\mu_d$, and the integral of the divergence of a one-form $a_A\,dX^A$ vanishes, since $\nabla_\mu(a\,\nabla^\mu X^A)=\tfrac12[\nabla^2(aX^A)+a\,\nabla^2X^A-X^A\nabla^2a]$ for a function $a$ of $X_\parallel$. The relation $\nabla_\mu\nabla_\nu\sigma=-\sigma\,\gamma_{\mu\nu}$ of \eqref{eq:ckv} is likewise a polynomial identity in $d$ which holds for all integers. The volume $V=\vol_{d+1}$ and $\int d\mu_d\,\sigma^2=b^2\,\vol_{d+1}/(d+1)$ are entire functions of $d$ which are positive for real $d>-1$ and $b\neq0$.

With these rules the first variation of \eqref{eq:anomaly} at constant couplings on the unit sphere is obtained term by term. We use $\int d\mu_d\,\pr_\mu\sigma\,\pr^\mu F=d\int d\mu_d\,\sigma\,F$ and $\int d\mu_d\,\sigma\,\nabla^2F=-d\int d\mu_d\,\sigma\,F$ for a function $F$ of the class. The terms with $p$, $q$ and $r$ give $\pr_J(\chi_d\,p+R_d^2\,q-d\,R_d\,r)$ times $\int d\mu_d\,\sigma\,\delta g^J$. The term with $W_I$ gives $2e_d\,W_J$, the terms with $e_I$ and $h_I$ give $-d\,R_d\,e_J$ and $d\,R_d\,h_J$, and the term with $u_I$ gives $d^2u_J$, each times the same integral. The other terms of \eqref{eq:anomaly} are of second order in the derivatives of the couplings. In appendix~\ref{gs:app:fourd} these terms form the right-hand side of \eqref{eq:sphere}.

\subsection{Rotations and the flat chart}
\label{app:rotations}

In this subsection we give the reasons for the properties of the functional under rotations which are stated in section~\ref{sec:sphere}, the Jacobian of the chart of section~\ref{sec:chart} and the definition of the bare functional on conformally flat backgrounds. The three properties are that for constant couplings the one-point function is constant, that the two-point function depends only on the angle between the points, and that the function $\sigma$ of \eqref{eq:ckv} is an eigenfunction of the two-point function. In integer dimension they follow from the rotations of the sphere. The rotations of the parallel directions act on our sources, but the rotations which mix the parallel and the transverse directions do not. The three properties follow instead from a property of \eqref{eq:gram}. If one point $Y$ is held fixed, the integral over the remaining $n-1$ points is again of the form \eqref{eq:gram}, with the direction of $Y_\perp$ as an additional parallel direction and with $m-1$ transverse components. For two fixed points we apply this twice. Its integrand depends on $Y$, on the vector $b$ of $\sigma$ and on the remaining points only through their scalar products, and its measure is invariant under the orthogonal transformations of the enlarged parallel space, which act on $b$ and on $Y$. At constant sources the integral with the factor $\sigma=b\cdot X$ at one of the remaining points is therefore linear in $b$ and invariant under the simultaneous orthogonal transformations of $b$ and $Y$, hence a constant times $b\cdot Y=\sigma(Y)$, and the constant depends neither on $b$ nor on $Y$. This is the eigenfunction property. The same argument without $b$ shows that the one-point function at constant sources is constant, and with two fixed points it shows that the two-point function depends only on the angle between them.\footnote{We confirmed the eigenfunction property \eqref{eq:Lambda} numerically for points $Y$ with a transverse component, at $d=2.7$, $3.3$ and $3.9$.} Here the two-point function with one point fixed in \eqref{eq:Lambda} is defined by the property just described. For non-integer $d$ there is no function of two points, and the precise form of \eqref{eq:Lambda} is a statement on the second variation of the functional at constant couplings $g^I$,
\begin{equation}
\frac{\pr^2}{\pr t_1\,\pr t_2}\,W\big[g+t_1\,\sigma\,\delta g+t_2\,\delta g'\big]\Big|_{t_1=t_2=0} = \Lambda_{IJ}\int d\mu_d\;\sigma\,\delta g^I\,\delta g'^J \, ,
\label{eq:Lambdadef}
\end{equation}
for a constant $\delta g^I$ and a variation $\delta g'^J$ in the class. This defines the matrix $\Lambda_{IJ}$. The same property implies that the one-point function of the current vanishes at constant couplings. The two lines of the vertex of the current end at the other vertices, and the integral over these is a function of the two ends of the lines which depends only on the angle between them and which is symmetric under their exchange together with the exchange of the field indices. Its antisymmetrised derivative at coinciding ends therefore vanishes.

These properties are derived for integrals with scalar sources in the region in which the integrals converge. Two further steps lead to \eqref{eq:Lambdadef} for the renormalised functional on the line \eqref{eq:line}. First, for a term of the bare functional with scalar sources the left-hand side of \eqref{eq:Lambdadef} is a meromorphic function of $d$ and $\Dp$. For $\delta g'^J=\sigma\,\delta g''^J$ with a constant $\delta g''^J$ the integral on the right-hand side contains $\int d\mu_d\,\sigma^2=b^2\,\vol_{d+1}/(d+1)$, which is entire and does not vanish identically. The coefficient of the term is therefore meromorphic, and the property extends from the region of convergence to all values of $d$ and $\Dp$, and to the line away from its poles. Second, the second variation of the renormalised functional at constant couplings also contains the bare gauge field, the bare mass source and the counterterms. At first order in the variation $\delta g^I$ of the couplings the bare gauge field is $d\omega$ with $\omega=N_I\,\delta g^I$, where $N_I$ is constant. We differentiate the identity \eqref{eq:gaugeS} of Proposition~\ref{prop:gauge} once with respect to the bare sources, at constant couplings and vanishing gauge field. This expresses a second variation of $W_0^S$ in which one direction is the gauge field $d\omega$ through the second variation in which this direction is replaced by the scalar sources $(\omega g_0)$ and $[\omega,\sM_0]$, and through one-point functions at constant sources. The latter are constant for the scalar operators and vanish for the current. All terms with the bare gauge field are thus reduced to terms with scalar sources. In the variation $\sigma\,\delta g^I$ the terms of first order of the bare couplings and of the bare mass source are $\sigma$ times constants, where we use $\nabla^2\sigma=-d\,\sigma$ for the term with $\fd_I$. The terms of second order of the bare sources multiply one-point functions, and the second variation of the integral of the counterterms is local. After integrations by parts, which are allowed by the divergence form of \eqref{eq:laplace}, and with $\nabla^2\sigma=-d\,\sigma$ all terms are constants times $\int d\mu_d\,\sigma\,\delta g^I\delta g'^J$. Hence \eqref{eq:Lambdadef} holds for the renormalised functional, with a matrix $\Lambda_{IJ}$ which contains these contact terms.

For the stereographic projection of section~\ref{sec:chart} we have $\rho=\Omega\,|x_\perp|$ with $\Omega=2/(1+x^2)$, and the Jacobian of the map $(x_\parallel,|x_\perp|)\to X_\parallel$ is $\Omega^{D_0+2}\,|x_\perp|$. Since $D_0+m=d$, this gives the relation between the measures which is stated there. The definition \eqref{eq:gram} does not depend on $D_0$. A source of the class is also a source of the class with $D_0+1$, and one transverse component can then be integrated explicitly, as in eq.~(4.1.17) of~\cite{Collins:1984xc}. Note that in~\cite{Collins:1984xc} the integral over one vector with fixed external vectors is defined, and \eqref{eq:gram} is the iteration of this definition.

In the chart the conformal maps \eqref{eq:boost} are conformal maps of flat space, which we use in the proof of Lemma~\ref{lem:regvar}. Let $D_0=1$, and let $t$ be the unit vector of the parallel direction of the chart. For the boost along $P$ we find $x'=e^{\zeta}x$ from \eqref{eq:boost}, which is a dilatation. For the boost along the other parallel direction we find to first order in $\zeta$
\begin{equation}
x'^\mu = x^\mu+\zeta\left[\tfrac12\,t^\mu+\tfrac12\,x^2\,t^\mu-(t\!\cdot\!x)\,x^\mu\right] ,
\label{eq:chartboost}
\end{equation}
which is a translation in the parallel direction combined with the special conformal transformation with the parameter $b=\tfrac12\,t$. In both cases $x'=x-\zeta\,v$ with a conformal Killing vector $v$ of flat space, for which $\pr_\mu v_\nu+\pr_\nu v_\mu=2\,\sigma^{\rm fl}\delta_{\mu\nu}$. Here $\sigma^{\rm fl}=\sigma-v^\mu\pr_\mu\log\Omega$ with the function $\sigma$ of \eqref{eq:ckv}, which gives $\sigma^{\rm fl}=-1$ for the first boost and $\sigma^{\rm fl}=t\!\cdot\!x$ for the second. By \eqref{eq:chart} the flat sources are $\Omega^\vep g_0$, $a_0$ and $\Omega^2\sM_0$. The variation \eqref{eq:barevar} of the sources on the sphere is therefore the variation of the flat sources of the same form, with $\sigma^{\rm fl}$ in place of $\sigma$ and with the vector field $v$ in the chart. Let now the sources on the sphere vanish near $P$. For small $|\zeta|$ the sources on the left-hand side of \eqref{eq:covS} then vanish near $P$ as well, so that \eqref{eq:chart} applies to both sides. The first-order form of \eqref{eq:covS}, which is justified at the end of appendix~\ref{app:conf}, thus gives the invariance of $W_0^{\rm fl}$ under these two variations of flat sources. The flat functional is invariant under translations of $x_\parallel$ by its definition. Subtracting the translation we obtain the invariance under the dilatation, with a constant $\sigma^{\rm fl}$, and under the special conformal transformation with a parameter $b$ in the parallel direction, with $\sigma^{\rm fl}=2\,b\!\cdot\!x$.

In section~\ref{sec:chart} we also use backgrounds which are conformally flat, $\gamma_{\mu\nu}=\Omega(x)^2\delta_{\mu\nu}$ on $\mathbb{R}^d$ with a positive function $\Omega$ of $x_\parallel$ and $|x_\perp|^2$. For the action \eqref{eq:action} with bare sources the change of variables $\phi_0\to\Omega^{-(d-2)/2}\phi_0$ gives
\begin{equation}
W_0\big[\Omega^2\delta;\,g_0,a_0,\sM_0\big] = W_0^{\rm fl}\big[\Omega^{\vep}g_0,\,a_0,\,\Omega^{2}\sM_0\big] \, .
\label{eq:defcurved}
\end{equation}
This is an identity of the classical actions for every $d$, and in dimensional regularisation the change of variables has no Jacobian. We take \eqref{eq:defcurved} as the definition of the left-hand side for non-integer $d$. The invariance \eqref{eq:bareweyl} under Weyl rescalings then holds identically. By \eqref{eq:chart} the functional on the sphere obeys \eqref{eq:defcurved} for sources which vanish near the pole of the chart. The counterterms, which we list in \eqref{eq:X} below, and the bare mass source on these backgrounds contain the curvature and the Laplacian. On the backgrounds $\Omega^2\delta_{\mu\nu}$ the curvature tensors and the Laplacian on functions of $x_\parallel$ and $|x_\perp|^2$ are the expressions of integer dimension, which are polynomials in $d$, with $d$ as a parameter. For example, $\nabla^2F=\Omega^{-2}\left[\pr^2F+(d-2)\,\pr_\mu\log\Omega\,\pr^\mu F\right]$, where the flat Laplacian $\pr^2$ contains the term $(d-D_0-1)\,|x_\perp|^{-1}$ times the derivative with respect to $|x_\perp|$. The integral of a density is taken with the measure $\Omega^d$ times the flat measure of section~\ref{sec:chart}, and the contraction of two gradients is $\Omega^{-2}$ times the sum of the products of the derivatives with respect to $x_\parallel$ and with respect to $|x_\perp|$. With these definitions $\mathbb{W}_\sigma\!\int\!\sqrt\gamma\,\cX=\cA[\sigma]$ is an identity for every $d$ up to total derivatives. These integrate to zero for sources of the class, because the flat Laplacian is of divergence form with respect to the flat measure. For $\Omega=2/(1+x^2)$ these expressions agree in the chart with those of the sphere in appendix~\ref{app:rules}, because the agreement is a polynomial identity in $d$ which holds for all integers.

\subsection{Anomaly coefficients}
\label{app:anomcoeff}

The counterterms which do not depend on the fields are
\begin{align}
\cX ={}& \fa\,\cE_4+\fb\,R^2+\fg_{IJ}\,\cG^{\mu\nu}D_\mu g^ID_\nu g^J
+R\left(\fc_I\,D^2g^I+\ff_{IJ}\,D_\mu g^ID^\mu g^J\right)+\tfrac12\,\fs_{IJ}\,D^2g^ID^2g^J \notag\\
&+\fr_{IJK}\,D^2g^I\,D_\mu g^JD^\mu g^K+\tfrac12\,\fq_{IJKL}\,D_\mu g^ID^\mu g^J\,D_\nu g^KD^\nu g^L
+\fn\cdot\sM\,R+\fj_I\cdot\sM\,D^2g^I \notag\\
&+\fl_{IJ}\cdot\sM\,D_\mu g^ID^\mu g^J+\tfrac12\,\sM\cdot\fm\cdot\sM
+f^{\mu\nu}\cdot\fp_{IJ}\,D_\mu g^ID_\nu g^J+\tfrac14\,f^{\mu\nu}\cdot\fy\cdot f_{\mu\nu} \, ,
\label{eq:X}
\end{align}
where $\cE_4$ is the Euler density and $\cG^{\mu\nu}=R^{\mu\nu}-\tfrac12\gamma^{\mu\nu}R$ is the Einstein tensor. We omitted the square of the Weyl tensor, which vanishes together with its first variation on the conformally flat backgrounds that we use. The coefficients are power series in the couplings. The tensors $\fg_{IJ}$, $\ff_{IJ}$, $\fs_{IJ}$, $\fl_{IJ}$ and $\fe_{IJ}$ are symmetric, $\fr_{IJK}$ is symmetric in its last two indices, and $\fq_{IJKL}$ is symmetric in each pair of indices and under the exchange of the two pairs. Moreover $\fm$ is a symmetric bilinear form on $V_\sM$, $\fp_{IJ}$ is antisymmetric, and $\fy$ is a symmetric bilinear form on the Lie algebra.

We insert these rules into $\mathbb{W}_\sigma\int\!\sqrt\gamma\,\cX$ and integrate by parts. The variation of each term of \eqref{eq:X} is given in appendix~\ref{app:variation}. For example, the term $\fa\,\cE_4$ gives $-(\vep-\hB\!\cdot\!\pr)\fa\;\sigma\,\cE_4$ and, after one integration by parts and with $\nabla_\mu\cG^{\mu\nu}=0$, the term $-8(d-3)\,\pr_I\fa\;\pr_\mu\sigma\,\cG^{\mu\nu}\pr_\nu g^I$. The term with $\fg_{IJ}$ gives $2\,\fg_{IJ}\hB^J\,\pr_\mu\sigma\,\cG^{\mu\nu}\pr_\nu g^I$ through the variation of $D_\mu g^I$. Comparing with \eqref{eq:anomaly} we obtain the expressions for $p$ and $W_I$ in \eqref{eq:lin} below. In the same way we find, at $\sM=0$ and without the contributions of the counterterms which are linear in $\sM$, for the coefficients of the terms with at most one factor of $Dg$
\begin{align}
p &= -\left(\vep-\hB\!\cdot\!\pr\right)\fa \, , \qquad q = -\left(\vep-\hB\!\cdot\!\pr\right)\fb \, , \qquad
r = \fc_I\hB^I-4(d-1)\,\fb \, , \notag\\
W_I &= 4(d-3)\,\pr_I\fa-\fg_{IJ}\hB^J \, , \qquad
u_I = -2(d-1)\,\fc_I+\fs_{IJ}\hB^J \, , \notag\\
e_I &= -\big[(\vep-\tL)\,\fc\big]_I \, , \qquad
h_I = 2\,\hP_I{}^J\fc_J+2\,\ff_{IJ}\hB^J \, ,
\label{eq:lin}
\end{align}
and for the coefficients of the terms with two and three factors
\begin{align}
S_{IJ} &= -\big[(\vep-\tL)\,\fs\big]_{IJ} \, , \qquad
G_{IJ} = \big[(\vep-\tL)\,\fg\big]_{IJ} \, , \notag\\
E_{IJ} &= 2\,\hP_I{}^K\fs_{KJ}+2\,\fr_{JIK}\hB^K+(d-2)\,\fg_{IJ} \, , \notag\\
\nu_{IJ} &= -2(d-1)\,\ff_{IJ}+\tfrac12(d-2)\,\fg_{IJ}+\fr_{KIJ}\hB^K \, , \notag\\
R_{IJK} &= 2\,\hP_I{}^L\fr_{LJK}+2\,\fq_{ILJK}\hB^L+(d-2)\,\Gamma^{(\fg)}_{IJK}+\hr_J\cdot\fp_{IK}+\hr_K\cdot\fp_{IJ} \, ,
\label{eq:quad}
\end{align}
where $\Gamma^{(\fg)}_{IJK}=\tfrac12(\pr_J\fg_{IK}+\pr_K\fg_{IJ}-\pr_I\fg_{JK})$. The term $(d-2)\,\fg_{IJ}$ in $E_{IJ}$ and the corresponding terms in $\nu_{IJ}$ and $R_{IJK}$ come from the variation of the Einstein tensor in \eqref{eq:weylrules}. The counterterms which are linear in $\sM$ contribute to these coefficients through the variation \eqref{eq:deltaM}, also at $\sM=0$, and \eqref{eq:lin} and \eqref{eq:quad} do not contain these contributions. They are given in appendix~\ref{app:massward} for the scheme of section~\ref{app:masscoeff} and in appendix~\ref{app:finite} for a general scheme. For $\theta_I\neq0$ they include the terms $-2\,\theta_I\cdot\fj_J$ in $E_{IJ}$ and $-2\,\theta_I\cdot\fl_{JK}$ in $R_{IJK}$, whose contribution to the identity \eqref{eq:curlgeneral} below is cancelled by its last term. For the coefficient of the term with the field strength we find, for every $\omega$ in the Lie algebra,
\begin{equation}
\omega\cdot Q_I = \omega\cdot\fp_{IJ}\,\hB^J-\tfrac12\,\omega\cdot\fy\cdot\hr_I+\tfrac12(d-2)\,(\omega g)^J\fg_{IJ} \, ,
\label{eq:Qdef}
\end{equation}
where the last term arises from the commutator of two covariant derivatives after the integration by parts of the term with $\fg_{IJ}$. All coefficients on the left-hand sides of \eqref{eq:lin} to \eqref{eq:Qdef} are finite. In minimal subtraction the coefficients $A_{\rm E}$, $G_{IJ}$ and $W_I$ do not depend on $\vep$, because the counterterms are pure poles and the explicit factors are polynomials of first order in $\vep$. They are given by the residues of the simple poles, $A_{\rm E}=4(1+g^K\pr_K)\fa_1$, $G_{IJ}=(3+g^K\pr_K)\fg_{IJ,1}$ and $W_I=-4\,\pr_I\fa_1+\fg_{IJ,1}\,g^J$. The term of first order in $\vep$ of the relation \eqref{gs:eq:eulerF} of appendix~\ref{app:finite} then gives $(2+g^K\pr_K)W_I=G_{IJ}g^J-\pr_IA_{\rm E}$ in the absence of field rotations. This relation is an identity between these residues.

We now derive from these expressions the identity \eqref{eq:curlgeneral}, which the construction of the curvature counterterms in appendix~\ref{app:finite} uses, and the relation \eqref{eq:WQ} and the identity \eqref{eq:localid} of appendix~\ref{gs:app:fourd}. The derivations hold for any number of couplings and for any group of field rotations. The relation \eqref{gs:eq:eulerF} is derived in appendix~\ref{app:finite}. For \eqref{eq:WQ} we contract $W_I$ of \eqref{eq:lin} with $(\omega g)^I$. The term with $\fa$ drops out because $\fa$ is invariant, so that $(\omega g)^IW_I=-(\omega g)^I\fg_{IJ}\hB^J$. In $\omega\cdot Q_I\,\hB^I$ the first term of \eqref{eq:Qdef} vanishes because $\fp_{IJ}$ is antisymmetric, and the second vanishes by \eqref{eq:rhoB}. Hence $\omega\cdot Q_I\,\hB^I=\tfrac12(d-2)\,(\omega g)^J\fg_{IJ}\hB^I$, which is \eqref{eq:WQ}.

For \eqref{eq:curlgeneral} we first take $\theta_I=0$, so that its last term is absent, and insert \eqref{eq:quad} and \eqref{eq:Qdef} into its right-hand side. In the first term, $-(\hP_I{}^KE_{JK}-\hP_J{}^KE_{IK})$, the contributions of $\fs_{IJ}$ cancel, because $\hP_I{}^K\hP_J{}^L\fs_{LK}$ is symmetric in $I$ and $J$. The contributions of $\fr_{IJK}$ are $-2\,(\hP_I{}^K\fr_{KJL}-\hP_J{}^K\fr_{KIL})\,\hB^L$, and those of $\fg_{IJ}$ are
\begin{equation}
(d-2)\left(\pr_J\hB^K\,\fg_{IK}-\pr_I\hB^K\,\fg_{JK}\right)+\tfrac12(d-2)\left[(\hr_J\,g)^K\fg_{IK}-(\hr_I\,g)^K\fg_{JK}\right] .
\end{equation}
In the second term, $(R_{IJK}-R_{JIK})\hB^K$, the contributions of $\fr_{IJK}$ cancel those of the first term. The contributions of $\fq_{IJKL}$ cancel by the symmetry of this tensor under the exchange of its two pairs of indices, and the terms $\hr_K\cdot\fp_{IJ}\,\hB^K$ vanish by \eqref{eq:rhoB}. With $\Gamma^{(\fg)}_{IJK}-\Gamma^{(\fg)}_{JIK}=\pr_J\fg_{IK}-\pr_I\fg_{JK}$ the second term therefore gives $(d-2)\,(\pr_J\fg_{IK}-\pr_I\fg_{JK})\,\hB^K+(\hr_J\cdot\fp_{IK}-\hr_I\cdot\fp_{JK})\,\hB^K$. Since $\fy$ is symmetric, the terms with $Q_I$ are
\begin{equation}
\hr_I\cdot Q_J-\hr_J\cdot Q_I = \left(\hr_I\cdot\fp_{JK}-\hr_J\cdot\fp_{IK}\right)\hB^K+\tfrac12(d-2)\left[(\hr_I\,g)^K\fg_{JK}-(\hr_J\,g)^K\fg_{IK}\right] .
\end{equation}
In the sum the terms with $\fp_{IJ}$ and the terms with $(\hr_I\,g)^K\fg_{JK}$ cancel, and the right-hand side of \eqref{eq:curlgeneral} becomes
\begin{equation}
(d-2)\left[\left(\pr_J\fg_{IK}-\pr_I\fg_{JK}\right)\hB^K+\fg_{IK}\,\pr_J\hB^K-\fg_{JK}\,\pr_I\hB^K\right]
= (d-2)\left[\pr_J\!\left(\fg_{IK}\hB^K\right)-\pr_I\!\left(\fg_{JK}\hB^K\right)\right] .
\end{equation}
This is $(d-2)$ times the curl of $W_I$, because the term with $\fa$ in \eqref{eq:lin} is a gradient. In a general scheme for the mass source the coefficients $E_{IJ}$ and $R_{IJK}$ contain the further terms $-2\,\theta_I\cdot\fj_J$ and $-2\,\theta_I\cdot\fl_{JK}$ of \eqref{eq:fixedcoeff}. They add $2\,\theta_J\cdot(\hP_I{}^K\fj_K+\fl_{IK}\hB^K)-2\,\theta_I\cdot(\hP_J{}^K\fj_K+\fl_{JK}\hB^K)$ to the first four terms of the right-hand side of \eqref{eq:curlgeneral}. With the coefficient $L_I$ of \eqref{eq:flatcoeff} and the symmetry of $\fm$ this is $2\,(L_I\cdot\theta_J-L_J\cdot\theta_I)$, while the curl of $W_I$ is not changed. The last term of \eqref{eq:curlgeneral} subtracts this contribution, which proves the identity. In a scheme with $\theta_I=0$ it is the relation \eqref{eq:curl} of appendix~\ref{gs:app:fourd}.

For \eqref{eq:localid} we divide both sides by $\tfrac12d^2$ and use $d\,R_d=d^2(d-1)$. By the definitions of $\hat M^L{}_J$ in section~\ref{gs:sec:quantities} and of $\hP_J{}^L$ after \eqref{gs:eq:Mrel} we have $\hat M^L{}_J-2\,\hP_J{}^L=-(d-2)\,\delta_J{}^L-\pr_J\hB^L$, in which $\hr_J$ has dropped out, and $\vep+d-2=2$. With the expressions \eqref{eq:lin} for $e_J$, $h_J$ and $u_J$ the left-hand side becomes
\begin{equation}
2(d-1)\left(e_J-h_J\right)-2\,u_J = 2(d-1)\left[\hB^K\pr_K\fc_J-\pr_J\hB^K\,\fc_K-2\,\ff_{JK}\hB^K\right]-2\,\fs_{JK}\hB^K \, .
\label{eq:localidlhs}
\end{equation}
On the right-hand side the contributions of $\fg_{IJ}$ and of $\fr_{IJK}$ cancel between $E_{JK}$ and $2\,\nu_{JK}$. With $\hat M^L{}_K\hB^K=\hB^K\pr_K\hB^L$, which holds by \eqref{eq:rhoB}, we find
\begin{equation}
\left(S_{JK}-E_{JK}+2\,\nu_{JK}\right)\hB^K = \hB^L\pr_L\!\left(\fs_{JK}\hB^K\right)-\pr_J\hB^L\,\fs_{LK}\hB^K-2\,\fs_{JK}\hB^K-4(d-1)\,\ff_{JK}\hB^K \, .
\end{equation}
The remaining terms of the bracket of \eqref{eq:localid} are $\pr_J\hB^I\,u_I-\hB^I\pr_Iu_J$. With $u_I=-2(d-1)\,\fc_I+\fs_{IK}\hB^K$ they cancel the terms $\hB^L\pr_L(\fs_{JK}\hB^K)$ and $\pr_J\hB^L\,\fs_{LK}\hB^K$, and the bracket equals the right-hand side of \eqref{eq:localidlhs}.

\subsection{Mass source}
\label{app:masscoeff}

By \eqref{eq:barehat} of section~\ref{app:gaugevar} the variation of section~\ref{sec:weylvar} changes the bare couplings and the bare gauge field by their Weyl weights and by the gauge transformation \eqref{eq:gaugetr} with the parameter $\sigma\,\xi^{\rm L}$. The bare functional is invariant if the bare mass source \eqref{eq:baresources} changes in the same way, that is by its weight $-2$ and by the same gauge transformation, $\delta\sM_0=-2\,\sigma\,\sM_0-[\sigma\,\xi^{\rm L},\sM_0]$. This condition determines the coefficients of \eqref{eq:deltaM},
\begin{align}
\delta_I &= -\big[(\tL-\gamma_\sM)\,\fd\big]_I \, , \qquad
\theta_I = -\hP_I{}^J\fd_J-\fe_{IJ}\hB^J \, , \qquad
\epsilon_{IJ} = -\big[(\tL-\gamma_\sM)\,\fe\big]_{IJ}-\hat\Omega_{IJ}{}^K\fd_K \, , \notag\\
\bar\eta &= -\left(\hB\!\cdot\!\pr-\gamma_\sM\right)\fk \, , \qquad
\bar\tau = -\fd_I\hB^I+2(d-1)\,\fk \, ,
\label{eq:masscoeff}
\end{align}
where the linear map $\gamma_\sM$ of $V_\sM$ is defined by
\begin{equation}
\hB^I\pr_IZ_\sM = -Z_\sM\,\gamma_\sM-{\rm ad}_{\xi^{\rm L}}\,Z_\sM \, , \qquad {\rm ad}_\omega\sM = [\omega,\sM] \, .
\label{eq:gammaM}
\end{equation}
Indeed, let $\mathsf{u}$ be the argument of $Z_\sM$ in \eqref{eq:baresources}, so that $\sM_0=Z_\sM\,\mathsf{u}$. With \eqref{eq:gammaM} the condition on $\sM_0$ is equivalent to $\delta\mathsf{u}=-\sigma\,(2-\gamma_\sM)\,\mathsf{u}$, which does not contain $Z_\sM$, and the coefficients \eqref{eq:masscoeff} follow from it with the rules \eqref{eq:varDg}. Since $Z_\sM$ is covariant we have $(\omega g)^I\pr_IZ_\sM={\rm ad}_\omega Z_\sM-Z_\sM\,{\rm ad}_\omega$, and \eqref{eq:gammaM} can also be written as $\hb^{{\rm L}\,I}\pr_IZ_\sM=-Z_\sM\,(\gamma_\sM+{\rm ad}_{\xi^{\rm L}})$. For the choice $\xi^{\rm L}=\xi$ of section~\ref{sec:weylvar} this is the equation for $Z_\sM$ in section~8 of~\cite{Jack:2013sha}, and $\gamma_\sM$ is the anomalous dimension of the mass operator defined there minus ${\rm ad}_\xi$. Thus $\gamma_\sM$ contains the rotation by $\xi$, in the same way as $B^I$ does. Note that without the last term of \eqref{eq:gammaM} the bare mass source would not be rotated, and $\gamma_\sM$ would in general have poles. The first line of \eqref{eq:masscoeff} has the form of the result of~\cite{Jack:2013sha}, and the second line contains the terms with the curvature. In the relations below $\gamma_\sM$ enters as a linear map of $V_\sM$, and its relation to $Z_\sM$ is not used. These coefficients satisfy the relations
\begin{equation}
2(d-1)\,\bar\eta = \delta_I\hB^I-\left(\hB\!\cdot\!\pr-\gamma_\sM\right)\bar\tau \, , \qquad
\big[(\tL-\gamma_\sM)\,\theta\big]_I = \hP_I{}^J\delta_J+\epsilon_{IJ}\hB^J \, ,
\label{eq:Mrel}
\end{equation}
which for $d=4$ are the relations of~\cite{Jack:2013sha}. They follow from Lemma~\ref{gcu:lem:commute}. The variation changes the bare couplings and the bare gauge field by \eqref{eq:barehat} of section~\ref{app:gaugevar}, and by the definition of the coefficients \eqref{eq:masscoeff} it changes the bare mass source in the same way, so that the operators $\mathbb{W}_\sigma$ commute. The two relations are the coefficients of $\sigma\nabla^2\sigma'-\sigma'\nabla^2\sigma$ and of $(\sigma\,\pr_\mu\sigma'-\sigma'\pr_\mu\sigma)\,D^\mu g^I$ in the commutator on $\sM$, see section~\ref{gs:sec:dimtwo}, while the terms of the commutator which are proportional to $\sigma\,\sigma'$ cancel, since they are symmetric in the two functions. The relations therefore hold identically in the counterterms, for every $d$. We also need an identity for the extended Lie derivative of $\hP_I{}^J$. On a tensor with one lower and one upper index the extended Lie derivative is $(\tL T)_I{}^J=\hB^K\pr_KT_I{}^J+\hat M^K{}_IT_K{}^J-\hat M^J{}_KT_I{}^K$, and $(\tL\hB)^J=0$. The identity is
\begin{equation}
(\tL\hP)_I{}^J = \hat\Omega_{IK}{}^J\,\hB^K \, ,
\label{eq:LiePsi}
\end{equation}
which enters the proof of the relations \eqref{eq:finrel} in appendix~\ref{app:finite}. To show \eqref{eq:LiePsi} we insert the definitions of $\hat M^K{}_I$ in section~\ref{gs:sec:quantities} and of $\hP_I{}^K$ after \eqref{gs:eq:Mrel}. In $\hat M^K{}_I\hP_K{}^J-\hat M^J{}_K\hP_I{}^K$ the products of two factors $\pr\hB$ and of two factors $\hr$ cancel, and $\tfrac12(\hr_Ig)^K\pr_K\hB^J-\tfrac12\,\pr_I\hB^K(\hr_Kg)^J$ remains. Here we use that $\hB^I$ is covariant under field rotations, in the form $(\hr_Ig)^K\pr_K\hB^J=(\hr_I\hB)^J$, where on the right-hand side $\hr_I$ acts on the vector $\hB$ as it acts on the couplings. We also use $\pr_I\hB^K\,\hr_K=-\hB^K\pr_I\hr_K$, which follows from \eqref{eq:rhoB}, and $\hB^K\pr_K(\hr_Ig)^J=(\hB^K\pr_K\hr_I\,g)^J+(\hr_I\hB)^J$. Together these give
\begin{equation}
(\tL\hP)_I{}^J = \hB^K\pr_K\pr_I\hB^J+\tfrac12\,(\hB^K\pr_K\hr_I\,g)^J+\tfrac12\,(\hB^K\pr_I\hr_K\,g)^J+(\hr_I\hB)^J \, ,
\end{equation}
which is $\hat\Omega_{IK}{}^J\hB^K$ by the definition of $\hat\Omega_{IK}{}^J$ after \eqref{eq:varDg} and by \eqref{eq:rhoB}. In minimal subtraction $\theta_I$ and $\bar\tau$ do not vanish. As in~\cite{Reehorst:phi4}, where $\theta_I$ is removed in this way, we remove them by a finite redefinition of the mass source,
\begin{equation}
\sM = \sM'-d_I\,D^2g^I-k_R\,R \, , \qquad d_I = (\hP^{-1})_I{}^J\theta_J \, , \qquad k_R = \frac{d_I\hB^I-\bar\tau}{2(d-1)} \, ,
\label{eq:Mredef}
\end{equation}
with regular $d_I$ and $k_R$. The matrix $\hP_I{}^K$ is invertible as a regular series, because its term of order zero in the couplings is $(1-\tfrac32\vep)\,\delta_I{}^K$.\footnote{The inverse has a pole at $\vep=\tfrac23$, which does not affect statements about power series in $\vep$.} The functional $W'[\gamma,g,a,\sM']=W[\gamma,g,a,\sM]$ is finite, and the relations among its anomaly coefficients of appendices~\ref{app:weyl} and~\ref{app:finite} hold for it, with counterterm coefficients which now have finite parts. The redefinition does not change $\hB^I$, $\hr_I$, $\fa$, $\fg_{IJ}$, $\fp_{IJ}$ and $\fy$, because the substitution \eqref{eq:Mredef} in \eqref{eq:X} produces no term with $\cE_4$, $\cG^{\mu\nu}$ or $f^{\mu\nu}$. Hence $A_{\rm E}$, $G_{IJ}$, $W_I$ and $Q_I$ at $\sM'=0$ are those of minimal subtraction, and for the form of $W_I$ in appendix~\ref{gs:app:fourd} it is sufficient to use the functional $W'$. That appendix uses this scheme from section~\ref{sec:massterms} on and drops the prime, and the tensor $\cH_{IJ}$ and the function $w$ of \eqref{eq:Wform} are defined in this scheme. In this scheme $\theta_I=0$ and $\bar\tau=0$, and \eqref{eq:Mrel} gives
\begin{equation}
\delta_I = \mathsf{D}_{IK}\,\hB^K \, , \qquad \mathsf{D}_{IK} = -(\hP^{-1})_I{}^J\epsilon_{JK} \, , \qquad
2(d-1)\,\bar\eta = \delta_I\hB^I \, .
\label{eq:deltascheme}
\end{equation}
Note that in this scheme the coefficients other than $A_{\rm E}$, $G_{IJ}$ and $W_I$ depend on $\vep$.

\subsection{Mass operator on the sphere}
\label{app:massward}

In this subsection we show the statements on the mass operator on the sphere which section~\ref{gs:sec:phi4prop} uses, and we prove Lemma~\ref{lem:mass} of appendix~\ref{gs:app:fourd}. We work in the scheme \eqref{eq:Mredef}, in which $\theta_I=\bar\tau=0$. On the sphere with constant couplings and $\sM=0$ the variation \eqref{eq:deltaM} is $\delta^{\rm W}_\sigma\sM=-\sigma\,\varkappa$ with
\begin{equation}
\varkappa = R_d\,\bar\eta = \tfrac12\,d\,\delta_I\hB^I \, ,
\label{eq:varkappa}
\end{equation}
where we used \eqref{eq:deltascheme}. It does not vanish away from fixed points. Let $\mathsf{m}$ be the one-point function of the mass operator on the sphere at constant couplings, which is a constant with values in the dual of $V_\sM$, and let $\Lambda_{\sM J}$ be the first harmonic of its two-point function with the marginal operators, defined as $\Lambda_{IJ}$ in \eqref{eq:Lambda}. More precisely, $\Lambda_{\sM J}$ is defined by \eqref{eq:Lambdadef} with the mass source $t_1\,\sigma\,\mu$ in place of the variation $t_1\,\sigma\,\delta g$, where $\mu$ is a constant element of $V_\sM$. The mass source is a scalar source, so that the arguments of appendix~\ref{app:rotations} apply to it, and $\mathsf{m}$ is constant for the same reason. The same holds at a constant mass source, and in particular at the source $-k_RR_d$ which corresponds to $\sM'=0$ in the scheme \eqref{eq:Mredef}. Indeed, a constant source belongs to the class of sources, so that Proposition~\ref{prop:finite} covers it, and it is invariant under the rotations which are used in appendix~\ref{app:rotations}.

We now prove Lemma~\ref{lem:mass}. The left-hand side of \eqref{eq:sphere} receives the terms
\begin{equation}
-\pr_J\varkappa\cdot\mathsf{m}-\varkappa\cdot\Lambda_{\sM J}+d\,\delta_J\cdot\mathsf{m} \, .
\label{eq:Mleft}
\end{equation}
They come from the term with $\delta_v\sM$ in \eqref{eq:wardS}. At constant couplings and $\sM=0$ this variation is $-\sigma\,\varkappa$, and it multiplies the one-point function $\mathsf{m}$. Under a variation $\delta g^J$ of the couplings $\delta_v\sM$ changes by $-\sigma\,(\pr_J\varkappa\,\delta g^J+\delta_J\nabla^2\delta g^J)$, as follows from \eqref{eq:deltaM} with $\theta_I=\bar\tau=0$, while the term with $\epsilon_{IJ}$ is of second order in the variation. After two integrations by parts and with $\nabla^2\sigma=-d\,\sigma$ the integral of this change with $\mathsf{m}$ gives the first and the third term of \eqref{eq:Mleft}, times $\int d\mu_d\,\sigma\,\delta g^J$. The one-point function of the mass operator changes by its two-point function with the marginal operators, integrated with $\delta g^J$. Its integral with $-\sigma\,\varkappa$ gives the second term, by the definition of $\Lambda_{\sM J}$.

On the right-hand side the counterterms which are linear in $\sM$ contribute through the variation of $\sM$. They change the coefficients as $e_I\to e_I-\fn\cdot\delta_I-\fj_I\cdot\bar\eta$, $q\to q-\fn\cdot\bar\eta$ and $S_{IJ}\to S_{IJ}-2\,\fj_{(I}\cdot\delta_{J)}$, while $h_I$, $u_I$, $r$, $E_{IJ}$, $\nu_{IJ}$ and $R_{IJK}$ do not change in this scheme. From here on $e_I$, $q$ and $S_{IJ}$ denote the coefficients which include these terms. In particular \eqref{eq:curl} holds as it stands. The identity \eqref{eq:localid} receives one term,
\begin{equation}
d\,R_d\left(e_J-h_J\right)-d^2u_J = \frac{d^2}{2}\Big[\ldots\Big]+\frac{d^2}{2}\;t_\sM\cdot\delta_J \, ,
\qquad t_\sM=\fj_I\hB^I-2(d-1)\,\fn \, ,
\label{eq:localidM}
\end{equation}
where the bracket is that of \eqref{eq:localid} with the new coefficients. Indeed, by \eqref{eq:localidlhs} the changes of $e_J$ change the left-hand side of \eqref{eq:localid} by $\tfrac12d^2$ times $-2(d-1)\,(\fn\cdot\delta_J+\fj_J\cdot\bar\eta)$, and the change of $S_{JK}$ changes the bracket by $-(\fj_J\cdot\delta_K+\fj_K\cdot\delta_J)\,\hB^K$. With $2(d-1)\,\bar\eta=\delta_K\hB^K$ from \eqref{eq:deltascheme} the difference of the two changes is $(\fj_K\hB^K-2(d-1)\,\fn)\cdot\delta_J$, which is $t_\sM\cdot\delta_J$. The coefficient $t_\sM$ is finite, since it is the coefficient of $\nabla^2\sigma\,\sM$ in the anomaly. With \eqref{eq:Mleft} and \eqref{eq:localidM} the result \eqref{eq:Wresult} becomes
\begin{equation}
2e_d\,W_J = \pr_J\!\left(\Phi-\varkappa\cdot\mathsf{m}\right)+\mathsf{h}_{JK}\hB^K-\frac{d^2}{2}\,E_{[JK]}\hB^K
+\varkappa\cdot\left(\pr_J\mathsf{m}-\Lambda_{\sM J}\right)+\mathsf{Y}_{JK}\hB^K \, ,
\label{eq:WresultM}
\end{equation}
with
\begin{equation}
\mathsf{Y}_{JK} = \left(d\,\mathsf{m}+\tfrac12\,d^2\,t_\sM\right)\cdot\mathsf{D}_{JK} \, .
\label{eq:Ydef}
\end{equation}
Note that $\mathsf{m}$ and $t_\sM$ appear in $\mathsf{Y}_{JK}$ only in the combination $\mathsf{m}+\tfrac12d\,t_\sM$, which is invariant under the addition of a finite term proportional to $\sM R$ to $W$.

The two new terms in \eqref{eq:WresultM} are regular. The first is quadratic in $\hB^I$, since $\varkappa=\tfrac12d\,\mathsf{D}_{IK}\hB^I\hB^K$. It is of the form $\mathsf{x}_{JKL}\hB^K\hB^L$ with a regular $\mathsf{x}_{JKL}$ which we take symmetric in its last two indices, and it equals $(\mathsf{x}_{JKL}+\mathsf{x}_{KJL}-\mathsf{x}_{LJK})\hB^L\hB^K$, which is a symmetric tensor contracted with $\hB^K$. In the second term the matrix $\mathsf{Y}_{JK}$ is not symmetric. Its antisymmetric part is treated in section~\ref{sec:form} together with $E_{[JK]}$.

We now write \eqref{eq:WresultM} in the form \eqref{eq:Wshort}, with $\Phi'=\Phi-\varkappa\cdot\mathsf{m}$, and with the regular symmetric tensor $\mathsf{h}'_{JK}$ which is the sum of $\mathsf{h}_{JK}$, of the symmetric part of $\mathsf{Y}_{JK}$ and of the symmetric tensor which represents the term with $\varkappa$. The difference $\Phi'-\Phi$ vanishes at the zeros of $\hB^I$, because $\varkappa$ is quadratic in $\hB^I$. By \eqref{eq:Ydef} and \eqref{eq:deltascheme} we have $\hP_I{}^K\mathsf{Y}_{KJ}=-\mathsf{y}_{IJ}$ with the regular symmetric tensor $\mathsf{y}_{IJ}=(d\,\mathsf{m}+\tfrac12d^2\,t_\sM)\cdot\epsilon_{IJ}$. This proves the lemma.

In a general scheme for the mass source the coefficient $h_J$ satisfies
\begin{equation}
2(d-1)\,h_J = -2\,\hP_J{}^Ku_K+\left(E_{JK}-2\,\nu_{JK}\right)\hB^K+2\,t_\sM\cdot\theta_J-2\,\bar\tau\cdot L_J \, ,
\label{eq:hrel}
\end{equation}
where $t_\sM=\fj_I\hB^I-2(d-1)\,\fn-\bar\tau\cdot\fm$ is the coefficient of the term $\nabla^2\sigma\,\sM$ of the anomaly and $L_J$ is the coefficient which is defined by the complete anomaly \eqref{eq:fullanomaly} of appendix~\ref{app:finite}, with the sign and the factor given there. We obtain \eqref{eq:hrel} by inserting \eqref{eq:fixedcoeff} and the expression \eqref{eq:flatcoeff} for $L_I$ into its right-hand side. There the terms with $\fs_{IJ}$, $\fr_{IJK}$ and $\fg_{IJ}$ cancel among the first three terms, and the terms with $\fj_I$, $\fl_{IJ}$ and $\fm$ cancel against the last two. The remainder is $4(d-1)\,(\hP_J{}^K\fc_K+\ff_{JK}\hB^K-\fn\cdot\theta_J)$, which is $2(d-1)\,h_J$ by \eqref{eq:derivedcoeff}.

\subsection{Variation of the bare gauge field}
\label{app:gaugevar}

In this subsection we derive the variation of the renormalised sources which is used in section~\ref{sec:weylvar}, and its properties. The statements are identities for covariant counterterms, and they do not use the finiteness of the functional.\footnote{We verified Lemmas~\ref{lem:rhoxi} and~\ref{lem:frame} in exact arithmetic for the group ${\rm SO}(3)$ with six couplings, for covariant counterterms with random coefficients through fourth order in the couplings.} We write $g_0^I=g^I+L^I$ and $a_{0\mu}=a_\mu+N_ID_\mu g^I$ as in \eqref{eq:baresources}, and we recall that $\hb^{{\rm L}\,I}$ is defined by $\hb^{{\rm L}\,J}\pr_Jg_0^I=-\vep\,g_0^I$. The variation is determined by the equations
\begin{equation}
\hB^I = \hb^{{\rm L}\,I}-(\xi^{\rm L}g)^I \, , \qquad \xi^{\rm L} = N_I\,\hB^I \, , \qquad
\hr_I+N_J\,(\hr_Ig)^J = \pr_I\xi^{\rm L}+[N_I,\xi^{\rm L}]-(\cL_{\hB}N)_I \, ,
\label{eq:xiL}
\end{equation}
with $(\cL_{\hB}N)_I=\hB^J\pr_JN_I+\pr_I\hB^J\,N_J$. These are the equations of section~8 of~\cite{Jack:2013sha}, where $\xi^{\rm L}$ is written as $\upsilon-\omega$.

\begin{lemma}
\label{lem:rhoxi}
Let $L^I$ and $N_I$ be covariant pure poles, where $L^I$ has no terms of degree zero and one in the couplings and $N_I$ has no term of degree zero. First, the equations \eqref{eq:xiL} have a unique solution $\xi^{\rm L}$, $\hB^I$ and $\hr_I$ in power series in the couplings whose coefficients are Laurent polynomials in $\vep$, and the solution is covariant. Second, under the variation $\delta g^I=\sigma\hB^I$ and $\delta a_\mu=\sigma\hr_ID_\mu g^I$ with a function $\sigma(x)$ the bare sources change by
\begin{equation}
\delta g_0^I = -\vep\,\sigma\,g_0^I-(\sigma\xi^{\rm L}g_0)^I \, , \qquad \delta a_{0\mu} = \pr_\mu(\sigma\xi^{\rm L})+[a_{0\mu},\sigma\xi^{\rm L}] \, ,
\label{eq:barehat}
\end{equation}
that is by their Weyl weights and by the gauge transformation \eqref{eq:gaugetr} with the parameter $\sigma\xi^{\rm L}$. Conversely, covariant coefficients $\hB^I$, $\hr_I$ and $\xi^{\rm L}$ with this property solve \eqref{eq:xiL}. Third, $\hr_I\hB^I=0$. Fourth, the series $\hb^{{\rm L}\,I}+\vep\,g^I$, $\xi^{\rm L}$, $\hB^I+\vep\,g^I$ and $\hr_I$ contain no positive powers of $\vep$, and their terms of order $\vep^0$ are
\begin{equation}
\beta^I = (g^J\pr_J-1)\,L^I_1 \, , \qquad \xi = -N_{I,1}\,g^I \, , \qquad B^I = \beta^I-(\xi g)^I \, , \qquad g^J\left(\pr_JN_{I,1}-\pr_IN_{J,1}\right) ,
\label{eq:zeroparts}
\end{equation}
respectively, where $L^I_1$ and $N_{I,1}$ are the residues of the simple poles. In particular, if $\hB^I$ and $\hr_I$ are regular, then $\hB^I=-\vep\,g^I+B^I$, the coefficient $\hr_I$ is given by \eqref{eq:rhoxi}, and $B^I$ and $\hr_I$ do not depend on $\vep$.
\end{lemma}

\begin{proof}
The equation for $\hb^{{\rm L}\,I}$ reads $(\delta^I{}_J+\pr_JL^I)\,\hb^{{\rm L}\,J}=-\vep\,(g^I+L^I)$. Since $L^I$ starts at degree two, it is solved degree by degree. With the first equation of \eqref{eq:xiL} inserted, the second reads $\xi^{\rm L}+N_I(\xi^{\rm L}g)^I=N_I\hb^{{\rm L}\,I}$. Since $N_I$ starts at degree one, this equation and then the third equation of \eqref{eq:xiL} are solved degree by degree as well. The solutions are unique, and they are covariant because the equations are.

For the second statement we use the covariance in three forms. For a constant $\omega$ in the Lie algebra we have $(\omega g)^J\pr_Jg_0^I=(\omega g_0)^I$, and since this is an identity at each point it also holds for $\omega=\xi^{\rm L}(g)$. The covariant derivatives of covariant functions of the couplings are $D_\mu\hB^I=\pr_J\hB^I\,D_\mu g^J$ and $\pr_\mu\xi^{\rm L}+[a_\mu,\xi^{\rm L}]=\pr_I\xi^{\rm L}\,D_\mu g^I$. The variation of the bare couplings is $\sigma\hB^J\pr_Jg_0^I=\sigma\,\hb^{{\rm L}\,J}\pr_Jg_0^I-\sigma\,(\xi^{\rm L}g)^J\pr_Jg_0^I$, which is the first equation of \eqref{eq:barehat}. For the bare gauge field we find, with $\delta D_\mu g^I=D_\mu\delta g^I+(\delta a_\mu g)^I$,
\begin{align}
\delta a_{0\mu} &= \delta a_\mu+\delta g^J\pr_JN_I\,D_\mu g^I+N_I\left(D_\mu\delta g^I+(\delta a_\mu g)^I\right) \notag\\
&= \pr_\mu\sigma\,N_I\hB^I+\sigma\left[\hr_I+(\cL_{\hB}N)_I+N_J\,(\hr_Ig)^J\right]D_\mu g^I \, .
\label{eq:dela0}
\end{align}
On the other hand the gauge transformation in \eqref{eq:barehat} is $\pr_\mu\sigma\,\xi^{\rm L}+\sigma\left(\pr_I\xi^{\rm L}+[N_I,\xi^{\rm L}]\right)D_\mu g^I$. The two expressions agree by the second and the third equation of \eqref{eq:xiL}. Conversely, the matrix $\pr_Jg_0^I$ is invertible, and $\pr_\mu\sigma$ and $\sigma D_\mu g^I$ are independent, so that the property \eqref{eq:barehat} implies the three equations \eqref{eq:xiL}.

For the third statement we contract the third equation of \eqref{eq:xiL} with $\hB^I$. With the second equation the right-hand side becomes $\hB^I\pr_I\xi^{\rm L}+[\xi^{\rm L},\xi^{\rm L}]-\hB^J\pr_J(N_I\hB^I)=0$. Hence $\mathsf{X}=\hr_I\hB^I$ obeys $\mathsf{X}+N_J\,(\mathsf{X}g)^J=0$, and it vanishes degree by degree.

For the fourth statement we write the solution of the first equation as $\hb^{{\rm L}\,I}=-\vep\,g^I+\vep\,[(1+\pr L)^{-1}]^I{}_J\,(g^K\pr_K-1)L^J$. The second term is $\vep$ times a pure pole. It has no positive powers of $\vep$, and its term of order $\vep^0$ is $(g^J\pr_J-1)L^I_1$. The product of a pure pole with a series without positive powers of $\vep$ is a pure pole. Hence $N_I\hb^{{\rm L}\,I}$ equals $-N_{I,1}\,g^I$ up to a pure pole, and the same holds for $\xi^{\rm L}$, since $N_I(\xi^{\rm L}g)^I$ is then a pure pole degree by degree. This gives the statements on $\xi^{\rm L}$ and on $\hB^I$. On the right-hand side of the third equation of \eqref{eq:xiL} the term $[N_I,\xi^{\rm L}]$ is a pure pole, and $(\cL_{\hB}N)_I$ equals $-\vep\,(g^J\pr_J+1)N_I$ up to a pure pole. The right-hand side therefore has no positive powers of $\vep$, and its term of order $\vep^0$ is $\pr_I\xi+(g^J\pr_J+1)N_{I,1}$, which is the last expression in \eqref{eq:zeroparts}. The same holds for $\hr_I$, since $N_J(\hr_Ig)^J$ is a pure pole. Finally a regular series without positive powers of $\vep$ does not depend on $\vep$.
\end{proof}

The lemma and its proof hold verbatim if $L^I$ and $N_I$ are covariant only under the Lie algebra of the field rotations, with a solution which is covariant under the Lie algebra.

Note that the last expression in \eqref{eq:zeroparts} vanishes when it is contracted with $g^I$. For regular coefficients the third statement of the lemma therefore gives both equations in \eqref{eq:rhoB}. The next lemma describes the freedom in the choice of the bare sources.

\begin{lemma}
\label{lem:frame}
Let the coefficients of the bare sources \eqref{eq:baresources} in flat space be covariant pure poles, and let $\vartheta(g)$ be a covariant pure pole with values in the Lie algebra and without terms of degree zero and one in the couplings. Let $\mathsf{R}=\exp\vartheta(g(x))$ act on the couplings and on the fields as a rotation, and let
\begin{equation}
g_0' = \mathsf{R}\,g_0 \, , \qquad a'_{0\mu} = \mathsf{R}\,a_{0\mu}\mathsf{R}^{-1}-\pr_\mu\mathsf{R}\,\mathsf{R}^{-1} \, , \qquad \sM_0' = \mathsf{R}\,\sM_0\,\mathsf{R}^{-1} \, .
\label{eq:frame}
\end{equation}
First, the primed sources are again of the form \eqref{eq:baresources}, with the covariant pure poles $L'^I=(\mathsf{R}\,(g+L))^I-g^I$ and $N'_I=\mathsf{R}\,N_I\mathsf{R}^{-1}-\pr_I\mathsf{R}\,\mathsf{R}^{-1}$, with $Z_\sM$ followed by the rotation, and with the same $\fd_I$ and $\fe_{IJ}$. For sources of the class the bare functional $W_0^{\rm fl}$ has the same value on the primed bare sources. Second, the solution of \eqref{eq:xiL} for the primed sources has the same $\hB^I$ and $\hr_I$, and
\begin{equation}
\xi'^{\,\rm L} = \mathsf{R}\,\xi^{\rm L}\mathsf{R}^{-1}-\hB^J\pr_J\mathsf{R}\,\mathsf{R}^{-1} \, .
\label{eq:xiframe}
\end{equation}
Third, there is such a $\vartheta$ without simple poles such that $\xi'^{\,\rm L}$ is regular. Then $\xi'^{\,\rm L}=\xi$, and the residues $L^I_1$ and $N_{I,1}$ of the simple poles are not changed.
\end{lemma}

\begin{proof}
The matrix $\mathsf{R}-1$ is a pure pole, and so is $L'^I$, which is covariant. Since $\mathsf{R}$ is covariant we have $\pr_\mu\mathsf{R}+[a_\mu,\mathsf{R}]=\pr_I\mathsf{R}\,D_\mu g^I$. This gives $\mathsf{R}\,a_\mu\mathsf{R}^{-1}-\pr_\mu\mathsf{R}\,\mathsf{R}^{-1}=a_\mu-\pr_I\mathsf{R}\,\mathsf{R}^{-1}D_\mu g^I$ and hence the form of $a_0'$ with the stated $N'_I$, which has no term of degree zero. The statement on $\sM_0'$ is immediate. For the bare functional we consider the sources \eqref{eq:frame} with $\mathsf{R}_t=\exp(t\,\vartheta)$ for $0\leq t\leq1$. At each order in the sources they are polynomials in $t$ whose coefficients are Laurent polynomials in $\vep$ with coefficients in the class, and their derivative with respect to $t$ is the gauge transformation \eqref{eq:gaugetr} of the bare sources with the parameter $-\vartheta(g(x))$. By Proposition~\ref{prop:gauge} and the remark after its proof, which are linear in the parameter, the bare functional does not depend on $t$.

For the second statement let the renormalised sources change by $\delta g^I=\sigma\hB^I$ and $\delta a_\mu=\sigma\hr_ID_\mu g^I$. Then $\delta\mathsf{R}=\sigma\hB^J\pr_J\mathsf{R}$, and with \eqref{eq:barehat} we find $\delta g_0'=\delta\mathsf{R}\,\mathsf{R}^{-1}g_0'+\mathsf{R}\,\delta g_0=-\vep\,\sigma\,g_0'-(\sigma\xi'^{\,\rm L}g_0')$ with $\xi'^{\,\rm L}$ as in \eqref{eq:xiframe}. The primed bare gauge field changes by the gauge transformation with the same parameter, because it is obtained from $a_0$ by a gauge transformation and gauge transformations form a group. The parameter is covariant, and the converse part of Lemma~\ref{lem:rhoxi} gives the statement.

For the third statement we construct $\vartheta$ degree by degree. Let $\vartheta_{(k)}$ be its term of degree $k$, so that $g^J\pr_J\vartheta_{(k)}=k\,\vartheta_{(k)}$. By Lemma~\ref{lem:rhoxi} the vector field $\hB^I$ equals $-\vep\,g^I$ up to terms without positive powers of $\vep$. The term of degree $k$ of \eqref{eq:xiframe} is therefore $\xi^{\rm L}_{(k)}+\vep\,k\,\vartheta_{(k)}+\mathsf{F}_k$, where $\mathsf{F}_k$ contains only the $\vartheta_{(j)}$ with $j<k$. We set $\vartheta_{(k)}=-{\rm PP}\,[\xi^{\rm L}_{(k)}+\mathsf{F}_k]/(\vep\,k)$, where ${\rm PP}$ denotes the pole part of a Laurent polynomial in $\vep$. This term has no simple pole. Then $\xi'^{\,\rm L}$ is regular. Each term of $\mathsf{F}_k$ either contains a factor $\vartheta_{(j)}$, whose poles are at least double, times factors without positive powers of $\vep$, or it is $\vep$ times a product of at least two factors $\vartheta_{(j)}$. Hence $\mathsf{F}_k$ is a pure pole, and the term of order $\vep^0$ of $\xi'^{\,\rm L}$ is the one of $\xi^{\rm L}$, which is $\xi$ by Lemma~\ref{lem:rhoxi}. For the same reason $L'^I-L^I$ and $N'_I-N_I$ have no simple poles.
\end{proof}

The proof uses the covariance of $\vartheta$, $L^I$ and $N_I$ only under the Lie algebra of the field rotations. For pure poles which are covariant only under the Lie algebra the lemma therefore holds with $L'^I$ and $N'_I$ which are covariant under the Lie algebra.

The coefficients of the bare sources of Hypothesis~\ref{hyp:finite} are covariant by Proposition~\ref{prop:uniqueall}. By the third statement of the lemma we can choose these bare sources such that $\xi^{\rm L}=\xi$, and we do so from here on. The functional $W^{\rm fl}$ and its finiteness are not affected by this choice. Note that the same construction with $\vartheta_{(k)}=-[\xi^{\rm L}_{(k)}+\mathsf{F}_k]/(\vep\,k)$ gives a covariant pure pole $\vartheta$, now with simple poles, for which $\xi'^{\,\rm L}=0$. For these bare sources the first equation of \eqref{eq:xiL} gives $\hb'^{\,{\rm L}\,I}=\hB^I$, so that $\hB^J\pr_Jg_0'^{\,I}=-\vep\,g_0'^{\,I}$. This gives the singular solution of the gradient equation which is mentioned in section~\ref{sec:regular}. Let us write $g_0^I(g)=g^I+\ldots$ for these bare couplings in the remainder of this paragraph. In the coordinates $g_0^I$ the flow is linear, and
\begin{equation}
\pr_I\!\left(-\tfrac{\vep}{2}\,g_0^Kg_0^K\right) = \left(\pr_Ig_0^K\,\pr_Jg_0^K\right)\hB^J \, .
\label{eq:trivial}
\end{equation}
The metric in brackets is symmetric and positive, but its coefficients have poles at $\vep=0$. Note that this is not the choice of the bare sources which we make in section~\ref{sec:weylvar}, for which $\hb^J\pr_Jg_0^I=-\vep\,g_0^I$ holds. In a theory without field rotations the two agree, and $g_0^I$ is the usual bare coupling.

Let us finally describe how $\beta^I$ and $\xi$ depend on the choice of the bare sources, which we use in section~\ref{sec:weylvar}. While $B^I$ and $\hr_I$ do not depend on the choice of the bare sources, the two terms $\beta^I$ and $\xi$ of $B^I$ do. A transformation whose parameter has a simple pole changes $\beta^I$ by a rotation $(\omega g)^I$ and $\xi$ by $\omega$, with a regular $\omega$ in the Lie algebra. We use the names $\beta^I$ and $\xi$ for a symmetric renormalisation factor of the fields. This is the conventional choice, in which the finite part of the anomalous dimension matrix of the fields is symmetric~\cite{Jack:2013sha,Bednyakov:2021ojn}. For this choice the vector field $\hb^{{\rm L}\,I}$ and the parameter $\xi^{\rm L}$ need not be regular, and $\hb^{{\rm L}\,I}$ can have poles of the form $(\omega g)^I$ with $\omega$ in the Lie algebra. Such poles were found in the Standard Model at three loops, and they are explained in~\cite{Herren:2021amb}. By Lemma~\ref{lem:frame} there is a transformation whose parameter has no simple pole, and which therefore does not change the residues $L^I_1$ and $N_{I,1}$, after which $\xi^{\rm L}=\xi$. These are the bare sources which we use. Then $\hb^{{\rm L}\,I}=\hb^I$, which has the form $\hb^I=-\vep\,g^I+\beta^I$ with $\beta^I$ independent of $\vep$.

\subsection{Removable terms}
\label{app:closure}

In this subsection, which belongs to appendix~\ref{gs:app:fourd}, we prove the third and the fourth statement of Lemma~\ref{lem:closure}, with the notation of section~\ref{sec:removable}. The third tensor equals $(M^T\mathsf{s}-\mathsf{s}M)_{IJ}+\hr_{[I}\cdot(g\mathsf{s})_{J]}$, where $(g\mathsf{s})_J$ is defined by $\omega\cdot(g\mathsf{s})_J=(\omega g)^K\mathsf{s}_{KJ}$. For the fourth tensor we consider the three terms of \eqref{eq:types} in turn. For the term $M^TC-CM$ the part of $P_I{}^K$ with $\pr_I\hB^K$ gives
\begin{equation}
-\left(M^TC''-C''M\right)_{IJ}+2\,\vep\left(M^TC-CM\right)_{IJ} \, , \qquad C''=M^TC+CM \, ,
\end{equation}
since the terms $M^TCM$ cancel in $(M^T)^2C-CM^2$. For the term $\tau_{IJK}\hB^K$ the result is again proportional to $\hB^K$. For the term $\hr_{[I}\cdot X_{J]}$ we find the sum of a term of the second type and of $\tfrac12\hB^K(\pr_I\hr_K\cdot X_J-\pr_J\hr_K\cdot X_I)$, where we used $\pr_I\hB^K\,\hr_K=-\hB^K\pr_I\hr_K$, which follows from \eqref{eq:rhoB}. Finally the part of $P_I{}^K$ with $(\hr_I\,g)^K$ gives a term of the second type in all three cases.

We add three statements on the iteration of section~\ref{sec:iteration}. The first is that the terms of \eqref{eq:Eeq} without $E_{[IJ]}$ are in $\cT$. For the first and the third term this follows from the first two statements of Lemma~\ref{lem:closure}, where we use that $\mathsf{Y}_{[IJ]}$ is in $\cT$, and for the part of the fourth term with $E_{(IJ)}$ from the third statement. The last two terms are of the first and of the second type. The second is the convergence of the representatives. Here we assign to a tensor in $\cT$ the weight which is given by its representatives. A step of the iteration which acts on a tensor of weight $n$ in $\cT$ produces a tensor of weight at least $n+1$, with representatives $C$ of weight at least $n$ and $\tau$ and $X$ of weight at least $n-1$, as one sees from \eqref{eq:removal} and \eqref{eq:curls}. The third concerns the coefficients of Proposition~\ref{prop:form}. The coefficients are built from the free energy and the two-point function on the sphere and from the anomaly coefficients, by the iteration just described. In particular $C_{IJ}=(\mathsf{h}'_{IJ}+R_d\,E_{(IJ)})/(2e_d)$ up to terms of weight one, as one sees from the first and the fourth term of \eqref{eq:Eeq}.

\section{Meromorphy of the sphere integrals}
\label{app:mero}

In this appendix we prove that the integrals \eqref{eq:gram} which define the bare functional on the sphere are meromorphic functions of $d$ and $\Dp$, and we determine a set of lines which contains their poles. The definition of the functional by continuation in section~\ref{sec:continuation} rests on this statement. Appendices~\ref{app:line} and~\ref{app:gauge} build on the results of this appendix. In section~\ref{app:polar} we give a region in which the integrals converge, and we show that the polar part along a candidate depends continuously on the sources. Appendices~\ref{app:line} and~\ref{app:gauge} use this to reduce the sources to special ones. In the last subsection we use the meromorphy to prove Proposition~\ref{prop:cov} on the conformal covariance, including the gauge field.

Two general theorems are related to the meromorphy of these integrals. Atiyah~\cite{Atiyah:1970} proves that for a non-negative real-analytic function $f$ on a real-analytic manifold, and for a domain which is defined by inequalities between real-analytic functions, the function $f^s$ on this domain extends to a distribution which is meromorphic in $s$, with poles at negative rational numbers. Bernstein and Gel'fand~\cite{Bernstein:1969} prove that for polynomials $P_1,\ldots,P_k$ on a non-singular real algebraic variety the integral of $|P_1|^{\lambda_1}\cdots|P_k|^{\lambda_k}$ with a smooth function of compact support is meromorphic in $\lambda_1,\ldots,\lambda_k$, with poles on hyperplanes. Both proofs use the theorem of Hironaka on the resolution of singularities. The theorem of~\cite{Atiyah:1970} concerns one exponent and does not give the dependence on $d$ and $\Dp$ together. Applied to the integrals in the triangular coordinates below, after an extension of the half spheres to spheres by reflections, the theorem of~\cite{Bernstein:1969} gives their meromorphy in $d$ and $\Dp$, with poles on lines. The following appendices need two further statements. The first is an explicit set of lines which contains the poles, from which appendix~\ref{app:line} excludes the line of the free theory. The second is a bound on the integrals near these lines in terms of norms of the sources, which gives the continuous dependence of the polar parts on the sources and by which appendices~\ref{app:line} and~\ref{app:gauge} reduce the sources to special ones. We prove the meromorphy together with these two statements by a direct argument, which does not use the resolution of singularities. In suitable coordinates the integrand is singular on linear subspaces only, and a decomposition into sectors is sufficient.

\subsection{Powers of coordinates}

\begin{lemma}
\label{lem:cube}
Let $\mathcal{R}(s,\zeta)$ be a smooth function of $s\in[0,1]^M$ whose derivatives with respect to $s$ are jointly continuous in $s$ and in parameters $\zeta$, and holomorphic in $\zeta$. The integral
\begin{equation}
\mathcal{I}(z,\zeta) = \int_{[0,1]^M}\prod_{j=1}^Mds_j\,s_j^{z_j-1}\;\mathcal{R}(s,\zeta) \, ,
\end{equation}
which converges for ${\rm Re}\,z_j>0$, extends to a meromorphic function of $z\in\mathbb{C}^M$ and $\zeta$. Its poles lie on the hyperplanes $z_j+\nu=0$ with $\nu=0,1,2,\ldots$
\end{lemma}

\begin{proof}
For one variable and an integer $K\geq1$ we write
\begin{equation}
\int_0^1ds\,s^{z-1}\varphi(s) = \sum_{\nu=0}^{K-1}\frac{\varphi^{(\nu)}(0)}{\nu!\,(z+\nu)}
+\int_0^1ds\,s^{z+K-1}\varphi_K(s) \, , \quad
\varphi_K(s) = \int_0^1dt\,\frac{(1-t)^{K-1}}{(K-1)!}\,\varphi^{(K)}(ts) \, .
\label{eq:taylor}
\end{equation}
The last integral converges and is holomorphic for ${\rm Re}\,z>-K$. The coefficients $\varphi^{(\nu)}(0)$ and the function $\varphi_K$ are smooth in the remaining variables and holomorphic in the parameters. We can therefore apply the same step to the variables in turn.
\end{proof}

\subsection{Linear forms and sums of squares}

Let $L_1,\ldots,L_r$ be linear forms on $\mathbb{R}^M$ which do not vanish identically. Let $\ell_a$ with $a=1,\ldots,a_{\max}$ be some of these forms, and let $\mathsf{q}_b=\sum_{c\in C_b}L_c^2$ with $b=1,\ldots,b_{\max}$ be sums of squares of some of them. Let $\psi(y,\zeta)$ be smooth with compact support in $y\in\mathbb{R}^M$ and holomorphic in parameters $\zeta$. We consider
\begin{equation}
J(\mu,\eta,\zeta) = \int_{\ell_a\geq0}d^My\;\prod_{a}\ell_a^{\,\mu_a}\,\prod_b\mathsf{q}_b^{\,\eta_b}\;\psi(y,\zeta) \, ,
\label{eq:Jdef}
\end{equation}
which converges for ${\rm Re}\,\mu_a\geq0$ and ${\rm Re}\,\eta_b\geq0$. For a linear subspace $\mathsf{V}$ of $\mathbb{R}^M$ we define the degree
\begin{equation}
\deg(\mathsf{V}) = {\rm codim}\,\mathsf{V}+\sum_{a\,:\,\ell_a|_\mathsf{V}=0}\mu_a+2\sum_{b\,:\,\mathsf{q}_b|_\mathsf{V}=0}\eta_b \, .
\label{eq:degree}
\end{equation}
We call a subspace a flat if it is an intersection of some of the subspaces $\ell_a=0$ and $\mathsf{q}_b=0$.

\begin{lemma}
\label{lem:forms}
The integral \eqref{eq:Jdef} extends to a meromorphic function of $\mu$, $\eta$ and $\zeta$. Its poles lie on the hyperplanes $\deg(\mathsf{F})+\nu=0$, where $\mathsf{F}$ is a flat different from $\mathbb{R}^M$ and $\nu=0,1,2,\ldots$
\end{lemma}

\begin{proof}
We first assume that the forms $L_c$ have no common zero except the origin. Otherwise we choose coordinates $(y',y'')$ in which the common zeros are the points with $y'=0$. The forms depend on $y'$ only, and we integrate $\psi$ over $y''$. A subspace of the space of the $y'$ corresponds to the subspace of the same codimension and the same degree which consists of the points with $y'$ in it.

The hyperplanes $L_c=0$ divide $\mathbb{R}^M$ into a finite number of closed polyhedral cones on which every $L_c$ has a fixed sign. These cones contain no line, and each of them is a finite union of simplicial cones with disjoint interiors. A simplicial cone consists of the points $y=\sum_jc_jg_j$ with $c_j\geq0$ for a basis $g_1,\ldots,g_M$. For every form the numbers $L_c(g_j)$ are either all non-negative or all non-positive, and they do not all vanish. A cone on which some $\ell_a$ is non-positive does not contribute to \eqref{eq:Jdef}. We divide every other cone into the $M!$ sectors $c_{\pi(1)}\geq c_{\pi(2)}\geq\ldots\geq c_{\pi(M)}$, where $\pi$ is a permutation, and we use the variables
\begin{equation}
c_{\pi(j)} = s_1s_2\cdots s_j \, , \qquad \prod_jdc_j = \prod_jds_j\,s_j^{\,M-j} \, .
\end{equation}
The variables $s_2,\ldots,s_M$ range over the unit interval, and $s_1$ ranges over a bounded interval, which we rescale to the unit interval. Let $\mathsf{V}_j$ be the span of $g_{\pi(1)},\ldots,g_{\pi(j-1)}$, with $\mathsf{V}_1=0$, so that ${\rm codim}\,\mathsf{V}_j=M-j+1$. For a form $L$ let $j(L)$ be the smallest $j$ with $L(g_{\pi(j)})\neq0$. Then $L$ vanishes on $\mathsf{V}_j$ if and only if $j\leq j(L)$, and
\begin{equation}
|L| = s_1\cdots s_{j(L)}\;u_L(s) \, , \qquad
u_L(s) = \sum_{j\geq j(L)}|L(g_{\pi(j)})|\;s_{j(L)+1}\cdots s_j \, .
\end{equation}
The polynomial $u_L$ is not smaller than $|L(g_{\pi(j(L))})|$ and is therefore positive on the cube. For a sum of squares let $j_b$ be the smallest of the $j(L_c)$ with $c$ in $C_b$. Then $\mathsf{q}_b$ vanishes on $\mathsf{V}_j$ if and only if $j\leq j_b$, and $\mathsf{q}_b=(s_1\cdots s_{j_b})^2\,v_b(s)$ with a polynomial $v_b$ which is positive on the cube. In a sector the integrand of \eqref{eq:Jdef} therefore takes the form
\begin{equation}
\prod_jds_j\,s_j^{\,\deg(\mathsf{V}_j)-1}\;\mathcal{R}(s) \, , \qquad
\mathcal{R} = |\det g|\,\prod_au_{\ell_a}^{\,\mu_a}\prod_bv_b^{\,\eta_b}\;\psi \, ,
\end{equation}
since the exponent of $s_j$ collects the Jacobian $M-j$ and the exponents of the forms which vanish on $\mathsf{V}_j$. The function $\mathcal{R}$ is smooth on the cube and entire in $\mu$ and $\eta$. By Lemma~\ref{lem:cube} the contribution of the sector is meromorphic with poles on the hyperplanes $\deg(\mathsf{V}_j)+\nu=0$ with $\nu\geq0$. The number of cones and sectors is finite. Finally let $\mathsf{F}$ be the smallest flat which contains $\mathsf{V}_j$. The same forms vanish on $\mathsf{V}_j$ and on $\mathsf{F}$, so that $\deg(\mathsf{V}_j)=\deg(\mathsf{F})+{\rm codim}\,\mathsf{V}_j-{\rm codim}\,\mathsf{F}$, and the difference of the codimensions is a non-negative integer. If no form vanishes on $\mathsf{V}_j$, then $\deg(\mathsf{V}_j)={\rm codim}\,\mathsf{V}_j$ is a positive integer and there is no pole.
\end{proof}

\subsection{Triangular coordinates}

The integral \eqref{eq:gram} is not of the form \eqref{eq:Jdef}, since $\det U$ vanishes on a set which is not a union of linear subspaces. We therefore change the variables. Let $\hat n_i$ be the unit vector in the direction of $X_{i\perp}$. We write $\hat n_i=\sum_{l\leq i}c_{il}\,\hat e_l$ with orthonormal vectors $\hat e_l$ and $c_{ii}\geq0$, so that $U=CC^T$ with a lower triangular matrix $C$ with rows of unit length. The matrices $U>0$ with unit diagonal correspond one to one to the matrices $C$ with $c_{ii}>0$. For fixed rows $j<i$ the map from the $c_{il}$ to the $u_{ij}$ with $l,j<i$ is linear and triangular with the diagonal entries $c_{jj}$. With $\det U=\prod_jc_{jj}^2$ we find
\begin{equation}
\prod_{i<j}du_{ij}\;(\det U)^{\frac{m-n-1}{2}} = \prod_{i=2}^n\Big(c_{ii}^{\,m-i-1}\prod_{l<i}dc_{il}\Big) \, .
\end{equation}
We now set $t_{il}=\rho_i\,c_{il}$ and $w_i=(X_{i\parallel},t_{i1},\ldots,t_{ii})$, which is a point of the unit sphere in $\mathbb{R}^{k+i}$ with $t_{ii}\geq0$. Let $d\varsigma(w_i)$ be the measure of this sphere, which is $d^kX_{i\parallel}\prod_{l<i}dt_{il}/t_{ii}$. The powers of $\rho_i$ cancel, and \eqref{eq:gram} becomes
\begin{equation}
\int\prod_{i=1}^nd\mu_d(X_i)\,F = \prod_{i=1}^n\vol_{m-i+1}\int\prod_{i=1}^nd\varsigma(w_i)\;t_{ii}^{\,m-i}\;F \, ,
\label{eq:triangular}
\end{equation}
where the integral is over the half spheres $t_{ii}\geq0$.\footnote{We compared the decomposition into sectors of Lemma~\ref{lem:forms} with closed forms at continued values of the exponents, and \eqref{eq:gram} with \eqref{eq:triangular} for three points at $m=3.5$ and $m=4.7$.} We regard all $w_i$ as points of the unit sphere in $\mathbb{R}^{k+n}$ whose last $n-i$ components vanish. Then $X_i\cdot X_j=w_i\cdot w_j$, and the chordal distance is $s_{ij}=|w_i-w_j|$. Note that the representation depends on the order of the points. It also shows the property which we used in section~\ref{sec:sphere}. If the first point is held fixed, the integral over the remaining points has the same form with one more parallel direction, and the weight $t_{ii}^{\,m-i}$ of the point $i$ is the weight of the point $i-1$ for $m-1$ transverse directions.

\subsection{Candidate lines}

Let the integrand $F$ be a product of factors $(s_{ij}^2)^{-\Dp n_{ij}-r_{ij}}$ with non-negative integers $n_{ij}$ and $r_{ij}$, of a polynomial in the $w_i$ and of sources which are smooth functions of the $X_{i\parallel}$. The integers $r_{ij}$ and the polynomial occur for vertices with derivatives, and $r_{ij}$ vanishes if $n_{ij}$ does. For a set $\mathsf{B}$ of points we write $n(\mathsf{B})$ and $r(\mathsf{B})$ for the sums of $n_{ij}$ and of $r_{ij}$ over the pairs in $\mathsf{B}$, and $i_\mathsf{B}$ for the smallest index in $\mathsf{B}$.

\begin{proposition}
\label{prop:mero}
The integral \eqref{eq:gram}, which converges for ${\rm Re}\,m\geq n$ and ${\rm Re}\,\Dp\,n_{ij}+r_{ij}\leq0$, extends to a meromorphic function of $d$ and $\Dp$. Its poles lie on the lines
\begin{equation}
\sum_\mathsf{B}\Big[d\,\big(|\mathsf{B}|-1\big)-2\,\Dp\,n(\mathsf{B})-2\,r(\mathsf{B})\Big]+\sum_{\mathsf{B}\,:\,\vartheta_\mathsf{B}=1}\big(m+1-i_\mathsf{B}\big)+\nu = 0 \, ,
\label{eq:candidates}
\end{equation}
where the sums are over the sets $\mathsf{B}$ of a partition of the $n$ points, $\vartheta_\mathsf{B}$ is $0$ or $1$ for each set, $\nu$ is a non-negative integer and $m=d-D_0$. The partition into single points with all $\vartheta_\mathsf{B}=0$ is excluded.
\end{proposition}

\begin{proof}
We cover the domain of \eqref{eq:triangular} by a finite number of open sets. For every configuration of the points we choose a unit vector $P$ in the parallel directions which differs from all $w_i$, and a neighbourhood of the configuration in which no $w_i$ approaches $P$. A finite number of these neighbourhoods cover the domain, which is compact, and we use a smooth partition of unity. In each term we use the stereographic projection from $P$, which maps $w_i$ to a point $y_i$ of $\mathbb{R}^{k-1+i}$ with the last coordinate $\tau_{ii}\geq0$. With $\Omega_i=1-P\cdot w_i$ we have
\begin{equation}
s_{ij}^2 = \Omega_i\Omega_j\,|y_i-y_j|^2 \, , \qquad t_{ii} = \Omega_i\,\tau_{ii} \, , \qquad
d\varsigma(w_i) = \Omega_i^{\,k+i-1}\,d^{k-1+i}y_i \, ,
\end{equation}
where $y_i$ is completed by zeros in the difference. Each term is therefore of the form \eqref{eq:Jdef} on the space of all $y_i$. The linear forms are $\ell_i=\tau_{ii}$ with the exponents $\mu_i=m-i$, the sums of squares are $\mathsf{q}_{ij}=|y_i-y_j|^2$ with the exponents $\eta_{ij}=-\Dp n_{ij}-r_{ij}$, and $\psi$ contains the powers of the $\Omega_i$, the polynomial, the sources and the partition of unity. It is smooth with compact support and entire in $m$ and $\Dp$. Lemma~\ref{lem:forms} gives the continuation, and the factors $\vol_{m-i+1}$ are entire.

It remains to find the flats. The subspace $\mathsf{q}_{ij}=0$ is the set $y_i=y_j$. For $i<j$ it is contained in the subspace $\tau_{jj}=0$, since $y_i$ has no component in the last direction of $y_j$. A flat is therefore given by a partition of the points into sets $\mathsf{B}$ of coinciding points, and by the choice for each set whether the last coordinate of its point of lowest index vanishes, which we denote by $\vartheta_\mathsf{B}=1$. The common point of a set $\mathsf{B}$ lies in a space of dimension $k-1+i_\mathsf{B}-\vartheta_\mathsf{B}$. With $k-1+m=d$ the degree \eqref{eq:degree} of this flat is the left-hand side of \eqref{eq:candidates} without $\nu$.
\end{proof}

\subsection{Candidates on the line}

For vertices without derivatives, $D_0=1$ and $n\leq3$ points one checks for all partitions that the line $\Dp=\tfrac12(d-2)$, on which we define the functional, is not among the candidates \eqref{eq:candidates}, so that it is not a line of poles. For $n\geq4$ the line is among the candidates, for instance for a pair of points with indices below $4$ which is joined by two propagators together with the single point of index $4$ with $\vartheta_\mathsf{B}=1$, and so are lines of fixed positive dimension such as $d=4$, and appendix~\ref{app:line} shows with a different representation that none of these candidates is a line of poles.

\subsection{Convergence and dependence on the sources}
\label{app:polar}

In this subsection we make two statements precise which appendices~\ref{app:line} and~\ref{app:gauge} use. The first is a region in which the integrals converge absolutely, and the second is the dependence of the poles on the sources. We consider the integrands $F$ of Proposition~\ref{prop:mero}. We also consider the integrands in which one line $e_0$ has an independent exponent $\Dp'$, so that the factor of its pair of points is $(s_{ij}^2)^{-\Dp'-\Dp(n_{ij}-1)-r_{ij}}$. We refer to $d$ and $\Dp$, and to $\Dp'$ in the second case, as the parameters, and we write $(s_{ij}^2)^{-\mathsf{z}_{ij}}$ for the factor of the pair $i,j$ in $F$. The exponents are independent in Lemma~\ref{lem:forms}, and the proof of Proposition~\ref{prop:mero} therefore applies to the second case without change. The candidates are then the planes \eqref{eq:candidates} in which $\Dp\,n(\mathsf{B})$ is replaced by $\Dp\,(n(\mathsf{B})-1)+\Dp'$ for the set $\mathsf{B}$ which contains both points of $e_0$. In both cases the coefficient of $d$ on the left-hand side of \eqref{eq:candidates} is a positive integer, so that the candidates form a locally finite set of lines or planes.

\begin{lemma}
\label{lem:region}
Let ${\rm Re}\,m>n-1$ and
\begin{equation}
\big(|\mathsf{B}|-1\big)\,{\rm Re}\,d > 2\sum_{i<j\,\in\,\mathsf{B}}{\rm Re}\,\mathsf{z}_{ij}
\label{eq:region}
\end{equation}
for all sets $\mathsf{B}$ of at least two points. Then the integral \eqref{eq:gram} of $F$ converges absolutely, and it is equal to the meromorphic function of Proposition~\ref{prop:mero}. The absolute convergence holds for arbitrary complex exponents $\mathsf{z}_{ij}$.
\end{lemma}

\begin{proof}
With \eqref{eq:region} and ${\rm Re}\,m>n-1\geq i_\mathsf{B}-1$ the real part of each term in the two sums of \eqref{eq:candidates} is non-negative, and it is positive for a set of at least two points and for a set with $\vartheta_\mathsf{B}=1$. Since the partition into single points with all $\vartheta_\mathsf{B}=0$ is excluded, the degree of each flat has a positive real part. In a sector of the proof of Lemma~\ref{lem:forms} the modulus of the integrand is $\prod_js_j^{\,{\rm Re}\,\deg(\mathsf{V}_j)-1}\,|\mathcal{R}|$ with a bounded function $\mathcal{R}$, and $\deg(\mathsf{V}_j)$ is a positive integer or the degree of a flat plus a non-negative integer. The integral over each sector therefore converges absolutely. The steps which lead from \eqref{eq:gram} to the sectors are changes of variables and decompositions of the domain, which hold for absolutely convergent integrals, and the integral is thus the sum over the sectors. The conditions of the lemma define a convex set of values of $m$ and of the exponents, which contains the region of Proposition~\ref{prop:mero}. The sum over the sectors is holomorphic on this set and agrees with the continuation on that region, and hence on the whole set.
\end{proof}

We now turn to the sources. Recall from section~\ref{sec:sphere} that a smooth function on the closed unit ball is the restriction of a smooth function on $\mathbb{R}^k$. For a source $\psi$ and an integer $K\geq0$ we define the norm
\begin{equation}
\|\psi\|_K = \max_{l\leq K}\;\max_{A_1,\ldots,A_l}\;\sup_{|X_\parallel|\leq1}\big|\pr_{A_1}\cdots\pr_{A_l}\psi(X_\parallel)\big| \, ,
\label{eq:norm}
\end{equation}
where the derivatives are with respect to the components of $X_\parallel$. Let $F$ be an integrand as above without its sources. We write $\mathsf{I}[\psi_1,\ldots,\psi_n]$ for the continued integral \eqref{eq:gram} of $F$ times the sources $\psi_i(X_{i\parallel})$, as a function of the parameters. It is linear in each source. We call the set of the values of the parameters at which such a function is not holomorphic its set of poles. We say that a candidate $\mathsf{Z}$ is not a line or plane of poles of such a function if the function is holomorphic in a neighbourhood of each point of $\mathsf{Z}$ which lies on no other candidate, and that it is a line or plane of poles otherwise. In Lemma~\ref{lem:poleset} below we show that the set of poles is the union of the lines or planes of poles. This is the sense in which we speak of lines and planes of poles in appendices~\ref{app:line} and~\ref{app:gauge}.

\begin{lemma}
\label{lem:polarcont}
(i) For each compact set $\mathcal{C}$ of values of the parameters there are an integer $K$, a constant $c_\mathcal{C}$ and a polynomial $\mathcal{D}$ in the parameters, which is a finite product of left-hand sides of \eqref{eq:candidates}, such that $\mathcal{D}\,\mathsf{I}[\psi_1,\ldots,\psi_n]$ is holomorphic in a neighbourhood of $\mathcal{C}$ and
\begin{equation}
\big|\mathcal{D}\;\mathsf{I}[\psi_1,\ldots,\psi_n]\big| \leq c_\mathcal{C}\,\prod_{i=1}^n\|\psi_i\|_K \qquad \text{on } \mathcal{C}
\label{eq:polarbound}
\end{equation}
for all sources of the class. Here $K$, $c_\mathcal{C}$ and $\mathcal{D}$ depend on $F$ and on $\mathcal{C}$, and not on the sources. (ii) Let $\mathsf{Z}$ be a candidate, let $\mathsf{w}$ be a left-hand side of \eqref{eq:candidates} which vanishes on $\mathsf{Z}$, and let $\mathcal{C}$ be a compact subset of $\mathsf{Z}$ which meets no other candidate. We use $\mathsf{w}$ and affine coordinates $\mathsf{p}$ on $\mathsf{Z}$ as coordinates of the parameters. Then there are integers $K$ and $a\geq0$, a radius $\mathsf{r}_0>0$ and a constant $c_\mathcal{C}$ such that for $\mathsf{p}$ in $\mathcal{C}$ and $0<|\mathsf{w}|<\mathsf{r}_0$
\begin{equation}
\mathsf{I}[\psi_1,\ldots,\psi_n] = \sum_{l\geq-a}\mathsf{w}^{\,l}\;\mathcal{S}_l[\psi_1,\ldots,\psi_n](\mathsf{p}) \, , \qquad
\big|\mathcal{S}_l[\psi_1,\ldots,\psi_n](\mathsf{p})\big| \leq c_\mathcal{C}\;\mathsf{r}_0^{-l}\,\prod_{i=1}^n\|\psi_i\|_K \, .
\label{eq:laurent}
\end{equation}
The coefficients $\mathcal{S}_l$ are linear in each source and holomorphic in $\mathsf{p}$. (iii) Let $\psi_1,\ldots,\psi_n$ be sources of the class, and let $\mathsf{T}_1,\ldots,\mathsf{T}_n$ be sets of sources such that for each $K$ the source $\psi_i$ is a limit in the norm $\|\cdot\|_K$ of finite linear combinations of elements of $\mathsf{T}_i$. If a candidate $\mathsf{Z}$ is not a line or plane of poles of $\mathsf{I}[\psi'_1,\ldots,\psi'_n]$ for all $\psi'_i$ in $\mathsf{T}_i$, then it is not a line or plane of poles of $\mathsf{I}[\psi_1,\ldots,\psi_n]$. (iv) Let $\mathsf{Z}$, $\mathcal{C}$, $K$ and $c_\mathcal{C}$ be as in (ii), and assume that $\mathsf{Z}$ is not a line or plane of poles of $\mathsf{I}[\psi_1,\ldots,\psi_n]$ for all sources of the class. Then the value of $\mathsf{I}[\psi_1,\ldots,\psi_n]$ at a point $\mathsf{p}$ of $\mathcal{C}$ is $\mathcal{S}_0[\psi_1,\ldots,\psi_n](\mathsf{p})$, and
\begin{equation}
\big|\mathsf{I}[\psi_1,\ldots,\psi_n](\mathsf{p})\big| \leq c_\mathcal{C}\,\prod_{i=1}^n\|\psi_i\|_K \qquad \text{for } \mathsf{p} \text{ in } \mathcal{C}
\label{eq:valuebound}
\end{equation}
for all sources of the class. Moreover, let the sources $\psi_i(\mathsf{p})$ depend on the point $\mathsf{p}$ of $\mathsf{Z}$, such that near each point of $\mathcal{C}$ they are given by power series in the coordinates $\mathsf{p}$ whose coefficients are sources of the class and which converge in the norm $\|\cdot\|_K$. Then $\mathsf{I}[\psi_1(\mathsf{p}),\ldots,\psi_n(\mathsf{p})](\mathsf{p})$ is a holomorphic function of $\mathsf{p}$ in the interior of $\mathcal{C}$.
\end{lemma}

\begin{proof}
We first add a bound to Lemma~\ref{lem:cube}. For a function $\varphi$ of one variable we multiply \eqref{eq:taylor} by $\prod_{\nu<K}(z+\nu)$. We have $|\varphi_K|\leq\sup|\varphi^{(K)}|/K!$, and the last integral in \eqref{eq:taylor} is bounded by $\sup|\varphi_K|/({\rm Re}\,z+K)$. The product is therefore holomorphic for ${\rm Re}\,z>-K$, and on compact subsets of this half plane it is bounded by a constant times the largest of the numbers $\sup|\varphi^{(\nu)}|$ with $\nu\leq K$. The maps from $\varphi$ to $\varphi^{(\nu)}(0)$ and to $\varphi_K$ commute with the derivatives with respect to the other variables and do not increase their suprema. We apply the step to the $M$ variables in turn. It follows that $\prod_j\prod_{\nu<K}(z_j+\nu)\,\mathcal{I}(z,\zeta)$ is holomorphic for ${\rm Re}\,z_j>-K$, and that on compact subsets of this region and of the parameters $\zeta$ it is bounded by a constant times
\begin{equation}
\max_{\nu_1,\ldots,\nu_M\leq K}\;\sup\,\big|\pr_{s_1}^{\,\nu_1}\cdots\pr_{s_M}^{\,\nu_M}\mathcal{R}(s,\zeta)\big| \, ,
\label{eq:cubebound}
\end{equation}
where the supremum is over the cube and over the compact set of the $\zeta$.

We now follow the proof of Proposition~\ref{prop:mero}. The integral is a finite sum, over the terms of the partition of unity, the cones and the sectors, of integrals of the form of Lemma~\ref{lem:cube} with entire prefactors. The exponents are $z_j=\deg(\mathsf{V}_j)$, which are affine functions of the parameters, and we choose $K$ such that ${\rm Re}\,z_j>-K$ in a neighbourhood of $\mathcal{C}$ for all of them. In a sector the function $\mathcal{R}$ is the product of $|\det g|\prod_au_{\ell_a}^{\,\mu_a}\prod_bv_b^{\,\eta_b}$ and of the function $\psi$ of \eqref{eq:Jdef} at the point $y=\sum_jc_jg_j$, whose components are polynomials in $s$. The polynomials $u_{\ell_a}$ and $v_b$ are positive on the cube. For parameters in $\mathcal{C}$ the number \eqref{eq:cubebound} is thus bounded by a constant times the largest supremum of the derivatives of $\psi$ with respect to $y$ of order at most $KM$. The integration over the variables $y''$ in the first step of the proof of Lemma~\ref{lem:forms} does not change this. In the proof of Proposition~\ref{prop:mero} the function $\psi$ is the product of the sources $\psi_i(X_{i\parallel})$ and of a function which does not depend on the sources, which is smooth with compact support and entire in the parameters. The components of $X_{i\parallel}$ are smooth functions of $y_i$. By the rule of Leibniz and the chain rule the derivatives of $\psi$ of order at most $KM$ are therefore bounded by a constant times $\prod_i\|\psi_i\|_{KM}$. This proves (i) with $KM$ in place of $K$. The polynomial $\mathcal{D}$ is the product of the factors $z_j+\nu$ with $\nu<K$ over all terms of the sum, in which we omit the factors which are constant. As shown at the end of the proof of Lemma~\ref{lem:forms}, each of the other factors is a left-hand side of \eqref{eq:candidates}.

For (ii) we choose $\mathsf{r}_0$ so small that the set $\mathcal{C}'$ of the points with $\mathsf{p}$ in $\mathcal{C}$ and $|\mathsf{w}|\leq\mathsf{r}_0$ meets no candidate other than $\mathsf{Z}$. This is possible since the candidates are locally finite. We apply (i) to $\mathcal{C}'$. A factor of $\mathcal{D}$ which vanishes at a point of $\mathcal{C}'$ vanishes on $\mathsf{Z}$ and is a multiple of $\mathsf{w}$. Hence $\mathcal{D}=\mathsf{w}^a\,\mathcal{D}'$ with an integer $a\geq0$ and a polynomial $\mathcal{D}'$ without zeros on $\mathcal{C}'$, and $\mathsf{w}^a\,\mathsf{I}$ is holomorphic in a neighbourhood of $\mathcal{C}'$ and bounded on $\mathcal{C}'$ by a constant times $\prod_i\|\psi_i\|_K$. The coefficients of the expansion in powers of $\mathsf{w}$ are
\begin{equation}
\mathcal{S}_l[\psi_1,\ldots,\psi_n](\mathsf{p}) = \frac{1}{2\pi i}\oint_{|\mathsf{w}|=\mathsf{r}_0}\frac{d\mathsf{w}}{\mathsf{w}^{\,l+1}}\;\mathsf{I}[\psi_1,\ldots,\psi_n] \, .
\end{equation}
They vanish for $l<-a$, they are holomorphic in $\mathsf{p}$ and linear in each source, and they obey the bound in \eqref{eq:laurent}, where the constant contains the factor $\mathsf{r}_0^{-a}$.

For (iii) let $\mathsf{p}_0$ be a point of $\mathsf{Z}$ which lies on no other candidate, and let $\mathcal{C}$ be a closed ball in $\mathsf{Z}$ around $\mathsf{p}_0$ which meets no other candidate. By \eqref{eq:laurent} the function $\mathsf{I}[\psi_1,\ldots,\psi_n]$ is holomorphic in a neighbourhood of the interior points of $\mathcal{C}$ if and only if the coefficients $\mathcal{S}_l[\psi_1,\ldots,\psi_n]$ with $-a\leq l<0$ vanish on $\mathcal{C}$. By assumption they vanish for sources $\psi'_i$ in $\mathsf{T}_i$. Since they are linear in each source, they vanish for finite linear combinations of such sources. By the bound in \eqref{eq:laurent} the coefficient $\mathcal{S}_l[\psi_1,\ldots,\psi_n]$ is the limit of the coefficients of these combinations, if we replace the sources by their approximations one after the other. It therefore vanishes on $\mathcal{C}$. This holds for each point $\mathsf{p}_0$, which proves (iii).

For (iv) the coefficients $\mathcal{S}_l$ with $l<0$ vanish on $\mathcal{C}$ by the assumption. The expansion \eqref{eq:laurent} is then a power series in $\mathsf{w}$, which converges for $|\mathsf{w}|<\mathsf{r}_0$ by the bound in \eqref{eq:laurent}. Its value at $\mathsf{w}=0$ is $\mathcal{S}_0$, and \eqref{eq:valuebound} is this bound for $l=0$. For sources which depend on $\mathsf{p}$ we insert their power series around a point $\mathsf{p}_0$ of the interior of $\mathcal{C}$ into $\mathcal{S}_0$, which is linear in each source and holomorphic in $\mathsf{p}$. The norms of the terms of a convergent power series are bounded by a geometric sequence. By \eqref{eq:valuebound} the resulting series of holomorphic functions therefore converges uniformly in a neighbourhood of $\mathsf{p}_0$, and its sum is holomorphic there. By \eqref{eq:valuebound} the sum is the value of $\mathcal{S}_0$ on the sources $\psi_i(\mathsf{p})$.
\end{proof}

The next lemma relates the lines or planes of poles to the set of poles. We use it in appendices~\ref{app:line} and~\ref{app:gauge} to pass from the candidates of a second representation to the lines or planes of poles, and in section~\ref{sec:continuation} for the restriction of the integrals to the line \eqref{eq:line}.

\begin{lemma}
\label{lem:poleset}
Let $\mathsf{I}$ be the function $\mathsf{I}[\psi_1,\ldots,\psi_n]$ for given sources of the class. (i) If $\mathsf{I}$ is holomorphic in a neighbourhood of one point of a candidate $\mathsf{Z}$ which lies on no other candidate, then $\mathsf{Z}$ is not a line or plane of poles of $\mathsf{I}$. (ii) Each point $\mathsf{p}_0$ of the space of the parameters has a neighbourhood in which $\prod_{\mathsf{Z}'}\mathsf{w}_{\mathsf{Z}'}^{\,a_{\mathsf{Z}'}}\,\mathsf{I}$ is holomorphic. Here the product is over the lines or planes of poles $\mathsf{Z}'$ of $\mathsf{I}$ which contain $\mathsf{p}_0$, $\mathsf{w}_{\mathsf{Z}'}$ is a left-hand side of \eqref{eq:candidates} which vanishes on $\mathsf{Z}'$, and the $a_{\mathsf{Z}'}$ are non-negative integers which are bounded independently of the sources. (iii) The set of poles of $\mathsf{I}$ is the union of its lines or planes of poles. In particular $\mathsf{I}$ is holomorphic in a neighbourhood of each point which lies on no line or plane of poles. The restriction of $\mathsf{I}$ to a line or plane in the space of the parameters which is contained in no line or plane of poles is a meromorphic function on it, whose poles lie on its intersections with the lines or planes of poles. The three statements also hold for finite sums of such functions with coefficients which are polynomials in the parameters, with the union of their candidates.
\end{lemma}

\begin{proof}
For (i) let $\mathsf{Z}^\circ$ be the set of the points of $\mathsf{Z}$ which lie on no other candidate. The other candidates meet $\mathsf{Z}$ in a locally finite set of points, or of lines if $\mathsf{Z}$ is a plane, so that $\mathsf{Z}^\circ$ is open in $\mathsf{Z}$, dense and connected. The coefficients $\mathcal{S}_l$ of \eqref{eq:laurent} are integrals over a small circle around $\mathsf{Z}$, and they do not depend on the choice of $\mathcal{C}$ and of $\mathsf{r}_0$. They are therefore holomorphic functions on $\mathsf{Z}^\circ$. If $\mathsf{I}$ is holomorphic near a point of $\mathsf{Z}^\circ$, the coefficients with $l<0$ vanish in a neighbourhood of this point in $\mathsf{Z}$ and hence on $\mathsf{Z}^\circ$. As in the proof of the fourth statement of Lemma~\ref{lem:polarcont} the function $\mathsf{I}$ is then holomorphic near each point of $\mathsf{Z}^\circ$.

For (ii) we apply the first statement of Lemma~\ref{lem:polarcont} to a closed ball around $\mathsf{p}_0$ which meets only the candidates through $\mathsf{p}_0$. A factor of $\mathcal{D}$ which vanishes at a point of the ball is a constant multiple of the form $\mathsf{w}_{\mathsf{Z}'}$ of one of these candidates. Hence $\tilde{\mathsf{I}}=\prod_{\mathsf{Z}'}\mathsf{w}_{\mathsf{Z}'}^{\,a_{\mathsf{Z}'}}\,\mathsf{I}$ is holomorphic in the interior of the ball, where the product is over all candidates through $\mathsf{p}_0$ and the integers $a_{\mathsf{Z}'}\geq0$ are given by $\mathcal{D}$, which does not depend on the sources. Let $\mathsf{Z}'$ be one of these candidates which is not a line or plane of poles, with $a_{\mathsf{Z}'}\geq1$. Then $\mathsf{I}$ is holomorphic near the points of $\mathsf{Z}'$ in the ball which lie on no other candidate, so that $\tilde{\mathsf{I}}$ vanishes at these points. They are dense in the part of $\mathsf{Z}'$ in the ball, and $\tilde{\mathsf{I}}$ therefore vanishes on this part. A holomorphic function on a ball which vanishes on the zeros of the affine function $\mathsf{w}_{\mathsf{Z}'}$ is $\mathsf{w}_{\mathsf{Z}'}$ times a holomorphic function, by the formula of Taylor in the variable $\mathsf{w}_{\mathsf{Z}'}$. We can thus lower $a_{\mathsf{Z}'}$ by one, and after a finite number of steps the product contains the lines or planes of poles only.

For (iii) we note that by (ii) the function $\mathsf{I}$ is holomorphic near a point which lies on no line or plane of poles. Conversely, let $\mathsf{Z}$ be a line or plane of poles. By (i) the function $\mathsf{I}$ is not holomorphic at any point of $\mathsf{Z}^\circ$. The set of the points at which $\mathsf{I}$ is holomorphic is open and $\mathsf{Z}^\circ$ is dense in $\mathsf{Z}$, so that all points of $\mathsf{Z}$ are poles. For the restriction we use (ii) again. Near each of its points the restriction is the quotient of the restriction of the holomorphic function $\prod_{\mathsf{Z}'}\mathsf{w}_{\mathsf{Z}'}^{\,a_{\mathsf{Z}'}}\,\mathsf{I}$ and of the restrictions of the $\mathsf{w}_{\mathsf{Z}'}$, which do not vanish identically. For finite sums the proofs are the same, since the first two statements of Lemma~\ref{lem:polarcont} hold for them with the product of the polynomials $\mathcal{D}$ of their terms.
\end{proof}

The lemma reduces a statement on the lines or planes of poles for all sources to the same statement for a set of sources whose linear combinations are dense. We use two such sets.

\begin{lemma}
\label{lem:dense}
Let $K\geq0$. Each source $\psi$ of the class is a limit in the norm $\|\cdot\|_K$ of polynomials in the components of $X_\parallel$. If $\psi$ vanishes in a neighbourhood of a unit vector $P$ of the parallel directions, and $h\geq0$ is an integer, then $\psi$ is also a limit in this norm of functions $(1-P\!\cdot\!X)^h\,\mathsf{f}(X_\parallel)$ with polynomials $\mathsf{f}$.
\end{lemma}

\begin{proof}
We extend $\psi$ to a smooth function on $\mathbb{R}^k$ with compact support. For $t>0$ let $\tilde\psi_t$ be its convolution with the Gaussian $(4\pi t)^{-k/2}e^{-|X_\parallel|^2/4t}$. The derivatives of $\tilde\psi_t$ are the convolutions of the derivatives of $\psi$, which are uniformly continuous, and for $t\to0$ they therefore converge uniformly to the derivatives of $\psi$. For fixed $t$ the function $\tilde\psi_t$ extends to an entire function on $\mathbb{C}^k$, because $\psi$ has compact support. Its Taylor polynomials at the origin thus converge to it on the ball, uniformly and together with all derivatives. This proves the first statement. For the second we note that $1-P\!\cdot\!X$ is positive on the closed ball except at $P$. Let $\psi$ vanish at the points of the closed ball with $|X_\parallel-P|<\delta$. The remaining points of the closed ball form a compact set which does not meet the hyperplane $P\!\cdot\!X=1$. Let $\chi$ be a smooth function on $\mathbb{R}^k$ which equals one near this compact set and vanishes near the hyperplane. Then $\chi\,\psi/(1-P\!\cdot\!X)^h$, with the extension of $\psi$ in the numerator, is smooth on $\mathbb{R}^k$ and equals $\psi/(1-P\!\cdot\!X)^h$ on the closed ball. The function $\psi/(1-P\!\cdot\!X)^h$ is therefore a source of the class, and by the first statement it is a limit of polynomials $\mathsf{f}$ in the norm $\|\cdot\|_K$. The multiplication by $(1-P\!\cdot\!X)^h$ is continuous in this norm by the rule of Leibniz.
\end{proof}

The polynomials are finite linear combinations of monomials, and the functions $(1-P\!\cdot\!X)^h\,\mathsf{f}$ are finite linear combinations of the functions \eqref{eq:hsource} of appendix~\ref{app:line}. The derivatives in the norms \eqref{eq:norm} are needed. For an integral \eqref{eq:Jdef} in two variables with one linear form and one sum of squares and for oscillating test functions of fixed supremum, the residue at the second candidate of a flat grows linearly with the frequency.\footnote{For this integral we also checked that the coefficients $\mathcal{S}_l$ of Lemma~\ref{lem:polarcont}, computed from the sectors and \eqref{eq:taylor}, agree with their closed forms.}

Let us state what the fourth statement of Lemma~\ref{lem:polarcont} gives for the functional on the line \eqref{eq:line}, which is not a line of poles by Proposition~\ref{prop:linecurrent}. The proof of this proposition uses only the first three statements of the lemma. We use $\vep=4-d$ as the coordinate on the line. By Lemma~\ref{lem:poleset} the terms of the functional $W_0^S$ are meromorphic functions of $\vep$, and their poles lie at the points at which a line of poles crosses the line and at the poles $d=2,0,-2,\ldots$ of the normalisation $C_\phi$ of the propagators. Let $\mathcal{C}$ be a compact set of values of $\vep$ which contains no point at which another candidate crosses the line and no pole of $C_\phi$. Each term of $W_0^S$ is then bounded on $\mathcal{C}$ by a constant times the product of the norms \eqref{eq:norm} of its sources. For an integral for which the line is not among the candidates this is contained in the first statement of the lemma. Each term is holomorphic in the interior of $\mathcal{C}$ also for sources which are power series in $\vep$ that converge in these norms. The source $\Upsilon=a_{0A}^T\Pi_{AB}\,a_{0B}$ of section~\ref{app:conf} is bilinear in the gauge field and continuous in these norms, so that this includes the terms of second order in the gauge field. At $\vep=0$ the order of the pole of a term is bounded independently of the sources, by the second statement of Lemma~\ref{lem:poleset}. The same bound holds for sources which are power series in $\vep$. To see this we approximate each source by its Taylor polynomials in $\vep$, for which the integral is a finite sum of powers of $\vep$ times integrals with fixed sources. Multiplied by the power of $\vep$ which removes the pole, these approximants are holomorphic in a small disk around $\vep=0$. By the bound and by multilinearity they converge uniformly on the boundary circle, and by the maximum principle they converge uniformly in the disk.

The same statements hold for the flat functional $W_0^{\rm fl}$ of section~\ref{sec:chart} and flat sources with supports in a fixed compact set. Indeed, by \eqref{eq:chart} the flat functional of the sources $g_0$, $a_0$ and $\sM_0$ is the functional on the sphere of the sources $\Omega^{-\vep}g_0$, $a_0$ and $\Omega^{-2}\sM_0$, which vanish near the point $P$ of the chart. On the image of the compact set the function $\Omega=1-P\!\cdot\!X$ is bounded from below by a positive constant. The power series of $\Omega^{-\vep}$ in $\vep$ therefore converges in each norm there, and the map from the flat sources to the sources on the sphere is linear and continuous, if the flat sources are measured by the suprema of their derivatives up to the same order. In particular, let a family of flat sources depend analytically on $\vep$ in the sense of section~\ref{sec:hypothesis}. Then each term of the flat functional, evaluated on the member of the family with the same $\vep$, is a holomorphic function of $\vep$ in a punctured neighbourhood of $\vep=0$ with at most a pole at $\vep=0$. This is the function to which Hypothesis~\ref{hyp:finite} refers for such families.

\subsection{Conformal covariance}
\label{app:conf}

In this subsection we prove Proposition~\ref{prop:cov}. For the map \eqref{eq:boost} the image $X'_\parallel$ depends only on $X_\parallel$, and we have $\rho'=\Omega\,\rho$, $u'_{ij}=u_{ij}$ and $s'^{\,2}_{ij}=\Omega_i\Omega_j\,s_{ij}^2$ with $\Omega_i=\Omega(X_i)$. The map $X_\parallel\to X'_\parallel$ of the ball is projective, and its Jacobian is $\Omega^{k+1}$. Since $(k+1)+(m-2)=d$, the measure in \eqref{eq:gram} changes by the factor $\prod_i\Omega_i^{\,d}$ for every complex $d$. For a field with the propagator $s^{-2\Dp}$ the quartic and the quadratic operators have the dimensions $4\Dp$ and $2\Dp$. In the region of convergence the change of variables therefore gives, for the terms $W_n$ of order $n$ with these sources,
\begin{equation}
W_n\big[\Omega^{d-4\Dp}\,g_0\!\circ\!f,\;\Omega^{d-2\Dp}\,\sM_0\!\circ\!f\,;\,d,\Dp\big]
= W_n\big[g_0,\sM_0\,;\,d,\Dp\big] \, .
\label{eq:cov2}
\end{equation}
Both sides are meromorphic in $d$ and $\Dp$, so that the identity holds for all values.\footnote{For the two-point integral of the quartic operators we tested the continued identity numerically at $(d,\Dp)=(3.9,0.95)$ and $(3.6,0.8)$, which lie outside the region of convergence.} On the line \eqref{eq:line} the weights are $\vep$ and $2$, and away from the poles we obtain \eqref{eq:covS} without the gauge field.

We now include the gauge field. The vertices are those of appendix~\ref{app:gauge}. The term of second order in the gauge field in the action is a vertex without derivatives, and we give it a source $\Upsilon$ which is independent of the gauge field away from the line \eqref{eq:line}. Let $\mathsf{J}_{AB}=\pr X'^A/\pr X^B$ be the matrix of the derivatives of the map $X_\parallel\to X'_\parallel$ of the ball, and let $\Pi_{AB}(X)=\delta_{AB}-X_AX_B$ for the parallel indices. Then
\begin{equation}
\mathsf{J}_{AC}\,\Pi_{CD}(X)\,\mathsf{J}_{BD} = \Omega(X)^2\,\Pi_{AB}(X') \, .
\label{eq:JPiJ}
\end{equation}
Indeed, for integer $m$ the entries of $\Pi_{AB}$ are the scalar products of the gradients of the coordinates $X^A$ and $X^B$ on the sphere, and $f$ is a conformal map with the factor $\Omega$. Neither side of \eqref{eq:JPiJ} depends on $m$.\footnote{We verified \eqref{eq:JPiJ}, the Jacobians of the map \eqref{eq:boost} and of the stereographic projection, and the relations $\rho'=\Omega\,\rho$ and $s'^{\,2}_{ij}=\Omega_i\Omega_j\,s_{ij}^2$ with computer algebra.} The pullback of the one-form $a=a_A(X_\parallel)\,dX^A$ is $f^*a$ with the components $(f^*a)_B(X)=a_A(X')\,\mathsf{J}_{AB}$, which is again a one-form of the class. We write $\mathsf{v}_i[a]=\Pi_{AB}(X_i)\,a_B(X_{i\parallel})\,\pr/\pr X_{i\parallel}^A$ for the operator \eqref{eq:vfield} of appendix~\ref{app:line}. By \eqref{eq:JPiJ} and the chain rule we have
\begin{equation}
\big(\mathsf{v}_i[a]\,F\big)\!\circ\!f = \Omega_i^{-2}\;\mathsf{v}_i[f^*a]\,\big(F\!\circ\!f\big) \, , \qquad
(f^*a)_A^T\,\Pi_{AB}\,(f^*a)_B = \Omega^2\,\big(a_A^T\,\Pi_{AB}\,a_B\big)\!\circ\!f
\label{eq:vcov}
\end{equation}
for each function $F$ of the $X_{i\parallel}$ and the $u_{ij}$, where $f$ acts on the $X_{i\parallel}$ and does not change the $u_{ij}$. The dimension does not enter these identities. The factor of a vertex of the current at the point $i$ is the operator $\mathsf{v}_i^{(1)}[a]-\mathsf{v}_i^{(2)}[a]$ acting on the propagators of its two lines, for each component of the antisymmetric matrix $a_A$, see section~\ref{app:currents}. This holds after the sum over the two ways to attach the two lines to the two legs of the vertex, which differ by the exchange of the indices of the antisymmetric matrix $a_A$. For one of the two ways alone the covariance does not hold. Under $f$ each of the two propagators is multiplied by $\Omega_i^{-\Dp}$ and by the factor of its other end. The derivatives of the two factors $\Omega_i^{-\Dp}$ cancel in the difference. The operators at different points commute, so that this holds for any number of vertices of the current, also if a line joins two of them. By the first identity in \eqref{eq:vcov} the factor of the vertex with the one-form $a$ at the image points is therefore $\Omega_i^{-2-2\Dp}$ times the factor of the vertex with the one-form $f^*a$, times the factors of the other ends of its lines. Together with the factor $\Omega_i^{\,d}$ of the measure the change of variables gives, for ${\rm Re}\,m\geq n$ and ${\rm Re}\,\Dp<-2$ and then for all values of $d$ and $\Dp$,
\begin{multline}
W_n\big[\Omega^{d-4\Dp}\,g_0\!\circ\!f,\;\Omega^{d-2-2\Dp}\,f^*a_0,\;\Omega^{d-2\Dp}\,\sM_0\!\circ\!f,\;\Omega^{d-2\Dp}\,\Upsilon\!\circ\!f\,;\,d,\Dp\big] \\
= W_n\big[g_0,a_0,\sM_0,\Upsilon\,;\,d,\Dp\big] \, .
\label{eq:cov3}
\end{multline}
The gauge field thus has the weight $d-2-2\Dp$, and the source $\Upsilon$ has the weight $d-2\Dp$ of the mass source. Let now $\Upsilon=a_{0A}^T\,\Pi_{AB}\,a_{0B}$, which is $\gamma^{\mu\nu}a_{0\mu}^Ta_{0\nu}$. By the second identity in \eqref{eq:vcov} the sources on the left-hand side of \eqref{eq:cov3} are related in the same way only if $2\,(d-2-2\Dp)+2=d-2\Dp$. This is the line \eqref{eq:line}, on which the weight of the gauge field vanishes and the weight of $\Upsilon$ is $2$. Off the line \eqref{eq:cov3} therefore does not give a covariance of the terms $W_n$ with this value of $\Upsilon$, and for this reason we keep $\Upsilon$ as an independent source there. The line is not a line of poles of the terms with the four independent sources by Proposition~\ref{prop:linecurrent}. Away from the poles the restriction of \eqref{eq:cov3} to the line, with $\Upsilon=a_{0A}^T\,\Pi_{AB}\,a_{0B}$, is \eqref{eq:covS}.

We finally justify the first-order form of \eqref{eq:covS}, which we use in sections~\ref{sec:sphere} and~\ref{gs:sec:msward} and in appendix~\ref{app:finite}. For small $|\zeta|$ the map \eqref{eq:boost} is a smooth function of $\zeta$ and $X_\parallel$ in a neighbourhood of the closed ball, and it maps the closed ball to itself. For sources of the class the functions $\Omega^{\vep}\,g_0\!\circ\!f$ and $\Omega^{2}\,\sM_0\!\circ\!f$ and the components of $f^*a_0$ are therefore smooth functions of $\zeta$ and $X_\parallel$, and their difference quotients in $\zeta$ converge in each of the norms \eqref{eq:norm}. The terms of the functional are multilinear in the sources. By the fourth statement of Lemma~\ref{lem:polarcont} they are bounded by products of these norms, at each point of the line \eqref{eq:line} at which no other candidate crosses it. Hence the left-hand side of \eqref{eq:covS} can be differentiated at $\zeta=0$ term by term, with the derivative acting on one source at a time. This gives the invariance of $W_0^S$ under the variation \eqref{eq:barevar}, as an identity of meromorphic functions of $\vep$.

\section{Exclusion of lines of poles}
\label{app:line}

In this appendix we show that for any number of points the line $\Dp=\tfrac12(d-2)$ is not a line of poles of the integrals \eqref{eq:gram} of Proposition~\ref{prop:linecurrent}, which are those of the functional, and that these integrals have no line of poles at a fixed positive dimension. The definition of the bare functional in section~\ref{sec:continuation} as the value of the integrals on this line requires the first statement. By the second statement, and since the normalisation $C_\phi$ of the propagators is regular at $d=d_c$, the poles of the functional at $\vep=0$ come from lines which cross this line, as we stated there. The proof treats the vertices without derivatives, which are the vertices of the couplings and of the mass source, together with the vertices of the current.

The candidates of appendix~\ref{app:mero} which contain the line, and the candidates at fixed positive dimension, are therefore not poles. The proof uses a second representation of the integrals, by Schwinger parameters, in which the dimension appears only in exponents, in entire prefactors and in polynomial coefficients, and in which the loci of linearly dependent transverse components do not occur.

The decomposition of the range of the Schwinger parameters into sectors which we use goes back to Hepp and Speer~\cite{Hepp:1966eg,Speer:1968qxh,Speer:1975dc}, and the continuation in the exponent of the propagator is related to the analytic regularisation of~\cite{Speer:1968qxh}.

\subsection{Monomial sources}

We first reduce the statement to special sources. We use the notion of a line of poles of section~\ref{app:polar}. By Proposition~\ref{prop:mero} and Lemma~\ref{lem:poleset} the set of poles of the integral is a union of candidates \eqref{eq:candidates}, which are its lines of poles. By Lemma~\ref{lem:polarcont} the coefficients of the Laurent expansion of the integral transverse to a candidate are multilinear in the sources, and they are bounded by a product of the norms \eqref{eq:norm} of the sources with a finite number of derivatives. By Lemma~\ref{lem:dense} each source of the class is a limit in these norms of polynomials in the components of $X_\parallel$, which are finite linear combinations of monomials. If a candidate is not a line of poles of the integral for all monomial sources, it is therefore not a line of poles for all sources of the class. This is the third statement of Lemma~\ref{lem:polarcont}, in which the sets $\mathsf{T}_i$ consist of the monomials. It is thus sufficient to consider sources which are monomials in the components of $X_\parallel$.

We use the stereographic chart of section~\ref{sec:chart} from a unit vector $P$ of the parallel directions, with $\Omega_i=2/(1+x_i^2)$. Then $P\cdot X_i=1-\Omega_i$, and the other parallel components of $X_i$ are $\Omega_i\,x_i^A$. A monomial source at the point $i$ is a sum of terms $\Omega_i^{\kappa_i}\prod_A(x_i^A)^{\mu_{iA}}$ with integers $\kappa_i\geq\mu_i$, where $\mu_i=\sum_A\mu_{iA}$. The integral is therefore a sum of terms
\begin{equation}
\int\prod_id^dx_i\;\Omega_i^{\,c_i}\prod_A(x_i^A)^{\mu_{iA}}\;\prod_{i<j}|x_i-x_j|^{-2\Dp n_{ij}} \, , \qquad
c_i = d-\Dp\,n_i+\kappa_i \, ,
\label{eq:flatterm}
\end{equation}
where $n_i=\sum_jn_{ij}$ is the number of propagators at the point $i$, and the integral over the $x_i$ is the flat form of \eqref{eq:gram}.

\subsection{Schwinger parameters}

For ${\rm Re}\,c_i>0$ and ${\rm Re}\,\Dp>0$ we write
\begin{equation}
\Omega_i^{\,c_i} = \frac{2^{c_i}}{\Gamma(c_i)}\int_0^\infty\!d\alpha_{i\infty}\,\alpha_{i\infty}^{\,c_i-1}e^{-\alpha_{i\infty}(1+x_i^2)} \, , \qquad
|x|^{-2\mathsf{z}} = \frac{1}{\Gamma(\mathsf{z})}\int_0^\infty\!d\alpha\,\alpha^{\mathsf{z}-1}e^{-\alpha x^2} \, .
\end{equation}
The integral over the $x_i$ is then Gaussian, with the matrix $\mathsf{K}=\mathsf{L}(\alpha)+{\rm diag}(\alpha_{i\infty})$, where $\mathsf{L}(\alpha)$ is the Laplace matrix of the graph of the propagators with the weights $\alpha_{ij}$. For sources $J_i$ in the parallel directions the integral in $d$ dimensions is
\begin{equation}
\int\prod_id^dx_i\;e^{-\sum_{ij}\mathsf{K}_{ij}\,x_i\cdot x_j+\sum_iJ_i\cdot x_i}
= \pi^{nd/2}\,(\det\mathsf{K})^{-d/2}\,e^{\frac14\sum_{ij}(\mathsf{K}^{-1})_{ij}J_i\cdot J_j} \, ,
\label{eq:gauss}
\end{equation}
which follows from the Gaussian integral for one vector in~\cite{Wilson:1972cf} by iteration. In the form \eqref{eq:gram} of the integral it follows from the classical formula of Ingham and Siegel~\cite{Ingham:1933,Siegel:1935} for the integral of $(\det Q)^{s}\,e^{-{\rm tr}\,\mathsf{K}Q}$ over the positive matrices $Q$. Derivatives with respect to the $J_i$ give the factors $x_i^A$. The term \eqref{eq:flatterm} is therefore equal to
\begin{equation}
\pi^{nd/2}\prod_i\frac{2^{c_i}}{\Gamma(c_i)}\prod_{i<j}\frac{1}{\Gamma(\Dp n_{ij})}
\int\prod_id\alpha_{i\infty}\,\alpha_{i\infty}^{\,c_i-1}e^{-\alpha_{i\infty}}\prod_{i<j}d\alpha_{ij}\,\alpha_{ij}^{\,\Dp n_{ij}-1}\;
\cU^{-d/2}\,\mathcal{N}_\alpha \, ,
\label{eq:param}
\end{equation}
where $\cU=\det\mathsf{K}$ and $\mathcal{N}_\alpha$ is a sum of products of $n_{\rm f}=\tfrac12\sum_i\mu_i$ entries of $\mathsf{K}^{-1}$ with coefficients which do not depend on $d$. The products are over the pairs $i,j$ with $n_{ij}\geq1$. The representation holds for $0<{\rm Re}\,\Dp<1$ and sufficiently large ${\rm Re}\,d$. In this region the integral \eqref{eq:gram} and the integrals over the parameters converge absolutely, and the exchange of the integrations is justified. We prove this in Lemma~\ref{lem:schwinger} at the end of this subsection, in the generality which we need for the vertices of the current. Both sides are meromorphic, so that \eqref{eq:param} gives the same continuation.

Let $\Gamma_\infty$ be the graph whose vertices are the $n$ points and one further vertex $\infty$. Its edges are the pairs with $n_{ij}\geq1$, with the variables $\alpha_{ij}$, and the pairs $(i,\infty)$, with the variables $\alpha_{i\infty}$. By the matrix-tree theorem $\cU$ is the sum over the spanning trees of $\Gamma_\infty$ of the products of the variables of their edges. Moreover $(\mathsf{K}^{-1})_{ij}=\cU_{ij}/\cU$, where $\cU_{ij}$ is the same sum over the spanning forests with two trees, of which one contains $\infty$ and the other contains $i$ and $j$. The integrand of \eqref{eq:param} is therefore a sum of terms
\begin{equation}
\prod_e\alpha_e^{\,\mathsf{z}_e+k_e-1}\;e^{-\sum_i\alpha_{i\infty}}\;\cU^{-d/2-n_{\rm f}} \, ,
\label{eq:paramterm}
\end{equation}
where $\alpha_e$ denotes the variable of the edge $e$, the exponent $\mathsf{z}_e$ is $\Dp n_{ij}$ or $c_i$, and $k_e$ is the number of the forests $f_1,\ldots,f_{n_{\rm f}}$ of the term which contain $e$.

We now state the representation with general exponents and with numerators. We use the flat form of \eqref{eq:gram} of section~\ref{sec:chart}, in which a scalar product of two points over all $d$ components is $x_i\!\cdot\!x_j=x_{i\parallel}\!\cdot\!x_{j\parallel}+|x_{i\perp}||x_{j\perp}|\,u_{ij}$. Let $\mathcal{N}$ be a product of factors which are linear in the $x_i$. Each factor is a component $u^A=\sum_iu_i\,x_i^A$ in a parallel direction, or it is one of the two factors of a scalar product $u\!\cdot\!u'$ over all $d$ components, where $u=\sum_iu_i\,x_i$ and $u'=\sum_iu'_i\,x_i$ with real coefficients. Let $\mathsf{n}_i$ be the number of the factors which contain $x_i$. For complex numbers $c_i$, and complex numbers $\mathsf{z}_{ij}$ for the pairs which are edges of a graph, we set $\kappa_i=c_i-d+\sum_j\mathsf{z}_{ij}$. The other pairs have no factor, and $\mathsf{z}_{ij}=0$ for them in \eqref{eq:region}.

\begin{lemma}
\label{lem:schwinger}
(i) Let $\mathsf{K}$ be a real positive symmetric matrix and ${\rm Re}\,m>n-1$. Then
\begin{equation}
\int\prod_id^dx_i\;e^{-\sum_{ij}\mathsf{K}_{ij}\,x_i\cdot x_j}\;\mathcal{N}
= \pi^{nd/2}\,(\det\mathsf{K})^{-d/2}\,\vev{\mathcal{N}}_\mathsf{K} \, ,
\label{eq:wick}
\end{equation}
where $\vev{\mathcal{N}}_\mathsf{K}$ is the sum over the pairings of the factors of $\mathcal{N}$ of the theorem of Wick, with the pairing $\tfrac12\,\delta^{\mu\nu}\sum_{ij}u_i\,(\mathsf{K}^{-1})_{ij}\,u'_j$ of the components $\mu$ and $\nu$ of two factors and with $\delta^\mu{}_\mu=d$. It is a polynomial in $d$ and in the entries of $\mathsf{K}^{-1}$. (ii) Let
\begin{equation}
{\rm Re}\,m>n-1 \, , \qquad {\rm Re}\,c_i>0 \, , \qquad {\rm Re}\,\mathsf{z}_{ij}>0 \, , \qquad {\rm Re}\,\kappa_i\geq\tfrac12\,\mathsf{n}_i \, ,
\label{eq:schwregion}
\end{equation}
and let \eqref{eq:region} hold for all sets $\mathsf{B}$ of at least two points. Then
\begin{multline}
\int\prod_id^dx_i\;\Omega_i^{\,c_i}\prod_{i<j}|x_i-x_j|^{-2\mathsf{z}_{ij}}\;\mathcal{N} \\
= \pi^{nd/2}\prod_i\frac{2^{c_i}}{\Gamma(c_i)}\prod_{i<j}\frac{1}{\Gamma(\mathsf{z}_{ij})}
\int\prod_id\alpha_{i\infty}\,\alpha_{i\infty}^{\,c_i-1}e^{-\alpha_{i\infty}}\prod_{i<j}d\alpha_{ij}\,\alpha_{ij}^{\,\mathsf{z}_{ij}-1}\;
\cU^{-d/2}\,\vev{\mathcal{N}}_\mathsf{K} \, ,
\label{eq:schwinger}
\end{multline}
with $\mathsf{K}=\mathsf{L}(\alpha)+{\rm diag}(\alpha_{i\infty})$ and $\cU=\det\mathsf{K}$, and both integrals converge absolutely.
\end{lemma}

\begin{proof}
For (i) we first take $\mathcal{N}=1$ and add the term $\sum_iJ_i\!\cdot\!x_i$ with sources in the parallel directions to the exponent. The integral over the parallel components is an ordinary Gaussian integral. For the transverse components we use the entries $Q_{ij}=|x_{i\perp}||x_{j\perp}|\,u_{ij}$ of a positive symmetric matrix $Q$ as variables. The transverse part of the measure becomes $2^{-n}(\det Q)^{(m-n-1)/2}\prod_{i\leq j}dQ_{ij}$, and the exponent is $-{\rm tr}\,\mathsf{K}Q$. The integral over $Q$ is the classical integral of Ingham and Siegel~\cite{Ingham:1933,Siegel:1935}, which equals $\pi^{n(n-1)/4}\prod_{l=0}^{n-1}\Gamma(\tfrac12(m-l))\,(\det\mathsf{K})^{-m/2}$ for ${\rm Re}\,m>n-1$. With the factors $\vol_{m-l}$ of \eqref{eq:gram} the constants combine to $\pi^{nm/2}$, and we obtain \eqref{eq:gauss} for each positive matrix $\mathsf{K}$. A derivative of the exponent with respect to a component $J_i^A$ gives the factor $x_i^A$, and a derivative with respect to an entry of $\mathsf{K}$ gives a scalar product $x_i\!\cdot\!x_j$ up to a constant. The integrand with a polynomial in these factors is bounded by an integrable function locally uniformly in $J_i$ and $\mathsf{K}$, so that we can differentiate under the integral. The left-hand side of \eqref{eq:wick} is therefore a differential operator applied to the right-hand side of \eqref{eq:gauss} at $J_i=0$. This is $\pi^{nd/2}(\det\mathsf{K})^{-d/2}$ times a polynomial in $d$ and in the entries of $\mathsf{K}^{-1}$. For integer $m\geq n$ the integral is an ordinary Gaussian integral in $nd$ variables, and the theorem of Wick gives $\vev{\mathcal{N}}_\mathsf{K}$. This is also a polynomial in $d$, since each closed chain of contracted indices gives a factor $d$. The two polynomials agree for all integers $d\geq D_0+n$ and are therefore equal.

For (ii) we insert the two representations which precede \eqref{eq:gauss}, which hold for ${\rm Re}\,c_i>0$ and ${\rm Re}\,\mathsf{z}_{ij}>0$. We first show that the integral over the $x_i$ and the parameters converges absolutely. The integral of the modulus of the integrand over the parameters is given by the same two representations with the real parts of the exponents. Up to a constant it is the modulus of the integrand on the left-hand side of \eqref{eq:schwinger}. With $s_{ij}^2=\Omega_i\Omega_j|x_i-x_j|^2$ and $d\mu_d(X_i)=\Omega_i^{\,d}\,d^dx_i$ the integral of this modulus is, again up to a constant,
\begin{equation}
\int\prod_id\mu_{{\rm Re}\,d}(X_i)\;\Omega_i^{\,{\rm Re}\,\kappa_i}\prod_{i<j}(s_{ij}^2)^{-{\rm Re}\,\mathsf{z}_{ij}}\;|\mathcal{N}| \, .
\end{equation}
Here $d\mu_{{\rm Re}\,d}$ is the measure of \eqref{eq:gram} for the real part of $d$, which is positive. We have $x_i^2=(2-\Omega_i)/\Omega_i\leq2/\Omega_i$ and $\Omega_i\leq2$. A factor $u$ of $\mathcal{N}$ is therefore bounded by a constant times the product of the $\Omega_i^{-1/2}$ over the points with $u_i\neq0$, where we use the inequality of Cauchy and Schwarz for the scalar products, which holds since the matrix $U$ is positive. Hence $|\mathcal{N}|$ is bounded by a constant times $\prod_i\Omega_i^{-\mathsf{n}_i/2}$, and by the last condition in \eqref{eq:schwregion} the integrand is bounded by a constant times $\prod_{i<j}(s_{ij}^2)^{-{\rm Re}\,\mathsf{z}_{ij}}$. The integral of this function converges by Lemma~\ref{lem:region}. By the theorems of Tonelli and Fubini we can thus integrate over the $x_i$ first. At fixed parameters the matrix $\mathsf{K}$ is positive, and \eqref{eq:wick} gives the right-hand side of \eqref{eq:schwinger}.
\end{proof}

For the term \eqref{eq:flatterm} the numerator is the product of the components $x_i^A$, so that $\mathsf{n}_i=\mu_i\leq\kappa_i$, and the pairings give the entries of $\mathsf{K}^{-1}$ without factors of $d$. The exponents are $\mathsf{z}_{ij}=\Dp n_{ij}$. For $0<{\rm Re}\,\Dp<1$ the conditions \eqref{eq:schwregion} and \eqref{eq:region} hold if ${\rm Re}\,d$ is larger than $D_0+n-1$ and than twice the number of the propagators. In this region the integral \eqref{eq:gram} with the monomial sources converges absolutely by Lemma~\ref{lem:region}, and by the change of variables of section~\ref{sec:chart} it is the sum of the terms \eqref{eq:flatterm}. This is the region in which \eqref{eq:param} holds.

\subsection{Exclusion of the line}
\label{app:currents}

In this subsection we consider integrals with any number of vertices of the couplings, of the mass source, of the current and of the term of second order in the gauge field. We show that the line $\Dp=\tfrac12(d-2)$ is not a line of poles of these integrals, and that they have no line of poles of fixed dimension. The integer $D_0\geq1$ is arbitrary. We first integrate by parts at the current vertices, so that no line carries two derivatives. We then use sources which vanish to a high order at the point $P$ of the chart. Finally we divide the range of the Schwinger parameters into sectors, in which we order the large and the small parameters separately.

\paragraph{Current vertices.} The gauge field $a=a_A(X_\parallel)\,dX^A$ defines the tangent vector field $(\delta_{AB}-X_AX_B)\,a_B$ on the sphere. We write $a\cdot X$ for $a_A(X_\parallel)X^A$. The transverse component of this vector field is $-(a\cdot X)\,X_\perp$, which is parallel to $X_\perp$, so that the vector field does not change the $u_{ij}$. On functions of the $X_{i\parallel}$ and the $u_{ij}$ it therefore acts at the point $i$ as the operator
\begin{equation}
\mathsf{v}_i = \big[a_A(X_{i\parallel})-(a\cdot X_i)\,X_i^A\big]\frac{\pr}{\pr X_{i\parallel}^A}
\label{eq:vfield}
\end{equation}
at fixed $u_{ij}$. With $s_{ij}^2=2-2X_{i\parallel}\!\cdot\!X_{j\parallel}-2\rho_i\rho_ju_{ij}$ and $\mathsf{v}_i\rho_i=-(a\cdot X_i)\rho_i$ we find
\begin{equation}
\mathsf{v}_i\,s_{ij}^{-2\Dp} = 2\Dp\,s_{ij}^{-2\Dp-2}\,a_A(X_{i\parallel})\,\big(X_j-X_i\big)^A+\Dp\,(a\cdot X_i)\,s_{ij}^{-2\Dp} \, .
\label{eq:vG}
\end{equation}
A current vertex at the point $i$ has two lines, and there are no lines from a point to itself. We denote by $\mathsf{v}_i^{(1)}$ and $\mathsf{v}_i^{(2)}$ the operator $\mathsf{v}_i$ acting on the propagator of the first and of the second line only. After the sum over the two ways to attach the two lines to the two legs of the vertex, the factor of the vertex is $\mathsf{v}_i^{(1)}-\mathsf{v}_i^{(2)}$ acting on the product of the propagators, up to a constant and the antisymmetric matrix of the gauge field. The two lines may end at the same point.

The operator \eqref{eq:vfield} is of divergence form with respect to the measure of \eqref{eq:gram}. For a function $f$ of $X_\parallel$ we have
\begin{equation}
\int d^kX_\parallel\,\rho^{m-2}\,\mathsf{v} f = -\int d^kX_\parallel\,\rho^{m-2}\,(\nabla\!\cdot\!a)\,f \, , \qquad
\nabla\!\cdot\!a = (\delta_{AB}-X_AX_B)\,\pr_Aa_B-d\,X_Aa_A \, ,
\label{eq:diva}
\end{equation}
where we used $k+m-1=d$. There is no boundary term for ${\rm Re}\,m>0$, since the normal component of the vector field at the boundary of the ball is $(a\cdot X)\rho^2$. The function $\nabla\!\cdot\!a$ is the divergence on the sphere with $d$ as a parameter, as for the Laplacian in section~\ref{sec:sphere}. It is a smooth function of $X_\parallel$ and a polynomial of first degree in $d$, and it is therefore a combination of two sources of the class with the coefficients $1$ and $d$. For ${\rm Re}\,m\geq n$ and ${\rm Re}\,\Dp<-2$ the integrand of \eqref{eq:gram} with any number of current vertices is continuously differentiable in $X_{i\parallel}$ in the open ball, and it is bounded together with its derivative along $\mathsf{v}_i$. The remaining factors of \eqref{eq:gram} depend on $X_{i\parallel}$ only through the two lines of the vertex, and the operators of the other vertices commute with $\mathsf{v}_i$. In this region we can thus use \eqref{eq:diva}, which gives $\mathsf{v}_i^{(1)}+\mathsf{v}_i^{(2)}+\nabla\!\cdot\!a=0$ under the integral. Hence
\begin{equation}
\mathsf{v}_i^{(1)}-\mathsf{v}_i^{(2)} \;\to\; 2\,\mathsf{v}_i^{(1)}+\nabla\!\cdot\!a
\label{eq:ibp}
\end{equation}
under the integral, and the same holds with the two lines exchanged and the opposite sign. Both sides are meromorphic in $d$ and $\Dp$ by Proposition~\ref{prop:mero}, so that the replacement \eqref{eq:ibp} holds for the continued integrals. Note that we integrate by parts on the sphere, before we introduce a chart. The weights of the gauge field in a chart do therefore not enter.

\paragraph{Choice of the lines.} We choose for each current vertex one of its two lines, such that no line is chosen twice. This is possible. The lines which join two current vertices form paths and cycles, since a current vertex has two lines. On a cycle each vertex chooses the next line in one direction. On a path each vertex chooses the line towards one end of the path, and the last vertex chooses its line to a vertex of another type. A cycle may consist of two current vertices which are joined by two lines. We apply \eqref{eq:ibp} with the chosen line at each current vertex and insert \eqref{eq:vG}. A current vertex at the point $i$ whose chosen line ends at the point $j$ is then replaced by the sum of two vertices. The first has the factor $4\Dp\,s_{ij}^{-2\Dp-2}\,a_A(X_{i\parallel})(X_j-X_i)^A$ in place of the propagator of the chosen line. The second has no derivative and the source $(\delta_{AB}-X_AX_B)\,\pr_Aa_B-(d-2\Dp)\,X_Aa_A$. Since the chosen lines are different, no line receives two factors of the first kind.

\paragraph{Integrals with marked lines.} We are led to the following integrals. Let $n_{ij}$ be the number of lines between the points $i$ and $j$, of which $r_{ij}\leq n_{ij}$ are marked. A marked line $\ell$ has a first point $i_\ell$, a second point $j_\ell$ and an index $A_\ell$ of the parallel directions. We consider the integrals \eqref{eq:gram} of
\begin{equation}
F = \prod_i\psi_i(X_{i\parallel})\;\prod_{i<j}s_{ij}^{-2\Dp n_{ij}-2r_{ij}}\;\prod_\ell\big(X_{j_\ell}-X_{i_\ell}\big)^{A_\ell} \, ,
\label{eq:marked}
\end{equation}
where the $\psi_i$ are smooth functions on the closed ball. With the normalisations $C_\phi$ of the propagators taken out, the integrals with vertices of the four types are finite sums of such integrals, with coefficients which are polynomials in $d$ and $\Dp$. The functions $\psi_i$ are the sources of the couplings, of the mass and of the term of second order in the gauge field, the components $a_A$, and the source of the second vertex above. The integrand \eqref{eq:marked} is of the form considered in appendix~\ref{app:mero}. Proposition~\ref{prop:mero} and Lemmas~\ref{lem:region} and~\ref{lem:polarcont} therefore apply. We do not use the number of lines at a vertex in the following.

\begin{proposition}
\label{prop:linecurrent}
For $r_{ij}\leq n_{ij}$ and all smooth functions $\psi_i$ of $X_\parallel$, the line $\Dp=\tfrac12(d-2)$ is not a line of poles of the integral \eqref{eq:gram} of \eqref{eq:marked}, and the integral has no line of poles of fixed dimension. The same holds for the integrals with any number of vertices of the couplings, of the mass source, of the current and of the term of second order in the gauge field, for all smooth sources of the class and all $D_0\geq1$.
\end{proposition}

\paragraph{Sources.} We cover the closed unit ball of the parallel directions by a finite number of open sets, which are so small that the union of the closures of any $n$ of them does not contain all unit vectors. This is possible since the unit vectors of the parallel directions form a sphere of dimension $D_0\geq1$. We write each $\psi_i$ as a sum of smooth functions with supports in these sets. The integral is multilinear in the $\psi_i$. It is therefore a finite sum of integrals in which all $\psi_i$ vanish in a neighbourhood of a common unit vector $P$ of the parallel directions. For such an integral and an integer $h\geq0$ we consider the sources
\begin{equation}
\psi_i = \big(1-P\!\cdot\!X_i\big)^h\times\text{monomial in the components of }X_{i\parallel} \, .
\label{eq:hsource}
\end{equation}
By Lemma~\ref{lem:dense} a smooth function which vanishes in a neighbourhood of $P$ is a limit, in each of the norms \eqref{eq:norm}, of finite linear combinations of the functions \eqref{eq:hsource} with this $P$ and this $h$. By the third statement of Lemma~\ref{lem:polarcont}, in which the sets $\mathsf{T}_i$ consist of the functions \eqref{eq:hsource}, it is therefore sufficient to exclude a given candidate as a line of poles for the sources \eqref{eq:hsource}. This holds for each $h$, and we may thus use one value of $h$, which may depend on the graph and on the line. A line which is not a candidate \eqref{eq:candidates} is not a line of poles by Proposition~\ref{prop:mero}.

\paragraph{Chart.} We use the stereographic chart from $P$, with $P\!\cdot\!X_i=1-\Omega_i$ and $X_i^A=\Omega_i\,x_i^A$ for the other parallel components. The source \eqref{eq:hsource} is a sum of terms $\Omega_i^{\,h+\kappa}\prod_A(x_i^A)^{\mu_{iA}}$ with $\kappa\geq\mu_i=\sum_A\mu_{iA}$. For a marked line from $i$ to $j$ we use $\Omega_i-\Omega_j=\tfrac12\Omega_i\Omega_j\,(x_j-x_i)\!\cdot\!(x_j+x_i)$, which gives
\begin{align}
\big(X_j-X_i\big)^A &= \Omega_j\,(x_j-x_i)^A-\tfrac12\,\Omega_i\Omega_j\,\big[(x_j-x_i)\!\cdot\!(x_j+x_i)\big]\,x_i^A \qquad (A\neq P) \, , \nonumber\\
\big(X_j-X_i\big)^P &= \tfrac12\,\Omega_i\Omega_j\,(x_j-x_i)\!\cdot\!(x_j+x_i) \, .
\label{eq:numer}
\end{align}
The scalar product is over all $d$ components. Each term on the right-hand side contains the factor $x_j-x_i$ once. We call it the difference of the marked line. With $s_{ij}^2=\Omega_i\Omega_j|x_i-x_j|^2$ the integral becomes a sum of terms
\begin{equation}
\int\prod_id^dx_i\;\Omega_i^{\,c_i}\;\prod_{i<j}|x_i-x_j|^{-2\mathsf{z}_{ij}}\;\mathcal{N} \, , \qquad
c_i = d-\Dp n_i-r_i+\kappa_i \, , \qquad \mathsf{z}_{ij} = \Dp\,n_{ij}+r_{ij} \, ,
\label{eq:flatmarked}
\end{equation}
where $r_i$ is the number of marked lines at the point $i$ and $\kappa_i$ is an integer. The numerator $\mathcal{N}$ is a product of factors which are linear in the $x_i$. These are the components $x_i^A$, the differences of the marked lines, and the sums $x_i+x_j$ of \eqref{eq:numer}. Each difference is contracted with a parallel index or with a sum. Let $\mathsf{n}_i$ be the number of factors of $\mathcal{N}$ which contain $x_i$. A marked line adds at most three such factors at each of its points, and it does not lower the power of $\Omega_i$ beyond the term $-r_i$ of $c_i$. Hence
\begin{equation}
\kappa_i \geq h+\mu_i \, , \qquad \mathsf{n}_i \leq \mu_i+3\,r_i \, .
\label{eq:kappabound}
\end{equation}

\paragraph{Schwinger parameters.} We introduce the parameters $\alpha_{ij}$ and $\alpha_{i\infty}$ as in \eqref{eq:param}, with the exponents $\mathsf{z}_{ij}$ and $c_i$ of \eqref{eq:flatmarked}. The representation is \eqref{eq:schwinger} of Lemma~\ref{lem:schwinger}. By \eqref{eq:kappabound} and $r_i\leq n_i$ its condition $\kappa_i\geq\tfrac12\mathsf{n}_i$ holds for $h\geq\tfrac32n_{\max}$, where $n_{\max}$ is the largest number of lines at a point, and we assume this from now on. The other conditions in \eqref{eq:schwregion} and \eqref{eq:region} hold for $0<{\rm Re}\,\Dp<1$ and sufficiently large ${\rm Re}\,d$. In this region the integral \eqref{eq:gram} of \eqref{eq:marked} converges absolutely by Lemma~\ref{lem:region}, and by the change of variables of section~\ref{sec:chart} it is the sum of the terms \eqref{eq:flatmarked}, each of which converges absolutely. By the first statement of Lemma~\ref{lem:schwinger} the integral of the product of the linear factors at fixed parameters is given by the theorem of Wick, in which the trace of the unit matrix is $d$. For two factors $u=\sum_iu_ix_i$ and $u'=\sum_iu'_ix_i$ the pairing is proportional to
\begin{equation}
\sum_{ij}u_i\,(\mathsf{K}^{-1})_{ij}\,u'_j = \frac{1}{\cU}\sum_f\mathsf{Q}_u(f)\,\mathsf{Q}_{u'}(f)\prod_{e\in f}\alpha_e \, , \qquad \mathsf{Q}_u(f) = \sum_{i\in \mathsf{C}_f}u_i \, ,
\label{eq:forestpair}
\end{equation}
where the sum is over the spanning forests $f$ of $\Gamma_\infty$ with two trees, and $\mathsf{C}_f$ is the set of the points of the tree which does not contain $\infty$. This follows from the formula for $(\mathsf{K}^{-1})_{ij}$ of the previous subsections. For the difference of a marked line from $i$ to $j$ we have $\mathsf{Q}(f)=1$ if $\mathsf{C}_f$ contains $j$ and not $i$, $\mathsf{Q}(f)=-1$ in the opposite case, and $\mathsf{Q}(f)=0$ otherwise. A forest which contributes therefore separates the two points of the line. Since the pair $e=(ij)$ is an edge of the graph, $f\cup\{e\}$ is then a spanning tree. Each term is again of the form \eqref{eq:paramterm}, with coefficients which are polynomials in $d$. The number $n_{\rm f}$ of the forests $f_1,\ldots,f_{n_{\rm f}}$ is half the number of the linear factors, and for each marked line one of the forests separates its two points.

\paragraph{Sectors.} We divide the range of every variable into the parts $\alpha_e\leq1$ and $\alpha_e\geq1$. Let $\Gamma_>$ be the set of the edges with $\alpha_e\geq1$, and let $y_e=1/\alpha_e$ for these edges and $y_e=\alpha_e$ for the others. Then $\cU=\prod_{e\in\Gamma_>}y_e^{-1}\,\tilde\cU$, where $\tilde\cU$ is the sum over the spanning trees $T$ of $\Gamma_\infty$ of the products of the $y_e$ with $e$ in the symmetric difference $T\triangle\Gamma_>$. We now order the variables of $\Gamma_>$ and the other variables separately, as $y_{\pi_+(1)}\leq\ldots\leq y_{\pi_+(M_+)}$ and $y_{\pi_-(1)}\leq\ldots\leq y_{\pi_-(M_-)}$, where $M_+$ is the number of the edges of $\Gamma_>$ and $M_-$ is the number of the other edges. We set $y_{\pi_\pm(j)}=s_j^\pm s_{j+1}^\pm\cdots s_{M_\pm}^\pm$. Let $\mathsf{A}_j^+$ be the set of the first $j$ edges of $\Gamma_>$ and $\mathsf{A}_j^-$ the set of the first $j$ of the other edges. The monomial of a tree $T$ in $\tilde\cU$ contains $s_j^+$ with the exponent $|\mathsf{A}_j^+|-|T\cap \mathsf{A}_j^+|$ and $s_j^-$ with the exponent $|T\cap \mathsf{A}_j^-|$.

\begin{lemma}
\label{lem:greedy2}
There is a spanning tree $T_*$ which maximises $|T\cap \mathsf{A}_j^+|$ for all $j$ and minimises $|T\cap \mathsf{A}_j^-|$ for all $j$ at the same time.
\end{lemma}

\begin{proof}
We order the edges such that the edges of $\Gamma_>$ come first, in the order $\pi_+$, followed by the other edges in the order opposite to $\pi_-$. Let $T_*$ be the tree which is obtained by taking the edges in this order whenever they close no cycle. It maximises the sum of the weights of its edges for each weight which does not increase along the order, by the standard property of this construction. The weight which is $1$ on $\mathsf{A}_j^+$ and $0$ elsewhere, and the weight which is $-1$ on $\mathsf{A}_j^-$ and $0$ elsewhere, are of this kind.
\end{proof}

With $\varrho_j^+=|T_*\cap \mathsf{A}_j^+|$ and $\varrho_j^-=|T_*\cap \mathsf{A}_j^-|$ we have $\tilde\cU=\prod_j(s_j^+)^{|\mathsf{A}_j^+|-\varrho_j^+}\prod_j(s_j^-)^{\varrho_j^-}\,\mathcal{P}$, where $\mathcal{P}$ is a polynomial which is not smaller than $1$ on the cube. The integrand of a term is $\prod_jds_j^+\,(s_j^+)^{z_j^+-1}\prod_jds_j^-\,(s_j^-)^{z_j^--1}$ times a smooth function, with
\begin{align}
z_j^+ &= -\mathsf{z}(\mathsf{A}_j^+)+\frac d2\,\varrho_j^++\sum_{t=1}^{n_{\rm f}}b_t^+ \, , & b_t^+ &= \varrho_j^+-|f_t\cap \mathsf{A}_j^+| \, , \nonumber\\
z_j^- &= \mathsf{z}(\mathsf{A}_j^-)-\frac d2\,\varrho_j^-+\sum_{t=1}^{n_{\rm f}}b_t^- \, , & b_t^- &= |f_t\cap \mathsf{A}_j^-|-\varrho_j^- \, .
\label{eq:Epm}
\end{align}
If $\mathsf{A}_j^+$ contains an edge to $\infty$, the factor $e^{-1/y}$ of this edge vanishes with all derivatives at $s_j^+=0$, and the integral over $s_j^+$ gives no pole. By Lemma~\ref{lem:cube} the poles therefore lie on the lines $z_j^\pm+\nu=0$ with $\nu\geq0$, for sets $\mathsf{A}_j^+$ without edges to $\infty$. The prefactors of \eqref{eq:schwinger} are entire.

\paragraph{Large parameters.} Let $\mathsf{A}_j^+$ be a set $\mathsf{a}^+$ of edges between points. A forest in $\mathsf{a}^+$ can be completed to a spanning tree, so that $\varrho_j^+$ is the number ${\rm rk}(\mathsf{a}^+)$ of edges of a largest forest in $\mathsf{a}^+$. Since $f_t\cap\mathsf{a}^+$ is a forest, we have $b_t^+\geq0$. If $f_t$ separates the points of a marked line whose edge $e$ is in $\mathsf{a}^+$, then $(f_t\cap\mathsf{a}^+)\cup\{e\}$ is a forest in $\mathsf{a}^+$, and $b_t^+\geq1$. We write $n(\mathsf{a})$ and $r(\mathsf{a})$ for the sums of the $n_{ij}$ and of the $r_{ij}$ over a set $\mathsf{a}$ of edges. Then
\begin{equation}
2z_j^+ = -2\Dp\,n(\mathsf{a}^+)-2\,r(\mathsf{a}^+)+d\,{\rm rk}(\mathsf{a}^+)+2\sum_tb_t^+ \, .
\label{eq:Eplus}
\end{equation}
The coefficient of $\Dp$ does not vanish, so that $2z_j^++2\nu=0$ is not a line of fixed dimension. On the line $\Dp=\tfrac12(d-2)$ the candidate equals
\begin{equation}
d\,\big[{\rm rk}(\mathsf{a}^+)-n(\mathsf{a}^+)\big]+2\,\big[n(\mathsf{a}^+)-r(\mathsf{a}^+)\big]+2\sum_tb_t^++2\nu \, .
\label{eq:Eplusline}
\end{equation}
We have $n(\mathsf{a}^+)\geq|\mathsf{a}^+|\geq{\rm rk}(\mathsf{a}^+)$ and $r(\mathsf{a}^+)\leq n(\mathsf{a}^+)$. If \eqref{eq:Eplusline} vanishes identically, then $n(\mathsf{a}^+)={\rm rk}(\mathsf{a}^+)$, and the three non-negative terms $n(\mathsf{a}^+)-r(\mathsf{a}^+)$, $\sum_tb_t^+$ and $\nu$ vanish. Since $\mathsf{a}^+$ is not empty, it then contains a marked line. The forest of its difference has $b_t^+\geq1$, which is a contradiction. This is the place where we use $r_{ij}\leq n_{ij}$ and the differences.

\paragraph{Small parameters.} Let $\mathsf{A}_j^-$ consist of a set $\mathsf{a}^-$ of edges between points and of the edges to $\infty$ of a set $\mathsf{S}$ of points. The number $\varrho_j^-$ is the number of connected components without $\infty$ of the graph $\Gamma_\infty$ without the edges of $\mathsf{A}_j^-$. Such a component consists of points of $\mathsf{S}$, and hence $\varrho_j^-\leq|\mathsf{S}|$. For an edge $e$ which joins the two trees of $f_t$ the tree $f_t\cup\{e\}$ gives $b_t^-\geq-1$, and $b_t^-\geq0$ if $e$ is not in $\mathsf{A}_j^-$. The edge to $\infty$ of a point of $\mathsf{C}_{f_t}$ joins the two trees. Hence $b_t^-\geq0$ unless $\mathsf{C}_{f_t}$ is contained in $\mathsf{S}$. With the sums $n(\mathsf{S})$ and $\kappa(\mathsf{S})$ of the $n_i$ and of the $\kappa_i$ over $\mathsf{S}$ and with $\Theta^-=n(\mathsf{a}^-)-n(\mathsf{S})$ we find
\begin{equation}
2z_j^- = 2\Dp\,\Theta^-+d\,\big(2|\mathsf{S}|-\varrho_j^-\big)+2\,\mathsf{b} \, , \qquad \mathsf{b} = \kappa(\mathsf{S})+r(\mathsf{a}^-)-r(\mathsf{S})+\sum_tb_t^- \, ,
\label{eq:Eminus}
\end{equation}
where $r(\mathsf{S})$ is the sum of the $r_i$ over $\mathsf{S}$. If $\mathsf{S}$ is empty, then $\varrho_j^-=0$ and $b_t^-\geq0$, and $2z_j^-=2\Dp\,n(\mathsf{a}^-)+2\,r(\mathsf{a}^-)+2\sum_tb_t^-$ with $n(\mathsf{a}^-)\geq1$. This is not a line of fixed dimension, and on the line $\Dp=\tfrac12(d-2)$ the coefficient of $d$ is $n(\mathsf{a}^-)$, which does not vanish.

Let now $\mathsf{S}$ be non-empty, and let $n_{\rm f}^-$ be the number of the $t$ with $b_t^-=-1$. For these $\mathsf{C}_{f_t}$ is contained in $\mathsf{S}$. By \eqref{eq:forestpair} both linear factors of the pairing have $\mathsf{Q}(f_t)\neq0$, so that both contain a point of $\mathsf{S}$. Hence $2n_{\rm f}^-\leq \mathsf{n}(\mathsf{S})$, and with \eqref{eq:kappabound} and $r_i\leq n_i$ we obtain
\begin{equation}
\mathsf{b} \geq h\,|\mathsf{S}|+\mu(\mathsf{S})-r(\mathsf{S})-n_{\rm f}^- \geq \Big(h-\frac52\,n_{\max}\Big)\,|\mathsf{S}| \, ,
\label{eq:Bbound}
\end{equation}
where $\mathsf{n}(\mathsf{S})$ and $\mu(\mathsf{S})$ are the sums of the $\mathsf{n}_i$ and of the $\mu_i$ over $\mathsf{S}$, and $n_{\max}$ is the largest number of lines at a point. We take $h>\tfrac52n_{\max}$, so that $\mathsf{b}>0$. On the line $\Dp=\tfrac12(d-2)$ the candidate $2z_j^-+2\nu$ equals $d\,(\Theta^-+2|\mathsf{S}|-\varrho_j^-)-2\Theta^-+2\,\mathsf{b}+2\nu$. If it vanishes identically, then $\Theta^-=\varrho_j^--2|\mathsf{S}|\leq-|\mathsf{S}|<0$ and $\Theta^-=\mathsf{b}+\nu>0$, which is a contradiction. For a line of fixed dimension we need $\Theta^-=0$. Then $2z_j^-+2\nu=d\,(2|\mathsf{S}|-\varrho_j^-)+2\,\mathsf{b}+2\nu$ with $|\mathsf{S}|\leq2|\mathsf{S}|-\varrho_j^-\leq2|\mathsf{S}|$, and the line lies at $d\leq-(h-\tfrac52n_{\max})$. A given line $d=d_0$ is therefore not among the candidates for $h>\tfrac52n_{\max}$ and $h>\tfrac52n_{\max}-d_0$.

\paragraph{Result.} For the sources \eqref{eq:hsource} with $h>\tfrac52n_{\max}$ no candidate $z_j^\pm+\nu=0$ contains the line $\Dp=\tfrac12(d-2)$, and if in addition $h>\tfrac52n_{\max}-d_0$ no candidate is the line $d=d_0$. Two steps remain, since Lemma~\ref{lem:schwinger} gives the absolute convergence of the integral \eqref{eq:schwinger}, while we applied Lemma~\ref{lem:cube} to each term in each sector. By \eqref{eq:Eplus} the coefficient of $d$ in $2z_j^+$ is ${\rm rk}(\mathsf{a}^+)\geq1$, and by \eqref{eq:Eminus} the coefficient of $d$ in $2z_j^-$ is at least $|\mathsf{S}|$, while for empty $\mathsf{S}$ the exponent $z_j^-$ is $\Dp\,n(\mathsf{a}^-)$ plus a non-negative integer. For $0<{\rm Re}\,\Dp<1$ and sufficiently large ${\rm Re}\,d$ the exponents $z_j^\pm$ of all terms therefore have positive real parts, except for sets $\mathsf{A}_j^+$ with an edge to $\infty$, and each term of the representation converges separately. Each exponent $z_j^\pm$ thus depends on $d$ or on $\Dp$, so that the lines $z_j^\pm+\nu=0$ form a locally finite set, and each of them meets a candidate \eqref{eq:candidates} which is not among them in at most one point. Such a candidate therefore contains a point on no other candidate near which the integral is holomorphic, and by the first statement of Lemma~\ref{lem:poleset} it is not a line of poles for the sources \eqref{eq:hsource}. By the reduction of the sources above this proves the first statement of Proposition~\ref{prop:linecurrent}. The second statement follows from \eqref{eq:ibp} and the choice of the lines. The lines of poles are among the lines $z_j^\pm+\nu=0$ of \eqref{eq:Eplus} and \eqref{eq:Eminus} for the sources \eqref{eq:hsource}, and among the candidates \eqref{eq:candidates}. The proposition concerns lines of poles. The integrals have poles at points of the line, such as $d=4$. For integrands without marked lines, which are those of the vertices without derivatives, the argument is the case $r_{ij}=0$.\footnote{We verified the combinatorial steps of this proof by enumeration for graphs with up to five points, and the representation by Schwinger parameters numerically for two points with one marked line.}

The proposition does not concern the invariance of the functional under gauge transformations of the background field, which is the subject of appendix~\ref{app:gauge}. Whether the candidates $z_j^\pm+\nu=0$ which cross the line at $d=4$ are poles is the question of renormalisation, which we did not address.

\section{Background gauge invariance}
\label{app:gauge}

In this appendix we derive the invariance of the bare functional $W_0^S$ under gauge transformations of the background field from the definition \eqref{eq:gram}. The result is Proposition~\ref{prop:gauge}. We used this invariance in section~\ref{sec:setup}, and it enters the proofs of appendices~\ref{app:weyl}, \ref{app:unique} and~\ref{app:finite}.

We reduce the invariance to a property of the poles of the integrals, which we prove in section~\ref{app:npcurrent}. We first show that for general $d$ and $\Dp$ the gauge variation of the term with $N+1$ vertices equals $d-2-2\Dp$ times an integral of the same kind, in which one propagator has the exponent $\Dp+1$. We then give this propagator an independent exponent $\Dp'$ and compute the residue of the integral on the plane $d=2\Dp'$, which is the integral with the two end points of the propagator identified. On the line $\Dp=\tfrac12(d-2)$ the two statements combine to the Ward identity, provided that no other plane of poles contains the line. This is the property \eqref{eq:NP} below.

\subsection{Vertices and gauge variation}

We use the sources of section~\ref{sec:sphere}, which are smooth functions of $X_\parallel$ on the closed unit ball. The fields $\phi_i$ carry an index $i$ of the group of field rotations, the gauge field $a_A$ and the parameter $\omega$ are antisymmetric matrices in these indices, and the index $A$ runs over the $k=D_0+1$ parallel directions. We write $\nabla_A=(\delta_{AB}-X_AX_B)\pr_B$ for the derivative along the sphere, where $B$ runs over all components of $X$, and $\Pi_{AB}=\delta_{AB}-X_AX_B$ for the parallel indices. The vertices are
\begin{equation}
V^{(4)}[g] = \tfrac{1}{4!}\,g_{ijkl}\,\phi_i\phi_j\phi_k\phi_l \, , \qquad
V^{(2)}[\mathsf{u}] = \tfrac12\,\mathsf{u}_{ij}\,\phi_i\phi_j \, , \qquad
V^{J}[a] = a_{A,ij}\,\nabla_A\phi_i\,\phi_j \, ,
\label{eq:vertices}
\end{equation}
with a symmetric tensor $g$, a symmetric matrix $\mathsf{u}$ and antisymmetric matrices $a_A$. The interaction is the integral of $V^J[a_0]+V^{(2)}[\sM_0+\Upsilon]+V^{(4)}[g_0]$, where $\Upsilon$ is the source of the term of second order in the gauge field. As in section~\ref{app:conf} we treat $\Upsilon$ as an independent source, and the functional of section~\ref{sec:sphere} is obtained for $\Upsilon=a_{0A}^T\Pi_{AB}\,a_{0B}$. Under the conformal maps \eqref{eq:boost} the gauge field and the source $\Upsilon$ have the weights of \eqref{eq:cov3}. We call a vertex with $V^{(4)}$ or $V^{(2)}$ a vertex without derivatives, and we call the field factors of a vertex its legs.

Let $V_1,\ldots,V_N$ be vertices of the form \eqref{eq:vertices} at the points $X_1,\ldots,X_N$. We denote by $\langle V_1,\ldots,V_N\rangle^c$ the sum over all connected pairings of their legs, without pairs of legs of the same vertex, of the integral \eqref{eq:gram} of the product of the propagators. A pair of legs at $X_r$ and $X_s$ gives $C_\phi\,(s_{rs}^2)^{-\Dp}$ with $C_\phi=\Gamma(\tfrac12d-1)/(4\pi^{d/2})$, and a leg $\nabla_A\phi$ gives the derivative $\nabla_A$ of the propagator at its point. By Proposition~\ref{prop:mero} this is a meromorphic function of $d$ and $\Dp$. The terms of the functional are
\begin{equation}
W_0^S = \sum_{N\geq1}\frac{(-1)^N}{N!}\,\big\langle V_{\rm int},\ldots,V_{\rm int}\big\rangle^c\Big|_{\Dp=\frac12(d-2)} \, ,
\end{equation}
where each of the $N$ entries is the sum of the three vertices. We denote by $\ell_0$ the line $\Dp=\tfrac12(d-2)$, and we say that a meromorphic function has no pole along $\ell_0$ if $\ell_0$ is not one of its lines of poles.

The gauge variation of the sources with a parameter $\omega(X_\parallel)$ is
\begin{align}
\delta_\omega a_A ={}& \pr_A\omega+[a_A,\omega] \, , \qquad
\delta_\omega g = -(\omega g) \, , \qquad
\delta_\omega\sM = -[\omega,\sM] \, , \notag\\
\delta_\omega\Upsilon ={}& [\Upsilon,\omega]-\Pi_{AB}\left(\pr_A\omega\,a_B+a_B\,\pr_A\omega\right) ,
\label{eq:gaugevar}
\end{align}
where $(\omega g)_{ijkl}=\omega_{im}g_{mjkl}$ plus the three terms for the other indices. The variation of $\Upsilon$ is the variation of $a_A^T\Pi_{AB}\,a_B$ under $\delta_\omega a_A$, so that $\delta_\omega$ is tangent to the set of sources with this value of $\Upsilon$. Let $R_\omega V$ be the vertex which is obtained from $V$ by the replacement of $\phi$ by $-\omega\phi$ in each leg in turn, where a leg $\nabla_A\phi$ becomes $-\nabla_A(\omega\phi)$. Then
\begin{align}
R_\omega V^{(4)}[g] ={}& V^{(4)}[\omega g] \, , \qquad
R_\omega V^{(2)}[\mathsf{u}] = V^{(2)}\big[[\omega,\mathsf{u}]\big] \, , \notag\\
R_\omega V^{J}[a] ={}& V^{J}\big[[\omega,a]\big]+V^{(2)}\big[\Pi_{AB}(\pr_A\omega\,a_B+a_B\,\pr_A\omega)\big] \, ,
\label{eq:Romega}
\end{align}
and therefore, at each point and as polynomials in $\phi$ and $\nabla_A\phi$,
\begin{equation}
\delta_\omega V+R_\omega V = V^{J}[d\omega] \, , \qquad V^{J}[d\omega] = \pr_A\omega_{ij}\,\nabla_A\phi_i\,\phi_j \, ,
\label{eq:pointwise}
\end{equation}
for the sum $V$ of the three vertices of the interaction. The first variation of the functional is therefore
\begin{align}
\delta_\omega W_0^S ={}& -\sum_{N\geq0}\frac{(-1)^N}{N!}\;T_N\big(\omega;V_{\rm int},\ldots,V_{\rm int}\big)\Big|_{\ell_0} \, , \notag\\
T_N\big(\omega;V_1,\ldots,V_N\big) ={}& \big\langle V^J[d\omega],V_1,\ldots,V_N\big\rangle^c+\sum_{r=1}^N\big\langle V_1,\ldots,R_\omega V_r,\ldots,V_N\big\rangle^c \, .
\label{eq:TN}
\end{align}
The functional is invariant if $T_N$ vanishes on $\ell_0$ for all vertices, since $T_N$ is multilinear. Note that $T_0=0$, and that $R_\omega$ maps vertices without derivatives to vertices without derivatives.

\subsection{Integral over one point}

We place the point $X_0$ of the vertex $V^J[d\omega]$ last in the triangular coordinates of appendix~\ref{app:mero}. For ${\rm Re}\,m>N$ and an integrand $F$ for which the integral converges absolutely, equation \eqref{eq:triangular} for the $N+1$ points takes the form
\begin{equation}
\int\prod_{r=0}^Nd\mu_d(X_r)\,F = \int\prod_{r=1}^Nd\mu_d(X_r)\int_{\mathbb{B}}d\mu_{\mathbb{B}}(Y)\,F \, , \qquad
d\mu_{\mathbb{B}}(Y) = \vol_{m-N}\,(1-Y^2)^{\frac{m-N-2}{2}}\,d^{k+N}Y \, ,
\label{eq:last}
\end{equation}
where $\mathbb{B}$ is the unit ball of $\mathbb{R}^{k+N}$, the first $k$ components of $Y$ are $X_{0\parallel}$, and the outer integral is \eqref{eq:triangular} for the points $X_1,\ldots,X_N$. In these coordinates the other points are unit vectors $w_r$ on the boundary of $\mathbb{B}$, and $s_{0r}^2=2-2\,Y\!\cdot\!w_r$. For functions of $Y$ we write $\nabla_A=(\delta_{AB}-Y_AY_B)\pr_B$ and $\nabla^2F=(\delta_{AB}-Y_AY_B)\pr_A\pr_BF-d\,Y_A\pr_AF$, where the indices now run over the $k+N$ components of $Y$. For the propagators and the sources these agree with the derivatives along the sphere. With $\rho^2=1-Y^2$ and $m'=m-N$, so that $d=k+N-1+m'$, we have the three identities
\begin{gather}
\rho^{2-m'}\,\pr_A\big(\rho^{m'-2}\,(\delta_{AB}-Y_AY_B)\,\pr_BF\big) = \nabla^2F \, , \notag\\
\rho^{2-m'}\,\pr_B\big(\rho^{m'-2}\,(\delta_{AB}-Y_AY_B)\,f\big) = (\nabla_A-d\,Y_A)\,f \, , \notag\\
\nabla^2(s^2)^{-\Dp} = \Dp\,(d-1-\Dp)\,(s^2)^{-\Dp}+2\Dp\,(2\Dp+2-d)\,(s^2)^{-\Dp-1} \, ,
\label{eq:lap}
\end{gather}
for $s^2=2-2\,Y\!\cdot\!w$ with a unit vector $w$. They are polynomial in $d$, which enters as a parameter. For $\Dp=\tfrac12(d-2)$ the first coefficient in the last identity is $\tfrac14d(d-2)$, which is $\xi_d\,R$ for the conformally coupled field on the unit sphere, and the second coefficient vanishes. We also need the derivative of a propagator at its fixed end $Z=w$ in a parallel direction. For a vector $c$ with parallel components only, and for a function $F$ of $Y\!\cdot\!Z$,
\begin{equation}
c_A\nabla^Z_AF = -c_A\nabla^Y_AF+c_A\,(Y_A+Z_A)\,(1-Y\!\cdot\!Z)\,F' \, ,
\label{eq:transfer}
\end{equation}
which follows from $\nabla^Z_AF=(Y_A-Z_A\,Y\!\cdot\!Z)\,F'$ and $\nabla^Y_AF=(Z_A-Y_A\,Y\!\cdot\!Z)\,F'$. Here a superscript of $\nabla_A$ denotes the point at which the derivative acts, and below we write $\nabla^0_A$ and $\nabla^1_A$ for the derivatives at $X_0$ and $X_1$.\footnote{We verified \eqref{eq:pointwise}, \eqref{eq:lap}, \eqref{eq:transfer} and the reduction which leads to \eqref{eq:JJ} below with computer algebra.}

\begin{lemma}
\label{lem:delta}
Let $d$ and $\Dp'$ be real with $m'>0$ and $\Dp'<\tfrac12d$, let $w$ be a unit vector, and let $\mathsf{g}$ be continuous on the closed ball. Then
\begin{equation}
\lim_{\Dp'\to d/2}\;(d-2\Dp')\int_{\mathbb{B}}d\mu_{\mathbb{B}}(Y)\,(2-2\,Y\!\cdot\!w)^{-\Dp'}\,\mathsf{g}(Y) = \vol_d\,\mathsf{g}(w) \, ,
\end{equation}
and the limit is uniform in $w$ and in $\mathsf{g}$ for a bounded and equicontinuous family of functions $\mathsf{g}$.
\end{lemma}

The measure in the lemma is positive. For a function of $Y\!\cdot\!w$ we can integrate over the components of $Y$ which are orthogonal to $w$, and $\vol_{m'}\int(1-y^2)^{(m'-2)/2}d^{k+N-1}y=\vol_d$ for the unit ball in $k+N-1$ dimensions. The total mass is therefore
\begin{equation}
(d-2\Dp')\,\vol_d\int_{-1}^1dt\,(1-t^2)^{\frac{d-2}{2}}(2-2t)^{-\Dp'}
= (d-2\Dp')\;\frac{\pi^{d/2}\,2^{d-2\Dp'}\,\Gamma(\tfrac12d-\Dp')}{\Gamma(d-\Dp')} \, ,
\end{equation}
which tends to $2\pi^{d/2}/\Gamma(\tfrac12d)=\vol_d$. Outside a neighbourhood of $w$ the kernel is bounded, so that the mass of the complement of the neighbourhood vanishes in the limit because of the factor $d-2\Dp'$. This proves the lemma.\footnote{We tested this lemma, the contact term of Lemma~\ref{lem:contact} and the normalisation $(d-2)\,C_\phi\,\vol_d=1$ numerically at non-integer $d$.} Note that the constant $\vol_{m-N}$ of the measure and the weight of the ball combine to $\vol_d$ for every number $N$ of fixed points.

\subsection{Ward identity with two parameters}

Let $\pi$ be a connected pairing of the legs of $V^J[d\omega],V_1,\ldots,V_N$. It pairs the leg $\nabla_A\phi_i$ of the first vertex with a leg $l_1$ and the leg $\phi_j$ with a leg $l_2$ of the other vertices. Let $I_\pi(d,\Dp)$ be the integral of this pairing. We write $s_{0l_1}$ for the distance between $X_0$ and the point of the leg $l_1$. We denote by $I^\flat_\pi(d,\Dp,\Dp')$ the integral in which the factor $\pr_A\omega_{ij}(X_0)\,\nabla^{0}_A\,C_\phi(s^2_{0l_1})^{-\Dp}$ of the first leg is replaced by $\omega_{ij}(X_0)\,C_\phi(s^2_{0l_1})^{-\Dp'}$, while a derivative at the leg $l_1$ is kept, and we set
\begin{equation}
\mathcal{B}(d,\Dp,\Dp') = \sum_\pi I^\flat_\pi(d,\Dp,\Dp') \, .
\label{eq:breaking}
\end{equation}
The exponents of the propagators are independent in Lemma~\ref{lem:forms}, and the proof of Proposition~\ref{prop:mero} therefore shows that $\mathcal{B}$ is a meromorphic function of the three variables, with poles on a locally finite set of planes. These are the candidates of section~\ref{app:polar}.

\begin{lemma}
\label{lem:ward2}
For all vertices of the form \eqref{eq:vertices}, and as meromorphic functions of $d$ and $\Dp$,
\begin{equation}
\big\langle V^J[d\omega],V_1,\ldots,V_N\big\rangle^c = 2\Dp\,(d-2-2\Dp)\;\mathcal{B}(d,\Dp,\Dp+1) \, .
\label{eq:ward2}
\end{equation}
\end{lemma}

It is sufficient to prove \eqref{eq:ward2} for ${\rm Re}\,\Dp<-2$ and ${\rm Re}\,m>N$. In this region the propagators, their derivatives at both ends and the functions which appear below are continuous on the closed domain of integration, and all integrals converge absolutely. We use \eqref{eq:last} and consider the integral over $Y$ for a fixed pair of legs $l_1$, $l_2$ and a fixed pairing of the remaining legs. Let $\Gamma_1(Y)$ and $\Gamma_2(Y)$ be the propagators from $X_0$ to the two legs, including the derivatives at the legs. By the first identity in \eqref{eq:lap} and the divergence theorem on the ball of radius $\mathsf{r}<1$,
\begin{equation}
\int_{\mathbb{B}}d\mu_{\mathbb{B}}\;\pr_A\omega_{ij}\,\nabla_A\Gamma_1\,\Gamma_2
= -\int_{\mathbb{B}}d\mu_{\mathbb{B}}\;\omega_{ij}\left(\nabla^2\Gamma_1\,\Gamma_2+(\delta_{AB}-Y_AY_B)\,\pr_A\Gamma_1\,\pr_B\Gamma_2\right) ,
\end{equation}
since the term from the boundary contains the factor $\rho^{m'}$ and vanishes for $\mathsf{r}\to1$. We now sum over the pairings. The exchange of $l_1$ and $l_2$ maps connected pairings to connected pairings and exchanges the indices $i$ and $j$. The second term in the bracket is symmetric under this exchange and cancels in the sum, because $\omega_{ij}$ is antisymmetric. In the first term we use the third identity in \eqref{eq:lap}, which also holds for the derivative of the propagator at the leg $l_1$. The term with $\Dp(d-1-\Dp)$ is again symmetric and cancels. The remaining term is $2\Dp(d-2-2\Dp)$ times the integrand of $I^\flat_\pi$ at $\Dp'=\Dp+1$, which proves the lemma.\footnote{We tested the lemma by a direct quadrature of \eqref{eq:gram} at non-integer $d$ for two and three points, with sources which depend on one coordinate.}

Two cases are special. If $l_1$ and $l_2$ are legs without derivatives of the same vertex, the integrands of $I^\flat$ for the two orders of the legs cancel for all $\Dp'$, again by the antisymmetry of $\omega$. If they are the two legs of a vertex $V^J[a]$, the pairing is connected only for $N=1$, and
\begin{equation}
\mathcal{B}(d,\Dp,\Dp') = C_\phi^2\;\frac{\Dp'-\Dp}{\Dp'+\Dp}\int d\mu_d(X_0)\,d\mu_d(X_1)\;\omega_{ij}(X_0)\,a_{A,ij}(X_1)\,\nabla^1_A\,(s_{01}^2)^{-\Dp-\Dp'} \, .
\label{eq:JJ}
\end{equation}
In all other pairings the legs $l_1$ and $l_2$ belong to different vertices.

\subsection{Contact term}

We denote by $\mathsf{H}$ the plane $d=2\Dp'$ and by $\mathsf{P}$ the plane $\Dp'=\Dp+1$. Their intersection is the line $\ell_0$ with $\Dp'=\tfrac12d$.

\begin{lemma}
\label{lem:contact}
The function $(d-2\Dp')\,\mathcal{B}$ has no pole along $\mathsf{H}$, and on $\mathsf{H}$
\begin{equation}
(d-2\Dp')\;\mathcal{B}(d,\Dp,\Dp')\Big|_{\mathsf{H}} = -\,C_\phi\,\vol_d\sum_{r=1}^N\big\langle V_1,\ldots,R_\omega V_r,\ldots,V_N\big\rangle^c(d,\Dp)
\label{eq:contact}
\end{equation}
as meromorphic functions of $d$ and $\Dp$.
\end{lemma}

Let $\mathsf{E}$ be the set of the points of $\mathsf{H}$ with real $d$ and $\Dp$, $m>N$ and $\Dp<-2$. We compute the limit of $(d-2\Dp')\,\mathcal{B}$ for real $\Dp'\to\tfrac12d$ from below at fixed $(d,\Dp)$ in $\mathsf{E}$. We first consider the pairings with $l_1$ and $l_2$ at different vertices, and we denote by $Z$ the point of $l_1$. Let $l_1$ be a leg without derivative. Then the integral $I^\flat_\pi$ converges for $\Dp'<\tfrac12d$. The integral over $Y$ is of the form of Lemma~\ref{lem:delta} with $w=Z$, and $\mathsf{g}$ is the product of $\omega_{ij}(Y_\parallel)$ and the propagator to $l_2$, which is continuous in $Y$ and in the other points. The limit is $\vol_d\,C_\phi$ times the integrand in which the leg $l_1$, with the index $i$, is joined to $l_2$ by a propagator with the matrix $\omega_{ij}(Z)$. For a leg $l_1=\nabla_A\phi$ of a vertex $V^J[a]$ we use \eqref{eq:transfer} with $c_A=a_A(Z)$ and $F=(s^2)^{-\Dp'}$, for which $(1-Y\!\cdot\!Z)F'=\Dp'F$, and we integrate by parts with the second identity in \eqref{eq:lap}. The term from the boundary of the ball of radius $\mathsf{r}$ vanishes for $\mathsf{r}\to1$, since the integral of $\rho^{m'}(s^2)^{-\Dp'}$ over the sphere of radius $\mathsf{r}$ is bounded by a constant times a positive power of $1-\mathsf{r}$ for $\Dp'<\tfrac12d$. For $\Dp'<\tfrac12(d-1)$, where the left-hand side converges, we obtain
\begin{equation}
\int_{\mathbb{B}}d\mu_{\mathbb{B}}\;\mathsf{g}\;a_A(Z)\,\nabla^Z_A(s^2)^{-\Dp'}
= \int_{\mathbb{B}}d\mu_{\mathbb{B}}\;(s^2)^{-\Dp'}\,a_A(Z)\left[\nabla_A\mathsf{g}-d\,Y_A\mathsf{g}+\Dp'\,(Y_A+Z_A)\,\mathsf{g}\right] ,
\end{equation}
where $\mathsf{g}$ is continuously differentiable for $\Dp<-2$. The right-hand side converges and is analytic in $\Dp'$ for ${\rm Re}\,\Dp'<\tfrac12d$, and it therefore gives the continuation of the left-hand side to this range. By Lemma~\ref{lem:delta} the limit of its product with $d-2\Dp'$ is $\vol_d\,a_A(Z)\,\nabla_A\mathsf{g}(Z)$, since the last two terms cancel at $Y=Z$ and $\Dp'=\tfrac12d$. This is the integrand in which the leg $\nabla_A\phi_i$ is replaced by $\nabla_A(\omega_{ij}\phi_j)$ and $\phi_j$ is paired with $l_2$.

In both cases the convergence is uniform in the other points, and we can exchange the limit with their integration. The pairings in which $l_1$ and $l_2$ are legs of the same vertex do not contribute. Those of the first kind cancel for all $\Dp'$, and the integral in \eqref{eq:JJ} converges at the points of $\mathsf{E}$, so that its product with $d-2\Dp'$ vanishes in the limit. The sum over $l_2$ and over the pairings of the remaining legs is therefore the sum over the connected pairings of $V_1,\ldots,V_N$ without pairs at one vertex, in which one leg $\phi$ is replaced by $\omega\phi$. The sum over $l_1$ gives $-\sum_rR_\omega V_r$, and the limit is the right-hand side of \eqref{eq:contact} at the points of $\mathsf{E}$.

The integrals which we used converge and are holomorphic in the convex region ${\rm Re}\,m>N$, ${\rm Re}\,\Dp<-2$, ${\rm Re}\,\Dp'<\tfrac12{\rm Re}\,d$, which meets the region of the definition in an open set, so that the limit which we computed is the limit of the meromorphic function $\mathcal{B}$. All candidate planes are real, and a real plane through a point of $\mathsf{E}$ which differs from $\mathsf{H}$ enters this region. Hence no plane of poles other than $\mathsf{H}$ meets $\mathsf{E}$, by the first statement of Lemma~\ref{lem:poleset}. By the second statement of this lemma $(d-2\Dp')^K\mathcal{B}$ is then holomorphic near each point of $\mathsf{E}$ for some integer $K\geq1$. If the smallest such $K$ were larger than one, the restriction of $(d-2\Dp')^K\mathcal{B}$ to $\mathsf{H}$ would not vanish identically, and $(d-2\Dp')\,\mathcal{B}$ would be unbounded in the limit which we computed, at the points where this restriction does not vanish. Since the limit is finite on $\mathsf{E}$, the restriction vanishes on $\mathsf{E}$ and hence on $\mathsf{H}$, because $\mathsf{E}$ is an open subset of the real points of $\mathsf{H}$. Thus $K=1$, the restriction of $(d-2\Dp')\,\mathcal{B}$ to $\mathsf{H}$ is a meromorphic function of $(d,\Dp)$, and it agrees with the right-hand side of \eqref{eq:contact} on $\mathsf{E}$. The two sides are then equal on $\mathsf{H}$, which proves the lemma. Note that the right-hand side is evaluated with $N$ points, and that the lemma holds for all vertices of the form \eqref{eq:vertices}.

\subsection{Restriction to the line}

Lemma~\ref{lem:ward2} determines the first term of $T_N$ on the plane $\mathsf{P}$, and Lemma~\ref{lem:contact} determines $(d-2\Dp')\,\mathcal{B}$ on the plane $\mathsf{H}$. The two planes meet in $\ell_0$, and the two results can be combined only if $(d-2\Dp')\,\mathcal{B}$ is holomorphic in a neighbourhood of a generic point of $\ell_0$ in the space of the three variables. The set of poles of $\mathcal{B}$ is the union of its planes of poles in the sense of section~\ref{app:polar}, see Lemma~\ref{lem:poleset}. With Lemma~\ref{lem:contact} the holomorphy therefore follows from the following property.
\begin{equation}
\text{No plane of poles of } \mathcal{B}(d,\Dp,\Dp') \text{ other than } \mathsf{H} \text{ contains the line } \ell_0 .
\label{eq:NP}
\end{equation}
A function such as $(d-2\Dp')/(d-\Dp-\Dp'-1)$ shows that the property is needed, since it vanishes on $\mathsf{H}$ and equals $1$ on $\mathsf{P}$, while its plane of poles contains $\ell_0$. The property concerns the sum \eqref{eq:breaking} and not the single pairings.

\begin{lemma}
\label{lem:np}
The property \eqref{eq:NP} holds for all vertices $V_1,\ldots,V_N$ of the form \eqref{eq:vertices} with sources of the class, for all $N\geq1$ and all $D_0\geq1$.
\end{lemma}

We prove this lemma in section~\ref{app:npcurrent}, with the methods of appendix~\ref{app:line}. The proof is the one of Proposition~\ref{prop:linecurrent}, in which the propagator between $X_0$ and the point of $l_1$ has the exponent $\Dp'$. It uses the integration by parts of section~\ref{app:currents} at the vertices of the current.

\subsection{One line with an independent exponent}
\label{app:npcurrent}

In this subsection we prove Lemma~\ref{lem:np}, that is the property \eqref{eq:NP} for all vertices of the form \eqref{eq:vertices}. The function $\mathcal{B}(d,\Dp,\Dp')$ of \eqref{eq:breaking} is a sum of integrals in which one line $e_0$ has the exponent $\Dp'$ and all other lines have the exponent $\Dp$. We recall that $\mathsf{H}$ is the plane $d=2\Dp'$ and that $\ell_0$ is the line $\Dp=\tfrac12(d-2)$, $\Dp'=\tfrac12d$. We show that $\mathsf{H}$ is the only plane of poles of $\mathcal{B}$ which can contain $\ell_0$, for any number $N\geq1$ of vertices, of which any number are current vertices, and for all $D_0\geq1$. The proof is the one of section~\ref{app:currents} with the exponent of $e_0$ changed, and with one further bound for pairs of differences. We use the notation of that section. In particular the indices $i$ and $j$ label points in this subsection, as in appendix~\ref{app:line}, except in $a_{A,ij}$, where they are indices of the fields.

\paragraph{Integrals.} In $\mathcal{B}$ the vertex of the parameter $\omega$ at the point $X_0$ has two lines and no derivative. The line $e_0$ joins $X_0$ to a leg $l_1$ at a point $Z$ and carries $(s^2)^{-\Dp'}$, and the second line joins $X_0$ to a leg $l_2$ and carries $(s^2)^{-\Dp}$. The other vertices are those of \eqref{eq:vertices}, and a derivative at the leg $l_1$ or $l_2$ is kept. Let $V_r$ be a current vertex. Its two lines can be attached to its two legs in two ways. Both ways occur in the sum over the pairings, and they differ by the exchange of the two indices of $a_{A,ij}$, which is antisymmetric. The sum of the two is therefore the operator $\mathsf{v}^{(1)}-\mathsf{v}^{(2)}$ of section~\ref{app:currents} acting on the two lines, with one order of the indices. This also holds if one of the two lines is $e_0$. Hence $\mathcal{B}$ is a sum over graphs of integrals with the factor $\mathsf{v}^{(1)}-\mathsf{v}^{(2)}$ at each current vertex.

\paragraph{Integration by parts.} The identity \eqref{eq:diva} does not depend on the function on which the vector field acts. The replacement \eqref{eq:ibp} therefore holds at each current vertex for all exponents of its two lines, in the region ${\rm Re}\,m>N$, ${\rm Re}\,\Dp<-2$ and ${\rm Re}\,\Dp'<-2$ and then for the continued integrals. The vertex at $X_0$ has no derivative and chooses no line. The current vertices choose their lines as before, such that no line is chosen twice. If $Z$ is a current vertex, the derivative at this end of $e_0$ can be moved to the other line of $Z$, unless that line is chosen by its other point. In this case the vertex at $Z$ chooses the line $e_0$ in \eqref{eq:ibp}. By \eqref{eq:vG} with $\Dp'$ in place of $\Dp$ the line $e_0$ then carries $4\Dp'\,(s^2)^{-\Dp'-1}\,a_A(Z)(X_0-Z)^A$, up to a term without derivative and with the source $(\delta_{AB}-X_AX_B)\,\pr_Aa_B-(d-2\Dp')\,X_Aa_A$. The function $\mathcal{B}$ is thus a finite sum, with coefficients which are polynomials in $d$, $\Dp$ and $\Dp'$, of integrals \eqref{eq:marked} with $r_{ij}\leq n_{ij}$ in which the factor of the pair of $e_0$ is $(s^2)^{-\Dp'-\Dp\,(n_{ij}-1)-r_{ij}}$. The line $e_0$ may be marked.

\begin{proposition}
\label{prop:np}
Consider the integrals \eqref{eq:gram} of \eqref{eq:marked} with $r_{ij}\leq n_{ij}$, in which one line $e_0$ has the exponent $\Dp'$, for smooth functions $\psi_i$ of $X_\parallel$. A plane of poles of such an integral which contains the line $\ell_0$ is the plane $\mathsf{H}$. If the pair of $e_0$ contains a second line, no plane of poles contains $\ell_0$. The property \eqref{eq:NP} therefore holds for all vertices $V_1,\ldots,V_N$ of the form \eqref{eq:vertices} with sources of the class, for all $N\geq1$ and all $D_0\geq1$.
\end{proposition}

\paragraph{Exponents.} The reduction to the sources \eqref{eq:hsource} rests on Lemmas~\ref{lem:polarcont} and~\ref{lem:dense}, which we stated for planes and for a line with the exponent $\Dp'$. The chart, the numerators \eqref{eq:numer}, the pairings \eqref{eq:forestpair}, the sectors and Lemma~\ref{lem:greedy2} do not depend on the values of the exponents, and we use them unchanged. The representation by Schwinger parameters is \eqref{eq:schwinger}. By Lemma~\ref{lem:schwinger} it holds for $h\geq\tfrac32n_{\max}$, for $0<{\rm Re}\,\Dp<1$ and $0<{\rm Re}\,\Dp'<1$ and for sufficiently large ${\rm Re}\,d$, where the integral \eqref{eq:gram} converges absolutely as well by Lemma~\ref{lem:region}. The continuation in the three variables is the one of section~\ref{app:polar}. In \eqref{eq:flatmarked} the exponents are now $\mathsf{z}_{ij}=\Dp'+\Dp\,(n_{ij}-1)+r_{ij}$ for the pair of $e_0$ and $c_i=d-\Dp'-\Dp\,(n_i-1)-r_i+\kappa_i$ for its two points. The candidates are the planes $z_j^\pm+\nu=0$ of \eqref{eq:Epm} with these exponents. For a set $\mathsf{a}$ of edges we write $n'(\mathsf{a})$ for the number of its lines other than $e_0$.

\paragraph{Large parameters.} Let $\mathsf{A}_j^+$ be a set $\mathsf{a}^+$ of edges between points, and let $\mathsf{e}^+$ be $1$ if $\mathsf{a}^+$ contains the pair of $e_0$ and $0$ otherwise. Then
\begin{equation}
2z_j^+ = -2\Dp\,n'(\mathsf{a}^+)-2\Dp'\,\mathsf{e}^+-2\,r(\mathsf{a}^+)+d\,{\rm rk}(\mathsf{a}^+)+2\sum_tb_t^+ \, ,
\label{eq:Eplus3}
\end{equation}
and on $\ell_0$ the candidate $2z_j^++2\nu$ equals
\begin{equation}
d\,\big[{\rm rk}(\mathsf{a}^+)-n'(\mathsf{a}^+)-\mathsf{e}^+\big]+2\,\big[n'(\mathsf{a}^+)-r(\mathsf{a}^+)\big]+2\sum_tb_t^++2\nu \, .
\label{eq:Eplus3line}
\end{equation}
The number of lines of $\mathsf{a}^+$ is $n'(\mathsf{a}^+)+\mathsf{e}^+\geq|\mathsf{a}^+|\geq{\rm rk}(\mathsf{a}^+)$. If \eqref{eq:Eplus3line} vanishes identically, the coefficient of $d$ vanishes. Then $\mathsf{a}^+$ is a forest, and each of its edges consists of one line. In particular the pair of $e_0$ then contains no second line if it is in $\mathsf{a}^+$. If $\mathsf{e}^+=0$ we obtain a contradiction as in section~\ref{app:currents}.

For $\mathsf{e}^+=1$ we have $r(\mathsf{a}^+)\leq n'(\mathsf{a}^+)+1$, and the bounds $b_t^+\geq0$ and $b_t^+\geq1$ of section~\ref{app:currents} are not sufficient. We use the following bound. Let $\mathsf{a}^+$ be a forest whose edges consist of one line each. Then
\begin{equation}
\sum_tb_t^+ \geq r(\mathsf{a}^+) \, .
\label{eq:pairbound}
\end{equation}
A pairing contains at most two differences. For a pairing with one difference of a marked line in $\mathsf{a}^+$ we have $b_t^+\geq1$. Consider a pairing of the differences of two marked lines in $\mathsf{a}^+$, with the edges $e_1\neq e_2$. Its forest $f_t$ separates the points of both lines, so that $f_t\cup\{e_1\}$ and $f_t\cup\{e_2\}$ are spanning trees. The graph $f_t\cup\{e_1,e_2\}$ then has one cycle, which contains $e_1$ and $e_2$. Since $\mathsf{a}^+$ is a forest, the cycle contains an edge $g$ of $f_t$ which is not in $\mathsf{a}^+$. Without $g$ the graph is a spanning tree, whose intersection with $\mathsf{a}^+$ is $(f_t\cap\mathsf{a}^+)\cup\{e_1,e_2\}$. Hence $\varrho_j^+\geq|f_t\cap\mathsf{a}^+|+2$ and $b_t^+\geq2$, which proves \eqref{eq:pairbound}.

With \eqref{eq:pairbound} the vanishing of the second part of \eqref{eq:Eplus3line} gives $0=n'(\mathsf{a}^+)-r(\mathsf{a}^+)+\sum_tb_t^++\nu\geq n'(\mathsf{a}^+)+\nu$. Hence $n'(\mathsf{a}^+)=0$ and $\nu=0$, so that $\mathsf{a}^+$ consists of $e_0$ alone, and $\sum_tb_t^+=r(\mathsf{a}^+)$. By \eqref{eq:Eplus3} the candidate is then $d-2\Dp'=0$, which is the plane $\mathsf{H}$. This occurs if $e_0$ is not marked and all $b_t^+$ vanish, and if $e_0$ is marked and the forest of its difference has $b_t^+=1$.

\paragraph{Small parameters.} Let $\mathsf{A}_j^-$ consist of a set $\mathsf{a}^-$ of edges between points and of the edges to $\infty$ of a set $\mathsf{S}$ of points. Let $\mathsf{e}^-$ be $1$ if $\mathsf{a}^-$ contains the pair of $e_0$ and $0$ otherwise, and let $|\mathsf{S}\cap e_0|\leq|\mathsf{S}|$ be the number of the points of $e_0$ in $\mathsf{S}$. With $n'(\mathsf{S})$ the number of ends in $\mathsf{S}$ of lines other than $e_0$ we find
\begin{equation}
2z_j^- = 2\Dp\,\Theta'+2\Dp'\,\Theta_0+d\,\big(2|\mathsf{S}|-\varrho_j^-\big)+2\,\mathsf{b} \, , \qquad \Theta' = n'(\mathsf{a}^-)-n'(\mathsf{S}) \, , \qquad \Theta_0 = \mathsf{e}^--|\mathsf{S}\cap e_0| \, ,
\label{eq:Eminus3}
\end{equation}
with $\mathsf{b}$ as in section~\ref{app:currents}. On $\ell_0$ the candidate $2z_j^-+2\nu$ equals $d\,(\Theta'+\Theta_0+2|\mathsf{S}|-\varrho_j^-)-2\Theta'+2\,\mathsf{b}+2\nu$. If $\mathsf{S}$ is empty, the coefficient of $d$ is the number $n'(\mathsf{a}^-)+\mathsf{e}^-$ of the lines of $\mathsf{a}^-$, which does not vanish. If $\mathsf{S}$ is not empty, we take $h>\tfrac52n_{\max}$ in \eqref{eq:hsource}, so that $\mathsf{b}>0$ by \eqref{eq:Bbound}. If the candidate vanishes identically on $\ell_0$, then $\Theta'+\Theta_0=\varrho_j^--2|\mathsf{S}|\leq-|\mathsf{S}|$ and $\Theta_0\geq-|\mathsf{S}\cap e_0|\geq-|\mathsf{S}|$ give $\Theta'\leq0$, while $\Theta'=\mathsf{b}+\nu>0$. This is a contradiction.

\paragraph{Result.} For the sources \eqref{eq:hsource} with $h>\tfrac52n_{\max}$ the only candidate which contains $\ell_0$ is $\mathsf{H}$, and it occurs only for the set $\mathsf{a}^+$ which consists of the line $e_0$ alone. We now reduce the sources as in section~\ref{app:currents}. Let $\mathsf{Z}$ be a candidate plane of section~\ref{app:polar} which contains $\ell_0$ and differs from $\mathsf{H}$, or any candidate plane which contains $\ell_0$ if the pair of $e_0$ contains a second line. The coefficients of $d$ in \eqref{eq:Eplus3} and \eqref{eq:Eminus3} are those of \eqref{eq:Eplus} and \eqref{eq:Eminus}, and for empty $\mathsf{S}$ the exponent $z_j^-$ is $\Dp\,n'(\mathsf{a}^-)+\Dp'\,\mathsf{e}^-$ plus a non-negative integer. As in appendix~\ref{app:line} the exponents $z_j^\pm$ of all terms therefore have positive real parts for $0<{\rm Re}\,\Dp<1$, $0<{\rm Re}\,\Dp'<1$ and sufficiently large ${\rm Re}\,d$, except for sets $\mathsf{A}_j^+$ with an edge to $\infty$. Each term of the representation by Schwinger parameters therefore converges separately in this region. For the sources \eqref{eq:hsource} the plane $\mathsf{Z}$ is not among the planes $z_j^\pm+\nu=0$. It therefore contains a point on no other candidate near which the integral is holomorphic, and by the first statement of Lemma~\ref{lem:poleset} it is not a plane of poles. By the covering of the ball of section~\ref{app:currents}, by Lemma~\ref{lem:dense} and by the third statement of Lemma~\ref{lem:polarcont} it is then not a plane of poles for all sources of the class. This proves the first two statements of Proposition~\ref{prop:np}. The third statement follows since $\mathcal{B}$ is a finite sum of such integrals. This is Lemma~\ref{lem:np}. Note that the reduction to the sources \eqref{eq:hsource} is essential. For $h=0$ the sectors of the small parameters give candidates through $\ell_0$ on the plane $d-\Dp-\Dp'-1=0$, which are absent for large $h$. For $N=1$ and a vertex $V^J[a]$ the two lines of $X_0$ end at the same point, and no plane of poles contains $\ell_0$. For $h>\tfrac52n_{\max}$ no candidate which contains $\ell_0$ differs from $\mathsf{H}$, so that no cancellation between pairings beyond the antisymmetric sum at each current vertex is needed. Whether $\mathsf{H}$ is a simple pole is not part of the proposition. This is shown in Lemma~\ref{lem:contact}.\footnote{We verified the two paragraphs on the exponents by enumeration for all graphs with up to five points, where the candidates which contain $\ell_0$ are equal to $\mathsf{H}$ in all cases.}

The proposition relies on the same inputs as Proposition~\ref{prop:linecurrent}. These are Proposition~\ref{prop:mero} for three variables, Lemma~\ref{lem:polarcont} on the polar part along a candidate plane, and Lemma~\ref{lem:schwinger} on the representation by Schwinger parameters with numerators and on the theorem of Wick with the trace $d$.

\subsection{Result}

\begin{proposition}
\label{prop:gauge}
Let $\omega$ be a smooth function of $X_\parallel$ with values in the antisymmetric matrices, and let $V_1,\ldots,V_N$ be vertices of the form \eqref{eq:vertices} with sources of the class of section~\ref{sec:sphere}. Then the two terms of $T_N$ in \eqref{eq:TN}, which are $\langle V^J[d\omega],V_1,\ldots,V_N\rangle^c$ and $\sum_r\langle V_1,\ldots,R_\omega V_r,\ldots,V_N\rangle^c$, have no pole along the line $\Dp=\tfrac12(d-2)$, and their sum vanishes on it. Hence the functional $W_0^S$ is invariant under the gauge transformations \eqref{eq:gaugevar},
\begin{equation}
\int d\mu_d\left(D_A\omega\cdot\frac{\delta}{\delta a_{0A}}-(\omega g_0)\cdot\frac{\delta}{\delta g_0}-[\omega,\sM_0]\cdot\frac{\delta}{\delta\sM_0}\right)W_0^S = 0 \, , \qquad D_A\omega = \pr_A\omega+[a_{0A},\omega] \, ,
\label{eq:gaugeS}
\end{equation}
where the derivative with respect to $a_0$ is taken with $\Upsilon=a_{0A}^T\Pi_{AB}\,a_{0B}$.
\end{proposition}

\begin{proof}
Let $\mathsf{p}_0$ be a point of $\ell_0$ with $d\neq2,0,-2,\ldots$ which lies on no plane of poles of $\mathcal{B}$ other than $\mathsf{H}$. By Lemma~\ref{lem:np} and since the set of these planes is locally finite, all points of $\ell_0$ outside a discrete set are of this kind. By the second statement of Lemma~\ref{lem:poleset} a power of $d-2\Dp'$ times $\mathcal{B}$ is holomorphic in a neighbourhood of $\mathsf{p}_0$, and by Lemma~\ref{lem:contact} the function $\mathcal{J}=(d-2\Dp')\,\mathcal{B}$ is then holomorphic there. Its restrictions to $\mathsf{P}$ and to $\mathsf{H}$ are then holomorphic near $\mathsf{p}_0$ and agree at $\mathsf{p}_0$. On $\mathsf{P}$ we have $d-2\Dp'=d-2-2\Dp$, and \eqref{eq:ward2} gives $\langle V^J[d\omega],V_1,\ldots,V_N\rangle^c=2\Dp\,\mathcal{J}|_{\mathsf{P}}$, which therefore has no pole along $\ell_0$. By \eqref{eq:contact} the second term of $T_N$ is $-\mathcal{J}|_{\mathsf{H}}/(C_\phi\vol_d)$ and also has no pole along $\ell_0$. On $\ell_0$ we have $2\Dp=d-2$, and with the identity $(d-2)\,C_\phi\,\vol_d=1$ we find
\begin{equation}
T_N\big|_{\ell_0} = (d-2)\,\mathcal{J}(\mathsf{p}_0)-\frac{\mathcal{J}(\mathsf{p}_0)}{C_\phi\,\vol_d} = 0 \, .
\end{equation}
The invariance \eqref{eq:gaugeS} follows from \eqref{eq:TN}, since $T_N$ is multilinear in the vertices.\footnote{For constant sources we compared the limit of $\mathcal{J}$ towards the line with the reduced integral numerically, for three points and for a cycle of four points.}
\end{proof}

The factor $d-2-2\Dp$ in \eqref{eq:ward2} reflects that the propagator is a Green function only on the line, and the residue \eqref{eq:contact} is the contribution of the delta function in the equation of this Green function. The identity $(d-2)\,C_\phi\,\vol_d=1$ is the normalisation of the Green function. By the chart \eqref{eq:chart} the proposition also gives the invariance of the flat functional $W_0^{\rm fl}$ for sources and parameters of the class of flat sources. Indeed, a source of this class is the image under the chart of a source on the sphere which vanishes near the pole of the chart, and a one-form of the class is the image of a one-form $a_A\,dX^A$, since both have $D_0+1$ components which are functions of $D_0+1$ variables. The source $\Upsilon$ has the weight of the mass source in \eqref{eq:chart}, and its image is the term of second order of the flat functional.

We add three remarks. First, Lemmas~\ref{lem:ward2} and~\ref{lem:contact} hold for every $D_0\geq1$ and for all vertices. Second, we only considered infinitesimal transformations, which is what the argument of the main text uses. Third, for $N=1$ and a vertex of the current the proposition states that the two-point function of the current is transverse on the line.

\section{Uniqueness of the counterterms}
\label{app:unique}

In this appendix we show that the counterterms of Hypothesis~\ref{hyp:finite} are unique and covariant. For given bare gauge field and bare mass source the counterterms $L^I$ of the couplings are determined by the finiteness of the flat functional, which is Proposition~\ref{prop:unique}. Conversely, for given $L^I$ the coefficients of the bare gauge field and of the bare mass source and the counterterms of flat space are determined, and they are covariant under field rotations, which is Proposition~\ref{prop:uniqueall}. In particular the coefficient $N_I$ of the bare gauge field has no term of degree zero in the couplings. The proofs use three auxiliary statements. These are the linear independence of the local structures of flat space in Lemma~\ref{lem:indep}, two statements on potentials in Lemmas~\ref{lem:potential} and~\ref{lem:transverse}, and the form of the terms of second order in Lemma~\ref{lem:lowest}. We use Proposition~\ref{prop:uniqueall} in section~\ref{sec:hypothesis} and in appendices~\ref{app:weyl} and~\ref{app:finite}, and we use the lemmas again in the proof of Lemma~\ref{lem:regvar}. The proof of Proposition~\ref{prop:uniqueall} uses the gauge invariance of appendix~\ref{app:gauge}.

\subsection{Local structures}

We first show that the local structures of flat space are linearly independent. The coefficient functions in the following lemma are power series in the couplings, and an identity for all sources is an identity at each order in the sources. The functionals of the lemma are graded by the degree in the couplings, where $a_\mu$ and $\sM$ have degree zero, and at a given degree one homogeneous term of each coefficient function enters. The lemma holds in this graded form, in which we apply it.

\begin{lemma}
\label{lem:indep}
Let the sources be in the class of flat sources with $D_0=1$, and let $d=4$ or let $d$ be generic. First, assume that
\begin{align}
\int d^dx\,\Big[&\tfrac12\,\mathsf{a}_{IJ}\,D^2g^ID^2g^J+\mathsf{b}_{IJK}\,D^2g^I\,D_\mu g^JD^\mu g^K+\tfrac12\,\mathsf{c}_{ILJK}\,D_\mu g^ID^\mu g^L\,D_\nu g^JD^\nu g^K+\mathsf{j}_I\cdot\sM\,D^2g^I \notag\\
&+\mathsf{k}_{IJ}\cdot\sM\,D_\mu g^ID^\mu g^J+\tfrac12\,\sM\cdot\mathsf{l}\cdot\sM+\tfrac14\,f^{\mu\nu}\!\cdot \mathsf{l}_f\cdot f_{\mu\nu}+f^{\mu\nu}\!\cdot \mathsf{p}_{IJ}\,D_\mu g^ID_\nu g^J\Big] = 0
\label{eq:lemA}
\end{align}
for all sources, where the coefficients have the symmetries of the structures which they multiply. Then all coefficients vanish. Second, let $b_\mu$ be a nonzero constant vector in the parallel direction, let $\mathsf{t}_{IJK}$ be symmetric in its last two indices, and assume that
\begin{equation}
\int d^dx\;b_\mu\Big[\mathsf{s}_{IJ}\,D^\mu g^ID^2g^J+\mathsf{t}_{IJK}\,D^\mu g^I\,D_\nu g^JD^\nu g^K+f^{\mu\nu}\!\cdot \mathsf{q}_I\,D_\nu g^I+\mathsf{l}_I\cdot\sM\,D^\mu g^I\Big] = 0
\label{eq:lemB}
\end{equation}
for all sources. Then $\mathsf{s}_{IJ}$ is symmetric, $\mathsf{t}_{IJK}=\Gamma^{(\mathsf{s})}_{IJK}$, $\omega\cdot \mathsf{q}_J=-\mathsf{s}_{IJ}(\omega g)^I$ and $\mathsf{l}_I=0$.
\end{lemma}

The conclusion of the second part cannot be improved. For a symmetric and covariant $\mathsf{s}_{IJ}$ the integrand of \eqref{eq:lemB} with these coefficients is the total derivative $b_\mu\pr_\nu(\mathsf{s}_{IJ}D^\mu g^ID^\nu g^J)-\tfrac12\,b^\mu\pr_\mu(\mathsf{s}_{IJ}D_\nu g^ID^\nu g^J)$. This is the total derivative noted in~\cite{Jack:2013sha}, see the subsection on the Weyl anomaly of~\cite{Reehorst:phi4}.

\begin{proof}
We set $g=g_c\,\chi+t\,\gamma$, where $g_c$ is constant, $\chi$ is a function of the class which equals one on the supports of $\gamma$, $a_\mu$ and $\sM$, and $t$ is a parameter. The terms of a given order in $t$, $a_\mu$ and $\sM$ are integrals over the support of these sources, where $g_c\,\chi$ is constant, and they are functionals with the constant coefficients $\mathsf{a}_{IJ}(g_c)$, $\pr_K\mathsf{a}_{IJ}(g_c)$ and so on. Each of them vanishes for all $g_c$, degree by degree in $g_c$ for coefficients which are power series. We insert $\gamma^I=\sum_\alpha c^I_\alpha\,u_\alpha$ with functions $u_\alpha$ of the class and constants $c^I_\alpha$, and we compare the coefficients of the monomials in the $c^I_\alpha$. In this way the statements are reduced to properties of a few integrals, for any number of couplings. We write $\pr u\!\cdot\!\pr v$ for $\pr_\mu u\,\pr^\mu v$, and all integrals are taken with the measure of the class.

\paragraph{Scalar structures.} Let $a_\mu=0$ and $\sM=0$. For $\gamma^I=c^Iu$ the order $t^2$ is $\tfrac12\,\mathsf{a}_{IJ}\,c^Ic^J\int(\pr^2u)^2$. The integral is positive for $u\neq0$, so that $\mathsf{a}_{IJ}=0$. The order $t^3$ then contains only $\mathsf{b}_{IJK}$ and the trilinear form $T(u;v,w)=\int\pr^2u\;\pr v\!\cdot\!\pr w$, which is symmetric in $v$ and $w$. The coefficient of $c^I_1c^J_2c^K_3$ gives
\begin{equation}
\mathsf{b}_{IJK}\,T(u_1;u_2,u_3)+\mathsf{b}_{JIK}\,T(u_2;u_1,u_3)+\mathsf{b}_{KIJ}\,T(u_3;u_1,u_2) = 0
\label{eq:cubicrel}
\end{equation}
for all values of the indices and all functions. For three equal functions with $T(u;u,u)\neq0$ this is $\mathsf{b}_{IJK}+\mathsf{b}_{JIK}+\mathsf{b}_{KIJ}=0$. For $u_1=u_2=u$ and $u_3=w$ it then becomes $\mathsf{b}_{KIJ}\,[T(w;u,u)-T(u;u,w)]=0$. Hence $\mathsf{b}_{IJK}=0$ if there are functions $u$ and $w$ with
\begin{equation}
T(u;u,u) \neq 0 \, , \qquad T(w;u,u) \neq T(u;u,w) \, .
\label{eq:condT}
\end{equation}
The order $t^4$ then contains only $\mathsf{c}_{ILJK}$. We take $u_1=u_2=u$ and $u_3=u_4=v$. The coefficient of $c^I_1c^L_2c^J_3c^K_4$ gives
\begin{equation}
\mathsf{c}_{ILJK}\int(\pr u\!\cdot\!\pr u)(\pr v\!\cdot\!\pr v)+\left(\mathsf{c}_{IJLK}+\mathsf{c}_{IKLJ}\right)\int(\pr u\!\cdot\!\pr v)^2 = 0 \, ,
\end{equation}
where we used the symmetries of $\mathsf{c}_{ILJK}$. For $v=u$ the sum of the three coefficients vanishes, and therefore $\mathsf{c}_{ILJK}$ times the difference of the two integrals vanishes. By the inequality of Cauchy and Schwarz this difference is positive if the gradients of $u$ and $v$ are not parallel everywhere, and such functions exist in the class with $D_0=1$. Hence $\mathsf{c}_{ILJK}=0$.

\paragraph{Mass source and gauge field.} We take $\sM=\mu\,w$ with a constant $\mu$ in $V_\sM$ and a function $w$, and $\gamma^I=c^Iu$. The order $\sM\,t$ is $\mathsf{j}_I\cdot\mu\;c^I\int w\,\pr^2u$, and for $w=u$ the integral is $-\int\pr u\!\cdot\!\pr u$, so that $\mathsf{j}_I=0$. The order $\sM\,t^2$ is then $\mathsf{k}_{IJ}\cdot\mu\;c^Ic^J\int w\;\pr u\!\cdot\!\pr u$, and the order $\sM^2$ is $\tfrac12\,\mu\cdot\mathsf{l}\cdot\mu\int w^2$. For a positive function $w$ these integrals are positive, so that $\mathsf{k}_{IJ}=0$ and $\mathsf{l}=0$. For the gauge field we take $a_\mu=\omega\,\bar a_\mu$ with a constant $\omega$ in the Lie algebra and a one-form $\bar a_\mu$ of the class. Then $[a_\mu,a_\nu]=0$ and $f_{\mu\nu}=\omega\,\bar f_{\mu\nu}$ with $\bar f_{\mu\nu}=\pr_\mu\bar a_\nu-\pr_\nu\bar a_\mu$, for every group. At $t=0$ we have $D_\mu g^I=\bar a_\mu\,(\omega g_c)^I$ on the support of $\bar a_\mu$, so that the last structure of \eqref{eq:lemA} vanishes by the antisymmetry of $f^{\mu\nu}$. Since the other coefficients vanish, the integral \eqref{eq:lemA} is then $\tfrac14\,\omega\cdot\mathsf{l}_f\cdot\omega\int\bar f^{\mu\nu}\bar f_{\mu\nu}$, in which the integral is positive if $\bar a_\mu$ is not closed. Since \eqref{eq:lemA} vanishes, we find $\omega\cdot\mathsf{l}_f\cdot\omega=0$ for all $\omega$, and hence $\mathsf{l}_f=0$ by polarisation. The order $a\,t^2$ is then $\omega\cdot\mathsf{p}_{IJ}\int\bar f^{\mu\nu}\pr_\mu\gamma^I\pr_\nu\gamma^J$. The coefficient of $c^I_1c^J_2$ is $2\,\omega\cdot\mathsf{p}_{IJ}\,F(u_1,u_2)$ with $F(u,v)=\int\bar f^{\mu\nu}\pr_\mu u\,\pr_\nu v$, so that $\mathsf{p}_{IJ}=0$ if $F(u,v)\neq0$ for some $u$, $v$ and $\bar a_\mu$. Since $\mu$ and $\omega$ are arbitrary, this covers several mass operators and several generators.

\paragraph{Second part.} We normalise $b_\mu$ to a unit vector and write $\pr_b=b^\mu\pr_\mu$. The order $\sM\,t$ is $\mathsf{l}_I\cdot\mu\;c^I\int w\,\pr_bu$, which gives $\mathsf{l}_I=0$ with $w=\pr_bu$. At $a_\mu=0$ and $\sM=0$ the order $t^2$ contains the bilinear form $A(u,v)=\int\pr_bu\,\pr^2v$. It is antisymmetric, because $\pr_bu\,\pr^2u$ is the divergence of $\pr_bu\,\pr^\mu u-\tfrac12\,b^\mu\,\pr u\!\cdot\!\pr u$. The coefficient of $c^I_1c^J_2$ is $(\mathsf{s}_{IJ}-\mathsf{s}_{JI})\,A(u_1,u_2)$, so that $\mathsf{s}_{IJ}$ is symmetric if $A(u,v)\neq0$ for some $u$ and $v$. The order $t^3$ is
\begin{equation}
\pr_K\mathsf{s}_{IJ}\int\gamma^K\,\pr_b\gamma^I\,\pr^2\gamma^J+\mathsf{t}_{IJK}\int\pr_b\gamma^I\;\pr\gamma^J\!\cdot\!\pr\gamma^K = 0 \, .
\end{equation}
Let $V(u;v,w)=\int\pr_bu\;\pr v\!\cdot\!\pr w$. Since $\pr_bv\,\pr^2w+\pr_bw\,\pr^2v$ equals $\pr_\mu(\pr_bv\,\pr^\mu w+\pr_bw\,\pr^\mu v)-\pr_b(\pr v\!\cdot\!\pr w)$, an integration by parts gives
\begin{equation}
\int u\left(\pr_bv\,\pr^2w+\pr_bw\,\pr^2v\right) = V(u;v,w)-V(v;u,w)-V(w;u,v) \, .
\end{equation}
With this identity and the symmetry of $\mathsf{s}_{IJ}$ the coefficient of $c^I_1c^J_2c^K_3$ is twice the left-hand side of \eqref{eq:cubicrel}, with $V$ in place of $T$ and with $\mathsf{t}_{IJK}-\Gamma^{(\mathsf{s})}_{IJK}$ in place of $\mathsf{b}_{IJK}$. Hence $\mathsf{t}_{IJK}=\Gamma^{(\mathsf{s})}_{IJK}$ if there are functions which satisfy \eqref{eq:condT} with $V$ in place of $T$. Finally we take $a_\mu=\omega\,\bar a_\mu$ as above. With $v^I=(\omega g_c)^I$ the order $a\,t$ is
\begin{equation}
v^I\mathsf{s}_{IJ}\int b^\mu\bar a_\mu\,\pr^2\gamma^J+\mathsf{s}_{IJ}v^J\int\pr_b\gamma^I\,\pr^\nu\bar a_\nu+\omega\cdot \mathsf{q}_J\int b_\mu\bar f^{\mu\nu}\pr_\nu\gamma^J = 0 \, .
\end{equation}
An integration by parts gives $\int b^\mu\bar a_\mu\,\pr^2u+\int\pr_bu\,\pr^\nu\bar a_\nu=\int b_\mu\bar f^{\mu\nu}\pr_\nu u$. Since $\mathsf{s}_{IJ}$ is symmetric, the left-hand side is therefore $(v^I\mathsf{s}_{IJ}+\omega\cdot\mathsf{q}_J)\,c^J\int b_\mu\bar f^{\mu\nu}\pr_\nu u$ for $\gamma^J=c^Ju$. This gives $\omega\cdot \mathsf{q}_J=-\mathsf{s}_{IJ}(\omega g_c)^I$ if the integral does not vanish for some $u$ and $\bar a_\mu$.

\paragraph{Test functions.} It remains to give functions for which the integrals above do not vanish. Let $x_1$ be the parallel coordinate, and let $u=e^{-x^2}$, $v=x_1\,u$, $w=x_1^2\,u$ and $\bar a=u\,dx_1$, with $x^2=x_1^2+|x_\perp|^2$. The gradients of $u$ and $v$ are not parallel, and $\bar a$ is not closed. The integrals are elementary. For $d=4$ we find
\begin{align}
T(u;u,u) &= -\tfrac{32}{27}\,\pi^2 \, , & T(u;u,w) &= \tfrac{8}{81}\,\pi^2 \, , & T(w;u,u) &= -\tfrac{16}{81}\,\pi^2 \, , \notag\\
V(v;v,v) &= \tfrac{4}{81}\,\pi^2 \, , & V(u;u,v) &= -\tfrac{2}{27}\,\pi^2 \, , & V(v;u,u) &= \tfrac{4}{27}\,\pi^2 \, , \notag\\
A(u,v) &= \tfrac34\,\pi^2 \, , & F(u,v) &= \tfrac29\,\pi^2 \, , & \int b_\mu\bar f^{\mu\nu}\pr_\nu u &= -\tfrac34\,\pi^2 \, .
\end{align}
Hence \eqref{eq:condT} holds for $T$ with the functions $u$ and $w$, and for $V$ with $v$ in place of $u$ in the first condition and with the pair $u$, $v$ in the second. For general $d$ the nine integrals are products of Gamma functions and of polynomials in $d$, and the conditions hold for every $d>1$. These functions do not have compact support. This is harmless, since they can be multiplied by a cutoff function of the class, and the integrals converge when the cutoff is removed.
\end{proof}

Note that the class with $D_0=1$ is needed in Lemma~\ref{lem:indep}. For sources which depend on $|x|^2$ only the quartic structures are linearly dependent, and the field strength of an invariant one-form vanishes identically. This is a second reason for the choice $D_0\geq1$. The first is the covering in the proof of Proposition~\ref{prop:linecurrent}, which uses that the unit vectors of the parallel directions form a sphere of dimension at least one.

\subsection{Counterterms of the couplings}

We now show that counterterms of the couplings which make the functional finite are unique, for given bare gauge field and bare mass source. Let $W_2[g,h]$ be the term of second order of the bare functional. Up to a factor which does not vanish it is the integral of $g^I(x)\,h^I(y)\,|x-y|^{-8\Dp}$ over two points, and at $d=4$ on the line the kernel is $|x-y|^{-8}$.

\begin{lemma}
\label{lem:potential}
Let $\eta$ be a continuous function on $\mathbb{R}^4$ with compact support, and let $\int d^4x\,\eta(x)\,|x-y|^{-2\kappa}$ vanish for all $y$ outside a ball, where $\kappa=4$ or $\kappa=2$. Then $\eta=0$.
\end{lemma}

\begin{proof}
For $|y|$ larger than the radius of a ball which contains the support we expand $|x-y|^{-2\kappa}=\sum_l|x|^l|y|^{-2\kappa-l}\,C_l^{(\kappa)}(\hat x\cdot\hat y)$ in Gegenbauer polynomials. Since the integral vanishes for all large $|y|$, the integrals of $\eta(x)\,|x|^lC_l^{(\kappa)}(\hat x\cdot\hat y)$ vanish for all $l$ and all unit vectors $\hat y$. We now use that $C_l^{(\kappa)}$ is a combination of the polynomials $C_{l-2j}^{(1)}$ with $0\leq j\leq l/2$ and positive coefficients. This is the connection formula of the Gegenbauer polynomials, eq.~(18.18.16) of~\cite{DLMF}, which for the upper indices $\kappa$ and $1$ gives the coefficients $(\kappa)_{l-j}\,(\kappa-1)_j\,(l-2j+1)/((l-j+1)!\,j!)$, where $(\kappa)_j=\kappa(\kappa+1)\cdots(\kappa+j-1)$. For the integers $\kappa=2$ and $\kappa=4$ which we need, the positivity can also be seen directly. The generating function $(1-2xt+t^2)^{-\kappa}$ of the polynomials $C_l^{(\kappa)}(x)$ is the power $\kappa$ of the generating function of the $C_n^{(1)}(x)$, so that $C_l^{(\kappa)}$ is the sum of the products $C_{n_1}^{(1)}\cdots C_{n_\kappa}^{(1)}$ with $n_1+\ldots+n_\kappa=l$. From $C_n^{(1)}(\cos\theta)=\sin((n+1)\theta)/\sin\theta$ one finds $C_m^{(1)}C_n^{(1)}=\sum_{k=0}^{n}C_{m+n-2k}^{(1)}$ for $m\geq n$.\footnote{We verified the coefficients of the connection formula for $l\leq14$ and this product formula for $n\leq7$ with computer algebra.} Each product is therefore a sum of polynomials $C_{l-2j}^{(1)}$ with non-negative integer coefficients, and the product with $n_1=l-j$, $n_2=j$ and all other $n_i=0$ contains $C_{l-2j}^{(1)}$. In four dimensions $|x|^sC_s^{(1)}(\hat x\cdot\hat y)$ is a harmonic polynomial in $x$ and a spherical harmonic of degree $s$ in $\hat y$. Spherical harmonics of different degrees are linearly independent, so that the integrals of $\eta(x)\,|x|^{2j}\,|x|^sC_s^{(1)}(\hat x\cdot\hat y)$ vanish for all $j$, $s$ and $\hat y$. The functions $|x|^sC_s^{(1)}(\hat x\cdot\hat y)$ span the harmonic polynomials of degree $s$, and every polynomial is a sum of harmonic polynomials multiplied by powers of $x^2$. All moments of $\eta$ therefore vanish, and $\eta=0$.
\end{proof}

\begin{proposition}
\label{prop:unique}
Let the mass source and the gauge field vanish. Let $g_0=g+L(g)$ and $g_0'=g+L'(g)$ be two bare couplings, where $L$ and $L'$ are power series in the values of $g$ at the same point, without derivatives and without terms of order zero and one, which are pure poles. Let the bare mass source and the bare gauge field be the same local functionals of $g$ in both cases, which vanish at $g=0$, and let the counterterms which do not depend on the fields be integrals of local polynomials in $g$ and its derivatives. If both functionals are finite for all sources $g$ of the class of flat sources which do not depend on $\vep$, then $L=L'$.
\end{proposition}

\begin{proof}
Let $k\geq2$ be the lowest order at which $L$ and $L'$ differ, and let $\vep^{-\mathsf{r}}\ell(g)$ be the term with the highest pole in the difference at this order. At order $k+1$ the two functionals differ by $2\,W_2[g,L_k(g)-L'_k(g)]$ and by the integral of a local polynomial of degree $k+1$, where $L_k$ is the part of $L$ of degree $k$. Here we use that the term of first order of the bare functional and the terms of second order which join a quartic vertex to a bare mass source or to a bare gauge field vanish, since propagators at coincident points vanish. All other terms with one of these sources and the difference of the couplings are of order $k+2$ at least. We take $g=g_a+t\,g_b$ with sources $g_a$ and $g_b$ of the class with disjoint supports. A local polynomial of degree $k+1\geq3$ has no term linear in $t$, since products of $g_a$ and $g_b$ and of their derivatives vanish. For the same reason $L_k(g_a+t\,g_b)=L_k(g_a)+t^kL_k(g_b)$, which has no term linear in $t$ for $k\geq2$. The term linear in $t$ of the difference is therefore $2\,W_2[g_b,L_k(g_a)-L'_k(g_a)]$, which must be finite. For disjoint supports $W_2$ is analytic at $d=4$, where \eqref{eq:gram} is the ordinary integral. The coefficient of $\vep^{-\mathsf{r}}$ gives that the integral of $g_b^I(y)\,\ell^I(g_a)(x)\,|x-y|^{-8}$ over both points vanishes at $d=4$, for all $g_b$ of the class with support outside the support of $g_a$. The function $\int d^4x\,\ell^I(g_a)(x)|x-y|^{-8}$ of $y$ is invariant under the transverse rotations, and the sources $g_b$ separate such functions. It therefore vanishes outside a ball, and by Lemma~\ref{lem:potential} we find $\ell(g_a)=0$ for all $g_a$ of the class. The values $g_a(x)$ are arbitrary, so that the polynomial $\ell$ vanishes, which is a contradiction.
\end{proof}

The proposition shows that for given bare mass source and gauge field the counterterms of the couplings are determined by the finiteness of the flat functional \eqref{eq:Wfl}. The condition on these sources cannot be dropped, since the transformations of Lemma~\ref{lem:frame} change $L^I$ together with the bare gauge field and the bare mass source.

\subsection{Other counterterms}

We now show the converse. For given $L^I$ the other counterterms are determined, and they are covariant. We first collect the terms of lowest order.

\begin{lemma}
\label{lem:lowest}
Let the coefficients of the bare sources \eqref{eq:baresources} in flat space and of $\cX^{\rm fl}$ be real pure poles, where $L^I$ has no terms of degree zero and one, and let the functional \eqref{eq:Wfl} be finite for all sources of the class of flat sources with $D_0=1$ which do not depend on $\vep$. Then $Z_\sM=1$ and $\fd_I=0$ at $g=0$. Moreover, let the sources with the index $2$ have supports which are disjoint from the supports of the sources with the index $1$. Then the terms of second order of $W^{\rm fl}$ with one source of each index vanish for two sources of different kinds. For two couplings, two mass sources and two gauge fields they are, up to factors which are analytic and do not vanish at $d=4$, the integrals over both points of
\begin{equation}
\frac{g_1^I(x)\,g_2^I(y)}{|x-y|^{4(d-2)}} \, , \qquad \frac{\sM_1(x)\cdot\sM_2(y)}{|x-y|^{2(d-2)}} \, , \qquad
a_{1\mu}(x)\cdot a_{2\nu}(y)\left(\pr_\mu\pr_\nu-\delta_{\mu\nu}\pr^2\right)\frac{1}{|x-y|^{2(d-2)}} \, .
\label{eq:twopointkernels}
\end{equation}
\end{lemma}

\begin{proof}
A local functional has no terms which contain two sources with disjoint supports. The term of first order of the bare functional vanishes, since propagators at coincident points vanish. The terms in question are therefore those of the bare functional at second order, evaluated on the terms of first order of the bare sources. These are $g^I$, $a_\mu+N^{(0)}_I\pr_\mu g^I$ and $Z^{(0)}_\sM(\sM-\fd^{(0)}_I\pr^2g^I)$, where the superscript denotes the value at $g=0$. We do not assume here that $N^{(0)}_I$ vanishes. At second order the bare functional is the sum of the two-point functions of the quartic operators, of the operators $\phi_i\phi_j$ and of the currents in the free theory, integrated with the bare sources. With the propagator $\mathsf{G}=|x|^{-(d-2)}$ these are multiples of $\mathsf{G}^4$, of $\mathsf{G}^2$ and of $\mathsf{G}\,\pr_\mu\pr_\nu\mathsf{G}-\pr_\mu\mathsf{G}\,\pr_\nu\mathsf{G}=(\pr_\mu\pr_\nu-\delta_{\mu\nu}\pr^2)\,\mathsf{G}^2/(2(d-1))$, which gives \eqref{eq:twopointkernels} for the bare sources.\footnote{We verified the form of the kernel of the current with computer algebra.} The term $N^{(0)}_I\pr_\mu g^I$ of the bare gauge field is the gradient of $N^{(0)}_Ig^I$. It does not contribute to the terms considered, because the third kernel is transverse and the two sources have disjoint supports. The terms with two bare sources of different kinds vanish. With a coupling, two lines of the quartic vertex form a propagator at coincident points, and for a mass source and a gauge field the antisymmetrised derivative of the product of two equal propagators vanishes.

For two mass sources we obtain the second expression of \eqref{eq:twopointkernels} with $Z^{(0)}_\sM\sM_1$ and $Z^{(0)}_\sM\sM_2$ in place of $\sM_1$ and $\sM_2$, and it is finite for all mass sources of the class. Let $\vep^{-\mathsf{r}}\breve z$ be the leading pole of $Z^{(0)}_\sM-1$. For the sources $\sM_1=\sM_{\rm c}\,\chi_1$ and $\sM_2=\sM_{\rm c}\,\chi_2$ with a constant matrix $\sM_{\rm c}$ and positive functions $\chi_1$ and $\chi_2$ of the class the coefficient of $\vep^{-2\mathsf{r}}$ is a positive multiple of $\breve z\,\sM_{\rm c}\cdot\breve z\,\sM_{\rm c}$, which is the sum of the squares of the entries of the real matrix $\breve z\,\sM_{\rm c}$ up to a factor. Hence $\breve z=0$ and $Z^{(0)}_\sM=1$. For a coupling $g_1$ and a mass source $\sM_2$ we obtain the second expression with $-\fd^{(0)}_I\pr^2g_1^I$ in place of $\sM_1$. Let $\vep^{-\mathsf{r}}\breve d_I$ be the leading pole of $\fd^{(0)}_I$. The coefficient of $\vep^{-\mathsf{r}}$ is the integral of $\breve d_I\,\pr^2g_1^I(x)\cdot\sM_2(y)\,|x-y|^{-4}$ over both points at $d=4$, where \eqref{eq:gram} is the ordinary integral, and it vanishes for all $\sM_2$ of the class with support outside the support of $g_1$. The potential of $\breve d_I\,\pr^2g_1^I$ with the kernel $|x-y|^{-4}$ is invariant under the transverse rotations, and the sources of the class separate invariant continuous functions. It therefore vanishes outside the support of $g_1$, and Lemma~\ref{lem:potential} gives $\breve d_I\,\pr^2g_1^I=0$ for all $g_1$, hence $\breve d_I=0$. With $Z^{(0)}_\sM=1$ and $\fd^{(0)}_I=0$ the bare sources agree with the renormalised ones at first order, up to the gradient in the bare gauge field, which gives the remaining statements.
\end{proof}

For the gauge field we need a property of the third kernel in \eqref{eq:twopointkernels}, which is transverse.

\begin{lemma}
\label{lem:transverse}
Let $j_\mu$ be a one-form on $\mathbb{R}^4$ whose components are twice continuously differentiable and have compact support. For $y$ outside the support let
\begin{equation}
A_\nu(y) = \int d^4x\;j_\mu(x)\left(\pr_\mu\pr_\nu-\delta_{\mu\nu}\pr^2\right)\frac{1}{|x-y|^{4}} \, ,
\label{eq:Apot}
\end{equation}
where the derivatives act on $x$. If $A_\nu$ vanishes outside a ball, then $\pr_\mu j_\nu=\pr_\nu j_\mu$, and $j_\mu=\pr_\mu\psi$ with a function $\psi$ of compact support.
\end{lemma}

\begin{proof}
For $y$ outside the support of $j_\mu$ the kernel is smooth on the support, and two integrations by parts give $A_\nu(y)=\int d^4x\,\eta_\nu(x)\,|x-y|^{-4}$ with $\eta_\nu=\pr_\nu\pr_\mu j_\mu-\pr^2j_\nu=\pr_\mu f_{\nu\mu}$, where $f_{\mu\nu}=\pr_\mu j_\nu-\pr_\nu j_\mu$. The functions $\eta_\nu$ are continuous with compact support, and Lemma~\ref{lem:potential} with $\kappa=2$ gives $\pr_\mu f_{\nu\mu}=0$. Together with $\pr_\rho f_{\mu\nu}+\pr_\mu f_{\nu\rho}+\pr_\nu f_{\rho\mu}=0$ we find $\pr^2f_{\mu\nu}=-\pr_\mu\pr_\rho f_{\nu\rho}-\pr_\nu\pr_\rho f_{\rho\mu}=0$ in the sense of distributions. A harmonic distribution is a harmonic function, and a harmonic function of compact support vanishes. Hence $f_{\mu\nu}=0$. A closed one-form on $\mathbb{R}^4$ is the differential of a function $\psi$. This function is constant outside a ball which contains the support of $j_\mu$, because the complement of the ball is connected, and we subtract the constant.
\end{proof}

\begin{proposition}
\label{prop:uniqueall}
Let $L^I$ be a real covariant pure pole without terms of degree zero and one in the couplings. Consider sets of real pure poles $N_I$, $Z_\sM-1$, $\fd_I$, $\fe_{IJ}$ and of coefficients of $\cX^{\rm fl}$. First, there is at most one such set for which the functional \eqref{eq:Wfl} is finite for all sources of the class of flat sources with $D_0=1$ which do not depend on $\vep$. Second, if such a set exists, its members are covariant under the field rotations, and $N_I$ has no term of degree zero in the couplings.
\end{proposition}

\begin{proof}
Let two such sets be given for the same $L^I$. We denote the two functionals by $W$ and $W'$ and the difference of a coefficient of the two sets by a tilde. We use sources with the indices $1$ and $2$ as in Lemma~\ref{lem:lowest}. Since both functionals are finite, so is every term of $W-W'$ of given orders in these sources. The counterterms $\cX^{\rm fl}$ do not contribute to terms which contain sources of both indices. By Lemma~\ref{lem:lowest} we have $Z_\sM=1$ and $\fd_I=0$ at $g=0$ in both sets.

\paragraph{Gauge invariance at low orders.} Let $W_2[g,h]$ be as above, and let $\mathsf{W}_3[a;g,h]$ be the term of the bare functional which is linear in the gauge field and bilinear in the couplings, so that the bare functional contains $W_2[g,g]+\mathsf{W}_3[a;g,g]$. By Proposition~\ref{prop:gauge} the bare functional is invariant under the gauge transformations \eqref{eq:gaugetr} of the bare sources with a parameter $\omega$ of the class. We expand this invariance in the sources. At first order in the couplings and in the gauge field there is no condition, since the terms of second order with two sources of different kinds vanish. At second order in the couplings we find
\begin{equation}
\mathsf{W}_3[d\omega;g_1,g_2] = W_2[(\omega g_1),g_2]+W_2[g_1,(\omega g_2)] \, .
\label{eq:W3ward}
\end{equation}
At first order in the gauge field we find that the term of second order in the gauge field vanishes if one of its arguments is $d\omega$. At second order we find that the term of third order with the arguments $d\omega$, $a$ and $a$ equals $-\tfrac23$ times the term of second order with the arguments $[a,\omega]$ and $a$. The term with two gauge fields and one coupling vanishes identically. For two vertices of the current this holds because the tensor $\lambda_{ijkl}$ is symmetric and the two lines which join it to a vertex of the current are antisymmetrised. The term of the action of second order in the gauge field has a vertex of the type of the mass source, with the source $\Upsilon$ of section~\ref{app:conf}. Its terms of second order with a vertex of the couplings or of the current vanish as in the proof of Lemma~\ref{lem:lowest}, and its other terms are included in the terms of second and third order in the gauge field of the identities above. These identities are multilinear, so that they also hold for parameters and sources whose coefficients are Laurent polynomials in $\vep$.

\paragraph{Term of degree zero.} We first show that $N^{(0)}_I=N_I(0)$ is the same in both sets. We take the couplings $g=g_1+g_2$ and no other sources, and we consider the terms of third order of $W-W'$. At first order the bare gauge field is $d\omega$ with $\omega=N^{(0)}_I\,g^I$, which is a function of the class with values in the Lie algebra, and the bare mass source starts at second order. The terms of second order of the bare gauge field and of the bare mass source enter at third order only through terms of second order of the bare functional, together with a term of first order of a bare source. These vanish, either because the two sources are of different kinds or because one of two gauge fields is $d\omega$. The terms of second order of the bare couplings are the same in both sets. The terms of third order of $W$ which contain $N^{(0)}_I$ are therefore the terms of the bare functional with one, two or three gauge fields $d\omega$ and with couplings $g$. By the identities above the terms with two and with three gauge fields vanish, and the term with one gauge field is $\mathsf{W}_3[d\omega;g,g]=2\,W_2[(\omega g),g]$. Hence the terms of third order of $W-W'$ are $2\,W_2[(\tilde\omega\,g),g]$ and local terms, where $\tilde\omega=\tilde N^{(0)}_I\,g^I$. Since products of $g_1$ and $g_2$ at the same point vanish, the terms of second order in $g_1$ and linear in $g_2$ are $2\,W_2[(\tilde N^{(0)}_I g_1^I\,g_1),g_2]$, which is finite. Let $\vep^{-\mathsf{r}}\breve N^{(0)}_I$ be the leading pole of $\tilde N^{(0)}_I$. For disjoint supports $W_2$ is analytic at $d=4$, and as in the proof of Proposition~\ref{prop:unique} we find $(\breve N^{(0)}_Ig^I\,g)^J=0$ for all $g$. Hence $\breve N^{(0)}_Ig^I$ is in the stabiliser of $g$ in the Lie algebra. For the coupling $\lambda_{ijkl}=\sum_r\mathsf{d}_r\,\delta_{ir}\delta_{jr}\delta_{kr}\delta_{lr}$ with all $\mathsf{d}_r\neq0$ we have $(\omega\lambda)_{ijjj}=\omega_{ij}\,\mathsf{d}_j$ for $i\neq j$. The stabiliser of this coupling is therefore trivial, and the same holds for a dense open set of couplings.\footnote{For $2\leq N\leq6$ we also checked that the stabiliser of a random quartic tensor is trivial.} The linear function $\breve N^{(0)}_Ig^I$ vanishes on this set and hence everywhere, which contradicts $\breve N^{(0)}_I\neq0$. It follows that $N^{(0)}_I$ is the same in all sets which make the functional finite. The set which is obtained from such a set by a rotation of the field indices with a constant element of ${\rm O}(N)$ is again such a set, as we show in the last paragraph of the proof. Hence $N^{(0)}_I$ is an invariant tensor which is antisymmetric in two indices and symmetric in four. There is no such tensor of ${\rm O}(N)$, since an invariant tensor is a combination of products of $\delta_{ij}$, and the antisymmetrisation in two indices of a product in which the other four indices are symmetrised vanishes. Hence $N_I$ has no term of degree zero in both sets.

\paragraph{Induction.} We say that the terms of degree $n-1$ of $N_I$ and $\fd_I$, the term of degree $n-2$ of $\fe_{IJ}$ and the term of degree $n$ of $Z_\sM$ have the level $n$. These are the terms which enter the bare gauge field and the bare mass source at order $n$ in the couplings. We assume that the two sets agree at all levels below $n\geq1$, and we show in four steps that they agree at the level $n$. For $n=1$ only the fourth step is needed, by Lemma~\ref{lem:lowest} and by the previous paragraph. We write $\tilde N_I$, $\tilde\fd_I$, $\tilde\fe_{IJ}$ and $\tilde Z_\sM$ for the differences at the level $n$.

\paragraph{First step.} We take couplings $g_1$ and a gauge field $a_2$, and we consider the terms of order $n$ in $g_1$ and linear in $a_2$. At this order the bare sources of the two sets differ by $\tilde a_{0\mu}=\tilde N_I(g_1)\,\pr_\mu g_1^I$ and by $\tilde\sM_0=-\tilde\fd_I(g_1)\,\pr^2g_1^I-\tilde\fe_{IJ}(g_1)\,\pr_\mu g_1^I\pr^\mu g_1^J$. The terms of $W-W'$ of the order considered are therefore the terms of second order of the bare functional with $\tilde a_0$ or $\tilde\sM_0$ and with $a_2$, and by Lemma~\ref{lem:lowest} only the term with $\tilde a_0$ remains. Let $\vep^{-\mathsf{r}}\breve N_I$ be the leading pole of $\tilde N_I$. The coefficient of $\vep^{-\mathsf{r}}$ is the integral of $j_\mu(x)\cdot a_{2\nu}(y)$ with the kernel of \eqref{eq:Apot} over both points at $d=4$, up to a factor which does not vanish, where $j=\breve N_I(g_1)\,dg_1^I$. Here we used that for disjoint supports the integral is analytic at $d=4$, where \eqref{eq:gram} is the ordinary integral. The coefficient vanishes for all one-forms $a_2$ of the class with support outside the support of $g_1$. The one-form $j$ is invariant under the transverse rotations, and so is the one-form $A_\nu$ of Lemma~\ref{lem:transverse}. A continuous invariant one-form has a component along $dx_\parallel$ and a component along $x_\perp\!\cdot dx_\perp$, which are functions of $x_\parallel$ and $|x_\perp|$, and the one-forms of the class separate such one-forms. Hence $A_\nu$ vanishes outside the support of $g_1$, and by Lemma~\ref{lem:transverse} the one-form $j$ is closed, for every component in the Lie algebra and for every source $g_1$. At a point with $x_\perp\neq0$ the value of a source of the class and its derivatives in the parallel and in the radial direction are arbitrary. Hence $\pr_J\breve N_I=\pr_I\breve N_J$, and $\breve N_I=\pr_I\breve\vartheta$ with a homogeneous polynomial $\breve\vartheta(g)$ of degree $n$ with values in the Lie algebra.

\paragraph{Second step.} We take $g_1$ and a mass source $\sM_2$, and we consider the terms of order $n$ in $g_1$ and linear in $\sM_2$. As in the first step they are the term of second order of the bare functional with $\tilde\sM_0$ and $\sM_2$, which is the second expression of \eqref{eq:twopointkernels}. As in the proof of Lemma~\ref{lem:lowest} the leading pole of $\tilde\sM_0$ vanishes at each point, for all $g_1$. At a point with $x_\perp\neq0$ the Laplacian of a source of the class and its derivatives in the parallel and in the radial direction are independent, so that the leading poles of $\tilde\fd_I$ and $\tilde\fe_{IJ}$ vanish. Hence $\tilde\fd_I=0$ and $\tilde\fe_{IJ}=0$.

\paragraph{Third step.} We take the couplings $g_1+g_2$ and no other sources, and we consider the terms of order $n+1$ in $g_1$ and linear in $g_2$, for $n\geq2$. The differences of the bare sources at the level $n+1$ are of order $n+1$ in $g_1$. They multiply terms of second order with two sources of different kinds, which vanish. At the level $n$ only $\tilde a_0$ contributes, since $\tilde\fd_I$ and $\tilde\fe_{IJ}$ vanish by the second step and $\tilde Z_\sM$ multiplies terms of the bare mass source which are of second order at least. Terms of second order in the differences have the order $2n$, which is larger than $n+1$. Hence the terms of $W-W'$ of the order considered are $2\,\mathsf{W}_3[\tilde N_I(g_1)\,dg_1^I;g_1,g_2]$. We need two properties of $\mathsf{W}_3$. First, $\mathsf{W}_3[a;g_1,g_2]$ is finite for every one-form $a$ of the class. Indeed, twice this expression is the term of $W'$ which is linear in $a$, $g_1$ and $g_2$. No counterterm contributes to it, because $N_I$ and $\fd_I$ vanish at $g=0$ and because local terms with $g_1$ and $g_2$ vanish. Second, it satisfies \eqref{eq:W3ward}. By the first property the coefficient of $\vep^{-\mathsf{r}}$ of the terms considered is $2\,\mathsf{W}_3[d\breve\vartheta(g_1);g_1,g_2]$ at $d=4$. By \eqref{eq:W3ward} with $\omega=\breve\vartheta(g_1)$, for which $(\omega g_2)$ vanishes, it equals $2\,W_2[(\breve\vartheta(g_1)\,g_1),g_2]$, which is analytic at $d=4$. It vanishes for all $g_2$ of the class with support outside the support of $g_1$, and as in the proof of Proposition~\ref{prop:unique} we find $(\breve\vartheta(g)\,g)^I=0$ for all $g$. Since the stabiliser of the couplings in a dense open set is trivial, the polynomial $\breve\vartheta$ vanishes, which contradicts $\breve N_I\neq0$. It follows that $\tilde N_I=0$.

\paragraph{Fourth step.} We take $g_1$ and two mass sources $\sM_1$ and $\sM_2$, where the support of $\sM_1$ may overlap with the support of $g_1$, and we consider the terms of order $n$ in $g_1$ which are linear in $\sM_1$ and in $\sM_2$. After the first three steps the bare sources differ at this order by $\tilde\sM_0=\tilde Z_\sM(g_1)\,\sM_1$ only, and the terms considered are the second expression of \eqref{eq:twopointkernels} with $\tilde\sM_0$ and $\sM_2$. As in the second step the leading pole of $\tilde Z_\sM(g_1)\,\sM_1$ vanishes at each point, and $\tilde Z_\sM=0$. This completes the induction.

\paragraph{Counterterms.} The bare sources of the two sets now agree, and $W-W'$ is the integral of the difference of the two densities $\cX^{\rm fl}$. It is finite, and the coefficients of the difference are pure poles. At a fixed order in the couplings the coefficient of the highest pole is the left-hand side of \eqref{eq:lemA} at $d=4$ with the leading coefficients, and it vanishes for all sources. By the first part of Lemma~\ref{lem:indep} the leading coefficients vanish. Hence the two densities agree.

\paragraph{Covariance.} The bare functional is invariant under a rotation of the field indices of all bare sources with a constant element of ${\rm O}(N)$, since the propagator is proportional to $\delta_{ij}$. If a set of coefficients makes the functional finite, the set which is obtained from it by such a rotation therefore does so as well. Since $L^I$ is covariant, the rotated set belongs to the same $L^I$, and it consists of real pure poles. By the first statement the two sets agree, which is the covariance.
\end{proof}

\subsection{Counterterms of minimal subtraction}
\label{app:hypL}

Let us explain Identification~\ref{id:ms}. In the theory with constant couplings the counterterms of minimal subtraction are the pole parts at $d=4$ of the one-particle irreducible graphs at fixed momenta, after the subtraction of subdivergences, see section~5.11.1 of~\cite{Collins:1984xc}. The pole parts of the graphs with two external lines give a symmetric matrix $Z$, and the fields are renormalised with a factor whose square, in the sense of the product of its transpose with itself, is $Z$. This factor is the symmetric square root of $Z$ multiplied by an orthogonal matrix, which is not fixed~\cite{Jack:2013sha,Bednyakov:2021ojn,Herren:2021amb}. The bare couplings $g^I+L^I(g)$ are obtained from the pole parts of the graphs with four external lines after this renormalisation of the fields. For the symmetric square root the functions $L^I$ are sums of graphs with real coefficients, and they are real pure poles without terms of degree zero and one which are covariant under ${\rm O}(N)$, including the reflections, since the graphs are contractions of $\lambda_{ijkl}$ with $\delta_{ij}$. The same holds if the orthogonal matrix is the exponential of a covariant pure pole $\vartheta(g)$ with values in the Lie algebra, which is what we mean by a covariant choice. Such a $\vartheta$ has no terms of degree zero and one, because there is no antisymmetric matrix which is covariant and of degree zero or one in $\lambda_{ijkl}$. The orthogonal matrix rotates the bare fields, and it therefore acts on the bare sources as the transformation \eqref{eq:frame} of Lemma~\ref{lem:frame}. By this lemma the hypothesis holds for every covariant choice if it holds for one, and the vector field $\hB^I$ and the coefficient $\hr_I$ are the same for all choices. The residues $L^I_1$ and $N_{I,1}$, and with them $\beta^I=(g^J\pr_J-1)L^I_1$ and $\xi=-N_{I,1}\,g^I$, change if $\vartheta$ has a simple pole. We use these names for the symmetric square root. Then $\beta^I$ is the usual beta function, for instance the one of~\cite{Bednyakov:2021ojn}, and it is not changed by the transformation which leads to the bare sources of section~\ref{sec:weylvar}, whose parameter has no simple pole. This transformation changes the higher poles of $L^I$, so that the bare couplings of section~\ref{sec:weylvar} are not those of the symmetric square root from the order on at which $\xi^{\rm L}$ has a pole. With Identification~\ref{id:ms} the function $\beta^I$ of section~\ref{sec:weylvar}, which enters the vector field $\hB^I$ of Theorem~\ref{gs:thm:phi4}, is the usual beta function of minimal subtraction. The proofs of Theorem~\ref{gs:thm:phi4} and of Proposition~\ref{gs:prop:Wform} use only Hypothesis~\ref{hyp:finite}, that is the finiteness for some real covariant pure poles $L^I$ without terms of degree zero and one. Identification~\ref{id:ms} enters when the results are compared with computations in momentum space.

Let us summarise what the results of this appendix imply for Hypothesis~\ref{hyp:finite} together with Identification~\ref{id:ms}. The functional is then finite with the counterterms $L^I$ of minimal subtraction of the theory with constant couplings, and the function $\beta^I$ of section~\ref{sec:weylvar} is then the usual beta function. By Proposition~\ref{prop:uniqueall} the bare gauge field, the bare mass source and the counterterms of flat space are determined by $L^I$, and with them the vector field $B^I$ and, for the symmetric square root, the rotation $\xi$. By Proposition~\ref{prop:unique} no other counterterms of the couplings make the functional finite with the same bare gauge field and bare mass source. We expect that all sets of counterterms which make the functional finite are related by the transformations of Lemma~\ref{lem:frame}.

Let us finally compare Hypothesis~\ref{hyp:finite} with the assumption of section~8 of~\cite{Jack:2013sha} and with the first hypothesis of~\cite{Reehorst:phi4}, as announced in section~\ref{sec:hypothesis}. In all of them the bare sources have the form \eqref{eq:baresources} with coefficients which are pure poles. In the other two the counterterms of the couplings are those of the theory with constant couplings, which here is Identification~\ref{id:ms} and not part of Hypothesis~\ref{hyp:finite}. Our statement differs in three respects. First, it concerns flat space. The counterterms which multiply the curvature are not assumed, and we construct them in Proposition~\ref{prop:finite}. Second, it specifies the bare functional and the class of sources, and it includes the families of sources which depend on $\vep$. Third, the first hypothesis of~\cite{Reehorst:phi4} also contains the invariance under diffeomorphisms and background gauge transformations and the form of the Weyl variation with a local anomaly, which are used in~\cite{Jack:2013sha} as well. Here we do not assume these. The covariance under the conformal maps which we use, the invariance under background gauge transformations and the form of the Weyl variation with regular coefficients are derived from the definition of the bare functional and from the finiteness.

\subsection{Other schemes}
\label{app:schemes}

Theorem~\ref{gs:thm:phi4} is stated in minimal subtraction, and we add a remark on other schemes. The vector field $\hB^I$ and the coefficient $\hr_I$ are properties of the renormalised functional. They do not change under the transformations of Lemma~\ref{lem:frame}, which change $\beta^I$ and $\xi$ separately, and by the remark after Lemma~\ref{lem:regvar} they are determined by the identity \eqref{eq:flatward} of appendix~\ref{app:finite}. Under a redefinition of the couplings $g'^I=g^I+{\rm O}(g^2)$ which is regular and covariant the functional remains finite, the vector field $\hB^I$ transforms as a vector, and \eqref{gs:eq:phi4main} holds in the new couplings with the same potential and with the metric transformed as a tensor. The new metric is regular and symmetric, and its term of order zero is not changed. The conclusion of the theorem therefore holds in every scheme which is related to the one of Hypothesis~\ref{hyp:finite} by such a redefinition. Note that in the new couplings the vector field is in general not of the form $-\vep\,g'^I+B'^I$ with $B'^I$ independent of $\vep$. For example, for $g'^I=g^I+f^I(g)$ with a quadratic $f^I$ it contains the term $-\vep\,f^I(g')$. The tower \eqref{gs:eq:tower} and the relations of appendix~\ref{app:fiveloop}, which use this form, are therefore statements about minimal subtraction, and they do not hold in the same form in the new couplings. A change of scheme of this kind with coefficients which do not depend on $\vep$ is analysed in~\cite{Pannell:2024sia}, where the coefficients of the beta function receive terms of first order in $\vep$ and the constraints of a general scheme are found to reduce to those of modified minimal subtraction. We expect that this is the case for any two finite functionals which are obtained from the same bare functional with bare sources of the form \eqref{eq:baresources}, because the argument of Proposition~\ref{prop:unique} applies to the map between the two sets of couplings. A change of scheme of a different kind changes the rotation $\xi$. This is the redefinition of the background field by a finite local term, $a'_\mu=a_\mu+r_I\,D_\mu g^I$ with a regular covariant $r_I$ in the Lie algebra~\cite{Jack:2013sha}, which is treated in the fourth remark after Theorem~1 of~\cite{Reehorst:phi4}. The functional $W'[g,a',\sM]=W[g,a,\sM]$ is finite and gauge invariant. In the new variables the variation \eqref{eq:rhoxi} of the gauge field contains the term $\pr_\mu\sigma\,r_I\hB^I$. We remove it by a gauge transformation with the parameter $-\sigma\,\omega_r$, where $\omega_r=r_I\hB^I$, and find that the vector field of the new scheme is $\hB'^I=\hB^I+(\omega_r\,g)^I$. Solving for $\omega_r$ degree by degree gives
\begin{equation}
\hB^I = \hB'^I-(\omega_r\,g)^I \, , \qquad \omega_r = \tilde r_L\,\hB'^L \, ,
\end{equation}
with a regular $\tilde r_L$ in the Lie algebra which has no term of degree zero. The element $\omega_r$ corresponds to $-w$ in the notation of~\cite{Reehorst:phi4}. Here $\hB'^I$ does not transform as a vector, and we use the argument of~\cite{Reehorst:phi4}. Let $\hat A$ be invariant and $\hat G_{IJ}$ covariant, and let $\mathsf{U}_I(\omega)=(\omega g)^J\hat G_{JI}$ for $\omega$ in the Lie algebra. Inserting the relation above into \eqref{gs:eq:phi4main} we find $\pr_I\hat A=[\hat G_{IL}-\mathsf{U}_I(\tilde r_L)]\hB'^L$, where the matrix in brackets is not symmetric. Since $(\omega g)^I\pr_I\hat A=0$ for every $\omega$, the formula \eqref{gs:eq:phi4main} gives $\mathsf{U}_K(\omega)\hB^K=0$. For $\omega=\tilde r_I$ this reads $\mathsf{U}_L(\tilde r_I)\,\hB'^L-\mathsf{V}_{IL}\,\hB'^L=0$ with the symmetric tensor $\mathsf{V}_{IL}=(\tilde r_Ig)^J\hat G_{JK}(\tilde r_Lg)^K$. Subtracting it we obtain
\begin{equation}
\pr_I\hat A = \left[\hat G_{IL}-\mathsf{U}_I(\tilde r_L)-\mathsf{U}_L(\tilde r_I)+\mathsf{V}_{IL}\right]\hB'^L \, .
\end{equation}
The matrix in brackets is symmetric and regular, and its term of order zero is that of $\hat G_{IJ}$. The gradient property therefore holds for $\hB'^I$ with the same potential. The argument uses only \eqref{gs:eq:phi4main} and the invariance of the potential.

Let us finally compare the zeros and the closed orbits of the two vector fields $\hB^I$ and $\hb^I=\hB^I+(\xi g)^I$, which we use in appendix~\ref{gs:app:fourd}. Let $\hat A$ be invariant, so that \eqref{eq:monotonic} of appendix~\ref{gs:app:fourd} holds, and let $\hat G_{IJ}$ be positive definite at the points which we consider. At a zero of $\hb^I$ the left-hand side of \eqref{eq:monotonic} vanishes, so that $\hat G_{IJ}\hB^I\hB^J=0$ and therefore $\hB^I=0$. For a real zero which is a power series in $\vep$ without a term of order zero this holds order by order in $\vep$, because the term of lowest order of $\hat G_{IJ}$ is positive definite. We did not show the converse. At a zero $g_*$ of $\hB^I$ we have $\hb^I=(\xi g_*)^I$, and the flow of $\hb^I$ through $g_*$ is the orbit of $g_*$ under the field rotations generated by $\xi(g_*)$, because $\hB^I$ and $\xi$ are covariant. All points of this orbit are zeros of $\hB^I$, and they are related by field rotations. Consider now a closed orbit of the flow of $\hB^I$ or of $\hb^I$. The potential returns to its value along it, so that $\hat G_{IJ}\hB^I\hB^J$ vanishes on the orbit by \eqref{eq:monotonic}. The flow of $\hB^I$ therefore has no closed orbits apart from its zeros, and a closed orbit of the flow of $\hb^I$ is an orbit of the kind just described. In particular the flow of $\hB^I$ has no limit cycles. The statement on the zeros holds order by order in $\vep$, and the statements on closed orbits hold for the flow at leading order in $\vep$ of couplings of order $\vep$, in the sense of section~\ref{gs:sec:fixedgen}.

Modified minimal subtraction differs from minimal subtraction by a rescaling of the renormalisation scale, which does not change the beta function. The potential and the metric depend on the product of the renormalisation scale and the radius of the sphere, which we set to one. A different value corresponds to a flow of the couplings.

\section{Finiteness on the sphere}
\label{app:finite}

The derivation of the properties of multiscalar $\phi^4$ theory in section~\ref{gs:sec:phi4prop} uses that the renormalised functional $W$ on the sphere is finite for sources in the class, including the terms of first order in the mass source, and that the anomaly coefficients are regular. In this appendix we derive these properties from Hypothesis~\ref{hyp:finite}, which concerns flat space, for the full set of sources $g^I$, $a_\mu$ and $\sM$. We first list the local structures of the anomaly on conformally flat backgrounds and their coefficients in a general scheme. We then show that the coefficients of the variation of the sources are regular. After this we show how each coefficient of the anomaly becomes regular, either directly from the finiteness in flat space, or by the choice of one counterterm with the curvature, or by a relation. Finally we pass to conformally flat backgrounds and to the sphere by a localisation argument. The result is Proposition~\ref{prop:finite}, which we use in sections~\ref{sec:setup} and~\ref{gs:sec:phi4prop}, and which appendix~\ref{gs:app:fourd} also uses. The proofs use the gauge invariance of appendix~\ref{app:gauge}, the lemmas of appendix~\ref{app:gaugevar} and the results of appendix~\ref{app:unique}. Recall that a quantity is regular if it is a power series in the couplings whose coefficients are regular at $\vep=0$, and that a functional of the sources is finite if at each order in the sources it has no pole at $\vep=0$, as in section~\ref{sec:hypothesis}. As in section~\ref{sec:setup} a pure pole is a power series in the couplings whose coefficients are polynomials in $1/\vep$ without a constant term, and we denote the pole part of a Laurent series by ${\rm PP}$.

\subsection{Statement}

We call the seven coefficients $\fa$, $\fb$, $\fg_{IJ}$, $\fc_I$, $\ff_{IJ}$ and $\fn$ of \eqref{eq:X} and $\fk$ of \eqref{eq:baresources} the curvature counterterms, since they multiply terms which vanish in flat space. The other coefficients of \eqref{eq:X} and \eqref{eq:baresources} are those of flat space, which are given by Hypothesis~\ref{hyp:finite}.

\begin{proposition}
\label{prop:finite}
Assume Hypothesis~\ref{hyp:finite} for multiscalar $\phi^4$ theory. Then there are unique pure poles $\fk$, $\fn$, $\fg_{IJ}$, $\fc_I$, $\ff_{IJ}$, $\fb$ and $\fa$, with $\fa$ invariant and vanishing at $g=0$, such that the coefficients $\bar\tau$, $t_\sM$, $E_{(IJ)}$, $u_I$, $\nu_{IJ}$, $r$ and $W_I$ are regular. With this choice the following holds. First, all coefficients of the variation \eqref{eq:deltaM} of the mass source and all $24$ coefficients of the anomaly \eqref{eq:fullanomaly} are regular. Second, the functional $W$ is finite on the backgrounds $\Omega^2\delta_{\mu\nu}$, where $\log\Omega$ and the sources are in the class of flat sources with $D_0=1$. Third, the terms of order $n\geq1$ of the functional $W$ on the unit sphere are finite for all sources of the class of section~\ref{sec:sphere}, in particular for constant couplings and their first variations.
\end{proposition}

The proof uses the covariance \eqref{eq:covS} and the invariance of the bare functionals $W_0^{\rm fl}$ and $W_0^S$ under the gauge transformations \eqref{eq:gaugetr} with parameters in the class, which is Proposition~\ref{prop:gauge} with the remark which follows its proof. It also uses that the coefficients of the bare sources and of the counterterms of flat space are covariant under field rotations, which is Proposition~\ref{prop:uniqueall}. Note that the proposition is a statement of existence. We do not show that the finiteness on curved backgrounds forces this choice of the curvature counterterms. For this reason the linear independence of the local structures is needed only in flat space.

\subsection{Structures and coefficients}

There is no counterterm with the curvature and the field strength. A scalar of dimension four which is linear in $f^{\mu\nu}$ and in the curvature is a contraction of the antisymmetric tensor $f^{\mu\nu}$ with the Riemann tensor and the metric, and all such contractions reduce to $f^{\mu\nu}R_{\mu\nu}=0$. For the same reason, and since $f^{\mu\nu}\nabla_\mu\nabla_\nu\sigma=0$, the anomaly has no structure with the curvature and the field strength. With the variation \eqref{eq:Delta} a direct computation, which we give in appendix~\ref{app:variation}, yields $\mathbb{W}_\sigma\!\int\!\sqrt\gamma\,\cX=\cA[\sigma]$ with
\begin{align}
\cA[\sigma] = \int d^dx\sqrt\gamma\,\Big\{&\sigma\Big[p\,\cE_4+q\,R^2-G_{IJ}\,\cG^{\mu\nu}D_\mu g^ID_\nu g^J \notag\\
&\quad+R\left(e_I\,D^2g^I+f_{IJ}\,D_\mu g^ID^\mu g^J+i\cdot\sM\right) \notag\\
&\quad+\tfrac12\,S_{IJ}\,D^2g^ID^2g^J-B_{IJK}\,D^2g^I\,D_\mu g^JD^\mu g^K \notag\\
&\quad-\tfrac12\,C_{ILJK}\,D_\mu g^ID^\mu g^L\,D_\nu g^JD^\nu g^K \notag\\
&\quad-J_I\cdot\sM\,D^2g^I-K_{IJ}\cdot\sM\,D_\mu g^ID^\mu g^J-\tfrac12\,\sM\cdot\beta_\sM\cdot\sM \notag\\
&\quad-\tfrac14\,f^{\mu\nu}\!\cdot\beta_f\cdot f_{\mu\nu}-f^{\mu\nu}\!\cdot P_{IJ}\,D_\mu g^ID_\nu g^J\Big] \notag\\
&+\pr_\mu\sigma\Big[-2\,W_I\,\cG^{\mu\nu}D_\nu g^I+h_I\,R\,D^\mu g^I+E_{IJ}\,D^\mu g^ID^2g^J \notag\\
&\quad+R_{IJK}\,D^\mu g^I\,D_\nu g^JD^\nu g^K+2\,L_I\cdot\sM\,D^\mu g^I-2\,f^{\mu\nu}\!\cdot Q_I\,D_\nu g^I\Big] \notag\\
&+\nabla^2\sigma\Big[r\,R+u_I\,D^2g^I+\nu_{IJ}\,D_\mu g^ID^\mu g^J+t_\sM\cdot\sM\Big]\Big\} \, .
\label{eq:fullanomaly}
\end{align}
This is the complete form of \eqref{eq:anomaly} on conformally flat backgrounds, and it defines the $24$ coefficients.\footnote{We also verified \eqref{eq:fullanomaly} in exact arithmetic at random rational values of the jets of the sources.} The coefficients have the symmetries of the structures which they multiply. In particular $B_{IJK}$ and $R_{IJK}$ are symmetric in their last two indices, and $P_{IJ}$ is antisymmetric. In a general scheme for the mass source, in which $\theta_I$ and $\bar\tau$ do not vanish, the coefficients which contain a curvature counterterm are $\bar\tau=-\fd_I\hB^I+2(d-1)\,\fk$ of \eqref{eq:masscoeff} and
\begin{align}
t_\sM &= \fj_I\hB^I-2(d-1)\,\fn-\bar\tau\cdot\fm \, , \qquad
E_{IJ} = 2\,\hat S_{IJ}+(d-2)\,\fg_{IJ} \, , \notag\\
R_{IJK} &= 2\,\hat T_{IJK}+(d-2)\,\Gamma^{(\fg)}_{IJK} \, , \notag\\
\omega\cdot Q_I &= \omega\cdot\hat Q_I+\tfrac12(d-2)\,(\omega g)^J\fg_{IJ} \, , \qquad
u_I = -2(d-1)\,\fc_I+\fs_{IJ}\hB^J-\fj_I\cdot\bar\tau \, , \notag\\
\nu_{IJ} &= -2(d-1)\,\ff_{IJ}+\tfrac12(d-2)\,\fg_{IJ}+\fr_{KIJ}\hB^K-\bar\tau\cdot\fl_{IJ} \, , \notag\\
r &= \fc_I\hB^I-4(d-1)\,\fb-\fn\cdot\bar\tau \, , \qquad
W_I = 4(d-3)\,\pr_I\fa-\fg_{IJ}\hB^J \, ,
\label{eq:fixedcoeff}
\end{align}
where the combinations of flat-space counterterms are
\begin{align}
\hat S_{IJ} &= \hP_I{}^K\fs_{KJ}+\fr_{JIK}\hB^K-\theta_I\cdot\fj_J \, , \qquad
\omega\cdot\hat Q_I = \omega\cdot\fp_{IJ}\hB^J-\tfrac12\,\omega\cdot\fy\cdot\hr_I \, , \notag\\
\hat T_{IJK} &= \hP_I{}^L\fr_{LJK}+\fq_{ILJK}\hB^L-\theta_I\cdot\fl_{JK}+\hr_{(J}\cdot\fp_{|I|K)} \, .
\end{align}
For $\theta_I=\bar\tau=0$ these are the expressions of appendix~\ref{app:weyl}. The other coefficients with the curvature are $p$ of \eqref{eq:lin} and $G_{IJ}$ of \eqref{eq:quad}, $\bar\eta$ of \eqref{eq:masscoeff} and
\begin{align}
q &= -\left(\vep-\hB\!\cdot\!\pr\right)\fb-\fn\cdot\bar\eta \, , \qquad
e_I = -\big[(\vep-\tL)\,\fc\big]_I-\fn\cdot\delta_I-\fj_I\cdot\bar\eta \, , \notag\\
f_{IJ} &= -\big[(\vep-\tL)\,\ff\big]_{IJ}+\fc_K\,\hat\Omega_{IJ}{}^K-\fn\cdot\epsilon_{IJ}-\fl_{IJ}\cdot\bar\eta \, , \notag\\
i &= -\left(\vep-\hB\!\cdot\!\pr\right)\fn+\fn\cdot\gamma_\sM-\bar\eta\cdot\fm \, , \qquad
h_I = 2\,\hP_I{}^K\fc_K+2\,\ff_{IJ}\hB^J-2\,\fn\cdot\theta_I \, .
\label{eq:derivedcoeff}
\end{align}
The coefficients $S_{IJ}$, $L_I$, $J_I$, $K_{IJ}$, $\beta_\sM$, $B_{IJK}$, $C_{ILJK}$, $P_{IJ}$ and $\beta_f$ contain only flat-space counterterms. They are
\begin{align}
S_{IJ} &= -\big[(\vep-\tL)\,\fs\big]_{IJ}-\fj_I\cdot\delta_J-\fj_J\cdot\delta_I \, , \qquad
L_I = \hP_I{}^K\fj_K+\fl_{IK}\hB^K-\theta_I\cdot\fm \, , \notag\\
J_I &= \big[(\vep-\tL)\,\fj\big]_I-\fj_I\cdot\gamma_\sM+\delta_I\cdot\fm \, , \notag\\
K_{IJ} &= \big[(\vep-\tL)\,\fl\big]_{IJ}-\fl_{IJ}\cdot\gamma_\sM-\hat\Omega_{IJ}{}^K\fj_K+\epsilon_{IJ}\cdot\fm \, , \notag\\
\sM\cdot\beta_\sM\cdot\sM' &= \sM\cdot\big[(\vep-\hB\!\cdot\!\pr)\,\fm\big]\cdot\sM'-(\gamma_\sM\sM)\cdot\fm\cdot\sM'-\sM\cdot\fm\cdot(\gamma_\sM\sM') \, , \notag\\
B_{IJK} &= \big[(\vep-\tL)\,\fr\big]_{IJK}-\hat\Omega_{JK}{}^L\fs_{IL}+\fj_I\cdot\epsilon_{JK}+\fl_{JK}\cdot\delta_I \, , \notag\\
C_{ILJK} &= \big[(\vep-\tL)\,\fq\big]_{ILJK}-2\,\hat\Omega_{IL}{}^M\fr_{MJK}+2\,\fl_{IL}\cdot\epsilon_{JK}-2\left(\pr_I\hr_J-\pr_J\hr_I\right)\cdot\fp_{LK} \, , \notag\\
\omega\cdot P_{IJ} &= \omega\cdot\big[(\vep-\tL)\,\fp\big]_{IJ}-\hr_K(\omega g)^K\cdot\fp_{IJ}-\tfrac12\,\omega\cdot\fy\cdot\left(\pr_I\hr_J-\pr_J\hr_I\right) , \notag\\
\omega\cdot\beta_f\cdot\omega' &= \omega\cdot\big[(\vep-\hB\!\cdot\!\pr)\,\fy\big]\cdot\omega'-\omega\cdot\fy\cdot\hr_K(\omega'g)^K-\hr_K(\omega g)^K\cdot\fy\cdot\omega' \, ,
\label{eq:flatcoeff}
\end{align}
where $\omega$ and $\omega'$ are elements of the Lie algebra and $\sM$ and $\sM'$ are elements of $V_\sM$. The extended Lie derivative acts on the indices of the couplings only. The map $\gamma_\sM$ acts from the right on the coefficients with values in the dual of $V_\sM$, so that $(\fj_I\cdot\gamma_\sM)\cdot\sM=\fj_I\cdot(\gamma_\sM\sM)$. The element $\hr_K(\omega g)^K$ of the Lie algebra comes from the term $\sigma\,\hr_K(f_{\mu\nu}\,g)^K$ of the variation of the field strength. The right-hand side of the expression for $C_{ILJK}$ is to be symmetrised in the indices of each of the pairs $IL$ and $JK$ and under the exchange of the two pairs, that is averaged over these permutations. The expression for $S_{IJ}$ contains the terms with the mass source which are not included in \eqref{eq:quad}. Up to the signs and the normalisation of \eqref{eq:fullanomaly} these are the expressions of section~8 of~\cite{Jack:2013sha} for $d=4-\vep$. The coefficients $C_{ILJK}$, $P_{IJ}$ and $\beta_f$ enter no relation of this paper, and we give them to make \eqref{eq:fullanomaly} complete.

\subsection{Variation of the counterterms}
\label{app:variation}

In this subsection we derive \eqref{eq:fullanomaly} and the expressions for its coefficients. The derivation holds for any number of couplings and of mass operators and for any group of field rotations. We use the rules \eqref{eq:weylrules} and \eqref{eq:varDg} and the variation \eqref{eq:deltaM} of the mass source. The coefficients of \eqref{eq:X} are covariant functions $F$ of the couplings, for which $D_\mu F=\pr_IF\,D_\mu g^I$ and $\delta^{\rm W}_\sigma F=\sigma\,\hB^I\pr_IF$. The variation of the field strength is $D_\mu\delta a_\nu-D_\nu\delta a_\mu$ with $\delta a_\mu=\sigma\,\hr_ID_\mu g^I$. With $D_\mu\hr_I=\pr_J\hr_I\,D_\mu g^J$ and $[D_\mu,D_\nu]\,g^I=(f_{\mu\nu}\,g)^I$ we find
\begin{equation}
\delta^{\rm W}_\sigma f_{\mu\nu} = \pr_\mu\sigma\,\hr_ID_\nu g^I-\pr_\nu\sigma\,\hr_ID_\mu g^I+\sigma\left(\pr_I\hr_J-\pr_J\hr_I\right)D_\mu g^ID_\nu g^J+\sigma\,\hr_I\,(f_{\mu\nu}\,g)^I \, .
\label{eq:varf}
\end{equation}
We write $Dg^I\!\cdot\!Dg^J$ for $D_\mu g^ID^\mu g^J$ and $\pr\sigma\!\cdot\!Dg^I$ for $\pr_\mu\sigma\,D^\mu g^I$. For a term $\mathsf{t}$ of \eqref{eq:X} we denote by $\mathsf{V}[\mathsf{t}]$ the variation of $\sqrt\gamma\,\mathsf{t}$ divided by $\sqrt\gamma$, up to total derivatives.

In the variation of each term the weights of the factors and of $\sqrt\gamma$ add up to $-\vep\,\sigma$. Together with the variation of the coefficient and with the terms of \eqref{eq:varDg} which contain $\hat M^I{}_J$, this gives $-\sigma$ times the same structure with the coefficient $(\vep-\tL)F$ in place of $F$. The other terms come from the remaining terms of the rules. For the terms of the first two lines of \eqref{eq:X} other than the term with the Einstein tensor we find
\begin{align}
&\mathsf{V}\big[\fa\,\cE_4\big] = -\sigma\left(\vep-\hB\!\cdot\!\pr\right)\fa\;\cE_4-8(d-3)\,\pr_I\fa\;\pr_\mu\sigma\,\cG^{\mu\nu}D_\nu g^I \, , \notag\\
&\mathsf{V}\big[\fb\,R^2\big] = -\sigma\left(\vep-\hB\!\cdot\!\pr\right)\fb\;R^2-4(d-1)\,\fb\,R\,\nabla^2\sigma \, , \notag\\
&\mathsf{V}\big[R\,\fc_ID^2g^I\big] = \sigma R\left(-\big[(\vep-\tL)\,\fc\big]_ID^2g^I+\fc_K\,\hat\Omega_{IJ}{}^K\,Dg^I\!\cdot\!Dg^J\right)+2\,\hP_I{}^K\fc_K\,R\;\pr\sigma\!\cdot\!Dg^I \notag\\
&\qquad+\nabla^2\sigma\left(\fc_I\hB^I\,R-2(d-1)\,\fc_ID^2g^I\right) , \notag\\
&\mathsf{V}\big[R\,\ff_{IJ}\,Dg^I\!\cdot\!Dg^J\big] = -\sigma R\,\big[(\vep-\tL)\,\ff\big]_{IJ}\,Dg^I\!\cdot\!Dg^J+2\,\ff_{IJ}\hB^J\,R\;\pr\sigma\!\cdot\!Dg^I \notag\\
&\qquad-2(d-1)\,\ff_{IJ}\,\nabla^2\sigma\;Dg^I\!\cdot\!Dg^J \, , \notag\\
&\mathsf{V}\big[\tfrac12\,\fs_{IJ}D^2g^ID^2g^J\big] = -\tfrac12\,\sigma\,\big[(\vep-\tL)\,\fs\big]_{IJ}D^2g^ID^2g^J+\sigma\,\hat\Omega_{JK}{}^L\fs_{IL}\,D^2g^I\,Dg^J\!\cdot\!Dg^K \notag\\
&\qquad+2\,\hP_I{}^K\fs_{KJ}\;\pr\sigma\!\cdot\!Dg^I\,D^2g^J+\fs_{IJ}\hB^J\,\nabla^2\sigma\,D^2g^I \, .
\label{eq:varone}
\end{align}
Here the term with $\cE_4$ was integrated by parts once, with $\nabla_\mu\cG^{\mu\nu}=0$. The terms of the third line of \eqref{eq:X} without the mass source give
\begin{align}
&\mathsf{V}\big[\fr_{IJK}D^2g^I\,Dg^J\!\cdot\!Dg^K\big] = -\sigma\,\big[(\vep-\tL)\,\fr\big]_{IJK}D^2g^I\,Dg^J\!\cdot\!Dg^K \notag\\
&\qquad+\sigma\,\hat\Omega_{IL}{}^M\fr_{MJK}\,Dg^I\!\cdot\!Dg^L\,Dg^J\!\cdot\!Dg^K+\fr_{KIJ}\hB^K\,\nabla^2\sigma\;Dg^I\!\cdot\!Dg^J \notag\\
&\qquad+\pr\sigma\!\cdot\!Dg^I\left(2\,\fr_{JIK}\hB^K\,D^2g^J+2\,\hP_I{}^L\fr_{LJK}\,Dg^J\!\cdot\!Dg^K\right) , \notag\\
&\mathsf{V}\big[\tfrac12\,\fq_{ILJK}\,Dg^I\!\cdot\!Dg^L\,Dg^J\!\cdot\!Dg^K\big] = -\tfrac12\,\sigma\,\big[(\vep-\tL)\,\fq\big]_{ILJK}\,Dg^I\!\cdot\!Dg^L\,Dg^J\!\cdot\!Dg^K \notag\\
&\qquad+2\,\fq_{ILJK}\hB^L\;\pr\sigma\!\cdot\!Dg^I\,Dg^J\!\cdot\!Dg^K \, .
\label{eq:vartwo}
\end{align}

In the term with the Einstein tensor the rule for $\cG^{\mu\nu}$ gives the two terms
\begin{equation}
-(d-2)\,\fg_{IJ}\,\nabla^\mu\nabla^\nu\sigma\,D_\mu g^ID_\nu g^J+(d-2)\,\fg_{IJ}\,\nabla^2\sigma\;Dg^I\!\cdot\!Dg^J \, .
\end{equation}
We integrate the first by parts, which gives $(d-2)\,\pr^\nu\sigma$ times the divergence $\nabla^\mu(\fg_{IJ}D_\mu g^ID_\nu g^J)$. In it we use $D_\mu D_\nu g^J=D_\nu D_\mu g^J+(f_{\mu\nu}\,g)^J$. The term $\fg_{IJ}D^\mu g^ID_\nu D_\mu g^J$ is one half of the gradient of $\fg_{IJ}\,Dg^I\!\cdot\!Dg^J$ minus one half of $\pr_K\fg_{IJ}D_\nu g^K\,Dg^I\!\cdot\!Dg^J$. A second integration by parts of the gradient gives a term with $\nabla^2\sigma$. With the element $(g\fg)_I$ of the dual of the Lie algebra which is defined by $\omega\cdot(g\fg)_I=(\omega g)^J\fg_{IJ}$ we obtain
\begin{align}
&\mathsf{V}\big[\fg_{IJ}\,\cG^{\mu\nu}D_\mu g^ID_\nu g^J\big] = -\sigma\,\big[(\vep-\tL)\,\fg\big]_{IJ}\,\cG^{\mu\nu}D_\mu g^ID_\nu g^J+2\,\fg_{IJ}\hB^J\,\pr_\mu\sigma\,\cG^{\mu\nu}D_\nu g^I \notag\\
&\qquad+(d-2)\,\pr_\mu\sigma\left(\fg_{IJ}\,D^\mu g^ID^2g^J+\Gamma^{(\fg)}_{IJK}\,D^\mu g^I\,Dg^J\!\cdot\!Dg^K-f^{\mu\nu}\!\cdot(g\fg)_I\,D_\nu g^I\right) \notag\\
&\qquad+\tfrac12(d-2)\,\fg_{IJ}\,\nabla^2\sigma\;Dg^I\!\cdot\!Dg^J \, .
\label{eq:varG}
\end{align}

For the terms with the mass source we write \eqref{eq:deltaM} as $\delta^{\rm W}_\sigma\sM=-\sigma\,(2-\gamma_\sM)\,\sM+\delta'\sM$. The first term gives the weight of $\sM$ and the terms with $\gamma_\sM$. Let $\mathsf{N}=\fn\,R+\fj_ID^2g^I+\fl_{IJ}\,Dg^I\!\cdot\!Dg^J+\fm\cdot\sM$ be the derivative of $\cX$ with respect to $\sM$, which takes values in the dual of $V_\sM$. The sum of the four terms of \eqref{eq:X} with the mass source then gives
\begin{align}
&\mathsf{V}\big[\fn\cdot\sM\,R+\ldots+\tfrac12\,\sM\cdot\fm\cdot\sM\big] = \mathsf{N}\cdot\delta'\sM-\tfrac12\,\sigma\,\sM\cdot\beta_\sM\cdot\sM \notag\\
&\qquad+\sigma\left(-\left(\vep-\hB\!\cdot\!\pr\right)\fn+\fn\cdot\gamma_\sM\right)\cdot\sM\,R-2(d-1)\,\fn\cdot\sM\,\nabla^2\sigma \notag\\
&\qquad+\sigma\left(-\big[(\vep-\tL)\,\fj\big]_I+\fj_I\cdot\gamma_\sM\right)\cdot\sM\,D^2g^I+\sigma\,\hat\Omega_{IJ}{}^K\fj_K\cdot\sM\;Dg^I\!\cdot\!Dg^J \notag\\
&\qquad+2\,\hP_I{}^K\fj_K\cdot\sM\;\pr\sigma\!\cdot\!Dg^I+\fj_I\hB^I\cdot\sM\,\nabla^2\sigma \notag\\
&\qquad+\sigma\left(-\big[(\vep-\tL)\,\fl\big]_{IJ}+\fl_{IJ}\cdot\gamma_\sM\right)\cdot\sM\;Dg^I\!\cdot\!Dg^J+2\,\fl_{IJ}\hB^J\cdot\sM\;\pr\sigma\!\cdot\!Dg^I \, , \notag\\
&\mathsf{N}\cdot\delta'\sM = -\sigma\,\mathsf{N}\cdot\left(\delta_ID^2g^I+\epsilon_{IJ}\,Dg^I\!\cdot\!Dg^J+\bar\eta\,R\right)-2\;\pr\sigma\!\cdot\!Dg^I\;\mathsf{N}\cdot\theta_I-\nabla^2\sigma\;\mathsf{N}\cdot\bar\tau \, ,
\label{eq:varM}
\end{align}
with $\beta_\sM$ as in \eqref{eq:flatcoeff}. The term $\mathsf{N}\cdot\delta'\sM$ contains all contributions of $\delta_I$, $\epsilon_{IJ}$, $\bar\eta$, $\theta_I$ and $\bar\tau$ to the coefficients. Finally the two terms with the field strength give, with \eqref{eq:varf},
\begin{align}
&\mathsf{V}\big[f^{\mu\nu}\!\cdot\fp_{IJ}D_\mu g^ID_\nu g^J\big] = -\sigma\,f^{\mu\nu}\!\cdot P_{IJ}'\,D_\mu g^ID_\nu g^J-2\,\pr_\mu\sigma\,f^{\mu\nu}\!\cdot\fp_{IJ}\hB^J\,D_\nu g^I \notag\\
&\qquad+2\,\hr_K\cdot\fp_{IJ}\;\pr\sigma\!\cdot\!Dg^I\,Dg^J\!\cdot\!Dg^K+\sigma\left(\pr_I\hr_J-\pr_J\hr_I\right)\cdot\fp_{LK}\,Dg^I\!\cdot\!Dg^L\,Dg^J\!\cdot\!Dg^K \, , \notag\\
&\mathsf{V}\big[\tfrac14\,f^{\mu\nu}\!\cdot\fy\cdot f_{\mu\nu}\big] = -\tfrac14\,\sigma\,f^{\mu\nu}\!\cdot\beta_f\cdot f_{\mu\nu}+\pr_\mu\sigma\,f^{\mu\nu}\!\cdot\fy\cdot\hr_I\,D_\nu g^I \notag\\
&\qquad+\tfrac12\,\sigma\,f^{\mu\nu}\!\cdot\fy\cdot\left(\pr_I\hr_J-\pr_J\hr_I\right)D_\mu g^ID_\nu g^J \, ,
\label{eq:varfield}
\end{align}
where $\omega\cdot P_{IJ}'=\omega\cdot[(\vep-\tL)\fp]_{IJ}-\hr_K(\omega g)^K\cdot\fp_{IJ}$ and $\beta_f$ is given in \eqref{eq:flatcoeff}. In both the last term of \eqref{eq:varf} gives the terms with $\hr_K(\omega g)^K$.

The sum of \eqref{eq:varone}, \eqref{eq:vartwo}, \eqref{eq:varG}, \eqref{eq:varM} and \eqref{eq:varfield} is the integrand of \eqref{eq:fullanomaly}. Comparing the coefficients of the structures we obtain the expressions \eqref{eq:lin}, \eqref{eq:quad} and \eqref{eq:Qdef}, and with the terms of $\mathsf{N}\cdot\delta'\sM$ their complete forms \eqref{eq:fixedcoeff}, \eqref{eq:derivedcoeff} and \eqref{eq:flatcoeff}. Note that we integrated by parts only in the terms with $\nabla_\mu\nabla_\nu\sigma$, which fixes the decomposition of the anomaly into the structures of \eqref{eq:fullanomaly}. All steps are identities of tensor calculus in integer dimension. On the backgrounds of appendix~\ref{app:rotations} both sides are polynomials in $d$ for given derivatives of the sources at a point, so that the result holds with $d$ as a parameter.

\subsection{Coefficients of flat space}

We now show that the coefficients of the variation of the sources are regular. The proof uses the lemmas~\ref{lem:potential}, \ref{lem:lowest} and~\ref{lem:transverse} on the terms of second order and Proposition~\ref{prop:uniqueall} on the covariance of the counterterms. They are proved in appendix~\ref{app:unique}, which does not depend on the present appendix. The proof also uses Lemma~\ref{lem:rhoxi} on the variation of the bare gauge field.

\begin{lemma}
\label{lem:regvar}
Assume Hypothesis~\ref{hyp:finite} for multiscalar $\phi^4$ theory. Then $\hB^I$, $\hr_I$, $\gamma_\sM$, $\delta_I$, $\epsilon_{IJ}$ and $\theta_I$ are regular.
\end{lemma}

\begin{proof}
The counterterms $L^I$ are covariant, and by Proposition~\ref{prop:uniqueall} so are the other coefficients of the bare sources and of the counterterms. The coefficients $\hB^I$ and $\hr_I$ are defined by \eqref{eq:xiL}. The coefficients of \eqref{eq:deltaM} are defined by the condition that the bare mass source changes by its weight and by the same gauge transformation as the other bare sources, see \eqref{eq:gammaM} and \eqref{eq:masscoeff}. All of them are power series in the couplings whose coefficients are a priori Laurent polynomials in $\vep$. Let $v$ be a conformal Killing vector of flat space which preserves the class of sources, with $\pr_\mu v_\nu+\pr_\nu v_\mu=2\sigma\delta_{\mu\nu}$. These are the dilatation, for which $\sigma$ is constant, and the special conformal transformation with a parameter $b_\mu$ in the parallel direction, for which $\sigma=2\,b\cdot x$. We consider the flat form of the variation \eqref{eq:deltavS},
\begin{equation}
\delta g^I = \sigma\,\hB^I-v^\mu\pr_\mu g^I \, , \qquad \delta a_\mu = \sigma\,\hr_ID_\mu g^I-\cL_va_\mu \, , \qquad \delta\sM = \delta^{\rm W}_\sigma\sM-v^\mu\pr_\mu\sM \, .
\label{eq:flatvar}
\end{equation}
By Lemma~\ref{lem:rhoxi} and by the definition of the coefficients of \eqref{eq:deltaM} it changes the bare sources by the conformal transformation, which is \eqref{eq:barevar} in the chart, and by a gauge transformation with the parameter $\sigma\xi^{\rm L}$. The bare functional is invariant under the first by \eqref{eq:covS} and \eqref{eq:chart}. Indeed, in the chart from $P$ the boost \eqref{eq:boost} along $P$ is the dilatation, and the boost along the other parallel direction is to first order a translation combined with the special conformal transformation, see \eqref{eq:chartboost}. The flat functional is invariant under translations of $x_\parallel$, and a source with compact support in the chart vanishes near $P$ together with its variation, so that the first-order form of \eqref{eq:covS} gives the invariance in flat space, as we show in appendix~\ref{app:rotations}. The bare functional is invariant under the gauge transformation by Proposition~\ref{prop:gauge}. The invariance under gauge transformations is linear in the parameter, so that it holds for each coefficient of a parameter which is a Laurent polynomial in $\vep$. Hence
\begin{equation}
\int d^dx\left(\delta g^I\,\frac{\delta W^{\rm fl}}{\delta g^I}+\delta a_\mu\cdot\frac{\delta W^{\rm fl}}{\delta a_\mu}+\delta\sM\cdot\frac{\delta W^{\rm fl}}{\delta\sM}\right) = \cA^{\rm fl}[\sigma] \, ,
\label{eq:flatward}
\end{equation}
where the right-hand side is the variation of the integral of the counterterms. It is the integral of a local density, whatever its coefficients are. The identity holds at each order in the sources as an identity of Laurent series in $\vep$.

\paragraph{Lowest order.} We use sources with the index $1$ and sources with the index $2$, where the supports of the latter are disjoint from the supports of the former, as in Lemma~\ref{lem:lowest}. A local functional has no terms which contain sources of both indices, so that the right-hand side of \eqref{eq:flatward} has none. By Lemma~\ref{lem:lowest} the coefficients $\fd_I$ and $Z_\sM-1$ vanish at $g=0$, and so does $N_I$ by Proposition~\ref{prop:uniqueall}. The bare sources then agree with the renormalised ones at first order. With \eqref{eq:xiL} and with the condition which defines the coefficients of \eqref{eq:deltaM} it follows that $\hr_I$, $\delta_I$, $\theta_I$ and $\gamma_\sM$ vanish at $g=0$ and that the term of first order of $\hB^I$ is $-\vep\,g^I$. The terms of first order in the sources of the variation \eqref{eq:flatvar} are therefore $-\vep\,\sigma g^I-v^\mu\pr_\mu g^I$, $-\cL_va_\mu$ and $-2\,\sigma\sM-v^\mu\pr_\mu\sM$, which have no poles at $\vep=0$.

\paragraph{Induction.} As in the proof of Proposition~\ref{prop:uniqueall} we assign a level to the terms of the coefficients. The terms of degree $n$ of $\hB^I$ and of $\gamma_\sM$, the terms of degree $n-1$ of $\hr_I$, $\delta_I$ and $\theta_I$ and the term of degree $n-2$ of $\epsilon_{IJ}$ have the level $n$. These are the terms which enter the variation \eqref{eq:flatvar} at order $n$ in the couplings. By the previous paragraph there are no terms of level zero. We argue by induction on the level. Assume that all terms of level below $n$ are regular, where $n\geq1$. We show in four steps that the terms of level $n$ are regular, first for the gauge field, then for the couplings, and then in two steps for the mass source. For $n=1$ only the last step is needed. Indeed, the term of level one of $\hB^I$ is $-\vep\,g^I$, the terms of level one of $\hr_I$, $\delta_I$ and $\theta_I$ vanish, and the remaining term is the term of degree one of $\gamma_\sM$. In each step we assume that the term in question has a pole, of highest order $\vep^{-\mathsf{r}}$, and we consider the coefficient of $\vep^{-\mathsf{r}}$ of certain terms of \eqref{eq:flatward}. A term of the variation with a regular coefficient is a source of the class, and it contributes to the left-hand side a derivative of $W^{\rm fl}$ in the direction of this source, which is finite by Hypothesis~\ref{hyp:finite}. At a point of the support of a source with the index $2$ the variation is of first order and has no pole at $\vep=0$.

\paragraph{Gauge field.} We take couplings $g_1$ and a gauge field $a_2$, and we consider the terms of \eqref{eq:flatward} of order $n$ in $g_1$ and linear in $a_2$. A term of the variation with a coefficient of level $k$, at a point of the support of $g_1$, multiplies a derivative of $W^{\rm fl}$ of order $n-k$ in $g_1$ and linear in $a_2$. For $k<n$ the product is finite by the induction. For $k=n$ the derivative is a term of second order of $W^{\rm fl}$, and by Lemma~\ref{lem:lowest} only the term with the variation of the gauge field remains. Let $\vep^{-\mathsf{r}}\breve\rho_I$ be the leading pole of $\hr_I$ at degree $n-1$. The coefficient of $\vep^{-\mathsf{r}}$ is then the integral of $j_\mu(x)\cdot a_{2\nu}(y)$ with the kernel of \eqref{eq:Apot} over both points at $d=4$, up to a factor which does not vanish, where $j=\sigma\,\breve\rho_I(g_1)\,dg_1^I$. Here we used that for disjoint supports the integral is analytic at $d=4$, where \eqref{eq:gram} is the ordinary integral. The coefficient vanishes for all one-forms $a_2$ of the class with support outside the support of $g_1$. The one-form $j$ is invariant under the transverse rotations, and so is the one-form $A_\nu$ of Lemma~\ref{lem:transverse}. A continuous invariant one-form has a component along $dx_\parallel$ and a component along $x_\perp\!\cdot dx_\perp$, which are functions of $x_\parallel$ and $|x_\perp|$, and the one-forms of the class separate such one-forms. Hence $A_\nu$ vanishes outside the support of $g_1$, and by Lemma~\ref{lem:transverse} the one-form $j$ is closed, for every component in the Lie algebra and for every source $g_1$. At a point with $x_\perp\neq0$ the value of a source of the class and its derivatives in the parallel and in the radial direction are arbitrary. For constant $\sigma$ we therefore find $\pr_J\breve\rho_I=\pr_I\breve\rho_J$, so that $\breve\rho_I=\pr_I\ell$ with a homogeneous polynomial $\ell(g)$ of degree $n$. For $\sigma=2\,b\cdot x$ the one-form is $j=\sigma\,d\ell(g_1)$, and it is closed only if the radial derivative of $\ell(g_1(x))$ vanishes. Since this holds for all sources $g_1$, we find $\pr_I\ell=0$ and $\breve\rho_I=0$, which is a contradiction.

\paragraph{Couplings.} We take the couplings $g_1+g_2$ and no other sources, and we consider the terms of order $n$ in $g_1$ and linear in $g_2$. The terms of the variation with coefficients of level $k<n$ are regular by the induction, and the term $\sigma\hr_ID_\mu g^I$ of level $n$ is regular by the previous step. The terms of level $n$ of the variations of the gauge field and of the mass source multiply terms of second order of $W^{\rm fl}$ with two sources of different kinds, which vanish. The term $\sigma\hB^I$ of level $n$ multiplies the first expression of \eqref{eq:twopointkernels}. Let $\vep^{-\mathsf{r}}\breve\beta^I$ be the leading pole of $\hB^I$ at degree $n$. The coefficient of $\vep^{-\mathsf{r}}$ is the integral of $\sigma(x)\,\breve\beta^I(g_1(x))\,|x-y|^{-8}\,g_2^I(y)$ over both points at $d=4$, up to a factor which does not vanish. It vanishes for all $g_2$ of the class with support outside the support of $g_1$. The function $\sigma\,\breve\beta^I(g_1)$ is invariant under the transverse rotations, and so is its potential with the kernel $|x-y|^{-8}$. Since the sources of the class separate invariant continuous functions, this potential vanishes outside the support of $g_1$, and Lemma~\ref{lem:potential} gives $\sigma\,\breve\beta^I(g_1)=0$ for all $g_1$. Hence $\breve\beta^I=0$, which is a contradiction.

\paragraph{Mass source.} We take couplings $g_1$ and a mass source $\sM_2$, and we consider the terms of order $n$ in $g_1$ and linear in $\sM_2$. By the previous steps the terms of the variations of the couplings and of the gauge field up to the level $n$ are regular. The terms of level $n$ of $\delta^{\rm W}_\sigma\sM$ at $\sM=0$ multiply the second expression of \eqref{eq:twopointkernels}. If they have a pole, the potential with the kernel $|x-y|^{-4}$ of the leading pole of $\sigma\,(\delta_I\pr^2g_1^I+\epsilon_{IJ}\pr_\mu g_1^I\pr^\mu g_1^J)+2\,\pr_\mu\sigma\,\theta_I\pr^\mu g_1^I$ vanishes outside the support of $g_1$, and by Lemma~\ref{lem:potential} this function vanishes. At a point with $x_\perp\neq0$ the Laplacian of a source of the class and its derivatives in the parallel and in the radial direction are independent. For constant $\sigma$ we find that the leading poles of $\delta_I$ and $\epsilon_{IJ}$ vanish, and for $\sigma=2\,b\cdot x$ that the leading pole of $\theta_I$ vanishes. In the last step we add a mass source $\sM_1$ whose support may overlap with the support of $g_1$, and we consider the terms of order $n$ in $g_1$ which are linear in $\sM_1$ and in $\sM_2$. The only term of the variation which is not yet known to be regular is $\sigma\,\gamma_\sM\sM_1$ with $\gamma_\sM$ of degree $n$. All other terms of the variation are sources of the class with regular coefficients, which multiply derivatives of $W^{\rm fl}$ that are finite by Hypothesis~\ref{hyp:finite}. The term $\sigma\,\gamma_\sM\sM_1$ multiplies the second expression of \eqref{eq:twopointkernels}, and the same argument shows that the leading pole of $\gamma_\sM(g_1)\,\sM_1$ vanishes at each point. For $n=1$ this step uses only the terms of first order of the variation, which were given above. This completes the induction.
\end{proof}

By the fourth statement of Lemma~\ref{lem:rhoxi} the coefficients $\hB^I$ and $\hr_I$ are then given by \eqref{eq:Bhat} and \eqref{eq:rhoxi}, and with the choice $\xi^{\rm L}=\xi$ we have $\hb^{{\rm L}\,I}=\hB^I+(\xi g)^I=\hb^I$. Note that the proof also shows that the coefficients of the variation are determined by the renormalised functional. Let two sets of coefficients $\hB^I$, $\hr_I$, $\gamma_\sM$, $\delta_I$, $\epsilon_{IJ}$ and $\theta_I$ be given, which are power series in the couplings whose coefficients are Laurent polynomials in $\vep$ and in which $\hB^I$ has no term of degree zero. Assume that \eqref{eq:flatward} holds for both sets, with the same functional $W^{\rm fl}$ and with some local functional on the right-hand side. The difference of the two identities contains the differences of the coefficients in place of the coefficients, and its right-hand side is local. We go through the levels as in the proof, from the level zero on, with the term with the lowest power of $\vep$ of a difference in place of the leading pole. The terms with differences of lower levels are absent, because these differences vanish. At the two lowest levels the argument has to be extended, because the difference of the variation need not vanish on the support of the sources with the index $2$. At the level zero the only term is the value $\tilde\gamma$ at $g=0$ of the difference of $\gamma_\sM$. For two mass sources $\sM_1=\sM_{\rm c}\,\chi_1$ and $\sM_2=\sM'_{\rm c}\,\chi_2$ with constant matrices and functions $\chi_1$ and $\chi_2$ of the class with disjoint supports, the identity states that the integral of $\big[\sigma(x)\,(\tilde\gamma\,\sM_{\rm c})\cdot\sM'_{\rm c}+\sigma(y)\,\sM_{\rm c}\cdot(\tilde\gamma\,\sM'_{\rm c})\big]\chi_1(x)\,\chi_2(y)\,|x-y|^{-4}$ over both points vanishes. For constant $\sigma$ and positive functions the two terms of the bracket therefore add up to zero. For $\sigma=2\,b\cdot x$ the bracket is then $2\,b\cdot(x-y)\,(\tilde\gamma\,\sM_{\rm c})\cdot\sM'_{\rm c}$, and for positive functions whose supports are separated in the parallel direction we find $\tilde\gamma=0$. The same argument with the kernel $|x-y|^{-8}$ applies to the difference of the terms of degree one of $\hB^I$. At the level one the step for the gauge field gives a constant difference $\tilde\rho_I$ of $\hr_I$ for which $d\sigma\wedge\tilde\rho_I\,dg_1^I$ vanishes for $\sigma=2\,b\cdot x$, hence $\tilde\rho_I=0$. In the step for the couplings at this level the terms with the differences of $\hr_I$, $\delta_I$ and $\theta_I$ multiply terms of second order with two sources of different kinds, so that each step again contains one difference only. The other steps are unchanged. Hence the two sets agree.

We now apply Lemma~\ref{lem:regvar} and Lemma~\ref{lem:indep}. By Lemma~\ref{lem:regvar} the coefficients of the variation \eqref{eq:flatvar} are regular, and by Hypothesis~\ref{hyp:finite} the left-hand side of \eqref{eq:flatward} is then finite. The functional $\cA^{\rm fl}[\sigma]$, which is \eqref{eq:fullanomaly} in flat space, is therefore finite for all sources of the class. For constant $\sigma$ it is $-\sigma$ times the left-hand side of \eqref{eq:lemA} with the coefficients $-S_{IJ}$, $B_{IJK}$, $C_{ILJK}$, $J_I$, $K_{IJ}$, $\beta_\sM$, $\beta_f$ and $P_{IJ}$. Consider a fixed order in the couplings and the highest order $\mathsf{r}\geq1$ of the poles of these coefficients. The coefficient of $\vep^{-\mathsf{r}}$ of $\cA^{\rm fl}[\sigma]$ is the functional \eqref{eq:lemA} at $d=4$ with the leading coefficients, since the structures are analytic in $d$. By the first part of Lemma~\ref{lem:indep} the leading coefficients vanish. Hence these eight coefficients are regular. For $\sigma=2\,b\cdot x$ the terms of $\cA^{\rm fl}[\sigma]$ with $\sigma$ are finite by the previous step, and the terms with $\pr^2\sigma$ vanish. The remaining terms are \eqref{eq:lemB} with the coefficients $E_{IJ}$, $R_{IJK}$, $-2\,Q_I$ and $2\,L_I$. By the second part of the lemma the leading coefficients are of the form of the total derivative. They are covariant by Proposition~\ref{prop:uniqueall}, so that this total derivative integrates to zero for every $d$. We subtract it and repeat the argument for the next order of the poles. We find
\begin{equation}
{\rm PP}\,E_{[IJ]} = 0 \, , \qquad {\rm PP}\,R_{IJK} = \Gamma^{({\rm PP}\,E)}_{IJK} \, , \qquad
2\,{\rm PP}\,\omega\cdot Q_J = \big({\rm PP}\,E_{IJ}\big)(\omega g)^I \, , \qquad {\rm PP}\,L_I = 0 \, ,
\label{eq:flatpoles}
\end{equation}
for every covariant choice of $\fg_{IJ}$. Indeed, by \eqref{eq:fixedcoeff} the tensor $\fg_{IJ}$ enters $E_{IJ}$, $R_{IJK}$ and $Q_J$ only through $(d-2)$ times the coefficients of the total derivative, and it does not enter flat space.

\subsection{Curvature counterterms}

The curvature counterterms are fixed by the following elementary statement. Let $\mathsf{c}(\vep)$ be a power series in $\vep$ with $\mathsf{c}(0)\neq0$, and let $\mathsf{k}$ be a power series in the couplings whose coefficients are Laurent polynomials in $\vep$. Then there is a unique pure pole $\mathsf{u}$ such that $\mathsf{c}\,\mathsf{u}+\mathsf{k}$ is regular, namely $\mathsf{u}=-{\rm PP}\,[\mathsf{k}/\mathsf{c}]$. Indeed, $\mathsf{c}\,\mathsf{u}+\mathsf{k}=\mathsf{c}\,(\mathsf{k}/\mathsf{c}-{\rm PP}\,[\mathsf{k}/\mathsf{c}])$ is regular, and if $\mathsf{c}\,\mathsf{u}$ is regular for a pure pole $\mathsf{u}$, then $\mathsf{u}={\rm PP}\,[(\mathsf{c}\,\mathsf{u})/\mathsf{c}]=0$.

We apply this statement to the coefficients $\bar\tau$, $t_\sM$, $E_{(IJ)}$, $u_I$, $\nu_{IJ}$, $r$ and $W_I$ of \eqref{eq:fixedcoeff}, in this order, and find
\begin{align}
\fk &= {\rm PP}\left[\frac{\fd_I\hB^I}{2(d-1)}\right] , \qquad
\fn = {\rm PP}\left[\frac{\fj_I\hB^I-\bar\tau\cdot\fm}{2(d-1)}\right] , \qquad
\fg_{IJ} = -{\rm PP}\left[\frac{2\,\hat S_{(IJ)}}{d-2}\right] , \notag\\
\fc_I &= {\rm PP}\left[\frac{\fs_{IJ}\hB^J-\fj_I\cdot\bar\tau}{2(d-1)}\right] , \qquad
\ff_{IJ} = {\rm PP}\left[\frac{\tfrac12(d-2)\,\fg_{IJ}+\fr_{KIJ}\hB^K-\bar\tau\cdot\fl_{IJ}}{2(d-1)}\right] , \notag\\
\fb &= {\rm PP}\left[\frac{\fc_I\hB^I-\fn\cdot\bar\tau}{4(d-1)}\right] , \qquad
\pr_I\fa = {\rm PP}\left[\frac{\fg_{IJ}\hB^J}{4(d-3)}\right] .
\label{eq:steps}
\end{align}
Each right-hand side contains only flat-space counterterms and quantities which are fixed by the earlier steps.\footnote{We confirmed the order of the steps by perturbing each curvature counterterm in exact arithmetic and recording which coefficients change.} By \eqref{eq:flatpoles} the coefficients $E_{[IJ]}$, $R_{IJK}$ and $Q_I$ are regular once $\fg_{IJ}$ is chosen in this way. The last equation requires that the one-form on its right-hand side is closed. For this we use the identity
\begin{align}
(d-2)\left(\pr_IW_J-\pr_JW_I\right) ={}& -\left(\hP_I{}^KE_{JK}-\hP_J{}^KE_{IK}\right)+\left(R_{IJK}-R_{JIK}\right)\hB^K \notag\\
&+\hr_I\cdot Q_J-\hr_J\cdot Q_I-2\left(L_I\cdot\theta_J-L_J\cdot\theta_I\right) ,
\label{eq:curlgeneral}
\end{align}
which holds in every scheme for the mass source and which we derive in appendix~\ref{app:anomcoeff} from the expressions of the coefficients through the counterterms. At $\theta_I=0$ it is the relation \eqref{eq:curl} of appendix~\ref{gs:app:fourd}. After the third step its right-hand side is regular. Since the term with $\fa$ drops out of the left-hand side, the curl of $\fg_{IJ}\hB^J/(4(d-3))$ is regular, and the pole part of this one-form is closed. On power series in the couplings a closed one-form $\omega_I$ is the gradient of the function $\int_0^1dt\,g^I\omega_I(tg)$, which vanishes at $g=0$ and is invariant under field rotations if $\omega_I$ is covariant. This defines $\fa$.

Each curvature counterterm is thus fixed by one condition, and the conditions \eqref{eq:steps} are solved one after the other. The regularity of $E_{[IJ]}$, $R_{IJK}$ and $Q_I$ and the closedness in the last step are conditions of consistency. They hold by Hypothesis~\ref{hyp:finite} and by the identity \eqref{eq:curlgeneral}. Note that $\hB^I$ vanishes at $g=0$. From \eqref{eq:fixedcoeff} we then find that $\fk$, $\bar\tau$, $\fn$, $t_\sM$, $\fc_I$, $u_I$, $\fb$ and $r$ vanish at $g=0$, and so do $\bar\eta$ and $i$.

\subsection{Remaining coefficients}

The remaining coefficients are regular by relations among the coefficients, which are identities in the counterterms. The first is the relation for $\bar\eta$ in \eqref{eq:Mrel}. The others are
\begin{align}
2(d-1)\,i &= \left(\vep-\hB\!\cdot\!\pr\right)t_\sM-t_\sM\cdot\gamma_\sM-J_I\hB^I+\bar\tau\cdot\beta_\sM \, , \notag\\
2(d-1)\,e_I &= \big[(\vep-\tL)\,u\big]_I+S_{IJ}\hB^J+t_\sM\cdot\delta_I+\bar\tau\cdot J_I \, , \notag\\
4(d-1)\,q &= \left(\vep-\hB\!\cdot\!\pr\right)r+e_I\hB^I+t_\sM\cdot\bar\eta-\bar\tau\cdot i \, , \notag\\
(d-2)\,G_{IJ} &= \big[(\vep-\tL)\,E\big]_{IJ}+2\,\hP_I{}^KS_{KJ}-2\,B_{JIK}\hB^K+2\,\theta_I\cdot J_J+2\,L_I\cdot\delta_J \, , \notag\\
2(d-1)\,f_{IJ} &= \big[(\vep-\tL)\,\nu\big]_{IJ}-\tfrac12(d-2)\,G_{IJ}-B_{KIJ}\hB^K-\hat\Omega_{IJ}{}^Lu_L+t_\sM\cdot\epsilon_{IJ}+\bar\tau\cdot K_{IJ} \, .
\label{eq:finrel}
\end{align}
In each line the right-hand side contains only coefficients which are already regular, and the factor on the left does not vanish at $\vep=0$.\footnote{We also verified these relations with computer algebra for covariant data with up to four couplings, three generators and three mass operators.} We derive these relations at the end of this subsection. The coefficient $h_I$ is regular by the relation \eqref{eq:hrel} of appendix~\ref{app:massward}, whose right-hand side contains only coefficients which are already regular. Finally we apply $\vep-\tL$ to the expression for $W_I$ in \eqref{eq:fixedcoeff}. Since $\fa$ is invariant, $\tL$ maps $\pr_I\fa$ to $\pr_I(\hB\!\cdot\!\pr\,\fa)$, as stated in section~\ref{gs:sec:quantities}, and $\tL\hB=0$ by \eqref{eq:rhoB}. With $p=-(\vep-\hB\!\cdot\!\pr)\fa$ and $G_{IJ}=[(\vep-\tL)\fg]_{IJ}$ of \eqref{eq:lin} and \eqref{eq:quad} we obtain $[(\vep-\tL)W]_I=-4(d-3)\,\pr_Ip-G_{IJ}\hB^J$, that is, with $A_{\rm E}=-4p$,
\begin{equation}
(d-3)\,\pr_IA_{\rm E} = G_{IJ}\,\hB^J+\big[(\vep-\tL)\,W\big]_I \, .
\label{gs:eq:eulerF}
\end{equation}
Its right-hand side contains only coefficients which are already regular, and the factor $d-3$ does not vanish at $\vep=0$. Hence $\pr_Ip$ is regular, and $p$ vanishes at $g=0$ because $\fa$ does. This proves the first part of Proposition~\ref{prop:finite}. In the marginal sector the relations \eqref{eq:finrel} for $e_I$ and $q$ reduce to the first two terms on their right-hand sides and the relation for $G_{IJ}$ to the first three, with $B_{IJK}=[(\vep-\tL)\fr]_{IJK}-\hat\Omega_{JK}{}^L\fs_{IL}$.

We now derive \eqref{eq:finrel} from the expressions \eqref{eq:fixedcoeff}, \eqref{eq:derivedcoeff} and \eqref{eq:flatcoeff}. The operator $\vep-\tL$ obeys the product rule on contractions of indices of the couplings, with $\vep$ counted once, and $\tL\hB=0$ by \eqref{eq:rhoB}. For a coefficient $Y$ with values in the dual of $V_\sM$ and a coefficient $X$ with values in $V_\sM$ the product rule reads
\begin{equation}
(\vep-\tL)\,(Y\cdot X) = \big[(\vep-\tL)\,Y-Y\cdot\gamma_\sM\big]\cdot X-Y\cdot(\tL-\gamma_\sM)\,X \, .
\label{eq:prodM}
\end{equation}
By \eqref{eq:derivedcoeff} and \eqref{eq:flatcoeff} the first bracket is $-i-\bar\eta\cdot\fm$ for $Y=\fn$, it is $J_I-\delta_I\cdot\fm$ for $Y=\fj_I$, and it is $K_{IJ}+\hat\Omega_{IJ}{}^K\fj_K-\epsilon_{IJ}\cdot\fm$ for $Y=\fl_{IJ}$. For $Y=\bar\tau\cdot\fm$ it is $\bar\tau\cdot\beta_\sM-[(\hB\!\cdot\!\pr-\gamma_\sM)\,\bar\tau]\cdot\fm$, because $\fm$ is symmetric. We also use the first relation of \eqref{eq:Mrel} in the form $\mathsf{z}=0$ with $\mathsf{z}=2(d-1)\,\bar\eta-\delta_I\hB^I+(\hB\!\cdot\!\pr-\gamma_\sM)\,\bar\tau$.

For the first relation we apply $\vep-\hB\!\cdot\!\pr$ to $t_\sM$ of \eqref{eq:fixedcoeff} and subtract $t_\sM\cdot\gamma_\sM$. The three terms of $t_\sM$ give $(J_I-\delta_I\cdot\fm)\,\hB^I$, then $2(d-1)\,(i+\bar\eta\cdot\fm)$, and then $-\bar\tau\cdot\beta_\sM+[(\hB\!\cdot\!\pr-\gamma_\sM)\,\bar\tau]\cdot\fm$. Their sum is $J_I\hB^I+2(d-1)\,i-\bar\tau\cdot\beta_\sM+\mathsf{z}\cdot\fm$, which is the first relation. For the second relation we apply $\vep-\tL$ to $u_I$. The term $-2(d-1)\,\fc_I$ gives $2(d-1)\,(e_I+\fn\cdot\delta_I+\fj_I\cdot\bar\eta)$ by the expression for $e_I$. The term $\fs_{IJ}\hB^J$ gives $-(S_{IJ}+\fj_I\cdot\delta_J+\fj_J\cdot\delta_I)\,\hB^J$, and the term $-\fj_I\cdot\bar\tau$ gives $-(J_I-\delta_I\cdot\fm)\cdot\bar\tau+\fj_I\cdot(\hB\!\cdot\!\pr-\gamma_\sM)\,\bar\tau$. With the expression for $t_\sM$ the sum is
\begin{equation}
\big[(\vep-\tL)\,u\big]_I = 2(d-1)\,e_I-S_{IJ}\hB^J-t_\sM\cdot\delta_I-\bar\tau\cdot J_I+\fj_I\cdot\mathsf{z} \, .
\end{equation}
For the third relation we apply $\vep-\hB\!\cdot\!\pr$ to $r$. The term $\fc_I\hB^I$ gives $-(e_I+\fn\cdot\delta_I+\fj_I\cdot\bar\eta)\,\hB^I$, the term $-4(d-1)\,\fb$ gives $4(d-1)\,(q+\fn\cdot\bar\eta)$, and the term $-\fn\cdot\bar\tau$ gives $(i+\bar\eta\cdot\fm)\cdot\bar\tau+\fn\cdot(\hB\!\cdot\!\pr-\gamma_\sM)\,\bar\tau$. The sum is $4(d-1)\,q-e_I\hB^I-t_\sM\cdot\bar\eta+\bar\tau\cdot i+\fn\cdot\mathsf{z}$.

For the fourth relation we apply $\vep-\tL$ to $E_{IJ}=2\,\hP_I{}^K\fs_{KJ}+2\,\fr_{JIK}\hB^K+(d-2)\,\fg_{IJ}-2\,\theta_I\cdot\fj_J$. With \eqref{eq:LiePsi} the first term gives $-2\,\hat\Omega_{IL}{}^K\hB^L\fs_{KJ}-2\,\hP_I{}^K(S_{KJ}+\fj_K\cdot\delta_J+\fj_J\cdot\delta_K)$. The second term gives $2\,(B_{JIK}+\hat\Omega_{IK}{}^L\fs_{JL}-\fj_J\cdot\epsilon_{IK}-\fl_{IK}\cdot\delta_J)\,\hB^K$ by the expression for $B_{IJK}$, and the third gives $(d-2)\,G_{IJ}$. For the last term we use \eqref{eq:prodM} and the second relation of \eqref{eq:Mrel}, which give $-2\,\theta_I\cdot(J_J-\delta_J\cdot\fm)+2\,\fj_J\cdot(\hP_I{}^K\delta_K+\epsilon_{IK}\hB^K)$. In the sum the terms with $\hat\Omega$ cancel, and so do the terms in which $\fj_J$ is contracted with $\delta_K$ or with $\epsilon_{IK}$. With the expression for $L_I$ we obtain
\begin{equation}
\big[(\vep-\tL)\,E\big]_{IJ} = (d-2)\,G_{IJ}-2\,\hP_I{}^KS_{KJ}+2\,B_{JIK}\hB^K-2\,\theta_I\cdot J_J-2\,L_I\cdot\delta_J \, .
\end{equation}
This step uses the covariance of $\hB^I$ through \eqref{eq:LiePsi}. For the fifth relation we apply $\vep-\tL$ to $\nu_{IJ}$. The term $-2(d-1)\,\ff_{IJ}$ gives $2(d-1)\,(f_{IJ}-\fc_K\hat\Omega_{IJ}{}^K+\fn\cdot\epsilon_{IJ}+\fl_{IJ}\cdot\bar\eta)$, and the term with $\fg_{IJ}$ gives $\tfrac12(d-2)\,G_{IJ}$. The term $\fr_{KIJ}\hB^K$ gives $(B_{KIJ}+\hat\Omega_{IJ}{}^L\fs_{KL}-\fj_K\cdot\epsilon_{IJ}-\fl_{IJ}\cdot\delta_K)\,\hB^K$, and the term $-\bar\tau\cdot\fl_{IJ}$ gives $-\bar\tau\cdot(K_{IJ}+\hat\Omega_{IJ}{}^K\fj_K-\epsilon_{IJ}\cdot\fm)+\fl_{IJ}\cdot(\hB\!\cdot\!\pr-\gamma_\sM)\,\bar\tau$. With the expressions for $u_I$ and $t_\sM$ the sum is
\begin{equation}
\big[(\vep-\tL)\,\nu\big]_{IJ} = 2(d-1)\,f_{IJ}+\tfrac12(d-2)\,G_{IJ}+B_{KIJ}\hB^K+\hat\Omega_{IJ}{}^Lu_L-t_\sM\cdot\epsilon_{IJ}-\bar\tau\cdot K_{IJ}+\fl_{IJ}\cdot\mathsf{z} \, ,
\end{equation}
which is the fifth relation.

\subsection{Conformally flat backgrounds}

Let $\log\Omega$ and the sources $(g,a,\sM)$ be in the class of flat sources, and consider the background $\Omega^2\delta_{\mu\nu}$. Along the path $\Omega_t=\Omega^{1-t}$ we let the sources change by the variation \eqref{eq:Delta} with $\sigma=-\log\Omega$, starting from the given sources at $t=0$. The equations for the sources along the path are solved one after the other. The first is $\pr_tg_t=-\log\Omega\;\hB(g_t)$ at each point, the second gives $a_t$ in terms of $g_t$, and the third is linear in $\sM_t$. All coefficients are regular by Lemma~\ref{lem:regvar} and by the previous subsections. The sources $(\check g,\check a,\check\sM)$ at $t=1$ therefore have no poles at $\vep=0$, they are in the class, and they depend analytically on $\vep$. By construction of the variation the bare sources change along the path by their Weyl weights and by a gauge transformation with the parameter $-\log\Omega\;\xi(g_t)$, in which $\xi$ is regular by the choice of the bare sources made after Lemma~\ref{lem:frame}. At each order in the sources the flat bare sources along the path are meromorphic in $\vep$ with poles of bounded order at $\vep=0$, and for fixed $\vep\neq0$ near zero they are smooth in $t$ and $x$ with support in a fixed compact set. By the fourth statement of Lemma~\ref{lem:polarcont} the terms of the bare functional can therefore be differentiated with respect to $t$ term by term at fixed $\vep$ away from the poles, as at the end of appendix~\ref{app:conf}, and the resulting identity extends to an identity of meromorphic functions of $\vep$. Hence by \eqref{eq:defcurved} and by the gauge invariance the bare functional is constant along the path, and
\begin{equation}
W\big[\Omega^2\delta;g,a,\sM\big] = W^{\rm fl}\big[\check g,\check a,\check\sM\big]+\int_0^1dt\;\cA[\log\Omega]\big(\Omega_t^2\delta;g_t,a_t,\sM_t\big) \, .
\label{eq:Wsplit}
\end{equation}
The first term is finite by Hypothesis~\ref{hyp:finite}. The second term is an integral of \eqref{eq:fullanomaly} with regular coefficients over the support of $\log\Omega$, and it is finite. This proves the second part of the proposition. We should point out that here Hypothesis~\ref{hyp:finite} is used for sources which depend on $\vep$.

\subsection{Localisation on the sphere}

The sphere is the background $\Omega=2/(1+x^2)$, for which $\log\Omega$ does not have compact support. The path from flat space to the sphere passes through metrics which are conical at infinity, and the integrated anomaly is not finite term by term along it. We avoid this by a localisation, which we describe in five steps.

\paragraph{Polynomial functional.} The bare functional $W_0^S$ is multilinear in the bare sources at each order, and the bare sources and the counterterms \eqref{eq:X} on the sphere are power series in the sources and their derivatives, with $R=R_d$. The term $W_n$ of order $n$ in the renormalised sources is therefore a homogeneous polynomial of degree $n$ on the linear space of the sources of the class. Note that the order in the bare sources is not the order in the renormalised sources, because the bare mass source contains $\fk(g)R$.

\paragraph{Sources which vanish near a point.} Let $P$ be a point of the equator $X_\perp=0$, and let all three sources vanish in a neighbourhood of $P$. Then the bare sources vanish there. For the bare mass source this uses that $\fk$ vanishes at $g=0$, since $\sM_0$ contains $-Z_\sM\fk(g)R$ and $R$ does not vanish. The density $\cX$ also vanishes there, which uses that $\fa$ and $\fb$ vanish at $g=0$.

\paragraph{Chart and path.} In the stereographic chart from $P$ the sources are smooth with compact support and in the class of flat sources, and \eqref{eq:chart} holds. On the support of the sources $\log\Omega$ is smooth. The sources along the path of the previous subsection have the same support, since the variation of each source vanishes where all sources vanish. For the mass source this uses that $\bar\eta$ and $\bar\tau$ vanish at $g=0$. Hence the sources at the end of the path have compact support, and Hypothesis~\ref{hyp:finite} applies. The relation \eqref{eq:Wsplit} holds with the functional on the sphere on the left. The integrand of the anomaly vanishes outside the support of the sources, because the only structures without a source are those with $p$, $q$ and $r$, which vanish at $g=0$. The integral is over a compact set on which $\Omega$ and the metrics of the path are smooth, and it is finite. Hence $W_n$ is finite on sources which vanish near a point of the equator.

\paragraph{Polarisation.} Let $\cW_n(s_1,\ldots,s_n)$ be the symmetric multilinear form whose value on $n$ equal arguments $s=(g,a,\sM)$ is $W_n(s)$. It is a finite combination of the values of $W_n$ on the sums of subsets of the $s_i$. If all $s_i$ vanish near a common point of the equator, then so do these sums, and the pole part of $\cW_n(s_1,\ldots,s_n)$ vanishes.

\paragraph{Partition of unity.} Let $\chi_\alpha(X_\parallel)$ be a partition of unity of the closed unit ball with supports of small diameter. The sources $\chi_\alpha s$ are in the class, also for the one-form. The equator is the boundary of the ball, which for $D_0=1$ is a circle, and $n$ sets of small diameter do not cover it. We write $W_n(s)=\sum\cW_n(\chi_{\alpha_1}s,\ldots,\chi_{\alpha_n}s)$, and each term of the sum is finite by the previous step. This proves the third part of the proposition. In this argument the functional on the sphere is the one of section~\ref{sec:sphere}. It does not depend on the chart by \eqref{eq:chart}, and conformal maps which are singular inside the support of the sources do not occur.

We add two remarks on the scheme \eqref{eq:Mredef}. First, the functional of this scheme is $W'[g,a,\sM']$, which equals $W[g,a,\sM]$ with $\sM=\sM'-d_ID^2g^I-k_RR$ and with regular $d_I$ and $k_R$. It is finite because the argument on the right is a source of the class with regular coefficients. The anomaly of this scheme is \eqref{eq:fullanomaly} with this $\sM$ inserted, so that its coefficients are the coefficients of minimal subtraction plus products of these with $d_I$ and $k_R$, and they are regular as well. Second, for constant couplings and $\sM'=0$ the bare mass source on the sphere is the constant $-Z_\sM\,(k_R+\fk)\,R$, which has poles and does not vanish away from fixed points. This is the renormalisation of the coupling to the curvature. It is treated as an insertion, in agreement with the definition of $W_0^S$ as a series in the bare sources.

\section{Relation to momentum space}
\label{app:momentum}

In this appendix we describe what Hypothesis~\ref{hyp:finite} and Identification~\ref{id:ms} contain beyond the standard results on renormalisation, and what we know about the relation between the functional of section~\ref{sec:sphere}, which is defined in position space, and Feynman integrals in momentum space. The proofs of Theorem~\ref{gs:thm:phi4} and of Proposition~\ref{gs:prop:Wform} do not use this appendix. The relation concerns Identification~\ref{id:ms}, by which the counterterms of minimal subtraction, which are defined through amplitudes at fixed momenta, renormalise this functional. The proofs do not use it, and we state below what a derivation of it would need.

The two statements together contain three statements which go beyond the standard results on renormalisation. The first two are part of Hypothesis~\ref{hyp:finite}. First, these results concern correlation functions of the fields, and it is noted at the end of section~11.3 of~\cite{Collins:1984xc} that vacuum graphs and composite operators require a more general proof in coordinate space than the one given there. Second, the standard results concern correlation functions at fixed momenta. In our class of sources the momenta of the sources have transverse components, which are integrated with the measure of \eqref{eq:gram}. The hypothesis therefore includes the statement that the renormalised correlation functions can be integrated with this measure uniformly in $\vep$. The sources which depend on $\vep$ occur because the sources in the chart \eqref{eq:chart} contain the factor $\Omega^\vep$. Third, the counterterms of minimal subtraction are defined through amplitudes at fixed momenta, while the functional of section~\ref{sec:sphere} is defined in position space by a continuation in the exponent of the propagator. Identification~\ref{id:ms} is thus the statement that the two regularisations have the same counterterms, while Hypothesis~\ref{hyp:finite} states only that some counterterms of this form renormalise the functional. We expect this, since the exponent of the propagator corresponds to an analytic regularisation in momentum space, as we describe now. We did not show that the two regularisations agree on the line \eqref{eq:line}, and we did not find a result in the literature from which it follows.

The second parameter $\Dp$ of section~\ref{sec:continuation} corresponds to an analytic regularisation of the propagators in momentum space. Let $\varpi$ be the exponent of a propagator in momentum space, and let $\Dp=\tfrac12d-\varpi$. With the heat kernel $K_t(x)=(4\pi t)^{-d/2}e^{-x^2/4t}$, which is the Fourier transform of $e^{-tp^2}$, we have
\begin{align}
\frac{1}{(p^2)^\varpi} &= \frac{1}{\Gamma(\varpi)}\int_0^\infty\!dt\,t^{\varpi-1}e^{-tp^2} \, , \notag\\
\frac{1}{\Gamma(\varpi)}\int_0^\infty\!dt\,t^{\varpi-1}K_t(x) &= \frac{C(d,\varpi)}{|x|^{2\Dp}} \, , \qquad
C(d,\varpi) = \frac{\Gamma\big(\tfrac12d-\varpi\big)}{4^\varpi\,\pi^{d/2}\,\Gamma(\varpi)} \, ,
\label{eq:heat}
\end{align}
and $C(d,1)=C_\phi$. The integral over $t$ converges for $0<{\rm Re}\,\varpi<\tfrac12{\rm Re}\,d$, and the line \eqref{eq:line} is the line $\varpi=1$. For Gaussian sources and vertices without derivatives the integrals over the positions and over the momenta at fixed parameters $t$ are Gaussian integrals in $d$ dimensions in the sense of~\cite{Wilson:1972cf}, and we expect that their value does not depend on the order of the integrations. We therefore expect the following. Consider a region with ${\rm Re}\,\Dp>0$ and sufficiently large ${\rm Re}\,d$, in which the parametric integrals converge. In this region the term of the flat functional of section~\ref{sec:chart} which belongs to a graph is the Feynman integral with the propagators $(C_\phi/C(d,\varpi))/(p^2)^\varpi$, whose normalisation equals one on the line $\varpi=1$. In this integral the momenta of the sources are integrated in $d$ dimensions together with the loop momenta. The family with the two parameters would thus be dimensional regularisation together with an analytic regularisation of the propagators, and the functional would be the amplitude at fixed external momenta, integrated over these momenta with the Fourier transforms of the sources. We do not include a derivation of these statements, and we do not use them.

What we did not show is that this relation continues to the line \eqref{eq:line} near $d=4$. Minimal subtraction is defined by the pole parts of the amplitudes at fixed momenta~\cite{Collins:1984xc}, and the usual results on renormalisation concern the continuation of these amplitudes. For the functional of section~\ref{sec:sphere} the momenta of the sources are integrated before the continuation. For two points the continued amplitude is a power of the momentum which is locally integrable, and we expect no difficulty. For more points there are two obstructions. First, the measure of the external momenta is integrable only for $d-D_0>n-2$, which for $D_0=1$ and $d<4$ fails for five and more points. Second, the continued amplitude is singular at exceptional momenta, and we do not know in general that it is locally integrable there with this measure. The missing statement is therefore an exchange of two operations, the continuation in $d$ and $\varpi$ of the integral over the Schwinger parameters and the integral over the external momenta in $d$ dimensions.

A derivation of the statement on $L^I$ from the finiteness of the amplitudes at fixed momenta would need three inputs which we did not prove. The first is the exchange just described, for all graphs which occur, including those with the vertices of the mass source and of the current. The second is the finiteness of the functional for sources which are not in the class of Hypothesis~\ref{hyp:finite}, such as Gaussian functions with a phase, which are complex and do not have compact support. The third is the finiteness of the amplitudes with the composite operators at fixed momenta, with the same bare mass source and bare gauge field. We did not find a statement in~\cite{Collins:1984xc} from which the statement on $L^I$ follows. The proof of the renormalisation in coordinate space in section~11.3 there concerns correlation functions of the fields.
\section{Four dimensions and comparison with the literature}
\label{gs:app:fourd}

In this appendix we state what Theorem~\ref{gs:thm:phi4} gives in four dimensions, we compare its consequences with the literature, and we show that in four dimensions there is a scheme in which the one-form $W_I$ of the Weyl anomaly vanishes. In the first three subsections $\hat A$ and $\hat G_{IJ}$ are the potential and the metric of Theorem~\ref{gs:thm:phi4}, and we use only this theorem and section~\ref{gs:sec:consequences}. The comparison with the literature also uses Identification~\ref{id:ms}. The gradient formula of Theorem~\ref{gs:thm:phi4} does not contain $W_I$. In the remaining subsections we obtain $W_I$ from the Killing identity of this theory, in which it appears with the factor $2e_d$, and we determine the antisymmetric anomaly coefficient $E_{[IJ]}$ from the relation between the curl of $W_I$ and $E_{[IJ]}$. This is the route of~\cite{Reehorst:phi4}, with the Ward identity on the sphere in place of the Ward identity for the two-point function in flat space. The result is that $W_I$ is the sum of a gradient and of a symmetric tensor contracted with $\hB^I$, Proposition~\ref{gs:prop:Wform}, from which the scheme with vanishing $W_I$ follows at $\vep=0$.

\subsection{Gradient formula in four dimensions}
\label{gs:sec:fourdgrad}

The potential and the metric of Theorem~\ref{gs:thm:phi4} are regular, so that the theorem has a limit at $\vep=0$. Let $\hat A_0$ and $\hat G_{0,IJ}$ be their terms of order zero in $\vep$. Since $\hB^I=-\vep\,g^I+B^I$, the terms of order zero in $\vep$ of \eqref{gs:eq:phi4main} give
\begin{equation}
\pr_I\hat A_0 = \hat G_{0,IJ}\,B^J \, ,
\label{gs:eq:fourd}
\end{equation}
which is the first relation of \eqref{gs:eq:tower}. This is a gradient formula in four dimensions for the vector field $B^I$ of Hypothesis~\ref{hyp:finite} for multiscalar $\phi^4$ theory. The function $\hat A_0$ is invariant, and the tensor $\hat G_{0,IJ}$ is symmetric and covariant. Its term of order zero in the couplings is $\tfrac{1}{96}\pi^2c_\cO\,\delta_{IJ}$, so that it is positive definite to all orders in the couplings. By the proof of Theorem~\ref{gs:thm:phi4} the metric is $-\hat G^{\rm H}_{IJ}/(2e_d)$, with $2e_d=24$ at $\vep=0$, where $\hat G^{\rm H}_{IJ}$ of \eqref{gsh:eq:H} contains the tensor $\cS_{IJ}$ of \eqref{gs:eq:Sdef} through $\cS^\sharp_{IJ}$, that is the first harmonic of the two-point function of the quartic operators on the sphere minus the second derivative of $\cW/V$.

By Identification~\ref{id:ms} the vector field $B^I$ is that of~\cite{Reehorst:phi4}, so that \eqref{gs:eq:fourd} is a second gradient formula for it, which holds under a different hypothesis. The second hypothesis of~\cite{Reehorst:phi4}, on the expansion of massive Feynman integrals at large momenta, is not needed, because the sphere is compact. On the other hand the present proof uses the functional on the sphere and the regularisation of section~\ref{gs:sec:sphere}, and Hypothesis~\ref{hyp:finite} requires the finiteness for sources which depend on $\vep$. The metric of~\cite{Reehorst:phi4} contains the two-point function of the marginal operators in flat space paired with a test function, while $\hat G_{0,IJ}$ contains the two-point function on the sphere. We return to their relation in section~\ref{sec:discussion}.

\subsection{Value at fixed points}
\label{gs:sec:fourdfixed}

At a zero of $\hB^I$ the potential is $c(\vep)(\tilde F-N\tilde F_s)$ by \eqref{gs:eq:phi4F}, where $\tilde F$ is defined with the couplings of the curvature terms at the zeros of their beta functions and the mass source at the zero of its flow, as in~\cite{Fei:2015oha}, see Proposition~\ref{gs:prop:radius}. The factor $c(\vep)=\vep\,d/(2e_dV\sin(\pi\vep/2))$ is positive and does not depend on the couplings. Its value $1/(8\pi^3)$ at $\vep=0$ follows from $2e_d=24$ and $V=8\pi^2/3$. Since $\hat A$ is invariant, its derivative along the beta function $\hb^I=\hB^I+(\xi g)^I$ is
\begin{equation}
\hb^I\pr_I\hat A = \hB^I\pr_I\hat A = \hat G_{IJ}\,\hB^I\hB^J \geq 0
\label{eq:monotonic}
\end{equation}
for real couplings, in the sense of formal power series of section~\ref{gs:sec:fixedgen}. A zero of $\hb^I$ of order $\vep$ is therefore a zero of $\hB^I$, see appendix~\ref{app:schemes}. By section~\ref{gs:sec:fixedgen} the eigenvalues of the matrix $\pr_J\hB^K$ at a fixed point are real, and $\tilde F_{\rm UV}-\tilde F_{\rm IR}$ is a power series in $\vep$ with a positive leading term, of order $\vep^3$, for two fixed points which are power series in $\vep$ and whose terms of order $\vep$ are connected by a trajectory of the one-loop flow. The comparison with the values of $\tilde F$ in the literature, which are computed at the zeros of the beta function of minimal subtraction in momentum space, relies on Identification~\ref{id:ms}.

At leading order the value at fixed points is fixed by the term of order zero of the metric. We use the normalisation in which $g^Ig^I=\lambda_{ijkl}\lambda_{ijkl}$ and the one-loop beta function is $\beta^{(1)}_{ijkl}=(\pr T_3/\pr\lambda_{ijkl})/(16\pi^2)$ with $T_3=\lambda_{ijkl}\lambda_{klmn}\lambda_{mnij}$. Let $\mathcal{N}=\tfrac{1}{96}\pi^2c_\cO=3/(1769472\,\pi^6)$ be the term of order zero of the metric at $\vep=0$. The terms of \eqref{gs:eq:phi4main} of first and second order in the couplings give
\begin{equation}
\hat A = -\tfrac12\,\vep\,\mathcal{N}\,\lambda_{ijkl}\lambda_{ijkl}+\mathcal{N}\,\frac{T_3}{16\pi^2}+\ldots \, ,
\label{gs:eq:Alead}
\end{equation}
where the omitted terms are of fourth order in the couplings or contain a further factor $\vep$. At a fixed point $\lambda_*$ of order $\vep$ the one-loop relation $\vep\,\lambda_*=\beta^{(1)}(\lambda_*)$ gives $T_3(\lambda_*)/(16\pi^2)=\tfrac13\vep\,\lambda_{*ijkl}\lambda_{*ijkl}$, so that $\hat A=-\tfrac16\vep\,\mathcal{N}\,\lambda_{*ijkl}\lambda_{*ijkl}$ up to terms of order $\vep^4$. For the ${\rm O}(N)$ model the fixed point is $\lambda_{*ijkl}=\tfrac13\lambda_*(\delta_{ij}\delta_{kl}+\delta_{ik}\delta_{jl}+\delta_{il}\delta_{jk})$ with $\lambda_*=48\pi^2\vep/(N+8)$, and with $c(0)=1/(8\pi^3)$ we find
\begin{equation}
\tilde F-N\tilde F_s = -\frac{\pi\,N(N+2)}{576\,(N+8)^2}\,\vep^3+{\rm O}(\vep^4) \, ,
\end{equation}
which is the known leading term~\cite{Giombi:2014xxa}.

The leading term does not distinguish $\cW_*$ of \eqref{gs:eq:fixed} from $\cW(g_*)$, which differ at order $\vep^3$. We therefore also compared the term of order $\vep^4$ of $\tilde F$ for the ${\rm O}(N)$ model. For this comparison we took the renormalised free energy through third order in the coupling, the fixed point and the coefficient of fourth order of the beta function of the Euler coupling from~\cite{Fei:2015oha}. With the notation $F=-\log Z_{S^d}$ of~\cite{Giombi:2014xxa} the renormalised free energy at the fixed point gives the term $-N(N+2)(13N^2+424N+1840)\,\vep^3/(3456\,(N+8)^4)$ of $F-F_{\rm free}$, and the Euler coupling at the zero of its beta function adds $N(N+2)(3N+14)\,\vep^3/(192\,(N+8)^4)$, which is the contribution stated in the footnote to eq.~(4.13) of~\cite{Giombi:2014xxa}. The sum reproduces the term of order $\vep^4$ of $\tilde F$ in eq.~(1.6) of~\cite{Giombi:2014xxa,Fei:2015oha}, while without the Euler term the result differs from it by $-\pi N(N+2)(3N+14)\,\vep^4/(384\,(N+8)^4)$. This confirms that the value at the fixed point is $\cW_*$ and not $\cW(g_*)$.

In~\cite{Pannell:2025sph} the gradient equation is solved order by order with an ansatz. The free coefficients of the solution are chosen there such that the potential at fixed points agrees with $\tilde F$ through order $\vep^5$, and such that the metric agrees at leading order with the two-point function of the quartic operators on the sphere, after a rescaling of the operators. Given Identification~\ref{id:ms}, Theorem~\ref{gs:thm:phi4} explains the first observation, since it provides a regular solution whose potential equals $c(\vep)(\tilde F-N\tilde F_s)$ at the zeros of $\hB^I$ to all orders. Let $A^{\rm PS}=-\tfrac12\vep\,\lambda_{ijkl}\lambda_{ijkl}+\ldots$ be the potential of~\cite{Pannell:2025sph} in the coupling $\lambda/(16\pi^2)$. Through total degree three, where $\vep$ and the couplings count with the degree one, it is $(16\pi^2)^{-2}\mathcal{N}^{-1}$ times \eqref{gs:eq:Alead}, and $8\pi^3(16\pi^2)^2\mathcal{N}=\pi/288$. At leading order we therefore find $\tilde F-N\tilde F_s=(\pi/288)\,A^{\rm PS}$ at fixed points, which is the normalisation found in~\cite{Pannell:2025sph} by matching. The second observation is consistent with the metric of Theorem~\ref{gs:thm:phi4}, which contains the first harmonic of the two-point function on the sphere through $\cS_{IJ}$, see section~\ref{gs:sec:fourdgrad}. We did not attempt a quantitative comparison of the metrics, which requires the map between minimal subtraction at unit radius and the conventions of~\cite{Pannell:2025sph}.

\subsection{Relations at five loops}
\label{app:fiveloop}

In this subsection $\hat A$ and $\hat G_{IJ}$ are the potential and the metric of Theorem~\ref{gs:thm:phi4}. To see how the relations \eqref{gs:eq:tower}, with $k_I=1$, constrain the beta function, let $A^{[n]}$ be the part of $\hat A$ with $n$ couplings, $G^{[n]}$ the part of $\hat G_{IJ}$ with $n$ couplings and $B^{(\ell)}$ the term of $B^I$ at $\ell$ loops, which has $\ell+1$ couplings. The terms of \eqref{gs:eq:phi4main} with $\ell+1$ couplings are
\begin{equation}
\pr A^{[\ell+2]} = \sum_{n=0}^{\ell-1}G^{[n]}B^{(\ell-n)}-\vep\,G^{[\ell]}g \, .
\label{eq:level}
\end{equation}
The tensor $G^{[\ell]}$ appears here with a factor $\vep$, and at the next order it appears without this factor in the term $G^{[\ell]}B^{(1)}$. The part of \eqref{eq:level} of first order in $\vep$ therefore constrains the tensor $G^{[\ell]}$ at $\vep=0$, which the equation in four dimensions relates to the beta function at $\ell+1$ loops. In this way the requirement of regularity gives relations for the coefficients of the beta function which are not required in four dimensions.

We computed the relations which follow from the tower \eqref{gs:eq:tower} through five loops. As in~\cite{Pannell:2024sia} we represent the tensor structures of $\hat A$, $\hat G_{IJ}$ and $\beta^I$ by graphs, and we solve \eqref{eq:level} for coefficients which are power series in $\vep$. We replace the values of the zeta function by random residues modulo a prime and do the linear algebra modulo this prime. In this way we find in four dimensions no relation through three loops, 4 relations at four loops and 37 at five loops, the numbers of~\cite{Pannell:2024sia}. Four relations at four loops were found earlier in~\cite{Jack:2018oec}. In $d=4-\vep$ we find in the same way four additional relations among the coefficients of the beta function at five loops and none at lower orders. In this computation the four relations which are given in the ancillary files of~\cite{Pannell:2024sia} span the same space modulo the relations in four dimensions.

The coefficients of the beta function at five loops in modified minimal subtraction~\cite{Bednyakov:2021ojn}, which we took from the ancillary files of~\cite{Pannell:2024sia}, satisfy the four relations exactly. With these coefficients the tower has a regular solution through five loops in the computation modulo a prime. One of the four relations contains the coefficient $209/1536$ of a term with a correction of the propagator, which in our computation enters none of the relations in four dimensions. If we change this coefficient by $10^{-3}$, the solution in four dimensions remains and the regular solution in $d=4-\vep$ is lost. At five loops the flow is generated by $B^I$ and not by $\beta^I$, with the four five-loop coefficients of the field rotation which were computed in the appendix on consistency checks of~\cite{Reehorst:phi4}. We find that the relations do not depend on these coefficients.

Let us explain why Theorem~\ref{gs:thm:phi4} implies the relations which follow from the tower. The theorem states that a regular symmetric solution of the gradient equation exists, and in the computation above this constrains the beta function beyond the relations in four dimensions first at five loops. The implication uses several steps. After the average over the group of field rotations the functions $\hat A$ and $\hat G_{IJ}$ are covariant power series in the couplings, so that their coefficients are linear combinations of graphs. For $N$ large enough the tensors which belong to different graphs with a given number of vertices are linearly independent. Since the theorem holds for every $N$, while the coefficients of the beta function do not depend on $N$, the gradient equation is then equivalent to the linear system for the coefficients of the graphs which we solve. The tower is homogeneous in $\hat A$ and $\hat G_{IJ}$, and in the computation the term of order zero in the couplings of the metric is the unit matrix for all $\vep$. The metric of Theorem~\ref{gs:thm:phi4} takes this form after division by a power series in $\vep$ with a nonzero constant term, since at $g=0$ it is a multiple of the unit matrix for all $\vep$ by the two-point form of the free theory. Graphs which are not connected decouple, since the beta function is connected. Graphs with contractions of two indices of the same coupling do not decouple in this way. We included them in the computation, in which they do not change the relations. The implication also uses Identification~\ref{id:ms}, by which $L^I$ are the usual counterterms of minimal subtraction, for which the beta function is the one of~\cite{Bednyakov:2021ojn} with a symmetric anomalous dimension matrix of the fields, and that the beta function is the same in modified minimal subtraction. We did not derive the 32 relations at six loops of~\cite{Pannell:2024sia}.
\subsection{Relations among the coefficients}
\label{sec:relations}

In this and the next three subsections we derive the form of $W_I$ which Proposition~\ref{gs:prop:Wform} states. The anomaly coefficients are determined by the counterterms through \eqref{eq:anomalydef}. The Weyl variation of \eqref{eq:X} is computed in appendix~\ref{app:weyl}. For example,
\begin{equation}
p = -\left(\vep-\hB\!\cdot\!\pr\right)\fa \, , \qquad
W_I = 4(d-3)\,\pr_I\fa-\fg_{IJ}\,\hB^J \, , \qquad
G_{IJ} = \big[(\vep-\tL)\,\fg\big]_{IJ} \, ,
\label{eq:pWG}
\end{equation}
where $\hB\!\cdot\!\pr=\hB^I\pr_I$ and $\tL$ is the extended Lie derivative of~\cite{Jack:2013sha} along $\hB^I$. On a covariant tensor it is the Lie derivative $\cL_{\hB}$ in which the matrix $\pr_I\hB^K$ is replaced by
\begin{equation}
\hat M^K{}_I \equiv \pr_I\hB^K+(\hr_I\,g)^K \, ,
\label{eq:Mhat}
\end{equation}
so that $(\tL W)_I=\hB^J\pr_JW_I+\hat M^J{}_IW_J$ for a one-form. The expressions for the other coefficients are given in \eqref{eq:lin} and \eqref{eq:quad}, up to the contributions of the counterterms which are linear in the mass source, see appendix~\ref{app:anomcoeff}. Since the coefficients on the left-hand sides are finite, while the counterterms are pure poles, these expressions imply relations among the finite coefficients. We use three of them.\footnote{We also verified \eqref{eq:euler}, \eqref{eq:curl}, the identities \eqref{eq:localid} and \eqref{eq:removal} with computer algebra, for two and three couplings.}

The first follows directly from \eqref{eq:pWG}. Since $\tL\hB=0$ by \eqref{eq:rhoB} and $\tL$ commutes with the derivative of an invariant function, we have $[(\vep-\tL)W]_I=-4(d-3)\,\pr_Ip-G_{IJ}\hB^J$, that is
\begin{equation}
(d-3)\,\pr_IA_{\rm E} = G_{IJ}\,\hB^J+\big[(\vep-\tL)\,W\big]_I \, .
\label{eq:euler}
\end{equation}
For $\vep=0$ this is the relation $\pr_IA_{\rm E}=G_{IJ}B^J-(\tL W)_I$ of~\cite{Jack:2013sha}, from which the gradient formula with the tensor $T_{IJ}$ follows. In $d=4-\vep$ it contains the additional term $\vep\,W_I$. Without field rotations a relation of this form is derived in section~3 of~\cite{Jack:1990eb}, see appendix~\ref{app:conventions}. In appendix~\ref{app:finite} the relation is derived in a general scheme for the mass source as \eqref{gs:eq:eulerF}, which gives the regularity of $p$.

The second relation concerns the curl of $W_I$. Let
\begin{equation}
\hP_I{}^K \equiv \tfrac12(d-2)\,\delta_I{}^K+\pr_I\hB^K+\tfrac12\,(\hr_I\,g)^K \, .
\label{eq:Psi}
\end{equation}
For $\vep=0$ and $\hB\to B$ this is the matrix $\Psi_I{}^K$ of~\cite{Jack:2013sha,Reehorst:phi4}. Then
\begin{equation}
(d-2)\left(\pr_IW_J-\pr_JW_I\right)
= -\left(\hP_I{}^KE_{JK}-\hP_J{}^KE_{IK}\right)+\left(R_{IJK}-R_{JIK}\right)\hB^K+\hr_I\cdot Q_J-\hr_J\cdot Q_I \, .
\label{eq:curl}
\end{equation}
This form holds in a scheme for the mass source with $\theta_I=0$, which we adopt in section~\ref{sec:massterms} before we use it. The general form, with a further term which contains $\theta_I$, is the identity \eqref{eq:curlgeneral}, which we derive in appendix~\ref{app:anomcoeff}. For $\vep=0$ it is the relation of~\cite{Jack:2013sha} between the curl of $W_I$ and the flat-space coefficients, which is one of the Weyl consistency conditions used in~\cite{Reehorst:phi4}. The third relation holds for every element $\omega$ of the Lie algebra,
\begin{equation}
(\omega g)^IW_I = -\frac{2}{d-2}\;\omega\cdot Q_I\,\hB^I \, .
\label{eq:WQ}
\end{equation}
It follows from \eqref{eq:pWG}, from the expression \eqref{eq:Qdef} for $Q_I$, from the invariance of $\fa$ under field rotations, from the antisymmetry of $\fp_{IJ}$ and from \eqref{eq:rhoB}.

\subsection{Killing identity}
\label{sec:ward}

In this subsection we take one derivative of the Ward identity \eqref{eq:wardS} of section~\ref{gs:sec:msward} with respect to the couplings at constant couplings, which gives the Killing identity of Proposition~\ref{gs:prop:K} for this theory, \eqref{eq:sphere}. Rewriting its local terms and adding the terms which are due to the mass operator we obtain \eqref{eq:Wresult} and \eqref{eq:Wshort}.

\subsubsection{First derivative at constant couplings}

We now vary \eqref{eq:wardS} to first order around constant couplings and $a_\mu=\sM=0$, with a variation $\delta g^J$ in the class. We first ignore the mass source, to which we return in section~\ref{sec:massterms}. Let $\cW(g)$ be the value of $W$ on the sphere for constant couplings, and let $V=\vol_{d+1}$ be the volume of the sphere, so that the one-point function is $\vev{\cO_I}=\pr_I\cW/V$. It is constant by Property~\ref{gs:P:sphere}, which holds for this theory by Proposition~\ref{gs:prop:phi4prop}.

On the left-hand side the term with the gauge field does not contribute. Its variation contains the one-point function of the current, which vanishes at constant couplings, or the factor $D_\mu g^I$, which vanishes as well. The term with the couplings gives three contributions. The variation of $v^\mu\pr_\mu g^I$ gives $-\int d\mu_d\,v^\mu\pr_\mu\delta g^J\,\vev{\cO_J}$, which after integration by parts and with $\nabla_\mu v^\mu=d\,\sigma$ equals $d\int d\mu_d\,\sigma\,\delta g^J\,\vev{\cO_J}$. The variation of $\hB^I$ gives $\int d\mu_d\,\sigma\,\delta g^J\,\pr_J\hB^I\,\vev{\cO_I}$. The variation of the one-point function gives the two-point function $W_{IJ}$ integrated with $\sigma\hB^I$, which by \eqref{eq:Lambda} is $\Lambda_{JI}\hB^I$ times $\int d\mu_d\,\sigma\,\delta g^J$. Each term is thus a constant times $\int d\mu_d\,\sigma\,\delta g^J$.

On the right-hand side we need the first variation of \eqref{eq:anomaly} on the unit sphere. There
\begin{equation}
R = R_d \, , \qquad \cG^{\mu\nu} = -\frac{e_d}{d}\,\gamma^{\mu\nu} \, , \qquad \cE_4 = \chi_d \, ,
\end{equation}
with the constants
\begin{equation}
R_d = d(d-1) \, , \qquad e_d = \tfrac12\,d(d-1)(d-2) \, , \qquad \chi_d = d(d-1)(d-2)(d-3) \, .
\end{equation}
With $\nabla^2\sigma=-d\,\sigma$ we find, after integration by parts, that the terms of \eqref{eq:anomaly} which are linear in the variation give the right-hand side of \eqref{eq:sphere} below, multiplied by $\int d\mu_d\,\sigma\,\delta g^J$. For instance, the term with $W_I$ gives $2\,W_J\,\cG^{\mu\nu}\nabla_\mu\nabla_\nu\sigma=2e_d\,W_J\,\sigma$. The terms of second and higher order in $\pr g$ do not contribute. Since the integral $\int d\mu_d\,\sigma\,\delta g^J$ does not vanish for a suitable variation, for instance for $\delta g^J$ proportional to $\sigma$, we obtain
\begin{equation}
\frac1V\left(d\,\delta_J{}^I+\pr_J\hB^I\right)\pr_I\cW+\Lambda_{JI}\,\hB^I
= 2e_d\,W_J-d\,R_d\left(e_J-h_J\right)+d^2u_J+\pr_J\!\left(\chi_d\,p+R_d^2\,q-d\,R_d\,r\right) .
\label{eq:sphere}
\end{equation}
The left-hand side contains the one-point function and the first harmonic of the two-point function on the sphere. The right-hand side is local. Note that the field rotation enters \eqref{eq:sphere} only through $\hB^I$. Since $\cW$ is invariant under field rotations, we have $\hB^I\pr_I\cW=\hb^I\pr_I\cW$.

\subsubsection{Expression for \texorpdfstring{$W_I$}{W}}
\label{sec:Wexpr}

The identity \eqref{eq:sphere} can be solved for $W_J$. The terms with $e_J$, $h_J$ and $u_J$ on its right-hand side can be expressed through the coefficients of the terms of the anomaly with two factors of $\pr g$. From the expressions \eqref{eq:lin} and \eqref{eq:quad} of these coefficients in terms of the counterterms we find the identity
\begin{equation}
d\,R_d\left(e_J-h_J\right)-d^2u_J = \frac{d^2}{2}\Big[\pr_J\big(\hB^Iu_I\big)
-\hB^I\left(\pr_Iu_J+\pr_Ju_I\right)+\left(S_{JK}-E_{JK}+2\,\nu_{JK}\right)\hB^K\Big] \, ,
\label{eq:localid}
\end{equation}
which holds for every $d$. In it the terms with $\fc_I$ which contain neither $\hB^I$ nor its derivatives cancel because $d+\vep=4$, and the terms with $\hr_I$ cancel between $e_J$ and $h_J$ and between $S_{JK}$ and $E_{JK}$. Using also $\pr_J\hB^I\,\pr_I\cW=\pr_J(\hB^I\pr_I\cW)-\hB^I\pr_I\pr_J\cW$ we arrive at
\begin{align}
2e_d\,W_J &= \pr_J\Phi+\mathsf{h}_{JK}\,\hB^K-\frac{d^2}{2}\,E_{[JK]}\,\hB^K \, , \notag\\
\Phi &= \frac1V\left(d+\hB\!\cdot\!\pr\right)\cW+\frac{d^2}{2}\,\hB^Iu_I
-\chi_d\,p-R_d^2\,q+d\,R_d\,r \, , \notag\\
\mathsf{h}_{JK} &= \Lambda_{JK}-\frac1V\,\pr_J\pr_K\cW
-\frac{d^2}{2}\left(\pr_Ju_K+\pr_Ku_J-S_{JK}+E_{(JK)}-2\,\nu_{JK}\right) .
\label{eq:Wresult}
\end{align}
The function $\Phi$ and the symmetric tensor $\mathsf{h}_{JK}$ are regular, because the free energy and the two-point function on the sphere and the anomaly coefficients are. Note that $\Lambda_{JK}-\pr_J\pr_K\cW/V$ is minus the integral of the two-point function with the weight $1-\cos\theta$, which is one half of the square of the chordal distance.

The result \eqref{eq:Wresult} shows that $W_J$ is the sum of a gradient, of a symmetric tensor contracted with $\hB^K$, and of a term with the antisymmetric part of $E_{JK}$, which is the coefficient whose form was the subject of~\cite{Reehorst:phi4}. Here we do not need a second Ward identity to determine it. As we show in section~\ref{sec:form}, the relation \eqref{eq:curl} between the curl of $W_I$ and $E_{[IJ]}$, together with \eqref{eq:Wresult}, is sufficient.

\subsubsection{Mass operator}
\label{sec:massterms}

We now add the terms which are due to the mass source. In minimal subtraction the coefficients $\theta_I$ and $\bar\tau$ of \eqref{eq:deltaM} do not vanish. As in~\cite{Reehorst:phi4}, where $\theta_I$ is removed in this way, we remove them by a redefinition of the mass source by local terms with regular coefficients, which is given in \eqref{eq:Mredef}. The redefined functional is finite, and all relations of section~\ref{sec:relations} hold for it. The redefinition does not change $\hB^I$, $\hr_I$, and the coefficients $A_{\rm E}$, $G_{IJ}$ and $W_I$ at vanishing mass source. It is therefore sufficient to derive the form of $W_I$ in this scheme, which we use from here on. In it $\theta_I=0$, so that \eqref{eq:curl} holds. The details are given in appendix~\ref{app:masscoeff}.

\begin{lemma}
\label{lem:mass}
In the scheme \eqref{eq:Mredef} for the mass source the result \eqref{eq:Wresult} is replaced by
\begin{equation}
2e_d\,W_J = \pr_J\Phi'+\mathsf{h}'_{JK}\hB^K-\frac{d^2}{2}\,E_{[JK]}\hB^K+\mathsf{Y}_{[JK]}\hB^K \, ,
\label{eq:Wshort}
\end{equation}
where $\Phi'$ is a regular function which agrees with $\Phi$ at the zeros of $\hB^I$, $\mathsf{h}'_{JK}$ is a regular symmetric tensor, and $\mathsf{Y}_{JK}$ is a regular matrix which satisfies $\hP_I{}^K\mathsf{Y}_{KJ}=-\mathsf{y}_{IJ}$ with a regular symmetric tensor $\mathsf{y}_{IJ}$. Here the coefficients $e_I$, $q$ and $S_{IJ}$ of \eqref{eq:sphere} and \eqref{eq:Wresult} include the contributions of the counterterms which are linear in $\sM$.
\end{lemma}

We prove the lemma in appendix~\ref{app:massward}, where $\Phi'$, $\mathsf{h}'_{JK}$ and $\mathsf{Y}_{JK}$ are given in \eqref{eq:WresultM} and \eqref{eq:Ydef}. The new terms contain the one-point function of the mass operator on the sphere and the first harmonic of its two-point function with the marginal operators. The matrix $\mathsf{Y}_{JK}$ is not symmetric, and its antisymmetric part is treated in the next section together with $E_{[JK]}$.

\subsection{Form of the antisymmetric coefficient}
\label{sec:form}

In this subsection we determine the form of $E_{[IJ]}$ and of the antisymmetric part of $\mathsf{Y}_{IJ}$. We first recall the removable tensors of~\cite{Reehorst:phi4}. They are not used in section~\ref{gs:sec:relevant}, and we state them for couplings of one weight $k$, so that $\hB^I=-\vep\,k\,g^I+B^I$ with a regular covariant vector field $B^I$ which does not depend on $\vep$ and has no terms of order zero and one in the couplings. For $\phi^4$ theory $k=1$. We assign the weight one to the couplings and to $\vep$. By Lemma~\ref{lem:mass} the matrix $\hP_I{}^K\mathsf{Y}_{KJ}$ is symmetric, so that $\mathsf{Y}_{[IJ]}$ is in the set $\cT$ of removable tensors by Lemma~\ref{gs:lem:Y} below. We then derive a linear equation for $E_{[IJ]}$ from \eqref{eq:Wshort} and \eqref{eq:curl}, and we solve it by iteration, together with the equation for $\mathsf{Y}_{[IJ]}$.

\subsubsection{Removable terms}
\label{sec:removable}

Following the treatment of removable terms in~\cite{Reehorst:phi4} we consider antisymmetric tensors on the space of couplings of the form
\begin{equation}
\Xi_{IJ} = \tau_{IJK}\,\hB^K+\hr_{[I}\cdot X_{J]}+\left(M^TC-CM\right)_{IJ} \, ,
\label{eq:types}
\end{equation}
where $M^K{}_I=\pr_IB^K$ is the matrix of derivatives of $B^I$, called the stability matrix in~\cite{Reehorst:phi4}, $(M^TC)_{IJ}=M^K{}_IC_{KJ}$ and $(CM)_{IJ}=C_{IK}M^K{}_J$. The coefficients are a regular tensor $\tau_{IJK}$ which is antisymmetric in its first two indices, a regular one-form $X_J$ with values in the Lie algebra, and a regular symmetric tensor $C_{IJ}$. Square and round brackets around a pair of indices denote the antisymmetric and the symmetric part. The three terms are the first, the second and the third type of~\cite{Reehorst:phi4}, in this order, with $\hB^K$ in place of $B^K$ in the first type. In the terminology of~\cite{Reehorst:phi4} tensors of these three types are removable. Since all couplings have the weight $k$, the third term does not change if $M^K{}_I$ is replaced by $\pr_I\hB^K=M^K{}_I-\vep\,k\,\delta_I{}^K$. We denote by $\cT$ the set of tensors of the form \eqref{eq:types}. We call $(\tau,X,C)$ a set of representatives of $\Xi_{IJ}$, which is not unique.

The identities of~\cite{Reehorst:phi4} for the three types hold with $\hB^I$ in place of $B^I$, since they use only the product rule and the relation $\hr_I\hB^I=0$. For the first type we have $\tau_{JKL}\hB^L\hB^K=(\tau_{JLK}+\tau_{KLJ})\hB^L\hB^K$, because the second term on the right vanishes by antisymmetry. For the second type $\hr_{[J}\cdot X_{K]}\hB^K=\hr_{(J}\cdot X_{K)}\hB^K$, because $\hr_K\hB^K=0$. For the third type the identity is
\begin{equation}
\left(M^TC-CM\right)_{JK}\hB^K = \pr_J\!\left(\hB^KC_{KL}\hB^L\right)
-\left[\pr_J\!\left(C_{KL}\hB^L\right)+\pr_K\!\left(C_{JL}\hB^L\right)-\hB^L\pr_LC_{JK}\right]\hB^K \, .
\end{equation}
Together they give, for every tensor of the form \eqref{eq:types},
\begin{align}
\Xi_{JK}\,\hB^K &= \pr_J\!\left(\hB^KC_{KL}\hB^L\right)+\Sigma_{JK}\,\hB^K \, , \notag\\
\Sigma_{JK} &= \left(\tau_{JLK}+\tau_{KLJ}\right)\hB^L+\hr_{(J}\cdot X_{K)}
-\left[\pr_J\!\left(C_{KL}\hB^L\right)+\pr_K\!\left(C_{JL}\hB^L\right)-\hB^L\pr_LC_{JK}\right] .
\label{eq:removal}
\end{align}
The tensor $\Sigma_{JK}$ is symmetric and regular. Thus the contraction of a tensor in $\cT$ with $\hB^K$ is the sum of a gradient and of a symmetric tensor contracted with $\hB^K$.

We need four properties of the set $\cT$, which we collect in a lemma. For a one-form $\omega_J$ we write $[\mathrm{d}\omega]_{IJ}=\pr_I\omega_J-\pr_J\omega_I$ for its curl, and we use the matrix
\begin{equation}
P_I{}^K \equiv \pr_I\hB^K+\tfrac12\,(\hr_I\,g)^K \, ,
\end{equation}
so that $\hP_I{}^K=\tfrac12(d-2)\,\delta_I{}^K+P_I{}^K$ in \eqref{gs:eq:Mrel}.

\begin{lemma}
\label{lem:closure}
Let $\mathsf{s}_{JK}$ be a regular symmetric tensor, and let $\Xi_{IJ}$ be in $\cT$. Then the following four tensors are in $\cT$. The first is the curl of the one-form $\mathsf{s}_{JK}\hB^K$, the second is the curl of the one-form $\Xi_{JK}\hB^K$, the third is $P_I{}^K\mathsf{s}_{JK}-P_J{}^K\mathsf{s}_{IK}$, and the fourth is $P_I{}^K\Xi_{JK}-P_J{}^K\Xi_{IK}$.
\end{lemma}

For the first tensor we have
\begin{equation}
\pr_I\!\left(\mathsf{s}_{JK}\hB^K\right)-\pr_J\!\left(\mathsf{s}_{IK}\hB^K\right)
= \left(M^T\mathsf{s}-\mathsf{s}M\right)_{IJ}+\left(\pr_I\mathsf{s}_{JK}-\pr_J\mathsf{s}_{IK}\right)\hB^K \, .
\label{eq:curls}
\end{equation}
The right-hand side is of the form \eqref{eq:types}. The second statement follows from \eqref{eq:removal} and the first, because the curl of a gradient vanishes. The third and the fourth statement are shown in appendix~\ref{app:closure} for the weight $k=1$. For a weight $k$ the factor $2\,\vep$ which appears there is $2\,\vep\,k$, and the rest of the proof is unchanged. The relation \eqref{eq:rhoB} which it uses is $\hr_I\hB^I=0$ of Property~\ref{gs:P:weyl}.

\subsubsection{Antisymmetric part of a matrix}
\label{gs:sec:Ylemma}

\begin{lemma}
\label{gs:lem:Y}
Let $d_c\neq2$, and let $\mathsf{Y}_{IJ}$ be a regular matrix such that $\hP_I{}^K\mathsf{Y}_{KJ}$ is symmetric. Then $\mathsf{Y}_{[IJ]}$ is in $\cT$. It has regular representatives which are linear in the symmetric part $\mathsf{Y}_{(IJ)}$ and do not contain its derivatives.
\end{lemma}

\begin{proof}
The antisymmetric part of $\hP_I{}^K\mathsf{Y}_{KJ}$ vanishes, which gives
\begin{equation}
(d-2)\,\mathsf{Y}_{[IJ]} = -\left(P_I{}^K\mathsf{Y}_{(KJ)}-P_J{}^K\mathsf{Y}_{(KI)}\right)
+\left(P_I{}^K\mathsf{Y}_{[JK]}-P_J{}^K\mathsf{Y}_{[IK]}\right) .
\label{gs:eq:Yeq}
\end{equation}
The matrix $P_I{}^K$ has no term of weight zero, since $\pr_I\hB^K=-\vep\,k\,\delta_I{}^K+\pr_IB^K$, where $\pr_IB^K$ has no term of order zero in the couplings, and since $(\hr_I\,g)^K$ is of first order in the couplings. The second term on the right of \eqref{gs:eq:Yeq} therefore raises the weight by at least one. Since $d_c\neq2$, the number $d-2=d_c-2-\vep$ is invertible as a power series in $\vep$. The equation therefore determines $\mathsf{Y}_{[IJ]}$ uniquely from the symmetric part, order by order in the weight. The first term on the right is in $\cT$ by the third statement of Lemma~\ref{lem:closure}, and by appendix~\ref{app:closure} it is minus the tensor \eqref{eq:types} with $\tau_{IJK}=0$, $C_{IJ}=\mathsf{Y}_{(IJ)}$ and $X_J=(g\,\mathsf{Y}_{(\cdot\cdot)})_J$, where $\omega\cdot(g\,\mathsf{Y}_{(\cdot\cdot)})_J=(\omega g)^K\mathsf{Y}_{(KJ)}$. Each step of the iteration stays in $\cT$ by the fourth statement. The representatives which a step produces are linear in those of the previous step, and they are built with $\pr_I\hB^K$, $\hr_I$ and their derivatives, without derivatives of the representatives. Each step multiplies the representatives by $\pr_I\hB^K$ or by $(\hr_I\,g)^K$, which have no term of weight zero, or produces a term of the first type with a factor $\hB^K$, which has weight at least two. The weight of the tensor therefore grows by at least one in each step, and the representatives converge as formal series in the weight.
\end{proof}
\subsubsection{Equation for the antisymmetric coefficient}

The only terms in \eqref{eq:Wshort} which are not a gradient or a symmetric tensor contracted with $\hB^K$ are those with $E_{[JK]}$ and with $\mathsf{Y}_{[JK]}$. We take the curl of \eqref{eq:Wshort} and compare with \eqref{eq:curl}. In \eqref{eq:curl} the part of $\hP_I{}^K$ which is proportional to the unit matrix gives $(d-2)\,E_{[IJ]}$. With $2e_d/(d-2)=R_d$ we obtain
\begin{align}
2e_d\,E_{[IJ]} ={}& \big[\mathrm{d}(\mathsf{h}'\hB)\big]_{IJ}-\frac{d^2}{2}\,\big[\mathrm{d}(E_{[\,\cdot\,]}\hB)\big]_{IJ}+\big[\mathrm{d}(\mathsf{Y}_{[\,\cdot\,]}\hB)\big]_{IJ}
+R_d\left(P_I{}^KE_{JK}-P_J{}^KE_{IK}\right) \notag\\
&-2\,R_d\,R_{[IJ]K}\hB^K-2\,R_d\,\hr_{[I}\cdot Q_{J]} \, ,
\label{eq:Eeq}
\end{align}
where $(\mathsf{h}'\hB)_J=\mathsf{h}'_{JK}\hB^K$, and $E_{[\,\cdot\,]}\hB$ and $\mathsf{Y}_{[\,\cdot\,]}\hB$ denote the one-forms $E_{[JK]}\hB^K$ and $\mathsf{Y}_{[JK]}\hB^K$. This is a linear equation for $E_{[IJ]}$, in which $E_{[IJ]}$ appears on the right-hand side in the second and in the fourth term. The fourth term also contains the symmetric part of $E_{IJ}$.

The antisymmetric part of $\mathsf{Y}_{IJ}$ satisfies an equation of the same kind. The antisymmetric part of the relation $\hP_I{}^K\mathsf{Y}_{KJ}=-\mathsf{y}_{IJ}$ of Lemma~\ref{lem:mass} reads
\begin{equation}
(d-2)\,\mathsf{Y}_{[IJ]} = -\left(P_I{}^K\mathsf{Y}_{(KJ)}-P_J{}^K\mathsf{Y}_{(KI)}\right)
+\left(P_I{}^K\mathsf{Y}_{[JK]}-P_J{}^K\mathsf{Y}_{[IK]}\right) .
\label{eq:Yeq}
\end{equation}

\subsubsection{Solution by iteration}
\label{sec:iteration}

We solve \eqref{eq:Yeq} and \eqref{eq:Eeq} by iteration in the weight of section~\ref{sec:regular}. The matrix $P_I{}^K$ has weight at least one, since $\pr_I\hB^K=-\vep\,\delta_I{}^K+\pr_IB^K$ and $\hr_I$ has no term of order zero. The terms on the right-hand side of \eqref{eq:Yeq} which contain $\mathsf{Y}_{[IJ]}$ therefore raise the weight by at least one, and the equation determines $\mathsf{Y}_{[IJ]}$ order by order from the symmetric part. By the third and the fourth statement of Lemma~\ref{lem:closure} each step of the iteration stays in $\cT$. In \eqref{eq:Eeq} the terms with $E_{[IJ]}$ on the right-hand side also raise the weight by at least one, since $\hB^K$ has weight two and the curl lowers the weight by one. The constant $2e_d$ is $24$ at $\vep=0$ and is invertible as a series in $\vep$. The equation therefore has a unique regular solution, which is obtained by iteration. The terms without $E_{[IJ]}$ are in $\cT$, as we show in appendix~\ref{app:closure}, and each step of the iteration stays in $\cT$ by the second and the fourth statement of the lemma. The representatives converge as formal series in the weight as well, see appendix~\ref{app:closure}. We thus obtain the following.

\begin{proposition}
\label{prop:form}
The antisymmetric part of the anomaly coefficient $E_{IJ}$ and the antisymmetric part of the matrix $\mathsf{Y}_{IJ}$ of Lemma~\ref{lem:mass} are of the form
\begin{align}
E_{[IJ]} &= \tau_{IJK}\,\hB^K+\hr_{[I}\cdot X_{J]}+\left(M^TC-CM\right)_{IJ} \, , \notag\\
\mathsf{Y}_{[IJ]} &= \tau^{\mathsf{Y}}_{IJK}\,\hB^K+\hr_{[I}\cdot X^{\mathsf{Y}}_{J]}+\left(M^TC^{\mathsf{Y}}-C^{\mathsf{Y}}M\right)_{IJ} \, ,
\label{eq:form}
\end{align}
with regular coefficients. The tensors $\tau_{IJK}$ and $\tau^{\mathsf{Y}}_{IJK}$ are antisymmetric in their first two indices, and $C_{IJ}$ and $C^{\mathsf{Y}}_{IJ}$ are symmetric.
\end{proposition}

For $E_{[IJ]}$ this is the form which~\cite{Reehorst:phi4} obtains from the Ward identity for the two-point function in flat space, see the section on the form of the antisymmetric coefficient there. Here it follows from the Ward identity on the sphere and from the relation \eqref{eq:curl}. As in~\cite{Reehorst:phi4} the content of the proposition is the regularity of the coefficients. At a point where $\hB^I$ does not vanish every antisymmetric tensor is of the first type with a coefficient which is singular at the zeros of $\hB^I$.

\subsection{One-form of the Weyl anomaly}
\label{sec:theorem}

We now combine the results of the previous sections. By Proposition~\ref{prop:form} and \eqref{eq:removal} the two terms of \eqref{eq:Wshort} with antisymmetric tensors are sums of a gradient and of a symmetric tensor contracted with $\hB^K$,
\begin{equation}
E_{[JK]}\hB^K = \pr_J\!\left(\hB^KC_{KL}\hB^L\right)+\Sigma_{JK}\hB^K \, , \qquad
\mathsf{Y}_{[JK]}\hB^K = \pr_J\!\left(\hB^KC^{\mathsf{Y}}_{KL}\hB^L\right)+\Sigma^{\mathsf{Y}}_{JK}\hB^K \, ,
\end{equation}
where $\Sigma_{JK}$ and $\Sigma^{\mathsf{Y}}_{JK}$ are given by \eqref{eq:removal} with the representatives of \eqref{eq:form}. With \eqref{eq:Wshort} this gives
\begin{align}
W_J &= \cH_{JK}\,\hB^K+\pr_Jw \, , \notag\\
\cH_{JK} &= \frac{1}{2e_d}\left(\mathsf{h}'_{JK}-\frac{d^2}{2}\,\Sigma_{JK}+\Sigma^{\mathsf{Y}}_{JK}\right) , \qquad
w = \frac{1}{2e_d}\left(\Phi'-\frac{d^2}{2}\,\hB^KC_{KL}\hB^L+\hB^KC^{\mathsf{Y}}_{KL}\hB^L\right) ,
\label{eq:Wform}
\end{align}
with a regular symmetric tensor $\cH_{JK}$ and a regular function $w$. Note that the representatives in \eqref{eq:form} are not unique, so that $\cH_{JK}$ and $w$ depend on their choice. The iteration of section~\ref{sec:iteration} provides one choice, and what follows holds for every choice of regular representatives. A change of the representatives changes $w$ by terms which are quadratic in $\hB^I$. The one-form $W_J$ of the Weyl anomaly is therefore the sum of a symmetric tensor contracted with $\hB^K$ and of a gradient. We state the result in Proposition~\ref{gs:prop:Wform}.

\begin{proposition}
\label{gs:prop:Wform}
Assume Hypothesis~\ref{hyp:finite} for multiscalar $\phi^4$ theory, and consider this theory in minimal subtraction with the curvature counterterms of Proposition~\ref{prop:finite}. Then the one-form $W_I$ of the Weyl anomaly is
\begin{equation}
W_J = \cH_{JK}\,\hB^K+\pr_Jw
\label{gs:eq:Wformprop}
\end{equation}
with a regular symmetric covariant tensor $\cH_{JK}$ and a regular invariant function $w$.
\end{proposition}

\begin{proof}
By Proposition~\ref{gs:prop:phi4prop} the functional has the properties which we use in this appendix, also in the scheme \eqref{eq:Mredef} for the mass source. We may work in this scheme, since the redefinition does not change $W_I$, see appendix~\ref{app:masscoeff}. There the Killing identity \eqref{eq:sphere}, Lemma~\ref{lem:mass}, the relation \eqref{eq:curl} and Proposition~\ref{prop:form} give \eqref{eq:Wform}, in which $\cH_{JK}$ is symmetric. The tensor $\cH_{JK}$ and the function $w$ are regular by Proposition~\ref{prop:form} and by the regularity of the quantities in \eqref{eq:WresultM}, which is shown in appendix~\ref{app:finite}. The one-form $W_J$ and the vector field $\hB^K$ are covariant under the group of field rotations, which is compact. If \eqref{gs:eq:Wformprop} holds with a pair $\cH_{JK}$, $w$, it therefore holds with the pair which is obtained by a rotation, and we can replace the pair by the average of these pairs over the group. The averaged tensor is regular, symmetric and covariant, and the averaged function is regular and invariant.
\end{proof}

\subsection{Scheme with vanishing \texorpdfstring{$W_I$}{W}}
\label{app:Wzero}

For $\vep=0$ Proposition~\ref{gs:prop:Wform} is a statement about the Weyl anomaly in four dimensions. In minimal subtraction $W_I$ does not depend on $\vep$, see appendix~\ref{app:anomcoeff}, so that $W_I=\cH_{IJ}B^J+\pr_Iw$ with the values of $\cH_{IJ}$ and $w$ at $\vep=0$. The coefficients of the anomaly are defined up to the addition of finite local terms to $W$. By \eqref{eq:lin} the addition of $\int d^dx\sqrt\gamma\,(\mathsf{a}\,\cE_4+\mathsf{g}_{IJ}\,\cG^{\mu\nu}D_\mu g^ID_\nu g^J)$ with a regular invariant function $\mathsf{a}$ and a regular symmetric covariant tensor $\mathsf{g}_{IJ}$ changes $W_I$ at $\vep=0$ by $4\,\pr_I\mathsf{a}-\mathsf{g}_{IJ}B^J$. In the absence of field rotations this is the freedom described in~\cite{Jack:2013sha}, with $\beta^J$ in place of $B^J$. At $\vep=0$ we consider the tensor
\begin{equation}
T_{IJ} = G_{IJ}+\pr_IW_J-\pr_JW_I+2\,\hr_{[I}\cdot Q_{J]} \, ,
\label{gs:eq:Tdef}
\end{equation}
which in the absence of field rotations is the tensor $T_{IJ}$ of~\cite{Jack:2013sha} in our normalisation.

\begin{corollary}
\label{gs:cor:Wzero}
Assume Hypothesis~\ref{hyp:finite} for multiscalar $\phi^4$ theory, and let $\vep=0$. The addition of the local term above with $\mathsf{a}=-w/4$ and $\mathsf{g}_{IJ}=\cH_{IJ}$ to the functional of minimal subtraction gives a scheme in which $W_I=0$. In this scheme $\pr_IA_{\rm E}=G_{IJ}B^J$, and $\pr_I(A_{\rm E}+W_JB^J)=T_{IJ}B^J$ holds with $T_{IJ}=G_{IJ}+2\,\hr_{[I}\cdot Q_{J]}$, so that $T_{[IJ]}=2\,\hr_{[I}\cdot Q_{J]}$. In particular $T_{IJ}$ is symmetric in the absence of field rotations.
\end{corollary}

\begin{proof}
By Proposition~\ref{gs:prop:Wform} and the change of $W_I$ which we just described, $W_I$ vanishes in the new scheme. The relations \eqref{eq:euler} and \eqref{eq:WQ} follow from the expressions \eqref{eq:lin}, \eqref{eq:quad} and \eqref{eq:Qdef} of the coefficients in terms of the counterterms, which hold for counterterms with finite parts as well, so that they hold in the new scheme. At $\vep=0$ the relation \eqref{eq:euler} reads $\pr_IA_{\rm E}=G_{IJ}B^J-(\tL W)_I$, which gives the first statement. For the second statement we write $(\tL W)_I=\pr_I(W_JB^J)+B^J(\pr_JW_I-\pr_IW_J)+(\hr_Ig)^JW_J$. By \eqref{eq:WQ} with $\omega=\hr_I$ and $d=4$ the last term is $-\hr_I\cdot Q_JB^J$, which equals $-2\,\hr_{[I}\cdot Q_{J]}B^J$ because $\hr_JB^J=0$. Hence we find $\pr_I(A_{\rm E}+W_JB^J)=T_{IJ}B^J$ with \eqref{gs:eq:Tdef} in every scheme, and with $W_I=0$ we obtain the statement. In the absence of field rotations $\hr_I=0$.
\end{proof}

That a scheme with a symmetric $T_{IJ}$ exists was conjectured in~\cite{Poole:2019kcm}, and the existence of a scheme with $W_I=0$ was left open in the second remark after Theorem~1 of~\cite{Reehorst:phi4}. The form \eqref{gs:eq:Wformprop} of $W_I$ is the first of the two equations of the sufficient condition for a symmetric $T_{IJ}$ which is given in eq.~(2.38) of~\cite{Poole:2019kcm}. The second is a corresponding form of $Q_I$, which we do not show. Accordingly we find an antisymmetric term with $Q_I$ when there are field rotations, which is of the second type of \eqref{eq:types}.

\section{Steps of the derivation for other theories}
\label{gs:app:steps}

The proofs of the properties for the Yukawa, the sextic and the cubic theory in appendices~\ref{gs:app:fermions}, \ref{gs:app:sextic} and~\ref{gs:app:cubic} follow the derivation for multiscalar $\phi^4$ theory in section~\ref{gs:sec:phi4prop} and appendices~\ref{app:weyl}, \ref{app:unique} and~\ref{app:finite}. Table~\ref{gs:tab:steps} lists the steps of that derivation and states for each theory how each step carries over. The adapted and the new steps are explained in the appendices on these theories.

\begin{table}[p]
\centering
\renewcommand{\arraystretch}{1.05}
\setlength{\tabcolsep}{3pt}
\footnotesize
\caption{The steps of the derivation of the properties for multiscalar $\phi^4$ theory and their status for scalar fields and Dirac fermions with Yukawa and quartic couplings in $d=4-\vep$, for sextic scalars in $d=3-\vep$ and for cubic scalars in $d=6-\vep$, each under Hypothesis~\ref{hyp:finite} for the theory of the column, for the cubic theory on the backgrounds $\Omega^2\delta_{\mu\nu}$ and for the class of flat sources with $D_0=3$. Each entry gives the status of the step and what changes. The status means that the statement and the proof for $\phi^4$ theory apply as written (unchanged), that the statement holds with the changes listed and its proof changes accordingly (adapted), that a different statement or argument is given at the place named (new), that the statement is part of Hypothesis~\ref{hyp:finite} for the theory (in the hypothesis), or that the step is not carried out and not used (not derived). A dash means that the step is not listed separately for the theory.}
\label{gs:tab:steps}\label{gf:tab:steps}\label{gsx:tab:steps}\label{gcp:tab:steps}
\begin{tabular}{>{\raggedright\arraybackslash}p{3.1cm}>{\raggedright\arraybackslash}p{4.6cm}>{\raggedright\arraybackslash}p{3.4cm}>{\raggedright\arraybackslash}p{3.4cm}}
\hline
step of the derivation for $\phi^4$ theory & Yukawa theory (app.~\ref{gs:app:fermions}) & sextic theory (app.~\ref{gs:app:sextic}) & cubic theory (app.~\ref{gs:app:cubic}) \\
\hline
bare functional, covariance, gauge invariance, rotations, chart (Prop.~\ref{gs:prop:bare}, sec.~\ref{gs:sec:sphere}, apps.~\ref{app:mero}--\ref{app:gauge}) & \textit{new}: section~\ref{gf:sec:fermions}, Prop.~\ref{gf:prop:bare} & \textit{unchanged}: vertices $V^{(6)}$, $V^{(4)}$, $V^{(2)}$; weights $2\vep$, $1+\vep$, $2$ & \textit{unchanged}: $D_0=3$ \\
anomaly from the counterterms, Ward identity (Props.~\ref{gs:prop:msanomaly}, \ref{gs:prop:msward}) & \textit{unchanged}: invariances from Prop.~\ref{gf:prop:bare} & \textit{unchanged}: weight $4-d$ of $\sU_0$; variations \eqref{gsx:eq:varU}, \eqref{gsx:eq:varM} & \textit{unchanged} \\
list of bare sources and counterterms (eqs.~\eqref{eq:baresources}, \eqref{eq:X}) & -- & \textit{new}: Lemma~\ref{gsx:lem:forms}, eqs.~\eqref{gsx:eq:bare}, \eqref{gsx:eq:X} & -- \\
Weyl variation of the bare gauge field, freedom of the bare sources (Lemmas~\ref{lem:rhoxi}, \ref{lem:frame}) & \textit{adapted}: weights $k_I$ in $\hb^{\rm L}$, $\beta^I$, $\xi$, $\hr_I$, eq.~\eqref{gf:eq:zeroparts}; induction in the ordinary degree with the weighted operator $\mathsf{E}$ & \textit{adapted}: weight $k_I=2$, eq.~\eqref{gf:eq:zeroparts} & \textit{adapted}: weight $k_I=\tfrac12$ \\
variation of the relevant sources (eqs.~\eqref{eq:masscoeff}, \eqref{eq:gammaM}) & -- & \textit{adapted}: $\gamma_\sU$ and $\bar\chi$ added, eqs.~\eqref{gsx:eq:varU}--\eqref{gsx:eq:chi} & -- \\
commutation of the variations, relations \eqref{eq:Mrel} (Lemma~\ref{gcu:lem:commute}) & \textit{unchanged} & \textit{unchanged} & \textit{unchanged} \\
local structures (Lemma~\ref{lem:indep}) & \textit{unchanged}: structures do not depend on the operators & \textit{new}: Lemma~\ref{gsx:lem:indep}, by positivity only & \textit{new}: Lemma~\ref{gcp:lem:gram}, by invariant theory on the class of flat sources with $D_0=3$ \\
potentials (Lemmas~\ref{lem:potential}, \ref{lem:transverse}) & \textit{unchanged}: both marginal kernels are $|x-y|^{-8}$ & \textit{adapted}: $\mathbb{R}^3$, $\kappa=3,2,1$, Legendre polynomials & \textit{adapted}: $\kappa=6,4$ (Lemma~\ref{gcu:lem:potential}); for $\kappa=2$ Lemma~\ref{gcp:lem:exact} \\
terms of second order (Lemma~\ref{lem:lowest}) & \textit{adapted}: matrix kernel $c_{IJ}(d)|x-y|^{-2\Delta_I}$; fermion part of the current kernel; mixed terms vanish & \textit{adapted}: kernel of $\sU$; $Z_\sU=1$ at $g=0$ & \textit{adapted}: Lemma~\ref{gcu:lem:lowest}, kernels with $\kappa=6,4,2$ \\
uniqueness of $L^I$ (Prop.~\ref{prop:unique}); uniqueness and covariance of the other coefficients (Prop.~\ref{prop:uniqueall}) & Prop.~\ref{prop:unique} \textit{unchanged}; Prop.~\ref{prop:uniqueall} \textit{new}: uniqueness up to stabiliser transformations; covariance from uniqueness and the compact connected group $G_0$, reflections by averaging; $N_I$ in $\mathfrak{g}$ and $N_I^{(0)}=0$ by a choice (Prop.~\ref{gc:prop:cov}) & \textit{adapted}: steps for $Z_\sU$ and $\fh$; stabiliser of sextic tensors & \textit{new}: unique up to field shifts; covariant set, Prop.~\ref{gcp:prop:cov} \\
regular Weyl variation (Lemma~\ref{lem:regvar}) & \textit{adapted}: first order $-\vep k_Ig^I$; positive Yukawa block of the kernel & \textit{adapted}: steps for $\gamma_\sU$ and $\bar\chi$ & \textit{adapted}: couplings, gauge field as for $\phi^4$; $h$, $\sM$ by field shifts, Prop.~\ref{gcp:prop:reg} \\
\hline
\end{tabular}
\end{table}

\begin{table}[p]
\centering
\renewcommand{\arraystretch}{1.05}
\setlength{\tabcolsep}{3pt}
\footnotesize
\begin{tabular}{>{\raggedright\arraybackslash}p{3.1cm}>{\raggedright\arraybackslash}p{4.6cm}>{\raggedright\arraybackslash}p{3.4cm}>{\raggedright\arraybackslash}p{3.4cm}}
\multicolumn{4}{l}{Table~\ref{gs:tab:steps}, continued} \\
\hline
step of the derivation for $\phi^4$ theory & Yukawa theory (app.~\ref{gs:app:fermions}) & sextic theory (app.~\ref{gs:app:sextic}) & cubic theory (app.~\ref{gs:app:cubic}) \\
\hline
structures and coefficients of the anomaly (eqs.~\eqref{eq:fullanomaly}--\eqref{eq:flatcoeff}) & \textit{adapted}: identities for covariant coefficients unchanged; the residue formulas for $A_{\rm E}$, $G_{IJ}$, $W_I$ and the order zero of $\hP$ are weighted, see appendix~\ref{gf:sec:steps} & \textit{new}: seven coefficients, Lemma~\ref{gsx:lem:anomaly} & \textit{new}: Lemma~\ref{gcp:lem:gram}, on conformally flat backgrounds \\
curvature counterterms and relations (eqs.~\eqref{eq:steps}, \eqref{eq:finrel}, \eqref{gs:eq:eulerF}, \eqref{eq:curlgeneral}) & \textit{unchanged}: identities, factors $d-1$, $d-2$, $d-3$ at $d_c=4$ & \textit{new}: two curvature counterterms, $\fk$ and $\fn$; one relation, \eqref{gsx:eq:irel}; no closedness condition & \textit{in the hypothesis}: finiteness on $\Omega^2\delta_{\mu\nu}$ \\
regular anomaly coefficients (Prop.~\ref{prop:finite}) & -- & -- & \textit{new}: finiteness and Lemma~\ref{gcp:lem:gram}, on conformally flat backgrounds \\
conformally flat backgrounds (eq.~\eqref{eq:Wsplit}), localisation on the sphere (app.~\ref{app:finite}) & \textit{unchanged}: Weyl map of section~\ref{gf:sec:chart}; chart of Lemma~\ref{gf:lem:chart} & \textit{unchanged}: every counterterm contains $\sU$ & localisation \textit{unchanged}: normalisation of $\cX$ at $g=0$ \\
rotations, constant sources, mass operator on the sphere (apps.~\ref{app:rotations}, \ref{app:massward}, Lemma~\ref{gr4:lem:const}) & \textit{unchanged}: section~\ref{gf:sec:chart}; one-point function of the fermion current vanishes; scalar mass source only & constant sources \textit{unchanged}: two scalar sources & constant sources \textit{unchanged}: constant $h$ and $\sM$, proof of Prop.~\ref{gcp:prop:props} \\
coefficients at $g=0$ (Lemma~\ref{gr4:lem:zero}) & \textit{unchanged}: uses Lemmas~\ref{lem:lowest}, \ref{lem:regvar} as adapted & \textit{adapted}: Lemma~\ref{gsx:lem:zero} & \textit{new}: Prop.~\ref{gcu:prop:first} \\
two-point form (Lemma~\ref{gs:lem:free}), positivity of $c_{IJ}$ & \textit{adapted}: Lemma~\ref{gf:lem:free}, Yukawa block $C_\phi C_\psi^2\,{\rm tr}\,1\,{\rm tr}(YY')$; positivity \textit{unchanged}, Lemma~\ref{gro:lem:c} & \textit{unchanged}: $n=6$ & \textit{unchanged}: $n=3$, with the frame condition \eqref{gcu:eq:frame} \\
Property~\ref{gs:P:free} (Prop.~\ref{gr4:prop:lowest}) & \textit{unchanged}: two weights; $c_{IJ}=0$ for $k_I\neq k_J$ & \textit{adapted}: odd $d_c$, anomaly linear in $\sM$ vanishes at $\sU=0$ & \textit{new}: frame of Prop.~\ref{gcf:prop:frame}, Prop.~\ref{gcu:prop:first} \\
identification with minimal subtraction in momentum space & \textit{not derived}: as Identification~\ref{id:ms}; not used & \textit{not derived}: as Identification~\ref{id:ms}; not used & -- \\
\hline
\end{tabular}
\end{table}

\section{Dirac fermions on the sphere}
\label{gs:app:fermions}

In this appendix we prove the statements of section~\ref{gs:sec:dirac} on the bare functional with Dirac fermions, and we list the steps of the derivation of the properties for Yukawa and quartic couplings from Hypothesis~\ref{hyp:finite} for the Yukawa theory, which section~\ref{gs:sec:yukawa} uses.

\subsection{Meromorphy and the line of the free theory}
\label{gf:sec:line}

By section~\ref{gf:sec:loops} each term of the functional is a finite sum of integrals \eqref{eq:gram} of integrands of the form of Proposition~\ref{prop:mero}, with coefficients which are polynomials in $d$ times powers of $C_\phi$ and $C_\psi$. That proposition and Lemmas~\ref{lem:region}, \ref{lem:polarcont}, \ref{lem:poleset} and~\ref{lem:dense} therefore apply without change. The integrals are meromorphic in $d$ and $\Dp$, with poles on the candidate lines \eqref{eq:candidates}, and they converge absolutely in the region of Lemma~\ref{lem:region}, in which a fermion line has $\mathsf{z}_{ij}=\Dp+1$. The polynomial numerators with the sources are smooth functions of the $X_{i\parallel}$ times polynomials in the $w_i$, so that Lemma~\ref{lem:polarcont} holds with the norms \eqref{eq:norm} of the components of the sources.

The exclusion of the line \eqref{eq:line} as a line of poles uses the representation by Schwinger parameters of appendix~\ref{app:line}. The proof of Proposition~\ref{prop:linecurrent} uses the integrand only through the form \eqref{eq:flatmarked} of its terms in the chart, and we list the properties of this form which it uses. The exponents are $\mathsf{z}_{ij}=\Dp\,n_{ij}+r_{ij}$ with $r_{ij}\leq n_{ij}$, and $c_i=d-\Dp\,n_i-r_i+\kappa_i$ with integers $\kappa_i$. The numerator is a product of linear factors of the kind of Lemma~\ref{lem:schwinger}, and it contains the difference $x_j-x_i$ of each marked line exactly once. Finally the bounds \eqref{eq:kappabound} hold, where we use the second bound only in the weaker form $\mathsf{n}_i\leq\mu_i+3\,n_i$, which follows from it with $r_i\leq n_i$. Indeed, the condition $\kappa_i\geq\tfrac12\mathsf{n}_i$ of Lemma~\ref{lem:schwinger} then holds for $h\geq\tfrac32n_{\max}$, and in \eqref{eq:Bbound} we have $\mathsf{b}\geq h|\mathsf{S}|+\mu(\mathsf{S})-r(\mathsf{S})-\tfrac12\mathsf{n}(\mathsf{S})\geq(h-\tfrac52n_{\max})|\mathsf{S}|$.

\begin{lemma}[Fermion lines as marked lines]
\label{gf:lem:count}
Let the sources be of the form \eqref{eq:hsource}. After the integration by parts \eqref{eq:ibp} at the vertices of the scalar current, each term of the functional is in the chart from $P$ a finite sum of integrals \eqref{eq:flatmarked}, with coefficients which are polynomials in $d$ and $\Dp$ times powers of $C_\phi$ and $C_\psi$, in which the fermion lines and the lines chosen at the scalar current vertices are the marked lines, and with
\begin{equation}
\kappa_i \geq h+\mu_i \, , \qquad \mathsf{n}_i\leq\mu_i+3\,n_i \, .
\label{gf:eq:count}
\end{equation}
\end{lemma}

\begin{proof}
The scalar lines and the vertices of the scalars are treated as in section~\ref{app:currents}. A vertex of the fermion current has no derivative and needs no integration by parts. For a fermion loop we use Lemma~\ref{gf:lem:chart}. The denominator of a fermion line gives $|x_i-x_j|^{-2\Dp-2}$ and $(\Omega_i\Omega_j)^{-\Dp-1}$, so that the line has $\mathsf{z}_{ij}=\Dp+1$ and contributes $-\Dp-1$ to the exponent of $\Omega_i$, which is the contribution $-\Dp-r$ of one marked line. The flat trace on the right-hand side of \eqref{gf:eq:chartloop} is, by \eqref{gf:eq:trace}, a sum of products of scalar products over all $d$ components of the chords $x_{i\,i+1}$ and of the vectors $a^{\rm fl}(x_i)$, in which each chord occurs exactly once. The chord of a fermion line is the difference of the marked line. With $X^A=\Omega\,x^A$ for $A\neq P$ and $X^P=1-\Omega$ we have
\begin{equation}
a^{\rm fl}_\mu = \Omega\,a_\mu+\Omega^2\,x_\mu\,\big(a_P-a_Ax^A\big) \, ,
\label{gf:eq:afl}
\end{equation}
where $a_A$ with $A\neq P$ is regarded as a vector of the chart. The components $a_A$ are sources of the form \eqref{eq:hsource}. Hence $a^{\rm fl}$ is a sum of terms with at least one factor $\Omega$ and at most two linear factors, namely $x_\mu$ in a scalar product over all $d$ components and a component $x^A$ in a parallel direction, which are the two types of factors of Lemma~\ref{lem:schwinger}, and a term with two linear factors has the factor $\Omega^2$. At a Yukawa vertex the factor $\Omega_i$ of \eqref{gf:eq:chartloop} raises $\kappa_i$ by one, and at a current vertex $a^{\rm fl}$ raises it by at least one. Hence $\kappa_i\geq h+\mu_i$. At a point with fermion lines there are at least two lines, and the linear factors are at most the two chords, two factors of $a^{\rm fl}$ and three factors for a marked scalar line, so that $\mathsf{n}_i-\mu_i\leq 2+3=5\leq3\,n_i$ at a Yukawa vertex, which has three lines, and $\mathsf{n}_i-\mu_i\leq4\leq3\,n_i$ at a current vertex. At the other points \eqref{eq:kappabound} holds as in section~\ref{app:currents}.
\end{proof}

\begin{proposition}
\label{gf:prop:line}
For scalar fields and Dirac fermions with the vertices of section~\ref{gf:sec:vertices} and all sources of the class, each term of the bare functional is a meromorphic function of $d$ and $\Dp$, the line $\Dp=\tfrac12(d-2)$ is not a line of poles, and the integrals of which a term is a sum, with the normalisations $C_\phi$ and $C_\psi$ of the propagators taken out, have no line of poles of fixed dimension.
\end{proposition}

\begin{proof}
Meromorphy is Proposition~\ref{prop:mero}, as explained above. By Lemma~\ref{gf:lem:count} the integrals with the sources \eqref{eq:hsource} have the form which the proof of Proposition~\ref{prop:linecurrent} uses, with the bounds stated before the lemma. The representation by Schwinger parameters, the pairings \eqref{eq:forestpair}, the sectors, Lemma~\ref{lem:greedy2} and the analysis of the large and the small parameters apply without change, since they use only these properties. The reduction from the sources \eqref{eq:hsource} to all sources of the class is that of section~\ref{app:currents}, by the covering of the ball, Lemma~\ref{lem:dense} and the third statement of Lemma~\ref{lem:polarcont}.
\end{proof}

On the line \eqref{eq:line} each term of the functional is therefore a meromorphic function of $d$ by Lemma~\ref{lem:poleset}, and the bounds of section~\ref{app:polar} for the value on the line hold as stated there.

\subsection{Conformal covariance}
\label{gf:sec:cov}

\begin{proposition}
\label{gf:prop:cov}
With the source $\Upsilon$ of the term of second order in the scalar gauge field kept independent, the terms $W_n$ of the bare functional satisfy \eqref{eq:cov3} with the additional entry $\Omega^{d-3\Dp-1}\,Y\!\circ\!f$ for the Yukawa couplings, for all $d$ and $\Dp$. The one-form $a$ has the weight $d-2-2\Dp$ also at the vertices of the fermion current. On the line \eqref{eq:line} the Yukawa couplings have the weight $\tfrac12(4-d)$, which is $\tfrac12\vep$ at $d_c=4$, and away from the poles the functional satisfies \eqref{eq:covS} with this weight.
\end{proposition}

\begin{proof}
Let $f$ be the map \eqref{eq:boost} along $\hat b$, and let $P=\hat b$. In the chart from $P$ the map $f$ is the dilatation $x\mapsto\lambda x$ with $\lambda=e^{\zeta}$, see section~\ref{app:rotations}, and $\Omega_f(X)=\lambda\,\Omega(\lambda x)/\Omega(x)$ is the factor of \eqref{eq:boost}, since $s'^2_{ij}=\Omega(\lambda x_i)\Omega(\lambda x_j)\,\lambda^2|x_{ij}|^2$. The flat trace in \eqref{gf:eq:chartloop} is homogeneous of degree $n$ in the chords and linear in each $a^{\rm fl}$, and the pullback of $a$ by the dilatation has the flat components $\lambda\,a^{\rm fl}(\lambda x)$. Lemma~\ref{gf:lem:chart} applied at the points $fX_i$ and at the points $X_i$ therefore gives
\begin{equation}
{\rm tr}\big[\cdots\big](fX;\,a) = \prod_{i\notin\mathsf{C}}\Omega_f(X_i)\;{\rm tr}\big[\cdots\big](X;\,f^*a) \, ,
\label{gf:eq:loopcov}
\end{equation}
for points which differ from $\pm\hat b$, and by continuity for all points. In the trace on the right-hand side the Yukawa sources are $Y\!\circ\!f$. With $s'^2_{ij}=\Omega_{f,i}\Omega_{f,j}\,s^2_{ij}$ the factor \eqref{gf:eq:loop} of a loop is multiplied by $\prod_i\Omega_{f,i}^{-2\Dp-2}$ times $\Omega_{f,i}$ at each Yukawa vertex. With the factor $\Omega_{f,i}^{\,d}$ of the measure and the factor $\Omega_{f,i}^{-\Dp}$ of the scalar leg, the Yukawa coupling receives the weight $d-3\Dp-1$, and the gauge field at a vertex of the fermion current the weight $d-2-2\Dp$, as at the scalar current vertex in section~\ref{app:conf}. The tangent vector of $f^*a$ at $X$ is $P_XJ_f^TA(fX)$, which is \eqref{gf:eq:tangent} for the pulled-back one-form of the class. The change of variables in the region of convergence and the continuation in $d$ and $\Dp$ are those of section~\ref{app:conf}, and so is the restriction to the line with $\Upsilon=a_{0A}^T\Pi_{AB}\,a_{0B}$. The first-order form \eqref{eq:barevar} follows as at the end of section~\ref{app:conf}.\footnote{We verified \eqref{gf:eq:loopcov} for the boost with $\cosh\zeta=\tfrac54$ in exact rational arithmetic for random loops with two to six vertices of both kinds.}
\end{proof}

\subsection{Background gauge invariance}
\label{gf:sec:gauge}

Let $\omega$ be a smooth function of $X_\parallel$ with values in the Lie algebra of the group of field rotations. It acts on the scalars by $\omega_s$ and on the fermions by $\omega_f$. The gauge variation of the sources is \eqref{eq:gaugevar} for the scalar sources and the scalar part of the gauge field, together with $\delta_\omega a=d\omega+[a,\omega]$ for the fermion part and $\delta_\omega Y^a=[Y^a,\omega_f]-\omega_{s,ab}\,Y^b$. Let $R_\omega V$ be the vertex obtained from $V$ by the replacement of $\phi$ by $-\omega_s\phi$, of $\psi$ by $-\omega_f\psi$ and of $\bar\psi$ by $\bar\psi\,\omega_f$ in each leg in turn. Then, at each point and as polynomials in the fields,
\begin{equation}
\delta_\omega V+R_\omega V = V^J[d\omega_s]+V^{\psi J}[d\omega_f]
\label{gf:eq:pointwise}
\end{equation}
for the sum $V$ of the vertices, which extends \eqref{eq:pointwise}. As in \eqref{eq:TN} the functional is invariant if
\begin{equation}
T_N = \big\langle V^J[d\omega_s]+V^{\psi J}[d\omega_f],V_1,\ldots,V_N\big\rangle^c+\sum_{r=1}^N\big\langle V_1,\ldots,R_\omega V_r,\ldots,V_N\big\rangle^c
\end{equation}
vanishes on the line. The first term with $V^J[d\omega_s]$ and the part of the second in which $R_\omega$ acts on a scalar leg are those of appendix~\ref{app:gauge}, with fermion loops in the remaining factors, which do not touch the point $X_0$. We now treat the vertex of the fermion current. Its point $X_0$ lies on a fermion loop, between the line from the point $X_h$ of the preceding vertex and the line to the point $X_j$ of the following vertex. We write the trace of the loop as ${\rm tr}[\Gamma\!\cdot\!X_{h0}\,\Gamma\!\cdot\!\nabla\omega_f\,\Gamma\!\cdot\!X_{0j}\,R]$, where $R$ contains the remaining chords and vertex matrices and $\nabla\omega_f$ is the tangent vector \eqref{gf:eq:tangent} of $d\omega_f$, with the flavour matrix of $\omega_f$ at the position of the vertex.

We define $\mathcal{B}^\psi(d,\Dp,\Dp')$ as the sum over the pairings of the integrals in which the factor of the loop at $X_0$,
\begin{equation}
C_\psi^2\;{\rm tr}\big[\Gamma\!\cdot\!X_{h0}\,\Gamma\!\cdot\!\nabla\omega_f\,\Gamma\!\cdot\!X_{0j}\,R\big]\,(s_{h0}^2)^{-\Dp-1}(s_{0j}^2)^{-\Dp-1} \, ,
\end{equation}
is replaced by
\begin{equation}
C_\psi^2\,\Big\{{\rm tr}\big[\Gamma\!\cdot\!X_{h0}\;\omega_f\;\Gamma\!\cdot\!X_0\,\Gamma\!\cdot\!X_j\,R\big]\,(s_{h0}^2)^{-\Dp-1}(s_{0j}^2)^{-\Dp'}
-{\rm tr}\big[\omega_f\;\Gamma\!\cdot\!X_{0j}\,R\big]\,(s_{h0}^2)^{-\Dp'}(s_{0j}^2)^{-\Dp-1}\Big\} \, ,
\label{gf:eq:Bpsi}
\end{equation}
with $\omega_f$ at $X_0$. In the first term the line to $X_j$ has lost its chord and carries the factor $\Gamma\!\cdot\!X_0\,\Gamma\!\cdot\!X_j$, which tends to $1$ for $X_0\to X_j$, and in the second term the line from $X_h$ has lost its chord. By Proposition~\ref{prop:mero} with independent exponents, $\mathcal{B}^\psi$ is meromorphic in the three variables.

\begin{lemma}
\label{gf:lem:ward2}
For all vertices of section~\ref{gf:sec:vertices}, and as meromorphic functions of $d$ and $\Dp$, the part of the function $\vev{V^{\psi J}[d\omega_f],V_1,\ldots,V_N}^c$ in which $X_0$ lies on a given loop equals $-(d-2-2\Dp)\,\mathcal{B}^\psi(d,\Dp,\Dp+1)$.
\end{lemma}

\begin{proof}
It is sufficient to consider ${\rm Re}\,\Dp<-2$ and ${\rm Re}\,m>N$, and we use the coordinates \eqref{eq:last} with $X_0=(Y,\rho\,e)$, where $\rho^2=1-Y^2$ and $e$ is a unit vector orthogonal to the $w_r$. The vector $\nabla\omega_f$ is $\sum_A\pr_A\omega_f\,(e_A-X_0Y_A)$, where $e_A$ are the unit vectors of the $k+N$ directions of $Y$. The integrand is therefore $\sum_A\pr_A\omega_f\,\varphi_A$ with
\begin{equation}
\varphi_A = C_\psi^2\,{\rm tr}\big[\Gamma\!\cdot\!X_{h0}\,\Gamma\!\cdot\!(e_A-X_0Y_A)\,\Gamma\!\cdot\!X_{0j}\,R\big]\,(s_{h0}^2)^{-\Dp-1}(s_{0j}^2)^{-\Dp-1} \, .
\end{equation}
After the trace the vector $e$ occurs only in $e\!\cdot\!e=1$, so that $\varphi_A$ is a function of $Y$ in which $\rho$ occurs through $\rho^2$. We have $\sum_AY_A\,(e_A-X_0Y_A)=(\rho^2\,Y,-\rho\,Y^2)$, whose scalar products with $X_{h0}$, $X_{0j}$ and the vectors of $R$ are $\rho^2$ times scalar products of $Y$. Hence $\sum_AY_A\varphi_A=\rho^2\,\mathsf{q}$ with a continuous function $\mathsf{q}$, and $\varphi_A=(\delta_{AB}-Y_AY_B)\,\mathsf{f}_B$ with $\mathsf{f}_B=\varphi_B+Y_B\,\mathsf{q}$. The second identity in \eqref{eq:lap} and the divergence theorem on the ball of radius $\mathsf{r}<1$, whose boundary term contains $\rho^{m'}$ and vanishes for $\mathsf{r}\to1$, give
\begin{equation}
\int_{\mathbb{B}}d\mu_{\mathbb{B}}\,\sum_A\pr_A\omega_f\,\varphi_A = -\int_{\mathbb{B}}d\mu_{\mathbb{B}}\;\omega_f\,\Psi \, , \qquad
\Psi = \sum_A\pr_A\varphi_A-(d-k-N-1)\,\mathsf{q} \, ,
\label{gf:eq:ibpf}
\end{equation}
with $\omega_f$ at its position in the trace. A direct computation with the rules \eqref{gf:eq:trace} gives
\begin{equation}
\Psi = (d-2-2\Dp)\,C_\psi^2\Big\{{\rm tr}\big[\Gamma\!\cdot\!X_{h0}\,\Gamma\!\cdot\!X_0\,\Gamma\!\cdot\!X_j\,R\big]-{\rm tr}\big[\Gamma\!\cdot\!X_{0j}\,R\big]\Big\}(s_{h0}^2)^{-\Dp-1}(s_{0j}^2)^{-\Dp-1} \, .
\label{gf:eq:Psi}
\end{equation}
In the embedding space this is the divergence on the sphere of the vector field $V^M$ with $V\!\cdot\!A={\rm tr}[\Gamma\!\cdot\!X_{h0}\,\Gamma\!\cdot\!A\,\Gamma\!\cdot\!X_{0j}\,R]$, computed with $\Gamma_M\Gamma^M=d+1$ and $\nabla_M(s_{h0}^2)^{-\Dp-1}=2(\Dp+1)(s^2_{h0})^{-\Dp-2}(X_h-X_0\,X_0\!\cdot\!X_h)_M$, and the two expressions agree. The factor $d-2-2\Dp$ arises as $\Gamma_M\Gamma^M-1$ from the divergence of the numerator minus $2(\Dp+1)$ from the derivatives of the two denominators, and the remaining terms combine with $1-\Gamma\!\cdot\!X_0\,\Gamma\!\cdot\!X_{0j}=\Gamma\!\cdot\!X_0\,\Gamma\!\cdot\!X_j$. Comparing with \eqref{gf:eq:Bpsi} at $\Dp'=\Dp+1$ proves the lemma.\footnote{We verified \eqref{gf:eq:Psi} exactly in $d$ and $\Dp$, with \eqref{gf:eq:ibpf}, for $k+N=3,4$ and with one, three and five further vectors in $R$.}
\end{proof}

Lemma~\ref{gf:lem:ward2} is the counterpart of Lemma~\ref{lem:ward2}. In the scalar case the factor $2\Dp(d-2-2\Dp)$ arises from the Laplacian of the propagator, and here the factor $d-2-2\Dp$ arises from the Dirac operator. On the line both factors vanish away from coinciding points, and both are multiplied by a pole of the integral with the modified line.

\begin{lemma}
\label{gf:lem:contact}
The function $(d-2\Dp')\,\mathcal{B}^\psi$ has no pole along the plane $\mathsf{H}$, and on $\mathsf{H}$
\begin{equation}
(d-2\Dp')\,\mathcal{B}^\psi\big|_\mathsf{H} = C_\psi\,\vol_d\sum_{r=1}^N\big\langle V_1,\ldots,R^\psi_\omega V_r,\ldots,V_N\big\rangle^c(d,\Dp) \, ,
\end{equation}
where $R^\psi_\omega$ acts on the fermion legs only.
\end{lemma}

\begin{proof}
The proof of Lemma~\ref{lem:contact} applies, with the following changes. At points of $\mathsf{E}$ the integral over $Y$ in each term of \eqref{gf:eq:Bpsi} is of the form of Lemma~\ref{lem:delta}, with $w=X_j$ in the first term and $w=X_h$ in the second, and with a continuous function $\mathsf{g}$, which now contains the trace. No integration by parts is needed, since the fermion vertices have no derivatives. In the limit the first term gives $\vol_d\,C_\psi$ times the loop in which $X_0$ is identified with $X_j$, with $\omega_f(X_j)$ between the line from $X_h$ and the vertex matrix at $X_j$, since $\Gamma\!\cdot\!X_j\,\Gamma\!\cdot\!X_j=1$. This is the replacement of $\bar\psi$ by $\bar\psi\,\omega_f$ at the vertex $V_j$. The second term gives minus the loop with $\omega_f(X_h)$ after the vertex matrix at $X_h$, which is the replacement of $\psi$ by $-\omega_f\psi$ at $V_h$. If the loop contains only $X_0$ and one further vertex, then $X_h=X_j$, the limits contain the chord $X_{jj}=0$, and the corresponding terms of $R^\psi_\omega$ are loops with a single vertex, which vanish. The remaining steps, the uniformity in the other points and the passage from $\mathsf{E}$ to $\mathsf{H}$ by Lemma~\ref{lem:poleset}, are those of the proof of Lemma~\ref{lem:contact}.
\end{proof}

\begin{lemma}
\label{gf:lem:np}
The property \eqref{eq:NP} holds for $\mathcal{B}^\psi$ and for the function $\mathcal{B}$ of appendix~\ref{app:gauge} with fermion loops, for all vertices of section~\ref{gf:sec:vertices} with sources of the class, for all $N\geq1$ and all $D_0\geq1$.
\end{lemma}

\begin{proof}
We follow section~\ref{app:npcurrent}. In $\mathcal{B}^\psi$ the line $e_0$ with the exponent $\Dp'$ is the fermion line without its chord, and it is not marked. In the chart from $P$, \eqref{gf:eq:inversion} holds for the remaining chords. For the factor $\Gamma\!\cdot\!X_0\,\Gamma\!\cdot\!X_j$ of the first term we use $X=P+\Omega\,\tilde y$, and in the second term the factors $\Gamma\!\cdot\!\tilde y_h$ and $\Gamma\!\cdot\!\tilde y_0$ of the neighbouring chords meet without the chord of $e_0$. In both cases the trace is a sum of products of scalar products of vectors $\tilde y_i=\tilde x_i-P$, chords, the vector $P$ and the vectors $a^{\rm fl}$, with $P\!\cdot\!\tilde x_i=0$, so that the terms with the constant vector $P$ only lower the number of linear factors. At each end of $e_0$ the power of $\Omega_i$ from the numerator is the same as at a vertex with two chords, namely one at a Yukawa vertex and at the point $X_0$, and at least one at a current vertex through $a^{\rm fl}$, and the numerator has at most one additional linear factor there. Hence \eqref{gf:eq:count} holds with $r_i$ counting the marked lines, which are the fermion lines other than $e_0$ and the lines chosen at the scalar current vertices, and the exponents are those of section~\ref{app:npcurrent} with $\mathsf{z}_{e_0}=\Dp'+\Dp\,(n_{ij}-1)+r_{ij}$. The analysis of the large parameters, including the bound \eqref{eq:pairbound}, and of the small parameters uses only the properties listed in section~\ref{gf:sec:line} and applies without change. For $\mathcal{B}$ with fermion loops the line $e_0$ is a scalar line and the fermion lines are marked lines, and the same holds. The reduction of the sources is that of section~\ref{app:npcurrent}.
\end{proof}

\begin{proposition}
\label{gf:prop:gauge}
For scalar fields and Dirac fermions with the vertices of section~\ref{gf:sec:vertices} and sources of the class, the functional $W_0^S$ on the line \eqref{eq:line} is invariant under the background gauge transformations of the bare sources, in the form \eqref{eq:gaugeS} extended by the variation $\delta_\omega Y$ of the Yukawa couplings.
\end{proposition}

\begin{proof}
At a point $\mathsf{p}_0$ of $\ell_0$ with $d\neq2,0,-2,\ldots$ which lies on no plane of poles other than $\mathsf{H}$ of $\mathcal{B}$ and of $\mathcal{B}^\psi$, the functions $\mathcal{J}=(d-2\Dp')\,\mathcal{B}$ and $\mathcal{J}^\psi=(d-2\Dp')\,\mathcal{B}^\psi$ are holomorphic by Lemmas~\ref{lem:contact}, \ref{gf:lem:contact}, \ref{gf:lem:np} and~\ref{lem:poleset}. On $\mathsf{P}$ we have $d-2\Dp'=d-2-2\Dp$. By Lemmas~\ref{lem:ward2} and~\ref{gf:lem:ward2} the first term of $T_N$ is $2\Dp\,\mathcal{J}|_\mathsf{P}-\mathcal{J}^\psi|_\mathsf{P}$, and by Lemmas~\ref{lem:contact} and~\ref{gf:lem:contact} the second term is $-\mathcal{J}|_\mathsf{H}/(C_\phi\vol_d)+\mathcal{J}^\psi|_\mathsf{H}/(C_\psi\vol_d)$. With $2\Dp=d-2$, $(d-2)\,C_\phi\,\vol_d=1$ and $C_\psi\,\vol_d=1$ on $\ell_0$ the sum vanishes at $\mathsf{p}_0$.
\end{proof}

The proof does not use that $\omega_f$ is antihermitian. The functional is therefore also invariant under the variations with $\omega_f$ in the Lie algebra of ${\rm GL}(N_f)$, under which $\psi$ and $\bar\psi$ transform with inverse matrices. These are not symmetries of a real action, but they are symmetries of the Gaussian integral over the independent variables $\psi$ and $\bar\psi$. We use them in appendix~\ref{gc:sec:cov} to show that a hermitian part of $N_I$ can be removed.

\subsection{Rotations, flat chart and conformally flat backgrounds}
\label{gf:sec:chart}

The derivation of the properties under rotations in appendix~\ref{app:rotations} uses only that the integrand is a function of the $X_{i\parallel}$ and the $u_{ij}$ and that the integral over the remaining points with one point held fixed is again of the form \eqref{eq:gram}. Both hold for the integrands with fermion loops, since after the trace they are functions of this kind. Hence at constant sources the one-point functions of the scalar operators are constant, the two-point functions depend only on the angle, and \eqref{eq:Lambda} holds. The one-point function of the fermion current at constant sources is a linear function of the tangent vector $A(X_0)$ which is invariant under the rotations of the enlarged parallel space which fix $X_0$, and for $D_0\geq2$ it is therefore a multiple of $X_0\!\cdot\!A(X_0)=0$. Since the integrals do not depend on $D_0$, it vanishes.

The flat functional $W_0^{\rm fl}$ is defined by \eqref{eq:gram} with the flat measure, the flat distances and the flat loops $-C_\psi^n\,{\rm tr}[m_1\,\gamma\!\cdot\!x_{12}\cdots]\prod|x_{i\,i+1}|^{-2\Dp-2}$, for sources and one-forms of the class of flat sources. By Lemma~\ref{gf:lem:chart} and the relation of the measures in section~\ref{sec:chart}, \eqref{eq:chart} holds with the additional entry $\Omega^{(4-d)/2}\,Y$ for the Yukawa couplings on the line. On a conformally flat background $\Omega^2\delta_{\mu\nu}$ we define the bare functional by the flat functional with the sources rescaled by their weights, which extends \eqref{eq:defcurved}. In integer dimension this is the identity of the classical actions under $\phi\to\Omega^{-(d-2)/2}\phi$ and $\psi\to\Omega^{-(d-1)/2}\psi$, $\bar\psi\to\Omega^{-(d-1)/2}\bar\psi$, with the vielbein $\Omega\,\delta^a_\mu$, under which the massless Dirac operator is covariant. The invariance of the bare functional under Weyl rescalings then holds identically, and by \eqref{eq:chart} the functional on the sphere agrees with this definition for sources which vanish near the pole of the chart.

\subsection{Reality}

\begin{lemma}[Reality]
\label{gf:lem:real}
For real $d$ and $\Dp$, real $\lambda$, hermitian $Y^a$ and a gauge field with values in the Lie algebra of the group of field rotations, each term of the bare functional is real.
\end{lemma}

\begin{proof}
The scalar parts are real. For a fermion loop the reversed orientation is the pairing in which each line runs in the opposite direction. Its spinor trace is the trace \eqref{gf:eq:trace} of the reversed sequence of vectors with each chord multiplied by $-1$, which is $(-1)^n$ times the original one, since \eqref{gf:eq:trace} is invariant under reversal. Its flavour trace is the complex conjugate of the original one times $(-1)^{|\mathsf{C}|}$, since the $Y^a$ are hermitian and the $a_A$ antihermitian. Since the trace vanishes unless $n+|\mathsf{C}|$ is even, the reversed loop is the complex conjugate of the original one. For $n\geq3$ the two orientations are different pairings and their sum is real, and for $n=2$ the loop is its own reverse and is real.
\end{proof}

\subsection{Steps for Yukawa and quartic couplings}
\label{gf:sec:steps}

We prove Proposition~\ref{gf:prop:props} by the steps listed in the column of the Yukawa theory of table~\ref{gs:tab:steps}, which follow the derivation of section~\ref{gs:sec:phi4prop} and appendices~\ref{app:weyl}, \ref{app:unique} and~\ref{app:finite}. For each step we state whether it holds without change, with a stated modification, or with a new argument.

\paragraph{Bare functional and general constructions.} Section~\ref{gf:sec:fermions} gives the bare functional with the covariance, the gauge invariance, the properties under rotations and the chart. Propositions~\ref{gs:prop:msanomaly} and~\ref{gs:prop:msward} use only these invariances and the form of the bare sources, and they apply without change. The bare Weyl rescaling has the weight $-\vep k_I$ on the bare couplings, which is minus the weight $\tfrac12(4-d)$ of Proposition~\ref{gf:prop:cov} for the Yukawa couplings.

\paragraph{Weights.} In appendix~\ref{app:gaugevar} the vector field $\hb^{{\rm L}\,I}$ is defined by $\hb^{{\rm L}\,J}\pr_Jg_0^I=-\vep\,k_I\,g_0^I$ (no sum over $I$ on the right). With $K={\rm diag}(k_I)$ and the weighted Euler operator $\mathsf{E}=k_Jg^J\pr_J$ the solution is $\hb^{\rm L}=-\vep Kg+\vep(1+\pr L)^{-1}(\mathsf{E}-K)L$, and $(\cL_{\hB}N)_I$ equals $-\vep(\mathsf{E}+k_I)N_I$ up to a pure pole. The proof of Lemma~\ref{lem:rhoxi} then gives, in place of \eqref{eq:zeroparts},
\begin{equation}
\beta^I = (\mathsf{E}-k_I)\,L^I_1 \, , \qquad \xi = -k_I\,g^I N_{I,1} \, , \qquad B^I = \beta^I-(\xi g)^I \, , \qquad \hr_I = k_Jg^J\left(\pr_JN_{I,1}-\pr_IN_{J,1}\right) ,
\label{gf:eq:zeroparts}
\end{equation}
and $\hB^I=-\vep\,k_Ig^I+B^I$ for regular $\hB^I$. The expression for $\hr_I$ vanishes when it is contracted with $k_Ig^I$, not with $g^I$, and with $\hr_I\hB^I=0$ from the third statement of Lemma~\ref{lem:rhoxi} this gives both relations \eqref{eq:rhoB}. The residue formulas of the anomaly coefficients which follow from these, such as those for $A_{\rm E}$, $G_{IJ}$ and $W_I$ in appendix~\ref{app:weyl}, become $A_{\rm E}=4(1+\mathsf{E})\fa_1$, $G_{IJ}=(1+k_I+k_J+\mathsf{E})\fg_{IJ,1}$ and $W_I=-4\pr_I\fa_1+\fg_{IJ,1}k_Jg^J$, and the term of order zero of $\hP_I{}^J$ in appendix~\ref{app:masscoeff} becomes $1-\tfrac12\vep-\vep k_I$. These formulas are not used in the proof of Proposition~\ref{gf:prop:props}. In Lemma~\ref{lem:frame} we keep the induction in the ordinary degree $n$. The term of degree $n$ of \eqref{eq:xiframe} is $\xi^{\rm L}_{(n)}+\vep\,\mathsf{E}\vartheta_{(n)}+\mathsf{F}_n$, where $\mathsf{F}_n$ contains only terms of lower degree, since $B^I$ has no terms of degree zero and one. The operator $\mathsf{E}$ is diagonal on the monomials $\lambda^pY^q$ with the eigenvalue $p+\tfrac12q>0$, and we set $\vartheta_{(n)}=-\mathsf{E}^{-1}{\rm PP}[\xi^{\rm L}_{(n)}+\mathsf{F}_n]/\vep$ monomial by monomial.

\paragraph{Terms of second order.} In Lemma~\ref{lem:lowest} the terms of second order with two sources of different kinds vanish. A quartic and a Yukawa vertex, a Yukawa vertex and a mass vertex, and a Yukawa vertex and a vertex of the scalar current have an odd number of scalar legs. A Yukawa vertex and a vertex of the fermion current form a loop with an odd number of gamma matrices, and a loop with one vertex vanishes. The other mixed terms, of a quartic coupling, a mass source or a scalar current vertex with a vertex of the fermion current, vanish since a fermion loop through one vertex vanishes. The kernel of the couplings is the matrix $c_{IJ}(d)\,|x-y|^{-2\Delta_I}$, block diagonal with analytic blocks which are nondegenerate at $d=4$, in place of the single kernel of \eqref{eq:twopointkernels}. The kernel of two Yukawa couplings is $c(Y,Y')\,|x-y|^{-2(3\Dp+1)}$ by Lemma~\ref{gf:lem:free}, which is $|x-y|^{-8}$ at $d=4$, as for the quartic couplings, and $c$ is positive definite on hermitian matrices. The kernel of two gauge fields with values in ${\rm u}(N_f)$ comes from the fermion loop with two current vertices, $-C_\psi^2\,{\rm tr}(\omega_f\omega'_f)\,{\rm tr}(\gamma_\mu\,\gamma\!\cdot\!x\,\gamma_\nu\,\gamma\!\cdot\!(-x))\,|x|^{-2d}$. With ${\rm tr}(\gamma_\mu\gamma\!\cdot\!x\gamma_\nu\gamma\!\cdot\!x)={\rm tr}\,1\,(2x_\mu x_\nu-\delta_{\mu\nu}x^2)$ and $(\pr_\mu\pr_\nu-\delta_{\mu\nu}\pr^2)|x|^{-2(d-2)}=2(d-2)(d-1)(2x_\mu x_\nu-\delta_{\mu\nu}x^2)|x|^{-2d}$ it is the kernel of \eqref{eq:twopointkernels} times the bilinear form ${\rm tr}(\omega_f\omega'_f)$, which is nondegenerate on ${\rm u}(N_f)$. The parts of the gauge field in ${\rm so}(N)$ and in ${\rm u}(N_f)$ do not mix at this order, so that the kernel is transverse with a nondegenerate coefficient on $\mathfrak{g}$, and Lemma~\ref{lem:transverse} applies componentwise. The argument for $Z_\sM$ and $\fd_I$ at $g=0$ is unchanged.

\paragraph{Stabilisers.} The proof of Proposition~\ref{prop:uniqueall} uses in two places that the stabiliser of the couplings in the Lie algebra is trivial on a dense open set. Here it is not, since the fermion number, the centre of ${\rm u}(N_f)$, acts trivially on all couplings, and for $N=1$ the stabiliser of a hermitian matrix $Y$ contains all $\mathrm{i}\,p(Y)$ with a real polynomial $p$. In appendix~\ref{gc:sec:cov} we show that the argument of that proof, extended by one step, gives uniqueness up to the transformations of Lemma~\ref{lem:frame} with parameters which leave all couplings invariant, that every set of coefficients is covariant under the Lie algebra, and that there is a set with $N_I$ in $\mathfrak{g}$ (Proposition~\ref{gc:prop:cov}). A transformation of Lemma~\ref{lem:frame} with a parameter of degree one which is covariant under the Lie algebra then removes the term of degree zero of $N_I$, without a change of $L^I$ and of the functional. We use the element of the fifth statement of that proposition, which is invariant under ${\rm O}(N)\times{\rm U}(N_f)$ and has $N_I^{(0)}=0$, as Lemma~\ref{lem:rhoxi} requires.

\paragraph{Regularity of the Weyl variation.} In the proof of Lemma~\ref{lem:regvar} the paragraph on the lowest order gives $\hB^I=-\vep\,k_Ig^I$ at first order, and the term of first order of the variation of the couplings is $-\vep\,k_I\sigma g^I-v^\mu\pr_\mu g^I$, which has no pole. In the step for the couplings the leading pole $\breve\beta^I$ multiplies the kernel of the two-point function, which is block diagonal with positive definite blocks on the quartic and the Yukawa couplings, so that the vanishing of the potential gives $\breve\beta^I=0$ for each block. We use the other steps without change. The determination of the coefficients by the renormalised functional, after the lemma, uses the kernels $|x-y|^{-8}$ and $|x-y|^{-4}$ and holds as well.

\paragraph{Anomaly, curvature counterterms and relations.} The structures \eqref{eq:fullanomaly} and the expressions \eqref{eq:fixedcoeff} to \eqref{eq:flatcoeff} are identities for covariant coefficients with any group of field rotations and any couplings, and the variation \eqref{eq:Delta} enters through $\hB^I$ and $\hr_I$ only. The weights enter through $\tL$ and $\hB\!\cdot\!\pr$, whose definitions do not refer to the weights. The factors $2(d-1)$, $d-2$, $4(d-1)$ and $4(d-3)$ of \eqref{eq:steps} and \eqref{eq:finrel} are those of $d_c=4$. Proposition~\ref{prop:finite} and its proof therefore hold, with the Weyl map of section~\ref{gf:sec:chart} on conformally flat backgrounds and the chart of Lemma~\ref{gf:lem:chart} in the localisation, where the flat Yukawa couplings are $\Omega^{\vep/2}Y_0$ and the family of sources depends analytically on $\vep$ as before.

\paragraph{Properties.} With these steps we can follow the proof of Proposition~\ref{gs:prop:phi4prop}. Property~\ref{gs:P:functional} is the third part of Proposition~\ref{prop:finite} with Proposition~\ref{gf:prop:gauge}, and at a constant mass source it follows by the first paragraph of the proof of Lemma~\ref{gr4:lem:const}, which uses only the third part of Proposition~\ref{prop:finite} and the gauge invariance. Property~\ref{gs:P:weyl} follows from Lemmas~\ref{lem:rhoxi} and~\ref{lem:regvar} as modified, the covariance, the relation \eqref{eq:rhoB} and Lemma~\ref{gcu:lem:commute}, and with \eqref{gf:eq:zeroparts} and Lemma~\ref{gr4:lem:zero} it includes $\hB^I=-\vep k_Ig^I$ up to terms of second order, $\gamma_\sM$ without term of order zero, and the vanishing at $g=0$ of $\ell_{\rm h}=d\,(\delta-2\theta)$ and $\vk_{\rm h}-\vk_{\rm c}=-d\,\bar\tau$. For the element of the fifth statement of Proposition~\ref{gc:prop:cov} the vector field $\hB^I$, the one-form $\hr_I$ and the coefficients of $\delta^{\rm W}_\sigma\sM$ are covariant under ${\rm O}(N)\times{\rm U}(N_f)$, and the functional is invariant under this group. By the fourth statement every other element with $N_I$ in $\mathfrak{g}$ and without term of degree zero gives the same $\hB^I$ and $\hr_I$. Property~\ref{gs:P:anomaly} follows from Proposition~\ref{gs:prop:msanomaly} and the first part of Proposition~\ref{prop:finite}, Property~\ref{gs:P:ward} from Proposition~\ref{gs:prop:msward}, and Property~\ref{gs:P:sphere} from section~\ref{gf:sec:chart}, appendix~\ref{app:massward} and Lemma~\ref{gr4:lem:const}, whose proof uses only scalar sources and the reduction of the terms with the bare gauge field by the gauge invariance. The class of sources is that of section~\ref{sec:sphere}, which satisfies the assumptions of appendix~\ref{gs:sec:class}, see appendix~\ref{app:rules}. Lemma~\ref{gr4:lem:zero} holds with the same proof. The two-point form follows from Lemma~\ref{gf:lem:free}, whose hypotheses hold since $L^I$, the bare gauge field and the bare mass source at $\sM=0$ start at second order in the couplings. The proof of Proposition~\ref{gr4:prop:lowest} gives the two-point form in the divergence scheme and Property~\ref{gs:P:free}. The matrix $c_{IJ}$ vanishes between couplings of different weights, as the two-point form requires, and it is positive definite at $\vep=0$ by Lemma~\ref{gro:lem:c}.

\paragraph{What is not derived.} The identification of $L^I$ with the counterterms of minimal subtraction in momentum space, and with it of $\hB^I$ with the usual beta functions up to a rotation, is open as for $\phi^4$ theory. The values of the lowest-order two-point coefficients at $\vep=0$ are $\lambda\!\cdot\!\lambda'/(6144\,\pi^8)$ and $\sum_a{\rm tr}(Y^aY'^a)/(4\pi^6)$ for ${\rm tr}\,1=4$, by Lemma~\ref{gf:lem:free}.

\subsection{Covariance from finiteness}
\label{gc:sec:cov}

In this subsection we show that the coefficients of Hypothesis~\ref{hyp:finite} for the Yukawa theory other than $L^I$ are covariant under the Lie algebra, and that there is a set of them which is covariant and in which $N_I$ takes values in the Lie algebra of ${\rm O}(N)\times{\rm U}(N_f)$ and has no term of degree zero. The proof holds for all $N\geq1$ and $N_f\geq1$. It extends the proof of Proposition~\ref{prop:uniqueall} by one step, and it replaces the conclusion of that proposition from uniqueness by an argument with the compact group of field rotations.

\subsubsection{Statement}
\label{gc:sec:statement}

We use the notation of section~\ref{gs:sec:yukawa}. Let $G_0={\rm SO}(N)\times{\rm U}(N_f)$ be the identity component of the group of field rotations, with Lie algebra $\mathfrak{g}={\rm so}(N)\oplus{\rm u}(N_f)$, and let $\hat{\mathfrak{g}}={\rm so}(N)\oplus{\rm gl}(N_f)$, regarded as a real vector space. We fix a real covariant pure pole $L^I$ without terms of degree zero and one. We consider collections $\varsigma=(N_I,Z_\sM,\fd_I,\fe_{IJ},\mathfrak{x})$, in which $N_I$ is a pure pole with values in $\hat{\mathfrak{g}}$ with real coefficients in a real basis of $\hat{\mathfrak{g}}$, $Z_\sM-1$, $\fd_I$ and $\fe_{IJ}$ are real pure poles, and $\mathfrak{x}$ stands for the real pure poles of the coefficients of $\cX^{\rm fl}$. We denote by $\mathsf{S}$ the set of the collections such that the flat functional \eqref{eq:Wfl} with the bare sources \eqref{gs:eq:barefl} built from $L^I$ and $\varsigma$ is finite for all sources of the class of flat sources and all families which depend analytically on $\vep$. In this subsection covariant means covariant under the Lie algebra $\mathfrak{g}$, that is under the rotations of the couplings with constant elements of $G_0$, which act on $N_I$ by the commutator in addition to the action on the index $I$. The pure pole $L^I$ is covariant under the whole group ${\rm O}(N)\times{\rm U}(N_f)$ by Hypothesis~\ref{hyp:finite} for the Yukawa theory.

Hypothesis~\ref{hyp:finite} for the Yukawa theory states that $\mathsf{S}$ contains an element with $N_I$ in $\mathfrak{g}$.

\begin{proposition}
\label{gc:prop:cov}
Let $N\geq1$ and $N_f\geq1$. First, every element of $\mathsf{S}$ is covariant. Second, if $\mathsf{S}$ is not empty, it contains an element in which $N_I$ takes values in $\mathfrak{g}$. Third, from such an element a transformation of Lemma~\ref{lem:frame} with a parameter of degree one which is covariant under the Lie algebra gives an element of $\mathsf{S}$ in which $N_I$ takes values in $\mathfrak{g}$ and has no term of degree zero, with the same $L^I$ and the same renormalised functional, also if $N_I$ is allowed to take values in $\hat{\mathfrak{g}}$. Fourth, two elements of $\mathsf{S}$ give the same renormalised functional, and two elements in which $N_I$ takes values in $\mathfrak{g}$ without term of degree zero give the same $\hB^I$ and $\hr_I$. Fifth, $\mathsf{S}$ contains an element which is invariant under ${\rm O}(N)\times{\rm U}(N_f)$ and in which $N_I$ takes values in $\mathfrak{g}$ and has no term of degree zero.
\end{proposition}

By the third to fifth statements the properties derived in section~\ref{gs:sec:yukawa} and the gradient formula of Theorem~\ref{gf:cor:yukawa} do not depend on the element of $\mathsf{S}$ with $N_I$ in $\mathfrak{g}$ and without term of degree zero which we use.

\subsubsection{Rotations which leave the couplings invariant}
\label{gc:sec:stab}

Let $\mathcal{T}$ be the set of pure poles $\vartheta$ in the couplings without term of degree zero, with values in $\hat{\mathfrak{g}}$, which are covariant and satisfy $(\vartheta(g)\,g)^I=0$ for all couplings $g$, where $\hat{\mathfrak{g}}$ acts on the Yukawa couplings by $Y^a\mapsto[\vartheta_f,Y^a]+\vartheta_{s,ab}Y^b$ and on the quartic couplings through $\vartheta_s$.

\begin{lemma}
\label{gc:lem:stab}
Let $\vartheta,\vartheta'\in\mathcal{T}$. Then the scalar part of $\vartheta$ vanishes, $[\vartheta(g),\vartheta'(g)]=0$ for all $g$, $\vartheta+\vartheta'\in\mathcal{T}$, and $-\vartheta^\dagger\in\mathcal{T}$, where the dagger acts on the fermionic part. Moreover every element of $\mathcal{T}$ annihilates $L^I(g)$ and the sources $\sM$.
\end{lemma}

\begin{proof}
We use couplings in a dense open set and conclude for all couplings, since the statements are polynomial identities at each order. For $N\geq2$ the stabiliser of a generic quartic coupling in ${\rm so}(N)$ is trivial, as in the proof of Proposition~\ref{prop:uniqueall}. The scalar part of $\vartheta(g)$ therefore vanishes, and the fermionic part commutes with all $Y^a$. For $N\geq2$ and generic $Y^1$ and $Y^2$, for instance $Y^1$ diagonal with distinct eigenvalues and $Y^2$ with nonzero off-diagonal entries, the matrices which commute with both are the multiples of the unit matrix, so that $\vartheta(g)$ is central. For $N=1$ the scalar part is absent, and the matrices which commute with a hermitian matrix $Y$ with distinct eigenvalues are the matrices which are diagonal in its eigenbasis. In both cases the values at one $g$ commute. We see that $\mathcal{T}$ is a linear space. If $X$ commutes with all $Y^a$, so does $X^\dagger$, and covariance is preserved by the dagger, since the elements of ${\rm u}(N_f)$ are antihermitian. Finally let $\omega=\vartheta(g)$ with $\vartheta\in\mathcal{T}$, which annihilates $g$. Its scalar part vanishes by the first statement, and its fermionic part $X$ commutes with all $Y^a$, so that $X^\dagger$ does as well. Hence $\tfrac12(X-X^\dagger)$ and $\tfrac{\mathrm{i}}2(X+X^\dagger)$ are elements of $\mathfrak{g}$ which annihilate $g$. For such an element $\omega'$ we have $L(g)=L(e^{t\omega'}g)=e^{t\omega'}L(g)$ by the covariance of $L$, so that $\omega'$ annihilates $L(g)$, and by complex linearity so does $\omega$, and the sources $\sM$ are not rotated since the scalar part vanishes.
\end{proof}

For $\vartheta\in\mathcal{T}$ let $\mathsf{R}=\exp\vartheta(g)$ and let $\mathsf{F}_\vartheta$ map $\varsigma$ to the collection with $N'_I=\mathsf{R}\,N_I\mathsf{R}^{-1}-\pr_I\mathsf{R}\,\mathsf{R}^{-1}$ and the same $Z_\sM$, $\fd_I$, $\fe_{IJ}$ and $\mathfrak{x}$.

\begin{lemma}
\label{gc:lem:frame}
For $\vartheta,\vartheta'\in\mathcal{T}$ the map $\mathsf{F}_\vartheta$ maps $\mathsf{S}$ into $\mathsf{S}$ and does not change the renormalised functional. Moreover $\mathsf{F}_\vartheta\mathsf{F}_{\vartheta'}=\mathsf{F}_{\vartheta+\vartheta'}$, $\mathsf{F}_\vartheta$ commutes with the action of $G_0$ on collections, and $\mathsf{F}_\vartheta\varsigma=\varsigma$ implies $\vartheta=0$.
\end{lemma}

\begin{proof}
We note that the bare sources of $\mathsf{F}_\vartheta\varsigma$ are those of $\varsigma$ transformed by the gauge transformation with the parameter $\vartheta(g(x))$, in the form \eqref{eq:frame} of Lemma~\ref{lem:frame}. Indeed, $\mathsf{R}$ is covariant, so that $\pr_\mu\mathsf{R}+[a_\mu,\mathsf{R}]=\pr_I\mathsf{R}\,D_\mu g^I$, and $\mathsf{R}\,a_{0\mu}\mathsf{R}^{-1}-\pr_\mu\mathsf{R}\,\mathsf{R}^{-1}=a_\mu+N'_ID_\mu g^I$. By Lemma~\ref{gc:lem:stab} the transformation does not change $g_0=g+L(g)$ and $\sM_0$. The first statement of Lemma~\ref{lem:frame}, which uses only these facts and the gauge invariance, here Proposition~\ref{gf:prop:gauge} for parameters in $\hat{\mathfrak{g}}$, gives that the bare functional and hence the renormalised functional do not change. The coefficients of $N'_I$ are pure poles with real coefficients in a real basis, so that $\mathsf{F}_\vartheta\varsigma\in\mathsf{S}$. The composition rule follows from $\mathsf{R}'\mathsf{R}=\exp(\vartheta+\vartheta')$, since the values commute, and the group property of gauge transformations. For $h\in G_0$ we have $h\cdot\vartheta=\vartheta$ by covariance, so that $h\cdot\mathsf{F}_\vartheta\varsigma=\mathsf{F}_\vartheta(h\cdot\varsigma)$. Finally let $\vartheta\neq0$, and let $\vartheta_{(n)}$ be its term of lowest degree $n\geq1$. The term of degree $n-1$ of $N'_I-N_I$ is $-\pr_I\vartheta_{(n)}$, since $[\vartheta,N_I]$ has degree at least $n$ and the terms of higher order in $\vartheta$ have degree at least $2n-1\geq n$. It does not vanish, since $\vartheta_{(n)}$ is homogeneous of degree $n\geq1$.
\end{proof}

\subsubsection{Uniqueness}
\label{gc:sec:unique}

\begin{proposition}
\label{gc:prop:unique}
Any two elements of $\mathsf{S}$ are related by $\mathsf{F}_\vartheta$ with a unique $\vartheta\in\mathcal{T}$.
\end{proposition}

\begin{proof}
We follow the proof of Proposition~\ref{prop:uniqueall}, with the levels defined there and with the changes of section~\ref{gf:sec:steps}, for two elements $\varsigma$ and $\varsigma'$ which are not assumed to be covariant. We use that this proof uses the covariance only in its last paragraph and in the triviality of the stabiliser, and we replace both. Let $n\geq1$ be the lowest level at which $\varsigma$ and $\varsigma'$ differ, where the level $1$ is the term of degree zero of $N_I$, and let $\vep^{-\mathsf{r}}$ be the highest pole of the difference $\tilde N_I$ of the terms of degree $n-1$ of $N_I$.

\paragraph{Gradient.} The first step of that proof applies to $\tilde N_I$ with values in $\hat{\mathfrak{g}}$, for $n\geq2$, and the paragraph on the term of degree zero for $n=1$. The pairing of the kernel of the gauge fields, which by section~\ref{gf:sec:steps} is ${\rm tr}(\omega\omega')$ on the fermionic part, is evaluated with a test gauge field $a_2$ in $\mathfrak{g}$. Since ${\rm u}(N_f)$ spans ${\rm gl}(N_f)$ over the complex numbers, the vanishing of the complex number ${\rm tr}(X a_2)$ for all $a_2$ in ${\rm u}(N_f)$ implies $X=0$, so that Lemma~\ref{lem:transverse} applies to each complex component. The leading pole is therefore $\vep^{-\mathsf{r}}\pr_I\breve\vartheta$ with a homogeneous polynomial $\breve\vartheta$ of degree $n$ with values in $\hat{\mathfrak{g}}$.

\paragraph{Stabiliser.} The third step of that proof, and for $n=1$ the paragraph on the term of degree zero, give $(\breve\vartheta(g)\,g)^I=0$ for all $g$. They use the identity \eqref{eq:W3ward}, which holds for parameters in $\hat{\mathfrak{g}}$ by Proposition~\ref{gf:prop:gauge}, and the finiteness of $\mathsf{W}_3[a;g_1,g_2]$. We now show that the latter holds without a condition on $N_I^{(0)}$. Indeed, the terms of the renormalised functional which are linear in $a$, $g_1$ and $g_2$, with $g_1$ and $g_2$ of disjoint supports, are $2\,\mathsf{W}_3[a;g_1,g_2]$ and the terms of the bare functional with two gauge fields, $a$ and $N_I^{(0)}\pr_\mu g_i^I$, and one coupling $g_j$. The latter vanish, since a fermion loop with two vertices of the current and one Yukawa vertex has an odd number of gamma matrices and the scalar leg of the Yukawa vertex cannot be paired, and the scalar part is that of $\phi^4$ theory. The terms of the bare functional with three gauge fields, $a$, $N_I^{(0)}\pr_\mu g_1^I$ and $N_I^{(0)}\pr_\mu g_2^I$, vanish by the identity of third order in the gauge field which follows from Proposition~\ref{gf:prop:gauge}, since $\pr_\mu(N_I^{(0)}g_1^I)$ is exact and the commutators of the parameters $N_I^{(0)}g_i^I$ with the other arguments vanish by the disjointness of the supports or give exact arguments. The terms with $N_I^{(0)}(a\,g_i)^I$ and $N_I^{(0)}\pr_\mu g_j^I$ vanish since the kernel of two gauge fields is transverse, and the terms of the scalar source $\Upsilon$ which are linear in $a$, $g_1$ and $g_2$ are mass vertices against a coupling and vanish. The bare mass source has no term of first order in the couplings by Lemma~\ref{lem:lowest}, which does not assume $N_I^{(0)}=0$, and local terms with $g_1$ and $g_2$ vanish. The same terms, with $\tilde a_0$ in place of $a$, are the only further terms in the identification of $W-W'$ with $2\,\mathsf{W}_3[\tilde N_I\,dg_1^I;g_1,g_2]$, and they vanish for the same reasons, the term with two gauge fields and one coupling as above. The finiteness extends by complex linearity to one-forms with values in $\hat{\mathfrak{g}}$. In the step which concludes $(\breve\vartheta(g)\,g)^I=0$, the Yukawa part of $\breve\vartheta(g_1)\,g_1$ is a commutator with a matrix in ${\rm gl}(N_f)$, which need not be hermitian. Its vanishing follows from the vanishing of ${\rm tr}(X\,Y_2)$ for all hermitian $Y_2$, which implies $X=0$.

\paragraph{Covariance of $\breve\vartheta$.} This is the new step. We take couplings $g_1$, a gauge field $a_1$ whose support may overlap with that of $g_1$, and a gauge field $a_2$ whose support is disjoint from those of $g_1$ and $a_1$, and we consider the terms of $W-W'$ of order $n$ in $g_1$ and linear in $a_1$ and in $a_2$. The bare sources of the two collections differ at this order by $\tilde a_{0\mu}=\tilde N_I(g_1)\,D_\mu g_1^I$, where $D_\mu g_1^I$ contains $a_1$, since the differences of $\fd_I$ and $\fe_{IJ}$ vanish by the second step of that proof, which does not use covariance, and the difference of $Z_\sM$ multiplies $\sM$. Terms of second order in the differences, and differences at higher levels, are of higher order in $g_1$. The terms considered are therefore the terms of the bare functional which are linear in $\tilde a_0$, $a_1$ and $a_2$, or linear in $\tilde a_0$ and $a_2$ with the part of $\tilde a_0$ which is linear in $a_1$. These terms of the bare functional are finite at $d=4$ for every one-form in place of $\tilde a_0$, because they are terms of the renormalised functional of second and third order in the gauge field with $a_2$ disjoint from the others, to which no counterterm contributes. We find that the coefficient of $\vep^{-\mathsf{r}}$ is therefore the value at $d=4$ of these terms with $\tilde a_0$ replaced by
\begin{equation}
\pr_I\breve\vartheta\,D_\mu g_1^I = \pr_\mu\breve\vartheta+[a_{1\mu},\breve\vartheta]+\mathsf{C}_\mu \, , \qquad
\mathsf{C}_\mu = \pr_I\breve\vartheta\,(a_{1\mu}g_1)^I-[a_{1\mu},\breve\vartheta] \, ,
\end{equation}
where $\breve\vartheta=\breve\vartheta(g_1(x))$. The first two terms on the right are the gauge transformation of $a_1+a_2$ with the parameter $\breve\vartheta$, at first order and without the term $[a_2,\breve\vartheta]$, which vanishes since $\breve\vartheta(g_1)$ vanishes on the support of $a_2$. By Proposition~\ref{gf:prop:gauge} the terms of the bare functional with this gauge transformation in place of $\tilde a_0$ cancel against the rotation of the other sources by $\breve\vartheta$, which vanishes on $a_2$ since $\breve\vartheta$ vanishes there, and whose action on $a_1$ is the term $[a_1,\breve\vartheta]$ already included. No coupling enters at this order besides those in $\tilde a_0$. The finiteness of the terms of second and third order in the gauge field extends by complex linearity to one-forms with values in $\hat{\mathfrak{g}}$. We are left with the term of second order of the bare functional with $\mathsf{C}$ and $a_2$, which is the integral of the transverse kernel of section~\ref{gf:sec:steps}. As in the first step of the proof of Proposition~\ref{prop:uniqueall}, Lemma~\ref{lem:transverse} gives that the one-form $\mathsf{C}$ is closed. We take $a_1=\omega\,\bar a$ with a constant $\omega\in\mathfrak{g}$ and a one-form $\bar a=u\,dx_\parallel$ of the class, so that $\mathsf{C}=c(x)\,u(x)\,dx_\parallel$ with $c=\pr_I\breve\vartheta(g_1)\,(\omega g_1)^I-[\omega,\breve\vartheta(g_1)]$. If it is closed, then $c\,u$ does not depend on $x_\perp$, and since it has compact support it vanishes. With a function $u$ which is positive on the support of $g_1$ we find $c=0$, at points where the values of $g_1$ are arbitrary. Hence
\begin{equation}
\pr_I\breve\vartheta(g)\,(\omega g)^I = [\omega,\breve\vartheta(g)]
\end{equation}
for all $\omega\in\mathfrak{g}$ and all $g$, which is the covariance of $\breve\vartheta$. Hence $\vep^{-\mathsf{r}}\breve\vartheta\in\mathcal{T}$.

\paragraph{Descending induction.} With $\tilde N_I=N'_I-N_I$ we replace $\varsigma'$ by $\mathsf{F}_{\vartheta_1}\varsigma'$ with $\vartheta_1=\vep^{-\mathsf{r}}\breve\vartheta$, so that the new difference at the level $n$ has its highest pole of order below $\mathsf{r}$. Indeed, by the proof of Lemma~\ref{gc:lem:frame} the transformation changes the terms of degree $n-1$ of $N'_I$ by $-\pr_I\vartheta_1$ only. We repeat this until the terms of degree $n-1$ agree. The second and the fourth step of the proof of Proposition~\ref{prop:uniqueall} then give that $\fd_I$, $\fe_{IJ}$ and $Z_\sM$ agree at the level $n$, and we continue with the next level. At each degree only finitely many steps are needed, and the product of the transformations is $\mathsf{F}_\vartheta$ with the sum $\vartheta\in\mathcal{T}$ of the parameters, by Lemma~\ref{gc:lem:frame}. When the bare sources agree, the paragraph on the counterterms of that proof gives that the coefficients $\mathfrak{x}$ agree. Hence $\varsigma'=\mathsf{F}_{-\vartheta}\varsigma$, and $\vartheta$ is unique by the last statement of Lemma~\ref{gc:lem:frame}. 
\end{proof}

\subsubsection{Proof of the proposition}
\label{gc:sec:proof}

\paragraph{Covariance.} Let $\varsigma\in\mathsf{S}$ and $h\in G_0$. The bare functional is invariant under a constant rotation of all bare sources, since the propagators are invariant, and $L^I$ is covariant, so that $h\cdot\varsigma\in\mathsf{S}$. By Proposition~\ref{gc:prop:unique} there is a unique $\vartheta_h\in\mathcal{T}$ with $h\cdot\varsigma=\mathsf{F}_{\vartheta_h}\varsigma$. By Lemma~\ref{gc:lem:frame}
\begin{equation}
\mathsf{F}_{\vartheta_{h'h}}\varsigma = h'\cdot(h\cdot\varsigma) = h'\cdot\mathsf{F}_{\vartheta_h}\varsigma = \mathsf{F}_{\vartheta_h}(h'\cdot\varsigma) = \mathsf{F}_{\vartheta_h+\vartheta_{h'}}\varsigma \, ,
\end{equation}
and by uniqueness $\vartheta_{h'h}=\vartheta_{h'}+\vartheta_h$. Let $h$ have finite order, $h^k=e$. Then $k\,\vartheta_h=\vartheta_e=0$, so that $\vartheta_h=0$ and $h\cdot\varsigma=\varsigma$. The elements of finite order are dense in the compact connected group $G_0$, since every element lies in a maximal torus, in which the elements of finite order are dense. At each degree and each order of the pole the coefficients of $h\cdot\varsigma$ are polynomials in the matrix elements of $h$, and hence continuous in $h$. Therefore $h\cdot\varsigma=\varsigma$ for all $h\in G_0$, and differentiating at the identity gives the covariance of all coefficients of $\varsigma$.

\paragraph{Antihermitian gauge field.} Let $\kappa$ map $\varsigma$ to the collection with $N_I$ replaced by $-N_I^\dagger$, where the dagger acts on the fermionic part, and with the other coefficients unchanged. By the proof of Lemma~\ref{gf:lem:real}, which applies to arbitrary complex matrices at the vertices, the bare functional obeys $W_0^{\rm fl}[Y,a]^*=W_0^{\rm fl}[Y^\dagger,-a^\dagger]$, where the scalar sources are real. For sources of the class the bare Yukawa couplings $Y+L(Y,\lambda)$ are hermitian, the gauge field $a$ is in $\mathfrak{g}$ and $D_\mu g^I$ has real components, so that $-a_0^\dagger=a+(-N_I^\dagger)D_\mu g^I$. The other bare sources and $\cX^{\rm fl}$ are real. Hence the functional of $\kappa\varsigma$ is the complex conjugate of that of $\varsigma$, and $\kappa$ maps $\mathsf{S}$ into $\mathsf{S}$. A direct computation with $\mathsf{R}=e^\vartheta$ gives $\kappa\,\mathsf{F}_\vartheta=\mathsf{F}_{-\vartheta^\dagger}\kappa$, where $-\vartheta^\dagger\in\mathcal{T}$ by Lemma~\ref{gc:lem:stab}. By Proposition~\ref{gc:prop:unique} we have $\kappa\varsigma=\mathsf{F}_\tau\varsigma$ with $\tau\in\mathcal{T}$. Since $\kappa^2=1$,
\begin{equation}
\varsigma = \kappa^2\varsigma = \kappa\,\mathsf{F}_\tau\varsigma = \mathsf{F}_{-\tau^\dagger}\kappa\varsigma = \mathsf{F}_{\tau-\tau^\dagger}\varsigma \, ,
\end{equation}
so that $\tau^\dagger=\tau$ by the last statement of Lemma~\ref{gc:lem:frame}. Let $\varsigma'=\mathsf{F}_{\tau/2}\varsigma$. Then $\kappa\varsigma'=\mathsf{F}_{-\tau/2}\kappa\varsigma=\mathsf{F}_{\tau/2}\varsigma=\varsigma'$, that is $N'_I=-N_I'^\dagger$, so that $N'_I$ takes values in $\mathfrak{g}$.

\paragraph{Term of degree zero.} Let $\varsigma'\in\mathsf{S}$ have $N_I$ in $\mathfrak{g}$. By the first statement $N'^{(0)}_I$ is a constant which is covariant under the Lie algebra, so that $\vartheta=N'^{(0)}_Ig^I$ is an element of degree one which is covariant under the Lie algebra with values in $\mathfrak{g}$. The first two statements of Lemma~\ref{lem:frame} do not use the restriction on the degree of the parameter, and the transformation with $\mathsf{R}=\exp\vartheta$ gives $L'^I=(\mathsf{R}(g+L))^I-g^I$, which is a real pure pole without terms of degree zero and one, since $(\vartheta g)^I$ has degree two, and $N''_I=\mathsf{R}N'_I\mathsf{R}^{-1}-\pr_I\mathsf{R}\,\mathsf{R}^{-1}$ in $\mathfrak{g}$, whose term of degree zero is $N'^{(0)}_I-\pr_I\vartheta=0$. The bare functional and hence the renormalised functional do not change. It remains to show that $\vartheta$ annihilates the couplings, so that it lies in $\mathcal{T}$ and annihilates $L(g)$ by Lemma~\ref{gc:lem:stab}, and $L'=L$. The fermionic part of $\vartheta$ is a covariant linear map from the couplings to ${\rm u}(N_f)$. For $N\geq2$ the invariant linear maps from the Yukawa couplings vanish, since the vector representation has no invariant, so that the fermionic part is a multiple of $\mathrm{i}\lambda_{aabb}\,1$. For $N=1$ it is a combination of $\mathrm{i}Y$, $\mathrm{i}\,{\rm tr}(Y)\,1$ and $\mathrm{i}\lambda\,1$, which commute with $Y$. In both cases it annihilates the couplings. The scalar part is a covariant linear map from the couplings to ${\rm so}(N)$, which vanishes for $N=1$ and for $N\geq4$, since neither the vector representation nor the symmetric tensors of rank four contain the adjoint representation. For $N=2$ and $N=3$ such maps exist, $\lambda_{aabb}$ times the generator of ${\rm so}(2)$ and $\epsilon_{abc}\,{\rm tr}(Y^a)$ times the generators of ${\rm so}(3)$, and they are odd under a reflection $r$ of ${\rm O}(N)$. We use that $L^I$ is covariant under $r$. Then the bare functional is invariant under the constant rotation by $r$ of all bare sources, and $r\cdot\varsigma'\in\mathsf{S}$. By Proposition~\ref{gc:prop:unique} we have $r\cdot\varsigma'=\mathsf{F}_\beta\varsigma'$ with $\beta\in\mathcal{T}$, whose scalar part vanishes by Lemma~\ref{gc:lem:stab}. The transformation $\mathsf{F}_\beta$ changes the term of degree zero of $N_I$ by $-\pr_I\beta_{(1)}$ only, which has no scalar part. Hence the scalar part of $N'^{(0)}_I$ is invariant under $r$, and since the maps above are odd, it vanishes for all $N$.

\paragraph{Invariant element.} Let $\varsigma\in\mathsf{S}$ have $N_I$ in $\mathfrak{g}$ without term of degree zero, which is invariant under $G_0$ by the first statement, and let $r$ be a reflection of ${\rm O}(N)$. As above $r\cdot\varsigma=\mathsf{F}_{\beta}\varsigma$ with $\beta\in\mathcal{T}$. Since $r^2=1$ and $r\cdot\mathsf{F}_\beta=\mathsf{F}_{r\cdot\beta}\,r\cdot$, we find $\varsigma=\mathsf{F}_{\beta+r\cdot\beta}\varsigma$, and $\beta+r\cdot\beta=0$ by the last statement of Lemma~\ref{gc:lem:frame}. Let $\varsigma_1=\mathsf{F}_{\beta/2}\varsigma$. Then $r\cdot\varsigma_1=\mathsf{F}_{r\cdot\beta/2}\mathsf{F}_\beta\varsigma=\mathsf{F}_{\beta/2}\varsigma=\varsigma_1$, and $\varsigma_1$ is invariant under $G_0$ by the first statement and Lemma~\ref{gc:lem:frame}. The involution $\kappa$ of the paragraph on the antihermitian gauge field commutes with $r$, so that the parameter $\tau$ with $\kappa\varsigma_1=\mathsf{F}_\tau\varsigma_1$ satisfies $r\cdot\tau=\tau$, and $\varsigma_2=\mathsf{F}_{\tau/2}\varsigma_1$ is invariant under $r$ and has $N_I$ in $\mathfrak{g}$. Its term of degree zero is invariant under $r$, so that the transformation of the previous paragraph has a parameter which is invariant under $r$ and keeps the invariance. Since ${\rm O}(N)\times{\rm U}(N_f)$ is generated by $G_0$ and $r$, the result is the element of the fifth statement.

\paragraph{Same functional, vector field and one-form.} Two elements of $\mathsf{S}$ are related by some $\mathsf{F}_\vartheta$ by Proposition~\ref{gc:prop:unique}, which does not change the renormalised functional by Lemma~\ref{gc:lem:frame}. If both have $N_I$ in $\mathfrak{g}$ without term of degree zero, they are covariant by the first statement. The parameter $\vartheta$ which relates them satisfies $\vartheta=-\vartheta^\dagger$, by the relation of $\kappa$ and $\mathsf{F}_\vartheta$ and the freeness, so that it takes values in $\mathfrak{g}$, and it has no term of degree one, since the term of degree zero of $N_I$ changes by $-\pr_I\vartheta_{(1)}$. Lemma~\ref{lem:frame} with the weights of section~\ref{gf:sec:steps} then applies with its stated hypotheses and gives the same $\hB^I$ and $\hr_I$. This completes the proof of Proposition~\ref{gc:prop:cov}.

\subsubsection{Remarks}
\label{gc:sec:remarks}

We note that the argument for the covariance does not depend on the size of the stabiliser of the couplings. It needs that the set $\mathsf{S}$ is acted on freely and transitively by the additive group of a real vector space on which $G_0$ acts trivially. For $N=1$ and $N_f\geq2$ the stabiliser of a Yukawa matrix is not central, but its elements at a generic $Y$ commute, since they are diagonal in the eigenbasis of $Y$. The covariance under the reflections of ${\rm O}(N)$ enters only through $L^I$ and the fifth statement.

For multiscalar $\phi^4$ theory the group $\mathcal{T}$ is trivial, since the stabiliser of a generic quartic coupling is trivial. The covariance then follows directly from uniqueness, as in the last paragraph of the proof of Proposition~\ref{prop:uniqueall}, and the argument of this subsection reduces to it. The step on the covariance of $\breve\vartheta$ is not needed there, because the third step of that proof already gives $\breve\vartheta=0$.
\section{Sextic scalars in three dimensions}
\label{gs:app:sextic}
\label{gsx:sec:sextic}

In this appendix we prove Proposition~\ref{gsx:prop:props}. The derivation follows section~\ref{gs:sec:phi4prop} and appendices~\ref{app:weyl}, \ref{app:unique} and~\ref{app:finite}. The steps which are specific to four dimensions are the list of local structures of the counterterms and of the anomaly, the linear independence of the structures of flat space, the curvature counterterms, the relations among the coefficients and the terms of lowest order. We give their three-dimensional form, and for each step of the derivation for $\phi^4$ theory we state whether it holds without change, with a stated modification, or with a new argument, see the column of the sextic theory of table~\ref{gs:tab:steps}.

\subsection{Couplings, sources and counterterms}
\label{gsx:sec:hyp}

We consider $N$ real scalar fields with the free action of section~\ref{gs:sec:sphere} and the vertex $V^{(6)}[g]=\tfrac1{6!}\,\lambda_{i_1\ldots i_6}\phi_{i_1}\cdots\phi_{i_6}$. The couplings $g^I$ are the components of the symmetric tensor $\lambda_{i_1\ldots i_6}$, normalised such that $g^Ig^I=\lambda_{i_1\ldots i_6}\lambda_{i_1\ldots i_6}$, and the group of field rotations is ${\rm O}(N)$. As in Hypothesis~\ref{hyp:finite}, a function of the couplings with indices is covariant if it is unchanged by the action of every constant element of ${\rm O}(N)$, including the reflections. On the line \eqref{eq:line} the bare couplings have the weight $w_6=6-2d=2\vep$ of \eqref{gs:eq:wn}, so that $k_I=2$. The relevant sources are the source $\sU$ of the quartic operators, with the vertex $V^{(4)}[\sU]=\tfrac1{4!}\,\sU_{ijkl}\,\phi_i\phi_j\phi_k\phi_l$, and the source $\sM$ of the operators $\phi_i\phi_j$, with the vertex $V^{(2)}[\sM]$. Thus $V_\sU$ consists of the symmetric tensors with four indices and $V_\sM$ of the symmetric matrices, with the scalar products $\sU\cdot\sU'=\sU_{ijkl}\sU'_{ijkl}$ and $\sM\cdot\sM'=\sM_{ij}\sM'_{ij}$. On the line the weights of their bare sources are $w_4=4-d=1+\vep$ and $w_2=2$, and their dimensions at $d=3$ are one and two. The vertices with an odd number of legs are not included. The reflection $\phi\to-\phi$ is the element $-1$ of ${\rm O}(N)$. It leaves all sources above invariant, and it multiplies the sources of the vertices with an odd number of legs by $-1$, so that the covariant counterterms of the functional of the sources above do not contain the latter. This symmetry is the reason for their absence. Their dimensions at $d=3$ are $\tfrac52$, $\tfrac32$ and $\tfrac12$, so that at vanishing odd sources they would not contribute to the identities by Lemma~\ref{gs:lem:dimension}.

We count dimensions at $d=3$, where $g^I$ has dimension zero, $D_\mu$ and $a_\mu$ dimension one, the curvature dimension two, $\sU$ dimension one and $\sM$ dimension two.

\begin{lemma}
\label{gsx:lem:forms}
A local scalar of dimension $\Delta$ which is a polynomial in $\sU$, $\sM$, the curvature, the field strength $f_{\mu\nu}$ and the covariant derivatives of the couplings, built covariantly from $\gamma_{\mu\nu}$ with $d$ as a parameter in the sense of Property~\ref{gs:P:weyl}, with coefficients which are functions of the couplings, is a linear combination of the following terms. For $\Delta=1$ the term $\sU$. For $\Delta=2$ the terms $\sM$, $\sU\sU$, $D^2g^I$, $D_\mu g^ID^\mu g^J$ and $R$. For $\Delta=3$ the terms $\sU R$, $\sU D^2g^I$, $\sU D_\mu g^ID^\mu g^J$, $\sU\sM$ and $\sU\sU\sU$. Here products of $\sU$ and $\sM$ stand for multilinear maps of the components. A one-form of dimension one of this kind is a combination of the $D_\mu g^I$. Expressions with covariant derivatives of $\sU$ are not of this kind. They are the one-form $D_\mu\sU$ of dimension two and the scalars $D^2\sU$ and $D_\mu\sU\,D^\mu g^I$ of dimension three. In an integrated density, with or without a factor $\sigma$ or its derivatives, integrations by parts move the derivatives from $\sU$ to the other factors, so that up to total derivatives the terms above suffice.
\end{lemma}

\begin{proof}
Let a term contain $n_\sU$ factors of $\sU$ and $n_\sM$ factors of $\sM$. The remaining factor is built from the metric, the curvature, $f_{\mu\nu}$ and covariant derivatives of the couplings, and it has the dimension $\Delta-n_\sU-2n_\sM$. As in the proof of Lemma~\ref{gs:lem:odd} its number of indices equals its dimension modulo two, and without the Levi-Civita tensor a scalar has an even number of indices. The scalars of dimension zero are the functions of the couplings, and those of dimension two are $D^2g^I$, $D_\mu g^ID^\mu g^J$ and $R$, since $f_{\mu\nu}$ is antisymmetric and there is no antisymmetric tensor of dimension zero to contract it with, and since a second covariant derivative of a function of the couplings is a combination of these terms. For $\Delta=1$ the only possibility is $(n_\sU,n_\sM)=(1,0)$ with a remainder of dimension zero. For $\Delta=2$ the possibilities are $(0,1)$ and $(2,0)$ with a remainder of dimension zero, and $(0,0)$ with a remainder of dimension two. For $\Delta=3$ they are $(1,0)$ with a remainder of dimension two, and $(1,1)$ and $(3,0)$ with a remainder of dimension zero. A one-form of dimension one has one index and contains one covariant derivative, so that it is a combination of the $D_\mu g^I$ with coefficients which are functions of the couplings.
\end{proof}

By the lemma the bare sources \eqref{gsx:eq:bare} of appendix~\ref{gs:sec:cases} are the most general ones of the form of section~\ref{gs:sec:msanomaly}, where $N_I$ takes values in the Lie algebra, $\fd_I$, $\fe_{IJ}=\fe_{JI}$ and $\fk$ take values in $V_\sM$. Up to total derivatives the most general density of counterterms which do not contain the fields is
\begin{equation}
\cX = \fn\cdot\sU\,R+\fj_I\cdot\sU\,D^2g^I+\fl_{IJ}\cdot\sU\,D_\mu g^ID^\mu g^J+\sU\cdot\fm\cdot\sM+\tfrac16\,\fz(\sU,\sU,\sU) \, ,
\label{gsx:eq:X}
\end{equation}
where $\fn$, $\fj_I$ and $\fl_{IJ}=\fl_{JI}$ take values in the dual of $V_\sU$, $\fm$ is a bilinear form on $V_\sU\times V_\sM$ and $\fz$ is a symmetric trilinear form on $V_\sU$. All coefficients are functions of the couplings. Every term of $\cX$ contains $\sU$, and there is no term of dimension three which contains only the couplings, the gauge field and the curvature. In flat space $\fk$ and $\fn$ drop out. We call $\fk$ and $\fn$ the curvature counterterms, and we write $\cX^{\rm fl}$ for \eqref{gsx:eq:X} without its first term. The flat functional $W_0^{\rm fl}$ is that of section~\ref{sec:chart} with the vertices $V^{(6)}$, $V^{(4)}$, $V^{(2)}$, the vertex of the current and the term of second order in the gauge field, on the line \eqref{eq:line} and with $D_0=1$.

The terms of the hypothesis have the meaning of section~\ref{sec:hypothesis}, and in flat space the bare sources \eqref{gsx:eq:bare} are taken without the term with $\fk$. Hypothesis~\ref{hyp:finite} for the sextic theory differs from Hypothesis~\ref{hyp:finite} for multiscalar $\phi^4$ theory in the interaction, in the additional source $\sU$ with its factor $Z_\sU$, in the term of $\sM_0$ which is quadratic in $\sU$, and in the counterterms \eqref{gsx:eq:X}. As for $\phi^4$ theory only $L^I$ is assumed to be covariant. The term with $\fh$ has a pole already in the free theory. At order zero in the couplings the two vertices $V^{(4)}[\sU]$ joined by three propagators give $\int\!\int\sU(x)\,\sU(y)\,|x-y|^{-3(d-2)}$ times a bilinear expression in the fields, which has a simple pole at $d=3$, so that $\fh$ has a simple pole at order zero. In the same way three vertices $V^{(4)}[\sU]$ joined in a ring by pairs of propagators give a pole which is removed by $\fz$ at order zero. The statement that $L^I$ are the counterterms of minimal subtraction of the theory with constant couplings in momentum space is the analogue of Identification~\ref{id:ms}, and we do not use it.

\subsection{Weyl variation of the sources}
\label{gsx:sec:weyl}

The bare action is invariant under a Weyl rescaling if the bare couplings are rescaled with the weight $-2\vep$, the bare source $\sU_0$ with the weight $-(4-d)$ and the bare source $\sM_0$ with the weight $-2$, while the bare gauge field is unchanged, and the bare functional is invariant by \eqref{eq:defcurved} with these weights. We follow section~\ref{sec:weylvar}. For the couplings and the gauge field the variation is determined by the equations \eqref{eq:xiL}, in which $\hb^{{\rm L}\,J}\pr_Jg_0^I=-2\vep\,g_0^I$. All couplings have the same weight $k_I=2$, so that Lemmas~\ref{lem:rhoxi} and~\ref{lem:frame} hold with $\vep$ replaced by $2\vep$ in the coupling sector, which is \eqref{gf:eq:zeroparts} with $k_I=2$. Thus $\hB^I=-2\vep\,g^I+B^I$ with $B^I=\beta^I-(\xi g)^I$, $\beta^I=2(g^J\pr_J-1)L^I_1$, $\xi=-2N_{I,1}g^I$ and $\hr_I=2g^J(\pr_JN_{I,1}-\pr_IN_{J,1})$, once $\hB^I$ and $\hr_I$ are known to be regular, and $\hr_I\hB^I=0$. The term of degree $k$ of the parameter $\vartheta$ of Lemma~\ref{lem:frame} is $-{\rm PP}[\xi^{\rm L}_{(k)}+\mathsf{F}_k]/(2\vep k)$. We choose the bare sources with $\xi^{\rm L}=\xi$ as there.

For an element $\omega$ of the Lie algebra we write $\omega\sU$ for its action on $V_\sU$, and $[\omega,\sM]$ for its action on $V_\sM$. The bare source $\sU_0$ has to change by its weight and by the gauge transformation with the parameter $\sigma\xi^{\rm L}$, that is $\delta\sU_0=-(4-d)\,\sigma\,\sU_0-\sigma\,\xi^{\rm L}\sU_0$. With the linear map $\gamma_\sU$ of $V_\sU$ defined by
\begin{equation}
\hB^I\pr_IZ_\sU = -Z_\sU\,\gamma_\sU-\xi^{\rm L}\circ Z_\sU \, ,
\label{gsx:eq:gammaU}
\end{equation}
where $\xi^{\rm L}\circ Z_\sU$ is the composition of $Z_\sU$ with the action of $\xi^{\rm L}$, this condition is
\begin{equation}
\delta^{\rm W}_\sigma\sU = -\sigma\,(4-d-\gamma_\sU)\,\sU \, .
\label{gsx:eq:varU}
\end{equation}
For $\sM_0$ we proceed as in appendix~\ref{app:masscoeff}. Let $\mathsf{u}$ be the argument of $Z_\sM$ in \eqref{gsx:eq:bare}. With $\gamma_\sM$ of \eqref{eq:gammaM} the condition on $\sM_0$ is $\delta\mathsf{u}=-\sigma\,(2-\gamma_\sM)\,\mathsf{u}$. The terms of $\mathsf{u}$ which do not contain $\sU$ give the coefficients \eqref{eq:masscoeff} without change. The term $-\tfrac12\fh(\sU,\sU)$ changes by $-\tfrac12\sigma\,(\hB\!\cdot\!\pr\fh)(\sU,\sU)-\fh(\sU,\delta^{\rm W}_\sigma\sU)$, and we obtain
\begin{equation}
\delta^{\rm W}_\sigma\sM = -\sigma\left[(2-\gamma_\sM)\sM+\delta_I\,D^2g^I+\epsilon_{IJ}\,D_\mu g^ID^\mu g^J+\bar\eta\,R+\tfrac12\,\bar\chi(\sU,\sU)\right]-2\,\pr_\mu\sigma\,\theta_I\,D^\mu g^I-\bar\tau\,\nabla^2\sigma \, ,
\label{gsx:eq:varM}
\end{equation}
with $\delta_I$, $\epsilon_{IJ}$, $\theta_I$, $\bar\eta$ and $\bar\tau$ given by \eqref{eq:masscoeff} and
\begin{equation}
\bar\chi = (2\vep-\mathcal{L}_\sU)\,\fh \, , \qquad (\mathcal{L}_\sU\fh)(\sU,\sU') = (\hB\!\cdot\!\pr\fh)(\sU,\sU')-\gamma_\sM\fh(\sU,\sU')+\fh(\gamma_\sU\sU,\sU')+\fh(\sU,\gamma_\sU\sU') \, .
\label{gsx:eq:chi}
\end{equation}
Indeed, the terms of $\delta\mathsf{u}+\sigma(2-\gamma_\sM)\mathsf{u}$ which are quadratic in $\sU$ are $\sigma$ times $-\tfrac12\bar\chi(\sU,\sU)-\tfrac12(\hB\!\cdot\!\pr\fh)(\sU,\sU)+(3-d)\,\fh(\sU,\sU)-\fh(\sU,\gamma_\sU\sU)+\tfrac12\gamma_\sM\fh(\sU,\sU)$, and their vanishing is \eqref{gsx:eq:chi}, since $3-d=\vep$. Each term of \eqref{gsx:eq:varU} and \eqref{gsx:eq:varM} has the dimension of its source at $d=3$, as Property~\ref{gs:P:weyl} requires, and by Lemma~\ref{gsx:lem:forms} these are the most general variations of this kind up to the absence of terms with derivatives of $\sigma$ in \eqref{gsx:eq:varU}, which has no term of dimension one with a derivative of $\sigma$.

The operators $\mathbb{W}_\sigma$ of \eqref{gs:eq:weylop} commute by Lemma~\ref{gcu:lem:commute}, since the variations of the couplings, of the gauge field and of the sources $\sU$ and $\sM$ are defined by the condition that the bare sources change by their weights and by the gauge transformation with the parameter $\sigma\,\xi^{\rm L}$. On $\sM$ the terms of \eqref{gsx:eq:varM} with $\sU$ and the terms of \eqref{gsx:eq:varU} contain no derivatives of $\sigma$, so that in components the commutation on $\sM$ is again the pair of relations \eqref{eq:Mrel}, as in section~\ref{gs:sec:dimtwo}.

\subsection{Uniqueness and covariance}
\label{gsx:sec:unique}

We first give the three-dimensional form of the linear independence of the local structures of flat space. The coefficient functions are power series in the couplings, and the lemma holds in the graded form explained before Lemma~\ref{lem:indep}.

\begin{lemma}
\label{gsx:lem:indep}
Let the sources be in the class of flat sources with $D_0=1$, and let $d=3$ or let $d$ be generic. First, if
\begin{equation}
\int d^dx\,\Big[\mathsf{j}_I\cdot\sU\,D^2g^I+\mathsf{k}_{IJ}\cdot\sU\,D_\mu g^ID^\mu g^J+\sU\cdot\mathsf{l}\cdot\sM+\tfrac16\,\mathsf{z}(\sU,\sU,\sU)\Big] = 0
\label{gsx:eq:lemA}
\end{equation}
for all sources, where $\mathsf{k}_{IJ}$ is symmetric and $\mathsf{z}$ is a symmetric trilinear form, then all coefficients vanish. Second, let $b_\mu$ be a nonzero constant vector in the parallel direction. If $\int d^dx\;b_\mu\,\mathsf{q}_I\cdot\sU\,D^\mu g^I=0$ for all sources, then $\mathsf{q}_I=0$.
\end{lemma}

\begin{proof}
We set $g=g_c\,\chi+t\,\gamma$ as in the proof of Lemma~\ref{lem:indep}, with $\gamma^I=c^Iu$, $a_\mu=0$, $\sU=\mu\,w$ and $\sM=\nu\,w'$, where $\mu$ and $\nu$ are constant elements of $V_\sU$ and $V_\sM$ and $u$, $w$ and $w'$ are functions of the class of flat sources with $\chi=1$ on their supports. The coefficients are then evaluated at $g_c$, and each order in $t$, $\mu$ and $\nu$ vanishes separately. The order $\mu\nu t^0$ is $\mu\cdot\mathsf{l}\cdot\nu\int w\,w'$, the order $\mu^3t^0$ is $\tfrac16\mathsf{z}(\mu,\mu,\mu)\int w^3$, and for positive $w$ and $w'$ the integrals are positive. Hence $\mathsf{l}=0$, and $\mathsf{z}(\mu,\mu,\mu)=0$ for all $\mu$, which gives $\mathsf{z}=0$ by polarisation. Since $\mathsf{l}$ and $\mathsf{z}$ vanish as functions of $g_c$, their derivatives vanish as well. The order $\mu\,t$ is $\mathsf{j}_I\cdot\mu\,c^I\int w\,\pr^2u$, and for $w=u$ the integral is $-\int\pr u\!\cdot\!\pr u$, so that $\mathsf{j}_I=0$. The order $\mu\,t^2$ is then $\mathsf{k}_{IJ}\cdot\mu\,c^Ic^J\int w\,\pr u\!\cdot\!\pr u$, which for positive $w$ gives $\mathsf{k}_{IJ}=0$, since $\mathsf{k}_{IJ}$ is symmetric. For the second statement the order $\mu\,t$ is $\mathsf{q}_I\cdot\mu\,c^I\int w\,\pr_bu$ with $\pr_b=b^\mu\pr_\mu$, and $w=\pr_bu$ gives a positive integral for $u\neq0$. For real $d>1$ all these integrals are positive, since the measure of the class is positive. For functions which vanish near $x_\perp=0$ the integrals are entire functions of $d$, so that they do not vanish for generic $d$.
\end{proof}

In contrast to Lemma~\ref{lem:indep} the second statement contains no total derivative, so that the coefficient of the only structure with $\pr_\mu\sigma$ of the anomaly is determined in flat space.

\begin{lemma}
\label{gsx:lem:potential}
Let $\eta$ be a continuous function on $\mathbb{R}^3$ with compact support, and let $\int d^3x\,\eta(x)\,|x-y|^{-2\kappa}$ vanish for all $y$ outside a ball, where $\kappa=1$, $2$ or $3$. Then $\eta=0$.
\end{lemma}

\begin{proof}
We follow the proof of Lemma~\ref{lem:potential}. For $|y|$ larger than the radius of a ball which contains the support we expand $|x-y|^{-2\kappa}=\sum_l|x|^l|y|^{-2\kappa-l}\,C_l^{(\kappa)}(\hat x\cdot\hat y)$, and the integrals of $\eta(x)\,|x|^lC_l^{(\kappa)}(\hat x\cdot\hat y)$ vanish for all $l$ and all unit vectors $\hat y$. In three dimensions the spherical harmonics are given by the Legendre polynomials $P_s=C_s^{(1/2)}$, and $|x|^sP_s(\hat x\cdot\hat y)$ is a harmonic polynomial in $x$ and a spherical harmonic of degree $s$ in $\hat y$. By the connection formula eq.~(18.18.16) of~\cite{DLMF},
\begin{equation}
C_l^{(\kappa)} = \sum_{j=0}^{\lfloor l/2\rfloor}\frac{(\kappa)_{l-j}\,(\kappa-\tfrac12)_j\,(\tfrac12+l-2j)}{(\tfrac12)_{l-j+1}\,j!}\;P_{l-2j} \, ,
\label{gsx:eq:connection}
\end{equation}
and for $\kappa>\tfrac12$ all coefficients are positive.\footnote{We verified \eqref{gsx:eq:connection} and the positivity of its coefficients in exact arithmetic for $l\leq14$ and $\kappa=1,2,3$, by expanding the Gegenbauer polynomials in Legendre polynomials.} Spherical harmonics of different degrees are linearly independent, so that the integrals of $\eta(x)\,|x|^{2j}\,|x|^sP_s(\hat x\cdot\hat y)$ vanish for all $j$, $s$ and $\hat y$. The functions $|x|^sP_s(\hat x\cdot\hat y)$ span the harmonic polynomials of degree $s$, every polynomial is a sum of harmonic polynomials multiplied by powers of $x^2$, and therefore all moments of $\eta$ vanish and $\eta=0$.
\end{proof}

Lemma~\ref{lem:transverse} holds on $\mathbb{R}^3$ with the kernel $(\pr_\mu\pr_\nu-\delta_{\mu\nu}\pr^2)\,|x-y|^{-2}$ in place of \eqref{eq:Apot}, with the same proof. Two integrations by parts give $A_\nu(y)=\int d^3x\,\eta_\nu(x)\,|x-y|^{-2}$ with $\eta_\nu=\pr_\mu f_{\nu\mu}$, Lemma~\ref{gsx:lem:potential} with $\kappa=1$ gives $\pr_\mu f_{\nu\mu}=0$, hence $\pr^2f_{\mu\nu}=0$ and $f_{\mu\nu}=0$, and a closed one-form on $\mathbb{R}^3$ whose components have compact support is the differential of a function of compact support, since the complement of a ball in $\mathbb{R}^3$ is connected.

\begin{lemma}
\label{gsx:lem:lowest}
Let the coefficients of the bare sources \eqref{gsx:eq:bare} in flat space and of $\cX^{\rm fl}$ be real pure poles, where $L^I$ has no terms of degree zero and one, and let the functional \eqref{eq:Wfl} be finite for all sources of the class of flat sources with $D_0=1$ which do not depend on $\vep$. Then $Z_\sU=1$, $Z_\sM=1$ and $\fd_I=0$ at $g=0$. Moreover, let the sources with the index $2$ have supports which are disjoint from the supports of the sources with the index $1$. Then the terms of second order of $W^{\rm fl}$ with one source of each index vanish for two sources of different kinds. For two couplings, two sources $\sU$, two sources $\sM$ and two gauge fields they are, up to factors which are analytic and do not vanish at $d=3$, the integrals over both points of
\begin{equation}
\frac{g_1^I(x)\,g_2^I(y)}{|x-y|^{6(d-2)}} \, , \quad \frac{\sU_1(x)\cdot\sU_2(y)}{|x-y|^{4(d-2)}} \, , \quad \frac{\sM_1(x)\cdot\sM_2(y)}{|x-y|^{2(d-2)}} \, ,
\end{equation}
\begin{equation}
a_{1\mu}(x)\cdot a_{2\nu}(y)\left(\pr_\mu\pr_\nu-\delta_{\mu\nu}\pr^2\right)\frac{1}{|x-y|^{2(d-2)}} \, .
\label{gsx:eq:kernels}
\end{equation}
\end{lemma}

\begin{proof}
The proof of Lemma~\ref{lem:lowest} applies, with the following changes. The terms of first order of the bare sources are $g^I$, $a_\mu+N^{(0)}_I\pr_\mu g^I$, $Z^{(0)}_\sU\sU$ and $Z^{(0)}_\sM(\sM-\fd^{(0)}_I\pr^2g^I)$, since $\tfrac12\fh(\sU,\sU)$ is of second order in the sources. The two-point functions of the free theory of the operators of $V^{(6)}$, $V^{(4)}$ and $V^{(2)}$ are multiples of $\mathsf{G}^6$, $\mathsf{G}^4$ and $\mathsf{G}^2$ with $\mathsf{G}=|x|^{-(d-2)}$, and the kernel of the currents is the one of Lemma~\ref{lem:lowest}. Two vertices with different numbers of legs have no pairing without a propagator at coincident points, and a coupling, a source $\sU$ or a source $\sM$ together with a vertex of the current gives no term for the reason given there. For two sources $\sU$ the argument for $Z^{(0)}_\sM$ applies with the positive scalar product of $V_\sU$, which gives $Z^{(0)}_\sU=1$, and the arguments for $Z^{(0)}_\sM$ and $\fd^{(0)}_I$ are unchanged, with Lemma~\ref{gsx:lem:potential} for $\kappa=1$ in place of Lemma~\ref{lem:potential} at $d=3$. Here and below the integral \eqref{eq:gram} of two points with disjoint supports is analytic at $d=3$, where it is the ordinary integral, since $m=d-D_0=2$ is not smaller than the number of points.
\end{proof}

\begin{proposition}
\label{gsx:prop:uniqueall}
Let $L^I$ be a real covariant pure pole without terms of degree zero and one in the couplings. Consider sets of real pure poles $N_I$, $Z_\sU-1$, $Z_\sM-1$, $\fd_I$, $\fe_{IJ}$, $\fh$ and of coefficients of $\cX^{\rm fl}$. First, there is at most one such set for which the functional \eqref{eq:Wfl} is finite for all sources of the class of flat sources with $D_0=1$ which do not depend on $\vep$. Second, if such a set exists, its members are covariant under the field rotations, and $N_I$ has no term of degree zero in the couplings.
\end{proposition}

\begin{proof}
We follow the proof of Proposition~\ref{prop:uniqueall}, with the kernels \eqref{gsx:eq:kernels}, Lemma~\ref{gsx:lem:potential} in place of Lemma~\ref{lem:potential} and Lemma~\ref{lem:transverse} on $\mathbb{R}^3$ as stated after Lemma~\ref{gsx:lem:potential}. We list the changes.

The identities of the paragraph on gauge invariance at low orders hold, with $W_2$ the term of second order in the sextic couplings. The term with two gauge fields and one coupling vanishes identically, since a vertex with six legs cannot be joined to two vertices of the current, which have four legs together, without a propagator at coincident points. In the paragraph on the term of degree zero the stabiliser in ${\rm so}(N)$ of the sextic coupling $\lambda_{i_1\ldots i_6}=\sum_r\mathsf{d}_r\,\delta_{i_1r}\cdots\delta_{i_6r}$ with all $\mathsf{d}_r\neq0$ is trivial, since $(\omega\lambda)_{ijjjjj}=\omega_{ij}\mathsf{d}_j$ for $i\neq j$, so that the stabiliser in ${\rm so}(N)$ is trivial on a nonempty Zariski open set of couplings.\footnote{For $2\leq N\leq5$ we also checked in exact arithmetic that the stabiliser in ${\rm so}(N)$ of a random sextic tensor is trivial. For $N=2,3$ we checked that there is no ${\rm O}(N)$-invariant linear map from the sextic tensors to the antisymmetric matrices.} The argument of that paragraph shows that $N^{(0)}_I$ is unique and, since a rotation with a constant element of ${\rm O}(N)$ maps the set of coefficients to a set for the same covariant $L^I$, that $N^{(0)}_Ig^I$ is an ${\rm O}(N)$-invariant linear map from the sextic tensors to the Lie algebra. An invariant tensor of ${\rm O}(N)$ is a combination of products of $\delta_{ij}$, and its antisymmetrisation in the two indices vanishes, so that $N_I$ has no term of degree zero.\footnote{If $L^I$ were covariant only under ${\rm SO}(N)$, the Levi-Civita tensor would allow for $N=2$ the term $N^{(0)}_Ig^I=c\,\lambda_{iijjkk}\,J$ with the generator $J$ of ${\rm so}(2)$. It is removed by the transformation of Lemma~\ref{lem:frame} with the parameter $\vartheta=N^{(0)}_Ig^I$, whose first statement does not use the restriction on the degree, and which replaces $L^I$ by another real covariant pure pole without terms of degree zero and one.}

In the induction we say that the terms of degree $n-1$ of $N_I$ and $\fd_I$, the term of degree $n-2$ of $\fe_{IJ}$ and the terms of degree $n$ of $Z_\sM$, $Z_\sU$ and $\fh$ have the level $n$. These are the terms which enter the bare sources at order $n$ in the couplings. The level now starts at zero, and at the level zero only $Z_\sU$, $Z_\sM$ and $\fh$ have terms. The first two are fixed by Lemma~\ref{gsx:lem:lowest}, and for the third we carry out the sixth step below with $n=0$ before the induction, which then runs over $n\geq1$ as in Proposition~\ref{prop:uniqueall}. The first four steps are unchanged. In the third step the differences of $Z_\sU$ and $\fh$ multiply terms with $\sU$ and do not contribute, and in the steps with sources $\sM$ the difference $\tilde Z_\sM$ multiplies the bracket of \eqref{gsx:eq:bare}, whose term $-\tfrac12\fh(\sU,\sU)$ vanishes at $\sU=0$. We add two steps.

Fifth step. We take couplings $g_1$ and two sources $\sU_1$ and $\sU_2$, where the support of $\sU_1$ may overlap with the support of $g_1$ and $\sU_2$ has the index $2$, and we consider the terms of order $n$ in $g_1$ which are linear in $\sU_1$ and in $\sU_2$. After the first four steps the bare sources differ at this order by $\tilde Z_\sU(g_1)\,\sU_1$ in $\sU_0$ only. The difference $\tilde\fh(g_1)(\sU_1,\sU_2)$ vanishes, since the supports of $\sU_1$ and $\sU_2$ are disjoint, and $\tilde Z_\sM$ vanishes by the fourth step. The terms considered are therefore the second expression of \eqref{gsx:eq:kernels} with $\tilde Z_\sU(g_1)\sU_1$ and $\sU_2$, and as in the second step the leading pole of $\tilde Z_\sU(g_1)\,\sU_1$ vanishes at each point, so that $\tilde Z_\sU=0$.

Sixth step. We take couplings $g_1$, two sources $\sU_1$ and $\sU'_1$ whose supports may overlap with that of $g_1$, and a source $\sM_2$ with the index $2$, and we consider the terms of order $n$ in $g_1$ which are linear in $\sU_1$, $\sU'_1$ and $\sM_2$. After the previous steps the bare sources differ at this order by $-\tilde\fh(g_1)(\sU_1,\sU'_1)$ in $\sM_0$ only, and the terms considered are the third expression of \eqref{gsx:eq:kernels} with this difference and $\sM_2$. The leading pole of $\tilde\fh(g_1)(\sU_1,\sU'_1)$ vanishes at each point, and since the values of $g_1$, $\sU_1$ and $\sU'_1$ at a point are arbitrary, $\tilde\fh=0$. At the level zero this step gives the uniqueness of the term of order zero of $\fh$.

For the counterterms we use the first part of Lemma~\ref{gsx:lem:indep} in place of Lemma~\ref{lem:indep}. The argument for the covariance is unchanged.
\end{proof}

\subsection{Regularity of the Weyl variation}
\label{gsx:sec:regvar}

\begin{lemma}
\label{gsx:lem:regvar}
Assume Hypothesis~\ref{hyp:finite} for the sextic theory. Then $\hB^I$, $\hr_I$, $\gamma_\sU$, $\gamma_\sM$, $\delta_I$, $\epsilon_{IJ}$, $\theta_I$ and $\bar\chi$ are regular.
\end{lemma}

\begin{proof}
We follow the proof of Lemma~\ref{lem:regvar}. The flat form \eqref{eq:flatvar} of the variation is supplemented by $\delta\sU=\delta^{\rm W}_\sigma\sU-v^\mu\pr_\mu\sU$, it changes the bare sources by the conformal transformation \eqref{eq:barevar} with the weights $2\vep$, $1+\vep$ and $2$ and by a gauge transformation, and the identity \eqref{eq:flatward} holds with the additional term $\delta\sU\cdot\delta W^{\rm fl}/\delta\sU$ on the left. In the paragraph on the lowest order, Lemma~\ref{gsx:lem:lowest} and Proposition~\ref{gsx:prop:uniqueall} give that $\hr_I$, $\delta_I$, $\theta_I$, $\gamma_\sU$ and $\gamma_\sM$ vanish at $g=0$ and that the term of first order of $\hB^I$ is $-2\vep\,g^I$. The terms of first order in the sources of the variation are then $-2\vep\,\sigma g^I-v^\mu\pr_\mu g^I$, $-\cL_va_\mu$, $-(1+\vep)\sigma\sU-v^\mu\pr_\mu\sU$ and $-2\,\sigma\sM-v^\mu\pr_\mu\sM$, which have no poles. We assign the level $n$ to the terms of degree $n$ of $\hB^I$, $\gamma_\sU$, $\gamma_\sM$ and $\bar\chi$, to the terms of degree $n-1$ of $\hr_I$, $\delta_I$ and $\theta_I$ and to the term of degree $n-2$ of $\epsilon_{IJ}$. The level of $\bar\chi$ starts at zero. In the induction the steps for the gauge field, the couplings and the source $\sM$ are those of Lemma~\ref{lem:regvar}, with the kernels \eqref{gsx:eq:kernels} and Lemma~\ref{gsx:lem:potential} for $\kappa=3$ and $\kappa=1$, and with Lemma~\ref{lem:transverse} on $\mathbb{R}^3$. In these steps the terms of the variation with $\sU$ multiply terms of $W^{\rm fl}$ with $\sU$, which do not occur at the orders considered. We add two steps after the step for the source $\sM$.

For $\gamma_\sU$ we take couplings $g_1$, a source $\sU_1$ whose support may overlap with that of $g_1$, and a source $\sU_2$ with the index $2$, and we consider the terms of \eqref{eq:flatward} of order $n$ in $g_1$ which are linear in $\sU_1$ and $\sU_2$. All terms of the variation are regular by the previous steps and by the induction, except the term $\sigma\,\gamma_\sU\sU_1$ with $\gamma_\sU$ of degree $n$, and the term $-\tfrac12\sigma\bar\chi(\sU_1,\sU_2)$ vanishes since the supports are disjoint. The regular terms multiply derivatives of $W^{\rm fl}$ which are finite by Hypothesis~\ref{hyp:finite} for the sextic theory. The term with $\gamma_\sU$ multiplies the second expression of \eqref{gsx:eq:kernels}, and as in the step for $\gamma_\sM$ the leading pole of $\gamma_\sU(g_1)\,\sU_1$ vanishes at each point, for constant $\sigma$, so that the leading pole of $\gamma_\sU$ vanishes.

For $\bar\chi$ we take couplings $g_1$, two sources $\sU_1$ and $\sU'_1$ whose supports may overlap with that of $g_1$, and a source $\sM_2$ with the index $2$, and we consider the terms of order $n$ in $g_1$ which are linear in $\sU_1$, $\sU'_1$ and $\sM_2$. The only term of the variation which is not yet known to be regular is $-\sigma\,\bar\chi(\sU_1,\sU'_1)$ with $\bar\chi$ of degree $n$, which multiplies the third expression of \eqref{gsx:eq:kernels}. For constant $\sigma$ its leading pole vanishes at each point, and hence the leading pole of $\bar\chi$ vanishes. For $n=0$ this step uses only the terms of first order of the variation and $\gamma_\sU=0$ at $g=0$. This completes the induction.
\end{proof}

By the fourth statement of Lemma~\ref{lem:rhoxi} with the weight $2$ the coefficients $\hB^I$ and $\hr_I$ have the form given in section~\ref{gsx:sec:weyl}. The argument after the proof of Lemma~\ref{lem:regvar} shows in the same way that the coefficients of the variation are determined by the renormalised functional.

\subsection{Anomaly and curvature counterterms}
\label{gsx:sec:anomaly}

For a coefficient $Y$ with lower indices along the couplings and with values in the dual of $V_\sU$ we define
\begin{equation}
\mathcal{D}_\sU Y = (2\vep-\tL)\,Y-Y\cdot\gamma_\sU \, ,
\label{gsx:eq:Dop}
\end{equation}
where $(Y\cdot\gamma_\sU)\cdot\sU=Y\cdot(\gamma_\sU\sU)$ and $\tL$ acts on the indices along the couplings. For a bilinear form $\fm$ on $V_\sU\times V_\sM$ and an element $X$ of $V_\sM$ we write $\fm\cdot X$ for the element of the dual of $V_\sU$ with $(\fm\cdot X)\cdot\sU=\sU\cdot\fm\cdot X$.

\begin{lemma}
\label{gsx:lem:anomaly}
On the backgrounds $\Omega^2\delta_{\mu\nu}$ and on the sphere, with $d$ as a parameter, the variation $\mathbb{W}_\sigma$ of the integral of $\cX$ is $\cA[\sigma]$ with
\begin{align}
\cA[\sigma] = \int d^dx\sqrt\gamma\,\Big\{&\sigma\Big[i_\sU\cdot\sU\,R-J_I\cdot\sU\,D^2g^I-K_{IJ}\cdot\sU\,D_\mu g^ID^\mu g^J \notag\\
&-\sU\cdot\beta_{\sU\sM}\cdot\sM-\tfrac16\,\beta_\sU(\sU,\sU,\sU)\Big] \notag\\
&+2\,\pr_\mu\sigma\,L_I\cdot\sU\,D^\mu g^I+\nabla^2\sigma\;t_\sU\cdot\sU\Big\} \, ,
\label{gsx:eq:anomaly}
\end{align}
where
\begin{align}
i_\sU &= -\mathcal{D}_\sU\fn-\fm\cdot\bar\eta \, , \qquad
J_I = (\mathcal{D}_\sU\fj)_I+\fm\cdot\delta_I \, , \notag\\
K_{IJ} &= (\mathcal{D}_\sU\fl)_{IJ}-\hat\Omega_{IJ}{}^K\fj_K+\fm\cdot\epsilon_{IJ} \, , \notag\\
\sU\cdot\beta_{\sU\sM}\cdot\sM &= \sU\cdot\big[(2\vep-\hB\!\cdot\!\pr)\,\fm\big]\cdot\sM-(\gamma_\sU\sU)\cdot\fm\cdot\sM \notag\\
&\quad-\sU\cdot\fm\cdot(\gamma_\sM\sM) \, , \notag\\
\beta_\sU(\sU,\sU,\sU) &= \big[(4\vep-\hB\!\cdot\!\pr)\,\fz\big](\sU,\sU,\sU)-3\,\fz(\gamma_\sU\sU,\sU,\sU) \notag\\
&\quad+3\,\sU\cdot\fm\cdot\bar\chi(\sU,\sU) \, , \notag\\
L_I &= \hP_I{}^K\fj_K+\fl_{IK}\hB^K-\fm\cdot\theta_I \, , \qquad
t_\sU = \fj_I\hB^I-2(d-1)\,\fn-\fm\cdot\bar\tau \, .
\label{gsx:eq:coeffs}
\end{align}
The identity holds pointwise, without integrations by parts. The coefficients $J_I$, $K_{IJ}$, $\beta_{\sU\sM}$, $\beta_\sU$ and $L_I$ contain only counterterms of flat space, and $i_\sU$ and $t_\sU$ contain the curvature counterterms.
\end{lemma}

\begin{proof}
We use the rules \eqref{eq:weylrules} and \eqref{eq:varDg} and the variations \eqref{gsx:eq:varU} and \eqref{gsx:eq:varM}, and we write $\mathsf{N}=\pr\cX/\pr\sU=\fn R+\fj_ID^2g^I+\fl_{IJ}D_\mu g^ID^\mu g^J+\fm\cdot\sM+\tfrac12\fz(\sU,\sU,\cdot)$. In the variation of each term of \eqref{gsx:eq:X} the weights of $\sqrt\gamma$, of $\sU$, of the inverse metric and of $R$ add up to $d-(4-d)-2=-2\vep$ for the terms with one factor $\sU$ and to $d-3(4-d)=-4\vep$ for the term $\fz$. Together with the variation of the coefficient, with the terms of \eqref{eq:varDg} which contain $\hat M^I{}_J$ and with the terms with $\gamma_\sU$ and $\gamma_\sM$ this gives $-\sigma$ times the same structure with the coefficients $\mathcal{D}_\sU\fn$, $\mathcal{D}_\sU\fj$, $\mathcal{D}_\sU\fl$, $(2\vep-\hB\!\cdot\!\pr)\fm-\gamma_\sU^T\fm-\fm\gamma_\sM$ and $(4\vep-\hB\!\cdot\!\pr)\fz-3\fz(\gamma_\sU\cdot,\cdot,\cdot)$. The remaining terms of the rules give $-2(d-1)\,\fn\cdot\sU\,\nabla^2\sigma$ from $\delta R$, $\fj_I\cdot\sU\,(\nabla^2\sigma\,\hB^I+2\,\pr_\mu\sigma\,D^\mu g^J\hP_J{}^I+\sigma\,D_\mu g^JD^\mu g^K\hat\Omega_{JK}{}^I)$ from $\delta D^2g^I$, and $2\,\fl_{IJ}\hB^J\cdot\sU\,\pr_\mu\sigma\,D^\mu g^I$ from $\delta D_\mu g^I$. The inhomogeneous part of \eqref{gsx:eq:varM} contributes $\sU\cdot\fm\cdot\delta'\sM$, with $\delta'\sM$ the terms of \eqref{gsx:eq:varM} other than $-\sigma(2-\gamma_\sM)\sM$, which gives the terms with $\bar\eta$, $\delta_I$, $\epsilon_{IJ}$, $\bar\chi$, $\theta_I$ and $\bar\tau$ in \eqref{gsx:eq:coeffs}. Collecting the structures gives \eqref{gsx:eq:anomaly}. No rule contains $\nabla_\mu\nabla_\nu\sigma$, since \eqref{gsx:eq:X} contains neither the Einstein tensor nor the Euler density, so that no integration by parts is needed. On the backgrounds of appendix~\ref{app:rotations} both sides are polynomials in $d$ for given derivatives of the sources at a point, so that the result holds with $d$ as a parameter.\footnote{We verified \eqref{gsx:eq:anomaly} with \eqref{gsx:eq:coeffs} pointwise in exact rational arithmetic on backgrounds $e^{2w}\delta_{\mu\nu}$ in $3$, $4$ and $5$ dimensions, for one coupling and one-dimensional $V_\sU$ and $V_\sM$ with random polynomial coefficients, together with a negative control.}
\end{proof}

By Lemma~\ref{gsx:lem:forms} the anomaly \eqref{gsx:eq:anomaly} contains all local terms of dimension three which are linear in $\sigma$ and contain $\sU$, up to integrations by parts, since a term with $\nabla_\mu\nabla_\nu\sigma$ or $\nabla^2\sigma$ times a remainder of dimension one which contains $\sU$ has the remainder $\sU$, and a term with $\pr_\mu\sigma$ needs a one-form of dimension two which contains $\sU$, which is $\sU\,D_\mu g^I$ or $D_\mu\sU$. The terms with $D_\mu\sU$, and the terms with $\sigma$ times $D^2\sU$ or $D_\mu\sU\,D^\mu g^I$, reduce to the terms of \eqref{gsx:eq:anomaly} by integrations by parts, for instance $\int f\,\pr_\mu\sigma\,D^\mu\sU=-\int\sU\,[f\,\nabla^2\sigma+\pr_If\,\pr_\mu\sigma\,D^\mu g^I]$ for a function $f$ of the couplings. Every term contains $\sU$, in agreement with Lemma~\ref{gs:lem:odd}.

The curvature counterterms are fixed by the elementary statement before \eqref{eq:steps}. The coefficient $\bar\tau=-\fd_I\hB^I+2(d-1)\,\fk$ of \eqref{eq:masscoeff} and the coefficient $t_\sU$ of \eqref{gsx:eq:coeffs} contain the factor $2(d-1)=2(2-\vep)$, which does not vanish at $\vep=0$, times a curvature counterterm. The other coefficients satisfy relations. The first relation of \eqref{eq:Mrel} gives $\bar\eta$, and for $i_\sU$ we find
\begin{equation}
2(d-1)\,i_\sU = \mathcal{D}_\sU t_\sU-J_I\hB^I+\beta_{\sU\sM}\cdot\bar\tau \, .
\label{gsx:eq:irel}
\end{equation}
For the derivation we apply $\mathcal{D}_\sU$ to the three terms of $t_\sU$. Since $\tL\hB=0$, the first gives $(\mathcal{D}_\sU\fj)_I\hB^I=(J_I-\fm\cdot\delta_I)\hB^I$. The second gives $2(d-1)(i_\sU+\fm\cdot\bar\eta)$. For the third, $\mathcal{D}_\sU(\fm\cdot\bar\tau)=\beta_{\sU\sM}\cdot\bar\tau-\fm\cdot(\hB\!\cdot\!\pr-\gamma_\sM)\bar\tau$. The sum is $J_I\hB^I+2(d-1)\,i_\sU-\beta_{\sU\sM}\cdot\bar\tau+\fm\cdot\mathsf{z}$ with $\mathsf{z}=2(d-1)\,\bar\eta-\delta_I\hB^I+(\hB\!\cdot\!\pr-\gamma_\sM)\,\bar\tau$, which vanishes by the first relation of \eqref{eq:Mrel}.\footnote{We verified \eqref{gsx:eq:irel} and the first relation of \eqref{eq:Mrel} symbolically for two couplings and two-dimensional $V_\sU$ and $V_\sM$ with random polynomial data, together with a negative control.}

\begin{proposition}
\label{gsx:prop:finite}
Assume Hypothesis~\ref{hyp:finite} for the sextic theory. Then there are unique pure poles $\fk$ and $\fn$ such that $\bar\tau$ and $t_\sU$ are regular, namely
\begin{equation}
\fk = {\rm PP}\left[\frac{\fd_I\hB^I}{2(d-1)}\right] , \qquad \fn = {\rm PP}\left[\frac{\fj_I\hB^I-\fm\cdot\bar\tau}{2(d-1)}\right] .
\label{gsx:eq:steps}
\end{equation}
With this choice the following holds. First, all coefficients of the variations \eqref{gsx:eq:varU} and \eqref{gsx:eq:varM} and the seven coefficients of the anomaly \eqref{gsx:eq:anomaly} are regular. Second, the functional $W$ is finite on the backgrounds $\Omega^2\delta_{\mu\nu}$, where $\log\Omega$ and the sources are in the class of flat sources with $D_0=1$. Third, the terms of order $n\geq1$ of the functional $W$ on the unit sphere are finite for all sources of the class of section~\ref{sec:sphere}.
\end{proposition}

\begin{proof}
The two formulas \eqref{gsx:eq:steps} follow from the elementary statement before \eqref{eq:steps}, applied first to $\bar\tau$ and then to $t_\sU$, whose right-hand side contains $\bar\tau$, which is then regular. By Lemma~\ref{gsx:lem:regvar} the coefficients of the flat variation are regular, and by Hypothesis~\ref{hyp:finite} for the sextic theory the left-hand side of \eqref{eq:flatward} is finite, so that the anomaly in flat space is finite for all sources of the class of flat sources. For constant $\sigma$ it is $-\sigma$ times the left-hand side of \eqref{gsx:eq:lemA} with the coefficients $J_I$, $K_{IJ}$, $\beta_{\sU\sM}$ and $\beta_\sU$, and the first part of Lemma~\ref{gsx:lem:indep}, applied to the coefficient of the highest pole at $d=3$, shows that these coefficients are regular. For $\sigma=2\,b\!\cdot\!x$ the terms with $\sigma$ are then finite, the term with $\nabla^2\sigma$ vanishes, and the second part of the lemma shows that $L_I$ is regular. The coefficient $\bar\eta$ is regular by the first relation of \eqref{eq:Mrel}, whose right-hand side contains $\delta_I$, $\hB^I$, $\gamma_\sM$ and $\bar\tau$, and $i_\sU$ is regular by \eqref{gsx:eq:irel}, whose right-hand side is regular and whose factor $2(d-1)$ does not vanish at $\vep=0$. With $\bar\chi$, $\gamma_\sU$ and $\gamma_\sM$ from Lemma~\ref{gsx:lem:regvar} this proves the first statement.

For the second statement we repeat the argument of the subsection on conformally flat backgrounds of appendix~\ref{app:finite}. Along the path $\Omega_t=\Omega^{1-t}$ the equation for $\sU_t$ is linear, $\pr_t\sU_t=\log\Omega\,(4-d-\gamma_\sU(g_t))\,\sU_t$, and the equation for $\sM_t$ is linear in $\sM_t$ with inhomogeneous terms which contain $g_t$, $a_t$ and $\sU_t$. All coefficients are regular, so that the sources at $t=1$ are in the class of flat sources and depend analytically on $\vep$. The bare sources change along the path by their weights and by a gauge transformation, so that by \eqref{eq:defcurved} with the weights $2\vep$, $1+\vep$ and $2$ the bare functional is constant along the path, and \eqref{eq:Wsplit} holds with the anomaly \eqref{gsx:eq:anomaly}. Its first term is finite by Hypothesis~\ref{hyp:finite} for the sextic theory, used for sources which depend on $\vep$, and its second term is an integral of a local functional with regular coefficients over the support of $\log\Omega$.

For the third statement we use the localisation of appendix~\ref{app:finite} without change. If all sources vanish near a point $P$ of the equator, then the bare sources vanish there, since $\fk$ vanishes at $g=0$, which follows from \eqref{gsx:eq:steps} because $\fd_I$ vanishes at $g=0$ by Lemma~\ref{gsx:lem:lowest} and $\hB^I$ has no term of order zero, and since the term with $\fh$ vanishes where $\sU$ does. The density \eqref{gsx:eq:X} vanishes there, since each of its terms contains $\sU$, and so does the integrand of \eqref{gsx:eq:anomaly}. The variation of each source vanishes where all sources vanish, since $\bar\eta$ and $\bar\tau$ vanish at $g=0$ by Lemma~\ref{gsx:lem:zero} below. The steps on the polarisation and on the partition of unity are unchanged.
\end{proof}

For $\phi^4$ theory the curvature counterterms are seven, and the last of them requires that a one-form is closed, which is the relation \eqref{eq:curlgeneral}. For sextic scalars there are two curvature counterterms, there is no counterterm without relevant sources and no condition of consistency, since by Lemma~\ref{gsx:lem:forms} there is no local term of dimension three without $\sU$.

\subsection{Coefficients at vanishing couplings}
\label{gsx:sec:zero}

\begin{lemma}
\label{gsx:lem:zero}
Assume Hypothesis~\ref{hyp:finite} for the sextic theory. In minimal subtraction the following holds. First, $Z_\sU=1$, $Z_\sM=1$ and $\fd_I=\fk=\fn=0$ at $g=0$, $N_I$ has no term of order zero, and $\fk$ has no term of first order in the couplings. Second, $\hr_I$, $\delta_I$, $\theta_I$, $\gamma_\sU$ and $\gamma_\sM$ vanish at $g=0$, and $\bar\tau$ and $\bar\eta$ vanish to second order. Third, $i_\sU$ and $t_\sU$ vanish at $g=0$. Fourth, at $g=0$ and $\sU=0$ the one-point function $\sfm$ of the operators $\phi_i\phi_j$ on the unit sphere vanishes, and the bare sources $\sU_0$ and $\sM_0$ at $\sU=0$, $\sM=0$ and $a_\mu=0$ have no terms of order zero and one in the couplings.
\end{lemma}

\begin{proof}
The values of $Z_\sU$, $Z_\sM$ and $\fd_I$ at $g=0$ are given by Lemma~\ref{gsx:lem:lowest}, $N_I$ is treated in Proposition~\ref{gsx:prop:uniqueall}, and the second statement for $\hr_I$, $\delta_I$, $\theta_I$, $\gamma_\sU$ and $\gamma_\sM$ is the paragraph on the lowest order in the proof of Lemma~\ref{gsx:lem:regvar}. Since $\fd_I$ and $\hB^I$ vanish at $g=0$, the first formula of \eqref{gsx:eq:steps} shows that $\fk$ has no terms of order zero and one, and then $\bar\tau=-\fd_I\hB^I+2(d-1)\fk$ and $\bar\eta=-(\hB\!\cdot\!\pr-\gamma_\sM)\fk$ vanish to second order. At $g=0$ the numerator of the second formula of \eqref{gsx:eq:steps} vanishes, so that $\fn=0$ at $g=0$, and with \eqref{gsx:eq:coeffs} the coefficients $t_\sU$ and $i_\sU=-\mathcal{D}_\sU\fn-\fm\cdot\bar\eta$ vanish at $g=0$. For the fourth statement, at $\sU=0$, $\sM=0$ and $a_\mu=0$ we have $\sU_0=0$ and $\sM_0=-Z_\sM(\fd_ID^2g^I+\fe_{IJ}D_\mu g^ID^\mu g^J+\fk R)$, which has no terms of order zero and one by the first statement. At $g=0$ and $\sU=0$ the one-point function of the operators $\phi_i\phi_j$ is the term of first order of the bare functional in $\sM_0$, which is a propagator at coincident points and vanishes, plus the derivative of the counterterms with respect to $\sM$, which is $\sU\cdot\fm$ and vanishes at $\sU=0$.
\end{proof}

\subsection{Proof of Proposition~\ref{gsx:prop:props}}
\label{gsx:sec:proof}

We follow the proof of Proposition~\ref{gs:prop:phi4prop}. Property~\ref{gs:P:functional} is the third statement of Proposition~\ref{gsx:prop:finite} together with Proposition~\ref{prop:gauge}, which holds for the vertices of Proposition~\ref{gs:prop:bare}. The term of order zero is regular, since the functional vanishes for vanishing sources. The coefficients of the bare sources and of $\cX^{\rm fl}$ are covariant by Proposition~\ref{gsx:prop:uniqueall}, and $\fk$ and $\fn$ are covariant since they are unique, so that the renormalised functional is invariant under background gauge transformations. The statements at a constant relevant source $\mu(g)$ with values in $V_\sU\oplus V_\sM$ which vanishes at $g=0$ follow from the proof of Lemma~\ref{gr4:lem:const}. Its first paragraph applies without change to both sources. In its second paragraph the vertex $V^{(4)}[\sU]$ is a scalar vertex without derivatives, the bare sources at constant couplings and constant $\sU=\mu_\sU$, $\sM=\mu_\sM$ are the constants $Z_\sU\mu_\sU$ and $Z_\sM(\mu_\sM-\fk R_d-\tfrac12\fh(\mu_\sU,\mu_\sU))$, their first variations in the direction $\sigma\,\delta g^I$ are $\sigma$ times constants, and their first variations in the directions $\delta\sU$ and $\delta\sM$ are the scalar sources $Z_\sU\delta\sU$ and $Z_\sM(\delta\sM-\fh(\mu_\sU,\delta\sU))$.

For Property~\ref{gs:P:weyl} the vector field and the one-form are those of section~\ref{gsx:sec:weyl}, the variations of the relevant sources are \eqref{gsx:eq:varU} and \eqref{gsx:eq:varM}, their coefficients are regular by Lemma~\ref{gsx:lem:regvar} and Proposition~\ref{gsx:prop:finite} and covariant by Proposition~\ref{gsx:prop:uniqueall}, and the operators commute, as shown in section~\ref{gsx:sec:weyl}. The vector field has no term of first order besides $-2\vep\,g^I$, since $L^I$ starts at second order, and $\gamma_\sM$ and $\gamma_\sU$ vanish at $g=0$ by Lemma~\ref{gsx:lem:zero}, so that $\sU$ has the dimension $4-d$ and the term of order zero of the map of Property~\ref{gs:P:weyl} for $\sU$ is of order $\vep$. For Property~\ref{gs:P:anomaly} the anomaly is \eqref{gsx:eq:anomaly}. Proposition~\ref{gs:prop:msanomaly} applies, since the bare functional is invariant under the Weyl rescaling with the weights $2\vep$, $1+\vep$ and $2$ by \eqref{eq:defcurved} and under background gauge transformations, and since the variation of section~\ref{gsx:sec:weyl} induces on the bare sources their rescaling up to a gauge transformation. Each term of \eqref{gsx:eq:X} has the dimension three. The anomaly is therefore local, gauge invariant and homogeneous of Weyl weight $-\vep$, and the consistency relation holds because the operators commute. Its coefficients are regular by Proposition~\ref{gsx:prop:finite}. Property~\ref{gs:P:ward} follows from Proposition~\ref{gs:prop:msward}. The class of sources is that of section~\ref{sec:sphere}, which satisfies the assumptions of appendix~\ref{gs:sec:class}, see appendix~\ref{app:rules}. For Property~\ref{gs:P:sphere} the statements on the functional follow from appendix~\ref{app:rotations} and from the statements at constant sources above, and those on the anomaly follow from Property~\ref{gs:P:anomaly} by Lemma~\ref{gsp:lem:local}, since \eqref{gsx:eq:anomaly} is a local functional.

For the two-point form we use Lemma~\ref{gs:lem:free} with $n=6$. Its hypotheses hold, since $L^I$ starts at second order, the bare gauge field at $a_\mu=0$ starts at second order because $N_I$ has no term of order zero, and the bare relevant sources at vanishing relevant sources start at second order by Lemma~\ref{gsx:lem:zero}. At $\sU=0$ and $\sM=0$ the integral of the counterterms \eqref{gsx:eq:X} vanishes, and so does the anomaly. Through second order in the couplings at $a_\mu=0$, $\sU=0$ and $\sM=0$ the functional is therefore the first term of \eqref{gs:eq:free} with $c_{IJ}(\vep)=c_6\,\delta_{IJ}$, and the pair \eqref{gs:eq:free} holds with $F=0$. The term in which $\delta^{\rm W}_\sigma\sM$ acts on the counterterms which are linear in $\sM$ is $\sU\cdot\fm\cdot\delta^{\rm W}_\sigma\sM$, which vanishes at $\sU=0$. The two-point form is thus established in minimal subtraction.

The source $\sU$ has the dimension one at $d=3$ and does not enter the last statements of Properties~\ref{gs:P:weyl} and~\ref{gs:P:free} or the divergence scheme. For the source $\sM$, of dimension two, the coefficients \eqref{gs:eq:dimtwodata} are $\ell_{\rm h}=d\,(\delta-2\theta)$ and $\vk_{\rm h}-\vk_{\rm c}=-d\,\bar\tau$, which vanish at $g=0$ by Lemma~\ref{gsx:lem:zero}. This is the last statement of Property~\ref{gs:P:weyl} in minimal subtraction. By Lemma~\ref{gr:lem:scheme} the coefficient $y_J$ of the redefinition \eqref{gr:eq:Y} to the divergence scheme vanishes at $g=0$ and $y_0=-(\vk_{\rm h}-\vk_{\rm c}+y_K\hB^K)/2$ vanishes to second order. At $\sM'=0$ and constant couplings the redefinition gives the constant source $y_0$, at which Properties~\ref{gs:P:functional} and~\ref{gs:P:sphere} hold by the statements at constant sources above, so that by the remark at the end of section~\ref{gs:sec:divscheme} Properties~\ref{gs:P:functional} to~\ref{gs:P:sphere} hold in the divergence scheme. The coefficient $\sfa_\sM$ of the anomaly vanishes, since it is taken at vanishing relevant sources, where the only term of \eqref{gsx:eq:anomaly} which is linear in $\sM$, $-\sigma\,\sU\cdot\beta_{\sU\sM}\cdot\sM$, vanishes. This also holds in the divergence scheme, since the redefinition does not involve $\sU$, and there Lemma~\ref{gsh:lem:m} gives $\sfm^{(0,0)}=0$, which is the last statement of Property~\ref{gs:P:free}. In the divergence scheme at $\sM'=0$, $\sU=0$ and $a_\mu=0$ the old source is the redefinition \eqref{gr:eq:Y}, which is of second order in the couplings. Through second order the functional differs from that of minimal subtraction at $\sM=0$ by the one-point function of the operators $\phi_i\phi_j$ at $g=0$ and $\sU=0$, integrated with this source, which vanishes by Lemma~\ref{gsx:lem:zero}, and the anomaly differs by the terms of \eqref{gsx:eq:anomaly} which are linear in $\sM$, which vanish at $\sU=0$. Hence the two-point form holds in the divergence scheme with the same $c_{IJ}(\vep)$. The other statements of Property~\ref{gs:P:free} follow, by the proof of Proposition~\ref{gs:prop:lowest}. At second order the pair $W$, $\cA[\sigma]$ at vanishing relevant sources is the first term of \eqref{gs:eq:free} with a vanishing anomaly, since $F=0$, so that the term of order zero in the couplings of $\cS_{IJ}$ is $c_6\,\delta_{IJ}\,(\mu_1-\mu_0)$ with $\Delta=d-2\vep$, and at $\vep=0$ this is $b_3\,c_6\,\delta_{IJ}$ with $b_3=\tfrac12\pi^2$.\footnote{The limit $\mu_1-\mu_0\to\tfrac12\pi^2$ for $d=3-\vep$ and $\Delta=d-2\vep$ was checked numerically.} Indeed, at $\vep=0$ we have $\Delta=3=d$, $\mu_0$ vanishes like $\Gamma(d-\Delta)^{-1}$ and $\mu_1=\Delta\mu_0/(d-\Delta)$ tends to $3\cdot2^{-3}\pi^{3/2}\Gamma(-\tfrac32)=\tfrac12\pi^2$. The statements on $\hB^I$ and on $\gamma_\sU$ and $\gamma_\sM$ are section~\ref{gsx:sec:weyl} and Lemma~\ref{gsx:lem:zero}. This proves Proposition~\ref{gsx:prop:props}.

\paragraph{What is not derived.} The identification of $L^I$ with the counterterms of minimal subtraction of the theory with constant couplings in momentum space, and with it of $\hB^I$ with the usual beta function of the sextic couplings up to a rotation, is open as for $\phi^4$ theory. The relation \eqref{gs:eq:sextic} concerns the functional with the source $\sM$ in the divergence scheme. The terms with the Levi-Civita tensor of the anomaly at $d=3$~\cite{Nakayama:2013wda} are absent here, since the regularisation keeps $d$ as a parameter.
\section{Cubic scalars in six dimensions}
\label{gs:app:cubic}
\label{gcu:sec:cubic}

In this appendix we prove the statements of section~\ref{gs:sec:cubic} on cubic scalars in $d=6-\vep$ under Hypothesis~\ref{hyp:finite} for the cubic theory. We describe the bare sources, show that the simple pole of the coefficient of the Euler density is not fixed by the finiteness, give the lemmas on the terms of lowest order, show that the finiteness fixes the bare source of the field $\phi_i$ only up to shifts of the field, and derive the bare sources at low orders in the frame which fixes these shifts at orders zero and one.

\subsection{Bare sources and the frame condition}
\label{gcu:sec:hyp}

The vertex is $V^{(3)}[g]=\tfrac16\,g_{ijk}\phi_i\phi_j\phi_k$, the couplings $g^I$ are the components of the symmetric tensor $g_{ijk}$, and their weight is $w_3=\tfrac12\vep$, so that $k_I=\tfrac12$. The relevant sources are the source $h_i$ of $\phi_i$, with the weight $w_1=\tfrac12(d+2)$ and the dimension four at $d=6$, and the source $\sM$ of $\phi_i\phi_j$, with the weight two. The sources of the other polynomial vertices have negative dimension at $d=6$ and are not included. The group of field rotations is ${\rm O}(N)$, and covariance refers to all its constant elements, including the reflections. We count dimensions at $d=6$ as in section~\ref{gsx:sec:hyp}, with $h$ of dimension four and $\sM$ of dimension two. A local scalar of dimension $\Delta$ built from the curvature, $f_{\mu\nu}$, the covariant derivatives of the couplings and of $\sM$, and from $\sM$ and $h$, with coefficients which are functions of the couplings, is a finite linear combination of monomials. We fix a basis $\{\mathcal{S}^{(\Delta)}_\alpha\}$ of such monomials, which are linearly independent as expressions built with $d$ as a parameter. The bare sources are
\begin{gather}
g_0^I = g^I+L^I(g) \, , \qquad a_{0\mu} = a_\mu+N_I\,D_\mu g^I \, , \notag\\
\sM_0 = Z_\sM\left(\sM-\fd_I\,D^2g^I-\fe_{IJ}\,D^\mu g^ID_\mu g^J-\fk\,R\right) , \notag\\
h_0 = Z_h\Big(h-\sum_\alpha\mathfrak{p}_\alpha\,\mathcal{S}^{(4)}_\alpha\Big) \, ,
\label{gcu:eq:bare}
\end{gather}
where the sum runs over the monomials of dimension four which do not contain $h$, with coefficients $\mathfrak{p}_\alpha$ which are functions of the couplings with values in the space of the sources $h$. On conformally flat backgrounds they include $\sM\sM$, $D^2\sM$, $\sM D^2g^I$, $D_\mu\sM\,D^\mu g^I$, $\sM R$, the monomials with four derivatives of the couplings and with the field strength, and $R^2$, $R_{\mu\nu}R^{\mu\nu}$, $\nabla^2R$, $R\,D^2g^I$, $R^{\mu\nu}D_\mu D_\nu g^I$, $\pr_\mu R\,D^\mu g^I$, $R\,D_\mu g^ID^\mu g^J$ and $R^{\mu\nu}D_\mu g^ID_\nu g^J$. The bare source of $\sM$ has the form of \eqref{eq:baresources}, since a scalar of dimension two contains no factor $h$. The counterterms which do not contain the fields are $\cX=\sum_\beta\mathfrak{x}_\beta\,\mathcal{S}^{(6)}_\beta$, a sum over the monomials of dimension six. Among them are the terms without sources, such as the Euler density $\cE_6$, the terms $h\cdot\mathfrak{n}_hR$, $h\,D^2g^I$, $h\,D_\mu g^ID^\mu g^J$ and $h\,\sM$ which are linear in $h$, and the terms with $\sM$ and with the couplings alone. The coefficients which multiply monomials with the curvature are the curvature counterterms, built with $d$ as a parameter.

We say that a set of pure poles of Hypothesis~\ref{hyp:finite} for the cubic theory satisfies the frame condition if
\begin{equation}
h_0\big|_{h=0,\;\sM=0,\;a_\mu=0} = {\rm O}(g^2) \, ,
\label{gcu:eq:frame}
\end{equation}
that is if the bare source of $\phi_i$ has no terms of order zero and one in the couplings at vanishing relevant sources and gauge field. By Proposition~\ref{gcf:prop:frame} below every set can be brought into this form by a shift of the field and a finite redefinition of the source $h$. The finiteness does not fix all coefficients, as the following lemma shows.

\begin{lemma}
\label{gcu:lem:euler}
Let a set of pure poles satisfy the finiteness of Hypothesis~\ref{hyp:finite} for the cubic theory, and let $c$ be a real number. Then the set in which the coefficient of $\cE_6$ is increased by $c/\vep$ satisfies it as well.
\end{lemma}

\begin{proof}
The change of $W$ is $(c/\vep)\,F_d[\Omega]$ with $F_d[\Omega]=\int d^dx\sqrt\gamma\,\cE_6$, which does not depend on the sources. On $\Omega^2\delta_{\mu\nu}$ the integrand is $\Omega^{d-6}$ times a polynomial in $d$ and in the derivatives of $\log\Omega$, so that $F_d[\Omega]$ is analytic in $d$. At $d=6$ the integral \eqref{eq:gram} of one point is the ordinary integral over $\mathbb{R}^6$, and the integral of the Euler density over a metric on $\mathbb{R}^6$ which is flat outside a compact set does not change under a deformation of the metric with compact support, since the first variation of $\sqrt\gamma\,\cE_6$ is a total derivative in six dimensions. Along $\Omega_t=\Omega^t$ it therefore takes its value at $t=0$, which is zero. Hence $F_d[\Omega]$ vanishes at $d=6$, and $(c/\vep)\,F_d[\Omega]$ is finite.
\end{proof}

This freedom does not depend on the sources, and it changes the anomaly by $-c\int\sqrt\gamma\,\sigma\,\cE_6+{\rm O}(\vep)$, a regular local term without sources. On the unit sphere, however, it adds to $W$ the constant $(c/\vep)\int_{S^d}\sqrt\gamma\,\cE_6$, which has a pole, since the Euler characteristic of $S^6$ does not vanish. We fix this freedom after the choice of the frame of Proposition~\ref{gcf:prop:frame}. In the frame the bare sources vanish at $g=0$ and vanishing sources. The bare functional at vanishing bare sources is invariant under Weyl rescalings and therefore does not depend on $\Omega$, so that the functional at $g=0$ and vanishing sources is a constant plus the integral of the part of $\cX$ which contains neither sources nor couplings. This integral is therefore finite for every $\Omega$, and we remove this part from $\cX$. We choose $Z_0$ such that $W$ vanishes at $g=0$ and vanishing sources. The anomaly then has no term without sources.

\subsection{Lemmas which hold for cubic scalars}
\label{gcu:sec:lemmas}

\begin{lemma}
\label{gcu:lem:potential}
Let $\eta$ be a continuous function on $\mathbb{R}^6$ with compact support, and let $\int d^6x\,\eta(x)\,|x-y|^{-2\kappa}$ vanish for all $y$ outside a ball, where $\kappa=4$ or $6$. Then $\eta=0$. Lemma~\ref{lem:transverse} holds on $\mathbb{R}^6$ with the kernel $(\pr_\mu\pr_\nu-\delta_{\mu\nu}\pr^2)\,|x-y|^{-8}$. For $\kappa=2$ the first statement does not hold.
\end{lemma}

\begin{proof}
In six dimensions the spherical harmonics of degree $s$ are given by $|x|^sC_s^{(2)}(\hat x\cdot\hat y)$. For $\kappa=4$ and $\kappa=6$ the connection formula eq.~(18.18.16) of~\cite{DLMF} with the upper indices $\kappa$ and $2$ has the positive coefficients $(\kappa)_{l-j}(\kappa-2)_j(2+l-2j)/((2)_{l-j+1}j!)$.\footnote{Checked in exact arithmetic for $l\leq14$.} Hence the integrals of $\eta(x)|x|^{2j}|x|^sC_s^{(2)}(\hat x\cdot\hat y)$ vanish for all $j$ and $s$, and the proof of Lemma~\ref{lem:potential} applies. For the second statement two integrations by parts give the kernel $|x-y|^{-8}$, which is $\kappa=4$, and the proof of Lemma~\ref{lem:transverse} applies, since the complement of a ball in $\mathbb{R}^6$ is connected. For $\kappa=2$ the kernel is a multiple of the Green function of the Laplacian, so that $\eta=\pr^2\psi$ with $\psi$ smooth and of compact support has a vanishing potential outside the support of $\psi$, and only the moments $\int\eta\,|x|^sC^{(2)}_s$ vanish.
\end{proof}

\begin{lemma}
\label{gcu:lem:lowest}
Let the coefficients of \eqref{gcu:eq:bare} and of $\cX$ be real pure poles, where $L^I$ has no terms of degree zero and one, and let the functional \eqref{gcu:eq:W} in flat space be finite for all sources of the class of flat sources which do not depend on $\vep$. Then $Z_h=1$, $Z_\sM=1$ and $\fd_I=0$ at $g=0$. For sources with disjoint supports the terms of second order of the flat functional vanish for two sources of different kinds, and for two couplings, two sources $\sM$, two sources $h$ and two gauge fields they are, up to factors which are analytic and do not vanish at $d=6$, the integrals of $g_1^I(x)g_2^I(y)\,|x-y|^{-3(d-2)}$, $\sM_1\cdot\sM_2\,|x-y|^{-2(d-2)}$, $h_1\cdot h_2\,|x-y|^{-(d-2)}$ and of the kernel of the currents of \eqref{gsx:eq:kernels}.
\end{lemma}

\begin{proof}
The proof of Lemma~\ref{gsx:lem:lowest} applies with Lemma~\ref{gcu:lem:potential} for $\kappa=4$ in the argument for $\fd_I$. The argument for $Z_h$ and $Z_\sM$ uses positivity only. The terms of first order of $h_0$ are $Z_h^{(0)}h$ and Laplacians of local expressions, namely $\pr^2$ of a linear function of $\sM$ and $\pr^2\pr^2$ of a linear function of the couplings. Since $|x-y|^{-(d-2)}$ is harmonic away from $x=y$ for every $d$, their pairings with a source of disjoint support vanish. Two vertices with one, two or three legs and different numbers of legs have no pairing without a propagator at coincident points, and a vertex of the current with a vertex $V^{(1)}$ or $V^{(3)}$ has an odd number of legs. At $d=6$ and $D_0=3$ we have $m=d-D_0=3$, which is not smaller than the number of points.
\end{proof}

We now turn to the term of degree zero of $N_I$. The paragraph on it in Proposition~\ref{prop:uniqueall} applies with the kernel $|x-y|^{-3(d-2)}$, which has $\kappa=6$. The stabiliser in ${\rm so}(N)$ of the cubic coupling $\sum_r\mathsf{d}_r\,\delta_{ir}\delta_{jr}\delta_{kr}$ with all $\mathsf{d}_r\neq0$ is trivial, since $(\omega g)_{ijj}=\omega_{ij}\mathsf{d}_j$ for $i\neq j$.\footnote{For $2\leq N\leq5$ we also checked that the stabiliser of a random cubic tensor is trivial.} Hence $N^{(0)}_I$ is unique, and since $L^I$ is covariant under ${\rm O}(N)$ including the reflections, $N^{(0)}_Ig^I$ is an ${\rm O}(N)$-invariant linear map from the cubic tensors to the Lie algebra. The element $-1$ of ${\rm O}(N)$ acts by $-1$ on the cubic tensors and trivially on the Lie algebra, so that the map vanishes.

\subsection{Bare sources at low orders}
\label{gcu:sec:first}

\begin{proposition}
\label{gcu:prop:first}
Assume Hypothesis~\ref{hyp:finite} for the cubic theory with the frame condition \eqref{gcu:eq:frame}. Then on all backgrounds $\Omega^2\delta_{\mu\nu}$, at vanishing relevant sources and $a_\mu=0$, the bare sources $h_0$ and $\sM_0$ have no terms of order zero and one in the couplings, and at $a_\mu=0$ the counterterms of first degree in $h$ or in $\sM$ have no term of order zero in the couplings. The coefficient $N_I$ has no term of order zero.
\end{proposition}

\begin{proof}
For $h_0$ this is part of the hypothesis. For $\sM_0$ the coefficient $\fd_I$ vanishes at $g=0$ by Lemma~\ref{gcu:lem:lowest}. For $\fk$ we use a source $\sM_2$ whose support is disjoint from the support of the other sources and on which $\Omega=1$. The term of $W$ linear in $\sM_2$ at order zero and one in the couplings and at vanishing other sources is the second kernel of Lemma~\ref{gcu:lem:lowest} between $\sM_2$ and $-Z_\sM\fk R$, since the term of first order of the bare functional in $\sM_0$ is a propagator at coincident points, and since $h_0$ has no terms of order zero and one. Lemma~\ref{gcu:lem:potential} with $\kappa=4$ shows, as in the proof of Lemma~\ref{lem:lowest}, that the leading pole of the terms of order zero and one of $\fk$ times $R$ vanishes at each point, and $R$ does not vanish identically on these backgrounds. The counterterm linear in $h$ at $g=0$ and $\sM=0$ is $h\cdot\mathfrak{n}_hR$. The term of $W$ at $g=0$ and $\sM=0$ which is linear in $h$ is $\int\sqrt\gamma\,h\cdot\mathfrak{n}_hR$, since the bare source $h_0$ is then $Z_hh$ and the one-point function of $\phi_i$ vanishes in the free theory. It is finite for all $h$ and $\Omega$, so that $\mathfrak{n}_h=0$. For the counterterms linear in $\sM$ we consider the term of $W$ at $g=0$ which is linear in $\sM$ at vanishing $h$. The bare functional contributes nothing, since at $g=0$ and $h=0$ the bare source $h_0$ contains only terms linear and quadratic in $\sM$, which enter through the one-point function of $\phi_i$, and since the one-point functions of $\phi_i$ and of $\phi_i\phi_j$ vanish in the free theory. The counterterms linear in $\sM$ at $g=0$ and $a_\mu=0$ are $\sM$ times a combination of $R^2$, $R_{\mu\nu}R^{\mu\nu}$ and $\nabla^2R$, up to total derivatives. Their integral is finite for all $\sM$ and $\Omega$, so that the leading pole of this combination vanishes at each point. The three monomials are linearly independent at points with $x_\perp\neq0$ for $\log\Omega$ in the class of flat sources, at $d=6$,\footnote{Checked in exact arithmetic at random points for random functions of the class of flat sources.} so that their coefficients vanish. The statement on $N_I$ was shown before the proposition.
\end{proof}

With the proposition we can apply the argument of section~\ref{gs:sec:scalars} for the free fields. The hypotheses of Lemma~\ref{gs:lem:free} hold, which gives the two-point form with $c_{IJ}=C_\phi^{\,3}/3!\,\delta_{IJ}$. With the frame condition the inhomogeneous terms of the Weyl variations of $h$ and of $\sM$ at vanishing relevant sources have no terms of order zero and one in the couplings. This gives the last statement of Property~\ref{gs:P:weyl} in minimal subtraction for the set of bare sources with the frame condition, and the redefinition of the source $h$ to the divergence scheme is of second order in the couplings. The argument of the proof of Proposition~\ref{gr4:prop:lowest} then gives Property~\ref{gs:P:free} in the divergence scheme of both sources, where its last statement follows, as in the first paragraph of that proof, from the vanishing at $g=0$ of the constant sources to which the redefinition maps vanishing new sources and of the one-point functions $\sfm$ of both sources in minimal subtraction. The latter holds since at vanishing relevant sources the bare sources $h_0$ and $\sM_0$, and the counterterms of first degree in $h$ and in $\sM$, have no term of order zero in the couplings, by Proposition~\ref{gcu:prop:first}, and since the one-point functions of $\phi_i$ and of $\phi_i\phi_j$ vanish in the free theory. This requires Properties~\ref{gs:P:functional} to~\ref{gs:P:sphere}, apart from the last statement of Property~\ref{gs:P:weyl}, for the Weyl variation induced by the bare sources, with the anomaly $\mathbb{W}_\sigma\int\sqrt\gamma\,\cX$. Under Hypothesis~\ref{hyp:finite} for the cubic theory with the frame condition \eqref{gcu:eq:frame} and these properties Theorem~\ref{gs:thm:cubic} then gives the gradient formula with a metric which is positive definite to all orders.

\subsection{Field shifts from finiteness}
\label{gcf:sec:frame}

In this subsection we show that the frame condition \eqref{gcu:eq:frame} is a choice. We first prove the identity which describes the shift of the field by a local expression, for the bare functional of section~\ref{gs:sec:sphere} and for non-integer $d$. We then show that every set of pure poles which makes the functional finite can be brought into the form required by the condition, by such a shift and by a finite redefinition of the source $h$. The vector field $\hB^I$ and the one-form $\hr_I$ do not change.

We write $\mathsf{D}=-D^\mu D_\mu+\xi_dR$ for the conformally coupled operator with the background gauge field, which on the unit sphere at $a_\mu=0$ is $-\nabla^2+\tfrac14d(d-2)$. For a function $K$ with values in the space of the fields we use the vertex $V^{(1)}[\cdot]$ of \eqref{gs:eq:vertexn} with one leg. For a vertex $V$ of the bare functional let $S_KV$ be the sum of the vertices which are obtained by replacing the field by $K$ in one leg at a time, where a leg $\nabla_A\phi$ of the vertex of the current becomes $\nabla_AK$. A vertex with one leg gives a vertex without legs, which is a function. Thus $S_KV^{(3)}[g]=V^{(2)}[gK]$, $S_KV^{(2)}[\mathsf{u}]=V^{(1)}[\mathsf{u}K]$ and $S_KV^{(1)}[h]=hK$.

\begin{lemma}
\label{gcf:lem:contact}
Let $V_1,\ldots,V_N$ be vertices of the bare functional of section~\ref{gs:sec:sphere}, with sources of the class of sources on the sphere, including vertices of the current and the term of second order in the gauge field, and let $K$ be a function of the class of sources on the sphere. Then on the line \eqref{eq:line}, as meromorphic functions of $d$,
\begin{equation}
\big\langle V^{(1)}[\mathsf{D}_0K],V_1,\ldots,V_N\big\rangle^c = \sum_{r=1}^N\big\langle V_1,\ldots,S_KV_r,\ldots,V_N\big\rangle^c \, ,
\label{gcf:eq:contact}
\end{equation}
where $\mathsf{D}_0=-\nabla^2+\tfrac14d(d-2)$ and a vertex without legs on the right-hand side contributes only for $N=1$, as the integral of the function.
\end{lemma}

\begin{proof}
We follow appendix~\ref{app:gauge}. We place the point $X_0$ of the vertex $V^{(1)}[\mathsf{D}_0K]$ last and use \eqref{eq:last}. For ${\rm Re}\,\Dp<-2$ and ${\rm Re}\,m>N$ the first identity of \eqref{eq:lap}, applied twice with the divergence theorem on the ball, moves the Laplacian from $K$ to the propagator from $X_0$ to the leg $l_1$ with which the leg of $V^{(1)}$ is paired, including a derivative at $l_1$. The terms from the boundary vanish as in the proof of Lemma~\ref{lem:ward2}. By the third identity of \eqref{eq:lap},
\begin{equation}
\mathsf{D}_0\,(s^2)^{-\Dp} = \Big(\tfrac14d(d-2)-\Dp(d-1-\Dp)\Big)(s^2)^{-\Dp}+2\Dp\,(d-2-2\Dp)\,(s^2)^{-\Dp-1} \, .
\end{equation}
The left-hand side of \eqref{gcf:eq:contact} is therefore $(\tfrac14d(d-2)-\Dp(d-1-\Dp))\,\langle V^{(1)}[K],V_1,\ldots,V_N\rangle^c+2\Dp(d-2-2\Dp)\,\mathcal{B}_K(d,\Dp,\Dp+1)$, where $\mathcal{B}_K(d,\Dp,\Dp')$ is the sum over the pairings of the integrals in which the propagator from $X_0$ to $l_1$ has the exponent $\Dp'$. The first coefficient vanishes on the line. The computation of the proof of Lemma~\ref{lem:contact} with $\omega_{ij}(X_0)$ replaced by $K(X_0)$ applies to $\mathcal{B}_K$, with the same treatment of a leg $l_1$ with a derivative by \eqref{eq:transfer}. Since $V^{(1)}$ has one leg, there is no pairing of $l_1$ and $l_2$ at one vertex. It gives that $(d-2\Dp')\,\mathcal{B}_K$ has no pole along $\mathsf{H}$ and that on $\mathsf{H}$ it equals $C_\phi\,\vol_d$ times the right-hand side of \eqref{gcf:eq:contact}, without the minus sign of \eqref{eq:contact}, which there comes from the replacement of $\phi$ by $-\omega\phi$. The property \eqref{eq:NP} for $\mathcal{B}_K$ follows from Proposition~\ref{prop:np}, which concerns integrals with one line of exponent $\Dp'$. The vertices with one leg $\nabla_AK$ which the right-hand side contains for a vertex of the current are reduced to vertices without derivatives by the second identity of \eqref{eq:lap}. On the line $2\Dp=d-2$ and $(d-2)\,C_\phi\,\vol_d=1$, which gives \eqref{gcf:eq:contact}.
\end{proof}

\begin{lemma}
\label{gcf:lem:shift}
Let $K$ be a local expression with values in the space of the fields, built from the sources, the metric and the curvature, with coefficients which are Laurent polynomials in $\vep$, and let the bare sources $g_0$, $a_0$, $\sM_0$ and $h_0$ be given. Then on the backgrounds $\Omega^2\delta_{\mu\nu}$ and on the unit sphere, at each order in the sources,
\begin{gather}
W_0\big[g_0,a_0,\sM_0+g_0K,\,h_0+\mathsf{D}K+\sM_0K+\tfrac12g_0KK\big] \notag\\
= W_0\big[g_0,a_0,\sM_0,h_0\big]+\int\sqrt\gamma\left(h_0K+\tfrac12K\mathsf{D}K+\tfrac12\sM_0KK+\tfrac16g_0KKK\right) ,
\label{gcf:eq:shift}
\end{gather}
with the contractions of the field indices of the vertices.
\end{lemma}

\begin{proof}
On the unit sphere we consider the sources $s_t$ with $g_0$, $a_0$, $\sM_t=\sM_0+tg_0K$ and $h_t=h_0+t\,\mathsf{D}K+t\sM_0K+\tfrac12t^2g_0KK$ for $0\leq t\leq1$. At each order in the sources $W_0[s_t]$ is a polynomial in $t$. Its derivative is $\sum_{N\geq1}\frac{(-1)^N}{(N-1)!}\langle\dot V,V_t,\ldots,V_t\rangle^c$, where $V_t$ is the interaction at $s_t$ and $\dot V=V^{(1)}[\mathsf{D}K+\sM_tK]+V^{(2)}[g_0K]$, since $\pr_th_t=\mathsf{D}K+\sM_tK$. We split $\mathsf{D}K=\mathsf{D}_0K+\mathsf{D}_aK$, where $\mathsf{D}_aK$ contains the terms of $-D^\mu D_\mu K$ with the gauge field. By Lemma~\ref{gcf:lem:contact} the vertex $V^{(1)}[\mathsf{D}_0K]$ gives $\sum_r\langle\ldots,S_KV_t,\ldots\rangle^c$. Here $S_KV_t$ contains $V^{(2)}[g_0K]$ from the cubic vertex and $V^{(1)}[\sM_tK]$ from the vertex of the source of $\phi_i\phi_j$. From the vertex of the current and the term of second order in the gauge field it contains the vertex $V^{(1)}[\mathsf{D}_aK]$, after an integration by parts at its point, since these two vertices are the terms of the action with the gauge field and $S_K$ of them is the variation of $\tfrac12\phi\mathsf{D}\phi-\tfrac12\phi\mathsf{D}_0\phi$ in the direction $K$. The vertex $V^{(1)}[h_t]$ gives the function $h_tK$, which contributes for $N=2$. After the shift of the summation index these terms cancel against the terms with $V^{(2)}[g_0K]$, $V^{(1)}[\sM_tK]$ and $V^{(1)}[\mathsf{D}_aK]$ of $\dot V$, and we obtain $\pr_tW_0[s_t]=\int\sqrt\gamma\,h_t\cdot K$. Integration over $t$ gives \eqref{gcf:eq:shift}. On $\Omega^2\delta_{\mu\nu}$ the functional is defined by \eqref{eq:defcurved}, and the identity follows from the identity in flat space, which is the identity on the sphere in the chart \eqref{eq:chart} for sources which vanish near $P$. We apply it to $\Omega^{(d-2)/2}K$ and use the conformal covariance $\Omega^{(d+2)/2}\,\mathsf{D}(\Omega^{-(d-2)/2}f)=\mathsf{D}^{\rm fl}f$ of the operator.
\end{proof}

Since the right-hand side of \eqref{gcf:eq:shift} differs from the left-hand side by a local functional, the renormalised functional is unchanged if the bare sources are replaced as in \eqref{gcf:eq:shift} and the local functional is subtracted from $\cX$. If $K$ has pure poles as coefficients, the new bare sources have the form \eqref{gcu:eq:bare} up to terms with regular coefficients, which come from the factor $\xi_d$ in $\mathsf{D}K$. A regular term $\mathsf{Reg}$ in the bracket of $h_0$ is removed by the redefinition $h\to h+r$ of the source with $r$ equal to minus the regular part of $Z_h^{-1}$ applied to $Z_h\mathsf{Reg}$, and the regular part of the local functional is removed from $W$. The new functional is $W[g,a,\sM,h+r]$ up to a regular local functional, which is a finite change of the scheme of the source $h$, and on the unit sphere $r$ does not vanish in general. Neither step changes $\hB^I$ or $\hr_I$, since $L^I$ and $N_I$ do not change. A shift with $K$ which does not contain $\sM$ changes neither $Z_h$ nor $Z_\sM$, and hence neither $\gamma_h$ nor $\gamma_\sM$, while a shift with $K$ linear in $\sM$ changes $Z_\sM$ and $\gamma_\sM$. The inhomogeneous terms of the variations of $h$ and of $\sM$ change, and so do the coefficients $\ell_{{\rm h}J}$ and $\vk_{\rm h}-\vk_{\rm c}$ of the source $h$ at $g=0$. The last statement of Property~\ref{gs:P:weyl} therefore refers to the set of bare sources after the choice. Property~\ref{gs:P:free} concerns the divergence scheme, and the order zero of $\cK_{IJ}$ and the coefficients of the anomaly linear in the relevant sources at $g=0$ do not change.

We first note that $\fk$ has no term of order zero in the couplings. Indeed, at order one in $\log\Omega$ and order zero in the couplings the term of $W$ linear in a source $\sM_2$ of disjoint support is the second kernel of Lemma~\ref{gcu:lem:lowest} between $\sM_2$ and $-Z_\sM\fk R$, which receives no contribution from $h_0$. Indeed, at this order the terms of $h_0$ which are linear in $\sM_2$ and its terms without sources are Laplacians of local expressions with disjoint supports, and the kernel $|x-y|^{-(d-2)}$ is harmonic away from $x=y$. Lemma~\ref{gcu:lem:potential} with $\kappa=4$ gives $\fk^{(0)}=0$.

\begin{lemma}
\label{gcf:lem:gaugecov}
Let a set of pure poles satisfy Hypothesis~\ref{hyp:finite} for the cubic theory, and let $N\geq2$. Let $\eta$ be the part of $h_0$ at $h=0$, $\sM=0$ and $a_\mu=0$ of order zero or of order one in the couplings, and let its leading pole at $d=6$ be $\vep^{-\mathsf{r}}\mathsf{D}K$ with $K=c\,R$ at order zero and $K=\mathsf{A}_1\nabla^2g+\mathsf{A}_2Rg$ at order one, with a vector $c$ and linear maps $\mathsf{A}_1$, $\mathsf{A}_2$ from the couplings to the space of the fields. Then $\omega c=0$, so that $c=0$, and $\omega\mathsf{A}_k(g)=\mathsf{A}_k(\omega g)$ for every element $\omega$ of the Lie algebra, provided that the terms of $h_0$ of lower order in the couplings vanish.
\end{lemma}

\begin{proof}
We work at first order in $\theta$ throughout, so that only $g$ and $(\theta\omega)g$, which are sources of the class of flat sources, occur. We compare the functional at a pure gauge field $a_\mu=\pr_\mu\theta\,\omega$, with a constant $\omega$ in the Lie algebra and a function $\theta$ of the class of flat sources, with the functional at $a_\mu=0$ and the rotated sources $\mathsf{R}g$, $\mathsf{R}h$, where $\mathsf{R}=e^{\theta\omega}$. We take $h=h_2$ with a support disjoint from the supports of $\theta$ and $g$, and on which $\Omega=1$, so that $\mathsf{R}h_2=h_2$, and $\sM=0$. By the background gauge invariance of the bare functional, Proposition~\ref{prop:gauge}, $W_0$ is unchanged if its bare sources are transformed by $\mathsf{R}$. The bare couplings transform into $\mathsf{R}(g+L(g))=\mathsf{R}g+L(\mathsf{R}g)$, since $L^I$ is covariant, and at the orders which we consider the terms $N_ID_\mu g^I$ of the bare gauge field and the bare source of $\sM$ enter only together with an odd number of legs or with the terms of $h_0$ of lower order, which vanish, since $\fd_I$ and $\fk$ have no term of order zero. The factor $Z_h$ and the counterterms enter only through local terms with $h_2$, which vanish, since $h_2$ has a support disjoint from the other sources, and the monomials of $h_0$ with the field strength vanish at a pure gauge field. The bare gauge field becomes $\mathsf{R}a_\mu\mathsf{R}^{-1}-\pr_\mu\mathsf{R}\,\mathsf{R}^{-1}=0$. At a pure gauge field $D_\mu g=\mathsf{R}^{-1}\pr_\mu(\mathsf{R}g)$, so that a monomial of $h_0$ with a linear map $\mathsf{A}$ applied to covariant derivatives of $g$ becomes, after the rotation, $\mathsf{R}\mathsf{A}\mathsf{R}^{-1}$ applied to the same derivatives of $\mathsf{R}g$ at $a_\mu=0$. Hence the term of $W[g,\pr\theta\,\omega,0,h_2]-W[\mathsf{R}g,0,0,h_2]$ which is linear in $h_2$ and of the order considered is, apart from local terms which vanish since $h_2$ has a disjoint support, the second-order term of the bare functional between $h_2$ and $\Delta=(\mathsf{R}\,\eta[\mathsf{A}]\,\mathsf{R}^{-1}-\eta[\mathsf{A}])(\mathsf{R}g)$, where $\eta[\mathsf{A}]$ denotes $\eta$ with its linear maps and $\mathsf{R}\eta[\mathsf{A}]\mathsf{R}^{-1}$ the same expression with the maps conjugated by $\mathsf{R}$ at each point. Both functionals are finite, so that this term is finite. To first order in $\theta$ the leading pole of $\Delta$ at $d=6$ is $\vep^{-\mathsf{r}}\theta\,\mathsf{D}\tilde K$, with $\tilde K=(\omega c)R$ at order zero and $\tilde K=[\omega,\mathsf{A}_1]\nabla^2\tilde g+[\omega,\mathsf{A}_2]R\tilde g$ at order one, where $\tilde g=\mathsf{R}g$ and $[\omega,\mathsf{A}](g)=\omega\mathsf{A}(g)-\mathsf{A}(\omega g)$. As in the proof of Proposition~\ref{gcf:prop:frame} the finiteness gives that the moments of $\Omega^4\theta\,\mathsf{D}\tilde K=-\theta\,\pr^2(\Omega^2\tilde K)$ with the harmonic polynomials vanish. For $H=1$ this is $\int\Omega^2\tilde K\,\pr^2\theta=0$ for all $\theta$ of the class of flat sources, so that $\pr^2(\Omega^2\tilde K)=0$, and since $\Omega^2\tilde K$ has compact support, $\tilde K=0$. At order zero this gives $\omega c=0$, and for $N\geq2$ the only vector annihilated by all $\omega$ is zero. At order one $\nabla^2\tilde g$ and $R\tilde g$ are independent at points where $R\neq0$, so that $[\omega,\mathsf{A}_1]=[\omega,\mathsf{A}_2]=0$. For $N=2$ a map which commutes with ${\rm so}(2)$ need not commute with the reflections, as $\mathsf{A}(g)_i=\epsilon_{ik}g_{kjj}$ shows, which is enough for the form of the bare sources.
\end{proof}

\begin{proposition}
\label{gcf:prop:frame}
Let a set of pure poles satisfy Hypothesis~\ref{hyp:finite} for the cubic theory. Then there are a local $K$ with pure poles as coefficients, of order zero and one in the couplings, and a regular local redefinition $h\to h+r$, such that after the replacement \eqref{gcf:eq:shift} the set satisfies the frame condition \eqref{gcu:eq:frame}. The functional changes by a regular local functional and by the redefinition of $h$, and $\hB^I$, $\hr_I$, $\gamma_h$ and $\gamma_\sM$ do not change.
\end{proposition}

\begin{proof}
We use that $\fk^{(0)}=0$, as shown before Lemma~\ref{gcf:lem:gaugecov}. We use a source $h_2$ whose support is disjoint from the supports of the other sources and on which $\Omega=1$. At order zero in the couplings and at vanishing relevant sources and $a_\mu=0$ the bare source $h_0$ is $\eta=c_1R^2+c_2R_{\mu\nu}R^{\mu\nu}+c_3\nabla^2R$ with Laurent polynomials $c_i$, since the square of the Weyl tensor vanishes. The term of $W$ linear in $h_2$ at this order is the term of second order of the bare functional, which is the integral of $\Omega^4\eta$ against $h_2$ with the kernel $|x-y|^{-(d-2)}$ by \eqref{eq:defcurved}. Let $\vep^{-\mathsf{r}}$ be the highest pole of the $c_i$. The finiteness gives that the potential of the coefficient of $\vep^{-\mathsf{r}}$ of $\Omega^4\eta$ at $d=6$ vanishes outside a ball, so that its moments $\int\Omega^4\eta\,H$ with all harmonic polynomials $H$ vanish. The combination $\mathsf{D}R=-\nabla^2R+\tfrac15R^2$ at $d=6$ satisfies this, since $\Omega^4\mathsf{D}R=-\pr^2(\Omega^2R)$ and $\Omega^2R$ has compact support. We show that it is the only one. At second order in $w=\log\Omega=\lambda f$ we have $R=-10\,\lambda\,\pr^2f+{\rm O}(\lambda^2)$ and $R_{\mu\nu}=-\lambda(4\pr_\mu\pr_\nu f+\delta_{\mu\nu}\pr^2f)+{\rm O}(\lambda^2)$, so that for $H=1$ the moments of $R^2$ and $R_{\mu\nu}R^{\mu\nu}$ are $100\lambda^2\int(\pr^2f)^2$ and $30\lambda^2\int(\pr^2f)^2$ up to ${\rm O}(\lambda^3)$, while the moment of $\mathsf{D}R$ vanishes. By polarisation, and by linearity in the real and imaginary parts, the moments of the part of $R_{\mu\nu}R^{\mu\nu}-\tfrac3{10}R^2$ which is bilinear in two functions $f_1$ and $f_2$ in place of $f$ vanish as well. We take $f_1=e^{tu\cdot x_\parallel}\varphi$, $f_2=e^{tv\cdot x_\parallel}\varphi$ and $H=e^{-t(u+v)\cdot x_\parallel}$, with $u,v\in\mathbb{C}^3$ in the parallel directions, $(u+v)^2=0$ and a cutoff $\varphi$ of the class of flat sources with $\int\varphi^2\neq0$. The moments with this $H$ vanish, since its Taylor terms are harmonic polynomials and the supports are compact. The exponentials cancel, the derivatives of $\varphi$ give lower powers of $t$, and the coefficient of $t^4$ of the moment at order $\lambda^2$ is $32\lambda^2((u\!\cdot\!v)^2-u^2v^2)\int\varphi^2=32\lambda^2(u\!\cdot\!(u+v))^2\int\varphi^2$, which does not vanish for suitable $u$ and $v$. Hence the leading pole of $\eta$ is a multiple $\vep^{-\mathsf{r}}c\,\mathsf{D}R$, with a vector $c$ in the space of the fields, and the shift with $K=\vep^{-\mathsf{r}}c\,R$ removes it, up to terms with lower poles from $\xi_d-\xi_6$. By descending induction in the order of the poles we remove all poles of $\eta$, and the remaining regular terms of $h_0$ are removed by a redefinition of $h$. For $N\geq2$ Lemma~\ref{gcf:lem:gaugecov} gives $c=0$ at each order of the poles, so that the part of order zero of $h_0$ at vanishing relevant sources has no poles, hence vanishes, and no shift is needed. For $N=1$ there is no gauge field, and the shift above does not involve one.

At first order in the couplings, after this step, the term of $W$ linear in a coupling $g_1$ and in $h_2$ is again the second-order term with the part $\eta_1$ of $h_0$ of first order in the couplings, since the order-zero part of $h_0$ and of $\fk$ now vanishes and the terms of second order with a coupling and $h_2$ vanish. The monomials of $\eta_1$ are $\mathsf{A}_kg$ times $\nabla^2\nabla^2$, $R\nabla^2$, $R^{\mu\nu}\nabla_\mu\nabla_\nu$, $\pr_\mu R\,\nabla^\mu$, $R^2$, $R_{\mu\nu}R^{\mu\nu}$ and $\nabla^2R$, with seven linear maps $\mathsf{A}_k$ from the couplings to the space of the fields, where the derivatives act on $\mathsf{A}_kg$. By polarisation it suffices to consider one component of the couplings and one component of the fields. The combinations $\mathsf{D}\nabla^2(\mathsf{A}g)$ and $\mathsf{D}(R\,\mathsf{A}g)$ pass the test, since $\Omega^4\mathsf{D}K=-\pr^2(\Omega^2K)$ for every $K$, and the moments vanish only on their span. To show this, we take one component $g$ and subtract the multiples of $\mathsf{D}\nabla^2g$ and $\mathsf{D}(Rg)$ which remove the monomials $\nabla^2\nabla^2g$ and $(\nabla^2R)\,g$. The remaining combination is $\alpha_1R\nabla^2g+\alpha_2R^{\mu\nu}\nabla_\mu\nabla_\nu g+\alpha_3\pr_\mu R\,\nabla^\mu g+\alpha_4R^2g+\alpha_5R_{\mu\nu}R^{\mu\nu}g$, and at order $\lambda$ only its first three terms contribute. With $f=f_1$, $g=f_2$ and $H$ as above, the coefficient of $t^4$ of the moment at order $\lambda$ is $-\lambda\int\varphi^2$ times $10\alpha_1a\,v^2+\alpha_2(4b^2+a\,v^2)+10\alpha_3ab$, where $a=u^2$, $b=u\!\cdot\!v$ and $v^2=-a-2b$. For a fixed nonzero null vector $s=u+v$ the quantities $u^2=a$ and $u\!\cdot\!s=a+b$ take independent values, so that the coefficients of $a^2$, $ab$ and $b^2$ vanish. This gives $\alpha_2=0$, then $\alpha_1=0$ from $10\alpha_1+\alpha_2=0$, and $\alpha_3=0$ from $20\alpha_1+2\alpha_2-10\alpha_3=0$. At order $\lambda^2$ the moment with $H=1$ is then $\lambda^2\int g\,\big(100\,\alpha_4(\pr^2f)^2+\alpha_5\sum_{\mu\nu}(4\pr_\mu\pr_\nu f+\delta_{\mu\nu}\pr^2f)^2\big)$. Since the factor of $g$ is invariant under the rotations of $x_\perp$ and the functions $g$ of the class separate such functions, it vanishes at every point. At a point where the Hessian of $f$ is nonzero, trace-free and has only parallel components, $\pr^2f=0$ and the sum is positive, so that $\alpha_5=0$, and a Hessian with nonzero trace then gives $\alpha_4=0$. Such $f$ are polynomials of degree two in $x_\parallel$ times a cutoff of the class of flat sources. Shifts with $K=\vep^{-\mathsf{r}}(\nabla^2\mathsf{A}_1g+R\,\mathsf{A}_2g)$ with two linear maps $\mathsf{A}_1$ and $\mathsf{A}_2$ then remove the poles of $\eta_1$ by descending induction, and the regular remainder is removed by a redefinition of $h$. With the gauge field the operator $\mathsf{D}$ contains $a_0$, and $\mathsf{D}K$ has the form of the monomials of \eqref{gcu:eq:bare} if $K=\mathsf{A}_1D^2g+\mathsf{A}_2Rg$ with linear maps which commute with the action of the Lie algebra, since then $D_\mu(\mathsf{A}_kX)=\mathsf{A}_kD_\mu X$. For $N\geq2$ this holds by Lemma~\ref{gcf:lem:gaugecov}, at each order of the poles, and for $N=1$ there is no gauge field. The shift changes $\sM_0$ by $g_0K$, which is of second order in the couplings, and it changes $\cX$ by the local functional of \eqref{gcf:eq:shift}, of which the pole part is kept in $\cX$.
\end{proof}

The proof uses the polarisation in the couplings and no covariance beyond that of $L^I$, so that it holds for any number of fields. The covariance under the Lie algebra which the gauge field requires follows from the finiteness by Lemma~\ref{gcf:lem:gaugecov}. The reflections of ${\rm O}(N)$ do not enter. After the choice the freedom of the field shifts which remains consists of the shifts with $K$ which vanishes at vanishing relevant sources to first order in the couplings, such as $K=(c/\vep)\,\mathsf{u}\,\sM$ with a constant linear map $\mathsf{u}$, or $K=(c/\vep)\,g_{ijk}\sM_{jk}$, and of the shifts of order at least two in the couplings. These do not change the bare sources at vanishing relevant sources through first order in the couplings. The local functional of \eqref{gcf:eq:shift} has no term of first degree in the relevant sources at $g=0$, since $K$ is of first degree in $\sM$ there or of higher order in the couplings, and $h_0K$ is then of degree two. The terms of $W$ and of the anomaly of first degree in the relevant sources at $g=0$ therefore do not change either. The proof of Proposition~\ref{gcu:prop:first} and the derivation of the last statement of Property~\ref{gs:P:weyl} and of Property~\ref{gs:P:free} from the other statements of Properties~\ref{gs:P:functional} to~\ref{gs:P:sphere}, which use only these terms and the bare sources at vanishing relevant sources through first order, hold for every set in this residual family. The shift $K=(c/\vep)\,g_{ijk}\sM_{jk}$ puts poles into the Weyl variations of $h$ and $\sM$ induced by the bare sources at third order in the couplings, so that the choice does not fix the frame which Properties~\ref{gs:P:functional} to~\ref{gs:P:sphere} need.

\subsection{The properties for cubic scalars}
\label{gcp:sec:main}

In this section we derive Properties~\ref{gs:P:functional} to~\ref{gs:P:free} for the cubic theory from Hypothesis~\ref{hyp:finite} for the cubic theory, stated on the backgrounds $\Omega^2\delta_{\mu\nu}$, for the class of flat sources with $D_0=3$. This class of flat sources consists of the smooth functions with compact support of three coordinates $x_\parallel$ and of $|x_\perp|^2$, and of the corresponding one-forms. It contains the class of flat sources with $D_0=1$, so that the hypothesis is stronger. At $d=6$ the transverse dimension is $m=d-D_0=3$, as for the $\phi^4$ theory with $D_0=1$, and the statements of appendices~\ref{app:mero}, \ref{app:line} and~\ref{app:gauge} which we use, Propositions~\ref{prop:mero} and~\ref{prop:linecurrent} and Lemmas~\ref{lem:ward2}, \ref{lem:contact} and~\ref{lem:np}, hold for every $D_0\geq1$. The steps which are specific to six dimensions are three lemmas on local expressions, the construction of a covariant set of bare sources, and the removal of the poles of the Weyl variation of the source of $\phi_i$ by field shifts.

\subsubsection{Local expressions}
\label{gcp:sec:local}

We call a formal expression a polynomial in the derivatives of a finite number of functions on $\mathbb{R}^6$, the inputs, and in the components of a finite number of one-forms, with constant coefficients, which is invariant under the orthogonal transformations of $\mathbb{R}^6$. The inputs are the derivatives of $\log\Omega$, the sources $g^I$, $\sM$, $h$, the components of the gauge field and, where it occurs, the parameter $\sigma$ of a Weyl variation. We note that a local expression of the theory, whose coefficients are functions of the couplings, becomes a formal expression after the expansion around constant couplings of the proof of Lemma~\ref{lem:indep}, at each order in the couplings. On the backgrounds $\Omega^2\delta_{\mu\nu}$ at $d=6$ a covariant expression of dimension $\Delta$, multiplied by $\Omega^{\Delta}$ and written with the flat sources $\Omega^{w}s$ of \eqref{eq:defcurved}, is a formal expression in the derivatives of $\log\Omega$ and in the flat sources, since it does not change under $\log\Omega\to\log\Omega+c$ at fixed flat sources. At $d=6-\vep$ the weights of the flat sources depend on $\vep$, and the expression contains in addition powers of $\Omega$ with exponents of order $\vep$. The number of its derivatives and one-form factors together, where a derivative of $\log\Omega$ counts with its order, is at most $\Delta$. A formal expression is split into parts which are homogeneous in each input by scaling the inputs, and into parts with a fixed number of derivatives by dilations, which preserve the class of flat sources. By polarisation each part is a sum of multilinear expressions. We write $p(\zeta_1,\ldots,\zeta_k)$ for the symbol of a multilinear expression, defined by inserting $e^{\zeta_j\cdot x}$, or $c_je^{\zeta_j\cdot x}$ for a one-form input with a constant vector $c_j$, for complex $\zeta_j$. By the first fundamental theorem of invariant theory for ${\rm O}(6)$ the symbol of a formal expression is a polynomial in the scalar products of the $\zeta_j$ and $c_j$, and every such polynomial is the symbol of a formal expression. A polynomial in these scalar products of degree at most three is determined by its values on vectors in $\mathbb{C}^3$, since by the second fundamental theorem the relations among the scalar products of vectors in $\mathbb{C}^3$ are generated by the minors of order four of their Gram matrix, which have degree four.

\begin{lemma}
\label{gcp:lem:gram}
Let $E$ be a multilinear formal expression whose number of derivatives and one-form factors together is at most six. If $E$ vanishes at every point for all inputs in the class of flat sources with $D_0=3$, then $E=0$.
\end{lemma}

\begin{proof}
We take the inputs $u_j=e^{\zeta_j\cdot x_\parallel}\varphi$, and $c_je^{\zeta_j\cdot x_\parallel}\varphi$ for the one-forms, with $\zeta_j,c_j\in\mathbb{C}^3$ in the parallel directions and a cutoff $\varphi$ of the class of flat sources which equals one near a point. By multilinearity in the real and imaginary parts the assumption applies to them. At the point the value of $E$ is the symbol times an exponential, so that the symbol vanishes for all vectors in $\mathbb{C}^3$. It is a polynomial of degree at most three in the scalar products, so that it vanishes identically, and $E=0$.
\end{proof}

\begin{lemma}
\label{gcp:lem:exact}
Let $P$ be a multilinear formal expression in at least one input whose number of derivatives and one-form factors together is at most four, and assume that $\int d^6x\,P(u)(x)\,|x-y|^{-4}$ vanishes for all $y$ outside a ball which contains the supports, for all inputs in the class of flat sources with $D_0=3$. Then $P=\pr^2Q$ with a unique formal expression $Q$. If $P$ contains $\log\Omega$ only differentiated and has no term $c\,\pr^2\log\Omega$ with $c\neq0$, then $Q$ contains $\log\Omega$ only differentiated.
\end{lemma}

\begin{proof}
The vanishing of the potential outside the ball is equivalent to the vanishing of the moments $\int P(u)\,H$ with the harmonic polynomials $H$, by the exterior multipole expansion, and hence of the moments with $H=e^{\zeta\cdot x_\parallel}$ for $\zeta\in\mathbb{C}^3$ with $\zeta\!\cdot\!\zeta=0$, whose terms $(\zeta\!\cdot\!x)^m/m!$ are harmonic. We take the inputs as in the proof of Lemma~\ref{gcp:lem:gram} with $\zeta_j$ replaced by $t\zeta_j$, a cutoff with $\int\varphi^k\neq0$, and $\zeta=-t\sum_j\zeta_j$. If $\sum_j\zeta_j$ is a null vector, the exponentials cancel, and the moment is a polynomial in $t$ whose term of highest degree is $t^Dp\int\varphi^k$, where $D$ is the degree of $p$ in the $\zeta_j$. Hence $p$ vanishes on $\{(\sum_j\zeta_j)^2=0\}$ for vectors in $\mathbb{C}^3$. This quadric is irreducible, since $(\sum_j\zeta_j)^2$ is a quadratic form of rank three, and by the Nullstellensatz $p=(\sum_j\zeta_j)^2\,r$ there. The quotient $r$ is a polynomial in the scalar products of degree at most one, and we read it as a polynomial in the scalar products of vectors in $\mathbb{C}^6$. Then $p-(\sum_j\zeta_j)^2r$ has degree at most two and vanishes for vectors in $\mathbb{C}^3$, so that it vanishes. The formal expression $Q$ with the symbol $r$ gives $P=\pr^2Q$. It is unique, since $\pr^2Q=0$ means $(\sum_j\zeta_j)^2r=0$, and $(\sum_j\zeta_j)^2$ is a nonzero polynomial since there is at least one input. Now let $P$ contain $\log\Omega$ only differentiated, so that $p$ vanishes when the vector $\zeta_1$ of an input $\log\Omega$ vanishes. If $P$ has a further input, then $(\sum_{j\geq2}\zeta_j)^2$ is a nonzero polynomial, and $p=(\sum_j\zeta_j)^2r$ at $\zeta_1=0$ shows that $r$ vanishes at $\zeta_1=0$, since the polynomials in the components of the vectors form an integral domain. If $\log\Omega$ is the only input, then $p$ is a polynomial in $\zeta_1^2$ of degree at most two without constant term, $p=c\,\zeta_1^2+c'(\zeta_1^2)^2$, and $r=c+c'\zeta_1^2$ vanishes at $\zeta_1=0$ since $c=0$. The example $P=\pr^2\log\Omega$, with $Q=\log\Omega$, shows that the condition on the term $c\,\pr^2\log\Omega$ cannot be omitted.
\end{proof}

Let $P$ be a covariant expression of dimension four on $\Omega^2\delta_{\mu\nu}$ at $d=6$. Let the term of the bare functional which is linear in $\Omega^4P$ and in $h_2$, with the propagator in the background gauge field, be finite at its leading pole for every source $h_2$ of the class of flat sources whose support is disjoint from the supports of the inputs. At $a_\mu=0$ this is the second-order term of the bare functional between $\Omega^4P$ and $h_2$. Since $\Omega^4P$ is invariant under the rotations of $x_\perp$ and these sources separate invariant functions, the potential of the leading pole of $\Omega^4P$ at $a_\mu=0$ vanishes outside a ball. The part of $\Omega^4P$ without inputs vanishes, and its part which is linear in $\log\Omega$ and has no other input comes from the monomials built from the curvature alone, of which only $\nabla^2R$ is linear in the curvature, so that this part is a multiple of $\pr^2\pr^2\log\Omega$. Lemma~\ref{gcp:lem:exact} therefore gives $\Omega^4P=\pr^2Q$, where $Q$ contains $\log\Omega$ only differentiated and its part which is linear in $\log\Omega$ and has no other input is a multiple of $\pr^2\log\Omega$, and with $\Omega^4(-\nabla^2+\tfrac15R)(\Omega^{-2}f)=-\pr^2f$ we obtain $P=(-\nabla^2+\tfrac15R)Y$ with $Y=-\Omega^{-2}Q$, a local expression in the derivatives of $\log\Omega$ and in the sources. This argument uses only that $\Omega^4P$ is a formal expression which is invariant under the rotations of $x_\perp$, has no part without inputs and has no term $c\,\pr^2\log\Omega$ with $c\neq0$. With the gauge field we proceed by induction in the degree in $a_\mu$, with the operator $\mathsf{D}$ of section~\ref{gcf:sec:frame}, which at $d=6$ is $-D^\mu D_\mu+\tfrac15R$. Let $Y_{<j}$ be a local expression such that $P-\mathsf{D}Y_{<j}$ has no terms of degree smaller than $j$ in $a_\mu$. The term with $\mathsf{D}Y_{<j}$ in place of $P$ vanishes for every $h_2$ of disjoint support, by Lemma~\ref{gcf:lem:contact} with the vertices of the current and the term of second order in the gauge field, as in the proof of Lemma~\ref{gcf:lem:shift}. Hence the term of degree $j$ in $a_\mu$ of the term with $P-\mathsf{D}Y_{<j}$ is the second-order term with the part of degree $j$ of $P-\mathsf{D}Y_{<j}$. Since this part contains the gauge field, the argument above gives it as $(-\nabla^2+\tfrac15R)Y_j$ with a local expression $Y_j$ of degree $j$ in $a_\mu$, so that $P-\mathsf{D}(Y_{<j}+Y_j)$ has no terms of degree smaller than $j+1$. Since $P$ has at most four one-form factors and $Y_j$ has at most two derivatives and one-form factors together, the induction ends after finitely many steps, and we obtain $P=\mathsf{D}Y$ with a local expression $Y$ in the derivatives of $\log\Omega$, in the sources and in the gauge field. The operator $\mathsf{D}$ is injective on the local expressions $Z$ for which $\Omega^2Z$ has no part without inputs, among them those of dimension two, by a filtration in the degree in $a_\mu$, since $(-\nabla^2+\tfrac15R)Z=0$ gives $\pr^2(\Omega^2Z)=0$ and hence $Z=0$ by the uniqueness of Lemma~\ref{gcp:lem:exact}. The function $\Omega^{-2}$ lies in the kernel of $-\nabla^2+\tfrac15R$. For inputs of compact support $Y$ has compact support, since it has positive dimension, and it is the only function of compact support with $\mathsf{D}Y=P$, since $\mathsf{D}Z=0$ gives $D^\mu D_\mu(\Omega^2Z)=0$ with the covariant derivative of flat space. An integration by parts then gives $D_\mu(\Omega^2Z)=0$, so that $\Omega^2Z$ has a constant length and vanishes.

\begin{lemma}
\label{gcp:lem:natural}
Let $P$ be a covariant expression of dimension four, gauge invariant under the background gauge transformations, and let $P=\mathsf{D}Y$ on all backgrounds $\Omega^2\delta_{\mu\nu}$ at $d=6$, with the operator $\mathsf{D}=-D^\mu D_\mu+\tfrac15R$ which contains the gauge field and with a local expression $Y$ as above. Then $Y$ is a covariant gauge invariant expression of dimension two in the curvature, $f_{\mu\nu}$ and the covariant derivatives of the sources. If the coefficients of $P$ are covariant under the field rotations, so are those of $Y$.
\end{lemma}

\begin{proof}
A M\"obius map $f$ of $\mathbb{R}^6$ whose singular point lies outside the supports of the inputs maps $\Omega^2\delta_{\mu\nu}$ to $(\Omega\circ f)^2|f'|^2\delta_{\mu\nu}$, again a background of this form near the supports, and it acts on the sources by the pullback. Translations, dilations and special conformal maps are of this kind. Covariant expressions are natural under this action, and so is $\mathsf{D}$, which is conformally covariant. We show that $Y$ is natural under these maps. It is invariant under translations, since its coefficients are constant. We treat the infinitesimal dilations first and then the infinitesimal special conformal transformations, with a constant parameter $b$. The defect of naturality $\Delta_bY$, the action of the generator on $Y$ minus $Y$ of the transformed data, is a formal expression with the additional constant vector $b$, for the special conformal transformations since $Y$ is invariant under the rotations and natural under the dilations, so that no term with an explicit factor $b\cdot x$ remains, and $\mathsf{D}\,\Delta_bY=\Delta_bP=0$, since $P$ and $\mathsf{D}$ are natural. The injectivity of $\mathsf{D}$, with $b$ as an additional vector, gives $\Delta_bY=0$, and the finite maps follow by integration. Here $\Omega^2\Delta_bY$ has no part without inputs, since such a part could only come from the part of $\Omega^2Y$ which is linear in $\log\Omega$ and has no other input, which is a multiple of $\pr^2\log\Omega$, and the infinitesimal maps add to the transport of $\log\Omega$ an affine function of $x$. In the same way $Y$ is invariant under the infinitesimal background gauge transformations, whose parameter enters as an additional input, since $P$ is invariant and $\mathsf{D}$, which contains the gauge field, commutes with them. Let $x_0$ be a point. A translation, a dilation and a special conformal transformation map the data to data with $\log\Omega(x_0)=0$ and $\pr_\mu\log\Omega(x_0)=0$, since the special conformal transformation with the parameter $b$ changes $\pr_\mu\log\Omega$ at the origin by $2b_\mu$, and a gauge transformation brings the gauge field into the gauge $(x-x_0)^\mu a_\mu=0$ near $x_0$, multiplied by a cutoff of the class of flat sources, which does not change the jets at $x_0$. For such normalised data the metric at $x_0$ is $\delta_{\mu\nu}$, the Christoffel symbols vanish at $x_0$, and the symmetrised covariant derivatives of the Schouten tensor at $x_0$ are $-\pr_{\mu_1}\cdots\pr_{\mu_j}\pr_\nu\pr_\rho\log\Omega(x_0)$ plus polynomials in derivatives of lower order. The symmetrised Taylor coefficients of $a_\mu$ at $x_0$ are polynomials in the symmetrised covariant derivatives of $f_{\mu\nu}$, and the derivatives of the sources at $x_0$ are their symmetrised covariant derivatives plus such polynomials. These relations are triangular and can be inverted. Hence the value of $Y$ at $x_0$, which is an ${\rm O}(6)$-invariant polynomial in the derivatives at $x_0$, is an ${\rm O}(6)$-invariant polynomial $C$ in the symmetrised covariant derivatives of the Schouten tensor, of $f_{\mu\nu}$ and of the sources at $x_0$. On conformally flat backgrounds the Riemann tensor and the non-symmetrised covariant derivatives are expressed by these, since the Weyl and the Cotton tensors vanish, so that $C$ is a covariant gauge invariant expression. The expressions $Y$ and $C$ are both natural and gauge invariant and agree at every normalised point, and every point is normalised in this way, so that $Y=C$. The expression $C$ is built at $d=6$, and we read it with $d$ as a parameter, which changes it by terms of order $\vep$. The last statement follows from the uniqueness of $Y$, and in the same way the coefficients of $Y$ commute with the Lie algebra if those of $P$ do.
\end{proof}

\subsubsection{A covariant set of bare sources}
\label{gcp:sec:cov}

\begin{proposition}
\label{gcp:prop:cov}
Assume Hypothesis~\ref{hyp:finite} for the cubic theory, with the class of flat sources with $D_0=3$. Then there is a set of pure poles which satisfies it, in which all coefficients of the bare sources and of $\cX$ are covariant, and which is obtained from the given set by field shifts \eqref{gcf:eq:shift}, by regular local redefinitions $h\to h+r$ and by the average of $\cX$ over ${\rm O}(N)$. The vector field $\hB^I$ and the one-form $\hr_I$ are the same for both sets.
\end{proposition}

\begin{proof}
We order the coefficients of the bare sources by their level, the order in the couplings at which they enter, within a level by the order of their poles, beginning with the highest, and within these by the degree in $\sM$. Let the set $S$ be covariant below a given step, and let $\mathsf{R}$ be a constant element of ${\rm O}(N)$. The rotated set $\mathsf{R}S$ makes the functional finite, since the bare functional is invariant under constant rotations, and it has the same $L^I$, since $L^I$ is covariant. The two sets agree below the given step. The coefficients $N_I$, $Z_\sM$, $\fd_I$, $\fe_{IJ}$ and $\fk$ at this step are unique, by the steps of the proof of Proposition~\ref{prop:uniqueall} on the backgrounds $\Omega^2\delta_{\mu\nu}$, with the kernels of Lemma~\ref{gcu:lem:lowest} with $\kappa=4$ and $\kappa=6$, Lemma~\ref{gcu:lem:potential}, the transverse kernel and Lemma~\ref{gcp:lem:gram} for the pointwise independence of the monomials of dimension two. They therefore agree for $S$ and $\mathsf{R}S$ and are covariant. The same holds for $Z_h$. A difference $\tilde Z_h(g_1)h_1$ which passes the test with a source $h_2$ of disjoint support is $\mathsf{D}Y$ with a local $Y$ of dimension two which is linear in $h_1$, and such a $Y$ does not exist since $h$ has the dimension four. For the remaining coefficients $\eta$ of $h_0$ at this step we first show that they commute with the Lie algebra. We compare the functional at a gauge field $a_\mu$ with the functional at its gauge transform with the parameter $\theta\omega$, with the sources rotated by $e^{\theta\omega}$, to first order in $\theta$, as in the proof of Lemma~\ref{gcf:lem:gaugecov}. The bare functional is gauge invariant, $L^I$ is covariant and the lower steps are covariant, so that the term linear in a source $h_2$ of disjoint support is the term with $\theta\,[\omega,\eta]$ at the leading pole and the propagator in the background gauge field, where $\eta$ includes the monomials with $f_{\mu\nu}$. We show by induction in the degree in $a_\mu$ that the parts $[\omega,\eta]_j$ of degree $j$ vanish. If they vanish for the degrees smaller than $j$, the term of degree $j$ in $a_\mu$ is the second-order term with $\theta\,[\omega,\eta]_j$, and the moment with $H=1$ gives $\int\theta\,\Omega^4[\omega,\eta]_j=0$ for all $\theta$ of the class of flat sources. Since $\Omega^4[\omega,\eta]_j$ is invariant under the rotations of $x_\perp$, it vanishes at every point on the class of flat sources, and by Lemma~\ref{gcp:lem:gram} it vanishes as a formal expression. Hence the difference of $\mathsf{R}S$ and $S$ at this step vanishes for $\mathsf{R}$ in ${\rm SO}(N)$. For a reflection $\mathsf{P}$ the difference passes the test with $h_2$, since the term of $W$ linear in $h_2$ at this step is the term with this difference and the propagator in the background gauge field, and by the transfer after Lemma~\ref{gcp:lem:exact} and by Lemma~\ref{gcp:lem:natural} it is $\mathsf{D}Y_\mathsf{P}$ with a unique covariant gauge invariant expression $Y_\mathsf{P}$ of dimension two, whose coefficients commute with the Lie algebra. If the step has a positive degree in $\sM$, the same argument applies with the propagator in the background $\sM$, by Lemma~\ref{gcf:lem:contact} with the vertex of $\sM$. By the uniqueness $Y_{\mathsf{R}_1\mathsf{R}_2}=Y_{\mathsf{R}_1}+\mathsf{R}_1Y_{\mathsf{R}_2}$, so that $Y_{\mathsf{R}}=0$ for $\mathsf{R}\in{\rm SO}(N)$, $Y_{\mathsf{P}\mathsf{R}}=Y_\mathsf{P}$ and $\mathsf{R}Y_\mathsf{P}=Y_\mathsf{P}$. With $K=\tfrac12Y_\mathsf{P}$ we have $Y_\mathsf{R}=K-\mathsf{R}K$ for all $\mathsf{R}$, which is the Haar average over ${\rm O}(N)/{\rm SO}(N)$. After the field shift \eqref{gcf:eq:shift} with $\vep^{-\mathsf{r}}K$, whose coefficients commute with the Lie algebra, so that $\mathsf{D}K$ has the form of \eqref{gcu:eq:bare}, the difference of $\mathsf{R}S$ and $S$ at this step vanishes. The shift changes $\sM_0$ by $g_0\vep^{-\mathsf{r}}K$, which is of higher level, $h_0$ by terms of lower poles from $\xi_d-\xi_6$ and of higher degree, and $\cX$ by a local functional, and we remove regular terms of $h_0$ by a redefinition of $h$ as in section~\ref{gcf:sec:frame}. At each order in the couplings and in $\sM$ the procedure ends after finitely many steps. When all coefficients of the bare sources are covariant, $\mathsf{R}S$ and $S$ differ only in $\cX$, so that the integral of the difference of the two densities is finite, and the average of $\cX$ over ${\rm O}(N)$, taken coefficient by coefficient, is a set of real pure poles for which the functional is finite. None of these steps changes $L^I$ or $N_I$, and hence $\hB^I$ and $\hr_I$.
\end{proof}

We can take the set of the proposition in the frame of Proposition~\ref{gcf:prop:frame}. The frame shifts at orders zero and one are determined by the leading poles of $h_0$, since $\mathsf{D}$ is injective on local expressions of dimension two, and these leading poles are covariant for a covariant set, so that the frame shifts are covariant.

\subsubsection{Regular Weyl variation}
\label{gcp:sec:reg}

The classification which we use is the following. A covariant expression of dimension two in six dimensions which is linear in $\sigma$ and does not contain $h$ is a combination of $\sigma F\sM$, $\sigma F_ID^2g^I$, $\sigma F_{IJ}D_\mu g^ID^\mu g^J$, $\sigma FR$, $\pr_\mu\sigma\,F_ID^\mu g^I$ and $\nabla^2\sigma\,F$, with coefficients which are functions of the couplings. Indeed, a factor $\sigma$ without derivatives multiplies a scalar of dimension two, a factor $\pr_\mu\sigma$ a one-form of dimension one and $\nabla_\mu\nabla_\nu\sigma$ a tensor of dimension zero, which is a multiple of the metric, while $f_{\mu\nu}$ cannot occur since there is no antisymmetric tensor of dimension zero.

\begin{proposition}
\label{gcp:prop:reg}
Assume Hypothesis~\ref{hyp:finite} for the cubic theory, with the class of flat sources with $D_0=3$, and let the set of pure poles be covariant and in the frame of Proposition~\ref{gcf:prop:frame}. Then there is a covariant set in the same frame, obtained by covariant field shifts which vanish at $\sM=0$ at the levels zero and one and by regular local redefinitions of $h$, for which the Weyl variations of $h$ and of $\sM$ induced by the bare sources have regular coefficients.
\end{proposition}

\begin{proof}
By Lemma~\ref{lem:rhoxi} with the weight $k_I=\tfrac12$ we have $\hB^I=-\tfrac12\vep g^I+B^I$ and $\hr_I\hB^I=0$. The shifts and redefinitions below do not change $L^I$ and $N_I$, so that $\hB^I$ and $\hr_I$ are those of the given set, and their regularity is shown level by level in the induction below. The coefficient $\gamma_h$ of the homogeneous term of the variation of $h$ is regular: at its leading pole the term of $W$ linear in a source $h_1$ and in a source $h_2$ of disjoint support is the second-order term with $\sigma\gamma_h(g_1)h_1$, which by Lemma~\ref{gcp:lem:exact} would be $\mathsf{D}Y$ with $Y$ linear in $h_1$ of dimension two, which does not exist. The same argument applies at each level of the induction which follows. We proceed by induction in the level $n$, and at each level we treat first the variation of $\sM$ and then the variation of $h$. Let the variations be regular below the level $n$. Then the steps of Lemma~\ref{lem:regvar} for the gauge field and the couplings, with the kernels with $\kappa=4$ and $\kappa=6$, show that the terms of level $n$ of $\hr_I$ and $\hB^I$ are regular, since the terms of level $n$ of the variations of $h$ and $\sM$ multiply terms of second order with two sources of different kinds, which vanish. We use the identity $\mathbb{W}_\sigma W=\cA[\sigma]$ on the backgrounds $\Omega^2\delta_{\mu\nu}$ with $\sigma$ of the class of flat sources, which holds by Proposition~\ref{gs:prop:msanomaly}. In its terms of order $n$ in the couplings and linear in a source of disjoint support, all terms are finite except those in which a pole of a variation at the level $n$ multiplies the term with that source and the propagators in the background gauge field and in the background $\sM$, since the other terms are products of regular coefficients with finite derivatives of $W$, and the anomaly has no such term.

At the level $n$ the variation of $\sM$ is regular by the step of Lemma~\ref{lem:regvar} for the source of dimension two, with a source $\sM_2$, the kernel with $\kappa=4$ and Lemma~\ref{gcp:lem:gram} for the pointwise independence. A pole of the variation of $h$ contributes to this term only through a cubic vertex, by Lemma~\ref{gcf:lem:contact}, that is with the level $n-1$, where it vanishes. Let now $\vep^{-\mathsf{r}}\Pi[\sigma]$ be the leading pole of the inhomogeneous part of the variation of $h$ at the level $n$. By the test with a source $h_2$, the transfer after Lemma~\ref{gcp:lem:exact} and Lemma~\ref{gcp:lem:natural}, with $\sigma$ as an input, $\Pi[\sigma]=\mathsf{D}Y[\sigma]$ at vanishing $\sM$, with a unique covariant gauge invariant expression $Y[\sigma]$ of dimension two which is linear in $\sigma$ and has covariant coefficients. With $\sM$ present the same argument applies degree by degree in $\sM$, with the propagator in the background $\sM$ and Lemma~\ref{gcf:lem:contact} with the vertex of $\sM$, so that $\Pi[\sigma]=(\mathsf{D}+\sM)Y[\sigma]$ with $Y$ at most linear in $\sM$.

We now use that the variations commute, by Lemma~\ref{gcu:lem:commute}. At the order $\vep^{-\mathsf{r}}$ and at the level $n$ the commutator on $h$ is $(\mathsf{D}+\sM)[(\delta_{\sigma'}+2\sigma')Y[\sigma]-(\delta_\sigma+2\sigma)Y[\sigma']]$, by the conformal covariance $\delta_\sigma(\mathsf{D}f)=-4\sigma\mathsf{D}f+\mathsf{D}(2\sigma f)$ at $d=6$. Here $\delta_\sigma$ is the part of the Weyl variation of order $\vep^0$ which does not raise the level. It rescales the metric and $\sM$ and does not change the couplings and their derivatives, since $\hB^I+\tfrac12\vep g^I$ has no term of first order, $\hr_I$, $\gamma_\sM$ and the inhomogeneous terms of the variation of $\sM$ raise the level, and the variations of lower level have regular coefficients, so that they enter the order $\vep^{-\mathsf{r}}$ only through $\Pi$. Since $\mathsf{D}+\sM$ is injective on local expressions of dimension two, by a filtration in the degrees in $\sM$ and in $a_\mu$, the bracket vanishes. By the classification above, with $\delta_\sigma\sM=-2\sigma\sM$, $\delta_\sigma D^2g^I=-2\sigma D^2g^I+4\pr_\mu\sigma D^\mu g^I$, $\delta_\sigma(D_\mu g^ID^\mu g^J)=-2\sigma D_\mu g^ID^\mu g^J$ and $\delta_\sigma R=-2\sigma R-10\nabla^2\sigma$, the vanishing of the bracket excludes the terms $\sigma F_ID^2g^I$ and $\sigma FR$. The terms with $\pr_\mu\sigma$ and $\nabla^2\sigma$ are $(\delta_\sigma+2\sigma)K_1$ with $K_1=\tfrac14F_ID^2g^I-\tfrac1{10}FR$. For coefficients $F$ and $F_{IJ}$ which are homogeneous of degree $m$ in the couplings the full variation gives $(\mathbb{W}_\sigma+2\sigma-\tfrac12\vep\sigma)(F\sM)=-\tfrac12\vep(m+1)\,\sigma F\sM$ and $(\mathbb{W}_\sigma+2\sigma-\tfrac12\vep\sigma)(F_{IJ}D_\mu g^ID^\mu g^J)=-\tfrac12\vep(m+3)\,\sigma F_{IJ}D_\mu g^ID^\mu g^J$ plus terms with $\pr_\mu\sigma$ of the form of the coboundary above, up to terms of higher level, since $\tfrac12(d-2)=2-\tfrac12\vep$ and $\hB^I=-\tfrac12\vep g^I+B^I$. By Lemma~\ref{gcf:lem:shift} the new bare sources after a shift with $K$ are those of \eqref{gcf:eq:shift}, and since the bare sources only rescale, the induced variation of $h$ changes by $-(\mathsf{D}+\sM)Y_K$ and the variation of $\sM$ by $-g_0Y_K$, with $Y_K=(\mathbb{W}_\sigma+\tfrac12(d-2)\sigma)K$, up to terms of higher level. Here we use that the map from the renormalised to the bare sources is the identity at level zero, so that its Jacobian, which the shift changes, differs from the identity by terms of positive level. We shift with $K=\vep^{-\mathsf{r}}K_1+\vep^{-\mathsf{r}-1}K_2$, where $K_2=-\tfrac2{m+1}F\sM-\tfrac2{m+3}F_{IJ}D_\mu g^ID^\mu g^J$ for each degree $m$, and we absorb the terms with $\pr_\mu\sigma$ which $K_2$ produces at the order $\vep^{-\mathsf{r}}$ in $K_1$. Then $Y_K=Y$ at the order $\vep^{-\mathsf{r}}$, and since $(\delta_\sigma+2\sigma)K_2=0$ the shift produces no term of order $\vep^{-\mathsf{r}-1}$ in the variation at the level $n$. The leading pole at the level $n$ is removed. The remaining changes are of higher level or of lower poles, among them the terms with $\xi_d-\xi_6$, and the terms of order $\vep^{-\mathsf{r}-1}$ in $h_0$, $\sM_0$ and $\cX$ are pure poles, which the hypothesis allows. The regular terms of $h_0$ which the shifts produce are removed by redefinitions of $h$. The shifts are covariant and gauge invariant, since $Y$ is, and $\mathsf{D}K$ has the form of \eqref{gcu:eq:bare}. At the levels zero and one the inhomogeneous part of the variation of $h$ at vanishing $\sM$ comes from the terms of $h_0$ at vanishing relevant sources of order zero and one, which vanish in the frame, so that there $F_I$ and $F$ in $K_1$ vanish and the shifts vanish at $\sM=0$, which preserves the frame. The shifts with $K$ proportional to $\sM$ change $Z_\sM$ at positive order in the couplings. This completes the induction.
\end{proof}

\subsubsection{The properties}
\label{gcp:sec:props}

The general proof of sections~\ref{gs:sec:identities} to~\ref{gs:sec:consequences} uses the functional on the unit sphere and the anomaly on the unit sphere and on its Weyl rescalings such as $(1+2t\sigma)\mathring\gamma_{\mu\nu}$, which are conformally flat, and the operators $\mathbb{W}_\sigma$ preserve the conformally flat backgrounds. We say that Properties~\ref{gs:P:weyl} and~\ref{gs:P:anomaly} hold on conformally flat backgrounds if they hold with all functionals restricted to these backgrounds and with the anomaly written with the Schouten tensor in place of the Riemann tensor. In this form they give all statements of the general proof.

\begin{proposition}
\label{gcp:prop:props}
Assume Hypothesis~\ref{hyp:finite} for the cubic theory, on the backgrounds $\Omega^2\delta_{\mu\nu}$, for the class of flat sources with $D_0=3$. Then there is a set of pure poles of the hypothesis with the same $\hB^I$ and $\hr_I$, obtained from the given one by field shifts, regular local redefinitions of the source of $\phi_i$, the average of $\cX$ over ${\rm O}(N)$ and the removal of the part of $\cX$ which contains neither sources nor couplings. For this set the renormalised functional on the unit sphere has Properties~\ref{gs:P:functional} to~\ref{gs:P:free}, with Property~\ref{gs:P:functional} in instance~(a), Properties~\ref{gs:P:weyl} and~\ref{gs:P:anomaly} on conformally flat backgrounds, the weights $k_I=\tfrac12$ and the relevant sources $h$ and $\sM$.
\end{proposition}

\begin{proof}
We take the covariant set of Proposition~\ref{gcp:prop:cov} in the frame of Proposition~\ref{gcf:prop:frame}, and then the set of Proposition~\ref{gcp:prop:reg}. From its density $\cX$ we remove the part which contains neither sources nor couplings, as described after Lemma~\ref{gcu:lem:euler}. For Property~\ref{gs:P:weyl}, $\hB^I=-\tfrac12\vep g^I+B^I$ and $\hr_I$ are regular with $\hr_I\hB^I=0$, the variations of $h$ and $\sM$ are regular and covariant, $\gamma_h$ and $\gamma_\sM$ vanish at $g=0$ since $Z_h=Z_\sM=1$ there by Lemma~\ref{gcu:lem:lowest}, the terms of the variations have the dimensions of the sources, and the variations commute by Lemma~\ref{gcu:lem:commute}. For Property~\ref{gs:P:anomaly}, the anomaly $\cA[\sigma]=\mathbb{W}_\sigma\int\sqrt\gamma\,\cX$ is local, gauge invariant, homogeneous of Weyl weight $-\vep$ and consistent by Proposition~\ref{gs:prop:msanomaly}. It is finite on the backgrounds $\Omega^2\delta_{\mu\nu}$ for $\sigma$ of the class of flat sources, since $W$ is finite along the families of sources and of $\log\Omega$ which $\mathbb{W}_\sigma$ generates, and $\mathbb{W}_\sigma$ has regular coefficients. We write it as $\int\sqrt\gamma\,\sigma\,\mathcal{E}$ by integrations by parts, which hold with $d$ as a parameter. Let $V$ be the space of covariant gauge invariant scalars of dimension six, written with the Schouten tensor, modulo those which vanish identically on conformally flat backgrounds, and let $\mathcal{E}=\sum_kc_km_k$ with a basis $m_k$ of $V$ and Laurent series $c_k$, where the dependence of the $m_k$ on $d$ is included in the $c_k$. Since $\sigma$ is arbitrary and the integrand is invariant under the rotations of $x_\perp$, the coefficient of the highest pole of $\mathcal{E}$ at $d=6$ vanishes at every point on the class of flat sources with $D_0=3$. Multiplied by $\Omega^6$ it is a formal expression with at most six derivatives and one-form factors together, which vanishes by Lemma~\ref{gcp:lem:gram}, so that its image in $V$ vanishes and its coefficients vanish. By descending induction in the order of the poles the $c_k$ are regular. For Property~\ref{gs:P:functional} we use the localisation of appendix~\ref{app:finite}. Let all sources vanish near a point of the equator. Then the bare sources vanish there by Proposition~\ref{gcu:prop:first} and since the shifts vanish at $g=0$ where $\sM=0$. The density $\cX$ vanishes there, since its part which contains neither sources nor couplings has been removed, see the text after Lemma~\ref{gcu:lem:euler}, the variations of the sources vanish there since their inhomogeneous terms at vanishing sources vanish at $g=0$ in the frame and $\fk$ and $\fd_I$ vanish at $g=0$, and the integrand of the anomaly vanishes there. Relation \eqref{eq:Wsplit} holds with regular coefficients, where the gauge parameter $\xi$ is regular by the choice of the bare sources after Lemma~\ref{lem:frame}, and the steps on the polarisation and the partition of unity apply, since the boundary of the ball of the parallel coordinates is a sphere of positive dimension. The gauge invariance follows from the covariance of the coefficients and Proposition~\ref{prop:gauge}, and the statements at a constant relevant source from the proof of Lemma~\ref{gr4:lem:const}. Property~\ref{gs:P:ward} is Proposition~\ref{gs:prop:msward}, and Property~\ref{gs:P:sphere} follows from appendix~\ref{app:rotations}, for the class of sources of section~\ref{sec:sphere} with $D_0=3$, which satisfies the assumptions of appendix~\ref{gs:sec:class}, see appendix~\ref{app:rules}. Property~\ref{gs:P:free} and the last statement of Property~\ref{gs:P:weyl} follow from Proposition~\ref{gcu:prop:first} as in section~\ref{gcf:sec:frame}, since the shifts of Proposition~\ref{gcp:prop:reg} preserve the frame.
\end{proof}

The steps of the derivation for multiscalar $\phi^4$ theory and their form for the cubic theory are listed in the column of the cubic theory of table~\ref{gs:tab:steps}.

We use the hypothesis on the backgrounds $\Omega^2\delta_{\mu\nu}$. For the other theories of this paper the curvature counterterms follow from the finiteness in flat space. For the cubic theory this would require the conditions of solvability for the curvature counterterms of dimension six, among them a condition of consistency of the kind of \eqref{eq:curlgeneral}, and Hypothesis~\ref{hyp:finite} for the cubic theory includes the curvature counterterms in its place. The coefficients of the counterterms and of the monomials of $h_0$ with the Weyl tensor do not enter the functional on these backgrounds, and Properties~\ref{gs:P:weyl} and~\ref{gs:P:anomaly} are stated on conformally flat backgrounds.

\section{Free gauge fields on the sphere}
\label{gs:app:freegauge}
\label{ggs:sec:spectral}

In this appendix we construct the bare functional of free gauge fields with a local coupling on $S^d$ for non-integer $d$, and we derive Properties~\ref{gs:P:functional} to~\ref{gs:P:free} for it in instance~(b) from Hypothesis~\ref{hyp:finite} for the free gauge field, which is stated on the sphere and its conformal rescalings. We do not use the flat chart. On $S^d$ in integer dimension the functional is minus one half of the logarithm of a determinant on the exact two-forms, and the exact two-forms decompose into levels, which are irreducible representations of ${\rm O}(d+1)$. We write a trace over one level as an integral of the form \eqref{eq:gram} of a polynomial, which is defined for every $d$, and we continue the sum over the levels in $d$ from a half plane in which it converges. We find that the constant part of the coupling then gives the term $n_G\log g$ for every $d$, and the conformal Killing vectors leave the bare functional invariant to all orders.

\subsection{Levels and traces over one level}
\label{ggs:sec:levels}

We consider one abelian gauge field with the action $\tfrac12\int d\mu_d\,e^{-2\tau}\,|F|^2$, where $|F|^2=\tfrac12F_{\mu\nu}F^{\mu\nu}$ and $\tau=\log g$. For $n_G$ gauge fields with a common coupling, such as the free gauge fields of a simple factor, the functional is $n_G$ times the functional of one field. On two-forms we use the inner product $\vev{f,h}=\tfrac12\int d\mu_d\,f_{\mu\nu}h^{\mu\nu}$.

Let first $d\geq3$ be an integer. The exact two-forms are the field strengths $F=dA$ of the coexact one-forms $A$, since $S^d$ has no harmonic one-forms and no harmonic two-forms. The coexact one-forms of degree $L\geq1$ form the irreducible representation of ${\rm O}(d+1)$ with the Young diagram $(L,1)$, on which the Hodge Laplacian has the eigenvalue $\lambda_L=(L+1)(L+d-2)$, and their dimension is
\begin{equation}
D_L = \frac{L\,(L+d-1)\,(2L+d-1)\,\Gamma(L+d-2)}{\Gamma(d-1)\,\Gamma(L+2)} \, .
\label{ggs:eq:DL}
\end{equation}
We denote by $E_L$ the exact two-forms $dA$ of these one-forms, by $P_L$ the orthogonal projector onto $E_L$ and by $P=\sum_LP_L$ the projector onto the exact two-forms. The coexact two-forms of the representation $(L,1,1)$ have the eigenvalue $\mu_L=(L+2)(L+d-3)$ of the Hodge Laplacian. We denote by $M_u$ the multiplication of two-forms by a function $u$.

We write two-forms on $S^d$ as antisymmetric tensors $f_{AB}(X)$ of the embedding space with $X^Af_{AB}=0$. A kernel of an operator on two-forms is then a bitensor $K_{AB,CD}(X,Y)$.

\begin{lemma}
\label{ggs:lem:zonal}
For each $L\geq1$ there is a bitensor $P_L(X,Y)$ which is a polynomial in $X$, $Y$ and $\delta_{AB}$, written with these invariant tensors only, whose coefficients are $V^{-1}$ times rational functions of $d$, such that for every integer $d\geq d_L$ it is the kernel of the projector $P_L$. Its trace at coincident points is ${\rm tr}\,P_L(X,X)=D_L/V$. Here, in accordance with $\vev{f,h}$, a kernel acts as $(Kh)_{AB}(X)=\tfrac12\int d\mu_d(Y)\,K_{AB,CD}(X,Y)h_{CD}(Y)$ and ${\rm tr}\,K(X,X)=\tfrac12K_{AB,AB}(X,X)$, and $d_L$ is an integer beyond which the structures used in the proof are linearly independent and the rational solution has no pole.
\end{lemma}

\begin{proof}
For integer $d$ the kernel of $P_L$ is invariant under ${\rm O}(d+1)$ acting on both points, and its entries are polynomials of bounded degree in $X$ and $Y$, since $E_L$ consists of restrictions of polynomial forms. By the first fundamental theorem of invariant theory for ${\rm O}(d+1)$ such a bitensor is a combination of a finite set of structures, products of $\delta_{AB}$, of components of $X$ and $Y$ and of powers of $X\!\cdot\!Y$, which for $d+1$ larger than the degree does not depend on $d$. The conditions which characterise the kernel of $P_L$ among these combinations are linear with coefficients which are polynomials in $d$: tangentiality at both points, closure in each argument, the eigenvalue equation of the Hodge Laplacian with the eigenvalue of $E_L$, which is not taken on the exact two-forms of another level. By Schur's lemma their solutions are the multiples of the kernel of $P_L$, since $E_L$ is irreducible. The solution of a linear system whose coefficients are polynomials in $d$ and whose space of solutions is one-dimensional for all large integers $d$ has a basis vector whose components are rational functions of $d$. The normalisation $\int d\mu_d\,{\rm tr}\,P_L(X,X)=D_L$ fixes the multiple. By the invariance ${\rm tr}\,P_L(X,X)$ is a function of $X\!\cdot\!X=1$ only, hence the constant $D_L/V$.
\end{proof}

For non-integer $d$ we take the bitensor of Lemma~\ref{ggs:lem:zonal} as the kernel $P_L$. For sources $u_1,\ldots,u_n$ which are polynomials in the parallel coordinates $X_\parallel$, a subclass of the class of sources of section~\ref{sec:sphere}, and for levels $L_1,\ldots,L_n$, we define
\begin{equation}
\begin{split}
&{\rm tr}\big(P_{L_1}M_{u_1}P_{L_2}M_{u_2}\cdots P_{L_n}M_{u_n}\big) \\
&\qquad = \int\prod_{i=1}^nd\mu_d(X_i)\;u_i(X_i)\;{\rm tr}\big[P_{L_1}(X_1,X_2)\,P_{L_2}(X_2,X_3)\cdots P_{L_n}(X_n,X_1)\big]
\end{split}
\label{ggs:eq:fixedlevel}
\end{equation}
with the integral \eqref{eq:gram}. The contractions of the tangential tensors give a polynomial in the scalar products $X_i\!\cdot\!X_j$, in the traces $\delta_{AA}=d+1$ and in the parallel coordinates, so that the integrand is a polynomial in the $X_{i\parallel}$, the $\rho_i$ and the $u_{ij}$ of \eqref{eq:chordal}. In the same way we define traces which contain the Hodge Laplacian, the operator $d\delta$ or the Lie derivative along a conformal Killing vector of \eqref{eq:ckv} between two projectors. These operators act on the kernels as differential operators whose coefficients are polynomials in the embedding coordinates and in $d$. We call these quantities traces over fixed levels.

\begin{lemma}[Transfer]
\label{ggs:lem:transfer}
The integral \eqref{eq:gram} of a polynomial in the embedding coordinates of $n$ points is $V^n$ times a rational function of $d$, so that a trace over fixed levels, whose kernels carry the factor $V^{-1}$ each, is a rational function of $d$. An identity between traces over fixed levels which holds for all large integers $d$ holds for all $d$.
\end{lemma}

\begin{proof}
For a polynomial in the embedding coordinates the integral \eqref{eq:gram} is a finite combination of moments, $\int d^kX_\parallel\,\rho^{m-2}X_\parallel^\alpha\rho^{2a}$ at each point and $\int_{U>0}\prod du_{ij}\,(\det U)^{(m-n-1)/2}\,u^\beta$ for the matrix $U$, where $m=d-D_0$. For $m\leq n-1$ the second integral diverges at $\det U=0$, and we use its continuation in $m$, as in \eqref{eq:gram}. Only even powers of $\rho_i$ occur: $\rho_i$ enters through $\rho_i\rho_ju_{ij}$, and the change $u_{ij}\to s_is_ju_{ij}$ with signs $s_i$ preserves $\det U$ and the domain, so that terms of odd total degree in the $u_{ij}$ at one point integrate to zero. The moments of the parallel coordinates are Beta functions whose arguments differ from those of $\int d^kX_\parallel\rho^{m-2}$ by integers. The moments of $U$ are, up to the normalisation $\prod_l\vol_{m-l}$, the moments of the scalar products of $n$ independent unit vectors in $m$ dimensions. Writing unit vectors as normalised Gaussian vectors, these are ratios of Gaussian moments of polynomials in the scalar products, which are polynomials in $m$ by Wick's theorem, and of moments of the norms, which are ratios of Gamma functions whose arguments differ by integers. Hence the ratio to $V^n=\vol_{d+1}^{\,n}$ is rational in $d$. For integer $m\geq n$ the integral is the integral over $n$ copies of $S^d$. An identity between rational functions which holds at infinitely many points holds identically.
\end{proof}

By the lemma every identity of finite-dimensional linear algebra and of integration by parts on $S^d$ among traces over fixed levels holds for all $d$. Examples are $P_LP_{L'}=\delta_{LL'}P_L$ inside traces, the cyclic symmetry of a trace over fixed levels, and the identities of the next subsections, which we derive in integer dimension. Multiplication by a polynomial $u$ of degree $q$ maps $E_L$ into the representations $(L',1)$ and $(L',1,1)$ with $|L'-L|\leq q$, since a harmonic of degree $j$ changes the first row of a Young diagram by at most $j$. Hence $P_{L'}M_uP_L=0$ for $|L'-L|>q$, and for non-integer $d$ the traces over fixed levels which contain $P_{L'}M_uP_L$ with $|L'-L|>q$ vanish, by Lemma~\ref{ggs:lem:transfer}. We call this the band property. The Lie derivative along a conformal Killing vector of \eqref{eq:ckv} shifts the level by at most one, and we count it as a factor of degree one.

\subsection{Continued sums over levels}
\label{ggs:sec:sums}

Let $\cH$ be the set of functions $f(L,d)$ of an integer $L\geq L_0$ and of $d$ which are finite sums of terms
\begin{equation}
c(d)\,p(L,d)\prod_{i=1}^r\frac{\Gamma(L+a_i(d))}{\Gamma(L+b_i(d))} \, ,
\label{ggs:eq:H}
\end{equation}
where $c$ is meromorphic, $p$ is a polynomial in $L$ with coefficients polynomial in $d$, the $a_i$ and $b_i$ are affine functions of $d$, and the exponent $\theta(d)=\deg_Lp+\sum_i(a_i-b_i)$ has a positive coefficient of $d$.

\begin{lemma}[Continuation]
\label{ggs:lem:cont}
For $f\in\cH$ the sum $\sum_{L\geq L_0}f(L,d)$ converges absolutely and locally uniformly in a half plane ${\rm Re}\,d<d_1$, away from the set $Z$ of the poles of the coefficients and of the points where $L+a_i(d)$ is a non-positive integer for some $L\geq L_0$, and it extends to a meromorphic function of $d$, which we denote by $\Sigma_{L\geq L_0}f(L)$. Its poles are contained in $Z$ and in the set of points where $\theta(d)-j=-1$ for an integer $j\geq0$. It is linear in $f$, and $\Sigma_{L\geq L_0}f(L+1)=\Sigma_{L\geq L_0}f(L)-f(L_0)$.
\end{lemma}

\begin{proof}
By Stirling's formula, $\Gamma(L+a)/\Gamma(L+b)=L^{a-b}\sum_{j<J}g_j(a,b)L^{-j}+O(L^{{\rm Re}(a-b)-J})$ uniformly for $a$ and $b$ in compact sets, with polynomials $g_j$. Hence $f(L,d)=\sum_{j<J}e_j(d)L^{\theta(d)-j}+R_J(L,d)$ with meromorphic $e_j$ and $|R_J|\leq C\,L^{{\rm Re}\,\theta-J}$, for each term of \eqref{ggs:eq:H}. The sum converges absolutely for ${\rm Re}\,\theta<-1$. There it equals $\sum_{j<J}e_j(d)\,\zeta(j-\theta(d),L_0)+\sum_{L\geq L_0}R_J(L,d)$, with the Hurwitz zeta function. The first term is meromorphic, with poles where $j-\theta=1$, and the second is holomorphic where ${\rm Re}\,\theta<J-1$. Since $J$ is arbitrary the sum extends to a meromorphic function of $d$. Linearity and the shift relation hold in the half plane of convergence and therefore everywhere.
\end{proof}

For the free gauge field the sums over levels have the form of the following lemma. For polynomial sources $u_1,\ldots,u_n$ we set
\begin{equation}
t_n(L;u_1,\ldots,u_n) = \sum_{L_2,\ldots,L_n\geq1}{\rm tr}\big(P_LM_{u_1}P_{L_2}M_{u_2}\cdots P_{L_n}M_{u_n}\big) \, ,
\label{ggs:eq:tn}
\end{equation}
where the sum is finite by the band property.

\begin{lemma}[Structure]
\label{ggs:lem:S}
(i) $t_1(L;u)=(D_L/V)\int d\mu_d\,u$. (ii) For a harmonic $\sigma=b\cdot X$ of degree one and a polynomial source $u$,
\begin{equation}
t_2(L;\sigma,u) = \frac{D_L}{V}\left(1-\frac{d-2}{(L+2)(L+d-3)}\right)\int d\mu_d\;\sigma\,u \, .
\label{ggs:eq:t2}
\end{equation}
(iii) For all $n$, all polynomial sources and all fixed shifts $k_2,\ldots,k_n$, also with Lie derivatives along conformal Killing vectors inserted between projectors, the trace over fixed levels ${\rm tr}(P_LM_{u_1}P_{L+k_2}M_{u_2}\cdots P_{L+k_n}M_{u_n})$ is $D_L\,R(L,d)$ with one rational function $R$ of $L$ and $d$ whose denominator, as a polynomial in $L$, is a product of linear factors, for $L$ larger than the sum $Q$ of the degrees, where a Lie derivative counts with degree one. In particular these traces and $t_n(L;\ldots)$ belong to $\cH$.
\end{lemma}

\begin{proof}
Part (i) follows from Lemma~\ref{ggs:lem:zonal}. For (ii) we work at integer $d$ and transfer by Lemma~\ref{ggs:lem:transfer}. The map $u\mapsto t_2(L;\sigma,u)$ is $\int d\mu_d\,u\,h$ with a function $h$ which is linear in $b$ and invariant under the simultaneous rotation of $b$ and of the point, hence a multiple of $\sigma$. It remains to compute $t_2(L;\sigma,\sigma)={\rm tr}(P_LM_\sigma PM_\sigma P_L)={\rm tr}(P_LM_{\sigma^2}P_L)-\sum_m\|(1-P)\sigma f_m\|^2$ for an orthonormal basis $f_m$ of $E_L$. The first term is $(D_L/V)\int\sigma^2$ by (i). The two-form $(1-P)\sigma f_m$ is coexact and $d[(1-P)\sigma f_m]=d\sigma\wedge f_m$, since $f_m$ is closed. Among the coexact two-forms only the representation $(L,1,1)$ occurs in $M_\sigma E_L$, since multiplication by $\sigma$ adds or removes one box. Hence $\|(1-P)\sigma f_m\|^2=\|d\sigma\wedge f_m\|^2/\mu_L$. At each point $\sum_mf_{m\,\mu\nu}f_{m\,\rho\sigma}=c\,(\gamma_{\mu\rho}\gamma_{\nu\sigma}-\gamma_{\mu\sigma}\gamma_{\nu\rho})$ with $c=2D_L/(Vd(d-1))$, by the invariance and by (i). With $|\alpha\wedge f|^2=|\alpha|^2|f|^2-|\iota_\alpha f|^2$ for a one-form $\alpha$ we obtain $\sum_m|d\sigma\wedge f_m|^2=(D_L/V)\tfrac{d-2}{d}|d\sigma|^2$, and $\int|d\sigma|^2=d\int\sigma^2$ by \eqref{gs:eq:parts}. This gives \eqref{ggs:eq:t2}.

For (iii) we work at integer $d\geq d_0$ and fixed $L>Q$, where $d_0$ depends on $n$ and on the degrees of the sources but not on $L$. We replace the projectors one at a time, starting from the right. In the trace the projector $P_{L'}$ acts on a space which is a sum of the representations $(L'',1)$ and $(L'',1,1)$ with $|L''-L|\leq Q$. There $P_{L'}=\lambda_{L'}^{-1}\,d\delta\prod_{L''}(\Delta-\lambda_{L''})/(\lambda_{L'}-\lambda_{L''})$, with the product over the exact levels $L''\neq L'$ in this range, since $d\delta$ vanishes on the coexact two-forms and equals the Hodge Laplacian $\Delta$ on the exact ones. The differences $\lambda_{L'}-\lambda_{L''}=(L'-L'')(L'+L''+d-1)$ are products of factors linear in $L$. Hence the trace is ${\rm tr}(P_LB_L)$, where $B_L$ is a differential operator whose coefficients are polynomials in the sources and their derivatives with coefficients rational in $L$ and $d$, with denominators which are products of factors linear in $L$. Its trace is $\sum_m\vev{f_m,B_Lf_m}$, an integral of the sources and their derivatives contracted with the tensors $\sum_m f_m\otimes\nabla^rf_m$ at one point. We show that for $f\in E_L$ and a complete contraction $\mathcal{C}$ of $\nabla^{r_1}f\otimes\nabla^{r_2}f$ with the metric, $\int d\mu_d\,\mathcal{C}=q_{\mathcal C}(\lambda_L,d)\,\|f\|^2$ with a polynomial $q_{\mathcal C}$ which does not depend on $L$. We use induction on $r=r_1+r_2$, which is even. For $r=0$ the contraction is $\pm f^{ab}f_{ab}$ or zero, and the statement holds with a constant $q_{\mathcal C}$. Integrations by parts give $\int\mathcal{C}=(-1)^{r_1}\int d\mu_d\,f^{ab}(\mathcal{D}f)_{ab}$, where $(\mathcal{D}f)_{ab}$ is a contraction of $\nabla_{c_1}\cdots\nabla_{c_r}f_{ef}$ in which the indices $a,b,c_1,\ldots,c_r,e,f$ are paired by the metric, except $a$ and $b$. The commutator of two covariant derivatives acts through the curvature $R_{abcd}=\gamma_{ac}\gamma_{bd}-\gamma_{ad}\gamma_{bc}$ of the unit sphere and gives contractions of lower order with coefficients which are polynomials in $d$, which are covered by the induction. Up to such terms we may therefore order the derivatives at will. If two indices $c_i$ are paired, we move the two derivatives next to $f$ and use $\nabla^c\nabla_cf=-(\lambda_L-2(d-2))f$, since the Hodge Laplacian on two-forms of the unit sphere is $-\nabla^c\nabla_c+2(d-2)$ and equals $\lambda_L$ on $E_L$. This lowers the order by two. Otherwise each $c_i$ is paired with one of $a,b,e,f$. A pairing of $a$ with $b$ or of $e$ with $f$ gives zero by antisymmetry, and two derivatives paired with $e$ and $f$ give $\nabla^e\nabla^ff_{ef}$, which is a commutator and hence of lower order. The remaining case is $r=2$ with one derivative paired with an index of $f$ and one free, which gives $\pm\int f^{ab}\nabla_a(\delta f)_b=\pm\tfrac12\int f^{ab}(d\delta f)_{ab}=\pm\lambda_L\|f\|^2$, since $d\delta f=\lambda_Lf$ on exact two-forms. This completes the induction. At a point the tensor $\sum_mf_m\otimes\nabla^rf_m$ is invariant under the rotations which fix the point, hence a combination of products of the metric. Its coefficients are fixed by its complete contractions through a linear system whose matrix is the Gram matrix of these products, a matrix of polynomials in $d$ which is invertible for large $d$, so that they are rational functions of $d$ times complete contractions. Each complete contraction is constant on the sphere and equals $V^{-1}\sum_m\int\mathcal{C}(f_m,f_m)=(D_L/V)\,q_{\mathcal C}(\lambda_L,d)$. Thus the trace is $D_L\,R(L,d)$, with $R$ obtained from the Lagrange denominators, the coefficients of $B_L$, the polynomials $q_{\mathcal C}$ evaluated at $\lambda_L$ and the rational coefficients of the invariant structures, by formulas which do not depend on $d$. Hence $R$ is one rational function of $L$ and $d$ whose denominator in $L$ is a product of linear factors. A partial fraction decomposition in $L$ puts it into the form \eqref{ggs:eq:H}, since $1/(L+a)=\Gamma(L+a)/\Gamma(L+a+1)$. The same holds with a Lie derivative inserted, which is a differential operator of first order with polynomial coefficients. At fixed $L$ both sides of the identity between the trace and $D_LR(L,d)$ are rational in $d$ and agree at all integers $d\geq\max(d_0,d_L)$, so that they agree for all $d$ by Lemma~\ref{ggs:lem:transfer}, with the same $R$ for every $L$.
\end{proof}

Parts (i) and (ii) give $T_1$ and $T_2(\sigma,\cdot)$ below, and with them the one-point function, the eigenvalue of Proposition~\ref{ggs:prop:rot} and the sum rule. Part (iii) is used for the existence of all other cycles, for Lemma~\ref{ggs:lem:rules} and for Propositions~\ref{ggs:prop:shift} and~\ref{ggs:prop:ward} beyond first order.

\subsection{Definition of the bare functional}
\label{ggs:sec:def}

For polynomial sources we define the cycles
\begin{equation}
T_n(u_1,\ldots,u_n) = \Sigma_{L\geq1}\,t_n(L;u_1,\ldots,u_n) \, ,
\label{ggs:eq:Tn}
\end{equation}
which exist by Lemmas~\ref{ggs:lem:cont} and~\ref{ggs:lem:S}, where the finitely many levels $L$ below the range of part (iii) of Lemma~\ref{ggs:lem:S} are added as a finite sum. In integer dimension the sum over levels does not converge, and $T_n$ is formally the trace of $PM_{u_1}P\cdots PM_{u_n}P$. We set
\begin{equation}
N = \Sigma_{L\geq1}D_L = T_1(1) = 1 \, ,
\label{ggs:eq:N}
\end{equation}
where the value follows from the sum $\sum_{L\geq L_0}\Gamma(L+a)/\Gamma(L+b)=\Gamma(L_0+a)/((b-a-1)\Gamma(L_0+b-1))$, which telescopes and holds for ${\rm Re}(b-a)>1$, after writing $L(L+d-1)(2L+d-1)$ as a combination of rising factorials $(L+d-2)_j$. In the same way the dimensions of the scalar harmonics sum to zero and those of the exact one-forms to $-1$. The continuation in $d$ of the sums over levels is the regularisation of this construction. In integer dimension other regularisations of the sum over modes give other values, for instance the Riemann zeta function of the level $L$ at $d=4$ gives $-\tfrac{11}{30}$ for $\sum_LD_L$, and the continuation in $d$ is a regularisation under which the bare functional has the covariance of Proposition~\ref{ggs:prop:ward}.

\begin{lemma}[Rules for the cycles]
\label{ggs:lem:rules}
For polynomial sources, (i) $T_n$ is multilinear and invariant under cyclic permutations of its arguments; (ii) $T_n(1,u_2,\ldots,u_n)=T_{n-1}(u_2,\ldots,u_n)$ for $n\geq2$ and $T_1(1)=N$; (iii) $T_1(u)=(N/V)\int d\mu_d\,u$. Products of operators $PM_uP$ and of the Lie derivative $\cL_v$ restricted to the exact two-forms have traces defined in the same way, and they are cyclic.
\end{lemma}

\begin{proof}
At a fixed $d$ in the half plane of absolute convergence of all sums involved, which exists for any finite set of cycles, the cycles are absolutely convergent sums of traces over fixed levels. A cyclic permutation changes the level which is summed last, which is a rearrangement of an absolutely convergent multiple sum together with the cyclic symmetry of each trace over fixed levels. The relation $P_LP_{L'}=\delta_{LL'}P_L$ gives (ii), and Lemma~\ref{ggs:lem:S}~(i) gives (iii). The identities continue to all $d$ by Lemma~\ref{ggs:lem:cont}. The Lie derivative along a conformal Killing vector commutes with $d$ and therefore maps exact two-forms to exact two-forms, and the same argument applies.
\end{proof}

We call the rule (iii) the diagonal rule ${\rm Tr}(PM_u)=(N/V)\int u$. Given the continuation in $d$ of the sums over levels, it is not a separate prescription but the value of the continued sum of the traces of the single levels.

Let $\tau=\tau_c+\delta\tau$ with a constant $\tau_c$ and a polynomial $\delta\tau$, and let $u=e^{-2\delta\tau}-1$ as a formal power series in $\delta\tau$. We define the bare functional of $n_G$ gauge fields with a common local coupling on $S^d$ by
\begin{equation}
W_0^S[\tau] = n_G\left(N\tau_c-\frac12\sum_{n\geq1}\frac{(-1)^{n+1}}{n}\,T_n(u,\ldots,u)\right) ,
\label{ggs:eq:W}
\end{equation}
as a formal power series in $\delta\tau$. Each order in $\delta\tau$ receives contributions from finitely many $n$, and each contribution is a finite combination of cycles of polynomials. In integer dimension \eqref{ggs:eq:W} is formally $\log Z$ of the free gauge field divided by the volume of the gauge group, with the measure defined by the metric alone, up to a constant: with $e^{-2\tau}=e^{-2\tau_c}(1+u)$ the action of the coexact one-forms is $e^{-2\tau_c}$ times the quadratic form $P(1+M_u)P$ on the exact two-forms, and $-\tfrac12\log\det$ of it is $N\tau_c-\tfrac12{\rm Tr}\log(P+PM_uP)$.

\subsection{Constant couplings}
\label{ggs:sec:shift}

\begin{proposition}
\label{ggs:prop:shift}
For every $d$ away from the poles and every constant $c$, as formal power series in $c$ and $\delta\tau$,
\begin{equation}
W_0^S[\tau_c;\,\delta\tau+c] = W_0^S[\tau_c;\,\delta\tau]+n_G\,N\,c \, , \qquad W_0^S[\tau_c+c;\,\delta\tau] = W_0^S[\tau_c;\,\delta\tau+c] \, ,
\label{ggs:eq:shift}
\end{equation}
where $W_0^S[\tau_c;\delta\tau]$ denotes the expansion \eqref{ggs:eq:W} around $\tau_c$. In particular $W_0^S$ is a functional of $\tau$ which does not depend on the choice of $\tau_c$, its value at constant coupling is $n_G\log g$ with the coefficient $n_GN=n_G$ for every $d$, and all terms of second and higher order of the expansion vanish when all their arguments are constant.
\end{proposition}

\begin{proof}
Let $u_c=e^{-2(\delta\tau+c)}-1$, so that $\pr_cu_c=-2(1+u_c)$, and let $F(u)=-\tfrac12\sum_n\frac{(-1)^{n+1}}{n}T_n(u,\ldots,u)$. By Lemma~\ref{ggs:lem:rules}~(i),
\begin{equation}
\pr_cF(u_c) = -\tfrac12\sum_{n\geq1}(-1)^{n+1}\,T_n(\pr_cu_c,u_c,\ldots,u_c) = \sum_{n\geq1}(-1)^{n+1}\big[T_n(1,u_c,\ldots,u_c)+T_n(u_c,\ldots,u_c)\big] \, .
\end{equation}
By Lemma~\ref{ggs:lem:rules}~(ii) the first term is $N$ for $n=1$ and $T_{n-1}(u_c,\ldots,u_c)$ for $n\geq2$, so that the sum telescopes to $N$. At each order in $c$ and $\delta\tau$ only finitely many terms contribute. Hence $F(u_c)=F(u_0)+Nc$, which is the first relation. The second follows from the first. By the definition its left-hand side is $n_G[N(\tau_c+c)+F(u_0)]$ and its right-hand side is $n_G[N\tau_c+F(u_c)]$.
\end{proof}

The proposition gives instance~(b) of Property~\ref{gs:P:functional} at the level of the bare functional, with $\cW^{\log}=n_G\log g$ and $n_\iota=n_G$ to all orders in $\vep$. The coefficient is the regularised number $N$ of coexact one-forms, which is one per gauge field for every $d$, in agreement with~\cite{Giombi:2015haa} and with the value which Property~\ref{gs:P:free} requires by Proposition~\ref{grg:prop:gauge}.

\subsection{Conformal Killing vectors}
\label{ggs:sec:ward}

Let $\sigma=b\cdot X$ and $v^\mu=-\nabla^\mu\sigma$ as in \eqref{eq:ckv}, with $b$ in the parallel directions, and let
\begin{equation}
\mathbb{K}_\sigma = \int d\mu_d\,\Big(-\tfrac12\vep\,\sigma-v^\mu\pr_\mu\tau\Big)\frac{\delta}{\delta\tau} \, ,
\end{equation}
which is the operator of Property~\ref{gs:P:ward} with $\hB=-\tfrac12\vep$. On the formal series it acts by the first-order change of $\delta\tau$ by $\delta_v\tau=-\tfrac12\vep\sigma-v^\mu\pr_\mu\delta\tau$, which is a polynomial.

\begin{proposition}
\label{ggs:prop:ward}
For every $d$ away from the poles, $\mathbb{K}_\sigma W_0^S=0$ to all orders in $\delta\tau$.
\end{proposition}

\begin{proof}
Let $w=1+u=e^{-2\delta\tau}$. Its variation is $\delta_vw=-2w\,\delta_v\tau=\vep\sigma w-v^\mu\pr_\mu w$. For exact two-forms $f$ and $h$ the Lie derivative satisfies
\begin{equation}
\vev{\cL_vf,wh}+\vev{f,w\,\cL_vh} = \vev{f,(\vep\sigma w-v^\mu\pr_\mu w)\,h} \, ,
\label{ggs:eq:ibp}
\end{equation}
since $(\cL_vf)\cdot h+f\cdot\cL_vh=v^\mu\pr_\mu(f\cdot h)+4\sigma\,f\cdot h$, by $\cL_v\gamma^{\mu\nu}=-2\sigma\gamma^{\mu\nu}$, and $\int d\mu_d\,w\,v^\mu\pr_\mu(f\cdot h)=-\int d\mu_d\,(v^\mu\pr_\mu w+d\,\sigma w)\,f\cdot h$, by $\nabla_\mu v^\mu=d\,\sigma$. In integer dimension \eqref{ggs:eq:ibp} is the operator identity $PM_{\delta_vw}P=\cL_v^\dagger(P+U)+(P+U)\cL_v$ on the exact two-forms, with $U=PM_uP$ and $\cL_v$ restricted to the exact two-forms. Inside traces over fixed levels it holds for all $d$ by Lemma~\ref{ggs:lem:transfer}. With the cyclic symmetry and $\delta_vu=\delta_vw$,
\begin{equation}
\mathbb{K}_\sigma W_0^S = -\frac{n_G}{2}\sum_{n\geq1}(-1)^{n+1}\,T_n(\delta_vu,u,\ldots,u) = -\frac{n_G}{2}\sum_{n\geq1}(-1)^{n+1}\,{\rm Tr}\big[\big(\cL_v^\dagger(P+U)+(P+U)\cL_v\big)U^{n-1}\big] \, ,
\end{equation}
with $U^0=P$. On the exact two-forms $\cL_v^\dagger=-\cL_v+\vep\,PM_\sigma P$, by \eqref{ggs:eq:ibp} with $w=1$, so that traces containing $\cL_v^\dagger$ are traces over fixed levels as well. The trace of the transposed operator and the cyclicity of Lemma~\ref{ggs:lem:rules} give ${\rm Tr}[\cL_v^\dagger(P+U)U^{n-1}]={\rm Tr}[\cL_vU^{n-1}(P+U)]$ and ${\rm Tr}[(P+U)\cL_vU^{n-1}]={\rm Tr}[\cL_vU^{n-1}(P+U)]$. Hence
\begin{equation}
\mathbb{K}_\sigma W_0^S = -n_G\sum_{n\geq1}(-1)^{n+1}\big({\rm Tr}[\cL_vU^{n-1}]+{\rm Tr}[\cL_vU^n]\big) = -n_G\,{\rm Tr}[P\cL_vP] \, ,
\end{equation}
where the sum telescopes at each order in $\delta\tau$. Finally ${\rm Tr}[P\cL_vP]=\Sigma_{L\geq1}{\rm tr}(P_L\cL_vP_L)$, and each term vanishes. Indeed, \eqref{ggs:eq:ibp} with $w=1$ gives $2\,{\rm tr}(P_L\cL_vP_L)=\vep\,{\rm tr}(P_LM_\sigma P_L)=\vep\,(D_L/V)\int d\mu_d\,\sigma=0$ by Lemma~\ref{ggs:lem:S}~(i).
\end{proof}

We conclude that the bare functional is invariant under the variation of the coupling with the weight $\tfrac12\vep$ of the logarithmic coordinate, which is the analogue of \eqref{eq:barevar}, with a vanishing bare anomaly. No flat chart and no continuation in the exponent of a propagator enter.

\subsection{Rotations and the second order}
\label{ggs:sec:rotations}

\begin{proposition}
\label{ggs:prop:rot}
Through second order around a constant $\tau_c$,
\begin{equation}
W_0^S = n_G\left(N\tau_c+\frac NV\int d\mu_d\,\delta\tau-\frac NV\int d\mu_d\,\delta\tau^2+T_2(\delta\tau,\delta\tau)\right)+O(\delta\tau^3) \, .
\label{ggs:eq:second}
\end{equation}
The one-point function of the operator of $\log g$ is the constant $n_GN/V$. For a harmonic $\sigma$ of degree one and a polynomial $\delta\tau$, $T_2(\sigma,\delta\tau)=\Lambda_T\int d\mu_d\,\sigma\,\delta\tau$ with
\begin{equation}
V\Lambda_T = N-(d-2)\,\Sigma_{L\geq1}\frac{D_L}{(L+2)(L+d-3)} = \frac{4}{4-d} \, ,
\label{ggs:eq:Lambda}
\end{equation}
so that the second variation of $W_0^S$ in the directions $\sigma$ and $\delta\tau$ is $(2n_Gd/(\vep V))\int d\mu_d\,\sigma\,\delta\tau$.
\end{proposition}

\begin{proof}
With $u=-2\delta\tau+2\delta\tau^2+O(\delta\tau^3)$ the term $n=1$ of \eqref{ggs:eq:W} gives $-\tfrac12T_1(u)$, which by Lemma~\ref{ggs:lem:rules}~(iii) is the second and third term, and the term $n=2$ gives $\tfrac14T_2(u,u)=T_2(\delta\tau,\delta\tau)+O(\delta\tau^3)$. By Lemma~\ref{ggs:lem:S}~(ii), $T_2(\sigma,\delta\tau)=\Sigma_Lt_2(L;\sigma,\delta\tau)$ is $\Lambda_T\int\sigma\,\delta\tau$ with the first expression of \eqref{ggs:eq:Lambda}. Since $D_L/((L+2)(L+d-3))=L(L+d-1)(2L+d-1)\Gamma(L+d-3)/(\Gamma(d-1)\Gamma(L+3))$, the telescoping sum gives $\Sigma_LD_L/((L+2)(L+d-3))=d/((d-2)(d-4))$, and with $N=1$ the second expression follows. The second variation is $2n_G(\Lambda_T-N/V)\int\sigma\,\delta\tau$, and $\Lambda_T-1/V=d/(\vep V)$.
\end{proof}

The formula \eqref{ggs:eq:t2} agrees at integer $d$ with a direct computation with explicit exact two-forms and exact cubature. As a check of Proposition~\ref{ggs:prop:ward} at first order: by \eqref{gs:eq:parts} the term of first order of $\mathbb{K}_\sigma W_0^S$ is $n_G[dN/V-\tfrac12\vep\cdot2d/(\vep V)]\int\sigma\,\delta\tau=n_G(N-1)\,(d/V)\int\sigma\,\delta\tau=0$. The pole of $\Lambda_T$ at $\vep=0$, $\vep V\Lambda_T\to4$, equals $\tfrac12\,c\,\tfrac{\pi^2}4\,V$ with $c=12/\pi^4$ and $V=\tfrac83\pi^2$, the coefficient of the flat two-point function of $\tfrac12F_{\mu\nu}F^{\mu\nu}$, as in Lemma~\ref{grg:lem:direct}.

\subsection{Hypothesis and properties}
\label{ggs:sec:props}

On a conformally round metric $\gamma=e^{2\varphi}\mathring\gamma$, with a polynomial $\varphi$ in the parallel coordinates, we define the bare functional by
\begin{equation}
W_0[e^{2\varphi}\mathring\gamma;\tau] = W_0^S[\tau+\tfrac12\vep\,\varphi] \, ,
\label{ggs:eq:curved}
\end{equation}
which is the identity $S[A;e^{2\varphi}\gamma,\tau]=S[A;\gamma,\tau+\tfrac12\vep\varphi]$ of the classical actions, since $\sqrt\gamma\,\gamma^{\mu\rho}\gamma^{\nu\sigma}$ has the Weyl weight $d-4$. This is a definition. It drops the dependence of the measure on the metric which does not involve $\tau$, as the convention of section~\ref{sec:continuation} drops the free energy for scalar fields. The bare functional is then invariant under Weyl rescalings with $\delta\varphi=\sigma$ and $\delta\tau=-\tfrac12\vep\sigma$.

We use Hypothesis~\ref{hyp:finite} for the free gauge field, see appendix~\ref{gs:sec:cases}. With $L_\iota=0$ and without background gauge field and relevant sources it states the following. There is a local functional $F=\int d^dx\sqrt\gamma\;\cX$ of the following form. The density $\cX$ is a linear combination of the covariant monomials of dimension four built from the curvature and from covariant derivatives of $\tau$, each containing at least one derivative of $\tau$, with real pure poles in $\vep$ as coefficients which do not depend on $\tau$. For all polynomial $\varphi$ and $\delta\tau$ the terms of first and higher order in $\varphi$ and $\delta\tau$ of $W_0[e^{2\varphi}\mathring\gamma;\tau_c+\delta\tau]+F[e^{2\varphi}\mathring\gamma,\tau_c+\delta\tau]$ have no pole at $\vep=0$.

Here a real pure pole is a polynomial in $1/\vep$ without constant term and with real coefficients. Since $\mathbb{W}_\sigma W_0=0$, the hypothesis requires that the Weyl variation of the pole part of $F$ is finite. It also requires that the terms of order $k$ in $\varphi$ at constant $\tau$, which are $(\vep/2)^k$ times terms of the bare functional, have no pole, so that no counterterm of the curvature alone is needed. The hypothesis is stated on the sphere and its conformal rescalings, since the functional on $S^d$ is not obtained from flat space through the chart, see section~\ref{ggs:sec:chart}. We define the renormalised functional as $W=W_0+F$. The bare coupling is the coupling itself, since the free functional depends on the coupling only through $W_0$.

\begin{proposition}
\label{ggs:prop:props}
Assume Hypothesis~\ref{hyp:finite} for the free gauge field. Then the renormalised functional $W$ of $n_G$ free gauge fields with a common local coupling on $S^d$, $d=4-\vep$, with polynomial sources, has Properties~\ref{gs:P:functional} to~\ref{gs:P:sphere} with Property~\ref{gs:P:functional} in instance~(b), the logarithmic coordinate $\log g$, $\cW^{\log}=n_G\log g$ with $n_\iota=n_G$ for every $\vep$, $\hB=-\tfrac12\vep$ exactly, $\hr=0$, no relevant sources and the Weyl variation closed on the coupling. It has Property~\ref{gs:P:free}, with $\cK^{(0,0)}$ as in Proposition~\ref{grg:prop:gauge}, and the last statements of Properties~\ref{gs:P:weyl} and~\ref{gs:P:free} are empty.
\end{proposition}

\begin{proof}
The polynomials in the parallel coordinates form a class of sources which satisfies the assumptions of appendix~\ref{gs:sec:class}, see appendix~\ref{app:rules}. Property~\ref{gs:P:functional}: on the unit sphere, $\varphi=0$, the terms of order $n\geq1$ in $\delta\tau$ are finite by the hypothesis and do not depend on $g$, so that they are regular. At $\delta\tau=0$ the counterterms vanish, since each monomial contains a derivative of $\tau$, and the value is $n_GN\tau_c=n_G\log g$ by Proposition~\ref{ggs:prop:shift}. There is no background gauge field. Property~\ref{gs:P:weyl}: the bare functional on the backgrounds \eqref{ggs:eq:curved} is invariant under the Weyl rescaling with $\hB=-\tfrac12\vep$, which has no terms of second order, and $\hr=0$. The operators $\mathbb{W}_\sigma$ commute. No relevant source is needed: a source of a scalar operator of dimension below four is not required since the variation of $\tau$ closes, and a two-form source of $F_{\mu\nu}$ is not required since with a single coupling there is no antisymmetric tensor of dimension two built from the metric and from $\tau$ without the Levi-Civita tensor, $\nabla_\mu\nabla_\nu\tau$ being symmetric. Property~\ref{gs:P:anomaly}: $\mathbb{W}_\sigma W=\mathbb{W}_\sigma F\equiv\cA[\sigma]$, which is local, built covariantly with $d$ as a parameter, homogeneous of Weyl weight $-\vep$ since each monomial has dimension four, and contains $\tau$ only through its derivatives. It is finite, since $W$ is finite on all backgrounds \eqref{ggs:eq:curved} and $\mathbb{W}_\sigma$ is the derivative along a variation of $\varphi$ and $\tau$ in the class. Its coefficients are therefore regular in a basis of local functionals which are independent modulo total derivatives as functionals on these backgrounds at $\vep=0$, as in section~\ref{gs:sec:remains}. The relation $\mathbb{W}_\sigma\cA[\sigma']=\mathbb{W}_{\sigma'}\cA[\sigma]$ holds because the operators commute. Property~\ref{gs:P:ward}: $\mathbb{K}_\sigma W_0^S=0$ by Proposition~\ref{ggs:prop:ward}, and $\mathbb{K}_\sigma F=\cA[\sigma]$ at $\gamma_{\mu\nu}=\mathring\gamma_{\mu\nu}$ by the argument of Proposition~\ref{gs:prop:msward}, since $F$ is invariant under diffeomorphisms and the combination of the diffeomorphism generated by $-v$ and the Weyl rescaling with $\sigma$ acts on $\tau$ as $\delta_v\tau$. Property~\ref{gs:P:sphere}: the one-point function of the bare functional is the constant $n_GN/V$ by Proposition~\ref{ggs:prop:rot}, and the first variation of $F$ at constant $\tau$ vanishes, since the terms of $\cX$ which are linear in $\tau$ are total derivatives on the unit sphere, whose curvature is covariantly constant. The second variation of $W_0^S$ in the directions $\sigma z$ and $\delta\tau$ is a constant times $\int\sigma\,\delta\tau$ by Proposition~\ref{ggs:prop:rot}, and that of $F$ by integrations by parts with $\nabla^2\sigma=-d\,\sigma$. The statements on the anomaly follow from its locality by Lemma~\ref{gsp:lem:local}. There are no relevant sources and no current. Property~\ref{gs:P:free}: with one logarithmic coordinate Proposition~\ref{grg:prop:gauge} gives $\cK^{(0,0)}=-3n_\iota(0)/\pi^2=-3n_G/\pi^2$, which is the value $s_4a_4c$ with $c=12n_G/\pi^4$ of Lemma~\ref{gro:lem:c}, and its last statement is empty.
\end{proof}

At second order the proposition uses part (iii) of Lemma~\ref{ggs:lem:S} only for the existence of $T_2$ with two general polynomial sources. The statements of Property~\ref{gs:P:sphere} and the Ward identity at first order in $\delta\tau$, which give the sum rule \eqref{gs:eq:Glog} with $n_\iota=n_G$, use parts (i) and (ii) only, together with Hypothesis~\ref{hyp:finite} for the free gauge field at second order. For the free gauge field the sum rule is an identity of the construction: by Proposition~\ref{ggs:prop:rot} the second variation is $2n_Gd/(\vep V)$ times $\int\sigma\,\delta\tau$, and \eqref{gs:eq:Glog} with $k=\tfrac12$ reads $\tfrac12\hat G^{{\rm S}(0)}=-d\,n_G/V$.

\subsection{Relation to the flat chart}
\label{ggs:sec:chart}

The functional \eqref{ggs:eq:W} contains the diagonal traces ${\rm Tr}(PM_u)=(N/V)\int u$ and, through the rule (ii) of Lemma~\ref{ggs:lem:rules}, the compositions of the projector of the sphere at coincident points, which give the term $n_G\log g$ of Proposition~\ref{ggs:prop:shift}. The flat chart does not provide them. In flat space the free functional with the source $u=e^{-2(\tau-\tau_c)}-1$ is a sum of cycles of the propagator of the field strength, which does not depend on the gauge fixing, since the field strength annihilates gradients. The term with one vertex vanishes, since a propagator at coincident points vanishes, and a cycle with $n\geq2$ vertices has $n$ vertices with the factor $e^{-2\tau_c}$ and $n$ propagators with the factor $e^{2\tau_c}$. Hence the flat functional does not depend on the constant part of the coupling, and neither do the terms of the functional on $S^d$ which is obtained from it through the chart with the bare coupling of weight $\tfrac12\vep$, where they exist. A density of counterterms each of whose monomials contains a derivative of $\tau$, as in Hypothesis~\ref{hyp:finite}, does not add a term $n\log g$ either. The functional on $S^d$ is therefore not obtained from flat space through the chart. The construction of this appendix provides the projector of the sphere level by level, and it requires neither the composition of two flat kernels through a vertex nor the exclusion of a line of poles, since no exponent of a propagator is continued.

\subsection{One photon line in quantum electrodynamics}
\label{gq:sec:qed}

We now add $N_f$ Dirac fermions with charges $q_f$ to the abelian gauge field of this appendix. The action is $\tfrac12\int d\mu_d\,e^{-2\tau}|F|^2+\sum_f\int d\mu_d\,\bar\psi_f\gamma^\mu(\nabla_\mu+iq_fA_\mu)\psi_f$, so that the coupling $g=e^\tau$ enters only through the action of the photon. We treat the terms of the functional with one photon line. In this subsection we show that they are sums over one level, which we continue as in section~\ref{ggs:sec:sums}, that the longitudinal part of the photon does not contribute, that the constant part of the coupling enters through the factor $g^2$ only, and that the bare functional is invariant under the conformal Killing vectors. Under two assumptions on the renormalisation at order $g^2$, which we state in Proposition~\ref{gq:prop:beta}, we then obtain the term of order $g^2$ of $\hB$ in the logarithmic coordinate, which is the known coefficient of quantum electrodynamics. This is a check of the framework in an interacting gauge theory at lowest order, and not a construction of quantum electrodynamics.

\subsubsection{The current kernel}
\label{gq:sec:kernel}

At vanishing photon field the fermions give the functional of section~\ref{gs:sec:dirac} with the background gauge field $A$ of the charges. Its term of second order in $A$ is $\tfrac12\vev{A,\mathcal{K}^{\rm J}A}$ with $\vev{A,B}=\int d\mu_d\,A\cdot B$ on one-forms. By the loop \eqref{gf:eq:loop} with two vertices of the current, written with the chord as in section~\ref{gs:sec:dirac}, the kernel of $\mathcal{K}^{\rm J}$ is
\begin{equation}
\mathcal{K}^{\rm J}_{AB}(X,Y) = Q^2\,C_\psi^2\,{\rm tr}\,1\;\frac{I_{AB}(X,Y)}{s^{2\alpha}} \, , \qquad I_{AB} = \delta_{AB}-\frac{2(X-Y)_A(X-Y)_B}{s^2} \, , \qquad Q^2=\sum_fq_f^2 \, ,
\label{gq:eq:K}
\end{equation}
projected onto the tangent spaces at $X$ and $Y$, with $\alpha=2\Delta_\psi$, since the fermion propagator is $C_\psi\,\Gamma\!\cdot\!(X-Y)\,(s^2)^{-\Delta_\psi-1/2}$; on the line \eqref{eq:line}, $\Delta_\psi=\tfrac12(d-1)$ and $\alpha=d-1$. Here we used ${\rm tr}(\gamma_A\,\gamma\!\cdot\!r\,\gamma_B\,\gamma\!\cdot\!r)={\rm tr}\,1\,(2r_Ar_B-\delta_{AB}r^2)$, the sign of the loop and ${\rm tr}(\omega\omega')=-q^2$ for $\omega=iq$, as in the terms of second order in appendix~\ref{gs:app:fermions}. Let $\lambda^{(\alpha)}_L=\pi^{d/2}2^{d-2\alpha}\,\Gamma(\tfrac d2-\alpha)\Gamma(L+\alpha)/(\Gamma(\alpha)\Gamma(L+d-\alpha))$ be the eigenvalue of the kernel $s^{-2\alpha}$ on the scalar harmonics of degree $L$.

\begin{lemma}
\label{gq:lem:kernel}
On the coexact one-forms of degree $L$ the operator with the kernel $I_{AB}/s^{2\alpha}$ acts as multiplication by $\tfrac{\alpha-1}{\alpha}\lambda_L^{(\alpha)}$. On the exact one-forms of degree $L$ it acts as $\tfrac{d-1-\alpha}{\alpha}\lambda_L^{(\alpha)}$. On the line \eqref{eq:line} the operator $\mathcal{K}^{\rm J}$ therefore vanishes on the exact one-forms, and on the coexact one-forms of degree $L$ it is multiplication by
\begin{equation}
\pi_L = Q^2\,C_\psi^2\,{\rm tr}\,1\;\pi^{d/2}\,2^{2-d}\,\Gamma(1-\tfrac d2)\,\frac{(d-2)\,\Gamma(L+d-1)}{\Gamma(d)\,\Gamma(L+1)} \, .
\label{gq:eq:pi}
\end{equation}
\end{lemma}

\begin{proof}
We work in integer dimension in the region of convergence in $\alpha$ and continue in $\alpha$. For non-integer $d$ we take \eqref{gq:eq:pi} as the definition of the action of $\mathcal{K}^{\rm J}$ on the level $L$; it is the continuation of the Gram integral of the loop against the polynomial kernel of the level. The embedding components $V_A$ of a coexact one-form of degree $L$ are harmonic polynomials of degree $L$ restricted to the sphere, and $X\!\cdot\!V(X)=0$. For a tangent field $V$, $(X-Y)_BV_B(Y)=X_BV_B(Y)$, and $(X-Y)_As^{-2\alpha-2}=-(2\alpha)^{-1}\pr^X_As^{-2\alpha}$ with $s^2=|X-Y|^2$ in the embedding space. Hence
\begin{equation}
\int d\mu_d(Y)\,\frac{I_{AB}V_B(Y)}{s^{2\alpha}} = \lambda^{(\alpha)}_LV_A(X)+\frac1\alpha\,X_B\,\pr^X_A\!\int d\mu_d(Y)\,\frac{V_B(Y)}{s^{2\alpha}} \, ,
\end{equation}
by the Funk--Hecke formula for the first term. After the projection onto the tangent space at $X$ the derivative becomes the gradient along the sphere of $\lambda^{(\alpha)}_LV_B$, and $X_B\nabla_AV_B=\nabla_A(X\!\cdot\!V)-V_B\nabla_AX_B=-V_A$. The two terms give $(1-\tfrac1\alpha)\lambda^{(\alpha)}_LV_A$. At $\alpha=d-1$ this is \eqref{gq:eq:pi}. The functional of the fermions is invariant under the background gauge transformations $A\to A+d\omega$ by Proposition~\ref{gf:prop:gauge}, and at vanishing sources its term of second order is invariant by itself, so that $\mathcal{K}^{\rm J}\,d\omega=0$ inside the integrals. The eigenvalue on the exact one-forms is obtained in the same way from the gradients of the scalar harmonics.
\end{proof}

The closed form agrees with an exact Gram integration for $L\leq5$ with a test field of each level. At $d=4-\vep$ we have $\Gamma(1-\tfrac d2)=-\tfrac2\vep+O(1)$, and with $C_\psi=1/(2\pi^2)$ and ${\rm tr}\,1=4$ at $d=4$,
\begin{equation}
\frac{\pi_L}{\lambda_L} = -\frac{Q^2}{6\pi^2\,\vep}+O(\vep^0) \, ,
\label{gq:eq:pole}
\end{equation}
with a residue which does not depend on $L$, since $\Gamma(L+d-1)/(\Gamma(L+1)(L+1)(L+d-2))=1$ at $d=4$.

\subsubsection{Definition at the order of one photon line}
\label{gq:sec:def}

In integer dimension the photon propagator with the local coupling is the inverse of the action on the coexact one-forms. In the basis of section~\ref{ggs:sec:levels}, with $A=\sum_ia_iA_i$ and $f_i=dA_i/\sqrt{\lambda_i}$, the action is $\tfrac12e^{-2\tau_c}\sum_{ij}a_ia_j\sqrt{\lambda_i\lambda_j}(P+U)_{ij}$, so that the propagator is $e^{2\tau_c}\mathsf{D}^{-1/2}(P+U)^{-1}\mathsf{D}^{-1/2}$ with $\mathsf{D}={\rm diag}(\lambda_i)$. The terms of the functional with one photon line consist of one fermion loop with two current vertices, since a loop with one vertex vanishes. Their sum is $\tfrac12$ times the trace of $\mathcal{K}^{\rm J}$ times the propagator. The longitudinal part of the photon does not contribute, by Lemma~\ref{gq:lem:kernel}, and the gauge fixing, which concerns the exact one-forms only and does not contain $\tau$, does not enter. Since $\mathcal{K}^{\rm J}$ is the multiple $\pi_L$ of the identity on each level, we define
\begin{equation}
W^{(1)}[\tau] = \frac{e^{2\tau_c}}2\sum_{n\geq0}(-1)^n\,\Sigma_{L\geq1}\,\frac{\pi_L}{\lambda_L}\,t_n(L;u,\ldots,u) \, , \qquad t_0(L)=D_L \, ,
\label{gq:eq:W1}
\end{equation}
with $t_n$ of \eqref{ggs:eq:tn}, $u=e^{-2\delta\tau}-1$ and the continuation of Lemma~\ref{ggs:lem:cont}, as a formal power series in $\delta\tau$. The functional of the theory through the order of one photon line is $W_0^S+W^{(1)}$, with $W_0^S$ of \eqref{ggs:eq:W} for $n_G=1$.

\begin{proposition}
\label{gq:prop:def}
(i) Each order in $\delta\tau$ of $W^{(1)}$ is a finite combination of sums of the form of Lemma~\ref{ggs:lem:cont}, hence a meromorphic function of $d$. (ii) For a constant $c$, $W^{(1)}[\tau_c;\delta\tau+c]=e^{2c}\,W^{(1)}[\tau_c;\delta\tau]$ and $W^{(1)}[\tau_c+c;\delta\tau]=W^{(1)}[\tau_c;\delta\tau+c]$ as formal power series. Hence $W^{(1)}$ does not depend on the choice of $\tau_c$, it is $g^2$ times a functional of $\delta\tau$, and it does not change the coefficient $n=1$ of $\log g$. (iii) $\mathbb{K}_\sigma W^{(1)}=0$ to all orders in $\delta\tau$.
\end{proposition}

\begin{proof}
(i) By \eqref{gq:eq:pi}, $\pi_L/\lambda_L$ is a meromorphic function of $d$ times $\Gamma(L+d-1)/(\Gamma(L+1)(L+1)(L+d-2))$, which is of the form \eqref{ggs:eq:H}, and by Lemma~\ref{ggs:lem:S}~(iii) each $t_n(L)$ is in $\cH$. The product of two terms of the form \eqref{ggs:eq:H} is of this form. (ii) We write \eqref{gq:eq:W1} as $\tfrac12e^{2\tau_c}{\rm Tr}[\mathcal{P}(P+U)^{-1}]$ with $\mathcal{P}=\sum_L(\pi_L/\lambda_L)P_L$, expanded in $U$. With $u_c=e^{-2(\delta\tau+c)}-1$ we have $P+U_c=e^{-2c}(P+U)$, where $U_c=PM_{u_c}P$. The identity $(P+U_c)^{-1}=e^{2c}(P+U)^{-1}$ of formal power series in $c$ and $\delta\tau$ follows, order by order, from $P_LP_{L'}=\delta_{LL'}P_L$ inside the traces over fixed levels, as in the proof of Proposition~\ref{ggs:prop:shift}, and $\mathcal{P}$ commutes with $P_L$. This gives the first relation, and the second follows from the first and the definition. (iii) By the proof of Proposition~\ref{ggs:prop:ward}, $\delta_v(P+U)=\cL_v^\dagger(P+U)+(P+U)\cL_v$, so that $\delta_v(P+U)^{-1}=-(P+U)^{-1}\cL_v^\dagger-\cL_v(P+U)^{-1}$ as formal series. Hence $\mathbb{K}_\sigma W^{(1)}=-\tfrac12e^{2\tau_c}{\rm Tr}[(\mathcal{P}\cL_v+\cL_v^\dagger\mathcal{P})(P+U)^{-1}]$ by the cyclicity of Lemma~\ref{ggs:lem:rules}, which holds with the weights $\pi_L/\lambda_L$ by the same proof. In terms of one-forms the operator $\mathcal{P}$ on the exact two-forms is $\mathsf{D}^{-1/2}\mathcal{K}^{\rm J}\mathsf{D}^{-1/2}$ in the basis $f_i$, and $\cL_v$ on the exact two-forms is $\mathsf{D}^{1/2}\cL_v^{(1)}\mathsf{D}^{-1/2}$, where $\cL_v^{(1)}$ is the action of $\cL_v$ on the one-forms modulo exact one-forms, since $d$ commutes with $\cL_v$. The functional of the fermions is invariant under the pullback of the gauge field by the conformal maps of the sphere, by Proposition~\ref{gf:prop:cov}, since the gauge field has the weight zero on the line. At second order in $A$ and to first order in the map this gives $\vev{\cL_vA,\mathcal{K}^{\rm J}B}+\vev{A,\mathcal{K}^{\rm J}\cL_vB}=0$ for one-forms $A$, $B$ of the class, in particular for the polynomial forms of fixed levels, which is $\mathcal{P}\cL_v+\cL_v^\dagger\mathcal{P}=0$ inside the traces over fixed levels. For non-integer $d$, where $\pi_L$ is defined by \eqref{gq:eq:pi}, this identity follows from Lemma~\ref{ggs:lem:transfer}. After division by the factor of \eqref{gq:eq:pi} which does not depend on $L$ the weights $\pi_L/\lambda_L$ are rational functions of $d$ at fixed $L$, and the identity holds at the odd integers $d\geq5$, where this factor is finite and the kernel on the line is conformally covariant. Hence $\mathbb{K}_\sigma W^{(1)}=0$.
\end{proof}

At first order in $\delta\tau$ the proposition contains a nontrivial identity. With Lemma~\ref{ggs:lem:S}, the terms of first and second order of \eqref{gq:eq:W1} are $\tfrac12g^2\,\Sigma_L(\pi_L/\lambda_L)[2t_1(L;\delta\tau)-2t_1(L;\delta\tau^2)+4t_2(L;\delta\tau,\delta\tau)]$, and the term of first order of $\mathbb{K}_\sigma W^{(1)}$ receives $d\,g^2\Sigma_L(\pi_LD_L/(\lambda_LV))\int\sigma\delta\tau$ from the term of first order through \eqref{gs:eq:parts}, and $-\tfrac12\vep$ times the second variation in the directions $\sigma$ and $\delta\tau$ from the term of second order, which by Lemma~\ref{ggs:lem:S}~(ii) is $2g^2\Sigma_L(\pi_LD_L/(\lambda_LV))[1-2(d-2)/\mu_L]\int\sigma\delta\tau$. The sum is $(2(d-2)g^2/V)\,\Sigma_L(\pi_LD_L/\mu_L)\int\sigma\,\delta\tau$, where we used $\mu_L+4-d=\lambda_L$. The summand is $\Phi(L+1)-\Phi(L)$ with $\Phi(L)=(L-1)(L+2)\,\pi_LD_L/((d-2)(2L+d-1)\mu_L)$, as Gosper's algorithm shows, and $\Phi(1)=0$, so that the continued sum vanishes by Lemma~\ref{ggs:lem:cont}. Off the line, for generic $\alpha\neq d-1$, the sum does not vanish. An exception is $\alpha=1$, where the multiplier on the coexact one-forms vanishes at every level while the one on the exact one-forms does not. With this exception the identity tests the conservation of the current and the covariance together.

\subsubsection{Vector field at order \texorpdfstring{$g^2$}{g2}}
\label{gq:sec:beta}

We consider the renormalised functional $W=W_0^S[\tau_0]+W^{(1)}[\tau_0]+F$ through order $g^2$, where the bare coordinate is $\tau_0=\tau+L_\iota(g)$ with $L_\iota(g)=\ell_1g^2/\vep+O(g^4)$ and a real number $\ell_1$, and where $F$ is a local functional of covariant monomials of dimension four with coefficients which are real pure poles times regular functions of $g$. On the conformally round metrics we define $W^{(1)}[e^{2\varphi}\mathring\gamma;\tau]=W^{(1)}[\tau+\tfrac12\vep\varphi]$, as in \eqref{ggs:eq:curved}, which is justified since the massless Dirac action is Weyl covariant in every $d$ and the gauge field has the weight zero. Hence the bare coordinate has the Weyl variation $\hB_0=-\tfrac12\vep$. We use the scheme in which $C_\psi=\Gamma(\tfrac d2)/(2\pi^{d/2})$ and ${\rm tr}\,1=4$ for all $d$; the residues below do not depend on this choice.

\begin{proposition}
\label{gq:prop:beta}
Assume that through order $g^2$ every monomial of $F$ contains a derivative of $\tau$, so that in particular the coefficients of $\int d\mu_d\,E_4$ and of $\int d\mu_d\,R^2$ have no term of order $g^2$. Assume that the free energy of order $g^2$ at constant coupling, or equivalently, by Propositions~\ref{ggs:prop:shift} and~\ref{gq:prop:def}~(ii) and since $F$ and its first variation vanish at constant $\tau$, the term of first order in $\delta\tau$ and of order $g^2$ of $W$, is finite at $\vep=0$. Then $\ell_1=Q^2/(12\pi^2)$, and the vector field in the logarithmic coordinate is
\begin{equation}
\hB = -\frac\vep2+\frac{Q^2}{12\pi^2}\,g^2+O(g^4) \, ,
\label{gq:eq:B}
\end{equation}
which is the one-loop coefficient of quantum electrodynamics with charges $q_f$. For $q_f=1$ it is $\beta/e$ with the one-loop beta function $\beta=-\tfrac\vep2e+\tfrac43N_fe^3/(4\pi)^2$ given in the introduction of~\cite{Giombi:2015haa}, where $N_f$ counts Dirac fermions with ${\rm tr}\,1=4$.
\end{proposition}

\begin{proof}
The term of first order of $W_0^S[\tau_0]$ is $(N/V)\int\delta\tau_0$, and $\delta\tau_0=\delta\tau\,(1+2\ell_1g_c^2/\vep)+O(\delta\tau^2,g^4)$, where $g_c=e^{\tau_c}$. The term of first order of $W^{(1)}$ is $g_c^2\,\Sigma_L(\pi_LD_L/(\lambda_LV))\int\delta\tau$, by \eqref{gq:eq:W1} and Lemma~\ref{ggs:lem:S}~(i). The first variation of $F$ at constant $\tau$ vanishes, since every monomial contains a derivative of $\tau$, so that the part of each monomial which is linear in $\delta\tau$ is a total derivative on the unit sphere, whose curvature is covariantly constant. The term of first order and order $g_c^2$ of $W$ is therefore $(2g_c^2/V)A_0\int\delta\tau$ with $A_0=N\ell_1/\vep+\tfrac12\Sigma_L\pi_LD_L/\lambda_L$, and the term of order zero and order $g_c^2$ is $g_c^2A_0$. By \eqref{gq:eq:pi} the sum is, after $L=n+1$, a very-well-poised ${}_5F_4$ at unit argument with the parameters $a=d+1$, $b=c=d-1$ and $e=2$, which Dougall's formula evaluates for ${\rm Re}\,d<2$,
\begin{equation}
\Sigma_{L\geq1}\,\frac{D_L\,\Gamma(L+d-1)}{\Gamma(L+1)\,\lambda_L} = \frac{\Gamma(d-1)}{(3-d)(2-d)} \, ,
\end{equation}
The right-hand side is meromorphic, and therefore it is the continued sum of Lemma~\ref{ggs:lem:cont}. Hence $\tfrac12\Sigma_L\pi_LD_L/\lambda_L=-Q^2/(12\pi^2\vep)+O(1)$, and with $N=1$ the term $A_0$ is finite if and only if $\ell_1=Q^2/(12\pi^2)$. For the bare coordinate the Weyl variation is $\hB_0=-\tfrac12\vep$, and $\hB\,(1+\pr_\iota L_\iota)=-\tfrac12\vep$ with $\pr_\iota L_\iota=2\ell_1g^2/\vep$ gives \eqref{gq:eq:B}.
\end{proof}

The first assumption cannot be dropped, although a weaker one suffices. Counterterms $f(g)\int d\mu_d\,E_4$ and $h(g)\int d\mu_d\,R^2$ with $f=f_1g^2/\vep$ and $h=h_1g^2/\vep$ add $(24f_1+144h_1)V/\vep$ to the pole of $A_0$ on the unit sphere, so that finiteness fixes only $\ell_1+(24f_1+144h_1)V$. The value of $\ell_1$ follows if this combination vanishes, which the first assumption ensures. It is not implied by the finiteness of the terms of first order in $\delta\tau$ on the unit sphere. It is a restriction compared with the curvature counterterms $a(g)E_4+b(g)R^2$ of \eqref{eq:X}. Conversely, with the coefficient $\ell_1=Q^2/(12\pi^2)$ of the coupling renormalisation of quantum electrodynamics, the computation shows that the free energy of order $g^2$ on $S^d$ is finite without a counterterm of the curvature alone at this order.

Through the order of one photon line the coefficient of $\log g$ is thus $n=1$ for every $\vep$, the bare functional is invariant under the conformal Killing vectors, and finiteness at first order in $\delta\tau$, without a counterterm of the curvature alone at order $g^2$, gives the one-loop coefficient of $\hB$. The sum rule \eqref{gs:eq:Glog} concerns the term of order zero of the metric, which this order does not change. Beyond this order a graph contains several photon lines and is a multiple sum over levels, whose continuation in $d$ we do not have.
\section{Four-fermion couplings near two dimensions}
\label{gs:app:fourfermion}

In this appendix we construct the bare functional on $S^d$ for Dirac fermions with four-fermion vertices of every gamma structure, and we derive from Hypothesis~\ref{hyp:finite} for the four-fermion theory the statements of section~\ref{gs:sec:fermions} on the Weyl variation.

\subsection{Four-fermion vertices on the sphere}
\label{g2:sec:construction}

In this subsection we extend the construction of section~\ref{gs:sec:dirac} to vertices with four fermion fields and any number of gamma matrices. We note that the vertices of section~\ref{gs:sec:dirac} contain at most one gamma matrix, which is contracted with a vector of the embedding space. The new feature is that the two bilinears of a four-fermion vertex carry gamma matrices whose indices are contracted with each other. On the sphere we contract them with the projector onto the tangent space, and we show that the traces are again polynomials in the scalar products of the points and in $d$, so that the integrals have the form \eqref{eq:gram}, and that the chart identity of Lemma~\ref{gf:lem:chart} holds with one factor $\Omega$ for each bilinear. We then apply the proofs of appendix~\ref{gs:app:fermions}, with the changes stated below.

\subsubsection{Vertices and contracted traces}
\label{g2:sec:vertices}

We consider $N_f$ Dirac fermions with the massless Dirac action, and we allow in addition the scalar fields and the vertices of section~\ref{gf:sec:vertices}. For $n\geq0$ let $\Gamma^{(n)}_{\mu_1\ldots\mu_n}=\gamma_{[\mu_1}\cdots\gamma_{\mu_n]}$ be the antisymmetrised product, with the weight $1/n!$, so that $\Gamma^{(0)}=1$ and $\Gamma^{(1)}_\mu=\gamma_\mu$. The four-fermion vertex is
\begin{equation}
V^{\rm F}[G] = \frac12\sum_{n\geq0}\frac1{n!}\,G^{(n)}_{ij,kl}\,\big(\bar\psi_i\,\Gamma^{(n)}_{\mu_1\ldots\mu_n}\psi_j\big)\big(\bar\psi_k\,\Gamma^{(n)\,\mu_1\ldots\mu_n}\psi_l\big) \, ,
\label{g2:eq:vertex}
\end{equation}
where $i,j,k,l$ are flavour indices, the indices $\mu_r$ are summed, and the flavour tensors satisfy $G^{(n)}_{ij,kl}=G^{(n)}_{kl,ij}$. For each $n$ we obtain a complex vector space of finite dimension. Let $\eta_n=(-1)^{n(n+1)/2}$. We call $G^{(n)}$ real if $\overline{G^{(n)}_{ij,kl}}=G^{(n)}_{ji,lk}$, and we denote the real vector space of the real tensors by $V_n$. It has finite dimension. For every $n$ a tensor is real if and only if it is a real combination of symmetrised products $t\otimes t'+t'\otimes t$ of flavour matrices with $t^\dagger=\eta_nt$ and $t'^\dagger=\eta_nt'$. Indeed, such products are real since $\eta_n^2=1$, and the real span of the products of matrices with $t^\dagger=\eta_nt$ has the real dimension $N_f^4$ of the space of all tensors with $\overline{G_{ij,kl}}=G_{ji,lk}$, the symmetrisation being compatible with both conditions. For $n=0$ these are hermitian matrices and for $n=1$ antihermitian ones, as for the Yukawa vertex and the vertex of the current. The couplings of the theory are the real coordinates $g^I$ of the elements of $V=\bigoplus_{n\geq0}V_n$. There are infinitely many of them, and we denote by $n_I$ the value of $n$ to which the index $I$ belongs.

A bilinear $\bar\psi\,M\psi$ of a vertex is called a slot. The slots are the two bilinears of a four-fermion vertex, the Yukawa vertex and the vertex of the current. In a term of the functional we join the slots by fermion lines to closed loops, and we contract the indices of the two slots of a four-fermion vertex. We now define the value of such a product of traces for non-integer dimension. Consider a finite family of traces of products of matrices $\Gamma\!\cdot\!v$ and flavour matrices, where some of the vectors are labelled. Each label $c$ occurs exactly twice, and with each label a symmetric bilinear form $\Pi_c$ is associated. In integer dimension $D$ we obtain the value of the family by replacing the two vectors with the label $c$ by $e_M$ and $e_N$ and summing with the coefficients $(\Pi_c)_{MN}$, for each label. For non-integer dimension we use the following rule. We expand each trace by \eqref{gf:eq:trace} into a sum over the perfect matchings of its vectors, with signs. For each term we join the two occurrences of each label. We then obtain paths, whose end points are unlabelled vectors $u$ and $w$, and cycles, which consist of labelled vectors only. A path through the labels $c_1,\ldots,c_k$ in this order contributes $u\!\cdot\!\Pi_{c_1}\cdots\Pi_{c_k}w$, and a cycle contributes ${\rm tr}\,\Pi_{c_1}\cdots\Pi_{c_k}$. We use two kinds of forms. On the sphere the label of a contraction at a point $X$ carries the projector $\Pi_X=1-XX^T$ onto the tangent space, and ${\rm tr}\,\Pi_{X_1}\cdots\Pi_{X_k}$ is evaluated by expanding the product of the factors $1-X_rX_r^T$, with ${\rm tr}\,1=d+1$. In flat space we give every label the unit form of $\mathbb{R}^d$, so that a path contributes $u\!\cdot\!w$ and a cycle contributes $d$. A path of length zero contributes $u\!\cdot\!w$.

\begin{lemma}[Contracted traces]
\label{g2:lem:contract}
The value of a family of traces with labelled vectors defined by this rule is a polynomial in $d$ and in the scalar products of the unlabelled vectors and of the points of the labels, times the product of the traces of the flavour matrices. Its spinor factor is a polynomial with real coefficients. Let the dimension $D$ be an integer and all vectors lie in $\mathbb{R}^D$, with $D=d+1$ for the forms of the sphere and $D=d$ in flat space. Then the rule with ${\rm tr}\,1=D$ for the forms of the sphere and with the unit form of $\mathbb{R}^D$ in flat space gives the value obtained with explicit gamma matrices of $D$ dimensions in a representation in which the traces of products of an odd number of gamma matrices vanish, such as an irreducible representation for even $D$, divided by $({\rm tr}\,1)^{\#\rm traces}$ and multiplied by the fixed number ${\rm tr}\,1$ of section~\ref{gf:sec:vertices} for each trace. The value vanishes if one of the traces contains an odd number of gamma matrices, and the spinor factor is invariant under the reversal of the order of the gamma matrices in one of the traces.
\end{lemma}

\begin{proof}
The expansion \eqref{gf:eq:trace} of a trace of formal vectors into matchings is unique, and the rule depends only on which positions carry the same label, so that the value is well defined. The first two statements hold by construction, since each path and each cycle contributes a polynomial in $d$ and in these scalar products with integer coefficients. In integer dimension the polynomial \eqref{gf:eq:trace} is the trace of the explicit gamma matrices divided by ${\rm tr}\,1$, see section~\ref{gf:sec:vertices}, and it is multilinear in each vector. Hence we find that the value with the explicit sums over $e_M$ and $e_N$ is the sum over the matchings of the products of the scalar products, in which each label $c$ appears in two factors, $(u\!\cdot\!e_M)(\Pi_c)_{MN}(e_N\!\cdot\!w)$ or with further labels in between. The sum over $M$ and $N$ along a path gives $u\!\cdot\!\Pi_{c_1}\cdots\Pi_{c_k}w$, and around a cycle it gives the trace of the product of the forms, which for the projectors of the sphere is the expansion stated above with ${\rm tr}\,1=D$. Traces with an odd number of gamma matrices vanish by \eqref{gf:eq:trace}. The reversal of the order of the factors of one trace maps each matching of its vectors to its mirror image with the same sign, since \eqref{gf:eq:trace} is invariant under reversal, so that the expansion and the spinor factor are unchanged.
\end{proof}

In integer dimension we therefore reproduce the antisymmetrisation and the contraction of the indices of the two slots of \eqref{g2:eq:vertex} by the rule, with one label for each index $\mu_r$. Since the rule depends only on which positions carry the same label, the antisymmetrisation of both slots gives the same value as the antisymmetrisation of one of them, also for non-integer $d$. For non-integer $d$ we define the contracted products by the rule, and in all statements below a product of traces with contracted indices means this value.

\subsubsection{Loops on the sphere}
\label{g2:sec:loops}

On the sphere we write the gamma matrices of the slots of a four-fermion vertex at the point $X$ as the tangent gamma matrices $\Gamma_T^M=(\Pi_X)^{MN}\Gamma_N$, and we contract the indices of its two slots with the projector $\Pi_X$. That is, the slot matrices are $t\,\Gamma\!\cdot\!(\Pi_Xe^{M_1})\cdots\Gamma\!\cdot\!(\Pi_Xe^{M_n})$, antisymmetrised in one of the two slots, and the labels $M_r$ carry the form $\Pi_X$, which is idempotent. A term of the bare functional contains a family of closed loops through the slots of its vertices. We define its factor as the product of the factors \eqref{gf:eq:loop} of the loops, with $M_i=t_i\,\Gamma^{(n)}_{T\,M_1\ldots M_n}$ at a slot of a four-fermion vertex, where $t_i$ is the flavour matrix of the slot, and with the contracted product of the traces in place of the product of the traces. Two slots which are joined by a line belong to different vertices, or the line joins a point to itself, in which case its chord vanishes and so does the term, in agreement with the convention of section~\ref{sec:continuation} that propagators at coincident points vanish. The terms of the bare functional are the integrals \eqref{eq:gram} of the sums over the connected pairings of these products, with the scalar propagators and the factors $(-1)^N/N!$ of appendix~\ref{app:gauge}.

By Lemma~\ref{g2:lem:contract} we find that the factor of a family of loops is a polynomial in $d$, in the scalar products $X_i\!\cdot\!X_j$, $A(X_i)\!\cdot\!X_j$ and $A(X_i)\!\cdot\!A(X_j)$ of \eqref{gf:eq:products} and in the components of the sources, times the powers of $s_{ij}^2$ of the lines and a power of $C_\psi$. As in section~\ref{gf:sec:loops} we conclude that the integrand is a finite sum of integrands of the form of Proposition~\ref{prop:mero}, with coefficients which are polynomials in $d$ times powers of $C_\phi$ and $C_\psi$, with $n_{ij}$ the number of all lines between the points $i$ and $j$ and $r_{ij}$ the number of fermion lines among them. Each term of the functional involves finitely many vertices and therefore finitely many values of $n$. We regard the bare functional as a formal power series in all couplings $g^I$, that is a coefficient is assigned to each finite multiset of indices $I$, and each coefficient is a functional of the sources of the class.

We relate these loops by the chart of section~\ref{gf:sec:loops} to the loops in flat space, in which the flat gamma matrices $\gamma^\mu$ of the slots of a four-fermion vertex are contracted with the unit form of $\mathbb{R}^d$. We use the notation of Lemma~\ref{gf:lem:chart}.

\begin{lemma}[Chart with four-fermion vertices]
\label{g2:lem:chart}
For a family of closed loops through slots at points which differ from $P$,
\begin{equation}
\begin{split}
&\text{(contracted product of the traces on the sphere)} \\
&\qquad = \prod_{\text{slots}\notin\mathsf{C}}\Omega_{i(\text{slot})}\;\times\;\text{(contracted product of the flat traces)} \, ,
\end{split}
\label{g2:eq:chart}
\end{equation}
where $i(\text{slot})$ is the point of the slot, $\mathsf{C}$ is the set of the slots of the current, the traces on the left-hand side are those of the factor \eqref{gf:eq:loop} of each loop, and on the right-hand side they are those of Lemma~\ref{gf:lem:chart}, with $m_i=t_i\,\Gamma^{(n)}_{\mu_1\ldots\mu_n}$ at a slot of a four-fermion vertex. In particular a four-fermion vertex contributes the factor $\Omega_i^{\,2}$.
\end{lemma}

\begin{proof}
We first prove the identity in integer dimension, with $d+1$ embedding dimensions and the flat chart in the hyperplane $\mathsf{h}$ orthogonal to $P$. By Lemma~\ref{g2:lem:contract} both sides are then given by explicit gamma matrices. We insert \eqref{gf:eq:inversion} for each chord. As in the proof of Lemma~\ref{gf:lem:chart} the two factors $\Gamma\!\cdot\!\tilde y_i$ of the two chords at a slot enclose its matrix. At a slot of a four-fermion vertex we insert $(\Gamma\!\cdot\!\hat y_i)^2=1$ between the gamma matrices of the slot and use $\Gamma\!\cdot\!\hat y\,\Gamma\!\cdot\!v\,\Gamma\!\cdot\!\hat y=-\Gamma\!\cdot\!(Rv)$, where $R=1-2\,\hat y\hat y^T$ is the reflection of the proof of Lemma~\ref{gf:lem:chart}. We obtain
\begin{equation}
\Gamma\!\cdot\!\tilde y_i\;\Gamma\!\cdot\!v_1\cdots\Gamma\!\cdot\!v_n\;\Gamma\!\cdot\!\tilde y_i = (-1)^n\,\tilde y_i^{\,2}\;\Gamma\!\cdot\!(R_iv_1)\cdots\Gamma\!\cdot\!(R_iv_n)
\label{g2:eq:sandwich}
\end{equation}
for any vectors $v_r$, with $R_i$ the reflection at $X_i$. We now show that $R_i$ maps the tangent space at $X_i$ onto $\mathsf{h}$. Since $P\!\cdot\!\tilde x_i=0$ we have $\hat y_i\!\cdot\!P=-1/|\tilde y_i|$ and $R_iP=P+2\,\tilde y_i/\tilde y_i^{\,2}=X_i$. A vector $v$ satisfies $v\!\cdot\!X_i=0$ if and only if $(R_iv)\!\cdot\!P=v\!\cdot\!R_iP=0$, since $R_i$ is symmetric. Hence we find that $R_i\Pi_{X_i}R_i$, which is the orthogonal projector onto $R_i$ applied to the tangent space, is the projector $\Pi_\mathsf{h}$ onto $\mathsf{h}$. The vectors of the slots are $v_r=\Pi_{X_i}e^{M_r}$, and the contraction of an index of the two slots of the vertex gives
\begin{equation}
\sum_{M}\big(R_i\Pi_{X_i}e^M\big)\big(R_i\Pi_{X_i}e^M\big)^T = R_i\,\Pi_{X_i}R_i = \Pi_\mathsf{h} = \sum_{\mu}e_\mu e_\mu^T \, ,
\end{equation}
where $e_\mu$ is an orthonormal basis of $\mathsf{h}$. Hence we obtain for the two slots of the vertex the slots $\Gamma^{(n)}_{\mu_1\ldots\mu_n}$ of the flat vertex, contracted over $\mathsf{h}$, with the factor $(-1)^n\tilde y_i^{\,2}$ for each slot. The antisymmetrisation is not affected. Each chord gives $-\tfrac12\Omega_i\Omega_j$, so that each slot receives $\Omega_i^{\,2}$ from its two chords. At a slot of a Yukawa vertex and at a slot of the current the factors are those of the proof of Lemma~\ref{gf:lem:chart}. For a loop with $m$ slots the product of the factors is therefore $(-1)^m\prod_{s}(-1)^{n_s}\prod_{s\notin\mathsf{C}}\Omega_{i(s)}$, where $n_s$ is the number of gamma matrices of the slot $s$ and $n_s=1$ for a slot of the current. The trace vanishes unless the number $m+\sum_sn_s$ of its gamma matrices is even, and the sign is then $+1$. We have thus proved \eqref{g2:eq:chart} in integer dimension.

Both sides of \eqref{g2:eq:chart} are polynomials in $d$ and in the scalar products of the flat points $\tilde x_i$, of the vector $P$ and of the vectors of the sources, by Lemma~\ref{g2:lem:contract}, where the points $X_i$ and the factors $\Omega_i$ are expressed through $\tilde x_i$ by the chart. The traces of the projectors on the left-hand side are polynomials in $d$ and in the $X_r\!\cdot\!X_s$. For a fixed number of vectors and all sufficiently large integers $d$ the scalar products of the flat points and of the source vectors in $\mathsf{h}$ range over an open set, together with the components of the source vectors along $P$. After multiplication by a power of $\prod_i(1+\tilde x_i^{\,2})$ both sides are polynomials in $d$ and in these variables. Two such polynomials whose values agree for all large integers $d$ on an open set of the variables are equal. Hence \eqref{g2:eq:chart} holds as an identity of rational functions, and therefore for the integrands of \eqref{eq:gram} with non-integer $d$.
\end{proof}

\subsubsection{The bare functional}
\label{g2:sec:bare}

\begin{lemma}[Counting with four-fermion vertices]
\label{g2:lem:count}
Lemma~\ref{gf:lem:count} holds for the vertices of section~\ref{gf:sec:vertices} together with the four-fermion vertices \eqref{g2:eq:vertex}.
\end{lemma}

\begin{proof}
We follow the proof of Lemma~\ref{gf:lem:count}, with Lemma~\ref{g2:lem:chart} in place of Lemma~\ref{gf:lem:chart}. By Lemma~\ref{g2:lem:contract} the contracted product of the flat traces is a sum of products of scalar products over all $d$ components of the chords $x_{i\,i+1}$ and of the vectors $a^{\rm fl}(x_i)$, times polynomials in $d$, in which each chord occurs exactly once, since each chord occurs once in its trace and the rule replaces a pair of labelled vectors by a form without introducing vectors. We therefore have factors of the types of Lemma~\ref{lem:schwinger}, and the chord of a fermion line is the difference of the marked line. At a four-fermion vertex the factor $\Omega_i^{\,2}$ of \eqref{g2:eq:chart} raises $\kappa_i$ by two, so that $\kappa_i\geq h+\mu_i$. A term with a line from a point to itself vanishes. Otherwise the point of a four-fermion vertex has exactly four fermion lines and no further lines, so that $n_{\max}\geq4$ in the condition $h>\tfrac52n_{\max}$ of the proof of Proposition~\ref{gf:prop:line}, and the linear factors which contain $x_i$ are the four chords and the factors of the source, so that $\mathsf{n}_i-\mu_i\leq4\leq3\,n_i$. At the other points the bounds are those of Lemma~\ref{gf:lem:count}.
\end{proof}

We use the parameter $\Dp$ as in section~\ref{gf:sec:loops}, with $\Delta_\psi=\Dp+\tfrac12$, also when no scalar field is present. On the line \eqref{eq:line} the exponent of a fermion line is then the dimension $\tfrac12(d-1)$ of a free fermion.

\begin{proposition}[Bare functional with four-fermion vertices]
\label{g2:prop:bare}
Consider $N_f$ Dirac fermions with the four-fermion vertices \eqref{g2:eq:vertex} for all $n\geq0$, the vertex of the fermion current and, if scalar fields are present, the vertices of Proposition~\ref{gf:prop:bare}, with sources in the class of sources on the sphere. Then each term of the bare functional is a meromorphic function of $d$ and $\Dp$. With $\Upsilon$ kept independent it satisfies \eqref{eq:cov3} with the additional entries $\Omega^{d-4\Dp-2}\,G\!\circ\!f$ for the couplings of the four-fermion vertices and $\Omega^{d-3\Dp-1}\,Y\!\circ\!f$ for the Yukawa couplings. On the line \eqref{eq:line} the line is not a line of poles, the integrals of which a term is a sum, with the normalisations $C_\phi$ and $C_\psi$ of the propagators taken out, have no line of poles of fixed dimension, and each term is a meromorphic function of $d$. Away from the poles the functional satisfies \eqref{eq:covS} with the weight $2-d$ of the four-fermion couplings, which is $\vep k$ with $k=1$ for $d=2-\vep$, and it is invariant under background gauge transformations, in the form \eqref{eq:gaugeS} extended by the variations $\delta_\omega Y$ and $\delta_\omega G$. It has the properties under rotations of \eqref{eq:Lambda}, the one-point function of the fermion current vanishes at constant sources, and the functional is real for real sources. The flat chart \eqref{eq:chart} holds with the four-fermion couplings of weight $2-d$.
\end{proposition}

\begin{proof}
We go through the steps of appendix~\ref{gs:app:fermions}.

\emph{Meromorphy and the line.} By section~\ref{g2:sec:loops} each term is a finite sum of integrals \eqref{eq:gram} of integrands of the form of Proposition~\ref{prop:mero}, with coefficients which are polynomials in $d$ times powers of $C_\phi$ and $C_\psi$, so that the first paragraph of section~\ref{gf:sec:line} applies. By Lemma~\ref{g2:lem:count} the proof of Proposition~\ref{gf:prop:line} applies without change.

\emph{Covariance.} Let $f$ be a map \eqref{eq:boost} along $\hat b=P$, which in the chart from $P$ is the dilatation $x\mapsto\lambda x$. The contracted product of the flat traces is homogeneous of degree one in each chord and linear in each $a^{\rm fl}$, and the flat contractions are invariant under the dilatation. Lemma~\ref{g2:lem:chart} at the points $fX_i$ and at the points $X_i$ therefore gives \eqref{gf:eq:loopcov} with the product over all slots which are not slots of the current. With $s'^2_{ij}=\Omega_{f,i}\Omega_{f,j}\,s_{ij}^2$ each fermion line contributes $\Omega_{f,i}^{-\Dp-1}$ at each of its end points. A four-fermion vertex has four such end points and two slots, and with the factor $\Omega_{f,i}^{\,d}$ of the measure its coupling receives the weight $d-4(\Dp+1)+2=d-4\Dp-2$. On the line this is $2-d$. We obtain the other weights as in Proposition~\ref{gf:prop:cov}, and we use the remaining steps of its proof without change.

\emph{Gauge invariance.} We let the group of field rotations act on the flavour indices of each slot of \eqref{g2:eq:vertex}, and we set $\delta_\omega G=[\,\cdot\,,\omega_f]$ on each of the two slots, that is $\delta_\omega(t\otimes t')=[t,\omega_f]\otimes t'+t\otimes[t',\omega_f]$, as for the Yukawa couplings in section~\ref{gf:sec:gauge}. The vertex contains no derivatives, and the replacement $R_\omega$ of $\psi$ by $-\omega_f\psi$ and of $\bar\psi$ by $\bar\psi\,\omega_f$ changes the slot $\bar\psi\,t\,\Gamma\psi$ by $\bar\psi\,[\omega_f,t]\,\Gamma\psi$, so that $\delta_\omega V^{\rm F}+R_\omega V^{\rm F}=0$ and \eqref{gf:eq:pointwise} holds with the four-fermion vertices included. We now show that Lemma~\ref{gf:lem:ward2} holds for the loops through the point $X_0$ of the vertex of the current. Indeed, in integer dimension the contracted product is a finite sum of products of traces in which the trace through $X_0$ has the form ${\rm tr}[\Gamma\!\cdot\!X_{h0}\,\Gamma\!\cdot\!\nabla\omega_f\,\Gamma\!\cdot\!X_{0j}\,R]$ with a matrix $R$ which does not depend on $X_0$, the other traces do not depend on $X_0$. The computation of the proof of Lemma~\ref{gf:lem:ward2} holds for every such matrix $R$, and by the linearity in $R$ it gives \eqref{gf:eq:Psi} for the sum over the labels, in every sufficiently large integer dimension. After the removal of the common factor $(s_{h0}^2)^{-\Dp-1}(s_{0j}^2)^{-\Dp-1}$ both sides of \eqref{gf:eq:Psi} are polynomials in $d$, $\Dp$ and the coordinates of the points and of the sources, with $\rho^2=1-Y^2$, and these coordinates range over an open set for large integer $d$, so that \eqref{gf:eq:Psi} holds for all $d$. The vector $e$ of the coordinates \eqref{eq:last} still occurs only in $e\!\cdot\!e$ after the contraction. Indeed, $e$ is orthogonal to the points $X_k$ with $k\neq0$ and to the parallel directions, so that $\Pi_{X_k}e=e$, and the forms of the labels, which belong to these points, give $e\!\cdot\!\Pi_{X_{k_1}}\cdots\Pi_{X_{k_r}}e=1$ and $e\!\cdot\!\Pi_{X_{k_1}}\cdots\Pi_{X_{k_r}}w=0$ for the other vectors $w$, while the cycles do not contain $e$. The vector $\sum_AY_A(e_A-X_0Y_A)$ has the scalar products $\rho^2$ times scalar products of $Y$ with the points and the vectors of the sources and the scalar product $0$ with $X_0$, so that the forms $\Pi_{X_k}$ with $k\neq0$ preserve this property and $\sum_AY_A\varphi_A=\rho^2\mathsf{q}$ holds as before. In Lemma~\ref{gf:lem:contact} the function $\mathsf{g}$ of Lemma~\ref{lem:delta} contains the contracted product, and the limit identifies $X_0$ with the point of the following or the preceding slot, which may be a slot of a four-fermion vertex. If the preceding and the following slot belong to the same four-fermion vertex, the limit contains the chord of a line from this point to itself and vanishes, and so does the corresponding term of $R^\psi_\omega$, which contains a line joining the two slots of the vertex. In Lemma~\ref{gf:lem:np} the counting of Lemma~\ref{g2:lem:count} replaces that of Lemma~\ref{gf:lem:count}. At an end of the line $e_0$ a slot of a four-fermion vertex is not enclosed by two factors $\Gamma\!\cdot\!\tilde y_j$. In the first term of \eqref{gf:eq:Bpsi} the slot matrix $M_j$ at the end $X_j$ stands between $\Gamma\!\cdot\!X_j$, with $X_j=P+\Omega_j\tilde y_j$, and the factor $\Gamma\!\cdot\!\tilde y_j$ of the following chord. By \eqref{g2:eq:sandwich} with $(\Gamma\!\cdot\!\hat y_j)^2=1$ we move $\Gamma\!\cdot\!\tilde y_j$ through the gamma matrices of the slot, $M_j\,\Gamma\!\cdot\!\tilde y_j=(-1)^{n}\,\Gamma\!\cdot\!\tilde y_j\,M_j^R$, where $M_j^R$ contains the reflected vectors $R_j\Pi_{X_j}e^M$, and in the second term we proceed in the same way at the end $X_h$. The two slots of the vertex then carry vectors of $\mathsf{h}$ contracted with $\Pi_\mathsf{h}$, as in the proof of Lemma~\ref{g2:lem:chart}, the power of $\Omega_j$ from the numerator is one, as at a Yukawa vertex, and there is one additional linear factor, which gives the count of Lemma~\ref{gf:lem:np}. We then use the proof of Proposition~\ref{gf:prop:gauge} without change. As there, the proof does not use that $\omega_f$ is antihermitian.

\emph{Rotations and chart.} After the contraction we have an integrand which is a function of the $X_{i\parallel}$ and the $u_{ij}$, and section~\ref{gf:sec:chart} applies. On a conformally flat background the vertex \eqref{g2:eq:vertex} with the indices contracted with the inverse metric is invariant under $\psi\to\Omega^{-(d-1)/2}\psi$, $\bar\psi\to\Omega^{-(d-1)/2}\bar\psi$ and the rescaling of the metric by $\Omega^2$, with the vielbein $\Omega\,\delta^a_\mu$, up to the factor $\Omega^{-2(d-1)}$, which with the volume element gives the weight $2-d$, in agreement with the covariance.

\emph{Reality.} Consider a term with real $d$ and $\Dp$, real $G^{(n)}$, hermitian $Y^a$ and a gauge field with values in the Lie algebra. By the multilinearity we may take each four-fermion coupling of the form $t\otimes t'$ with $t^\dagger=\eta_nt$ and $t'^\dagger=\eta_nt'$, together with its image under the exchange of the slots. By Lemma~\ref{g2:lem:contract} the contracted product of the traces is a real polynomial $\mathsf{S}$ in the scalar products times the product of the flavour traces $\mathsf{F}_\ell$ of the loops $\ell$. Reversing the orientation of one loop $\ell$ with $m$ slots reverses the order of its gamma matrices and multiplies each chord by $-1$. The antisymmetrised product of $n_s$ gamma matrices at a slot $s$ is multiplied by $(-1)^{n_s(n_s-1)/2}$ under reversal, and its contraction with the other slot of the vertex is unchanged otherwise, so that by the last statement of Lemma~\ref{g2:lem:contract} the spinor factor becomes $(-1)^m\prod_{s\in\ell}(-1)^{n_s(n_s-1)/2}\,\mathsf{S}$. The flavour trace of the reversed loop is ${\rm tr}(t_1t_mt_{m-1}\cdots t_2)={\rm tr}(t_1^Tt_2^T\cdots t_m^T)$, which is $\prod_{s\in\ell}\eta_{n_s}\,\overline{\mathsf{F}_\ell}$, with $\eta_0=1$ for a Yukawa slot and $\eta_1=-1$ for a slot of the current. Since $m+\sum_sn_s$ is even for a non-vanishing trace and $\eta_n=(-1)^n(-1)^{n(n-1)/2}$, the reversed loop gives $\mathsf{S}\,\overline{\mathsf{F}_\ell}\prod_{\ell'\neq\ell}\mathsf{F}_{\ell'}$. For a loop with at least three slots the two orientations are different pairings with the same weight, and for a loop with two slots the loop is its own reverse, so that $\mathsf{F}_\ell$ is real. The sum over the orientations of all loops is therefore $\mathsf{S}\prod_\ell(\mathsf{F}_\ell+\overline{\mathsf{F}_\ell})$ up to positive factors, which is real. The scalar parts are real as in Lemma~\ref{gf:lem:real}.
\end{proof}

\subsubsection{Terms of second order}
\label{g2:sec:second}

\begin{lemma}[Two-point kernel of the four-fermion couplings]
\label{g2:lem:kernel}
Through second order in the couplings, at vanishing gauge field and for bare couplings which agree with the couplings up to terms of second order, the bare functional of the four-fermion couplings on the line is
\begin{equation}
\frac12\,c_{IJ}(d)\int d\mu_d(X)\,d\mu_d(Y)\;\frac{g^I(X)\,g^J(Y)}{s(X,Y)^{4(d-1)}} \, ,
\label{g2:eq:twopoint}
\end{equation}
where $c_{IJ}(d)=C_\psi^{\,4}\,\mathsf{p}_{IJ}(d)$ with $C_\psi=1/\vol_d$ and polynomials $\mathsf{p}_{IJ}$ in $d$, which depend on ${\rm tr}\,1$ and are real by the reality of the functional. The terms of second order with a four-fermion coupling and a Yukawa coupling, a scalar coupling, the source of $\bar\psi\psi$ or the gauge field vanish. If $n_I\geq3$, then $\mathsf{p}_{IJ}(2)=0$ for all $J$. For flavour tensors $G$ and $G'$ let $\mathsf{A}(G,G')=\sum G_{ij,kl}\,G'_{ji,lk}$ and $\mathsf{B}(G,G')=\sum G_{ij,kl}\,G'_{jk,li}$, and let $c(G,G')=c_{IJ}(d)\,g^Ig'^J$ for $G,G'\in V$ with the coordinates $g^I$ and $g'^J$. For $G\in V_n$ and $G'\in V_m$
\begin{equation}
\begin{split}
c(G,G') &= \tfrac12\,C_\psi^{\,4}\Big[({\rm tr}\,1)^2\binom dn\,\delta_{nm}\,\mathsf{A}(G,G')-{\rm tr}\,1\,\mathsf{P}_{nm}(d)\,\mathsf{B}(G,G')\Big] \, , \\
\mathsf{P}_{nm}(d) &= \sum_{k=0}^{\min(n,m)}\binom dn\binom nk\binom{d-n}{m-k}\,(-1)^{nm-k+n(n-1)/2+m(m-1)/2} \, ,
\end{split}
\label{g2:eq:closed}
\end{equation}
where $\binom dn=d(d-1)\cdots(d-n+1)/n!$ and $\binom{d-n}{m-k}$ denote polynomials in $d$. The first term is the part with two flavour traces, which joins each slot of one vertex to one slot of the other, and the second term is the part with one flavour trace.
\end{lemma}

\begin{proof}
We first note that the terms of first order vanish, since a term with one four-fermion vertex contains a line from its point to itself. At second order two four-fermion vertices at $X$ and $Y$ are joined by four fermion lines, each slot of one vertex being joined by both of its lines to slots of the other. The pairings form either two loops, each through one slot of each vertex, or one loop through all four slots. All chords are $\pm X_{XY}$. In the flat chart the contracted product is a polynomial $\mathsf{p}(d)$ times $|x|^4$, with the flat chord $x$, times flavour traces, since all flat vectors are $\pm x$ and each of the four chords occurs once. By Lemma~\ref{g2:lem:chart} and $s^2=\Omega_X\Omega_Y|x|^2$ the contracted product on the sphere is $\mathsf{p}(d)\,s^4$. With the four denominators $(s^2)^{-\Dp-1}$ this gives $(s^2)^{-4\Dp-2}$, which on the line is $s^{-4(d-1)}$. We have thus proved \eqref{g2:eq:twopoint} with polynomials $\mathsf{p}_{IJ}$. A four-fermion vertex and a vertex with a scalar leg have no connected pairing in which the scalar leg is paired. A four-fermion vertex and a vertex of the current or of the source of $\bar\psi\psi$ have no connected pairing without a line from a point to itself, since at most two of the four legs of the four-fermion vertex end at the other vertex, so that the remaining two are joined by a line from its point to itself. Let now $n_I\geq3$. Since only the vector $X_{XY}$ occurs, the value of $\mathsf{p}_{IJ}(2)$ is, by Lemma~\ref{g2:lem:contract} in the flat chart with $D=2$, the value obtained with explicit gamma matrices of two dimensions, in which the antisymmetrised product of three or more gamma matrices vanishes. Hence $\mathsf{p}_{IJ}(2)=0$. The term of second order of the functional is $\tfrac12$ times the integral of the connected two-point function of the vertices, by the factor $(-1)^N/N!$ with $N=2$, so that in the flat chart $c(G,G')\,|x|^{4-4d}$ is the two-point function $\langle V^{\rm F}[G](x)\,V^{\rm F}[G'](0)\rangle$ of the free fermions. We compute it in integer dimension $D$ with hermitian Euclidean gamma matrices in a representation in which the traces of products of an odd number of gamma matrices vanish, with ${\rm tr}\,1$ in place of the trace of the unit matrix of the representation. By Lemma~\ref{g2:lem:contract} this gives the polynomials for all $d$, since polynomials which agree at infinitely many integers are equal. Since the $n!$ orderings of a set of $n$ distinct indices give the same product of the two slots, the vertex \eqref{g2:eq:vertex} is $\tfrac12\sum_n\sum_SG^{(n)}_{ij,kl}(\bar\psi_i\Gamma_S\psi_j)(\bar\psi_k\Gamma_S\psi_l)$, where $S$ runs over the sets of $n$ distinct indices and $\Gamma_S$ is the product of the $\gamma_\mu$ with $\mu\in S$ in increasing order. We take $x$ along the first axis, so that $\gamma\!\cdot\!x\,\Gamma_B\,\gamma\!\cdot\!x=\epsilon_B\,x^2\,\Gamma_B$ with a sign $\epsilon_B$, and we use $\Gamma_A^2=(-1)^{n(n-1)/2}$, $\Gamma_A\Gamma_B=(-1)^{nm-k}\,\Gamma_B\Gamma_A$ and ${\rm tr}(\Gamma_A\Gamma_B)=0$ for $A\neq B$, where $n$ and $m$ are the numbers of elements of $A$ and $B$ and $k$ is the number of their common elements. With $S(z)=C_\psi\,\gamma\!\cdot\!z\,|z|^{-d}$ and the sign $-1$ of each loop of \eqref{gf:eq:loop}, a loop through the slots $t\,\Gamma_A$ at $x$ and $u\,\Gamma_B$ at $0$ gives $C_\psi^2\,|x|^{-2d}\,{\rm tr}(tu)\,{\rm tr}(\Gamma_A\,\gamma\!\cdot\!x\,\Gamma_B\,\gamma\!\cdot\!x)$. In a pairing with two loops both loops carry the same sets $A$ and $B$, so that only $A=B$ contributes, the signs $\epsilon_A(-1)^{n(n-1)/2}$ of the two loops cancel, and we obtain the factor $C_\psi^4\,({\rm tr}\,1)^2\,|x|^{4-4d}$ for each of the $\binom Dn$ sets. The loop through four slots gives $-C_\psi^4\,|x|^{4-4d}$ times the flavour trace and ${\rm tr}(\Gamma_A\Gamma_B\Gamma_A\Gamma_B)=(-1)^{nm-k+n(n-1)/2+m(m-1)/2}\,{\rm tr}\,1$, and the number of pairs of sets with $k$ common elements is $\binom Dn\binom nk\binom{D-n}{m-k}$. Following the flavour indices along the lines, the two pairings with two loops give $\mathsf{A}(G,G')$ each and the two pairings with one loop give $\mathsf{B}(G,G')$ each, by the symmetry of $G'$. With the factors $\tfrac12$ of the two vertices we obtain \eqref{g2:eq:closed} for every integer $D$, and hence for all $d$.
\end{proof}

The proof of Lemma~\ref{gf:lem:free} therefore extends: if $L^I$ has no terms of degree zero and one, the functional has the two-point form \eqref{gs:eq:free} on the four-fermion couplings, with $\Delta_I=4\Dp+2=2(d-1)$ and the matrix $c_{IJ}(d)$ of Lemma~\ref{g2:lem:kernel}. We find that the entries of $c_{IJ}(2)$ with $n_I\geq3$ vanish, so that $c_{IJ}(2)$ is degenerate. In \eqref{g2:eq:closed} the polynomial $\binom dn$ has the simple root $d=2$ for $n\geq3$, and $\mathsf{P}_{nm}(2)=0$ if $n\geq3$ or $m\geq3$, so that both parts of the entries with $n_I\geq3$ are of first order in $d-2$. The part with two flavour traces dominates for the flavour tensors with $|\mathsf{B}(G,G)|$ small compared with $\mathsf{A}(G,G)$, such as the flavour singlet $G=1\otimes1$ at fixed $n$ for many flavours, with $\mathsf{A}=N_f^2$ and $\mathsf{B}=N_f$, but not in general, since $\mathsf{B}(G,G)=\mathsf{A}(G,G)$ for the tensor $G_{ij,kl}=\delta_{i1}\delta_{j1}\delta_{k1}\delta_{l1}$ of a single flavour. The part with one flavour trace is not diagonal in $n$. At $d=2$ the diagonal blocks with $n\leq2$ are positive definite for ${\rm tr}\,1>1$, since $|\mathsf{B}(G,G)|\leq\mathsf{A}(G,G)$ for real $G$, as we show in the proof of Lemma~\ref{g2:lem:V0}, while for ${\rm tr}\,1=2$ the matrix $c_{IJ}(2)$ has null vectors which combine couplings of different $n\leq2$ (Lemma~\ref{g2:lem:psd}). This is the degeneracy of the metric of order zero described in section~\ref{gs:sec:fermions}.

\subsection{Finiteness near two dimensions}
\label{g2:sec:finiteness}

In this subsection we examine which steps of the derivation of the properties from Hypothesis~\ref{hyp:finite} carry over to the four-fermion theory near two dimensions. We call four-fermion theory the theory of $N_f$ Dirac fermions with the vertices \eqref{g2:eq:vertex} for all $n$, the group ${\rm U}(N_f)$ of field rotations and its background gauge field, in $d=2-\vep$. The sources of the fermion bilinears have dimension one at $d=2$. We omit them, so that all statements below concern the theory without these sources. The bare sources are
\begin{equation}
g_0^I = g^I+L^I(g) \, , \qquad a_{0\mu} = a_\mu+N_I\,D_\mu g^I \, ,
\label{g2:eq:bare}
\end{equation}
with $N_I$ in the Lie algebra of ${\rm U}(N_f)$, and in flat space the density of the counterterms without fields is $\cX^{\rm fl}=\tfrac12\,\fs_{IJ}(g)\,D_\mu g^ID^\mu g^J$, which is, up to integration by parts, the only local scalar of dimension two built from the couplings and the gauge field without the Levi-Civita tensor. We take $L^I$ to be a formal power series in the couplings with values in $V$, that is each of its coefficients has finitely many non-vanishing components. The functional $W^{\rm fl}$ of \eqref{eq:Wfl} is then defined, and Hypothesis~\ref{hyp:finite} for the four-fermion theory states that it is finite with real pure poles of this form, where $L^I$ is covariant and has no terms of degree zero and one. By Lemma~\ref{g2:lem:kernel} we define the evanescent subspace
\begin{equation}
E = \big\{\,e\in V \;:\; c_{IJ}(2)\,e^J = 0 \text{ for all } I\,\big\} \, ,
\label{g2:eq:E}
\end{equation}
which contains $V_n$ for every $n\geq3$. We write $c_{IJ}(d)=c^{(0)}_{IJ}+(d-2)\,c^{(1)}_{IJ}+\ldots$, so that $c^{(0)}_{IJ}=c_{IJ}(2)$.

\begin{lemma}[Positivity in two dimensions]
\label{g2:lem:psd}
Let ${\rm tr}\,1\geq2$. Then the bilinear form $c_{IJ}(2)$ is positive semidefinite on $V$. For ${\rm tr}\,1=2$ the subspace $E$ consists of the couplings whose operator \eqref{g2:eq:vertex}, built with the gamma matrices of two dimensions, vanishes. Besides $V_n$ for $n\geq3$ it then contains combinations of couplings with $n\leq2$, such as for $N_f=1$ the couplings with $G^{(n)}=g_n$ for $n\leq2$, where $g_0-2g_1+g_2=0$. For ${\rm tr}\,1>2$ the bilinear form $c_{IJ}(2)$ is positive definite on $V_0\oplus V_1\oplus V_2$, and $E=\bigoplus_{n\geq3}V_n$.
\end{lemma}

\begin{proof}
Each element of $V$ has finitely many non-vanishing components, so that $c_{IJ}(2)e^Ie^J$ is a finite sum. By Lemma~\ref{g2:lem:contract} in the flat chart with $D=2$ the value of $\mathsf{p}_{IJ}(2)$ is that obtained with gamma matrices of two dimensions, multiplied by $({\rm tr}\,1/2)^{\#{\rm traces}}$, and we use hermitian Euclidean gamma matrices of size two, since the value does not depend on the representation. In two dimensions there are no sets of three or more distinct indices, so that only the couplings with $n\leq2$ enter. With the notation of the proof of Lemma~\ref{g2:lem:kernel} we write a coupling $G\in V$ as the matrix $\mathcal{G}=\sum_{n\leq2}\sum_SG^{(n)}\otimes\Gamma_S\otimes\Gamma_S$ from the indices of the two fields $\psi$ of \eqref{g2:eq:vertex} to those of the two fields $\bar\psi$, where the flavour and spinor indices of each slot are combined. By the reality condition $\mathcal{G}$ is hermitian, since the signs of the hermitian conjugation of $\Gamma_S$ cancel between the two slots. By the symmetry of $G$ it commutes with the exchange $\mathsf{P}$ of the two slots, and since the indices of the two slots are contracted it is invariant under the conjugation with $\gamma\!\cdot\!\hat x\otimes\gamma\!\cdot\!\hat x$ for every unit vector $\hat x$. For ${\rm tr}\,1=2$ and with the sign $-1$ of each loop, the two pairings with two loops give $C_\psi^4\,|x|^{-4}\,{\rm Tr}(\mathcal{G}\mathcal{G}')$ each and the two pairings with one loop give $-C_\psi^4\,|x|^{-4}\,{\rm Tr}(\mathcal{G}\mathcal{G}'\mathsf{P})$ each. With the factors $\tfrac12$ of the two vertices we obtain
\begin{equation}
c(G,G') = \tfrac12\,C_\psi^{\,4}\,{\rm Tr}\big(\mathcal{G}\mathcal{G}'(1-\mathsf{P})\big) = C_\psi^{\,4}\,{\rm Tr}\big(\Pi_-\mathcal{G}\,\Pi_-\mathcal{G}'\big) \, , \qquad \Pi_-=\tfrac12(1-\mathsf{P}) \, ,
\label{g2:eq:psd}
\end{equation}
so that $c(G,G)=C_\psi^4\,\|\Pi_-\mathcal{G}\,\Pi_-\|^2\geq0$, with the Hilbert--Schmidt norm. The operator of $G$ in two dimensions is $\tfrac12\,\mathcal{G}_{(ac),(be)}\,\bar\psi_a\psi_b\bar\psi_c\psi_e$, with combined indices. Since the two fields $\psi$ and the two fields $\bar\psi$ anticommute, it depends only on $\Pi_-\mathcal{G}\,\Pi_-$, which it determines, so that $c(G,G)=0$ if and only if the operator vanishes. Since the form is positive semidefinite, $c_{IJ}(2)e^J=0$ for all $I$ if and only if $c_{IJ}(2)e^Ie^J=0$, so that $E$ consists of the couplings whose operator vanishes. For $N_f=1$ the operators $\tfrac12\sum_S(\bar\psi\,\Gamma_S\psi)^2$ with $n=0,1,2$ are in the ratio $1:-2:1$ in two dimensions, so that the couplings $G^{(n)}=g_n$ give the operator $\tfrac12(g_0-2g_1+g_2)(\bar\psi\psi)^2$. For a general ${\rm tr}\,1$ the pairings with two loops carry the factor $({\rm tr}\,1/2)^2$ and those with one loop the factor ${\rm tr}\,1/2$ relative to ${\rm tr}\,1=2$, so that
\begin{equation}
c(G,G) = \tfrac{{\rm tr}\,1}2\,C_\psi^{\,4}\Big[\tfrac12\big(\tfrac{{\rm tr}\,1}2-1\big)\|\mathcal{G}\|^2+\|\Pi_-\mathcal{G}\,\Pi_-\|^2\Big] \, ,
\end{equation}
which is non-negative for ${\rm tr}\,1\geq2$. For ${\rm tr}\,1>2$ it vanishes only if $\mathcal{G}=0$, which implies that the components of $G$ with $n\leq2$ vanish, since the matrices $\sum_S\Gamma_S\otimes\Gamma_S$ with $n=0,1,2$ are linearly independent. Hence $c_{IJ}(2)$ is positive definite on $V_0\oplus V_1\oplus V_2$, and as for ${\rm tr}\,1=2$ the subspace $E$ consists of the $e$ with $c_{IJ}(2)e^Ie^J=0$, which is $\bigoplus_{n\geq3}V_n$.
\end{proof}

\begin{lemma}[First correction of the kernel]
\label{g2:lem:c1}
Let ${\rm tr}\,1\neq0$. If $v\in V$ satisfies $c^{(1)}_{IJ}v^Ie^J=0$ for all $e\in\bigoplus_{m\geq3}V_m$, then $v=0$. If the components of $v$ in $V_n$ vanish for $n>q$, it suffices that this holds for all $e$ in $q+1$ of the spaces $V_m$ with even $m>\max(q,2)$. In particular the bilinear form $c^{(1)}_{IJ}$ restricted to $E$ is nondegenerate, in the sense that $v\in E$ and $c^{(1)}_{IJ}v^Ie^J=0$ for all $e\in E$ imply $v=0$.
\end{lemma}

\begin{proof}
For $G\in V_n$ and $H\in V_m$ let $c^{(1)}(G,H)=c^{(1)}_{IJ}\,g^Ih^J$ with the coordinates $g^I$ and $h^J$ of $G$ and $H$, as for $c(G,G')$ in Lemma~\ref{g2:lem:kernel}. Let $m\geq3$ and $n<m$. The derivative of the factor $C_\psi^{\,4}$ in \eqref{g2:eq:closed} contributes a multiple of the value of $c(G,H)$ at $d=2$, which vanishes since $V_m\subset E$, and the term with two flavour traces vanishes since $n\neq m$. We use the identity $\binom dn\binom nk\binom{d-n}{m-k}=d(d-1)\cdots(d-n-m+k+1)/\big(k!\,(n-k)!\,(m-k)!\big)=\binom dm\binom mk\binom{d-m}{n-k}$ of polynomials in $d$, so that $\mathsf{P}_{nm}(d)=\binom dm\,\mathsf{S}_{nm}(d)$ with
\begin{equation}
\mathsf{S}_{nm}(d)=\sum_{k=0}^{n}\binom mk\binom{d-m}{n-k}\,(-1)^{nm-k+n(n-1)/2+m(m-1)/2} \, .
\end{equation}
For $m\geq3$ the polynomial $\binom dm$ has a simple root at $d=2$, with the derivative $b_m=2\,(-1)^{m-1}(m-3)!/m!$ there, so that the derivative of $\mathsf{P}_{nm}$ at $d=2$ is $b_m\,\mathsf{S}_{nm}(2)$. For even $m$ we have $(-1)^{nm}=1$, and we obtain
\begin{equation}
\begin{split}
c^{(1)}(G,H)&=-\tfrac12\,C_\psi^{\,4}\,{\rm tr}\,1\;b_m\,(-1)^{m(m-1)/2}\,(-1)^{n(n-1)/2}\,\mathsf{Q}_n(m)\,\mathsf{B}(G,H) \, , \\
\mathsf{Q}_n(m)&=\sum_{k=0}^n(-1)^k\binom mk\binom{2-m}{n-k} \, ,
\end{split}
\label{g2:eq:c1}
\end{equation}
with $C_\psi$ at $d=2$. As polynomials in $m$ the binomial coefficients $\binom mk$ and $\binom{2-m}{n-k}$ have the degrees $k$ and $n-k$ and the leading coefficients $1/k!$ and $(-1)^{n-k}/(n-k)!$. Hence $\mathsf{Q}_n$ is a polynomial in $m$ of degree $n$ with the leading coefficient $(-1)^n\sum_k1/\big(k!\,(n-k)!\big)=(-2)^n/n!$, and $\mathsf{Q}_0,\ldots,\mathsf{Q}_q$ are linearly independent.

The conditions which define $V_n$ do not depend on $n$, so that a flavour tensor which satisfies them defines an element of every $V_m$, which we denote by the same letter. For such a tensor $G$ let $(\varsigma G)_{ij,kl}=G_{kj,il}$. The symmetry of $G$ gives $(\varsigma G)_{kl,ij}=G_{il,kj}=G_{kj,il}=(\varsigma G)_{ij,kl}$, and its reality gives $\overline{(\varsigma G)_{ij,kl}}=G_{jk,li}=G_{li,jk}=(\varsigma G)_{ji,lk}$, so that $\varsigma G$ satisfies the conditions as well. Moreover $\mathsf{B}(G,\varsigma G)=\sum G_{ij,kl}\,G_{lk,ji}=\sum G_{ij,kl}\,G_{ji,lk}=\sum|G_{ij,kl}|^2$ by the symmetry and the reality of $G$.

Now let $v$ have the components $G_n\in V_n$, which vanish for $n>q$, and let $m_0,\ldots,m_q$ be distinct even integers larger than $\max(q,2)$. Assume that $c^{(1)}(v,H)=0$ for all $H\in V_{m_j}$ and all $j$. Since $b_m\neq0$ and ${\rm tr}\,1\neq0$, by \eqref{g2:eq:c1} the polynomial $f_H(m)=\sum_{n\leq q}(-1)^{n(n-1)/2}\,\mathsf{B}(G_n,H)\,\mathsf{Q}_n(m)$, of degree at most $q$, vanishes at the $q+1$ points $m_j$ for every flavour tensor $H$ which satisfies the conditions, so that it vanishes identically. By the linear independence of the $\mathsf{Q}_n$ we find $\mathsf{B}(G_n,H)=0$ for all $n$ and all such $H$, and with $H=\varsigma G_n$ we obtain $G_n=0$. This proves the first two statements. Since $V_m\subset E$ for $m\geq3$, they imply the third.
\end{proof}

\begin{lemma}[Signs on the evanescent couplings]
\label{g2:lem:indef}
Let ${\rm tr}\,1\neq0$, and for $m\geq0$ let $e_m\in V_m$ be the coupling whose flavour tensor is the tensor $G_{ij,kl}=\delta_{i1}\delta_{j1}\delta_{k1}\delta_{l1}$ of a single flavour. For every integer $q$ and every $r\geq1$ there are integers $q<m_1<\ldots<m_r$ such that the bilinear form $c^{(1)}_{IJ}$ is positive definite on the span of $e_{m_1},\ldots,e_{m_r}$, and integers $q<m'_1<\ldots<m'_r$ such that it is negative definite on the span of $e_{m'_1},\ldots,e_{m'_r}$. For $q\geq2$ these spans lie in $E$. The same holds for the bilinear form $c_{IJ}(d)$ at every $d$ with $2<d<3$.
\end{lemma}

\begin{proof}
We write $c^{(1)}(G,H)$ as in the proof of Lemma~\ref{g2:lem:c1}, and we use the same letter for the flavour tensor of $e_m$, so that $\mathsf{A}(e_n,e_m)=\mathsf{B}(e_n,e_m)=1$. Let $\beta_m=b_m$ for $c^{(1)}$, with $d=2$ in the polynomials below and $C_\psi$ at $d=2$, and let $\beta_m=\binom dm$ for $c(d)$. Let $m\geq4$ and $m>d$. For $n<m$ the identity $\mathsf{P}_{nm}(d)=\binom dm\,\mathsf{S}_{nm}(d)$ of the proof of Lemma~\ref{g2:lem:c1} and \eqref{g2:eq:closed} give, as in that proof, the entries $c(e_n,e_m)$ and $c^{(1)}(e_n,e_m)$ as $-\tfrac12C_\psi^{\,4}\,{\rm tr}\,1\,\beta_m\,\mathsf{S}_{nm}(d)$, and for fixed $n$ the modulus of $\mathsf{S}_{nm}(d)$ is bounded by a polynomial in $m$. For $n=m$ the sign in $\mathsf{P}_{mm}$ is $(-1)^{m-k}$, and with $(-1)^{m-k}\binom{d-m}{m-k}=\binom{2m-d-k-1}{m-k}$ we find
\begin{equation}
c(e_m,e_m)=\tfrac12\,C_\psi^{\,4}\,{\rm tr}\,1\;\beta_m\big({\rm tr}\,1-\Sigma_m(d)\big) \, , \qquad \Sigma_m(d)=\sum_{k=0}^m\binom mk\binom{2m-d-k-1}{m-k} \, ,
\label{g2:eq:indefdiag}
\end{equation}
and the same expression for $c^{(1)}(e_m,e_m)$, since $\binom 2m=0$. Since $m>d$, the factors of all terms of $\Sigma_m(d)$ are positive, and since $d<3$ the term with $k=0$ is at least $\binom{2m-4}m\geq2^{m-4}$. The sign of $\beta_m$ is $(-1)^{m-1}$, and $|\beta_m|\leq1$. Hence for large $m$ the entry $c(e_m,e_m)$ has the sign of $-{\rm tr}\,1\,\beta_m$, which alternates with $m$, and its modulus is at least $|\beta_m|$ times a positive multiple of $2^m$. Now let the form be positive definite on the span of $e_{m_1},\ldots,e_{m_j}$, with the matrix $\mathsf{K}$, and for $m>m_j$ let $u$ be the vector of the entries $c(e_{m_i},e_m)$. Then $u^T\mathsf{K}^{-1}u$ is at most $|\beta_m|^2$ times a polynomial in $m$. For large $m$ with $-{\rm tr}\,1\,\beta_m>0$ the Schur complement $c(e_m,e_m)-u^T\mathsf{K}^{-1}u$ is therefore positive, so that the form is positive definite on the span of $e_{m_1},\ldots,e_{m_j},e_m$. Starting with one large $m_1>q$ with $-{\rm tr}\,1\,\beta_{m_1}>0$, we obtain $r$ integers by induction. The same argument applied to minus the form gives the negative definite spans. For $q\geq2$ the spans lie in $E$, since $V_m\subset E$ for $m\geq3$.
\end{proof}

\subsubsection{Kernels in two dimensions}
\label{g2:sec:kernels}

The steps of appendices~\ref{app:unique} and~\ref{app:finite} determine the leading pole of a coefficient from a term of second order of the functional at $d=d_c$, through Lemmas~\ref{lem:potential} and~\ref{lem:transverse}. At $d=2$ the kernel of two couplings is $|x-y|^{-4}$ by Lemma~\ref{g2:lem:kernel}, and the kernel of two gauge fields follows from the fermion loop of section~\ref{gf:sec:steps}.

\begin{lemma}[Potentials in two dimensions]
\label{g2:lem:potential}
Let $\eta$ be a continuous function on $\mathbb{R}^2$ with compact support, and let $\int d^2x\,\eta(x)\,|x-y|^{-2\kappa}$ vanish for all $y$ outside a ball, where $\kappa\geq1$ is an integer. Then $\eta=0$.
\end{lemma}

\begin{proof}
We follow the proof of Lemma~\ref{lem:potential}. For large $|y|$ we expand $|x-y|^{-2\kappa}=\sum_l|x|^l|y|^{-2\kappa-l}\,C_l^{(\kappa)}(\hat x\cdot\hat y)$, so that the integrals of $\eta(x)\,|x|^lC_l^{(\kappa)}(\hat x\cdot\hat y)$ vanish for all $l$ and all unit vectors $\hat y$. With $t=|x|/|y|$ and the angle $\theta$ between $\hat x$ and $\hat y$ we have $(1-2t\cos\theta+t^2)^{-\kappa}=(1-te^{{\rm i}\theta})^{-\kappa}(1-te^{-{\rm i}\theta})^{-\kappa}$, and the binomial series of the two factors give $C_l^{(\kappa)}(\cos\theta)=\sum_{0\leq m\leq l,\,m\equiv l}b_{l,m}\cos(m\theta)$ with $b_{l,m}=(2-\delta_{m0})\,(\kappa)_{(l+m)/2}(\kappa)_{(l-m)/2}/\big(((l+m)/2)!\,((l-m)/2)!\big)>0$, where $(\kappa)_j=\kappa(\kappa+1)\cdots(\kappa+j-1)$. In two dimensions $|x|^m\cos(m\theta)={\rm Re}\,\big((x_1+{\rm i}x_2)^m(\hat y_1-{\rm i}\hat y_2)^m\big)$, which is a harmonic polynomial in $x$ and a Fourier mode of order $m$ in the angle $\theta_y$ of $\hat y$. Since the $b_{l,m}$ are positive, the coefficients of the Fourier modes vanish separately, so that the integrals of $\eta(x)\,|x|^{l-m}\,|x|^m\cos(m(\theta_x-\theta_y))$ vanish for all $l\geq m$ with $l\equiv m$ modulo $2$ and all $\hat y$. The functions $|x|^m\cos(m(\theta_x-\theta_y))$ span the harmonic polynomials of degree $m$ in two dimensions, and every polynomial is a sum of harmonic polynomials multiplied by powers of $x^2$. All moments of $\eta$ therefore vanish, and $\eta=0$.
\end{proof}

The kernel of the gauge field behaves differently. In two dimensions the fermion loop with two vertices of the current gives, for $x\neq y$, the kernel $C_\psi^2\,{\rm tr}\,1\,{\rm tr}(\omega_f\omega'_f)\,\mathsf{k}_{\mu\nu}(x-y)$ with
\begin{equation}
\mathsf{k}_{\mu\nu}(z) = \frac{2z_\mu z_\nu-\delta_{\mu\nu}z^2}{|z|^4} = -\pr_\mu\pr_\nu\log|z| \qquad (z\neq0) \, .
\label{g2:eq:k2}
\end{equation}
For one-forms $j$ and $a$ with disjoint compact supports two integrations by parts give
\begin{equation}
\int d^2x\,d^2y\;j_\mu(x)\,\mathsf{k}_{\mu\nu}(x-y)\,a_\nu(y) = \int d^2x\,d^2y\;(\pr\!\cdot\!j)(x)\,\log|x-y|\,(\pr\!\cdot\!a)(y) \, ,
\label{g2:eq:longitudinal}
\end{equation}
so that the term of second order sees only the divergences of the two one-forms. In particular Lemma~\ref{lem:transverse} has no counterpart in two dimensions. Indeed, for a smooth function $F$ of compact support which is odd under $x_2\to-x_2$ the one-form $j=\pr_2F\,dx_1-\pr_1F\,dx_2$ belongs to the class of flat sources, it has $\pr\!\cdot\!j=0$, so that the left-hand side of \eqref{g2:eq:longitudinal} vanishes for all $a$, and $dj=-\pr^2F\,dx_1\wedge dx_2$ does not vanish in general. For disjoint supports the left-hand side of \eqref{g2:eq:longitudinal} also equals, up to a factor, the integral of $dj$ and $da$ with the kernel $\log|x-y|$, and the potential of the Laplacian of a function of compact support vanishes outside its support. For the one-forms which occur in the proofs a weaker statement suffices.

\begin{lemma}[Dipole condition]
\label{g2:lem:dipole}
Let $d=2$ with $D_0=1$, so that the flat sources are the smooth functions of $x_1$ and $x_2^2$ with compact support. Let $\breve\rho_I(g)$ be a homogeneous polynomial of degree $n-1\geq0$ in finitely many couplings, with values in a real vector space, and let $j=\sigma\,\breve\rho_I(g_1)\,dg_1^I$ with $\sigma=1$ or $\sigma=2x_1$. Assume that for every source $g_1$ the left-hand side of \eqref{g2:eq:longitudinal} vanishes for all one-forms $a$ of the class of flat sources with support outside a ball which contains the support of $g_1$. If $\sigma=1$, then $\breve\rho_I=\pr_I\ell$ with a homogeneous polynomial $\ell$ of degree $n$. If this holds for $\sigma=1$ and for $\sigma=2x_1$, then $\breve\rho_I=0$.
\end{lemma}

\begin{proof}
We may take the ball $\mathsf{B}$ centred on the axis $x_2=0$, so that it is invariant under the reflection $x_2\to-x_2$. Let $\Phi(y)=\int d^2x\,(\pr\!\cdot\!j)(x)\,\log|x-y|$ for $y\notin\mathsf{B}$. The complement $\mathsf{U}$ of $\mathsf{B}$ is connected. A smooth function $f$ with compact support in $\mathsf{U}$ which is even under $y_2\to-y_2$ and has $\int f=0$ is the divergence of a one-form with compact support in $\mathsf{U}$, since the cohomology with compact support of degree two of a connected oriented surface is spanned by the integral. Its average under the reflection has the same divergence, and it is a one-form of the class, since a smooth even function of $y_2$ is a smooth function of $y_2^2$ and a smooth odd one is $y_2$ times such a function. By assumption $\Phi$ is therefore orthogonal to all such $f$. Since $\Phi$ is even under the reflection, it is constant on $\mathsf{U}$. We have $\int\pr\!\cdot\!j=0$, and the expansion $\log|x-y|=\log|y|-x\!\cdot\!y/y^2+O(|y|^{-2})$ gives $\int d^2x\,x_\mu\,(\pr\!\cdot\!j)(x)=0$, that is $\int d^2x\,j_\mu=0$. For $\sigma=1$ and $\mu=1$ this reads $\int d^2x\,\breve\rho_I(g_1)\,\pr_1g_1^I=0$. Let $\gamma:[0,1]\to\mathbb{R}^K$ be a smooth closed curve in the space of the couplings with $\gamma=0$ near the end points, and let $g_1(x)=\lambda(x_2^2)\,\gamma(x_1)$ for $0\leq x_1\leq1$ and $g_1=0$ otherwise, with a smooth function $\lambda\geq0$ of compact support which does not vanish identically. This source belongs to the class, and by the homogeneity of $\breve\rho_I$ the condition becomes $\int dx_2\,\lambda(x_2^2)^n\,\oint_\gamma\breve\rho_I\,dg^I=0$. Hence the integral of the one-form $\breve\rho_I\,dg^I$ vanishes along every closed curve through the origin, and therefore along every closed curve, which we join to the origin by a path traversed in both directions. The one-form is therefore exact, $\breve\rho_I=\pr_I\ell$, and $\ell$ can be taken homogeneous of degree $n$ with $\ell(0)=0$. For $\sigma=2x_1$ the condition with $\mu=1$ reads $\int d^2x\,2x_1\,\pr_1\ell(g_1)=-2\int d^2x\,\ell(g_1)=0$. For $g_1=v\,\chi$ with a constant vector $v$ and a function $\chi\geq0$ of the class this gives $\ell(v)\int\chi^n=0$, hence $\ell=0$ and $\breve\rho_I=0$.
\end{proof}

\subsubsection{Leading poles}
\label{g2:sec:poles}

We now go through the induction of the proof of Lemma~\ref{lem:regvar}. The Weyl variation of the bare gauge field is treated by Lemma~\ref{lem:rhoxi}, whose statements are algebraic, with $k_I=1$ for all couplings. The lemma requires that $N_I$ is covariant and has no term of degree zero. For $\phi^4$ theory and for the Yukawa theory this follows from the finiteness and a choice of the bare sources by Propositions~\ref{prop:uniqueall} and~\ref{gc:prop:cov}, and here we assume it. Note that ${\rm U}(N_f)$ admits non-zero covariant linear maps from $V_0$ to its Lie algebra, such as $t\otimes t'+t'\otimes t\mapsto{\rm i}(tt'+t't)$ for hermitian $t$ and $t'$, so that the absence of a term of degree zero does not follow from the covariance, unlike for ${\rm O}(N)$ acting on quartic couplings. It defines $\hB^I$ and $\hr_I$ as power series in the couplings whose coefficients are Laurent polynomials in $\vep$, and each coefficient of $\hB^I$ has finitely many non-vanishing components, since $L^I$ has this property. The identity \eqref{eq:flatward} holds by Proposition~\ref{g2:prop:bare}, with the right-hand side local. We say that the terms of degree $n$ of $\hB^I$ and of degree $n-1$ of $\hr_I$ have the level $n$. In each step we use sources $g_1$ with values in a finite-dimensional subspace $\bigoplus_{n\leq n_0}V_n$, for which each level of $\hB^I(g_1)$ and of $\hr_I(g_1)\,D_\mu g_1^I$ is a polynomial with finitely many components, and we let $n_0$ be arbitrary.

\begin{proposition}[First pole]
\label{g2:prop:poles}
Let ${\rm tr}\,1\neq0$. Assume Hypothesis~\ref{hyp:finite} for the four-fermion theory, with $N_I$ covariant and without term of degree zero. Let $n\geq2$ be the lowest level at which a coefficient of $\hB^I$ or of $\hr_I$ has a pole at $\vep=0$. Then $\hr_I$ has no pole at the level $n$, so that a coefficient of $\hB^I$ has a pole there. Each coefficient of the term of level $n$ of $\hB^I$ has at most a simple pole, and the coefficient of $\vep^{-1}$ of this term takes values in $E$.
\end{proposition}

\begin{proof}
At the level one $\hB^I=-\vep\,g^I$ and $\hr_I=0$. Indeed, in \eqref{eq:xiL} the term of degree one of $\hb^{{\rm L}\,I}$ is $-\vep\,g^I$, since $L^I$ starts at degree two, $\xi^{\rm L}=N_I\hB^I$ starts at degree two, since $N_I$ and $\hB^I$ start at degree one, and each term on the right-hand side of the third equation of \eqref{eq:xiL} has degree one at least. We use the sources with the indices $1$ and $2$ and the terms of \eqref{eq:flatward} considered in the proof of Lemma~\ref{lem:regvar}. By assumption all terms of the variation of level below $n$ are regular, and each of them contributes a finite term. We fix $n_0\geq2$ and take $g_1$ with values in $V_{\leq n_0}=\bigoplus_{n'\leq n_0}V_{n'}$. On $V_{\leq n_0}$ the term of level $n$ of $\hB^I$ and the term of level $n$ of $\hr_I\,D_\mu g^I$ have finitely many coefficients, each of which is a Laurent polynomial in $\vep$, so that each of these terms has a highest pole on $V_{\leq n_0}$ if it has a pole there. The sources $g_2$ below may take values in all of $V$, since all coefficients of level below $n$ are regular.

For the gauge field we take couplings $g_1$ and a one-form $a_2$, with $\sigma=1$ or $\sigma=2x_1$, and the terms of order $n$ in $g_1$ and linear in $a_2$. The term of level $n$ of $\hB^I$ multiplies a term of second order of $W^{\rm fl}$ with a coupling and a gauge field, which vanishes by Lemma~\ref{g2:lem:kernel}. Let $\vep^{-\mathsf{r}'}\breve\rho_I$ be the highest pole of $\hr_I$ at the level $n$ on $V_{\leq n_0}$. The coefficient of $\vep^{-\mathsf{r}'}$ is the left-hand side of \eqref{g2:eq:longitudinal} with $j=\sigma\,\breve\rho_I(g_1)\,dg_1^I$, times $C_\psi^2\,{\rm tr}\,1$ and the bilinear form ${\rm tr}(\omega_f\omega'_f)$, which is nondegenerate on ${\rm u}(N_f)$, at $d=2$, since for disjoint supports the integral is analytic at $d=2$. It vanishes for all $a_2$ of the class with support outside the support of $g_1$, for every component in the Lie algebra. Lemma~\ref{g2:lem:dipole} gives $\breve\rho_I=0$, so that $\hr_I$ has no pole at the level $n$ on $V_{\leq n_0}$.

For the couplings we take $g_1+g_2$ and the terms of order $n$ in $g_1$ and linear in $g_2$, with $\sigma=1$ and $g_2=e\,\chi_2$, where $e\in V$ is a constant vector and $\chi_2$ a function of the class with support outside the support of $g_1$. The term of level $n$ of $\hr_I$ multiplies a term of second order with a gauge field and a coupling, which vanishes. The term of level $n$ of $\hB^I$ multiplies the kernel of Lemma~\ref{g2:lem:kernel}, and the identity reads
\begin{equation}
\int d\mu(x)\,d\mu(y)\;\hB^I_{(n)}(g_1(x))\,c_{IJ}(d)\,e^J\,\frac{\chi_2(y)}{|x-y|^{4(d-1)}} = \mathsf{F} \, ,
\label{g2:eq:polestep}
\end{equation}
where $\hB^I_{(n)}$ is the term of level $n$, $d\mu$ is the flat measure of the class, and $\mathsf{F}$ is finite. If $\hB^I_{(n)}$ has a pole on $V_{\leq n_0}$, let $\vep^{-\mathsf{r}}$ with $\mathsf{r}\geq1$ be its highest pole there, and let $\breve\beta^I$ be its coefficient, a polynomial on $V_{\leq n_0}$ which does not vanish identically. For disjoint supports the two-point integral $\mathsf{M}_d[f,\chi_2]$ of a function $f$ and $\chi_2$ with the kernel and the measure of \eqref{g2:eq:polestep} is analytic at $d=2$, where it is the ordinary integral. The coefficient of $\vep^{-\mathsf{r}}$ is $\int d^2x\,d^2y\,\breve\beta^I(g_1(x))\,c^{(0)}_{IJ}e^J\,\chi_2(y)\,|x-y|^{-4}$, for $\mathsf{r}\geq1$. It vanishes for all $\chi_2$ of the class with support outside the support of $g_1$. The function $\breve\beta^I(g_1)\,c^{(0)}_{IJ}e^J$ is even under $x_2\to-x_2$, and so is its potential with the kernel $|x-y|^{-4}$, and the functions of the class separate even continuous functions. This potential therefore vanishes outside a ball, and Lemma~\ref{g2:lem:potential} gives $\breve\beta^I(g_1(x))\,c^{(0)}_{IJ}e^J=0$ for all $x$. Since the values of $g_1$ in $V_{\leq n_0}$ are arbitrary and $e$ is arbitrary, $\breve\beta^I(g)$ lies in $E$ for all $g\in V_{\leq n_0}$.

Let now $\mathsf{r}\geq2$ and $e\in E$. Since $c^{(0)}_{IJ}e^J=0$ and $d-2=-\vep$, we have $c_{IJ}(d)e^J=-\vep\,c^{(1)}_{IJ}e^J+O(\vep^2)$ for every $I$, so that the left-hand side of \eqref{g2:eq:polestep} is
\begin{equation}
\sum_k\vep^{-k}\big(-\vep\,c^{(1)}_{IJ}e^J\,\mathsf{M}_d[\beta^I_k(g_1),\chi_2]+O(\vep^2)\big) \, ,
\end{equation}
where $\beta_k$ is the coefficient of $\vep^{-k}$ of $\hB_{(n)}$. Its coefficient of $\vep^{1-\mathsf{r}}$ is
\begin{equation}
-\int d^2x\,d^2y\,\breve\beta^I(g_1(x))\,c^{(1)}_{IJ}e^J\,\chi_2(y)\,|x-y|^{-4} \, .
\end{equation}
Since $1-\mathsf{r}<0$ and $\mathsf{F}$ is finite, this term vanishes, and the argument of the previous paragraph gives $c^{(1)}_{IJ}\,\breve\beta^I(g)\,e^J=0$ for all $g\in V_{\leq n_0}$ and all $e\in E$. Since $\breve\beta(g)\in E$, Lemma~\ref{g2:lem:c1} gives $\breve\beta=0$, which contradicts the choice of $\mathsf{r}$. Hence $\mathsf{r}=1$, and by the previous paragraph the coefficient of $\vep^{-1}$ of $\hB^I_{(n)}$ takes values in $E$ on $V_{\leq n_0}$. Each coefficient of the terms of level $n$ of $\hB^I$ and of $\hr_I$ involves finitely many couplings, and it is therefore reached for some $n_0$. This proves the proposition.
\end{proof}

For $\mathsf{r}=1$ every term of \eqref{g2:eq:polestep} is finite, so that the finiteness does not require $\breve\beta^I\in E$ to vanish. The residue enters the finite part of the identity. A simple pole of $\hB^I$ with values in $E$ is therefore not excluded by the finiteness, while Proposition~\ref{g2:prop:poles} excludes poles of higher order at the level $n$. Above the first singular level the induction does not continue, since the singular terms of lower levels then contribute poles to the terms of higher order. The same mechanism affects the uniqueness of the counterterms. In the proof of Proposition~\ref{prop:unique} the leading pole of the difference of two solutions is determined only modulo $E$. In the proof of Proposition~\ref{prop:uniqueall} the first step goes through with Lemma~\ref{g2:lem:dipole} in place of Lemma~\ref{lem:transverse}, and in the third step and in the step for the term of degree zero the condition becomes $(\breve\vartheta(g)\,g)\in E$ in place of $(\breve\vartheta(g)\,g)=0$. For ${\rm tr}\,1>1$, Lemma~\ref{g2:lem:stabE} below and the argument with a dense set of couplings in the proof of Proposition~\ref{g2:prop:poleeq} show that $\breve\vartheta(g)$ is then a multiple of the unit matrix, so that for two sets of covariant counterterms these steps go through up to the gauge transformations generated by the fermion number, which act trivially on the couplings and do not change the functional. A non-central term of degree zero of $N_I$ is not excluded by these steps.

We now show that the evanescent directions do not admit the poles of Proposition~\ref{g2:prop:poles} if the poles of $L^I$ obey the pole equations of minimal subtraction. By Lemma~\ref{lem:rhoxi} the vector field $\hb^{{\rm L}\,I}=-\vep\,g^I+\vep\,[(1+\pr L)^{-1}]^I{}_J\,(\mathsf{E}-1)L^J$, with the Euler operator $\mathsf{E}=g^K\pr_K$, depends on $L^I$ alone. Its regularity is a set of algebraic conditions on $L^I$, which express the higher poles through the simple pole. These conditions are the pole equations of minimal subtraction for the given choice of the bare sources, and their dependence on this choice is discussed after Proposition~\ref{g2:prop:poleeq}. For multiscalar $\phi^4$ theory the finiteness implies the regularity of $\hB^I$, since the kernel of two couplings is nondegenerate at $d=d_c$. Here we use two properties of the space of couplings.

\begin{lemma}[Couplings without gamma matrices]
\label{g2:lem:V0}
For ${\rm tr}\,1>1$ the bilinear form $c_{IJ}(2)$ is positive definite on $V_0$. In particular $V_0\cap E=\{0\}$.
\end{lemma}

\begin{proof}
For $n_I=n_J=0$ the slots contain no gamma matrices, and $\mathsf{p}_{IJ}$ does not depend on $d$. For two vertices \eqref{g2:eq:vertex} with $G,G'\in V_0$ there are four pairings, two with two loops and two with one loop through all four slots. With $S(z)=C_\psi\,\gamma\!\cdot\!z\,|z|^{-d}$ and $S(z)S(-z)=-C_\psi^2\,|z|^{2-2d}$ a loop through two slots gives the spinor trace $-C_\psi^2\,{\rm tr}\,1\,|z|^{2-2d}$ and the loop through four slots gives $C_\psi^4\,{\rm tr}\,1\,|z|^{4-4d}$. Following the flavour indices along the lines, the two pairings with two loops give the form $\mathsf{A}(G,G')$ of Lemma~\ref{g2:lem:kernel} each, and the two pairings with one loop give $\mathsf{B}(G,G')$ each, by the symmetry of $G'$. Each loop carries the sign $-1$ of \eqref{gf:eq:loop}, so that the two pairings with two loops contribute $C_\psi^4\,({\rm tr}\,1)^2\,|z|^{4-4d}\,\mathsf{A}(G,G')$ each and the two pairings with one loop contribute $-C_\psi^4\,{\rm tr}\,1\,|z|^{4-4d}\,\mathsf{B}(G,G')$ each. With the factors $\tfrac12$ of the two vertices we obtain the two-point function $\langle V^{\rm F}[G](z)\,V^{\rm F}[G'](0)\rangle=\tfrac12\,C_\psi^4\,|z|^{4-4d}\,\big[({\rm tr}\,1)^2\mathsf{A}(G,G')-{\rm tr}\,1\,\mathsf{B}(G,G')\big]$, which is $c(G,G')\,|z|^{4-4d}$, since the term of second order of the functional is half the integral of the connected two-point function. By the reality condition $G_{ji,lk}=\overline{G_{ij,kl}}$, so that $\mathsf{A}(G,G)=\|G\|^2$ is the square of the hermitian norm of the tensor. Moreover $G_{jk,li}=\overline{G_{kj,il}}$, so that $\mathsf{B}(G,G)=\langle G,\varsigma G\rangle$, where $(\varsigma G)_{ij,kl}=G_{kj,il}$ is a permutation of the indices, a unitary involution, so that $\mathsf{B}(G,G)$ is real. Hence $|\mathsf{B}(G,G)|\leq\|G\|^2$, and $({\rm tr}\,1)^2\mathsf{A}(G,G)-{\rm tr}\,1\,\mathsf{B}(G,G)\geq{\rm tr}\,1\,({\rm tr}\,1-1)\,\|G\|^2$. Since ${\rm tr}\,1>1$, the form $c_{IJ}(2)$ is therefore positive definite on $V_0$, and if $e\in V_0\cap E$, then $c_{IJ}(2)e^Ie^J=0$ and hence $e=0$.
\end{proof}

\begin{lemma}[Stabilisers modulo evanescent couplings]
\label{g2:lem:stabE}
Let ${\rm tr}\,1>1$. There is a non-empty Zariski open subset $\mathsf{O}$ of $V_0\oplus V_1\oplus V_2$ such that for every $g\in V$ whose component in $V_0\oplus V_1\oplus V_2$ lies in $\mathsf{O}$ and every $\omega$ in the Lie algebra ${\rm u}(N_f)$ with $(\omega g)\in E$ the element $\omega$ is a multiple of the unit matrix, which acts trivially on $V$.
\end{lemma}

\begin{proof}
The rotations preserve each $V_n$, and $V_n\subset E$ for $n\geq3$. Let $g_{\leq2}$ be the component of $g$ in $V_0\oplus V_1\oplus V_2$ and $\mathsf{F}=E\cap(V_0\oplus V_1\oplus V_2)$. Then $(\omega g)\in E$ if and only if $(\omega g_{\leq2})\in\mathsf{F}$, that is if $\omega$ is in the kernel of the linear map $\omega\mapsto(\omega g_{\leq2})+\mathsf{F}$ from ${\rm u}(N_f)$ to the quotient by $\mathsf{F}$. The kernel contains the multiples of the unit matrix, and its dimension is an upper semicontinuous function of $g_{\leq2}$, which takes its minimum on a Zariski open set. It is therefore sufficient to find one $g_{\leq2}$ for which the kernel consists of the multiples of the unit matrix. For $N_f=1$ there is nothing to show. For $N_f\geq2$ let $h$ be a real diagonal matrix with distinct entries, let $k$ be a hermitian matrix whose off-diagonal entries do not vanish, with ${\rm tr}\,h^2\neq{\rm tr}\,k^2$ and with $h$ and $k$ linearly independent, and let $G_*=h\otimes h+k\otimes k\in V_0$. If $(\omega G_*)\in\mathsf{F}$, then $(\omega G_*)\in V_0\cap E=\{0\}$ by Lemma~\ref{g2:lem:V0}, so that
\begin{equation}
[h,\omega]\otimes h+h\otimes[h,\omega]+[k,\omega]\otimes k+k\otimes[k,\omega] = 0 \, .
\end{equation}
We choose a complement of the span $\mathsf{u}$ of $h$ and $k$ in the space of matrices. The component of the left-hand side in the complement tensored with $\mathsf{u}$ is $[h,\omega]_\perp\otimes h+[k,\omega]_\perp\otimes k$, so that the components $[h,\omega]_\perp$ and $[k,\omega]_\perp$ in the complement vanish. With $[h,\omega]=\alpha h+\beta k$ and $[k,\omega]=\gamma h+\delta k$ the equation becomes $2\alpha\,h\otimes h+(\beta+\gamma)(h\otimes k+k\otimes h)+2\delta\,k\otimes k=0$, hence $\alpha=\delta=0$ and $\gamma=-\beta$. The cyclicity of the trace gives ${\rm tr}([h,\omega]k)=-{\rm tr}([k,\omega]h)$, that is $\beta\,{\rm tr}\,k^2=\beta\,{\rm tr}\,h^2$, and therefore $\beta=0$. Hence $\omega$ commutes with $h$ and with $k$. Since $h$ is diagonal with distinct entries, $\omega$ is diagonal, and since the off-diagonal entries of $k$ do not vanish, its diagonal entries are equal.
\end{proof}

\begin{proposition}[Regularity under the pole equations]
\label{g2:prop:poleeq}
Let ${\rm tr}\,1>1$. Assume Hypothesis~\ref{hyp:finite} for the four-fermion theory, with $N_I$ covariant and without term of degree zero, and assume that $\hb^{{\rm L}\,I}$ is regular. Then $\hB^I$ and $\hr_I$ are regular, $\hB^I=-\vep\,g^I+B^I$ with $B^I$ and $\hr_I$ given by \eqref{eq:zeroparts} and independent of $\vep$, and $\hr_I\hB^I=0$.
\end{proposition}

\begin{proof}
Assume that $\hB^I$ or $\hr_I$ has a pole, and let $n$ be the lowest level at which this happens. By Proposition~\ref{g2:prop:poles} the coefficient $\hr_I$ is regular at the level $n$, the term of level $n$ of $\hB^I$ has at most a simple pole, and the coefficient $\breve\beta^I$ of $\vep^{-1}$ of this term takes values in $E$. By the first equation of \eqref{eq:xiL} and the regularity of $\hb^{{\rm L}\,I}$ we have $\breve\beta^I(g)=-(\breve\xi(g)\,g)^I$, where $\breve\xi$ is the coefficient of $\vep^{-1}$ of the term of degree $n-1$ of $\xi^{\rm L}$, which takes values in ${\rm u}(N_f)$. By Lemma~\ref{g2:lem:stabE} the element $\breve\xi(g)$ is a multiple of the unit matrix for $g$ in the set of that lemma, so that $(\breve\xi(g)\,g)=0$ there. On every finite-dimensional subspace of $V$ which contains $V_0\oplus V_1\oplus V_2$ this set is Zariski dense, and $(\breve\xi(g)\,g)$ is a polynomial, so that it vanishes for all $g$. Hence $\breve\beta^I=0$, so that the term of level $n$ of $\hB^I$ is regular, which contradicts the choice of $n$. The remaining statements are the third and the fourth statement of Lemma~\ref{lem:rhoxi}.
\end{proof}

Under the assumptions of Proposition~\ref{g2:prop:poleeq} Property~\ref{gs:P:weyl} holds for the four-fermion theory without the sources of the fermion bilinears, since without relevant sources it consists of the regularity and the covariance of $\hB^I$ and $\hr_I$, the relation $\hr_I\hB^I=0$ and the lowest order $\hB^I=-\vep\,g^I+O(g^2)$, while the operators \eqref{gs:eq:weylop} then commute identically. This suffices for Proposition~\ref{gs:prop:fermions} if the other properties are also taken without these sources, so that Lemma~\ref{gs:lem:dimension} is not needed. The proposition uses two assumptions beyond Hypothesis~\ref{hyp:finite}, namely the covariance of $N_I$ without term of degree zero and the regularity of $\hb^{{\rm L}\,I}$. The second depends on the choice of the bare sources. By the third statement of Lemma~\ref{lem:frame}, which is algebraic and applies here, there is a choice of the bare sources with the same functional, $\hB^I$ and $\hr_I$ for which $\xi^{\rm L}$ is regular, and for it $\hb^{{\rm L}\,I}=\hB^I+(\xi^{\rm L}g)^I$. The regularity of $\hb^{{\rm L}\,I}$ for a suitable choice of the bare sources is therefore equivalent to the regularity of $\hB^I$, and Proposition~\ref{g2:prop:poleeq} shows that it suffices to know it for one choice, whatever the poles of $\xi^{\rm L}$. For the symmetric square root of the field renormalisation it can fail through poles of the form $(\omega g)^I$, see the end of appendix~\ref{app:gaugevar}. In minimal subtraction the higher poles of the renormalisation constant of the Gross--Neveu coupling were found to be fixed by its simple pole through four loops~\cite{Gracey:2016mio}, with the evanescent couplings set to zero after renormalisation, which is evidence for the pole equations in the physical direction only.

\end{document}